\documentclass[structabstract]{aa}  
\usepackage{graphicx}
\usepackage{lscape}
\usepackage{txfonts}
\usepackage{caption}
\usepackage{color}
\usepackage{grffile}
\usepackage{afterpage}
\usepackage{amsmath}
\newcommand{\hii}{H\,{\scriptsize II}}
\newcommand{\uchii}{UC-H\,{\scriptsize II}}
\newcommand{\hi}{H\,{\scriptsize I}}

\newcommand{\kms}{km s$^{-1}$}

\newcommand{\at}{{ATLASGAL}}

\newcommand{\mum}{$\mu$m}
\newcommand{\msol}{M$_{\odot}$}
\newcommand{\lsol}{L$_{\odot}$}
\newcommand{\lbolm}{$L_{\rm bol}/M$}

\newcommand{\nhii}{78}
\newcommand{\nirdark}{217}
\newcommand{\nirbright}{135}

\newcommand{\nsoutot}{427} 

\newcommand{\emirwing}{160} 

\newcommand{\nsouapex}{128} 

\newcommand{\nsoufivedet}{117} 
\newcommand{\nsoufivedetfivesigma}{114} 

\begin{document}

 \title{ATLASGAL-selected massive clumps in the inner Galaxy}
   \subtitle{II: Characterisation of different evolutionary stages and their SiO emission}

   \author{T. Csengeri
          \inst{1}
          \and
           S. Leurini
          \inst{1}
          \and
           F. Wyrowski
          \inst{1}
         \and
           J. S. Urquhart
          \inst{1}
          \and
           K. M. Menten
          \inst{1}
          \and
          M. Walmsley
          \inst{2,3}
          \and
           S. Bontemps
          \inst{4}
          \and
           M. Wienen
          \inst{1}
          \and
            H. Beuther
          \inst{5}   
          \and
           F. Motte
          \inst{6}
          \and
          Q. Nguyen-Luong
          \inst{7,8}
          \and
           P. Schilke
           \inst{9}
          \and
            F. Schuller
          \inst{10}
          \and
          A. Zavagno
          \inst{11}
          \and
          C. Sanna
          \inst{12}
          }

   \institute{Max Planck Institute for Radio Astronomy,
              Auf dem H\"ugel 69, 53121 Bonn, Germany
              \email{csengeri@mpifr-bonn.mpg.de}
         \and
          INAF-Osservatorio Astrofisico di Arcetri, Largo E. Fermi 5, I-50125 Firenze, Italy
         \and
           Dublin Institute for Advanced Studies, Burlington Road 10, Dublin 4, Ireland
          \and
           OASU/LAB-UMR5804, CNRS, Universit\'e Bordeaux 1, 33270 Floirac, France    
        \and
          Max Planck Institute for Astronomy, K\"onigstuhl 17, 69117 Heidelberg, Germany
         \and
          Laboratoire AIM Paris Saclay, CEA-INSU/CNRS-Universit\'e Paris Diderot, IRFU/SAp CEA-Saclay, 91191 Gif-sur-Yvette, France
          \and
     National Astronomical Observatory of Japan, 2-21-1 Osawa, Mitaka, Tokyo 181-8588, Japan
     \and
Canadian Institute for Theoretical Astrophysics, University of Toronto, 60 St. George Street, Toronto, ON M5S~3H8, Canada
           \and
          I. Physikalisches Institut der Universit\"at zu K\"oln, Z\"ulpicher Str. 77, 50937, K\"oln, Germany
        \and
             European Southern Observatory, Alonso de Cordova 3107, Vitacura, Santiago, Chile
            \and
          Aix Marseille Universit\'e, CNRS, LAM (Laboratoire d'Astrophysique de Marseille), UMR 7326, 13388 Marseille, France
       \and
         Dipartimento di Fisica, Universit\'a di Cagliari, Strada Prov.le Monserrato-Sestu Km 0.700, I-09042 Monserrato (CA), Italy
             }

   \date{Received ; accepted }

 
  \abstract
   {   
The processes leading to the birth of high-mass stars are poorly understood. 
 The key
 first  step to reveal their formation processes is characterising the clumps and cores from which they form.
   }
   {
  We define a representative sample of massive clumps in different evolutionary 
  stages selected from the APEX Telescope Large Area Survey of the Galaxy  
({\at}), from which we aim to establish
  a census of molecular tracers of their evolution.
  As a first step, we  study the shock tracer, SiO, 
  mainly associated with shocks from jets probing
  accretion processes. In low-mass young stellar objects (YSOs), 
  outflow and jet activity decreases with time
  during the star formation processes. Recently, a
  similar scenario was suggested for massive clumps based on SiO observations.
   Here we analyse observations of the SiO ($2-1$) and ($5-4$) lines
   in a statistically significant sample to constrain
   the change of SiO abundance and the excitation conditions
   as a function of evolutionary stage of massive star-forming clumps.
   }
   {
    We performed an unbiased spectral line survey covering the 3-mm atmospheric window
   between 84$-$117~GHz with the IRAM~30m telescope 
   of a sample of $430$ sources of the \at\ survey, covering various evolutionary 
   stages of massive clumps.
    A smaller sample of \nsouapex\ clumps has been observed
   in the SiO ($5-4$)
   transition with the APEX telescope to complement the ($2-1$) line and 
   probe the excitation conditions of the emitting gas. We derived 
   detection rates to assess the star formation activity of the sample,   and we estimated the column density and abundance using both an LTE 
   approximation and non-LTE calculations for a smaller subsample, where
   both transitions have been observed.
   }
   {
   We characterise the physical properties of the selected sources, 
   which greatly supersedes
   the largest samples studied so far,
   and show that they are representative of different evolutionary stages.
   We report a high detection rate of $>75$\% of the SiO ($2-1$) line and  a $>90$\% detection rate from the 
   dedicated follow-ups in the ($5-4$) transition.
   Up to 25\% of the infrared-quiet clumps exhibit high-velocity line wings,
   suggesting that molecular tracers are more efficient tools to determine the level of
   star formation activity than infrared colour criteria.
   We also find infrared-quiet clumps that exhibit {only a low-velocity  
   component ($FWHM\sim5-6$~\kms)} 
   SiO emission in the ($2-1$) line. 
   In the current picture, where
   this is attributed to low-velocity shocks from cloud-cloud collisions, 
   this can be used to pinpoint the youngest, thus, likely prestellar massive 
   structures. Using the optically thin isotopologue ($^{29}$SiO), we estimate
   that  the ($2-1$) line is optically thin towards most of the sample. 
    Furthermore, based on the line ratio of the ($5-4$) to the ($2-1$) line,
   our study reveals a trend of changing excitation conditions that 
   lead to brighter emission in the ($5-4$) line towards more evolved sources.
   Our models show that a proper treatment of non-LTE effects and beam dilution is 
   necessary to constrain trends in the SiO column density and abundance. 
   }
    {
    We conclude that the SiO ($2-1$) line with broad line profiles and
    high detection rates is a powerful probe of star formation activity in the 
    deeply embedded phase of the evolution of massive clumps.
    The ubiquitous detection of SiO in all evolutionary stages
    suggests a continuous star formation process in massive clumps.
    Our analysis delivers a more robust estimate of SiO column density and abundance
    than previous studies and questions the decrease of jet activity in massive clumps
    as a function of age. The observed increase of excitation conditions towards the
    more evolved clumps suggests a higher pressure in the shocked gas towards
    more evolved or more massive clumps in our sample. 
    }


   \maketitle
%

\section{Introduction}\label{sec:intro-sio}

The origin of high-mass stars is still an enigma in modern
astrophysics. To reveal the processes playing a major role
in their formation, first the origin
of their mass reservoir, and, thus, the origin of massive clumps,
needs to be studied. The APEX Telescope Large Area Survey of the Galaxy  
({\at} survey), covering a
420~sq. deg area of the inner Galaxy~\citep{schuller2009,csengeri2013}, provides  
unprecedented statistics of massive clumps 
hosting various evolutionary stages of embedded
high-mass (proto-)stars.

Molecular tracers are useful tools
 to study physical conditions such as temperature and density within
massive clumps. The various species 
may provide zooming lenses, even if the internal
structure of the clump is observationally not resolved, molecular emission 
probes different layers
and physical components, 
revealing the properties
of both the diffuse and dense gas.
Spectral surveys have so far targeted a few selected
 sources in
mostly distinct evolutionary stages, such as
\uchii\ regions, high-mass protostellar objects (HMPOs; e.g.\,\citealp{Klaassen2012}),
and infrared dark clouds (IRDCs; e.g.\,\citealp{Beuther2007, Vasyunina2011}).
Larger samples covering a broad range 
of evolutionary stages are needed
for statistical analysis of their molecular properties.
Examples of these studies are e.g.\,\citet{Codella1999,Miettinen2006,M07, Sakai2010,LS2011,Sanhueza2012,Gerner2014} which 
cover and compare various evolutionary stages, with  
the largest sample of 159 clumps studied in 37 IRDCs by \citet{Sanhueza2012}.

Since complete samples drawn from Galaxy-wide surveys, such as
\at,\ are 
now available, we have undertaken
an unbiased spectral line survey 
with the IRAM~30m telescope. This survey
covers 32~GHz  of the 3-mm atmospheric
transmission window towards an unprecedentedly large sample of 430 clumps
selected from \at\ 
representing various stages of high-mass star formation (see example in Fig.\,\ref{fig:example}).
Tracing the low-J transitions of several species, this survey
aims to characterise the physical conditions and chemical imprints of the evolution of
massive clumps while forming high-mass stars.
In particular,
we present here a dedicated study focusing on the
low-J SiO emission of this survey to provide an observational framework to constrain shock-
and clump evolution models.  
Based on the \at\ survey and complementary stellar tracers,
we have selected the largest, and likely most 
complete, sample of massive clumps 
ranging in evolution from the starless, infrared-quiet 
 to the infrared-bright and {\uchii} stage for which the change of SiO
emission, and consequently the evolution of shocks and jets,
can be addressed.

Shocks seem to be a ubiquitous phenomenon in the process of
star formation. Fast shocks related to outflows have been regularly observed
at sites of massive protostars (e.g.\,\citealp{Beuther2004,Qiu2007,Duarte-Cabral2014}), 
while the role of low-velocity shocks
forming massive structures has just started to be explored~\citep{Csengeri2011a, Csengeri2011b, Quang2013}. 
A characteristic tracer of shocks is the SiO molecule, 
which has been observed in various conditions
such as (extremely) high-velocity shocks ($ v_{\rm s}\ge 15$ up to 100~\kms) related to powerful jets
from protostars, photon-dominated regions (PDRs) and low-velocity
shocks ($v_{\rm s}\leq10$~\kms).
Although it is
commonly used as an indirect tracer of jet activity revealing the
processes of material ejection
from young protostars \citep{Codella1999,Nisini2007,Duarte-Cabral2014},
the mechanisms responsible for the observed complex 
line profiles are not well constrained by shock models
(see also~\citealp{Lefloch1998, Jimenez-Serra2005, Anderl2013}).

Previous studies targeting smaller samples
report evolutionary trends seen
in the abundance of shock tracers, such as SiO.
 \citet{Codella1999} studied SiO emission towards both low- and high-mass 
 young stellar objects (YSOs) 
 and found a 
trend of brighter SiO emission from higher luminosity sources, suggesting
more powerful shocks in the vicinity of more massive YSOs.
The studies of \citet{M07} and \citet{Sakai2010}, 
focusing only on massive clumps and cores,
found a difference between mid-infrared dark
and bright sources, suggesting an evolutionary trend of shocks
from protostars in massive clumps. The colder, thus, younger sources showed 
brighter SiO emission in the $J= 2-1$ line, 
while mid-infrared bright sources were found to 
show weaker emission.  \citet{M07} and \citet{Sakai2010} interpreted this result as indicating that the younger sources are associated with
newly formed shocks compared to the older shocks
towards the more evolved objects.
Using the same transition, \citet{Miettinen2006}
also reports a decrease in SiO abundance in warmer, i.e.\,
likely more evolved sources. 
The study of \citet{Beuther2007} targets a larger sample of 43 IRDCs
in the $J= 2-1$ line, and finds that it is barely detected in quiescent clumps.
The largest statistical sample was
studied by \citet{LS2011}, who 
report a decrease of SiO ($2-1$) line luminosity as a function
of \lbolm. This finding was further supported by \citet{SM2013} on a smaller sample
and complemented with the $5-4$ transition.
However, the \citet{LS2011} study only considers
line luminosities without attempting to estimate the SiO abundance.

Other studies, however, such as \citet{Sanhueza2012}, \citet{Miettinen2014},
and \citet{Leurini2014}, find no trend of a 
decreasing SiO intensity; in particular, \citet{Miettinen2014}
reports increasing SiO abundance towards infrared-bright clumps, following the findings of \citet{Gerner2014}.
The aim of this study is therefore
to investigate the SiO abundance and excitation variations looking
for evolutionary trends
and thus monitor the shock activity in a Galactic-wide sample of massive clumps.
Taking this a step further,  we calculate the 
SiO abundance and use it as an independent value to investigate the statistical 
trends in the sample. 
In addition, we complement a large fraction of the sources
with the SiO ($5-4$) line, probing a higher energy transition 
and thereby allowing the investigation of excitation effects.
Using a larger sample,  
probing more transitions and performing
LTE and non-LTE abundance estimations, this study  
represents the most complete
approach in which single-dish telescopes have been used to establish a census
of SiO 
emission associated with massive clumps hosting a variety of evolutionary
stages. 
Based on this large sample of massive clumps, 
we derive statistical trends in SiO emission and thus characterise the
shock properties along the evolution of massive clumps.

The paper is summarised as follows: Sect.~\ref{sec:obs} describes the
observations and Sect.~\ref{sec:sample} describes the targeted sample.
In Sect.~\ref{sec:results} we discuss the results in terms of detection rates
and statistics, and we also discuss the line profiles.
The analysis of the derived abundances and the excitation conditions is
presented in Sect.~\ref{sec:analysis}, and the results are discussed in Sect.~
\ref{sec:discussion}. The conclusions are summarised in Sect.~\ref{sec:conclusion}. 


\section{Observations and data reduction}\label{sec:obs}

\subsection{IRAM~30m observations: Molecular fingerprints of massive clumps}\label{sec:obs-iram}

\begin{figure*}[!htpb]
\centering
\includegraphics[width=0.9\linewidth, angle=0]{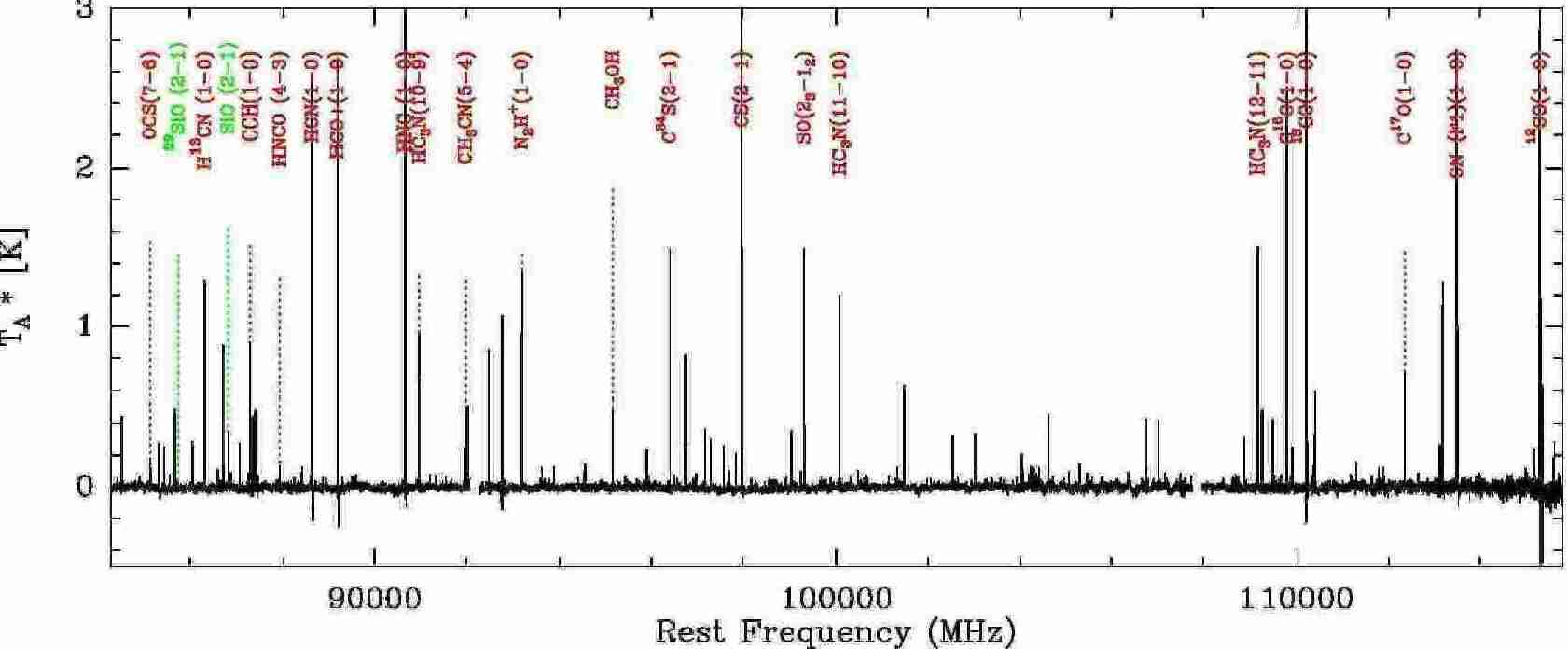}
\caption{
Complete frequency
coverage of the survey carried out with the EMIR instrument at the IRAM~30m telescope towards 
an example source, showing typical hot-core emission (G29.95--0.02). The SiO 2--1 transition and 
its isotopologue is labelled in green, other common transitions, which are not used in this paper, 
are labelled in red.
}\label{fig:example}
\end{figure*}

The present study is based on
an extensive spectroscopic follow-up project
of \at\ sources carried out with the IRAM~30m telescope\footnote{IRAM is supported by INSU/CNRS (France), MPG (Germany), and IGN (Spain).} and the EMIR receiver. Over 430~sources have been selected in various 
evolutionary stages to be observed 
as part of this unbiased spectral line survey 
covering almost the entire 3~mm atmospheric window, corresponding to 
a frequency range between 
$\sim$84~GHz$-115$~GHz (Fig.\,\ref{fig:example}).  

The observations have been carried
out in two sessions. First, in a pilot study, we targeted a smaller sample of the 36 most extreme
sources in terms of sub-millimeter peak flux density
from the \at\ survey between $83.8-115.7$~GHz. These observations
were carried out 
between the $8^{\rm th}$ and $11^{\rm th}$ of April 2011. The 
Fast Fourier Transform Spectrometer (FFTS) backend was 
used in 200~kHz spectral resolution giving a velocity resolution between 
$0.68$ and $0.51$~\kms\ across the band.
This frequency range was covered in 4~GHz blocks, using the actual capabilities 
of the receiver and backends. We also correlated the signal in eight different setups
with the VESPA backend giving a velocity resolution of 0.08~\kms.

We carried out the second part of the project  on several days between the 22$^{\rm nd}$ February and 13$^{\rm th}$
of March in the observing pool, 
as well as between $9^{\rm th}$  and  22$^{\rm nd}$ of October 2012\footnote{The corresponding IRAM project ids for this set of observations are: 181-10, 049-11, 037-12, respectively.}, 
similarly in the observing pool. We
used the FTS backends together with the 16~GHz 
instantaneous bandwidth of the EMIR receiver in dual polarisation. The frequency range
between $84.26 -115.73$~GHz was covered in two setups centred on 86 and 94~GHz, 
respectively, in the lower inner band of the receiver. 
The FTS backend was used with 200~kHz spectral resolution, giving the 
same velocity resolution as described above. A summary of the observations
is listed in Table\,\ref{tab:table-obs}.

Both observing campaigns used the same observing strategy: pointing and focus were checked
regularly and a reference spectrum was obtained at the beginning of the observations. 
The cross-calibration of each observing session is described in Sect.\,\ref{app:app-cal}.
During
the pilot study, several of the targets themselves served as reference, while in the second 
campaign the spectral line calibrator G34.26+0.15 was regularly observed.
The observations were carried out in position
switching mode with a constant offset for the reference position
of 10{\arcmin} in RA and DEC.
 
From this survey the current paper 
focuses only on the SiO ($2-1$) transition at 86.847~GHz,
as well as the $^{29}$SiO ($2-1$) line at 85.759~GHz. 
Analysis of the other transitions will be presented in a subsequent series of papers.
To go from $T_{\rm A}^{*}$ 
to $T_{\rm mb}$ temperature scales, 
we applied a  
beam efficiency correction of 0.81. 
After defining a window around the lines, 
a baseline of order 3 was removed from the spectra 
over a limited velocity range of $\sim$150\,km/s 
around the line. 
The $T_{\rm sys}$ varies between 97 and 222~K, and we measure
an $rms$ noise level of $\sim20$~mK averaged in a 2~\kms\ velocity bin 
on $T_{\rm A}^{*}$
scale.
The nominal velocity resolution and beam sizes are indicated in Table\,\ref{tab:table-obs}.

\subsection{APEX observations}

We followed up a smaller sample of \nsouapex\ sources, 
including all the  nearby massive clumps
and those exhibiting 
the brightest SiO ($2-1$) detections
 in the different source categories 
(see Sect.\,\ref{sec:class}) with the APEX telescope\footnote{APEX is a collaboration between the Max-Planck-Institut f\"ur Radioastronomie, the European Southern Observatory, and the Onsala Space Observatory.}, using the APEX-1 receiver centred on 217.2~GHz. 
The XFFTS backends  provided a {77}~kHz spectral resolution.
The corresponding velocity resolution and the beam size
 are indicated in Table\,\ref{tab:table-obs}.

Pointing and focus were checked regularly,
and the same observing strategy was adopted as for the observations with the IRAM~30m
telescope.
A baseline of
order 3 was removed from the spectra and a beam efficiency of 0.75 
has been applied. The $T_{\rm sys}$ varies between 143 and 280~K, and we measure
the noise between 15 and 30 mK on a $T_{\rm A}^{*}$ scale
at a velocity resolution smoothed to 2~\kms. 

\begin{table*}
\centering
\caption{Summary of observations.}\label{tab:table-obs}
\begin{tabular}{rrrrrrrrrrrr}
\hline\hline
Date & No. of sources & Instruments          & Frequency range & Velocity resolution & Beam\tablefootmark{a} & $rms$ [$T_{\rm A}^\star$]\tablefootmark{b}  \\
\hline 
$8^{\rm th}$ -- $11^{\rm th}$ April 2011 & 36 & IRAM~30m/EMIR & 84--117 GHz         & $0.68$ to $0.51$~\kms\ & 29\arcsec\ & 20 mK \\
February, {October} 2012 & 395& IRAM~30m/EMIR & 84--117 GHz         & $0.68$ to $0.51$~\kms\ & 29\arcsec\ & 20 mK \\
July -- October 2013& \nsouapex\ & APEX/SheFI & 215.2--219.2~GHz         & $0.105$~\kms\ & 28\arcsec\ & 15--30 mK \\
\hline
\end{tabular}
 \tablefoot{ 
 {
\tablefoottext{a}{The beam size is given at the frequencies of the SiO (2--1) and SiO (5--4) lines.}
\tablefoottext{b}{The $rms$ noise per channel is given for 2~\kms\ velocity bins.}
}}\end{table*}

\section{A sample of massive clumps from the \at\ survey}\label{sec:sample}

The targeted sources have been selected from the 
\at\ survey~\citep{schuller2009, csengeri2013}, where flux limited samples
corresponding to various evolutionary stages of star formation have been identified.
In the following, we describe the physical
properties of the sample starting with a discussion of their distances
(Sect.\,\ref{sec:dist}),
and then, we characterise their evolutionary stage using
ancillary tracers (Sect.\,\ref{sec:class}). We then estimate
their evolutionary stage assessing their 
mid-infrared monochromatic luminosity (Sect.\,\ref{sec:ir}). 
We discuss the dust properties
obtained from the \at\ survey (Sect.\,\ref{sec:mass}), and finally 
describe a mass- and distance-limited sample (Sect.\,\ref{sec:sample_sel}), 
which is used for a part of the analysis.

\subsection{Distances}\label{sec:dist}

Over 1000 sources from the \at\ survey have been surveyed in the NH$_3$ (1,1)
transition to determine kinematic distances \citep{Wienen2012,Wienen2015}.
We used {\hi} absorption against the background continuum and {\hi} emission spectra to distinguish between the near and far distance solution. Towards young sources,
where no background continuum is observed (i.e.\,prior to the appearance of ionizing emission associated with expanding {\hii} regions)
and the source is seen in absorption 
at 22~\mum\ with WISE, we systematically consider the source to be at the near distance.
Since several of our sources have been studied in detail as 
part of the Red MSX 
survey~(RMS; \citealp{Lumsden2013}),
we adopted their distance estimate where available~\citep{Urquhart2014}.

\begin{figure}[!htpb]
\centering
\includegraphics[width=7cm, angle=90]{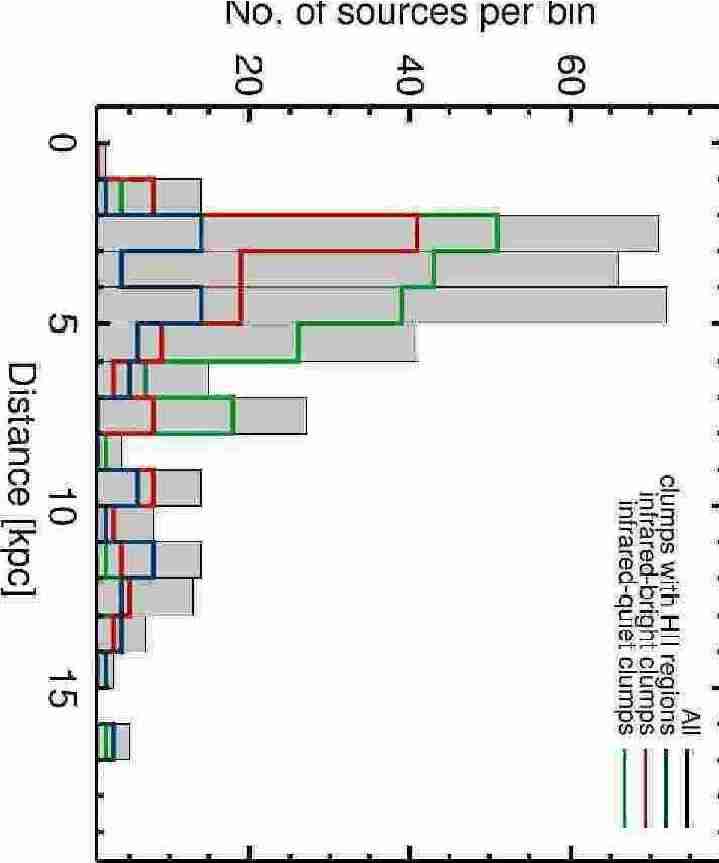}
\caption{Distance distribution of the whole 
              sample with the different source types
              indicated in colours. 
              Black line with grey filling corresponds to the whole sample,
              blue shows the genuine {\hii} regions, red corresponds to
              clumps with embedded protostars, and the green histogram shows
              the infrared-quiet clumps.}\label{fig:histo-dist}
\end{figure}

The distance distribution of the sample is shown in Fig.\,\ref{fig:histo-dist}.
We have a small fraction { ($\sim5$\%)} of sources at large distances ($>12$~kpc), which 
are mainly clumps with embedded {\hii} regions and/or bright massive young stellar objects (MYSOs) identified by the RMS survey. Massive clumps with {\hii} regions are more evenly distributed in distance, while both the infrared-bright
and infrared-quiet clumps peak at $3-4$~kpc (for a description of source classification, see Sect.\,\ref{sec:class}). This is 
a selection effect, since nearby sources appear brighter and have therefore
been preferred in the initially flux-limited sample selection 
(see also \citealp{Giannetti2014}).
Infrared-quiet clumps seem to have a similar
distribution as a function of distance as clumps hosting MYSOs. 
The mean and median values 
of the adopted distances 
of the sample (5.0 and 4.6\,kpc, respectively)
show that on average clumps associated with embedded {\uchii}
regions are more distant
with a broader distribution, while star-forming clumps without radio emission
are on average at a closer distance 
and show a smaller scatter. Infrared-quiet clumps are the most nearby
sources with an even smaller scatter.
The statistical properties of the sample are summarised
in Table\,\ref{tab:table-stat-class}.

\subsection{Classification of the sources}\label{sec:class}

The sources have been selected
to represent various evolutionary stages of high-mass star formation
(see also \citealp{Giannetti2014}).
After the original sample selection, new catalogues became available and therefore
we refine  this classification to
reflect  the physical properties of the dominating embedded source in distinct
evolutionary stages better. Below we provide the details of the classification:

\begin{itemize}
\item[(1)] Clumps hosting (UC-){\hii} regions 
have been identified by cross-correlating the dust peaks 
with the CORNISH catalogue \citep{Hoare2012,Purcell2013}, 
which were then further classified as (UC-){\hii} regions 
by \citet{UCornish2013}, and 
the extensive follow-up observations by the RMS Survey~\citep{Lumsden2013}. 
This category therefore consists  of genuine sources showing 
free-free emission from already formed high-mass star(s) (M$_\star>8\,$\msol).

\item[(2)] Clumps with embedded protostars and/or YSOs 
have been identified by 
cross-matching the dust peaks with the WISE point source catalogue 
(see also Sect.\,\ref{sec:ir}).
These infrared-bright, star-forming clumps 
host
a point source in the 22~\mum\ band of WISE, 
but lack detectable free-free emission at radio wavelengths. 
Part of this sample consists of genuine MYSOs, which
 have also been identified in the RMS Survey, while
the rest of the sample host likely a low- or intermediate-mass protostar at this,
mid-infrared bright stage of its evolution. The embedding clump is, however, in several
cases massive enough to form high-mass stars, 
and therefore even the weaker sources
are likely to be precursors of massive protostars.

\item[(3)] Infrared-quiet clumps have no embedded 
point source associated with the dust peak in the 22~\mum\ band
in the WISE point source catalogue.
These sources show no mid-infrared signatures of an embedded
protostar or protocluster coinciding with the dust peak.
They are expected to be the youngest in the sample, either hosting 
deeply embedded Class~0-like
high-mass protostars~\citep{Bontemps2010}, or an even earlier stage of
gravitationally bound structures on the verge of collapse, 
i.e. prestellar or starless clumps. Several of these clumps
appear in absorption against the mid-infrared background
and are associated with IRDCs.
We term these sources to be infrared-quiet, similar
to the terminology used by, 
for example \,\citet{M07,Bontemps2010,Csengeri2011a}
for massive dense cores.
\end{itemize}

\begin{figure*}[!htpb]
\centering
  \includegraphics[width=6.0cm,angle=0]{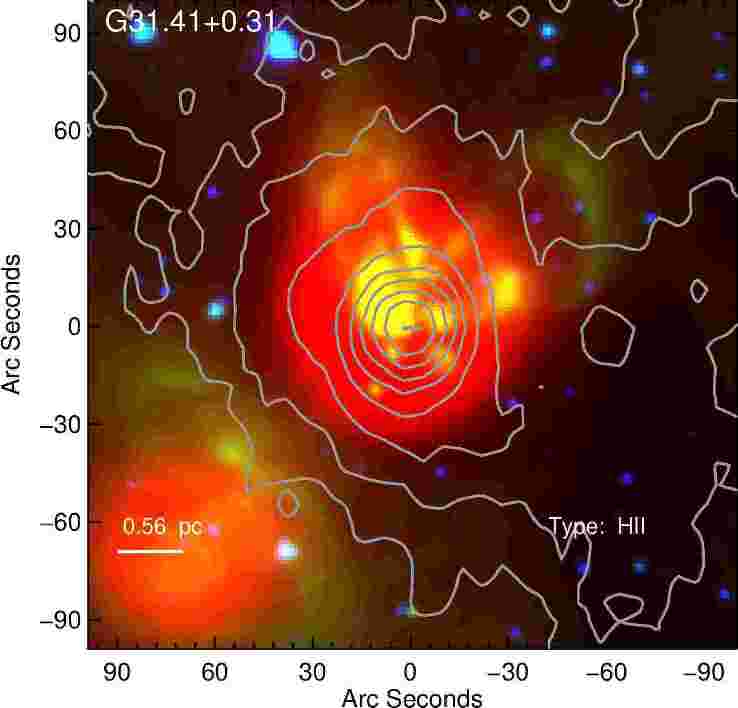}
  \includegraphics[width=6.0cm,angle=0]{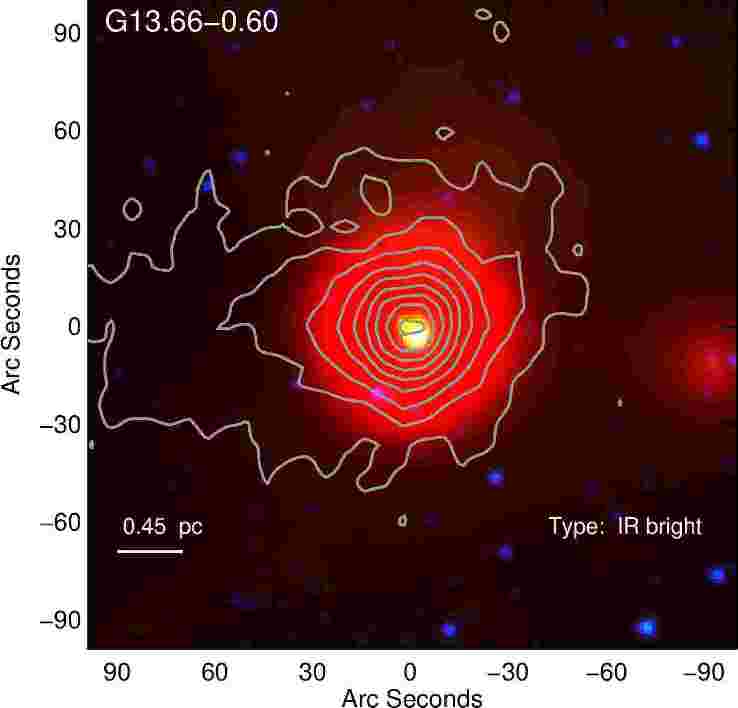}
  \includegraphics[width=6.0cm,angle=0]{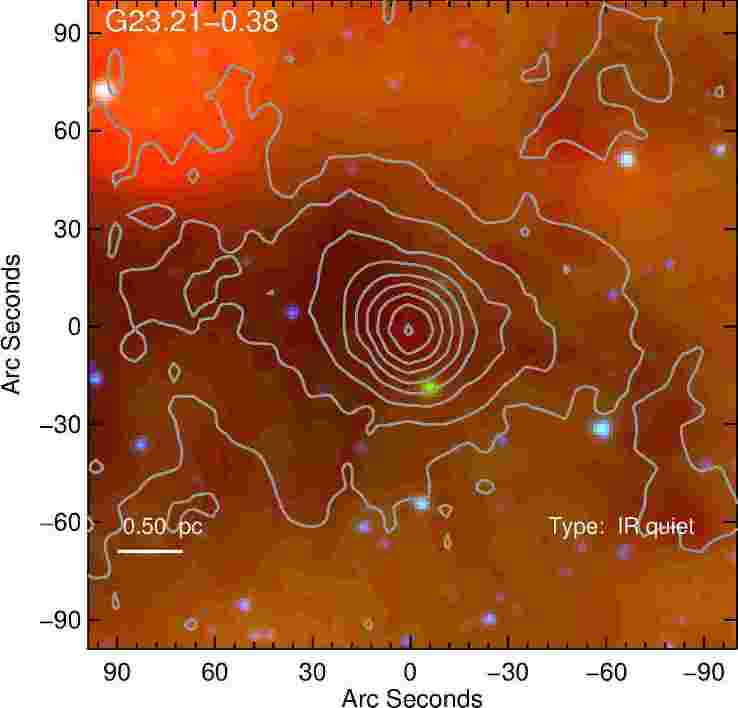}
 \caption{Examples of sources in different classes are shown from left to right: massive 
               clumps hosting embedded {\uchii} region, infrared-bright source, and
               an infrared-quiet massive clump. The background image is a three colour 
               composite image from Spitzer 3.6~\mum\ (blue), 8~\mum (green), and 
               WISE 22~\mum\ (red) band images. Grey contours show
               the 870~\mum\ emission from \at. White labels show the source name,
               the classification of the source, and a bar shows the physical
               scale considering the distance of the source.}\label{fig:examples}
 \end{figure*}

The first category with {\uchii} regions clearly corresponds to the latest
stage of the formation of high-mass stars when the objects have reached 
the main sequence while still accreting material from their surrounding clump.
The (M)YSOs are in a stage prior to that when a massive and luminous central object 
has formed already. 
The infrared-quiet clumps represent the earliest stages
at the onset of collapse, when the protostar is so deeply embedded that
it remains hidden at mid-infrared wavelengths and only the
high dust column density suggests the capability of the clump to
form high-mass stars.  
In fact, not all of these clumps may have high enough
 density to potentially form high-mass stars. 
 We investigate this
aspect in more detail in Sect.\,\ref{sec:sample_sel}.
Examples of sources according to this
classification are shown in Fig.\,\ref{fig:examples}. 

Altogether out of the  430 selected sources, 
 \nhii\ sources host embedded {\hii} regions, 
\nirbright\ sources embedded protostars, 
and \nirdark\ sources are  infrared-quiet, or starless.
Two sources turned out to be not related to star formation: 
the RMS survey identifies the source G50.05+0.77 as a planetary nebula 
and G59.18+0.11 as an evolved star~\citep{Lumsden2013}.
Therefore, these two sources are not be considered in the following analysis.

\begin{figure*}[!htpb]
\centering
\includegraphics[width=0.26\linewidth, angle=90]{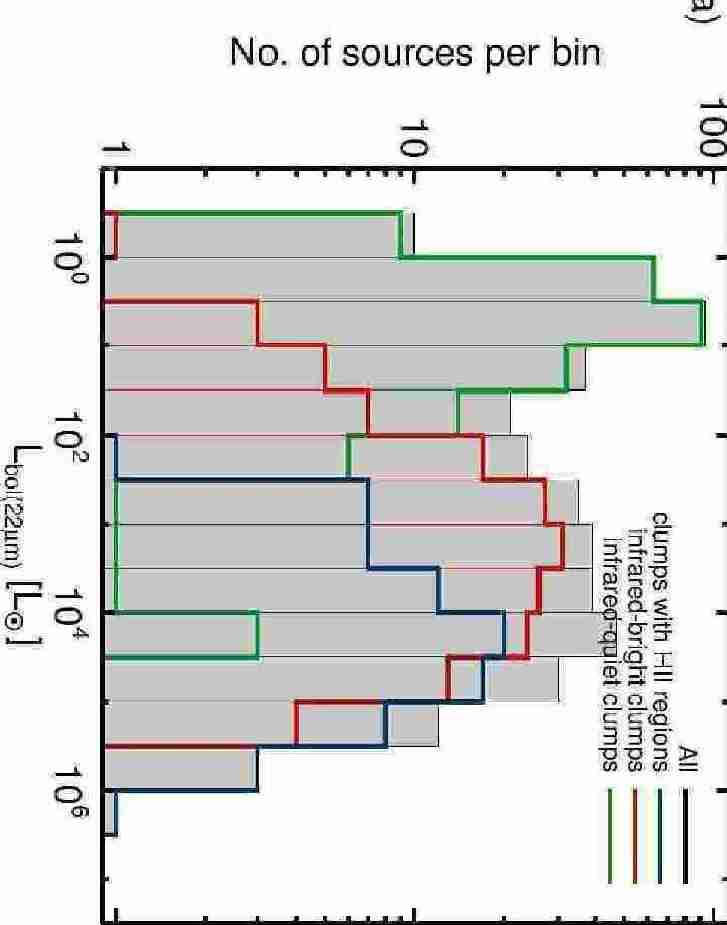}
\includegraphics[width=0.26\linewidth, angle=90]{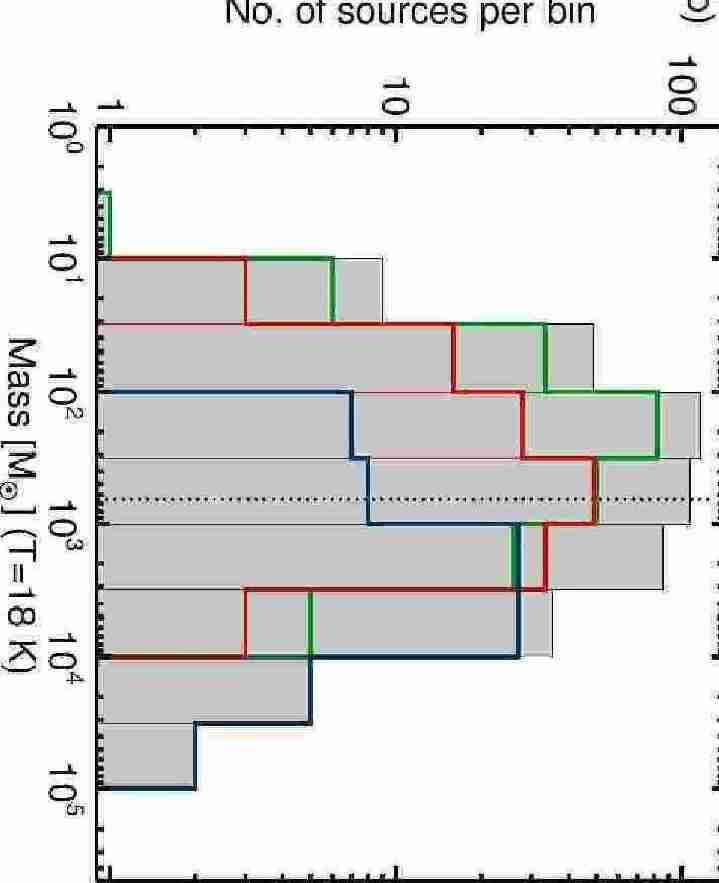}
\includegraphics[width=0.26\linewidth, angle=90]{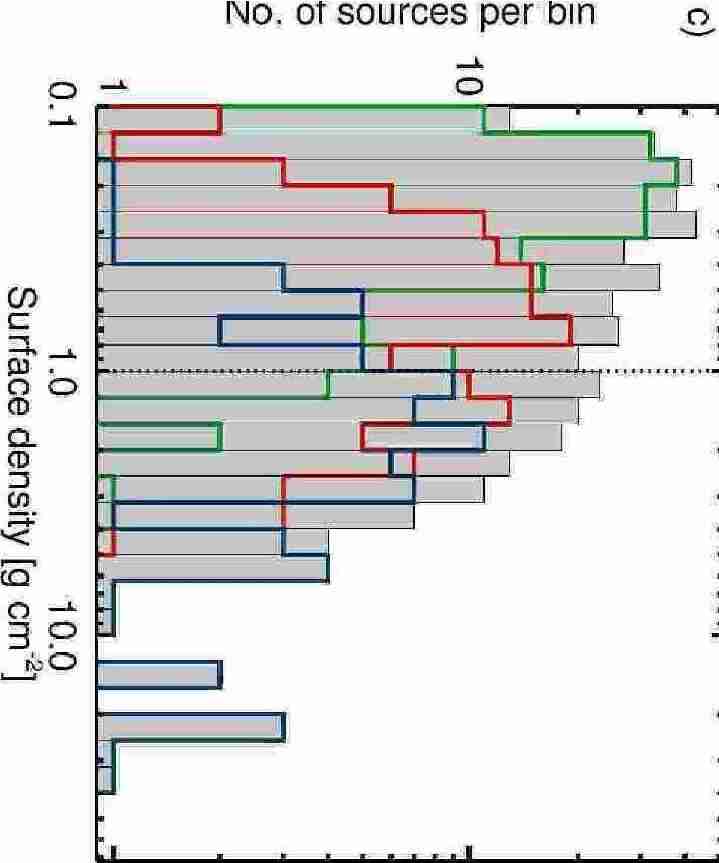}
\caption{Histograms of  
               the bolometric luminosity extrapolated from the monochromatic
              luminosity at 22~\mum\ from \citep{Mottram2011a}; we present
              ({\bf a}), the mass {\bf b}) and
              ({\bf c})  the surface density
  for the whole sample. 
              The dotted line in panel {\sl b)} shows the mass limit
              of 650~\msol.
              The dashed line in panel {\sl c)}
              show the theoretical value of 1~g~cm$^{-2}$ from
              \citep{Krumholz2008}.
              The colour scheme is the same as in Fig.\,\ref{fig:histo-dist} and is
              indicated in the legend of panel {\sl a)}.}
              \label{fig:histo-mass}
\end{figure*}

\subsection{Mid-infrared monochromatic luminosity}\label{sec:ir}

We calculate a monochromatic luminosity from the $21-22$~\mum\  flux densities as
\begin{equation}
      L_{\rm \nu}^{22~{\mu m}} = 4\pi \times d^2\times S_{\nu}^{22~{\mu m}},
\end{equation}
where $S_{\nu}^{22~\mu m}$ is flux per unit frequency and $d$ corresponds to
the distance to the source. 
The flux is taken from the band 4 
magnitude values from the WISE point source catalogue and converted to magnitudes\footnote{
The 22~\mum\ fluxes were taken from the WISE point source catalogue~\citep{Wright2010}. We considered an embedded mid-infrared source to be associated with the dust, when 
the angular offset was small ($\le10$\arcsec)
and required a good quality flag measurement for the 22~\mum\ band footnote,
and a maximum fraction of saturated pixels of 25\%. (Objects flagged 
as "D", "H", "O", or "P" are
likely to be spurious sources according to the WISE catalogue description). 
The magnitudes were 
then converted to fluxes using a zero point magnitude of 8.2839 at 22~\mum~\citep{Wright2010}, 
a colour correction factor of 1.0 and an additional correction factor of 0.9 from \citet{Cutri2012}.}.
Where such measurements from the WISE catalogue were not available because of saturation, 
we used the MSX
point source catalogue. 
In a consecutive step we
visually inspected
our positions with the \at\ maps for the dust, as well as Spitzer-GLIMPSE
images ($3.6-8$~{\mum}), Spitzer-MIPSGAL images (24~{\mum}) and
the WISE images (22~{\mum}) to check for spurious matching with  
foreground stars.
We compare
the $21-22$~\mum\ flux measurements between MSX and WISE in Appendix\,\ref{app:msx_wise}.

For sources without any embedded mid-infrared source, we 
estimate an upper limit from the publicly available 22~{\mum} WISE images.
The files have been reprojected with the Montage tool \citep{montage2010}
and the flux corresponding to the position of the continuum peak has been extracted.
No correction for the background has been applied. The map units 
from DN\footnote{http://wise2.ipac.caltech.edu/docs/release/allsky/expsup/}
have been converted to Jy  using a conversion factor of 
$5.27\times10^{-5}$~\citep{Cutri2012}.
The 22~\mum\ band is sensitive not only to emission from point sources, but 
also to bright and extended structures associated with PAH emission.
In particular,  
the typically complex sites of massive star formation can cause
signicant confusion due to extended PAH emission,
and the actual detection limit is higher than in fields with rather isolated embedded sources.

The remaining sources have been visually inspected and were found to be 
associated with very bright saturated emission in the 22~\mum\ band of WISE
and were likely also saturated in the MSX catalogue. We adopt a lower
limit for these sources of 330~Jy~\citep{Cutri2012}. 
The left panel of Fig.\,\ref{fig:histo-mass} shows the monochromatic luminosity
distribution of the sample. There is a trend for the more evolved
clumps to be more luminous at mid-infrared wavelengths, while the
infrared-quiet clumps are likely less evolved 
than clumps hosting {\uchii} regions or (M)YSOs.

From the monochromatic flux densities, we estimated 
the bolometric luminosity of the sources by applying
a scaling factor derived in \citet{Mottram2011a}. We used the
respective filter widths for both MSX and WISE and scaled
the WISE fluxes at 22~\mum\ to the 21~\mum\ wavelength of the
MSX E-band assuming $F_{\nu}\sim {\nu}^3$.
Towards the infrared-quiet clumps these 
estimates are based on upper limits (see Sect.\,\ref{sec:ir}). Although these estimates
are only good to an order of magnitude, we see a clear shift in the peak luminosity
of the different classes with the infrared-quiet clumps showing up as the 
lowest luminosity sources.

\begin{figure}[!htpb]
\centering
\includegraphics[width=0.7\linewidth, angle=90]{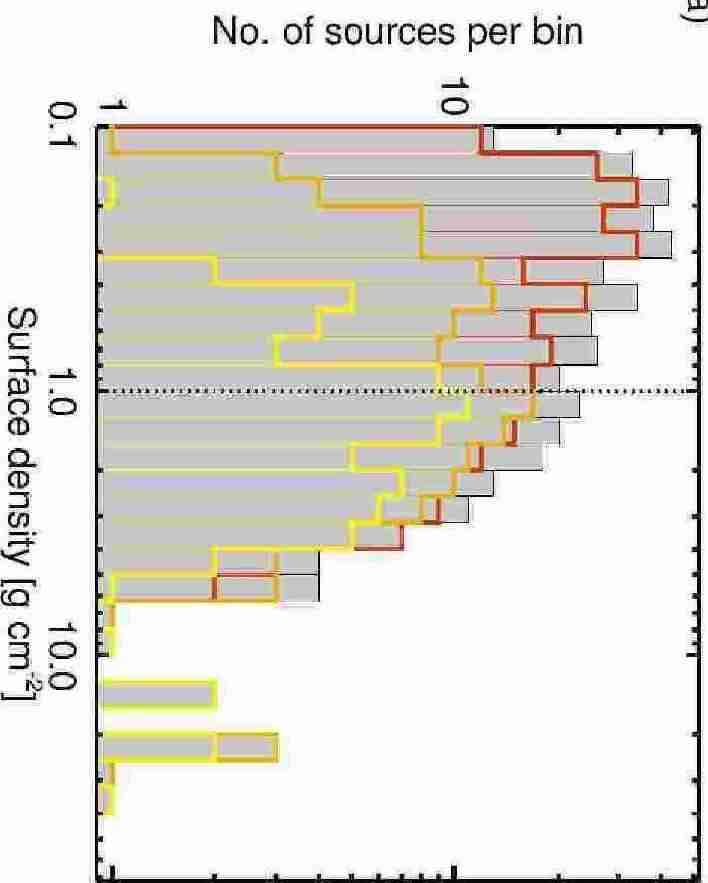}
\includegraphics[width=0.7\linewidth, angle=90]{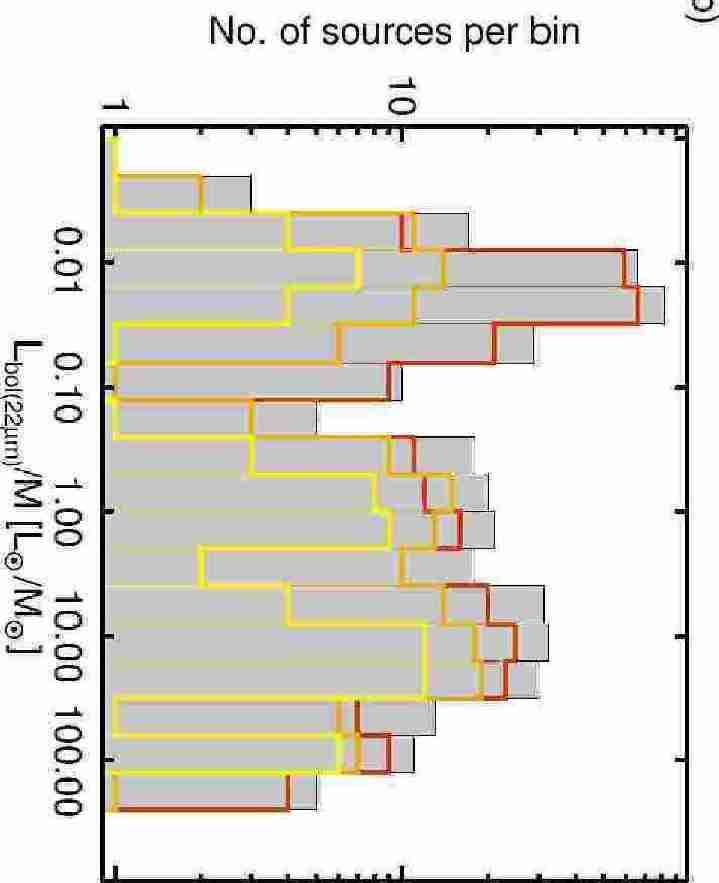}
\includegraphics[width=0.7\linewidth, angle=90]{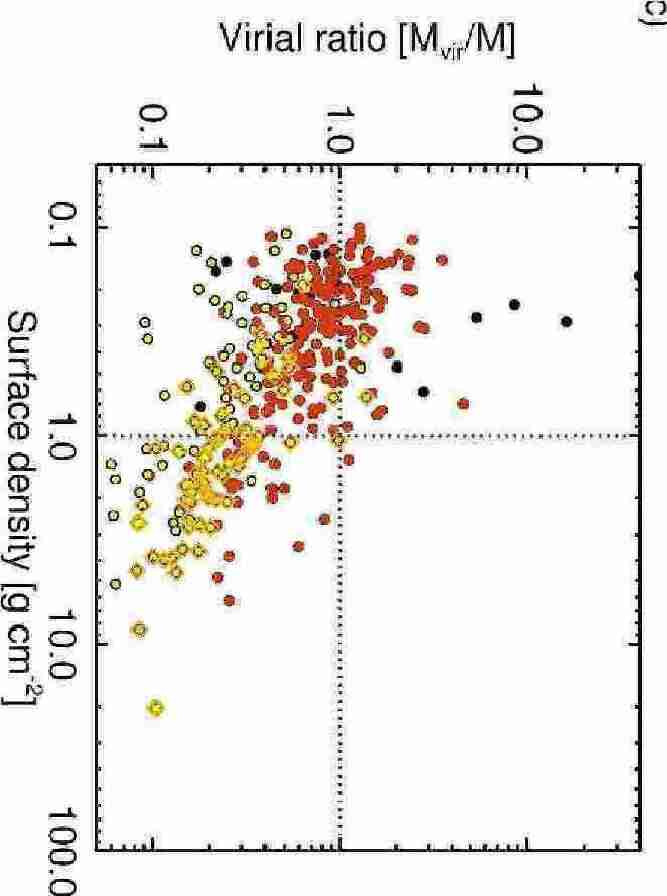}
\caption{{\bf a)} Histogram of the surface density of all sources, the
 dark orange,  orange, and yellow histograms correspond to the distance, 
  mass, and both mass- and distance-limited subsamples, respectively.
  {\bf b)} Histogram of  \lbolm\ [\lsol/\msol]. The lines are the same as on panel{ a)}.
    {\bf c)} 
    Plot of surface density versus the ratio of $M_{\rm vir}/M$. 
    The colour codes are the same as for panels a and b. Dotted lines
    show the line of $M_{\rm vir}=M$, and a surface density of 1~g~cm$^{-2}$.
     The properties of the selected subsamples are discussed in Sect.\,\ref{sec:sample_sel}.}\label{fig:histo-subsample}
\end{figure}
\begin{figure}[!htpb]
\centering
\includegraphics[width=0.7\linewidth, angle=90]{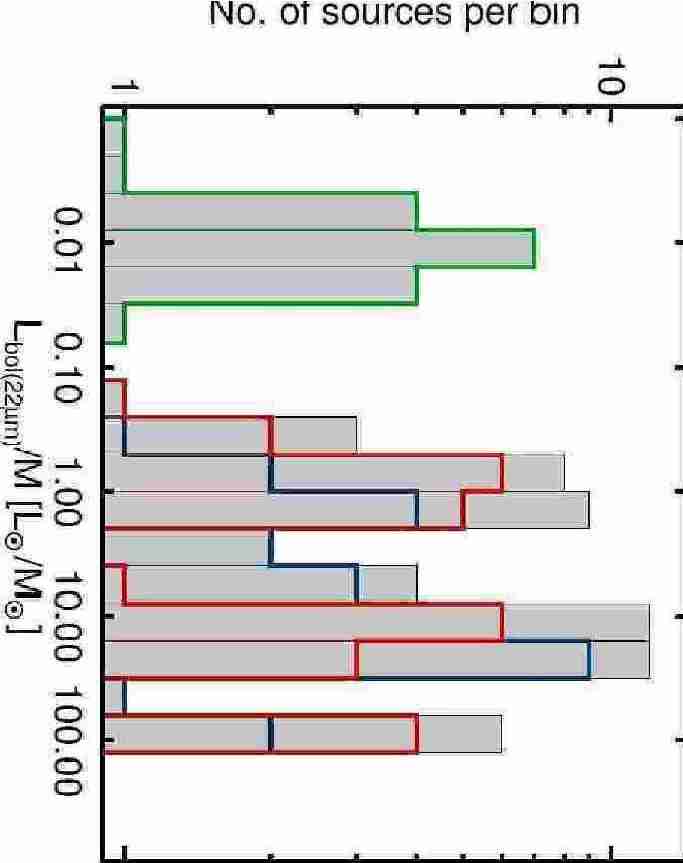}
\caption{
Histogram of the $L_{\rm bol}/M$ distribution of the distance- and mass-limited
subsample. Colours are the same as in Fig.\,\ref{fig:histo-mass}, and indicate sources in different
classification.
}\label{fig:histo-subsamplelm}
\end{figure}

\setlength{\tabcolsep}{3pt}
 \begin{table}
 \centering
 \caption{
 { 
 Classification of the sources and statistics of the subsamples defined in Sect.\,\ref{sec:sample_sel}.
 }}
 \begin{tabular}{rrrrrrrrrrrr}
 \hline\hline
 Class & No. (\%) & $1$~kpc$<d<7$~kpc& $M\ge650\,$\msol \\
& & &   $1$~kpc$<d<7$~kpc \\
 \hline
     all      &   430                  & 310           &  75 \\  
     {\hii}   &          78 (18\%) & 42 (14\%)  & 29 (40\%) \\    
     infrared-bright &  135 (31\%) & 99 (32\%)  &    28   (36\%) \\
     infrared-quiet &    217  (51\%) & 169 (54\%)& 18  (24\%) \\
\hline
 \end{tabular}\label{tab:table-stat-class}
 \tablefoot{ 
 {
 In parenthesis we show the fraction of sources with respect to the total number of objects within the given subsamples.
}}
 
 \end{table}
 \begin{table*}
 \centering
 \caption{{ Statistics of the physical} properties of the sample.}
 \begin{tabular}{lrrrrrrrrrrr}
 \hline\hline
Parameter & & Min. & Max. & Mean & Median & Standard~deviation \\
 \hline 
 \hline 
 Mass [\msol]: & all      & 4 & 9.13$\times10^4$ &  1555 &   408 &  5240
 \\
     & {\hii}   &   110 &    9.13$\times10^4$ &     5363 &     2709 &    11273 \\
     & infrared-bright &    17 &     7735 &      866 &      534 &     1138 \\
     & infrared-quiet &     4 &     8260 &      577 &      260 &      990 \\
\hline
 Distance [kpc]:& all      &     0.8 &    16.9 &     5.5 &     4.8 & 
    3.4 \\
     & {\hii}   &     1.12 &    16.5 &     7.5 &     6.5 &     4.2 \\
     & infrared-bright &     0.9 &    13.8 &     5.2 &     4.2 &     3.4 \\
     & infrared-quiet &     0.8 &    16.9 &     5.0 &     4.6 &     2.7 \\
\hline
 $\Sigma$ [g~cm$^{-3}$]: & all      &     0.10 &    33.84 &     1.18 &     0.43 & 
    3.01 \\
     & {\hii}   &     0.19 &    33.84 &     3.78 &     1.68 &     6.27 \\
     & infrared-bright &     0.11 &     4.83 &     0.92 &     0.64 &     0.83 \\
     & infrared-quiet &     0.10 &     3.90 &     0.38 &     0.22 &     0.52 \\
\hline
 \end{tabular}\label{tab:table-stat}
 \end{table*}

\subsection{Mass estimates}\label{sec:mass}

We estimate the 
gas
mass of the clumps using the same formula and numerical values
as in \citet{schuller2009} and \citet{csengeri2013}, i.e. 
\begin{equation}
 \frac{M}{[\rm M_{\odot}]}=\frac{S_\nu\,R\,d^2}{B_\nu(T_d)\,\kappa_{\nu}} 
\simeq 6.33 \times \frac{S_{\nu}^{870~\mu m}}{[\rm Jy]} \times \Big( \frac{d}{[\rm kpc]} \Big)^2
,\end{equation}
where $S_{\nu}^{870~\mu m}$ is the integrated flux from the \at\ survey from \citet{csengeri2013},
and $d$ corresponds to the 
heliocentric distance. To be consistent with our previous studies (e.g. \citealp{csengeri2013}),
we use a gas-to-dust mass ratio (R) of 100 and  $\kappa_{\nu}=1.85$ cm$^2$~g$^{-1}$. 
We derived the numerical factor  with the constants given above, 
$\nu=345$~GHz, and
a typical dust temperature dominated by the interstellar radiation
field of $T_{\rm d}=18$~K \citep{Bernard2010}.
This is close to
the average gas kinetic temperature determined from NH$_3$ measurements on large 
samples of massive clumps~\citep{Dunham2011, Wienen2012}.

The { middle} panel of Fig.\,\ref{fig:histo-mass} shows the mass distribution of the sources
with the different classifications indicated. Clumps hosting embedded {\hii}
regions are the most massive in the sample, which is also due 
to their larger distances (see Sect\,\ref{sec:dist}). Star-forming
clumps exhibit a peak mass higher than that of infrared-quiet clumps. 
Infrared-quiet clumps are on average a factor of more than two less massive than
star-forming clumps and an order of magnitude less massive 
than clumps with embedded
{\hii} regions. 
We note that the mass estimates
are subject to uncertainties due to variations in the dust temperature.
Massive protostars heat up their environment, which would lead 
to overestimating the mass, while 
the dust temperature may be lower
towards the infrared-quiet sources. Taking $T_{\rm d}=10$~K, we would underestimate
the masses  by a factor of three, while
with $T_{\rm d}=20$~K, we would measure 15\% less mass.
The uncertainty due to the dust temperature
 is, however, unlikely to introduce a significant bias 
because sources at larger distances cover larger physical regions as well, where the
dust temperature is expected to be in equilibrium with the interstellar radiation field~
\citep{Bernard2010}.

The right panel of Fig.\,\ref{fig:histo-mass}   shows the distribution of the surface density of
the clumps calculated by 
\begin{equation}
\frac{\Sigma}{[\rm g~cm^{-2}]}=6.65\times10^{-5}\times{{\frac{M}{{[\rm M_{\odot}]}}}\times\Big(\frac{1}{2}\frac{R}{[\rm pc]}\Big)^{-2}}~\rm ,
\end{equation}
where $R$ corresponds to the beam-deconvolved $FWHM$ size 
of the Gaussian 
fits to the sources from \citet{csengeri2013}. As this histogram shows,
a substantial fraction (25\%) of { all} 
the sources lie above the conservative, 
theoretical limit of 1~g~cm$^{-2}$ 
for massive cores to form stars by \citet{Krumholz2008}, 
suggesting that our sample selection is robust enough to study the
properties of massive clumps.
The derived physical parameters of the sources are listed in Table\,\ref{tab:table-dust}.

\begin{table*}
\caption{List of sources selected for the survey with the corresponding dust parameters from ATLASGAL.}\label{tab:table-dust}
\begin{center}
\begin{tabular}{lrrlrrrrrrrr}
\hline\hline
Source name\tablefootmark{a} & Ra    & Dec     &  ATLASGAL source\tablefootmark{b} & Offset \tablefootmark{c} & Distance & Mass & $N(\rm H_2)$\tablefootmark{d} & $\Sigma$ & Classification &  \\
            & J2000 [$^\circ$] & J2000 [$^\circ$]  &  &  [\arcsec] &[kpc] & [M$_{\odot}$] & [cm$^{-2}$] & [g\,cm$^{-2}$] & \\
\hline
G06.22-0.61   &  270.5120   &    -23.8869   &      G006.2159-0.6098   &     3.8   &  3.78   &    446   &                   6.7$\times10^{22}$   &      0.79   &       infrared-bright  \\
G08.05-0.24   &  271.1470   &    -22.1112   &      G008.0495-0.2433   &     3.4   &    --   &     --   &                   2.9$\times10^{22}$   &      0.45   &        infrared-quiet  \\
G08.68-0.37   &  271.5974   &    -21.6189   &      G008.6834-0.3675   &     0.0   &  12.0   &   7592   &                  13.4$\times10^{22}$   &      1.84   &       infrared-bright  \\
G08.71-0.41   &  271.6528   &    -21.6213   &      G008.7064-0.4136   &     0.0   &  11.9   &   3249   &                   3.0$\times10^{22}$   &      0.26   &        infrared-quiet  \\
G11.35+0.80   &  271.8970   &    -18.7276   &      G011.3441+0.7961   &     4.7   &    --   &     --   &                   1.8$\times10^{22}$   &      0.19   &       infrared-bright  \\
G11.38+0.81   &  271.9020   &    -18.6882   &      G011.3811+0.8103   &     3.6   &  3.37   &     90   &                   1.9$\times10^{22}$   &      0.25   &        infrared-quiet  \\
G10.47+0.03   &  272.1588   &    -19.8638   &      G010.4722+0.0277   &     0.0   &  11.0   &  32316   &                  97.8$\times10^{22}$   &     28.57   &            HII region  \\
G10.08-0.19   &  272.1610   &    -20.3155   &      G010.0791-0.1951   &     6.7   &  3.76   &    117   &                   2.1$\times10^{22}$   &      0.28   &        infrared-quiet  \\
G10.45-0.02   &  272.1869   &    -19.9100   &      G010.4446-0.0178   &     0.0   &  10.7   &   2730   &                   6.2$\times10^{22}$   &      0.87   &        infrared-quiet  \\
G10.28-0.12   &  272.1940   &    -20.0978   &      G010.2841-0.1134   &     5.2   &  2.20   &    269   &                   8.8$\times10^{22}$   &      0.82   &        infrared-quiet  \\
G10.29-0.12   &  272.2010   &    -20.0984   &      G010.2863-0.1240   &    10.1   &  2.20   &    280   &                   9.9$\times10^{22}$   &      0.97   &        infrared-quiet  \\
G10.66+0.08   &  272.2070   &    -19.6731   &      G010.6627+0.0811   &     7.3   &  2.96   &    112   &                   1.7$\times10^{22}$   &      0.15   &        infrared-quiet  \\
G10.34-0.14   &  272.2500   &    -20.0599   &      G010.3420-0.1426   &     1.1   &  2.00   &    191   &                  13.1$\times10^{22}$   &      1.99   &        infrared-quiet  \\
G10.32-0.14   &  272.2560   &    -20.0867   &      G010.3225-0.1611   &     5.2   &  3.55   &    747   &                  10.3$\times10^{22}$   &      1.03   &       infrared-bright  \\
G10.33-0.16   &  272.2580   &    -20.0834   &      G010.3225-0.1611   &    10.6   &  3.55   &    747   &                  10.3$\times10^{22}$   &      1.03   &       infrared-bright  \\
G10.62-0.03   &  272.2860   &    -19.7654   &      G010.6174-0.0304   &     5.2   &  5.53   &    219   &                   1.6$\times10^{22}$   &      0.19   &        infrared-quiet  \\
G10.75+0.02   &  272.3100   &    -19.6288   &      G010.7471+0.0161   &     5.1   &  2.88   &     91   &                   1.6$\times10^{22}$   &      0.14   &        infrared-quiet  \\
G10.32-0.23   &  272.3220   &    -20.1205   &      G010.3221-0.2304   &     1.4   &    --   &     --   &                   2.1$\times10^{22}$   &      0.20   &       infrared-bright  \\
G10.21-0.30   &  272.3350   &    -20.2506   &      G010.2144-0.3051   &     1.8   &  1.90   &     34   &                   4.0$\times10^{22}$   &      2.52   &        infrared-quiet  \\
G10.15-0.34   &  272.3380   &    -20.3244   &      G010.1506-0.3434   &     2.4   &  1.60   &    129   &                   8.0$\times10^{22}$   &      0.75   &       infrared-bright  \\
G10.21-0.32   &  272.3520   &    -20.2607   &      G010.2130-0.3234   &     0.4   &  3.55   &    814   &                  10.8$\times10^{22}$   &      1.05   &       infrared-bright  \\
G10.17-0.36A   &  272.3610   &    -20.3175   &      G010.1672-0.3624   &    17.0   &  3.55   &   1032   &                   8.2$\times10^{22}$   &      0.65   &       infrared-bright  \\
G10.17-0.36B   &  272.3650   &    -20.3181   &      G010.1672-0.3624   &     5.7   &  3.55   &   1032   &                   8.2$\times10^{22}$   &      0.65   &       infrared-bright  \\
G10.11-0.41   &  272.3820   &    -20.3994   &      G010.1065-0.4168   &    10.3   &  2.00   &     43   &                   1.7$\times10^{22}$   &      0.15   &        infrared-quiet  \\
G10.83-0.02   &  272.3860   &    -19.5746   &      G010.8276-0.0207   &     6.4   &  3.70   &    157   &                   1.6$\times10^{22}$   &      0.14   &        infrared-quiet  \\
G10.96+0.02   &  272.4140   &    -19.4412   &      G010.9576+0.0223   &     2.5   &  13.6   &   5911   &                   8.1$\times10^{22}$   &      1.12   &            HII region  \\
G11.03+0.06   &  272.4160   &    -19.3557   &      G011.0335+0.0615   &     2.8   &  14.2   &   4188   &                   4.9$\times10^{22}$   &      0.61   &            HII region  \\
G10.74-0.13   &  272.4400   &    -19.7021   &      G010.7418-0.1255   &     1.7   &  3.60   &    239   &                   3.2$\times10^{22}$   &      0.32   &        infrared-quiet  \\
G10.67-0.22   &  272.4920   &    -19.8111   &      G010.6700-0.2211   &     1.7   &  3.70   &    205   &                   2.5$\times10^{22}$   &      0.25   &        infrared-quiet  \\
G10.75-0.20   &  272.5130   &    -19.7275   &      G010.7528-0.1975   &     1.9   &  3.87   &    208   &                   1.6$\times10^{22}$   &      0.14   &        infrared-quiet  \\
G10.99-0.08   &  272.5270   &    -19.4628   &      G010.9916-0.0815   &     2.8   &  3.70   &    408   &                   2.7$\times10^{22}$   &      0.21   &        infrared-quiet  \\
\hline
\end{tabular}
 \tablefoot{The full table is available only in electronic form at the CDS via anonymous ftp to cdsarc.u-strasbg.fr (130.79.125.5) or via http://cdsweb.u-strasbg.fr/cgi-bin/qcat?J/A\&A/. 
 \tablefoottext{a}{We provide here the source names used for the observations.}
 \tablefoottext{b}{The corresponding ATLASGAL source name from \citet{csengeri2013}.
 \tablefoottext{c}{Offset in arcseconds from the ATLASGAL dust peak from the catalog of \citet{csengeri2013}.}
 \tablefoottext{d}{The $N(\rm{H_2})$ estimate corresponds to the peak column density and is calculated from Eq.\,\ref{eq:dust_cdens} in Sect.\,\ref{sec:sio_21_abundance}.}}
 } 
\end{center}
\end{table*}

\subsection{A distance- and mass-limited subsample}\label{sec:sample_sel}

Both the IRAM~30m and the APEX telescope have a 
beam of 29\arcsec\ at the frequency of the SiO ($2-1$) and ($5-4$) transitions, respectively, which translates to physical scales of $0.1-2.45$~pc
at the distance range of $0.4-17$~kpc.
SiO emission has been claimed to be extended in 
IRDCs, as well as towards massive ridges 
on these scales~(e.g.\,\citealp{Jimenez-Serra2010,Quang2013}).
\citet{SM2013} resolved the spatial distribution of some
of their sources, although with a nearly three-times smaller beam.
It is therefore reasonable to 
assume here that
the extension of bipolar outflows is likely
to be smaller
than the beam, which may be particularly true for the more distant sources.
Comparing the whole 
sample at different distances may consequently introduce a bias due to
different filling factors.
To obtain more robust statistical results, we define a 
distance-limited subsample with sources
located between $1-7$~kpc, where the angular resolution
corresponds to $0.2-$1~pc and a potential distance bias 
is minimised. 

In order to investigate the properties of strictly massive clumps,
we consider clumps above a threshold of
$\sim$\,$650$\,\msol\ ($\Sigma\ge 1$\,g\,cm$^{-2}$, for a $R$ of 0.4\,pc,
derived from the average beam-convolved 
angular size of 25\arcsec\ at 5\,kpc), 
which is taken as a rough criterion for massive 
clumps to form high-mass stars \citep{csengeri2013}. As a comparison,
this limit corresponds to that proposed by \citet{Krumholz2008},
and it is more {conservative} than that used by \citet{Tackenberg2012} as well as
in the study of \citet{LS2011}, where
all clumps above 100~\msol\ at distances of several kilo-parsecs 
were considered  capable of
forming high-mass stars. 
A total of 161 sources (37\%) fulfil our mass threshold, which
makes this a significantly richer sample of more massive clumps 
compared to other spectral line studies. 
As a comparison, only five of the  \citet{LS2011}
sources are above this mass threshold. 

Combining these criteria,
we find 310 sources (72\% of the complete sample) to 
fall within
the distance limit of $1-7$~kpc, 
from which 75 sources  (14\%) 
also fulfil the $M>650$~\msol\ 
limit. 
The latter are  
the most extreme sources of the sample.
Among these extreme sources, we find 18 infrared-quiet clumps (i.e.\,$\sim$25\%, as in \citealp{csengeri2013}),
thus potential sites 
to study the initial conditions for high-mass star formation. 

Even putting such a conservative limit to restrict the sample
to the most massive sources, our initial selection is rich enough
to provide better statistics compared to precedent studies
and is a representative sample of 
different evolutionary stages (Table~\ref{tab:table-stat-class}).
The corresponding distribution of surface density and
luminosity to mass ratio ($L_{\rm bol}/M$) of these subsamples is 
shown in Fig.\,\ref{fig:histo-subsample} (upper and middle panels, respectively).
The statistics of the physical properties of the different source classes
are summarised in Table~\ref{tab:table-stat}.

To demonstrate that the sources are 
likely to form stars, we
calculated the ratio of the mass estimate from dust
versus the virial mass of the sources 
 to investigate the stability of the clumps in 
Fig.\,\ref{fig:histo-subsample} (lower panel).
We used the following equation from $M_{\rm Vir}=3\times \frac{R_{\rm sph}\times\sigma^2}{G}$~\citep{BertoldiMcKee1992}, where $R_{\rm sph}$ corresponds to the radius of a sphere, $\sigma$ is the velocity dispersion, and $G$ is the gravitational constant. Assuming 
an inverse square density profile
\begin{equation}
\frac{M_{\rm Vir}}{[\rm M_{\odot}]}= 697 \times \frac{1}{2} \frac{R}{[\rm pc]} \times \frac{\sigma_{tot}^2}{[\rm km~s^{-1}]},
\end{equation}
where $\sigma_{tot}$ corresponds to the velocity dispersion of the gas
derived from a fit considering the hyperfine structure of 
the N$_2$H$^+$ ($J=1-0$) line extracted from this spectral survey with the
IRAM~30m telescope, and $R$
to the beam-deconvolved $FWHM$ sizes of the source. 
As Fig.\,\ref{fig:histo-subsample} lower panel shows, the mass and distance-limited selection
very likely corresponds to unstable, self-gravitating clumps.
We therefore consider this
subsample statistically significant
and free from distance or selection biases.
Furthermore, it has a representative fraction of sources in different evolutionary stages (see also Table\,\ref{tab:table-stat-class}). We show their distribution of $L_{\rm bol}/M$ in
Fig.\,\ref{fig:histo-subsamplelm}, which is used later in Sect.\,\ref{sec:results} and
\ref{sec:analysis}.

In summary, in the following analysis in Sect.\,\ref{sec:results}, we first analyse
the statistical properties of the entire sample 
 of 430 sources, out of which \nsoutot\ were observed in the 
SiO ($2-1$) transition. To study the line profiles, we minimise the distance bias by using the
distance-limited sample, and investigate evolutionary trends in the line profile for 
a distance- and mass-limited sample.
In Sect.\,\ref{sec:analysis} we again derive first statistics on the full sample,
and then compare these results to the mass- and distance-limited subsample.

\section{Results: SiO detection rates and line profiles}\label{sec:results}

Based on the \at\ survey, we identified a 
representative sample of massive clumps located throughout the inner Galaxy.
Here we focus on the properties of SiO emission. 
In Sect.\,\ref{sec:detrate} we investigate first
the detection rates to reveal 
the statistics of the presence
of shocked gas. 
We report the typical line profiles and the velocity integrated emission
in Sect.\,\ref{sec:int}. 
In Sect.\,\ref{sec:line_profiles}
we study the stacked SiO line profiles in the 
distance- and mass-limited sample, and then investigate the statistical properties of the 
velocity structure of the 2--1 line in Sect.\,\ref{sec:width}.

\subsection{Detection rates}\label{sec:detrate}
\subsubsection{The SiO ($2-1$) line}
\begin{figure*}[!htpb]
\centering
  \includegraphics[width=0.9\linewidth,angle=0]{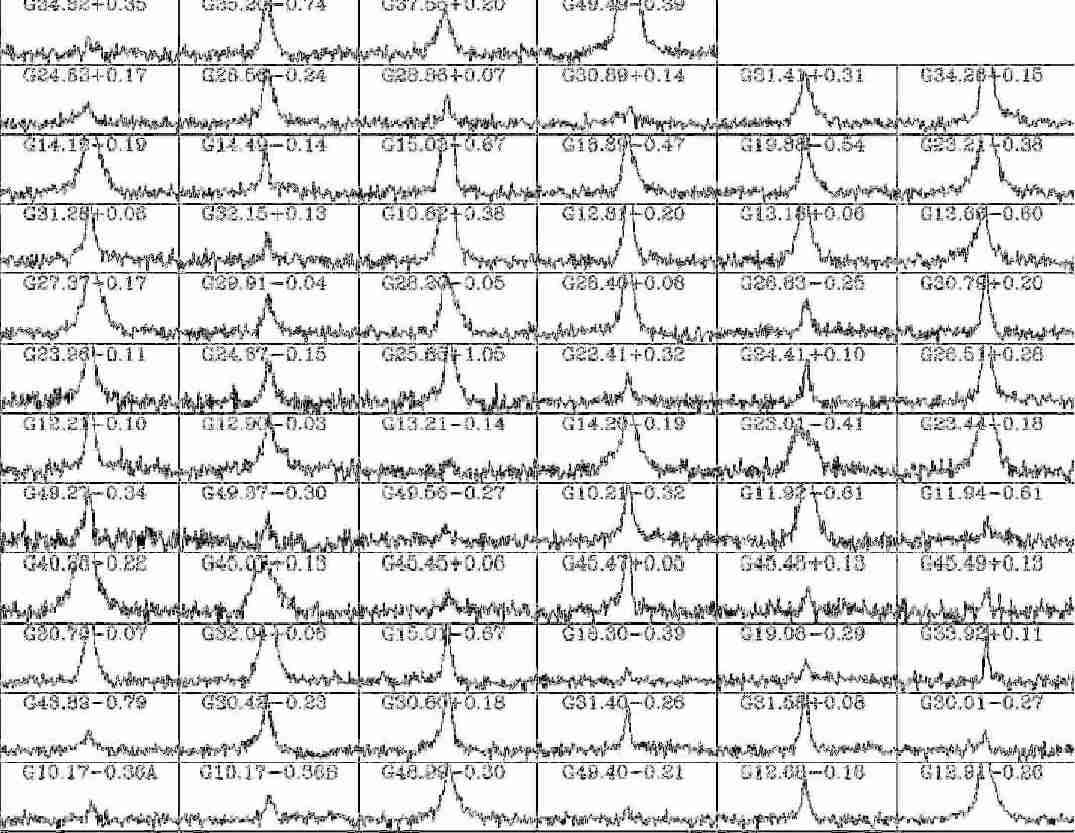}
 \caption{SiO ($2-1$) spectra of the mass- and distance-limited subsample.
These plots show a  $\pm70$~\kms\ velocity range and are centred on 
the rest velocity of the source,
while the y axis ranges from $-0.1$ to $0.5$~K on a $T_{\rm mb}$ scale.}
\label{fig:examples-sio}
 \end{figure*}

\begin{figure}[!htpb]
\centering
\includegraphics[width=6.5cm, angle=90]{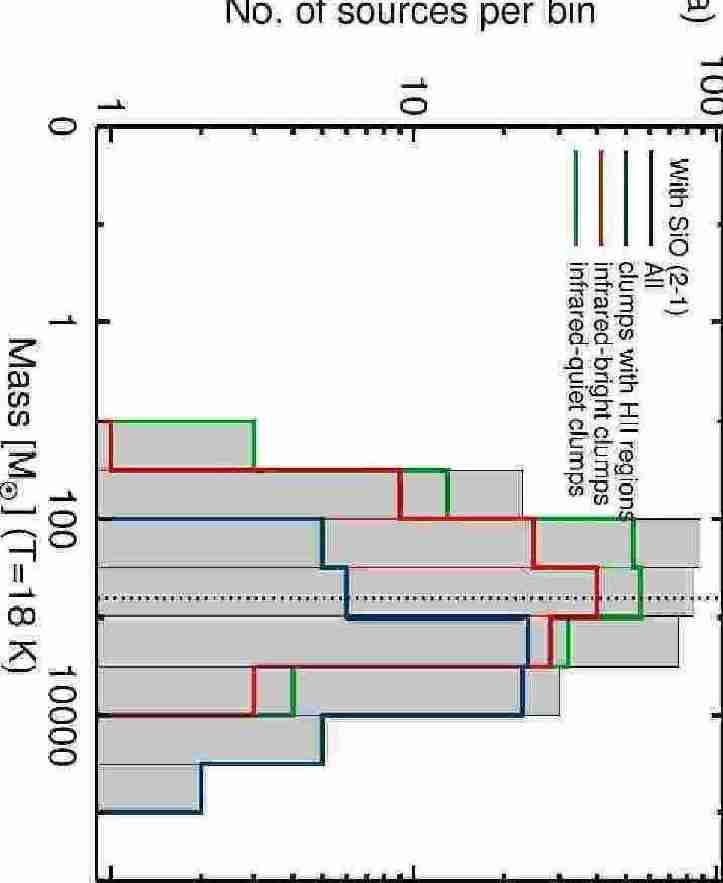}
\includegraphics[width=6.5cm, angle=90]{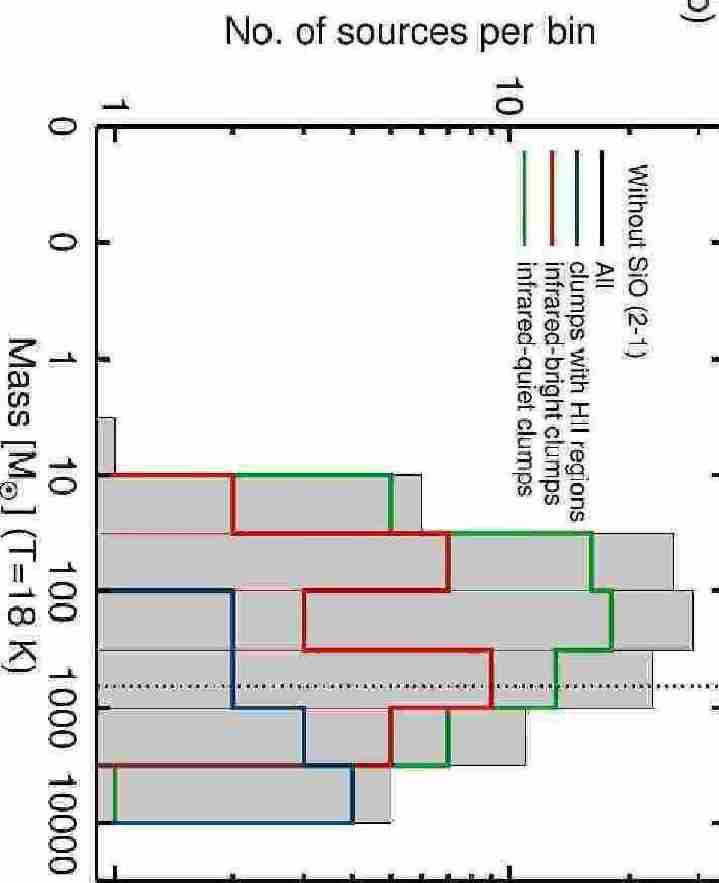}
\caption{Histogram of the massive clumps with ({\bf top}) and lacking ({\bf bottom}) SiO ($2-1$) detection
from the entire sample and the different categories labelled. The dashed line corresponds to 650~\msol, which is a crude limit for clumps to form massive stars. Colours are the same as in Fig.\,\ref{fig:histo-dist}.
}\label{fig:nondet}
\end{figure}

From the total of \nsoutot\ observed sources, our spectral line survey
detects 
molecular emission in all but one source.
In the SiO ($2-1$) line, 301 sources (70\%) show peak line intensities 
above $3\sigma$, where $\sigma$ corresponds to the $rms$ noise per 
velocity bin.
The detection rates according to source type
 are summarised in Table~\ref{tab:table-detection}, 
and  we note that the 
relative detection rate increases with sub-millimeter flux.
The SiO ($2-1$) line is  most frequently detected towards clumps
hosting {\hii} regions, and
a substantial fraction (61\%) of the infrared-quiet clumps also
exhibit SiO emission. 
Although with much higher sensitivity, already
 \citet{M07} found a high detection rate of SiO towards
infrared-quiet massive cores in the Cygnus-X molecular complex. 
However, 
such a high detection rate towards a large number of more distant
and infrared-quiet
massive clumps is intriguing since SiO is considered a typical tracer of
shocks~\citep{Schilke1997}, mostly related to protostellar activity.
Our {overall} detection rate is similar to that of \citet{LS2011}, who
had {at most} two times better sensitivity.
As shown in Table\,\ref{tab:table-detection},
the detection rates are similar for the distance-limited (1--7 kpc)
subsample as for the entire sample (see Sect.\,\ref{sec:sample_sel}).
Towards the mass and distance limited sample the
detection rates in each source type are similar, however, the overall
 detection rate is higher:  
out of the total of {75} sources
SiO ($2-1$), emission is detected towards {70} sources (93\%).
{The spectra of this subsample} are shown in Fig.\,\ref{fig:examples-sio}.
In Fig.\,\ref{fig:nondet} we show the histogram of the SiO 
detections and non-detections
for the different type of sources. 
While the detections are equally distributed in all categories for the whole sample, 
in the mass- and distance-limited sample  
practically all non-detections (5) are associated with infrared-bright embedded 
protostars or {\hii} regions that may have already started to disrupt their
natal cloud and where outflow activity may have ceased. 
The only infrared-quiet source with non-detection, G49.21-0.34, also  
shows a weak SiO emission, but 
below our detection
threshold.
All massive infrared-quiet clumps show evidence for emission
from shocked gas. Considering the full sample, 
the distribution of non-detections is clearly shifted to the lower mass regime
below the threshold for massive star formation.
These non-detections may therefore reflect the more diffuse
clumps without star formation.

For the reported detection rates, given the 3$\sigma$ limit, we expect 
$<2$ sources to be spurious detections
due to the statistical noise, which does not influence our results.
The resulting spectra, along with a three-colour image of the targeted clumps
with 870~\mum\ contours from \at,\ are shown as Online Material 
in Appendix\,\ref{app:both}. 
The sources where only the $2-1$ line has been observed
are shown in Appendix\,\ref{app:only21}.

Because of the non-Gaussian line profiles,
we first visually determined the 
velocity range where the emission reaches the noise level.
We used this velocity range ($FWZP$) 
 to 
 calculate the integrated intensity ($\int T_{\rm mb}(v)\, dv$). 
We calculated the errors on the line area  from the standard
formula of $\sigma_{area} = \sqrt{N_{ch}}\times \delta \, v \times \sigma$,
where $N_{\rm ch}$ is the number of channels with emission, $\delta \rm v$
is the velocity resolution and $\sigma$ is the $rms$ noise per channel
of each spectrum.
The extracted line properties, such as the 
$v_{\rm lsr}$ of the sources, the full width at zero power ($FWZP$), integrated intensity,
noise level, presence of wings and SiO column density estimates
(see also Sect.\,\ref{sec:int}, and Sect.\,\ref{sec:sio_21_coldens} )
are
summarised in Table~\ref{tab:table-large-sio21}.


\begin{table*}
\centering
\caption{Detection rates.}\label{tab:table-detection}
\begin{tabular}{rrrr|rrrrrrrrr}
\hline\hline
Source type & \multicolumn{3}{c}{SiO ($2-1$)} &  \multicolumn{3}{c}{SiO ($5-4$)} \\
                        &  \multicolumn{3}{c}{$>3\sigma$ detection} &  \multicolumn{3}{c}{$>3\sigma$ detection}\\
                       &  All distances & $1$~kpc$<d<7$~kpc&  $1$~kpc$<d<7$~kpc & All distances  & $1$~kpc$<d<7$~kpc&$1$~kpc$<d<7$~kpc\\
                       &                      &  &  M$>650$\,\msol\ &   & & M$>650$\,\msol\ \\

\hline
All sources &       301 (70\%)  &    226 (73\%)          &    70 (93\%) &      117    & 95    & 56   \\
{\hii} regions &      63 (80\%)   & 36 (85\%)               &   27 (93\%)       & 40     &  30   & 23    \\
infrared-bright sources & 106 (79\%) & 83 (84\%)     &   26 (93\%)      & 42     & 38    &  23   \\
infrared-quiet sources &  132 (61\%) & 106 (63\%)    &   17 (94\%)     &  35    & 27    & 10     \\
\hline
\end{tabular}
 \tablefoot{ 
  For the SiO (2--1) line, we indicate in parenthesis the percentage of the detection with respect to the total number of sources in the full sample (from Table\,\ref{tab:table-stat-class}) and the different source type, respectively. }
\end{table*}

\subsubsection{The SiO ($5-4$) line}\label{sec:detrate_sio54}

From the spectroscopic survey of the 430 sources targeted with 
the IRAM 30m telescope, we followed up \nsouapex\ sources
with the APEX telescope in the SiO~($5-4$) transition. 
Out of these sources  \nsoufivedet\ are detected
 with peak intensities above $3\sigma$ and
\nsoufivedetfivesigma\ detections  are above 5$\sigma$.
The line parameters of the SiO (5--4) emission are
summarised in Table~\ref{tab:table-large-sio54}. 
Altogether we have 112 sources with $>5\sigma$ detection in both transitions.
The statistics of the SiO (5--4) detection rates is shown in Table\,\ref{tab:table-detection}.
\begin{table*}
\caption{SiO ($2-1$) line parameters and column density.}\label{tab:table-large-sio21}
\begin{center}
\begin{tabular}{lrrrrrrrrrrr}
\hline\hline
Source name\tablefootmark{a} & Ra    & Dec     &  $v_{\rm lsr}\tablefootmark{b}$& $\Delta v$\tablefootmark{c} &$\int$T$_{mb}\,\rm dv$ & $\sigma_{21}$\tablefootmark{d}& Wings &  $N(\rm{SiO})$ \tablefootmark{e}\\
            & J2000 [$^\circ$] & J2000 [$^\circ$]  &  [km/s] &  [km/s] &[K km/s]  & [K km/s] & & $\times10^{12}$ [cm$^{-2}$] \\
\hline
G06.22-0.61   &  270.5120   &    -23.8869   &   18.46   &  [    0;   27] &    2.84   &    0.25   &  y   &  5.08  \\
G08.68-0.37   &  271.5974   &    -21.6189   &   38.03   &  [   32;   54] &    8.68   &    0.24   &  y   & 15.52  \\
G08.71-0.41   &  271.6528   &    -21.6213   &   39.64   &  [   28;   52] &    4.14   &    0.16   &  y   &  7.41  \\
G11.35+0.80   &  271.8970   &    -18.7276   &   27.59   &  [   12;   41] &    1.36   &    0.25   &  n   &  2.43  \\
G10.47+0.03   &  272.1588   &    -19.8638   &   67.05   &  [   56;   83] &   12.12   &    0.22   &  n   & 21.68  \\
G10.45-0.02   &  272.1869   &    -19.9100   &   76.06   &  [   66;   85] &    2.10   &    0.11   &  n   &  3.75  \\
G10.28-0.12   &  272.1940   &    -20.0978   &   14.29   &  [   -9;   47] &    6.67   &    0.25   &  y   & 11.92  \\
G10.29-0.12   &  272.2010   &    -20.0984   &   14.37   &  [    4;   29] &    2.84   &    0.12   &  y   &  5.08  \\
G10.34-0.14   &  272.2500   &    -20.0599   &   11.97   &  [   -9;   39] &    9.63   &    0.37   &  y   & 17.22  \\
G10.62-0.03   &  272.2860   &    -19.7654   &   64.19   &  [   59;   76] &    0.62   &    0.12   &  n   &  1.10  \\
G10.21-0.30   &  272.3350   &    -20.2506   &   12.50   &  [   -9;   47] &    3.58   &    0.25   &  y   &  6.40  \\
G10.15-0.34   &  272.3380   &    -20.3244   &    9.81   &  [   -1;   23] &    1.36   &    0.12   &  y   &  2.43  \\
G10.21-0.32   &  272.3520   &    -20.2607   &   10.19   &  [  -17;   25] &    5.43   &    0.25   &  y   &  9.72  \\
G10.17-0.36A   &  272.3610   &    -20.3175   &   14.02   &  [   -5;   23] &    1.48   &    0.12   &  n   &  2.65  \\
G10.17-0.36B   &  272.3650   &    -20.3181   &   15.21   &  [    8;   21] &    1.11   &    0.12   &  y   &  1.99  \\
G10.96+0.02   &  272.4140   &    -19.4412   &   21.36   &  [    8;   35] &    2.84   &    0.25   &  y   &  5.08  \\
G10.74-0.13   &  272.4400   &    -19.7021   &   29.00   &  [    4;   61] &    4.94   &    0.37   &  y   &  8.83  \\
G10.67-0.22   &  272.4920   &    -19.8111   &   29.78   &  [   20;   47] &    1.85   &    0.12   &  y   &  3.31  \\
G10.75-0.20   &  272.5130   &    -19.7275   &   31.86   &  [   18;   35] &    1.11   &    0.25   &  y   &  1.99  \\
G10.99-0.08   &  272.5270   &    -19.4628   &   29.36   &  [   20;   37] &    0.86   &    0.25   &  n   &  1.55  \\
\hline
\end{tabular}
 \tablefoot{The full table is available only in electronic form at the CDS via anonymous ftp to cdsarc.u-strasbg.fr (130.79.125.5) or via http://cdsweb.u-strasbg.fr/cgi-bin/qcat?J/A\&A/. 
 \tablefoottext{a}{Abbreviated source names used for the observations.}
 \tablefoottext{b}{The $v_{\rm lsr}$ of the sources is determined from optically thin tracers in the survey, such as N$_2$H$^+$ (1--0) and H$^{13}$CO$^+$ (1--0).}
 \tablefoottext{c}{The velocity range corresponding to the $FWZP$.}
 \tablefoottext{d}{Error in the integrated intensity of the SiO (2--1) transition.}
 \tablefoottext{e}{The $N(\rm{SiO})$ estimate based on Eq.\,\ref{eq:cdens} and LTE assumption.}
 } 
\end{center}
\end{table*}

\begin{table*}
\begin{center}
\caption{SiO ($5-4$) line parameters. }\label{tab:table-large-sio54}
\begin{tabular}{lrrrrrrrrrrr}
\hline\hline
Source name\tablefootmark{a}  & Ra      & Dec   & $\Delta v$\tablefootmark{b} &$\int$T$_{mb}\,\rm dv$ & $\sigma_{54}$\tablefootmark{c} & \\
            & J2000 [$^\circ$] & J2000 [$^\circ$]  &  [km/s] &[K km/s]  & [K km/s] & & \\
\hline
G08.68-0.37   &  271.5974   &  -21.6189   &  [    25;    46] &    1.31   &    1.27  \\
G08.71-0.41   &  271.6528   &  -21.6213   &  [    36;    41] &    0.28   &    0.01  \\
G10.47+0.03   &  272.1588   &  -19.8638   &  [    58;    81] &    4.77   &    0.01  \\
G10.45-0.02   &  272.1869   &  -19.9100   &  [    71;    84] &    0.27   &    0.01  \\
G10.28-0.12   &  272.1940   &  -20.0978   &  [    -3;    45] &    2.14   &    0.22  \\
G10.29-0.12   &  272.2010   &  -20.0984   &  [     6;    21] &    0.78   &    0.06  \\
G10.34-0.14   &  272.2500   &  -20.0599   &  [    -7;    49] &    3.85   &    0.13  \\
G10.21-0.32   &  272.3520   &  -20.2607   &  [   -17;    25] &    0.75   &    0.18  \\
G10.74-0.13   &  272.4400   &  -19.7021   &  [    16;    32] &    0.59   &    0.08  \\
G10.62-0.38   &  272.6193   &  -19.9300   &  [   -17;    12] &    8.79   &    0.07  \\
G12.42+0.51   &  272.7110   &  -17.9292   &  [     2;    31] &    0.37   &    0.07  \\
G11.08-0.53   &  272.9950   &  -19.6004   &  [    23;    46] &    1.17   &    0.10  \\
G11.90-0.14   &  273.0460   &  -18.6918   &  [    38;    46] &    0.16   &    0.19  \\
G12.20-0.03   &  273.0980   &  -18.3801   &  [    38;    58] &    1.38   &    0.19  \\
G12.21-0.10   &  273.1650   &  -18.4040   &  [     8;    43] &    1.84   &    0.01  \\
G12.90-0.03   &  273.4500   &  -17.7595   &  [    34;    83] &    1.48   &    0.10  \\
G12.68-0.18   &  273.4760   &  -18.0294   &  [    40;    63] &    0.31   &    0.11  \\
G11.92-0.61   &  273.4920   &  -18.9052   &  [    10;    57] &    8.80   &    0.10  \\
G11.94-0.61   &  273.5030   &  -18.8899   &  [    28;    51] &    0.50   &    0.08  \\
\hline
\end{tabular}
 \tablefoot{The full table is available only in electronic form at the CDS via anonymous ftp to cdsarc.u-strasbg.fr (130.79.125.5) or via http://cdsweb.u-strasbg.fr/cgi-bin/qcat?J/A\&A/.
 \tablefoottext{a}{Abbreviated source names used for the observations.}
 \tablefoottext{b}{The velocity range corresponding to the $FWZP$ determined from the SiO (2--1) observations.}
 \tablefoottext{c}{Error in the integrated intensity of the SiO (5--4) transition.}}
\end{center}
\end{table*}

\subsection{Line shape and velocity integrated emission}
\label{sec:int}

In both observed lines, towards the majority of the sources
we detect nearly Gaussian emission profiles centred on 
the rest velocity ($v_{\rm lsr}$). 
Towards several sources
sources, we detected additional emission in the high-velocity line wings.
Altogether we detect such wings in the SiO ($2-1$) transition
in \emirwing\ sources, which is 50\% of the total 
number of detections. 
The sources exhibit large variations in the wing profiles,
however, in most of the cases we find a symmetric broad component
centred on the line, while in a few cases we detect asymmetric
line profiles. We find altogether 63 of the infrared-quiet sources to exhibit high-velocity
wings; these likely indicate deeply embedded
high-mass protostars in a Class~0-like phase
~\citep{Bontemps2010,Duarte-Cabral2013}.
 They correspond to $\sim$30\% of the detections in this category of sources, which are higher than
for the more evolved objects
in agreement with the earlier findings of \citet{M07}.

\subsection{Averaged line profiles}\label{sec:line_profiles}
\begin{figure*}[!htpb]
\centering
\includegraphics[width=0.265\linewidth]{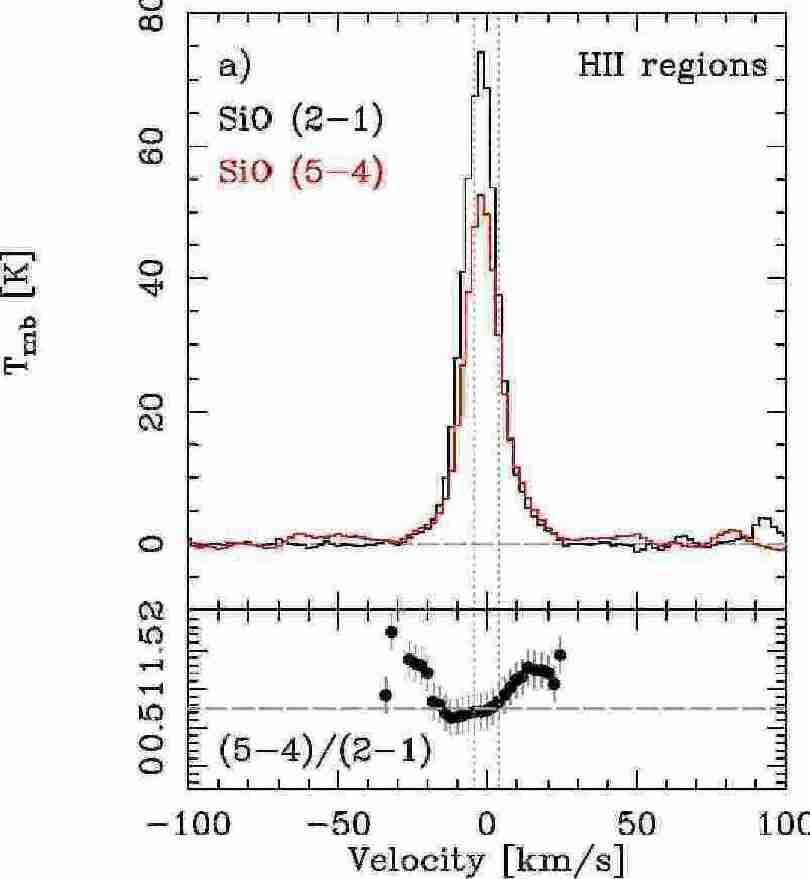}
\includegraphics[width=0.22\linewidth]{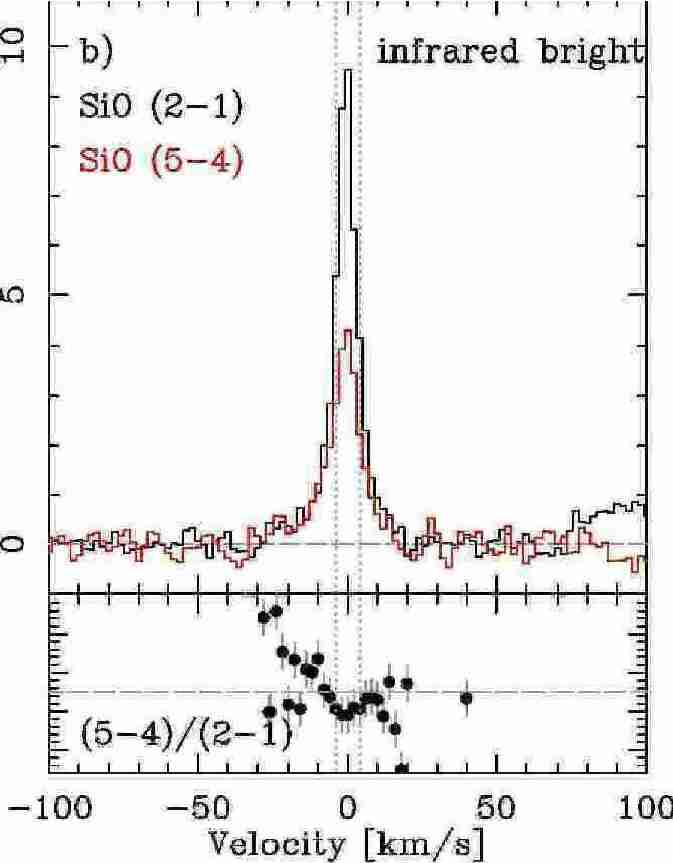}
\includegraphics[width=0.22\linewidth]{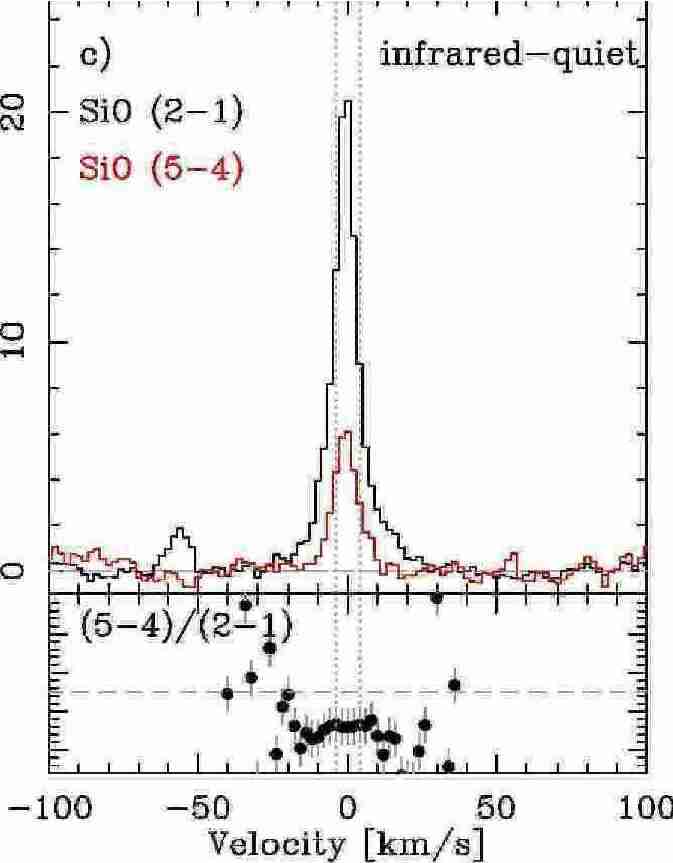}
\includegraphics[width=0.22\linewidth, angle=0]{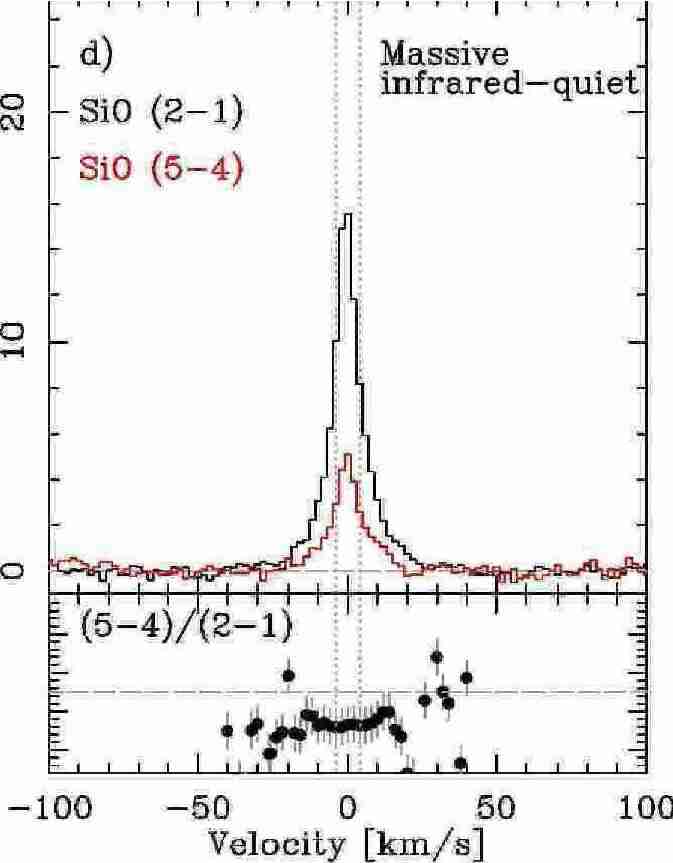}
\caption{Averaged, distance weighted line profiles. The SiO (2--1) line is shown in black, while the SiO (5--4) line is red. 
Dotted grey line indicates the velocity range $\Delta v=8$~\kms\
with respect to the averaged line centre. The panel \textbf{a)} corresponds to the {\hii} regions,  \textbf{b)} shows the infrared-bright sources,  \textbf{c)} shows the infrared-quiet sample, and  \textbf{d)} shows a subset of the infrared-quiet sample, where $\rm 1<d<7$~kpc and $\rm M>650$~\msol. 
The bottom panels show the ratio of the $5-4$ over
the $2-1$ line intensity per 2~\kms\ velocity bins. 
{ For a better comparison, a ratio of 0.75 is indicated
with dotted grey line.}
}\label{fig:line_profiles}
\end{figure*}

In Fig.\,\ref{fig:line_profiles} we show the
 averaged line profiles of both transitions from the
 different source classes scaled to a common distance
 of 1\,kpc.
Since  the beam sizes are similar for both lines, 
we assume that 
the emission originates from the same volume of gas and the excitation
conditions are the same.

We find that the ratio of the (5--4) to the (2--1) line intensities clearly
change with the respective evolutionary phase.
The smallest ratios are observed towards the
infrared-quiet clumps. 
High-velocity wings with
emission broader than 8\,\kms\ 
are observed towards all of these 
classes. 
For a detailed description of the velocity structure of the line,
see Sect.\,\ref{sec:width}.
We find equally high-velocity gas in the sample of
massive infrared-quiet clumps
as towards the clumps with embedded {\hii} regions, 
which is similar to the findings of \citet{M07}. 
These high-velocity wings are, therefore, confirmed in the early stage
of massive clumps 
on a  Galactic scale statistics. 

The signal-to-noise ratio of the individual spectra
does not allow us to study the excitation conditions in the
high-velocity wings, therefore, the averaged line profiles
are used here for a qualitative description. 
In the lower panel of Fig.\,\ref{fig:line_profiles}, we  
show the ratio of the (5--4/2--1) lines. 
We find that the line ratio is larger in the high-velocity wings
compared to the low-velocity Gaussian component, suggesting 
a change in the excitation conditions for the
low- versus the high-velocity components, and,
therefore, a different
origin for the low- and high-velocity component,
as also noted by \citet{Nisini2007, Leurini2013}. 
This trend is, however, less clear for the intermediate
velocities, in particular towards the infrared-quiet sample.
Using the line ratios, in Sect.\,\ref{sec:analysis} we provide 
a quantitative analysis of the velocity averaged 
excitation conditions for individual sources with 
the highest signal-to-noise detections in the low-velocity component.

\subsection{The velocity structure of the SiO ($2-1$) line}\label{sec:width}

Since we have a representative sample of the different
evolutionary stages, i.e.\,clumps with {\hii} regions,
star-forming, and infrared-quiet clumps, 
here we use the higher signal-to-noise ratio ($2-1$) transition
to statistically investigate the SiO line profiles over its entire velocity extent
across the sample and different categories.
In Sect.\,\ref{sec:nofit} we therefore first analyse   
the velocity extent of the flow with the $FWZP$ 
of the distance-limited
subsample to avoid bias due to  
lower sensitivity for the high-velocity wings towards
more distant sources.

Emission in high-velocity wings, corresponding to emission from shocked gas due to fast 
material ejection, is commonly observed in the SiO lines
(e.g.\,\citealp{Gueth1998,Codella1999, Nisini2007}). However, there
is no clear definition to distinguish the high-velocity component from a Gaussian profile 
commonly observed at the systemic velocity of the source.
Various definitions have been used so far in the literature. 
\citet{Quang2013} refer to a broad component with $FWZP> 25$\kms, 
while other studies are more conservative. For example, \citet{Duarte-Cabral2014}
refer to emission with $FWHM>9$~\kms\ as a  broad component, and 
define a low-velocity component with $FWHM$ of $2.5-6$~\kms.
\citet{Jimenez-Serra2010} refer to a low-velocity component with
$FWHM$ line-widths up to 3~\kms, and consider as a high-velocity
component $FWHM$ widths between 4--7 \kms.
In low-mass protostars \citet{Lefloch1998} report weak low-velocity
components with $FWHM$ of 0.4-1.2~\kms, while \citet{Codella1999}
identify low-velocity components with up to 3~\kms\ $FWHM$
line width. In Sect.\,\ref{sec:twocompfit}, 
we use a Gaussian fit to statistically disentangle the low- 
and high-velocity component. We stress at this point that throughout the sample the low-velocity component always peaks at the ambient velocity
of the clump.

\subsubsection{The high-velocity component}\label{sec:nofit}
\begin{figure}[!htpb]
\centering
\includegraphics[width=0.8\linewidth, angle=90]{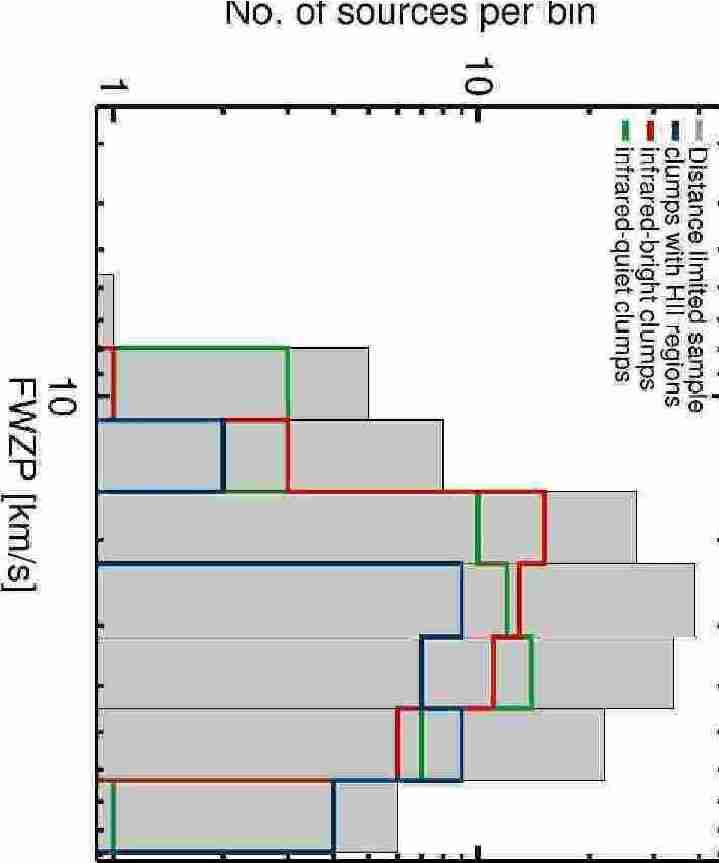}
\caption{
{
Histogram of the velocity range for the measured
$FWZP$ of the SiO ($2-1$) emission of 
the distance-limited subsample (see Sect\,\ref{sec:sample_sel}). 
The cutoff in velocity was determined by visual inspection
as described in Sect.\,\ref{sec:detrate}, where emission drops to 
zero.}
}\label{fig:width_histo}
\end{figure}

We have looked for trends 
as a function of the evolutionary stage and 
 velocity of the flow. 
To accomplish this, we analyse the statistical properties of
the $FWZP$, which also accounts for the
wings of the lines and was used 
for determining the total integrated emission.
Broad velocity ranges up to $70-76$~\kms\
are observed towards clumps hosting
several embedded {\hii}-regions, such as W51~Main, W49A,
as well as other mid-infrared bright clumps hosting YSOs.
The broadest infrared-quiet source shows high-velocity wings up to
a total velocity range of 65~\kms\ associated with the
star-forming complexes of W43~Main and South as well as W33.
To study the statistical properties of the highest velocities observed, 
we show the histogram of the $FWZP$ 
of the distance-limited subsample in Fig.\,\ref{fig:width_histo}. 
We find no significant difference
as a function of source type.

\begin{figure}[!htpb]
\centering
\includegraphics[width=5.cm, angle=90]{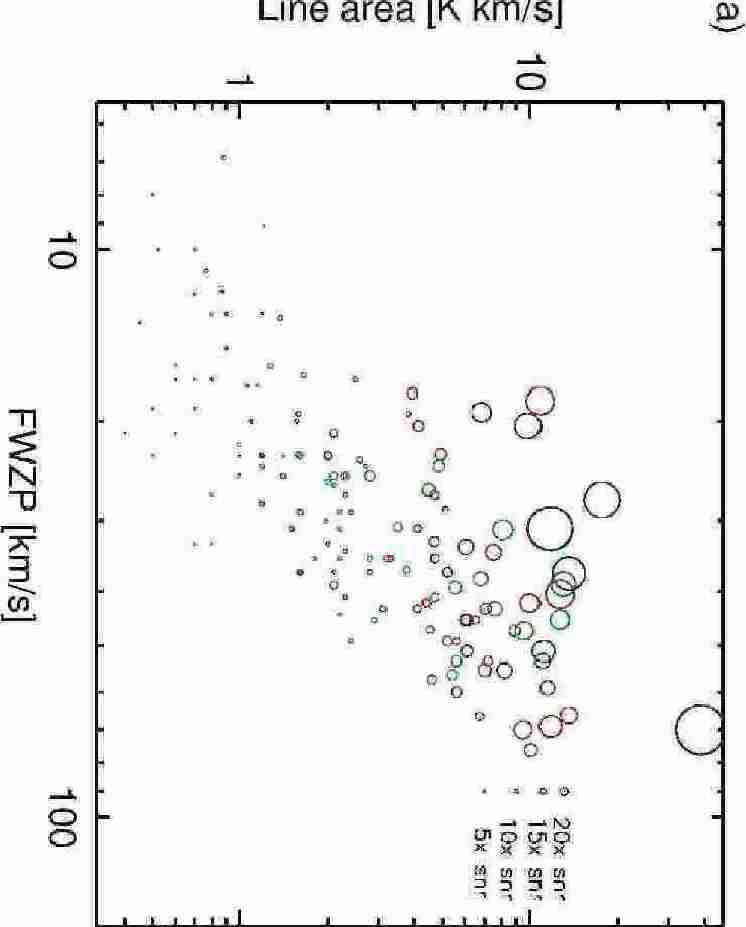}
\includegraphics[width=5cm, angle=90]{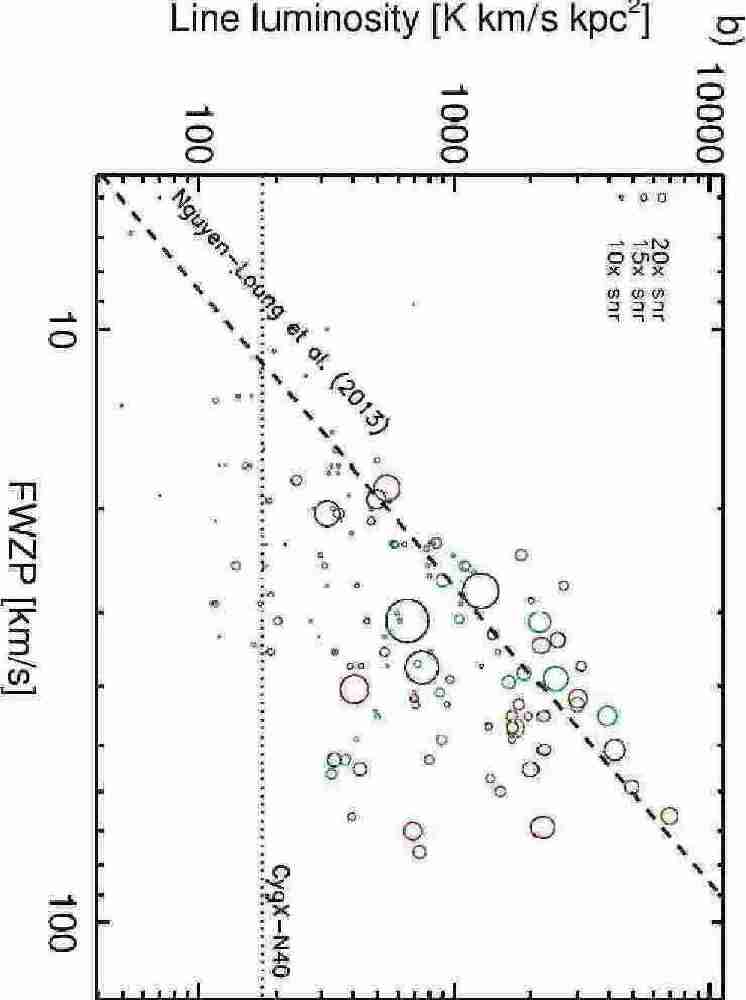}
\caption{Panel {\bf a)} shows the integrated line area as a function of
the $FWZP$ { for the distance- and mass-limited subsample}. 
The size of the open circles show the signal-to-noise of the line area. 
Blue circles show the clumps associated with {\hii} regions, red are associated with (M)YSOs, and green corresponds to the infrared-quiet clumps. Panel {\bf b)} shows the line
luminosity. 
The dashed line shows the $L_{\rm SiO(2-1)}=\alpha_{\rm L} \times FWZP_{\rm SiO(2-1)}-\beta_{\rm L}$ relation, where \citet{Quang2013} determine
 $\alpha_{\rm L}=144$, $\beta_{\rm L}=751$ for the
sample of objects in the W43 Main complex.
The SiO (2--1) line luminosity of the CygX-N40
MDC from \citet{Duarte-Cabral2014} is shown in dotted line.
}\label{fig:fwzp}
\end{figure}

In Fig.\,\ref{fig:fwzp} we show the SiO (2--1) line area and the line luminosity as a function
of the $FWZP$, following \citet{Quang2013}, for the distance- and mass-limited subsample. 
We find several sources that have high SiO luminosity 
comparable to the mean $L_{\rm SiO}\sim10^3$~K~\kms~kpc$^2$ 
found by \citet{Quang2013} towards the ridges of W43-Main. 
On panel b) we show their relations and find that most of our sources fall 
close to their fit. 
For comparison, we show two examples of SiO line luminosities
from shocks at low velocities,
which are possibly a fingerprint of small-scale converging flows (see also
\citealp{Csengeri2011a, Csengeri2011b}).
An example of a nearby massive dense core (Cyg-X N40) from 
\citet{Duarte-Cabral2014} exhibits $1.77\times10^2$~K~\kms~kpc$^2$
SiO (2--1), which is much lower compared to the more distant and massive ridge associated with W43-MM1.
Compared to the example of Cyg-X N40 
our sample probes on average more distant, thus 
larger scale structures, where higher line luminosities from low-velocity shocks are expected, as high 
as that of the W43 ridge.

\begin{figure}[!htpb]
\centering
\includegraphics[width=0.7\linewidth, angle=90]{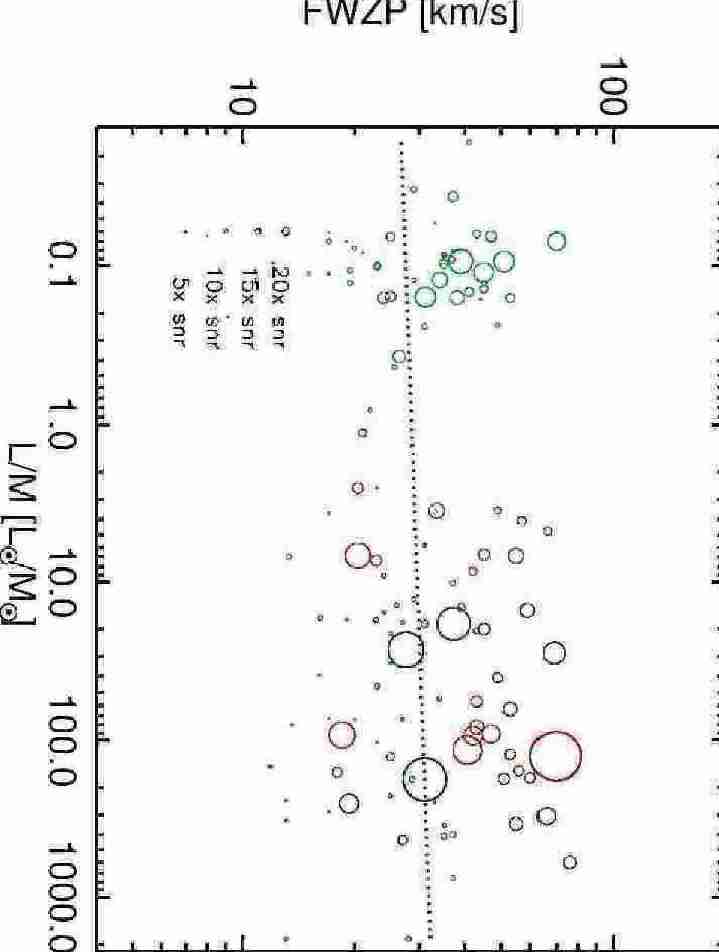}
\caption{
 $FWZP$ of SiO (2--1) emission for the most massive nearby ($1< \rm d<7$~kpc)
and massive ($\rm M>650$~\msol)
clumps. Dotted line shows a robust linear fit to the data points, which
results in a nearly constant relationship ($f(x)=1.5-0.01x$) 
between the $L_{\rm bol}/M$ and $FWZP$ 
quantities.
The colours are indicated in the legend.
}
\label{fig:width_comp}
\end{figure}

We correlated the velocity extent of the flow
of the distance-
and mass-limited sample with $L_{\rm bol}/M$,
but find no clear trend (Fig.\,\ref{fig:width_comp}), which is similar to what we qualitatively derived in Sect.\,\ref{sec:line_profiles}.
We find
no significant difference between the flow velocity associated with
infrared-bright and infrared-quiet clumps, 
only the dispersion of the observed velocities increases
with evolutionary stage. 
This could possibly hint 
at a more confused underlying
origin of the observed shocks 
driven by several protostars forming in a clustered environment.

\subsubsection{Two-component Gaussian fit}\label{sec:twocompfit}

To distinguish between the low- and high-velocity regimes, 
we use a two-component Gaussian fit: one centred on the ambient 
velocity with a narrower line width ($\Delta v\le8$~\kms, see below)
compared to the second component characterised 
by high-velocity line wings. 
These wings often show asymmetries
in which case the Gaussian fit does not reproduce  the line profile well.
For a statistical study, this approach is, however, a powerful tool that allows a systematic analysis, 
while this method is also
commonly used in the literature (e.g. \citealp{Duarte-Cabral2013}).

To obtain more robust results, for the two-component Gaussian fitting, we only
consider detections with a peak intensity above 
5$\sigma$. From the entire sample, 
76 sources can be fitted with two components. 
Only a broad component centred on the ambient velocity
was fitted to 34, while only a low-velocity component was fitted to 122 
sources. From these, 61 (50\%) are classified as infrared-quiet clumps. 
To exclude sensitivity limitation as a possible reason for the lack of 
detection of the high-velocity component, 
we checked the distance distribution of the sample with detection of
the high-velocity component and find a relatively 
homogenous distribution as a function of distance up to  
$\sim$14~kpc. 
{ Similarly, we find a homogenous distance distribution for the sources where
only the low-velocity component is detected.
This suggests that the lack of detection in the high-velocity regime
is rather the intrinsic property of the
shock in these sources.}

On average, we find $\sim$5--6~\kms\ $FWHM$ width for the 
low-velocity component centred at the ambient velocity. 
The high-velocity component is fitted on average with 
19~\kms\ towards the younger sources 
and we find that it decreases to 14~\kms\
for the more evolved clumps where the two components could be fitted. 
The values show, however, a large scatter
and do not show statistically robust trends, as also seen in
Sect\,\ref{sec:nofit}. 

The maximum $FWHM$ for the 
broad component is found to be 69~\kms\
with a mean of 18.4~\kms\ towards the entire sample.
The broad line profiles suggest already on-going star formation with significant
high-velocity gas likely associated with the jets emanating
from young protostars.

From the distance- and mass-limited subsample ,
34 sources show both a low- ($\Delta v\le8$~\kms) and a 
high-velocity component, while 25 are fitted with only a low-velocity component
roughly equally distributed in each category of sources. 
We find four sources that are fitted with only a high-velocity 
component ($\Delta v\sim14-20$~\kms).
As a comparison, 
our statistically derived line-width for the low-velocity regime is 
a factor of at most ten larger than found by \citet{Jimenez-Serra2010}
for spatially extended SiO emission (see Sect.\,\ref{sec:width} for more details), but are comparable to the findings
of \citet{Duarte-Cabral2014}. Our definition of narrow-velocity component is, however,
more conservative than \citet{Quang2013}.

\begin{figure}[!htpb]
\centering
\includegraphics[width=0.7\linewidth, angle=90]{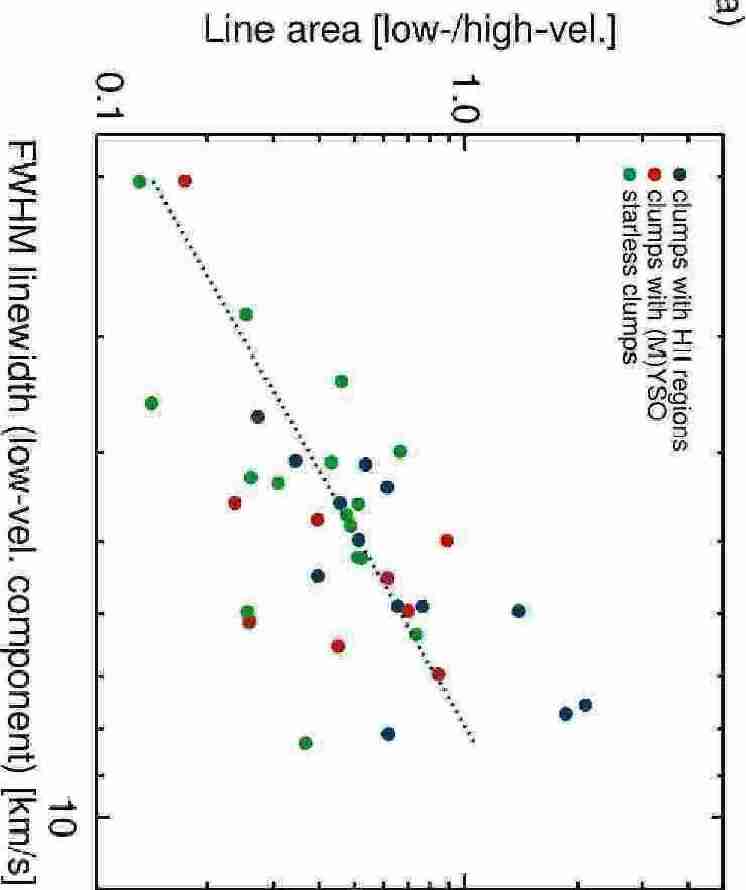}
\includegraphics[width=0.7\linewidth, angle=90]{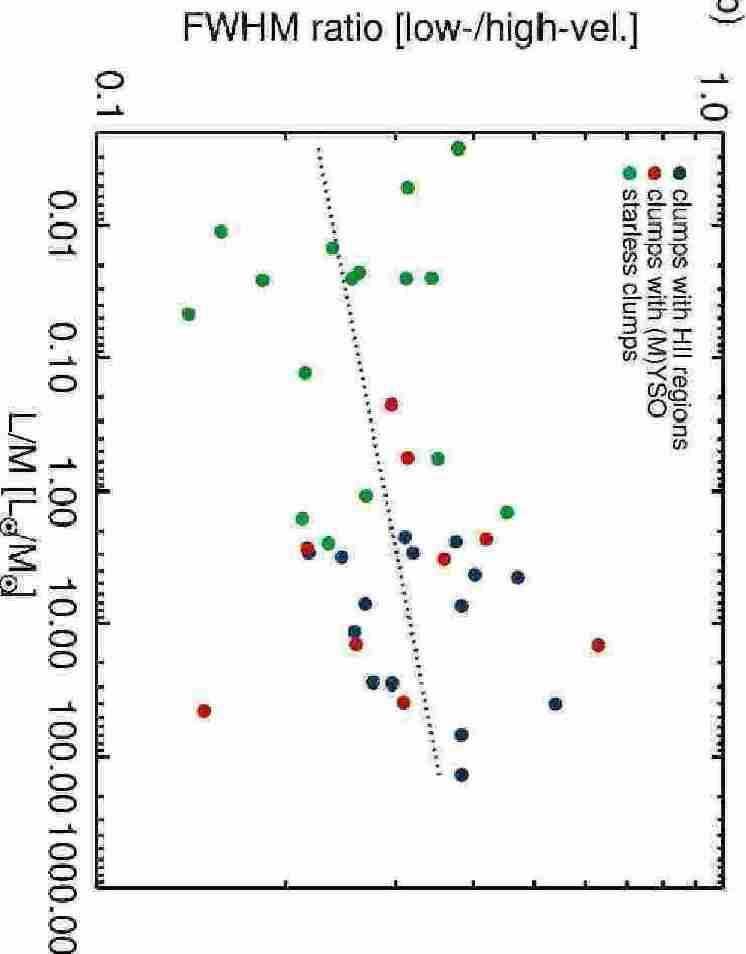}
\caption{
{
$FWHM$ width of the low-velocity Gaussian component versus the ratio of the line area 
of the low- and the high-velocity 
components of the distance-limited sample,} showing detections above $7\sigma$ ({\bf a}). 
The dotted line shows a robust linear fit ($f(x)=-1.3+1.4x$).
The $L_{\rm bol}/M$ ratio versus the ratio of the $FWHM$ of the narrow and 
broad velocity components {\bf b}). Dotted lines show
a robust linear fit to the data points ($f(x)=-0.5+0.04x$).
The colours are the same as in Fig.\,\ref{fig:fwzp}.}
\label{fig:width}
\end{figure}

To look for trends in the properties of the two components,
we investigate their ratio in Fig.\,\ref{fig:width} (upper panel)
towards the distance-limited subsample, where 
both components are detected with a high signal-to-noise ratio. 
We find a weak trend that the contribution of the high-velocity 
component decreases with an increasing width of the low-velocity component.
This may simply mean that  the distinction
between the two components becomes less clear towards these sources, or 
that the high-velocity component has a more important contribution in integrated emission
for the sources exhibiting a low-velocity component.
We find  three massive clumps (G10.62-0.38, G34.26+0.15, G49.49-0.39), 
where 
the high-velocity component may dominate the emission over the 
low-velocity component. 

In the lower panel of Fig.\,\ref{fig:width}, we show  the ratio of the 
$FWHM$ of the two components as a function of $L_{\rm bol}/M$. We find a weak trend
that the width of the high-velocity component decreases with $L_{\rm bol}/M$. 
This may reflect the evolution
of the shocks, where the high-velocity component is more associated with 
the jets emanating from the youngest protostars, although our results are statistically 
not robust enough to draw further conclusions.

Overall, we 
do not find robust trends for the properties of the
 line width as a function of evolution. This is further supported by 
Fig.\,\ref{fig:width_histo}, which shows the histogram of the
velocity range determined from the SiO ($2-1$) emission in
Sect.\,\ref{sec:nofit}. 
The peak of the distribution falls between
$10-50$~\kms, and
 there is no significant difference between 
 different source categories
confirming that high-velocity
shocks are present in all evolutionary stages.

To understand the origin of the SiO low-velocity, and relatively narrow component,
we correlated the line width with that of the N$_2$H$^+$ (1--0) transition
and found no correlation. This suggests that
the origin of the SiO narrow component is not related to the turbulent gas of the 
clump, and is likely to have a different origin.

Finally, we note that the detectability of the broad component depends not only
on the signal-to-noise of the spectra (and therefore its distance), 
but also on the inclination angle.  We have not considered this potential bias in the analysis
because in cluster forming clumps
the picture is even more complicated because of a distribution in number and strength
of outflows and their orientation. 



\section{SiO emission and excitation conditions in massive clumps}\label{sec:analysis}
In this section, using the two SiO transitions of (2--1) and (5--4),
we investigate the excitation conditions of the gas emitting these lines
and quantify the properties of the shocked gas towards this large sample
of massive clumps with well-characterised physical properties.
First, we derive the optical depth of the SiO (2--1) line (Sect.\,\ref{sec:coldens}). 
To estimate the SiO column densities, we first use a non-LTE approach for the subsample with 
observed 5--4 transitions
consisting of the most massive sources and the highest signal-to-noise detections
of the 2--1 line. Based on these results we then extend the column density estimates
to the full sample based on LTE assumption (Sect.\,\ref{sec:sio_21_coldens}). 
We then use these results to 
estimate SiO abundances (Sect.\,\ref{sec:sio_21_abundance}) 
and investigate possible trends in the different evolutionary phases.
However, the larger statistics and observations of the two transitions with an 
energy difference of $\sim 20$\,K allows us to go a step further than previous works
(e.g.\,\citealp{LS2011,SM2013,Leurini2014}). 
The ratio of the SiO ($5-4$) to SiO ($2-1$) line
is not affected by beam dilution under the assumption that the transitions originate in the same gas,
while other quantities such as the SiO column density and abundance are affected by crude assumptions
needed in the analysis of such a large sample (LTE conditions and a beam dilution factor of one, for example).
Therefore we base our discussion on the properties of SiO as a function of evolution in the trends derived from the ratio of the two lines (Sect.\,\ref{sec:area}).
\subsection{Optical depth}\label{sec:coldens}

Since the IRAM spectral survey (Sect.\,\ref{sec:obs-iram})
covers the ($2-1$) transition of the $^{29}$SiO line at 85.759~GHz,  
we use this information
to estimate the optical depth of the main isotopologue. 
Although the survey has not been designed for such weak lines,
 we still find  $\sim40$ detections above $3\sigma$. 
While this is  
a significantly lower detection rate compared to the main isotopologue,
it still provides useful upper limits of the optical depth of the main line. 
 
\begin{figure}[!htpb]
\centering
\includegraphics[width=6.5cm, angle=-90]{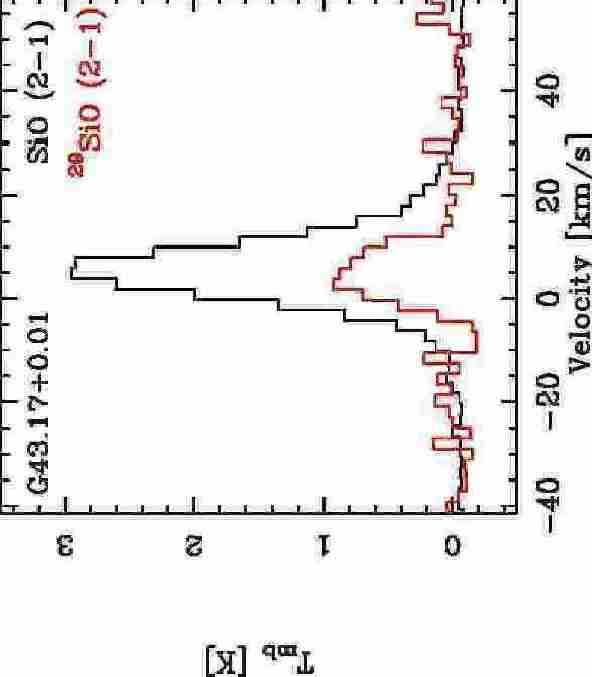}
\caption{Example of a source showing moderately optically thick emission
inferred from the intensity ratio of the $^{28}$SiO ($2-1$) and the  $^{29}$SiO lines
towards one of the brightest sources in the sample, known as W49N.
}\label{fig:tau}
\end{figure}

The terrestrial ratio of 
$X(^{28}$SiO)/$X(^{29}$SiO) is 19.6 and in the interstellar medium 
values between $10-20$ have been measured~\citep{Penzias1981},
hence, we adopt an isotopologue
ratio of 20. The ratio of the peak intensity of the two lines  
suggests that only in a few cases ($\sim8$, corresponding to $<20\%$, Fig.\,\ref{fig:tau}) the $^{28}$SiO ($2-1$) transition becomes saturated
suggesting  moderately optically thick emission. Therefore, we conclude
that towards
the majority of
the sample the emission is optically thin.

\subsection{SiO column density}\label{sec:sio_21_coldens}

Although the critical density of SiO transitions is high and
LTE conditions may not apply at typical conditions in shocks,
it is a commonly used approach to estimate the SiO column density assuming LTE.
(e.g.\,\citealp{SM2013,Quang2013}).
Based on the subsample 
of 112 sources
where both the ($2-1$) and ($5-4$) lines 
are detected (see Sect.\,\ref{sec:detrate_sio54}), in the following
we compare the column density estimates using a 
non-LTE (Sect\,\ref{sec:sio_21_coldens_nlte}) and LTE approach (Sect\,\ref{sec:sio_21_coldens_lte}).
We show that within an order of magnitude
the LTE approach gives a relatively good estimate for the column density 
considering the uncertainties such as the unknown source size.

\subsubsection{Column density estimates from RADEX}\label{sec:sio_21_coldens_nlte}
We first show column density estimates for the subsample of sources
where both the (2--1) and (5--4) transitions are detected with a high
signal-to-noise. We used RADEX~\citep{vanderTak2007} 
with a plane parallel slab geometry
to constrain the SiO column density from
the ratio of these two lines. 
We use these results to show the general trends of the
SiO column density and abundance in
non-LTE conditions (Sect.\,\ref{sec:sio_21_abundance}), and then use
the LTE approach to extend our statistics of column
density and abundance to 
the full sample.

\begin{figure}[!htpb]
\centering
\includegraphics[width=6.5cm, angle=90]{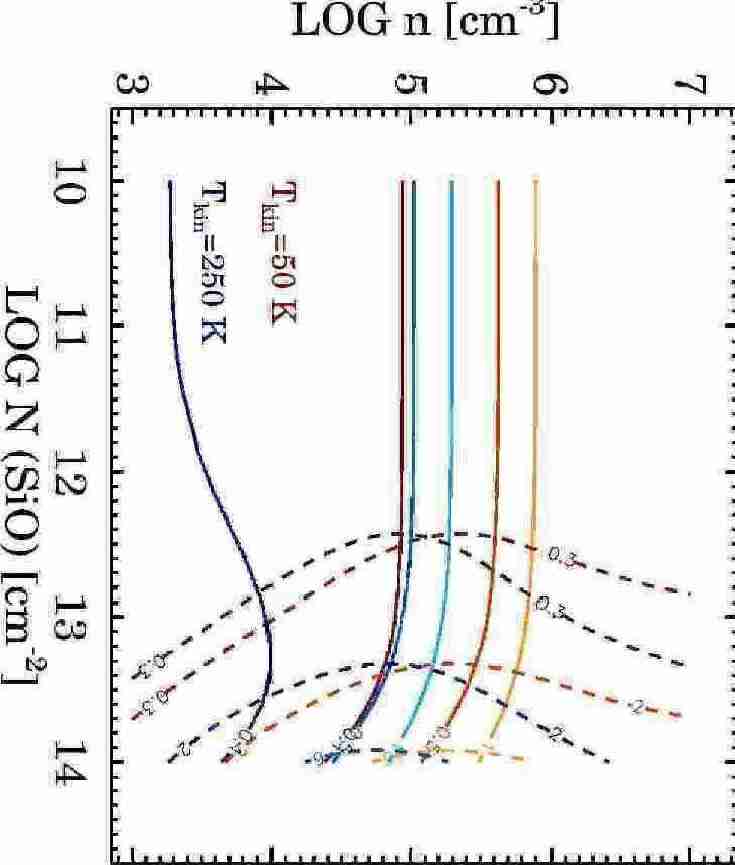}
\includegraphics[width=6.5cm, angle=90]{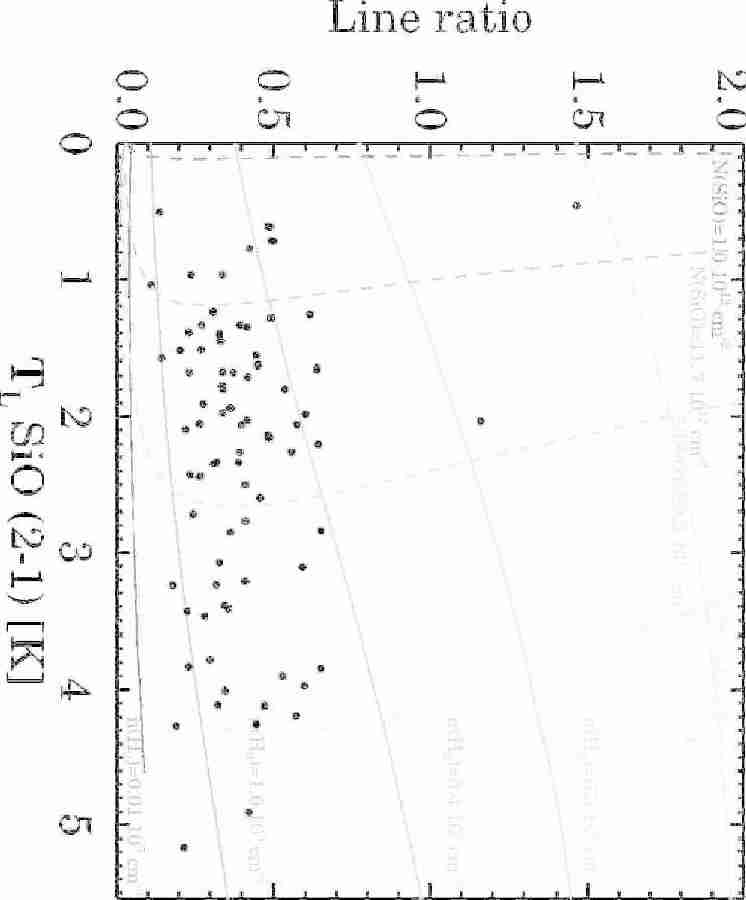}
\caption{{\bf Top:} Non-LTE calculations with RADEX of the ($5-4$) to ($2-1$) line ratios
for the typical conditions for $n(\rm H_2)$ and $N(\rm SiO)$ in massive clumps.
Solid lines show the estimated line ratios for $T_{\rm kin}=50$~K, and 250~K
in varying shades corresponding to different line area ratios,
which are comparable to the observed values.
Dashed lines show the estimated line temperatures for the (2--1) 
transition with a
beam filling factor of unity.
{\bf Bottom:}
The same calculations as above for the $T_{\rm kin}=50$~K
compared with the observed line temperatures for the (2--1)
transition and the line ratio, where both values are normalised to 
a line width of 5~\kms. Solid lines show the solutions for the given
values of H$_2$ density; dashed lines show the solutions
for fixed SiO column density.
}
\label{fig:ratio}
\end{figure}

We measure the line properties with a single-component Gaussian fit to both the
(5--4) and (2--1) lines, and then normalised the line area to
a typical width of 5~\kms\ (see Sect.\,\ref{sec:twocompfit})\footnote{In total, 225 sources are well fitted with the single Gaussian component corresponding to 75\% of the total detections, where the average line width is close to 5 km/s.}. 
Typical line ratios of the SiO (5--4) and (2--1) lines are found between
$0.1-1.5$, with a mean of 0.4. This is compared to non-LTE calculations with RADEX 
in Fig.\,\ref{fig:ratio}, which shows the
($5-4$) and ($2-1$) line ratio for a typical range of $n({\rm H_2})\sim10^3-
10^7$~cm$^{-3}$ and $N({\rm SiO})\sim10^{10}-10^{14}$~cm$^{-2}$
for $T_{\rm kin}=50$ and $250$~K.
We find typical values for $T_{\rm ex}=4$ to $10$~K (see also 
Sect.\,\ref{sec:sio_21_coldens_lte} and Appendix\,\ref{app:excitation}).
The upper panel shows
the observed values of the line ratios
in solid lines, and the line temperatures in dashed lines.

From the upper panel of Fig.\,\ref{fig:ratio}, 
we first establish that the different kinetic temperatures 
do not have a significant impact on the resulting column density, 
only a factor of up to 20-25\%. Therefore, in the subsequent analysis 
we rely on models with a single value of $T_{\rm kin}=50$~K.
We then estimated the column densities of SiO for each source with RADEX
using the ratio of the peak line intensity normalised to a common line width
of 5~\kms. Assuming a beam filling factor of unity, we find
$N({\rm SiO})$ between $9.6\times10^{11}-1.1\times10^{13}$\,cm$^{-2}$ 
(see Fig.\,\ref{fig:ratio}, lower panel). 
However,
as pointed out earlier, owing to the large dispersion in the distance
of the sample, beam dilution may play an important role.
We therefore tested this effect by assuming a common 8~\arcsec\ 
angular size for the emission at 1~kpc and scaled the measured
line temperatures with the distance of the source. This gives a factor
of few higher $N({\rm SiO})$ between $1.5\times10^{12}-1.4\times10^{14}$\,cm$^{-2}$.
In Fig.\,\ref{fig:sio21_abundance_radex}, we compare the result of this test with the SiO column density estimates without beam dilution.

In the lower panel of Fig.\,\ref{fig:ratio}, we compare
these calculations with the observed line parameters normalised
to a line width of 5~\kms. The plotted lines show
the model results for fixed column density and H$_2$ density for
$T_{\rm kin}=50$~K. The observed values correspond
to an H$_2$ density range of $\sim10^5$~cm$^{-3}$.

\subsubsection{Column density estimates from LTE}\label{sec:sio_21_coldens_lte}

To estimate column densities for the whole sample, we can only
rely on the ($2-1$) transition. 
We therefore have to assume LTE conditions, 
optically thin emission, and that the source fills
the main beam. 
An 
additional assumption is that the bulk emission originates
from the low-velocity component of the spectra (see also Sect.\,\ref{sec:twocompfit}), 
therefore,  including the high-velocity 
wings, we can still use a single excitation temperature ($T_{\rm ex}$).  
The total column density is then 
estimated by 
\begin{multline}\label{eq:cdens}
N_{\rm tot} = \frac{3\,k^2}{4\,\pi^3\,h\,\nu^2} \frac{1}{S\,\mu^2}\,T_{\rm ex} \,e^{\frac{E_u}{k\,T_{\rm ex}}}\,\int{T_{\rm mb}\,\rm dv}\,\frac{\tau}{1-e^{-\tau}} \simeq \\
\simeq 1.8\times10^{12}\,\int{T_{\rm mb}\,\rm dv} \rm\, [cm^{-2}],
\end{multline}
where $S=2$ is the statistical weight, $\rm \mu=3.1$~Debye is the electric dipole moment, $\nu=86.85$~GHz and we used $T_{\rm ex}=10$~K. 
To justify the choice of $T_{\rm ex}$ for the LTE calculations, 
we implicitly assumed optically thin emission for both transitions, 
which is a commonly used approximation 
and is valid in our large statistics as shown in Sect.\,\ref{sec:coldens}. 
First, based on the LTE approach and 
similar to the results of \citet{SM2013}, from the ratio of the
two transitions we find that the sources
have on average $T_{\rm ex}$ of $\sim 10$~K with 
extreme cases reaching $\sim 30$~K (Appendix\,\ref{app:excitation}). 
In the following, for the LTE calculations
we use a single value for the whole sample, 
since 
larger uncertainties are introduced by
relying on velocity and beam averaged estimates.
We adopt $T_{\rm ex}=10$~K for all sources,
which is similar to what is used by \citet{SM2013}, and 
also \citet{Leurini2014} show that 
this excitation temperature is a good assumption for the typical densities and
temperatures for such clumps.
Using $T_{\rm ex}=30$~K would increase
the estimated column densities by a factor of two,
while using a lower value of $T_{\rm ex}=5$~K,
we would only obtain less than ten per cent lower column densities.

We derive SiO column densities between 
$1.6\times10^{12}-7.9\times10^{13}$~cm$^{-2}$ for the whole sample
based on the total integrated intensity derived in Sect.\,\ref{sec:int}.
The highest column densities are observed towards the massive clumps
hosting {\uchii} regions, and we do not find significant differences between the
derived column densities for the infrared-quiet and infrared-bright clumps.

Comparing the column densities from the LTE assumption with the non-LTE RADEX
calculations with a beam filling factor of unity, 
we find that the LTE column densities are on average
a factor of two higher. In comparison with RADEX results, corrected for a source size and
different distances, we find that these estimates
are within a factor of eight, and the LTE
estimates give on average a factor of three lower column densities.
 Given the uncertainties in the assumptions,
these tests show that on average the column density estimates from
the LTE assumption are within an order of magnitude comparable
to that of the non-LTE models. However, systematic
biases may be introduced in large samples, and therefore this simple 
 analysis may give misleading results due to the combination of
non-LTE effects and varying source size, which may scale with the distance.

\subsection{SiO abundance}\label{sec:sio_21_abundance}

To estimate the SiO abundance,  
 we use the peak flux densities from \citet{csengeri2013} and estimate
 the H$_2$ column density from the dust
\begin{equation}\label{eq:dust_cdens}
N({\rm H_2})=\frac{F_\nu\,R}{B_\nu(T_d)\,\Omega\,\kappa_{\nu}\,\mu_{\rm H_2}\,m_{\rm H}},
\end{equation}
where {$F_\nu$ is the peak flux density}, $\Omega$  the beam solid angle, $\mu_{\rm H_2}$  
 the mean molecular weight of the interstellar medium
with respect to hydrogen molecules, which is equal to 2.8,
and $m_{\rm H}$ is the mass of an hydrogen atom. We adopt here the same assumptions as \citet{schuller2009}: a 
gas-to-dust mass ratio (R) of 100 and  $\kappa_{\nu}=1.85$ cm$^2$~g$^{-1}$, which is interpolated to 870~$\mu$m from Table 1, 
Col.\,5 of \citet{OH1994}. 
A more precise estimation of the H$_2$ column density would be given
 using a different molecular tracer with a similar line profile and a 
well-constrained abundance with respect to H$_2$.
Other species, such as HCO$^+$ \citep{SM2013}, $^{13}$CO ,  
or high-$J$ CO lines \citep{Leurini2014} have been commonly used to
estimate the H$_2$ column density in the high-velocity regime corresponding
to the line wings. In our survey the signal-to-noise ratio in the wings of 
the $^{13}$CO and HCO$^+$ ($1-0$) lines is 
typically low, therefore, we do not attempt  to estimate
the $\rm H_2$, and SiO column densities strictly arising from the high-velocity
wings, but estimate the velocity integrated column density. To test this, we
selected the H$^{13}$CO$^+$ ($1-0$) line
at 86.754~GHz from the line survey (see Sect.\,\ref{sec:obs-iram}),
which is expected to be an optically thin tracer of the gas.
Given the variation in the line profiles we selected
a handful of sources and estimated the H$_2$ column density 
from the  integrated emission in the same velocity range as used for the SiO line,
and using the formula from \citet{Schneider2010}.
This results at the order of magnitude similar
H$_2$ column density values based on the dust, 
between $1.4\times10^{-11}$ to $1.9\times10^{-10}$.
For the whole sample we therefore 
only provide abundance estimates based on the dust emission,
and point out that other studies, such as 
\citet{Miettinen2006, Sanhueza2013, Gerner2014}, also rely on the dust to estimate
$N(\rm H_2)$.
{ As SiO may not strictly follow the distribution of dust,}
our results of abundance estimates are lower limits
from which local deviations within the beam may be large.

\begin{figure}[!htpb]
\centering
\includegraphics[width=6.5cm, angle=90]{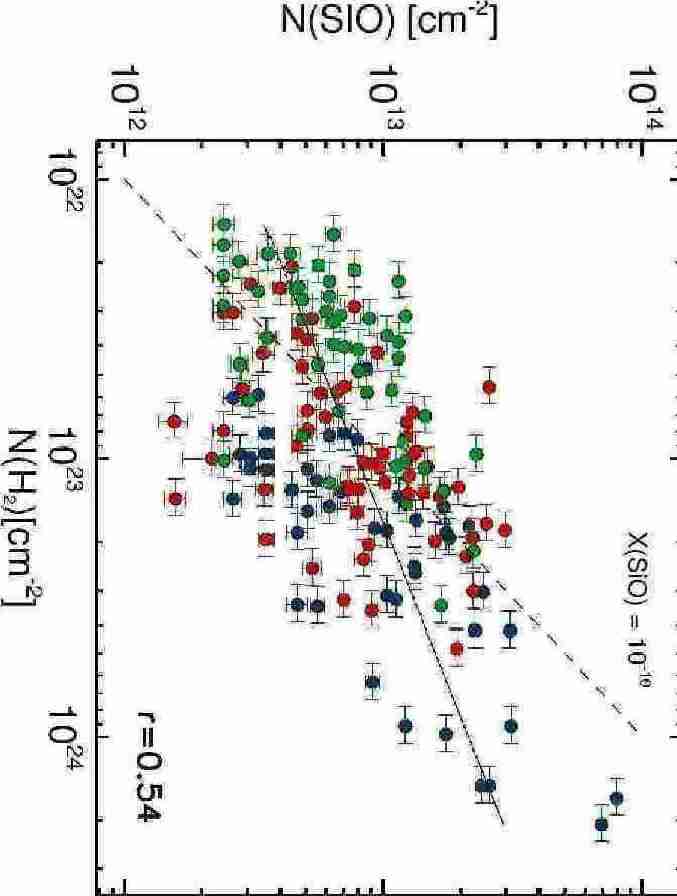}
\caption{Column density of SiO estimated from the (2--1) transition for sources with { 
$>10\sigma$} detection versus column density derived from the 870~\mum\ integrated flux density from \citet{csengeri2013}.
The Spearman rank correlation coefficient, $r$, is shown on the panel. In total
155 sources are shown. Sources classified as clumps with embedded {\hii} regions are
shown in blue, red circles show the infrared-bright clumps, while green corresponds to the infrared-quiet clumps. The black line shows 
a linear fit to the data in log-log space. Dashed line
corresponds to an SiO abundance of $10^{-10}$.
}\label{fig:sio21_abundance}
\end{figure}

First, in Fig.\,\ref{fig:sio21_abundance} we show the SiO column density
calculated based on the LTE approximation from the ($2-1$) transition in Sect\,\ref{sec:sio_21_coldens_lte}
versus the H$_2$ column density ($N_{\rm H_2}$) calculated from the 
870~\mum\ integrated flux densities from the \at\ survey,  
assuming a dust temperature (T$_d$) of 18~K\footnote{
We use the equations of \citet{schuller2009} and \citet{csengeri2013}.}. 
We perform a Spearman-rank correlation test to see 
if there is a monotonic relationship between the two variables. 
The rank correlation coefficient ($r$) indicates the strength of the correlation and we
also determine its significance level. 
We find a significant correlation between the SiO and H$_2$ column density
with a high significance (>99\%) for sources above a $10~\sigma$
detection. We performed the same tests for the distance- and mass-limited subsample
and find a very similar correlation.
More distant sources may be too small in angular scale to fill the beam, while 
the dust emission is always found to be extended, therefore, a different beam dilution 
may affect the more distant sources. To account for this, we have also performed
a partial Spearman-rank correlation test against the distance 
and found practically the same correlation coefficients, suggesting that distance bias does
not influence the determined correlations. 
{
This result is in agreement with our findings in Sect.\,\ref{sec:width}, where we did not
find any difference in the full and the distance-limited sample.
}
We performed the same test for the subsample with non-LTE column density estimates
and find a similar trend, but with
a weaker correlation coefficient (see also Sect.\,\ref{app:ls}). 
This suggests a correlation with an
 increasing trend between $N({\rm SiO})$ and $N({\rm H_2})$ in our sample.

Despite the uncertainties in the SiO column density estimations, 
we estimate here a beam and velocity averaged SiO abundance  
from the column density calculated from the SiO ($2-1$) transition above,
and the $\rm H_2$ column density of the clump derived from dust emission
using $X(\rm{SiO})$$ = N_{\rm SiO}/ N_{\rm H_2}$.
The estimated SiO abundances do not exhibit large variations in the sample,
ranging between $1.1\times10^{-11}-4.9\times10^{-10}$ 
 with a median of $9\times10^{-11}$. 
 This compares well with other studies in the literature,
which find variations in the SiO abundance by several orders
of magnitude between $10^{-12}$ to $10^{-7}$ (e.g.\,\citealp{Bachiller1997,Garay1998,
Codella2005,Nisini2007,Gerner2014}). 
 In contrast to the
 column densities, this simple analysis suggests higher SiO
 abundance towards the infrared-quiet clumps, and similarly high
abundances towards the infrared-bright sample. 
The lowest SiO abundance is seen towards 
 the clumps associated with {\hii} regions. These results are in agreement with
 the findings of \citet{Gerner2014}.
 We caution again that we derive here beam averaged abundances, 
 and local deviations from the reported values can be very large.
Because of the uncertainties of the abundance estimations,
we investigate this in more detail in the next section.

\begin{table*}
\centering
\caption{Column densities and abundances estimated from the LTE analysis SiO (2--1) line towards the whole sample.}\label{tab:table}
\begin{tabular}{rrrrrrrrrrrr}
\hline\hline
Source groups & Min. N(SiO)  & Max. N(SiO) & Median N(SiO) &  Average N(SiO) & Average X(SiO)  \\
  & [cm$^{-2}$]  & [cm$^{-2}$] & [cm$^{-2}$] & [cm$^{-2}$] &   \\
\hline
All sources &      $1.6\times10^{12}$  &  $7.9\times10^{13}$   & $7.4\times10^{12}$  & $1.0\times10^{13}$  & $1.1\times10^{-10}$  \\
{\hii} regions &   $2.6\times10^{12}$  &  $7.9\times10^{13}$   & $9.1\times10^{12}$   & $1.4\times10^{13}$  &  $5.2\times10^{-11}$ \\
infrared-bright sources & $1.6\times10^{12}$   & $3.0\times10^{13}$ &   $7.9\times10^{12}$  & $9.6\times10^{12}$ &  $1.0\times10^{-10}$ \\
infrared-quiet sources & $2.4\times10^{12}$ &  $2.3\times10^{13}$  & $6.4\times10^{12}$ &  $8.0\times10^{12}$  & $1.8\times10^{-10}$ \\
\hline
\end{tabular}
\end{table*}

\subsection{Variation in SiO emission as a function of $L_{\rm bol}/M$}\label{sec:area}

Based on a sample of 47 massive clumps,
the study of \citet{LS2011}
 revealed a correlation between
the SiO (2--1) line luminosity of clumps and their evolutionary stage  
(see also \citealp{M07}).
In particular, they find a decrease between $L_{\rm SiO}/L_{\rm bol}$
with $L_{\rm bol}/M$, that is interpreted as a decrease in the SiO outflow
energetics or a decrease of SiO abundance with time.
To further test this scenario, \citet{SM2013} extended this study
 using small maps and  the SiO ($5-4$) transition and
confirmed the previous findings of \citet{LS2011}.
 
Compared to these previous studies,  we present a richer 
statistics in terms of the number of
observed sources but also covering a much larger
range of $L_{\rm bol}/M$. Furthermore, we attempt 
to provide column density estimates with both non-LTE and LTE
methods. As a comparison, in Appendix\,\ref{app:ls} we show the same 
correlations as \citet{LS2011}, which reveals
a strong correlation between $L_{\rm SiO}/L_{\rm bol}$
and $L_{\rm bol}/M$,
similar to the study of \citet{SM2013}.
These parameters are, however, not
independent as $L_{b\rm ol}$ appears on both
axes. Therefore, in the following we base our analysis on correlations
of independent variables to obtain statistically robust trends. 

\begin{figure}[!htpb]
\centering
\includegraphics[width=6.5cm, angle=90]{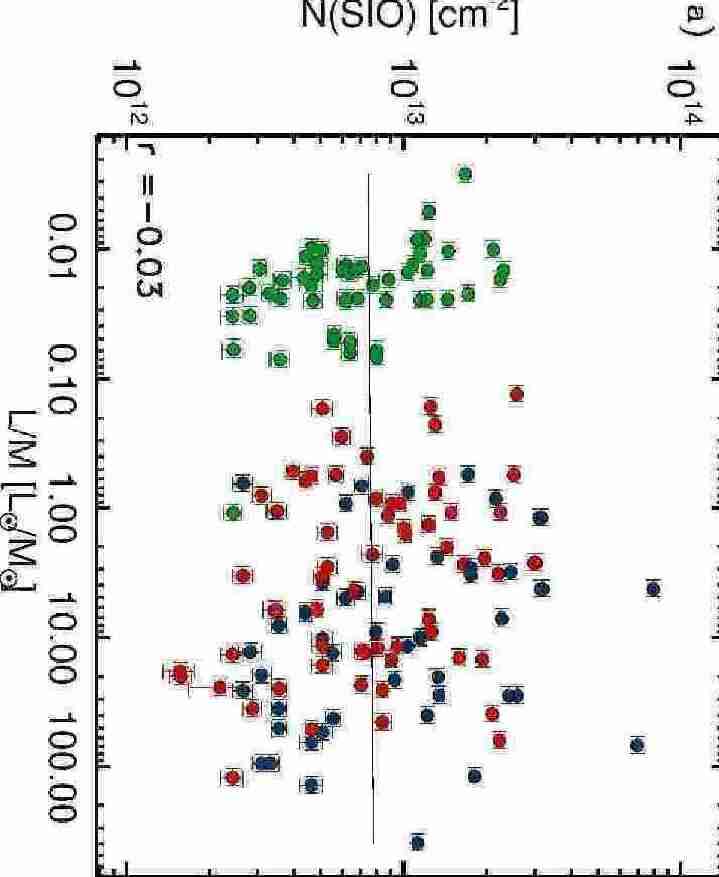}
 \includegraphics[width=6.5cm, angle=90]{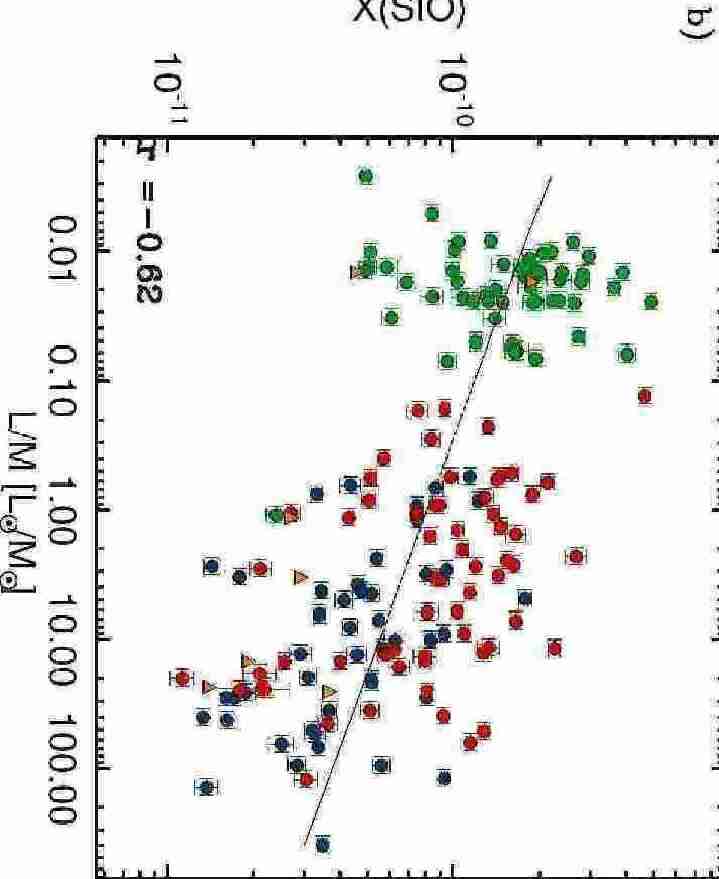}
\caption{{\bf a)} SiO column density estimated from the LTE assumption and the 
($2-1$) transition versus 
 $L_{\rm bol}/M$.
 {\bf b)} SiO abundance for the same sample versus 
$L_{\rm bol}/M$.
{
Triangles show abundance estimates for a few sources where the H$^{13}$CO$^+$
1--0 line could be used as an alternative estimate for H$_2$ column density. These 
points follow the same trend as the abundance estimate for the entire sample based on
dust. 
}
}
\label{fig:sio21_lm_column}
\end{figure}

We first use our SiO column density estimation and correlate it with
$L_{\rm bol}/M$ in Fig.\,\ref{fig:sio21_lm_column} (panel a), where 
$N(\rm SiO)$ is an LTE estimate from the (2--1) transition calculated using the total integrated intensity of the line. 
We find no correlation between
the two variables and the same holds for the subsample of sources analysed with RADEX.
On panel b we also show the LTE-estimated  
abundances as a function of $L_{\rm bol}/M$,
which in turn show a correlation with 
a coefficient of $r=-0.62$. 
This negative correlation is a robust, statistically significant result 
with a significance of $p\sim0.0001$. To derive the abundances, we carried out
a similar test, assuming a uniform, 8\arcsec\ source size at 1~kpc
and derive a significantly lower correlation coefficient of $r=-0.3$.  
Abundance estimates based on non-LTE calculations similarly show a 
much weaker correlation (see also Appendix\,\ref{app:ls}, Fig.\,\ref{fig:sio21_abundance_radex}). 
Although the first correlation is consistent with \citet{LS2011} and \citet{SM2013}, 
who also interpret their results as a decreasing 
trend between $X(\rm SiO)$ versus $L_{\rm bol}/M$,
our additional tests suggest that both non-LTE effects and beam filling
at various distances decreased the correlation between these parameters.

The results of these tests of abundance estimations are similar to that obtained by
\citet{Leurini2014}, 
who used a more reliable estimation of the SiO abundance
based on the SiO ($5-4$) and $(8-7)$ lines and on a dust independent estimate of $N(\rm H_2)$; they
 do not find a correlation between $X(\rm SiO)$ and $L_{\rm bol}/M$.
That study, however, focuses on a much smaller sample and consequently covers a 
considerably smaller range of $L_{\rm bol}/M$, covering a range of
 only a factor of 50.

\begin{figure}[!htpb]
\centering
\includegraphics[width=6.5cm, angle=90]{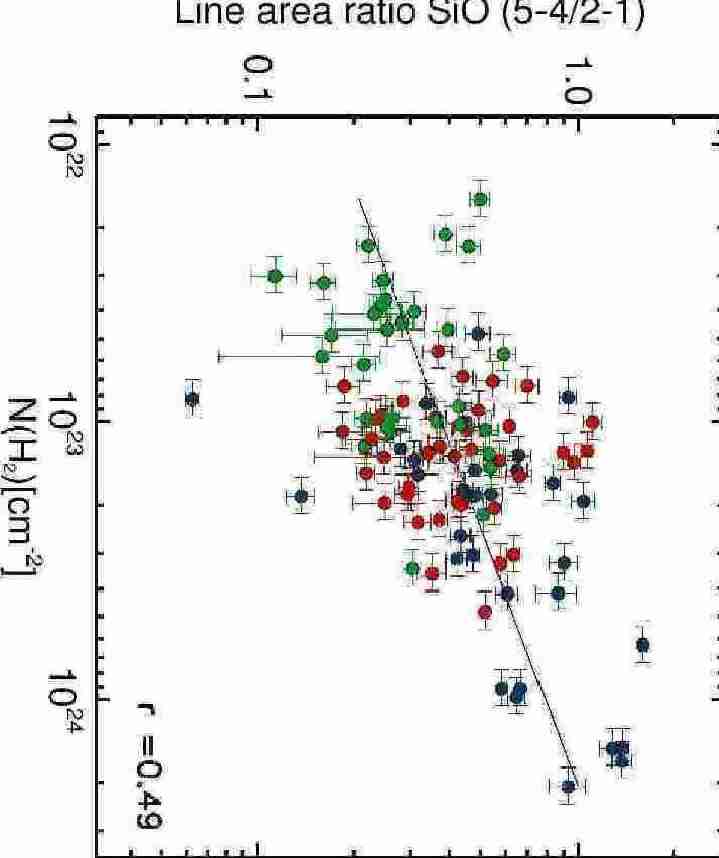}
\caption{Same as Fig.\,\ref{fig:sio21_abundance} for the SiO ($J= 5-4$)/($2-1$) ratio. The error is estimated to be 15\% for N(H$_2$), while for the line ratio the error is calculated from the noise in the two lines: $\sigma_{5-4/2-1}=area_{5-4}/area_{2-1}\,\sqrt{(\sigma_{5-4}/area_{5-4})^2+(\sigma_{2-1}/area_{2-1})^2}$. Altogether 100 sources are shown on the plot. 
Sources classified as clumps with embedded {\hii} regions are
shown in blue, red circles show the infrared-bright clumps, while green corresponds to the infrared-quiet clumps.
}\label{fig:area}
\end{figure}

\begin{figure}[!htpb]
\centering
\includegraphics[width=6.5cm, angle=90]{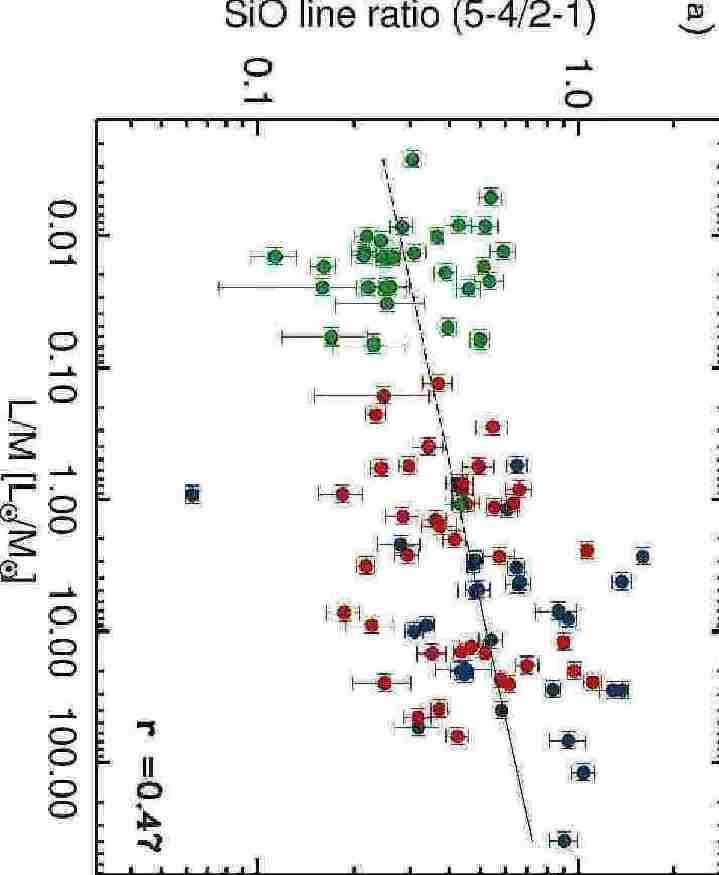}
\includegraphics[width=6.5cm, angle=90]{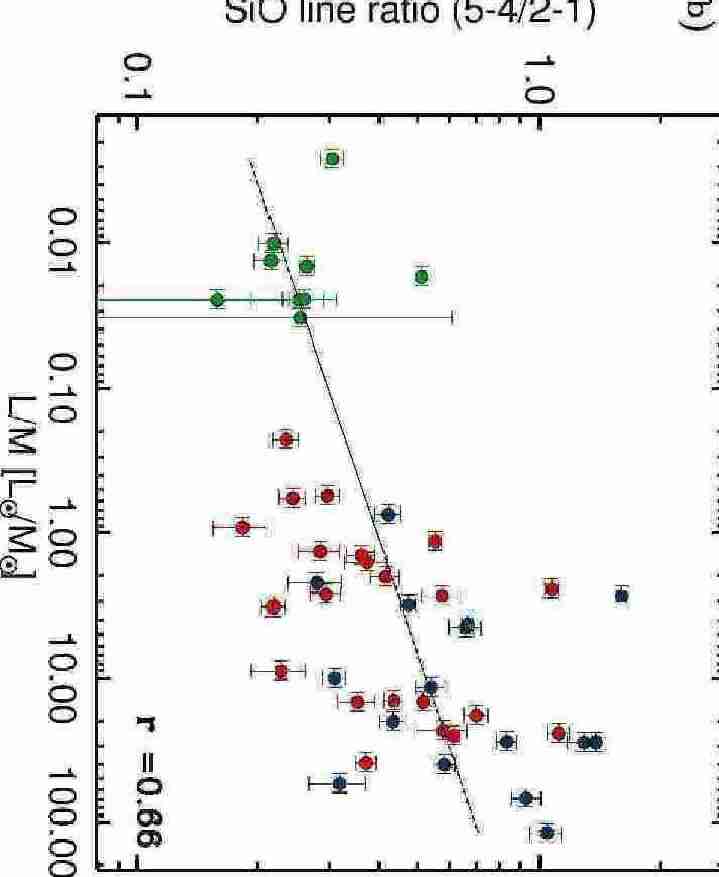}
\caption{Ratio of the integrated intensity of the SiO (5--4) and (2--1) transitions as a function
of $L_{\rm bol}/M$. { a)} All sources with $>5\sigma$ detection in both lines, altogether 100 sources. { b)} A subsample limited in distance and mass.
}\label{fig:sio21_lm}
\end{figure}

As pointed out earlier, 
more robust results are obtained when using 
the ratio between the SiO ($5-4$) to ($2-1$) 
transition to probe excitation effects.
Since the APEX and IRAM~30m telescopes have the same beam size at these frequencies, uncertainties due 
to the beam filling and source size affect both lines the same way, assuming that they originate from the same gas. 
Therefore, we investigate 
the change in the ratio of the integrated intensity of the two transitions 
as a function of  ${\rm H_2}$
column density (see Fig.\,\ref{fig:area}) and $L_{\rm bol}/M$ (Fig.\,\ref{fig:sio21_lm}).

In Fig.\,\ref{fig:area} the colour coding of the different classes suggests 
that a significantly higher ratio is measured
towards more evolved sources with higher H$_2$ column
density, suggesting that the emission in the $(5-4)$ transition is increasing 
as a function of H$_2$ column
density.
We derive
a correlation coefficient of 0.49 with
a significance of $p<0.001$, 
corresponding to the probability that
a random distribution would reproduce the observed coefficient. 
This suggests a moderate, but robust correlation. 
The correlation coefficient becomes slightly higher when considering only the
distance- and mass-limited sample.

Considering the highest
signal-to-noise detections ($>5\sigma$) we find
a robust, statistically significant positive correlation with $L_{\rm bol}/M$ ($r=0.47$) (Fig.\,\ref{fig:sio21_lm}).
This correlation gets considerably stronger ($r=0.66$) 
when taking the distance- and mass-limited subsample. 
The increasing trend of line ratios with the distance independent 
measure of age 
clearly points towards an evolutionary trend of changing excitation conditions
favouring the excitation of the 5--4 transition.

\section{Discussion}\label{sec:discussion}

We obtained SiO emission in two transitions (J$_{up}=2, 5$) 
with $>25$~K difference in energy
levels of a statistically significant sample of 
massive clumps in various evolutionary stages.
In the previous sections, we derived the statistical properties of this sample
in terms of detection rates, line profiles, 
SiO column densities, abundances, and line ratios of the two transitions as a function
of evolutionary stages. Here we discuss the implications of 
these statistical results in light of what is currently known about
the astrochemical origin of the SiO molecule (Sect.\,\ref{sec:sio_origin}), and then we report on the most recent observational results 
(Sect.\,\ref{sec:sio_shocks})
{ and compare our results to 
the reported evolutionary trends in the literature }
(Sect.\,\ref{sec:sio_lit}). Finally, we discuss the implication
of the observed trends in SiO chemistry as a function of
evolution (Sect.\,\ref{sec:shocks}).

\subsection{How well do we understand the origin of SiO emission?}
\label{sec:sio_origin}

After many years of theoretical and observational studies, the
astrochemical origin of SiO in the gas phase is still not fully constrained,
mainly because of a poor understanding of the Si chemistry
(e.g.\,\citealp{Mackay1996, Walmsley1999}). 
SiO is efficiently formed from Si atoms in the gas phase  
via the reactions of
$\rm Si + O_2 \rightarrow SiO + O$ and $\rm Si + OH \rightarrow SiO + H$~(e.g.\,\citealp{Hartquist1980,Herbst1989}). 
Theoretical studies conclude that, in fact, most of the gas-phase Si
ends up in SiO \citep{Herbst1989}, hence, it is a good tracer
 of Si in the gas phase.
{
Si atoms may be brought to the gas phase by various
physical processes, such as } dust destruction 
by sputtering~\citep{Schilke1997} or  direct photodesorption
\citep{Walmsley1999}, the latter of which  is a very inefficient process. 
In shocks, other processes, such as vaporisation
and shattering from grain-grain collisions,  
may also bring Si and SiO directly to the gas phase \citep{Guillet2009,Guillet2011,Anderl2013}. 
These processes are more important when the density is above 
$n_{\rm H_2}\geq10^4$~cm$^{-3}$, which is typical for shocks in dense clumps. 
Alternatively, SiO may also be directly present mixed with the grain
mantel ices and get released to the gas phase via 
sputtering (e.g.\,\citealp{Jimenez-Serra2009}).

From an observational point of view, SiO has been detected
in various astrophysical conditions with abundances of $\sim10^{-11}-10^{-10}$
towards translucent clouds~(e.g.\,\citealp{Turner1998}) 
in the Orion Bar PDR~(e.g.\,\citealp{Schilke2001}) with $\sim10^{-11}$.
In nearby dense clouds, however, the SiO abundance is very low with upper
limits of $<10^{-12}$ (e.g.\,\citealp{Ziurys1989}). While it generally shows  a low
abundance in the cold dense gas, in high-velocity gas from shocks
($>25$\,\kms) the SiO abundance increases by orders of magnitudes up to $10^{-7}$
(e.g.\,\citealp{Neufeld1989, Caselli1997, Schilke1997}, for theory; and, \citealp{ Bachiller1997,Garay1998,Nisini2007}, for observations).
Recent observations report
low-velocity SiO emission towards 
IRDCs~\citep{Beuther2007}, 
high-mass,
star-forming clumps~\citep{Miettinen2006, Sanhueza2013}, 
and widespread, low-velocity emission towards
IRDCs \citep{Jimenez-Serra2010} as well as 
 young massive ridges~\citep{Quang2013},
where typical abundances of $\sim10^{-11}-10^{-10}$
have been reported.

While classical shock models explain 
the observed high abundances and the high-velocity emission ($\sim10^{-9}-10^{-7}$ ) by 
fast ($v_{\rm s}>25$~\kms) shocks, 
the observed abundances
of $\sim10^{-11}$ in the low-velocity component pose more of a challenge to models.
Having SiO only in the grain cores,
fast shock velocities of $v_{\rm s}\sim30$~\kms\
are needed to produce the observed abundances.
However, \citet{Jimenez-Serra2008} show that 
an Si abundance as low as 10$^{-4}$
with respect to water in the mantles 
can produce narrow, low-velocity SiO line profiles with abundances of $10^{-11} - 10^{-10}$
by sputtering with shock velocities as low as $v_{\rm s}=12$~\kms.
Towards massive ridges other authors find that a larger fraction,  
up to 10\% \citep{Quang2013,Anderl2013}, of SiO in the
grain mantels is needed to explain its bright and extended emission.

Alternatively, photo-evaporation by an external UV field
could also lead to SiO emission at the ambient velocities \citep{Schilke2001}.
We find no clear evidence for a PDR origin for these detections because it
is mostly observed towards the youngest, infrared-quiet category sources without
a strong external UV field, and we find no correlation between the velocity extent of the SiO 
emission and tracers of the ambient gas. This suggests an upper limit 
on the amount of Si to be present in the grain mantle ices.
To unambiguously show that low-velocity
shocks releasing ice mantles are the origin of narrow, {low-velocity} SiO
emission, SiO would be expected to show a correlation with bona-fide
ice tracers, such as water (e.g.\,\citealp{vanDishoeck2011})
and methanol \citep{Menten1986,vanderTak2000}, which are beyond the scope of the current paper.

Based on these considerations and comparing the typical abundances
estimated
 for the sample, we attribute the high-velocity line component 
to fast shocks related to outflows driven by embedded protostars.
We detect, however, a large number of sources with only the low-velocity component
with similar abundances as derived from other studies of massive clumps~\citep{Miettinen2006, Sanhueza2013}. 
Towards low-mass protostars, the line profile and abundance of the
low-velocity component of the SiO emission is
found to be consistent with a magnetic precursor of C-type shocks
\citep{Jimenez-Serra2004, Jimenez-Serra2009}. 
However, to explain the extended low-velocity SiO emission reported
towards massive clumps, slow shocks related to cloud formation
processes through collision are invoked
(e.g.\,\citealp{Jimenez-Serra2010,Kauffmann2013,Quang2013,Miettinen2014}).
Alternatively, a population of outflows from low-mass stars, or
decelerated gas from high-velocity shocks,
could also explain the narrow, low-velocity SiO emission 
peaking at ambient velocities \citep{Lefloch1998, Jimenez-Serra2010}.
To put in context our observations with these studies, 
the low-velocity component of our sample
exhibits similar, or a factor of at most two larger line widths 
compared to the high angular resolution observations of
\citet{Duarte-Cabral2014}, 
but a factor of ten broader as typically observed
in the vicinity of low-mass protostars
and in the IRDC G035.39-00.33 by \citet{Lefloch1998,Jimenez-Serra2004,Jimenez-Serra2010}.
The low-velocity component is, however, smaller compared to the observations of \citet{Quang2013}
of the W43~Main star-forming complex, where its origin is explained
by large-scale cloud collision.

\subsection{SiO as a tracer of protostellar activity and shocks}\label{sec:sio_shocks}

Our sample is large enough to address the characteristic properties of SiO emission
in massive clumps as well as to put constraints on evolutionary trends of these
shocks. 
The high-velocity wings are excellent probes of 
fast shocks associated with protostellar jets~(e.g.\,\citealp{Schilke1997}). 
Our finding of a high detection rate of the broad component 
 in the SiO ($2-1$) line suggests the presence of fast shocks.
Line wings are also detected in the most evolved stages of massive clumps,
where {\uchii} regions are already formed. 
Although infall with high accretion rates 
has been reported towards these types of sources
(e.g.\,\citealp{Klaassen2012}), 
the detection of the broad outflow component suggests that there is 
accretion ongoing not only from the envelope, but also onto the (proto)stars leading to jets.
Considering that most of our clumps  eventually form not only a single star,
but a protocluster, these 
jets may originate from younger protostars nearby {\uchii} regions, which have
been frequently observed with high angular resolution data (e.g.\,\citealp{Leurini2009, Csengeri2011b}).
If so, the observed omni-presence of the broad component implies a continuous
star formation within massive protoclusters. 
Clearly, high angular resolution studies would be required
to pinpoint the origin of jet activity and its relation to the {\uchii} region.

We also find a large fraction of the infrared-quiet clumps exhibiting a broad
velocity component. Similar findings have been reported 
from case studies, such as a sample of massive dense cores in Cygnus-X by \citet{M07},
who reveal particularly bright SiO (2--1) emission towards their infrared-quiet sample.
Here we also find some extreme examples, which show the broad component 
despite being classified as infrared-quiet sources, and being completely dark at 22~\mum.
The detection of fast-shocks towards these sources suggests that they host deeply
embedded Class~0-like protostars \citep{Bontemps2010,Duarte-Cabral2013},
and also imply that molecular tracers are more efficient tools to determine the level of 
star formation activity than the infrared colour criteria due to large extinction.

Other detections towards the infrared-quiet class originate from
a low-velocity component SiO emission.
The non-detection of the wings may be due to sensitivity limitations in the survey.
However, since SiO is depleted towards low-mass dense cores { (with typical
abundances of $<10^{-12}$, \citealp{Ziurys1989}), 
it is intriguing that  SiO is detected towards massive clumps with at least  
an order of magnitude} higher abundance (see also e.g.\,\citealp{Miettinen2006,Sanhueza2013,Gerner2014}).
More recent models
explain its presence by low-velocity shocks with $v_{\rm s}<25$~\kms
 \citep{Jimenez-Serra2010,Quang2013}
due to cloud-cloud collisions, which are expected to lead to a narrow ($<10$~\kms),
low-velocity component at the ambient cloud velocity.
As also discussed in Sect.\,\ref{sec:sio_origin},
the most likely origin for this emission is related to low-velocity shocks
and cloud formation processes. If so, 
only a low-velocity SiO emission
towards massive clumps lacking any star formation activity
may be an indicator of youth and could pinpoint the earliest, likely prestellar phase of their evolution.

In particular,
high angular resolution studies by \citet{Duarte-Cabral2014} 
show that up to 90\% of the SiO emission could be associated
with the ambient gas, produced by cloud-cloud collision at the origin of dense 
structures, also claimed by \citet{Csengeri2011b}. 
They find SiO (2--1) line luminosities of the order of $\sim1-4.6\times10^3\,L_{\rm SiO}$
[K km/s kpc$^2$] for the narrow, low-velocity component at the ambient velocity 
with a $FWHM < 6$~\kms.
We indeed find several sources with similar line luminosity, however, they
exhibit on average larger line-widths up to 10~\kms.
However, since our
observations provide beam-averaged information on the shocked gas
in a forming cluster, we are not able to disentangle the contribution of the 
low-velocity component produced by the
high-velocity outflow shocks and only 
low-velocity shocks not related to outflow activity.

Shock models predict a rapid decrease of the SiO intensity and the line width
as the shock passed~\citep{Gusdorf2008a}. This is also supported by observations
in a sample of low- to high-mass cores in \citet{Codella1999}, as well as 
observations in a sample of massive clumps by \citet{Sakai2010}, who
report decreasing line width as a function of age. The latter sample 
is comparable to our distance-limited sample
with sources between 1.7--4.6~kpc.
It is therefore intriguing, that in our large sample we do not find any
robust trend in the line width of the SiO emission
as a function of evolutionary stage. 
This could  either be due to beam confusion from nearby sources, blending of different evolutionary stages in the beam,
or simply a limited sensitivity for the { high-velocity}
component.

A contribution from an additional, low-velocity shock component
to the high-velocity shocks produced by protostellar jets
may explain why in this large sample we find no evidence for
any robust trend between the line width and age of the clumps.

\subsection{SiO chemistry as a function of evolution}\label{sec:sio_lit}
One of the motivations for this study is to 
extend previous works to investigate the trends between
the SiO emission 
and the evolutionary stage. To accomplish this,
either mid-infrared flux, or more recently
$L_{\rm bol}/M,$ was used to distinguish between young
and/or evolved sources.
The $L_{\rm bol}/M$ ratio is commonly used as a
tracer of evolutionary stage 
(e.g.\,\citealp{Molinari2008, Ma2013, Urquhart2014, Leurini2014});
low values correspond to young cores, where the ratio is still dominated
by the envelope mass, while higher values are dominated by high $L_{\rm bol}$ 
 dominated by an accretion luminosity consuming the envelope mass.
It is not clear how this value characterises massive clumps forming a cluster,
however, following previous studies
we also rely on this value as an indicator of evolution.
Using $L_{\rm bol}$ extrapolated from monochromatic luminosity
at 22~\mum, our estimates here are good to an order of magnitude
and are sensitive to varying dust extinction along the line of sight 
and inclination angle. 
Nevertheless, since they have been derived systematically, we can rely on this measure
to study the statistical properties of the sample. Given the large number of
objects, in particular, that of massive clumps, our 
study increases the statistics of massive clumps in the SiO line
by a factor of six compared to previous surveys (e.g.\,\citealp{LS2011}).
Therefore, our study represents to date the most 
extensive study of SiO emission in massive clumps.

{\citet{Sakai2010} studies a sample of massive clumps and distinguishes between 
MSX bright and dark sources, while more recent studies on larger samples by
\citet{LS2011} and \citet{SM2013} use the $L_{\rm bol}/M$ quantity derived
from Herschel to distinguish between young and more evolved
sources. Both of these studies suggest 
a decrease of SiO line luminosity with age.
These findings were interpreted as
SiO being largely enhanced in the earliest evolutionary phases due to an
excess in the release of the grains from powerful jets emanating from young protostars.
As such, shocks are related to the jet activity age, and the SiO line intensity decreases on short
timescales~\citep{Gusdorf2008} leading to lower line luminosities.
This explanation is particularly interesting because it suggests that 
the intensity of the shocks and consequently 
the jet activity decreases with time. 
This would imply a picture very similar 
to low-mass protostars, 
where a decrease with jet and outflow activity with time has been shown
by several studies (e.g.\,\citealp{Bontemps1996}).

More recently, \citet{Sanhueza2013} and \citet{Miettinen2014} studied the SiO ($2-1$)
line in samples of IRDCs and found no trend of decreasing SiO abundance. 
Although this may be due to a small range
of evolutionary sequence covered 
or a different contribution of low- and
high-velocity shocks in IRDCs 
(see \citet{Duarte-Cabral2014} for a discussion),
similar results are found by \citet{Leurini2014} for
a subset of sources from \citet{LS2011}.
This study is also affected by a limited range of $L_{\rm bol}/M$.
However, for the first time the authors investigated the effects of excitation
by studying high-J SiO lines and derived the SiO abundance directly
through comparison with high-J CO transitions.
Clearly, a statistically significant sample 
covering a larger range of $L_{\rm bol}/M$ is essential, and
we cover at least four orders of magnitude of $L_{\rm bol}/M$ from
infrared-quiet sources to the luminous latest evolutionary stages of star formation
in two SiO transitions.

In Sect.\,\ref{sec:area}, we derive a crude estimation of the SiO abundance
from the ($2-1$) transition, which is subject to large uncertainties due to the LTE assumption, the unknown beam dilution factor, and the fact that the $N(\rm{H_2})$
was estimated from the dust. 
Using the LTE approximation, we find a decreasing trend between 
$X(\rm{SiO})$ and $L_{\rm bol}/M$. Although this result is similar to
that found by \citet{Miettinen2006,LS2011,SM2013}, it is statistically not robust. 
This is because these variables are not entirely independent as both quantities are derived from the dust. On the other hand, using the non-LTE column density estimates
this trend becomes even weaker (Fig.\,\ref{fig:sio21_abundance_radex}), suggesting
that the trend is an observational bias due to the LTE approximation.
This is further supported by the fact that plotting the SiO
 line luminosity as a function of $L_{\rm bol}/M$, following
 \citet{LS2011}, we do not find any trend in our sample (Fig.\,\ref{fig:ls2011}). 
Further supported by more recent studies, such as \citet{Leurini2014}, 
who find no decreasing trend as a function of age, and that of \citet{Gerner2014},
who report smaller abundances for the youngest sources in their sample
than for the more evolved sources, our results dispute an 
evolutionary trend in the SiO column density or abundance as a function of age.

To overcome the limitations set by the approximations for the SiO column
density and abundance calculations, the strength of our study is that we can go
a step further to investigate the excitation effects towards an unprecedentedly
large sample. These results in Fig.\,\ref{fig:sio21_lm} 
clearly reveal a robust trend of increasing ratio of the area of the
($2-1$) and the ($5-4$) transitions as a function of $L_{\rm bol}/M$.
This implies that towards sources with larger
$L_{\rm bol}/M$, the higher energy levels are more populated. 
This implicitly suggests a {change in the excitation conditions}
with increasing $L_{\rm bol}/M$, which we interpret here as an evolutionary 
tracer. This explains why the SiO abundance derived only from the ($2-1$) 
transition in previous studies suffers from a systematic underestimation
of the column density towards the more evolved sources. This is the case because the 
excitation conditions have not been properly accounted for. Furthermore,
the effects of beam dilution are more important when comparing lines with
different beam sizes such as in the study of \citet{LS2011}.
Even if \citet{LS2011} used the $J=3-2$ transition to probe the conditions of the
gas,  the difference in the upper state energy is larger for $5-4$ line 
and, consequently, the line ratio is more sensitive
to different conditions, and although \citet{SM2013} extended this initial study of
with the ($5-4$) transition, they suffered from poor statistics.

Using the $J=8-7$ and the $J=3-2$ transitions, which have a larger difference of $\sim$62.5~K 
in their upper energy levels, \citet{Leurini2014} also 
find a weak trend for which the more evolved
sources exhibit larger line ratios pointing to an increased population of the higher J 
transition. Our study confirms this trend with a  much larger statistical sample.

Based on our statistically robust results,
we therefore conclude
that the excitation conditions change
due to increasing temperature and/or density 
towards massive clumps as a 
function of $L_{\rm bol}/M$, which reflects different evolutionary stages.
In cluster-forming clumps, the high detection rate of SiO and broad wings
together with the continuously increasing excitation effects suggests 
a continuously on-going star formation process.

\subsection{Shock evolution in massive clumps}\label{sec:shocks}

The origin of this change in excitation conditions, 
however, needs to be understood.  
SiO is sensitive to the radiative pumping via its vibrational and bound electronic states. 
In PDR environments,
at low densities already weak near-infrared, optical, and UV-radiation field can be sufficient to modify the 
rotational level populations~\citep{Godard2013}, 
hence, the interpretation of the origin of low- and high-$J$ emission of the 
SiO in the vicinity of young massive stars may pose a more complex problem than previously 
thought (see also \citealp{Leurini2014}).
According to \citet{Godard2013}, the higher J transitions in the vicinity of bright protostars
may be affected by radiative pumping. 
Although we have these kinds of objects among the sources classified as star-forming and clumps associated with {\uchii} regions, this effect on $J<5$ transitions is marginal. 

We find that the observed correlation of the SiO (5--4)/(2--1) integrated intensity ratio with $L_{\rm bol}/M$
is stronger for considering only the distance- and mass-limited 
sample, however, the fact that the correlation is still statistically significant including all sources 
suggests that radiative pumping does not significantly influence our results. Based on non-LTE
calculations with RADEX \citep{vanderTak2007}, 
we see that the line ratios are more sensitive to a change in 
density rather than temperature. 
This has already been suggested by \citet{Schilke2001}, who
find that the ratio of the (5--4)/(2--1) line is more sensitive to the
thermal pressure, i.e. the product of $n(\rm H_2)T$. 
Here we find typical line area ratios of $\sim0.5$, which are
well reproduced by non-LTE calculations with a pressure term
of $\sim5\times10^6-1.7\times10^7$~K~cm$^{-3}$. 
Therefore, 
we arrive at the conclusion
that the observed trend is likely either 
due to an increase of pre-shock gas density
or is a signature of higher thermal pressure for the more evolved clumps.

The former could be explained by the fact that jets impact the already compressed
material along the outflow cavity or 
by an increasing mass accretion rate,  as
observed by e.g.~\citet{Klaassen2012}. Numerical simulations of a collapsing
massive core forming a $20-30$~\msol\ star report increasing mass
accretion rate in the early evolution of the protostar-core system \citep{Kuiper2013}. 
Collective effects and the strong gravitational potential
in massive clumps
may also explain such an increasing ambient gas density~\citep{Peters2014}.

Alternatively, the reported trend of change in excitation 
may simply reflect the type 
of the central object forming. 
\citet{Faundez2004} proposes an alternative scenario 
for the $L_{\rm bol}/M$ being an indicator of 
the most massive object forming in the clump.
\citet{csengeri2013} also finds a hint that there is a correlation
between the 22~\mum\ flux and the sub-millimeter emission,
further suggesting that this ratio may be rather determined
by the most massive object forming in the clump.
This kind of trend would then imply that the SiO emission
is dominated by the central, most massive object forming
in a clump, and, consequently, the observed trend would reflect
that the more massive the clumps are, the more massive  central object
(see also \citealp{Urquhart2014}).
As also discussed in \citet{Leurini2014}, the $L_{\rm bol}/M$,
the properties of the central
object and the age of the protocluster are not independent, and, hence, this
value could be used as an indicator of the evolutionary stage of the clump.
Our results in this context would imply that more massive stars
would form in a higher density and higher thermal
pressure environment on a clump scale.

\section{Conclusions}\label{sec:conclusion}
We present here a statistically significant, representative sample of massive clumps 
in different evolutionary stages selected from the \at\ survey. 
Towards these sources, an unbiased
spectral line survey covering 32~GHz from the 3mm atmospheric window
has been carried out with the IRAM~30m telescope. We focus here on the shock tracer,  
SiO, where the ($2-1$) line has been complemented with observations of 
the ($5-4$) transition with 
the APEX telescope. The presented study is 
the richest and most extensive statistical study of the properties of shocked gas 
towards massive clumps. 
The main results are as follows:
\begin{itemize}
\item We constrain the physical properties of a sample of massive clumps
         and classify the sources by assessing the presence of embedded 
         {\uchii} regions, infrared-bright sources, while the rest of the sources
         are classified as infrared-quiet clumps. In particular, we show that the distance-
         and mass-limited sample is a representative sample of 
         massive clumps in all evolutionary stages.
         
\item SiO ($2-1$) 
         emission has a high detection rate of 77\% above 3~$\sigma$ in the complete sample. From these detections, we find a 
 high detection rate, over 50\%, in the SiO ($2-1$) line
         towards infrared-quiet clumps. 
         Furthermore, we find 34 infrared-quiet sources
         that exhibit emission in the higher energy, ($5-4$) transition.
         All infrared-quiet sources in the mass- and distance-limited sample exhibit evidence 
         for emission from shocked gas traced by SiO.
         
\item {
         The line profiles often show two components, one centred at the ambient velocity (low-velocity component) and high-velocity line wings. 
         Towards the largest fraction of all infrared-quiet sources, we only detect the
         low-velocity component. 
         This may originate from fast shocks from deeply embedded, thus
         undetectable population of low-mass stars or low-velocity shocks originating from
         cloud-cloud collisions.}

\item The high-velocity SiO emission with broad component
         is seen through all evolutionary stages, suggesting that jet activity 
         is present throughout the evolution of massive clumps, 
         likely indicating a continuously on-going star formation in a clustered environment.

\item We find a weak hint that the broad component is less dominant for 
         more evolved sources, however, we do not find any evidence for a 
         robust trend between 
         the line width and source type,
         which could be due to source confusion in clusters.
         
\item Using the $^{29}$SiO isotopologue, we estimate that towards most of the sources
        the ($2-1$) emission is optically thin. 
        
\item Non-LTE models towards a subsample of sources with both the $2-1$
        and the $5-4$ transitions observed, indicate SiO column densities
        between $9.6\times10^{11}-1.1\times10^{13}$~cm$^{-2}$ in a density range of 
        $\sim10^5$~cm$^{-3}$ and $T_{\rm {kin}}=50$~K. 

\item We show that the commonly used LTE estimates
         of the SiO column density are biased due to
         a varying, distance dependent beam dilution and non-LTE effects.
         This questions the results of previous
         studies claiming a decrease of SiO abundance as a function of evolutionary stage
         and thus a decrease of jet activity.
         
\item Instead, we find a significant correlation between the
         ratio of the SiO ($5-4$) and ($2-1$) lines with increasing $L_{\rm bol}/M$, which 
         are independent quantities. This is interpreted as a change in excitation conditions,
         with an increasing product of $n(\rm H_2)T$, i.e. pressure
         as a function of evolutionary stage. 
         Our study rather suggests that the shock conditions change with 
         either the age of the clump 
         or as a function of the most massive object already formed in the clumps.
        
\end{itemize}

The origin of this
trend remains, however, unclear.  
To further pin down the origin of the SiO emission, multi-transition studies  constraining
shock models are the next step. In addition, sensitive high-angular resolution observations with
ALMA will unambiguously reveal the morphology of the SiO emission and
thereby allow us to distinguish between the proposed explanation if the observed shocks
originate from a single jet or a multiple system of material ejections.

\begin{acknowledgements}
{
We thank the referee for a careful reading of the manuscript.
}
This work was partially funded by the ERC Advanced Investigator Grant GLOSTAR (247078) and was partially carried out within the Collaborative Research Council 956, sub-project A6, funded by the Deutsche Forschungsgemeinschaft (DFG). 
This paper is based on data acquired with the Atacama Pathfinder EXperiment (APEX). APEX
is a collaboration between the Max Planck Institute for Radioastronomy, the European
Southern Observatory, and the Onsala Space Observatory.
This research made use of data products from the Midcourse Space Experiment. Processing of the data was funded by the Ballistic Missile Defense Organization with additional support from NASA Office of Space Science. This research has also made use of the NASA/ IPAC Infrared Science Archive, which is operated by the Jet Propulsion Laboratory, California Institute of Technology, under contract with the National Aeronautics and Space Administration. 
This publication makes use of data products from the Wide-field Infrared Survey Explorer, which is a joint project of the University of California, Los Angeles, and the Jet Propulsion Laboratory/California Institute of Technology, funded by the National Aeronautics and Space Administration.
This research made use of Montage, funded by the National Aeronautics and Space Administration's Earth Science Technology Office, Computation Technologies Project, under Cooperative Agreement Number NCC5-626 between NASA and the California Institute of Technology. Montage is maintained by the NASA/IPAC Infrared Science Archive.
\end{acknowledgements}
\bibliography{ag-sio.bib}
\bibliographystyle{aa}

\begin{appendix} 
\section{Calibration}\label{app:app-cal}

To monitor the system's performance on different days, we
observed a reference source,   mostly G34.26+0.15, at the beginning of each 
observing session and each frequency setup. 
We then performed a Gaussian fit to the brightest lines, e.g.\,
H$^{13}$CO$^+$ (1--0), SiO (2--1), OCS (7-6), HC$_3$N (7--6) lines.
Figure\,\ref{fig:app-cal} shows
the variation in the peak intensity
 obtained from  N$_2$H$^+$
line of the reference source. 
We monitor an $rms$
variation between the different observing days of 7\%.
The derived line intensities are therefore comparable between the different set of observations.

\begin{figure}[!htpb]
\centering
\includegraphics[width=0.95\linewidth, angle=0]{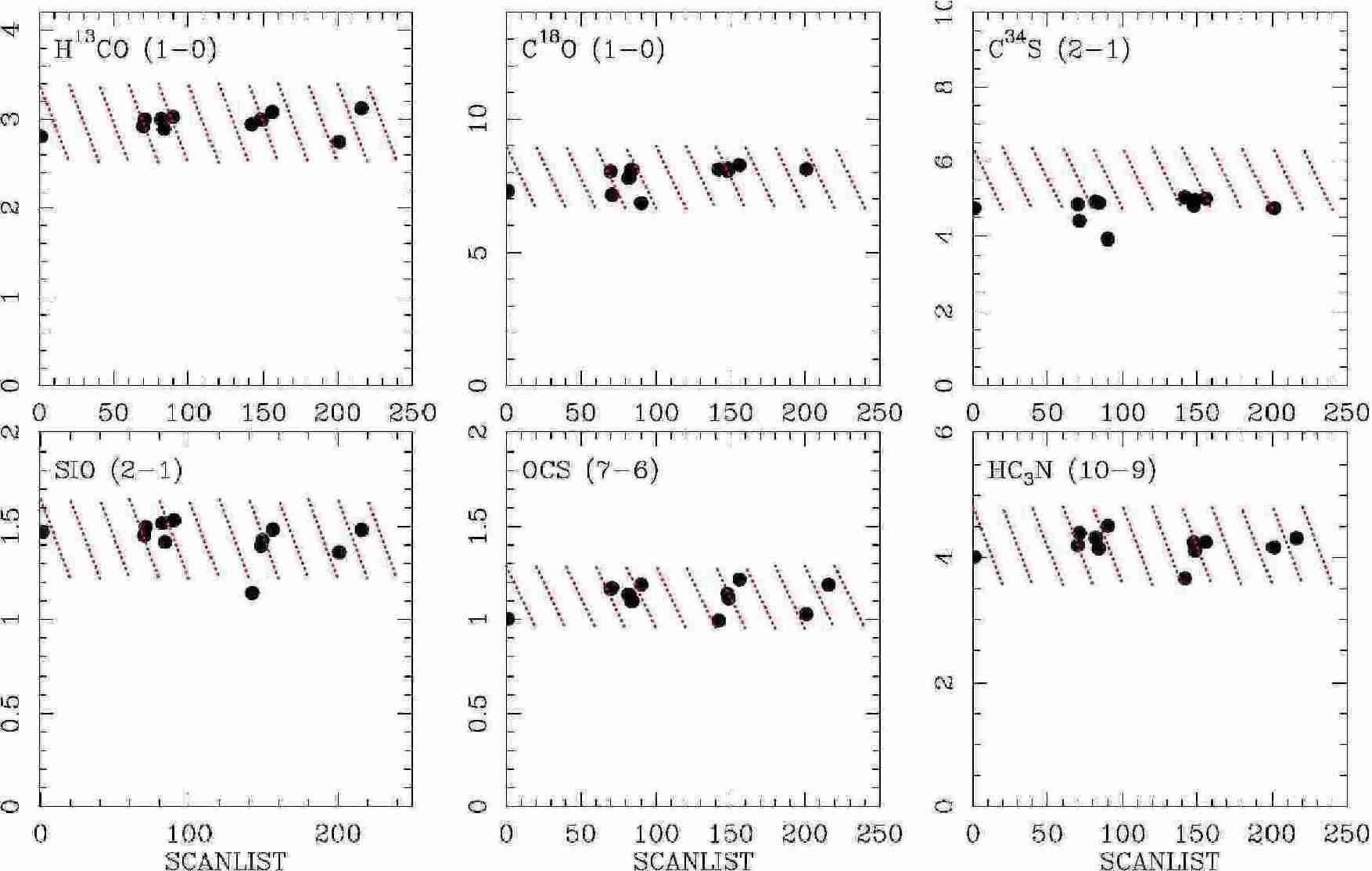}
\caption{Peak intensity of selected line measured on different observing days.
Red dashed area shows 15\% uncertainty around the mean.}
\label{fig:app-cal}
\end{figure}

\section{The 22~\mum\ flux density}\label{app:msx_wise}

We check the consistency of the various datasets used for determining
the luminosity of the sources. First we cross-correlate
the used 21~\mum\ fluxes from the MSX E-band with the WISE W4~band
photometry at 22~\mum. In Fig.\,\ref{fig:msx_wise}, we show
the correlation between the extracted flux densities from the two catalogues.
There is a substantial
overlap in the sensitivity of the two detectors and there is a good agreement between the two measurements,
as already discussed in \citet{csengeri2013}; see also \citealp{Lumsden2013}.
In Fig.\,\ref{fig:msx_wise}, we show the comparison between the MSX and
WISE fluxes.

\begin{figure}[!htpb]
\centering
\includegraphics[width=6.5cm, angle=90]{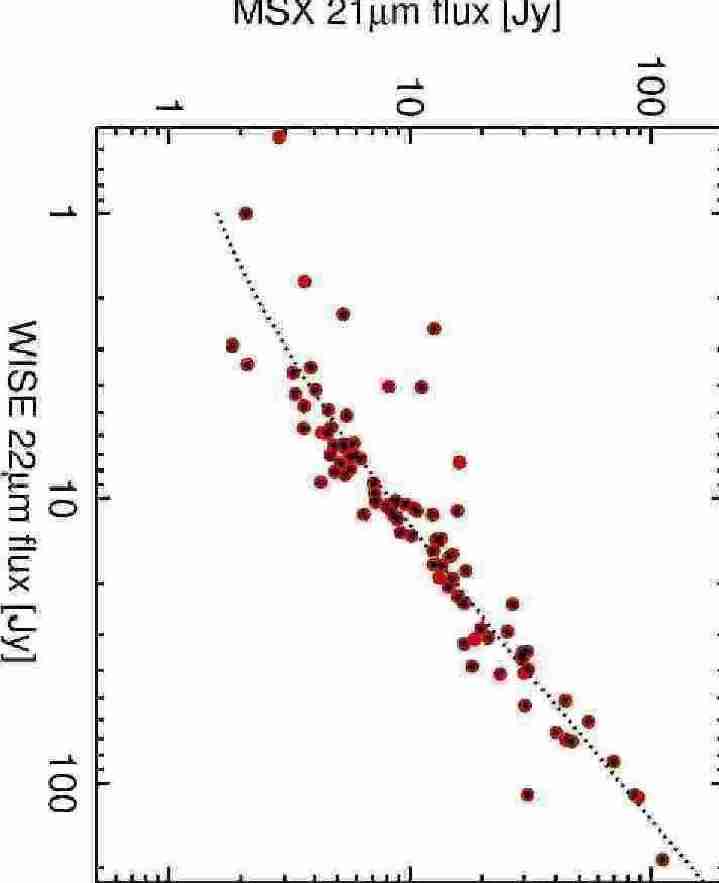}
\caption{Comparison of the extracted 21~\mum\ fluxes from the MSX catalogue
and the 22~\mum\ fluxes from the WISE catalogue.}\label{fig:msx_wise}
\end{figure}

\section{Excitation temperature}\label{app:excitation}
Fig.\,\ref{fig:tex} shows line ratio estimations from LTE calculations.
\begin{figure}[!htpb]
\centering
\includegraphics[width=6.5cm, angle=90]{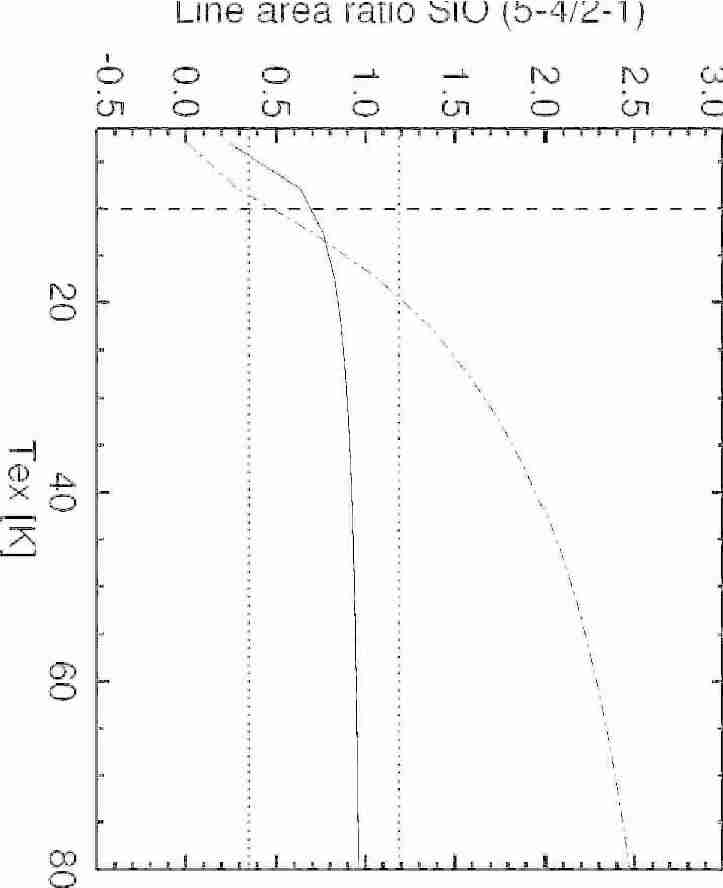}
\caption{Comparison of the $5-4$ to $2-1$ line ratios
in case of optically thin (dashed dotted line) 
and optically thick (solid line) emission. 
Vertical dashed line shows the adopted $T_{\rm ex}$ of 10\,K.
Dotted lines show the average and  maximum line ratio. 
}\label{fig:tex}
\end{figure}

\section{Additional correlations}\label{app:ls}

We show here the same correlations as in Fig.\,\ref{fig:sio21_abundance_radex}
{\sl a} panel,
but for the non-LTE SiO column density estimates with RADEX. Figure\,\ref{fig:sio21_abundance_radex} shows the SiO column densities from
two tests: the filled symbols and  black labels correspond to calculations
with a beam filling factor of unity, which show a similar, although weaker, correlation
as the abundances estimated from the LTE assumption. Open circles with grey
labels show models where the distance effect is taken into account with a fixed source
size. The correlation coefficient is smaller in this case.

As a comparison, in the Fig.\,\ref{fig:sio21_abundance_radex} {\sl b} panel we show the
correlation between $X$(SiO) and $L_{\rm bol}/M$ for the non-LTE models. This plot 
shows that in taking the distance effect into account, the decrease of the
SiO abundance as a function of age indicators becomes less significant.

Finally, to compare our statistical results with that of \citet{LS2011}, we 
show  the plots of the same quantities as in their paper.
For simplicity we show the sources presented in \citet{SM2013}
with revised distances. 
As Fig.\,\ref{fig:ls2011} { lower} panel shows, there is a strong correlation
when plotting the L(SiO $2-1$)/L$_{\rm bol}$ against $L_{\rm bol}/M$, while
we find no correlation between the SiO (2--1) line luminosity and $L_{\rm bol}/M$.
\begin{figure}[!htpb]
\centering
\includegraphics[width=6.5cm, angle=90]{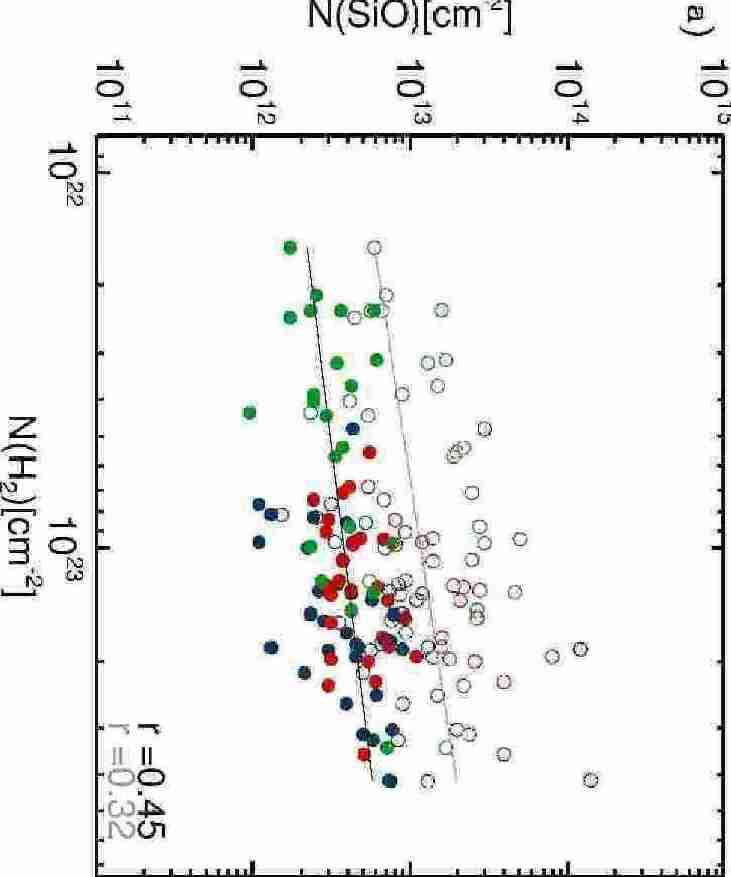}
\includegraphics[width=6.5cm, angle=90]{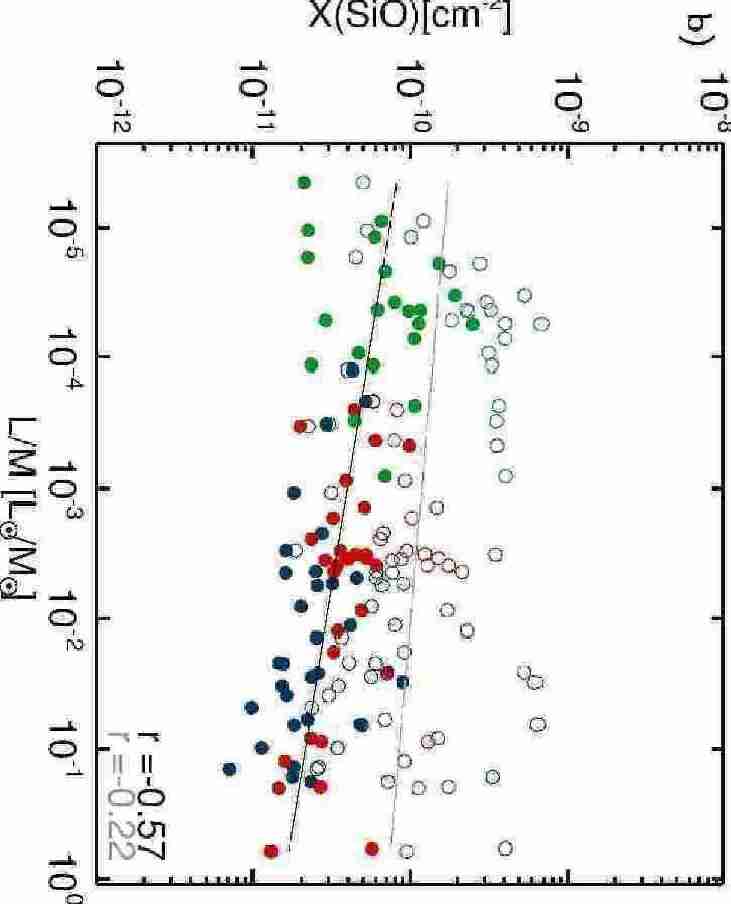}
\caption{{ a)} Column density of SiO estimated from the ratio of the (5--4) to (2--1) transitions with RADEX, versus column density derived from the 870~\mum\ integrated flux density from \citet{csengeri2013}. Filled symbols correspond to the calculations
assuming uniform beam filling, while circles show the calculations with a fixed
angular size of 8~\arcsec\ at 1~kpc, and scaled to the distance of the sources.
The fit to the data are shown in black and grey lines, respectively. The coefficient of the Spearman-rank correlation test is shown in the figure. The correlation is weaker
for the calculations taking the distances into account.
{ b)} Abundance of SiO estimated as described in Sect\,\ref{sec:sio_21_abundance}.
Symbols are the same as on panel { a)}.
}\label{fig:sio21_abundance_radex}
\end{figure}

\begin{figure}[!htpb]
\centering
\includegraphics[width=6.5cm, angle=90]{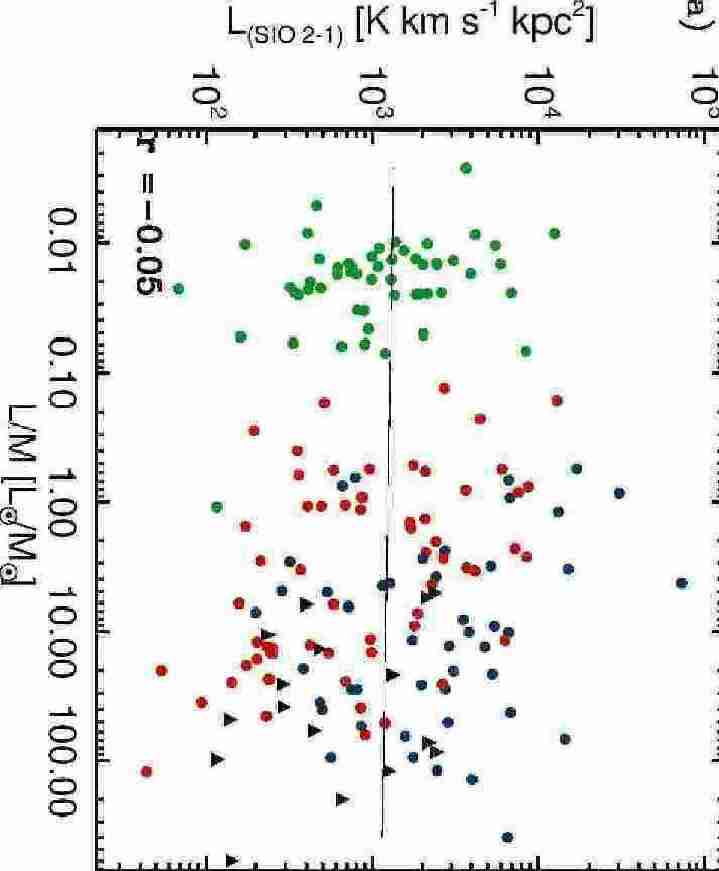}
\includegraphics[width=6.5cm, angle=90]{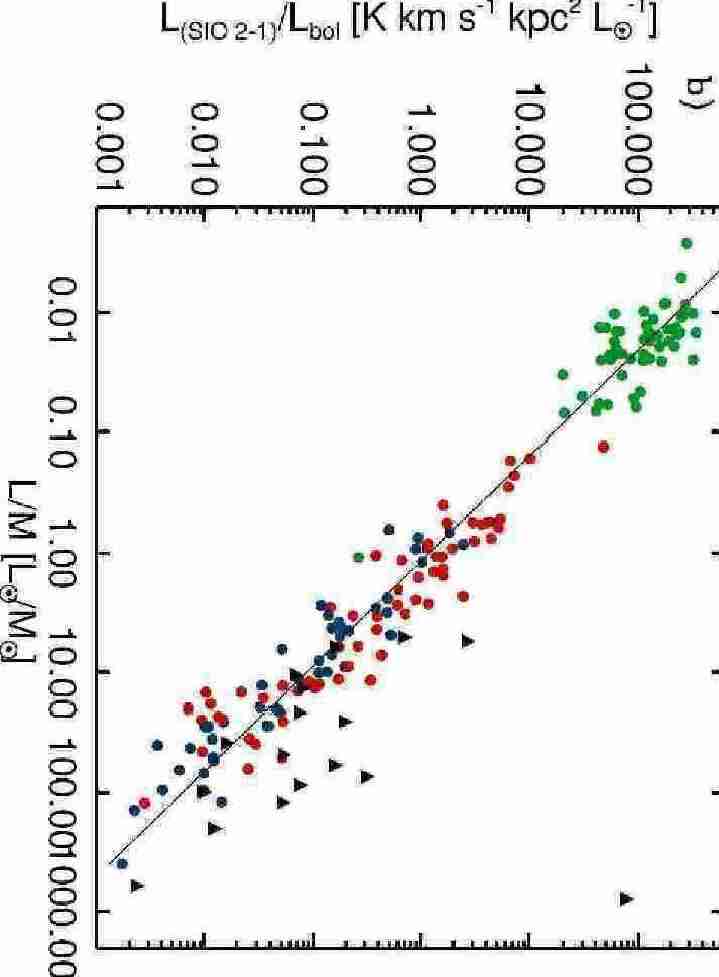}
\caption{The same plot for our sample as in \citet{LS2011}. Colour
symbols are the same as in Fig.\,\ref{fig:width_histo}, and used throughout the
paper. Black triangles show 
the sources from \citet{LS2011} with distance estimates listed in
 \citet{SM2013}.}\label{fig:ls2011}
\end{figure}

\section{The SiO ($2-1$) and ($5-4$) lines of the sample}\label{app:spectra}
\subsection{Spectra of sources observed in both the SiO ($2-1$) and ($5-4$) lines}\label{app:both}
\onlfig{
\begin{landscape}
\begin{figure}
  \includegraphics[width=6.cm]{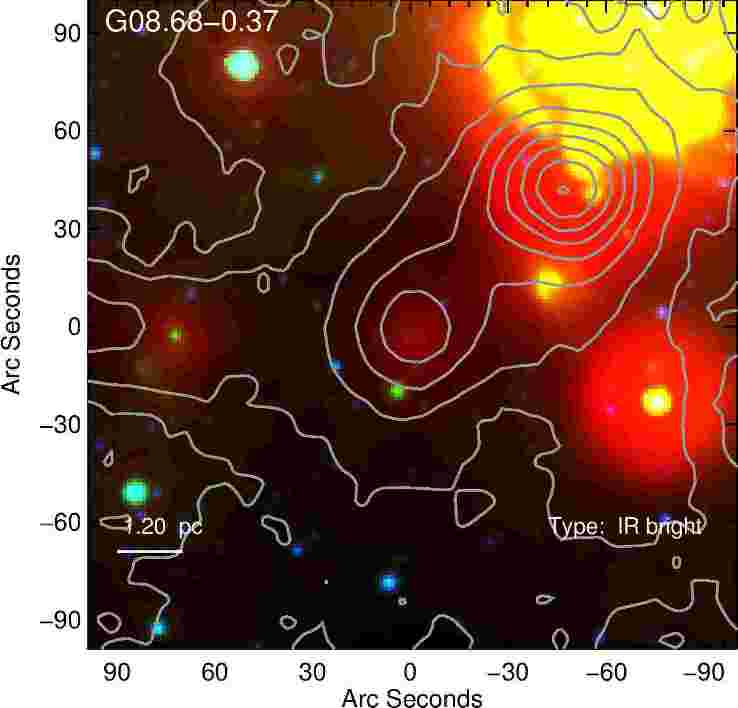}
  \includegraphics[width=6.0cm,angle=90]{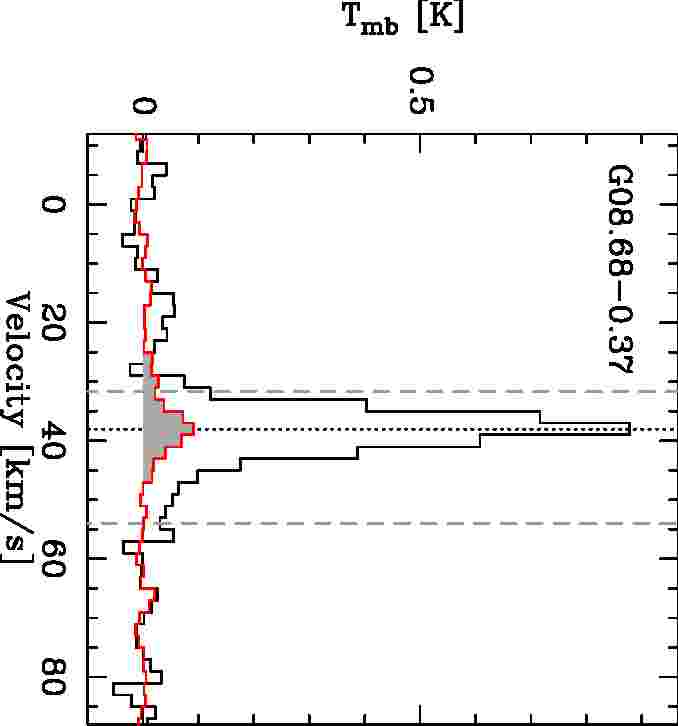}
  \includegraphics[width=6.0cm]{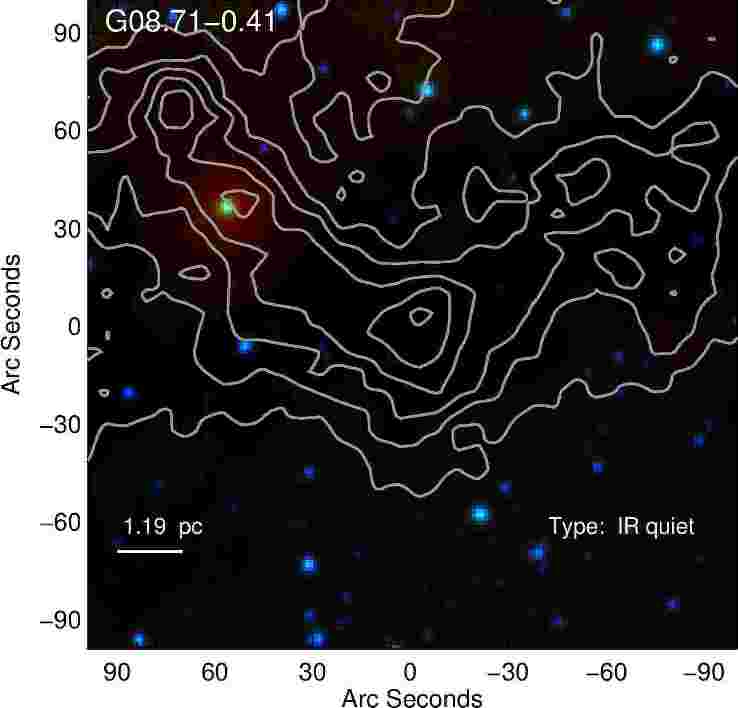}
  \includegraphics[width=6.0cm,angle=90]{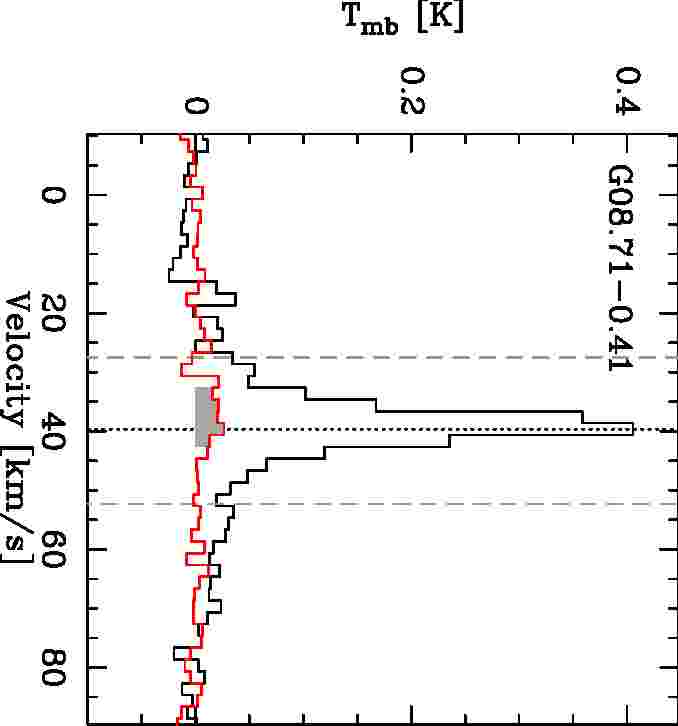}
  \includegraphics[width=6.0cm]{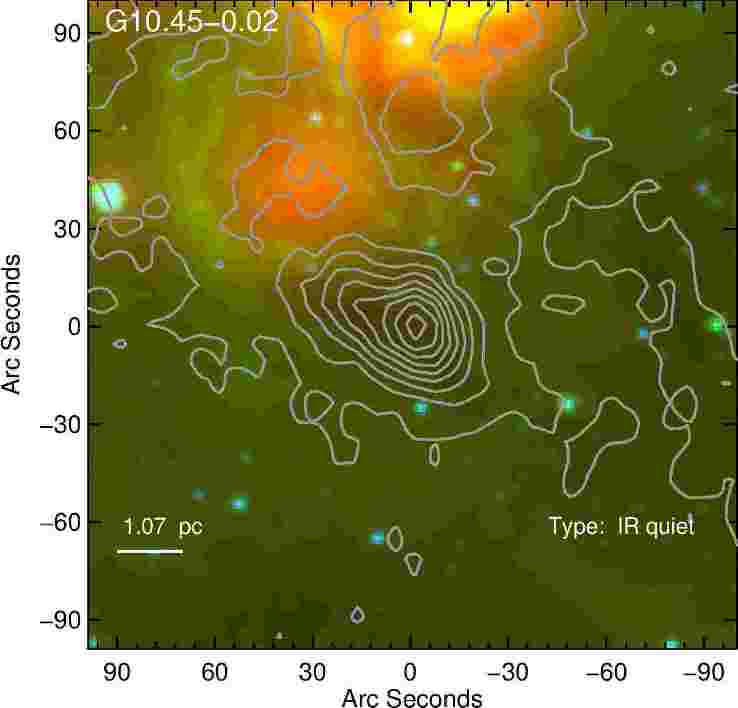}
  \includegraphics[width=6.0cm,angle=90]{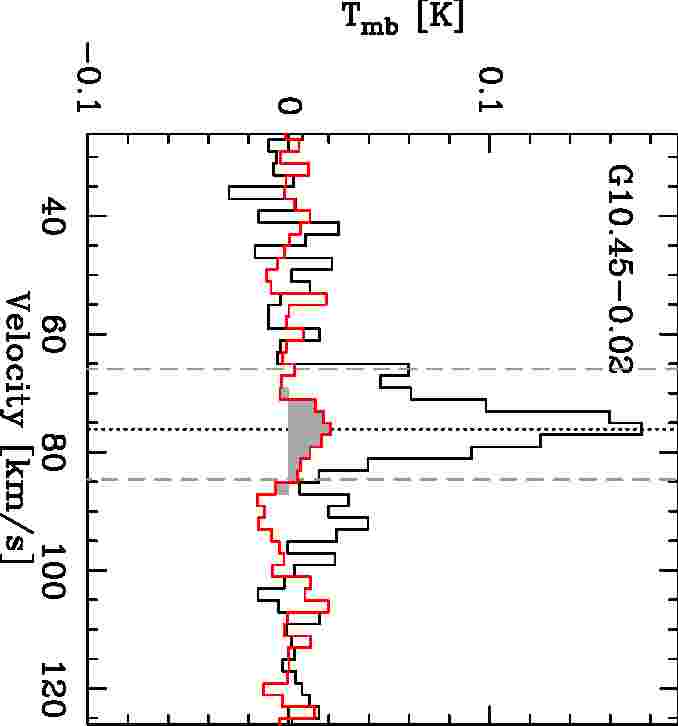}
  \includegraphics[width=6.0cm]{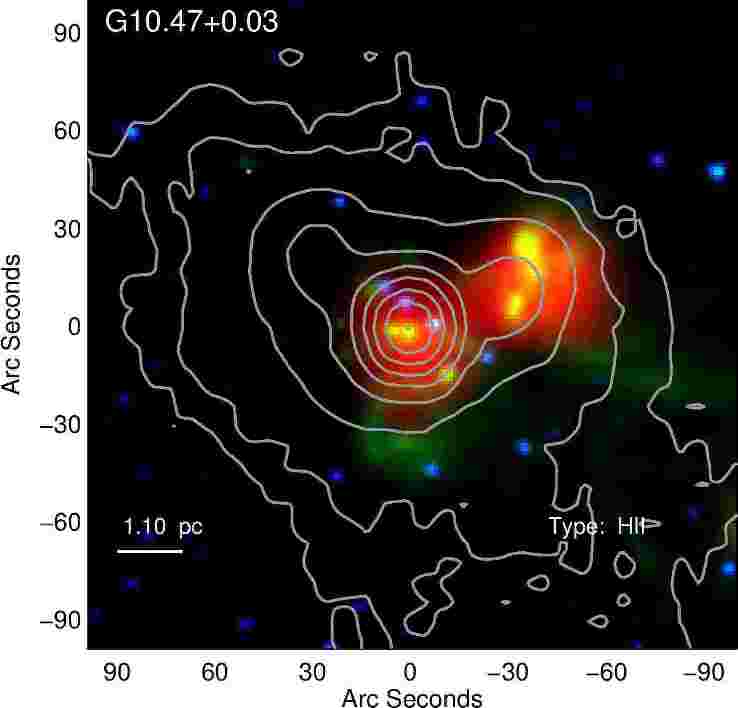}
  \includegraphics[width=6.0cm,angle=90]{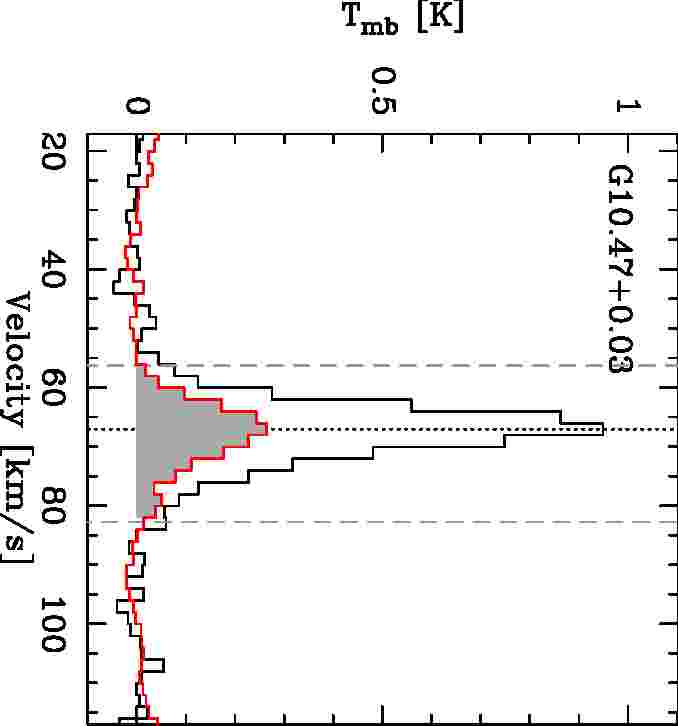}
 \caption{{ Left:} The colour plots show a three-colour 
               composite image from Spitzer 3.6~\mum\ (blue), 8~\mum(green), and 
               WISE 22~\mum\ (red) band images, similar to 
               Fig.\,\ref{fig:examples}. Grey contours show
               the 870~\mum\ emission from \at. White labels show the source name,
               the classification of the source, and a bar shows the physical
               scale considering the distance of the source. 
               { Right:} The spectra of the SiO ($2-1$) transition is shown in black
               and the $5-4$ in red. Grey shaded area shows the velocity
               range used to determine the integrated emission. Dotted line shows
               the systemic velocity of the source 
               (v$_{\rm lsr}$, see also Table\,\ref{tab:table-large-sio21}), dashed lines
               correspond to the velocity range determined from the SiO 
               ($2-1$) transition. Light grey dotted lines show the velocity range
               of the $5-4$ transition.
               These plots are first shown for the targets of 
               the first IRAM~30m observing campaign.}\label{app:fig1}
 \end{figure}
\end{landscape}

\begin{landscape}
\begin{figure}
\ContinuedFloat
  \includegraphics[width=6.0cm]{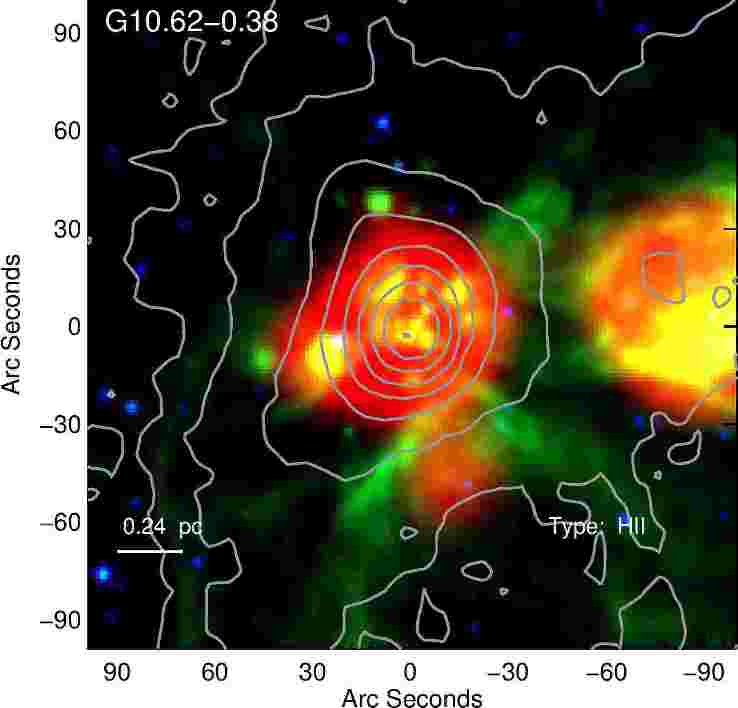}
  \includegraphics[width=6.0cm,angle=90]{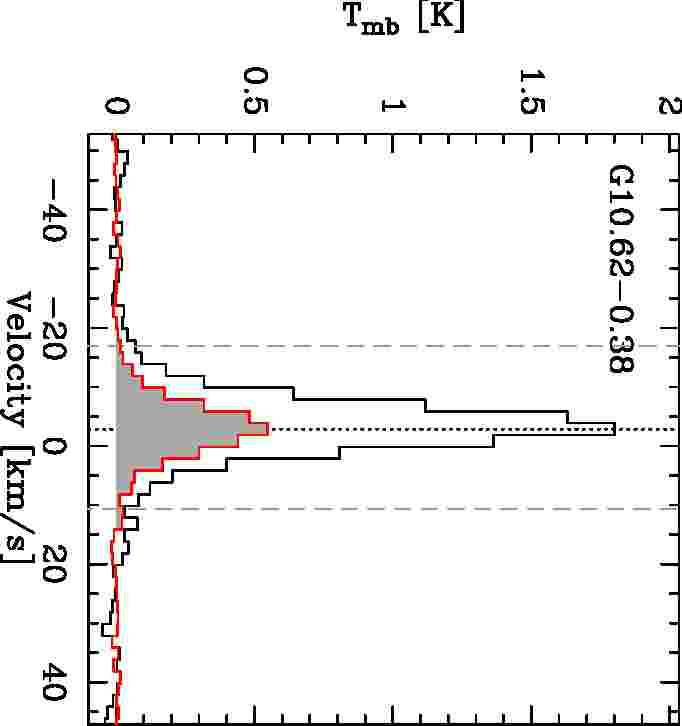}
  \includegraphics[width=6.0cm]{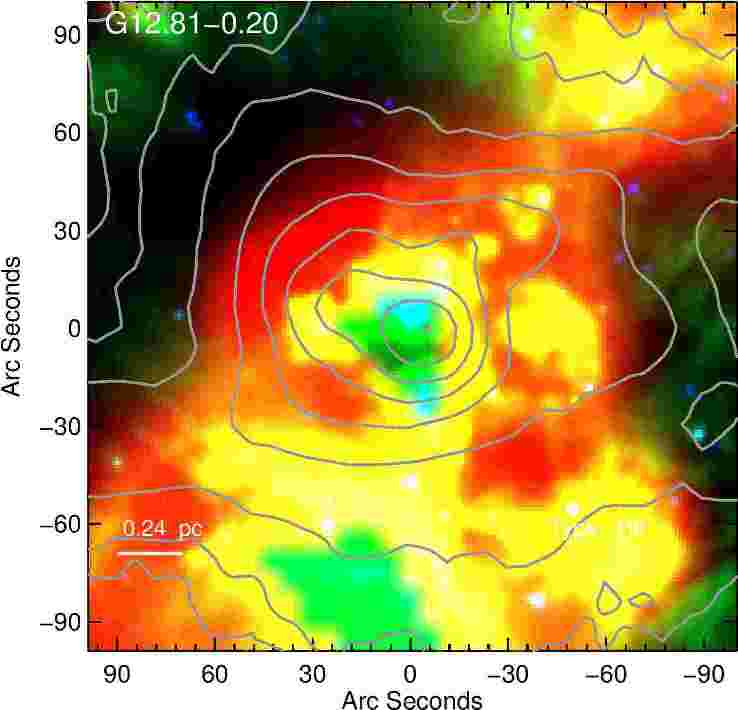}
  \includegraphics[width=6.0cm,angle=90]{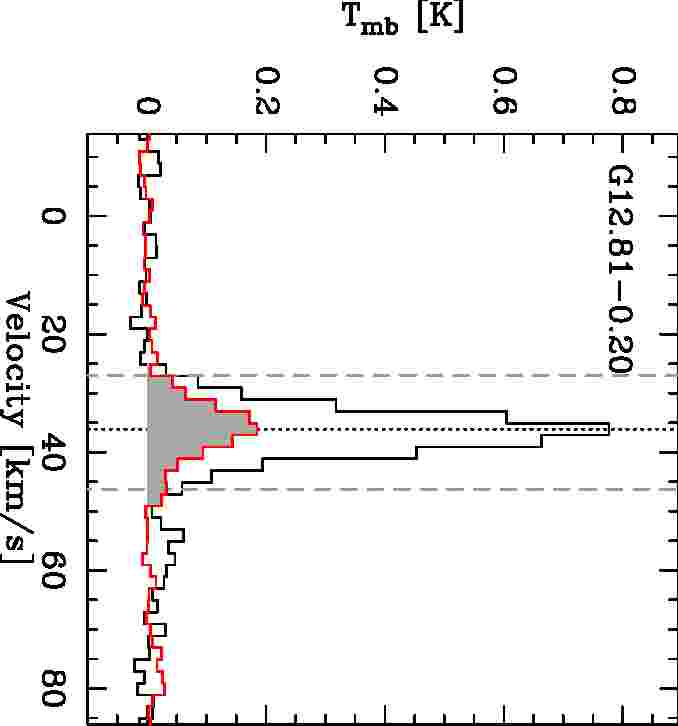}
  \includegraphics[width=6.0cm]{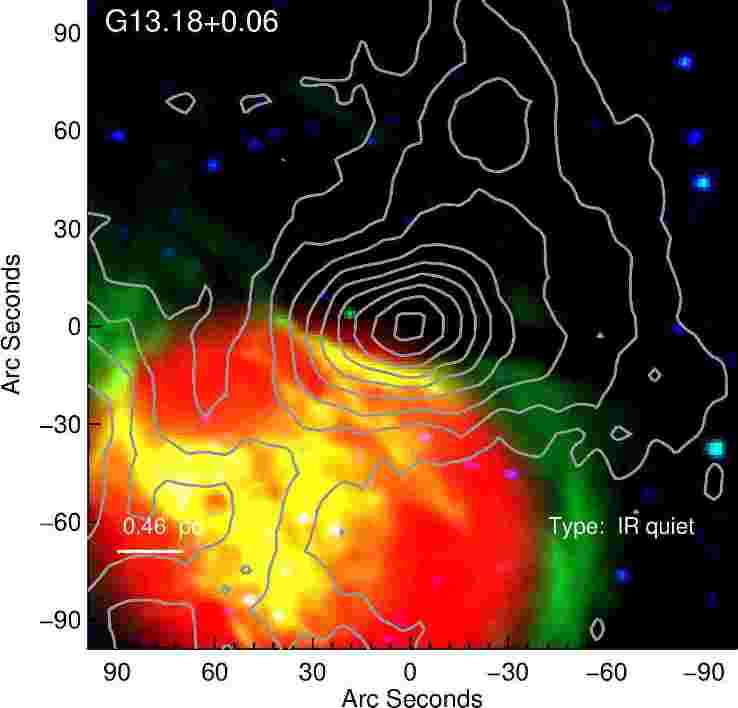}
  \includegraphics[width=6.0cm,angle=90]{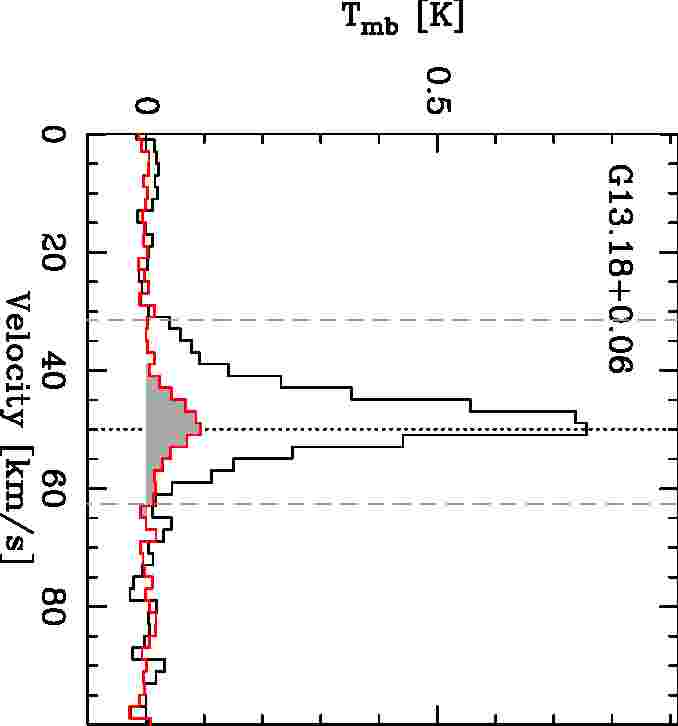}
  \includegraphics[width=6.0cm]{FIG_ASTRO_PH/SIO_PILOT_FINAL/G13.66-0.60_3_colour.eps.jpg}
  \includegraphics[width=6.0cm,angle=90]{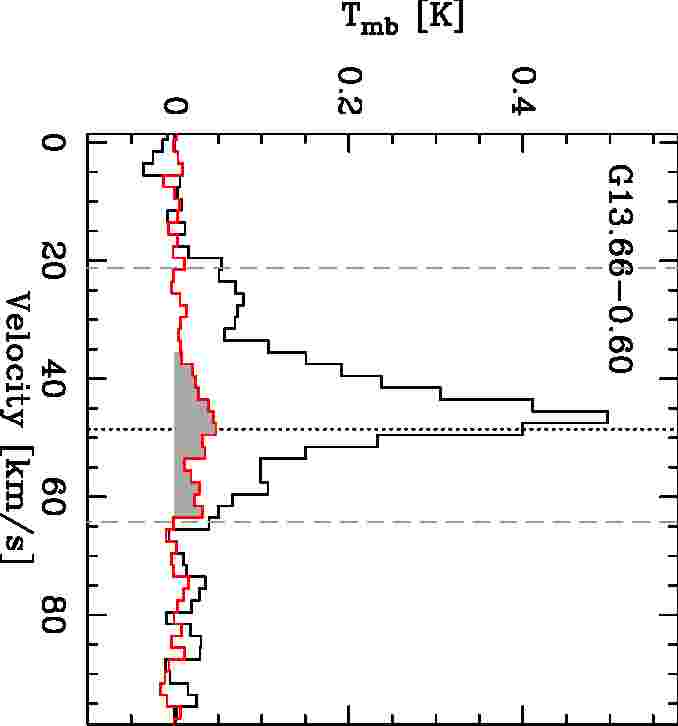}
  \includegraphics[width=6.0cm]{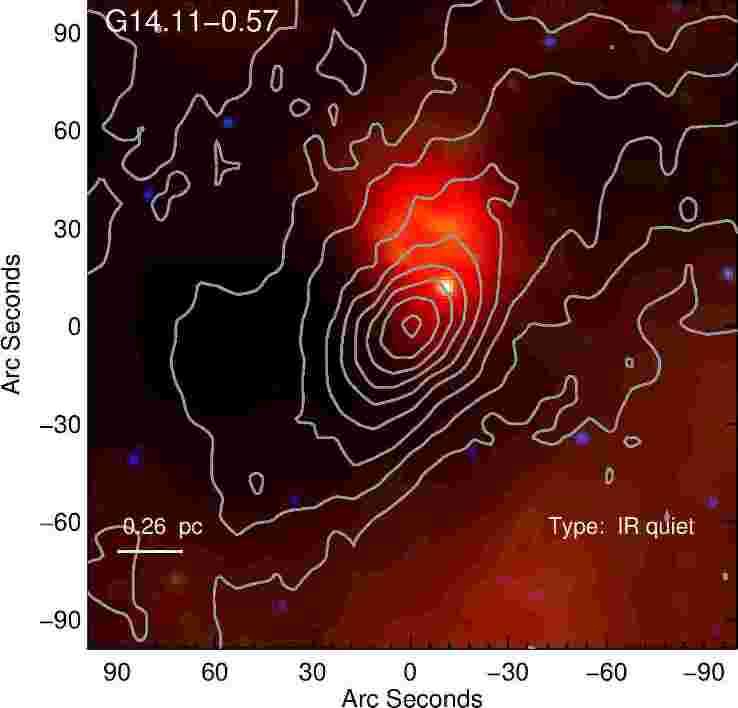}
  \includegraphics[width=6.0cm,angle=90]{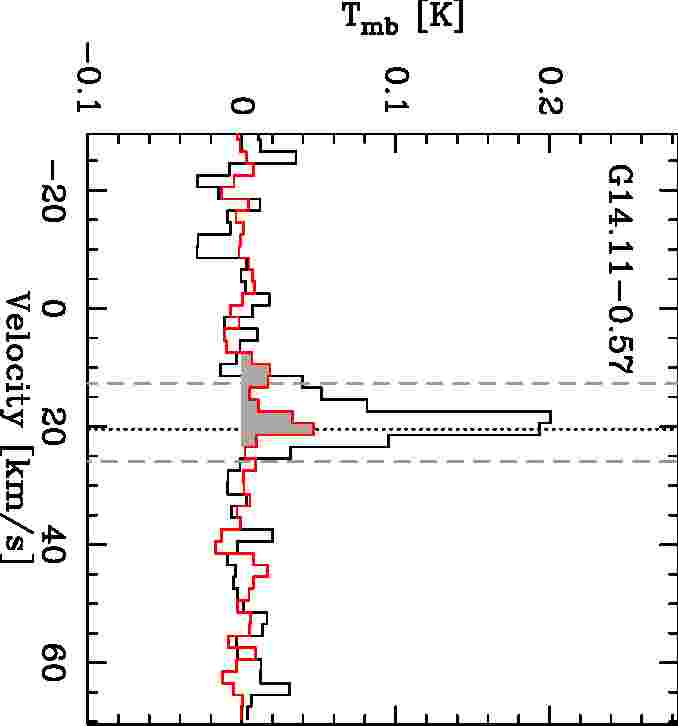}
  \includegraphics[width=6.0cm]{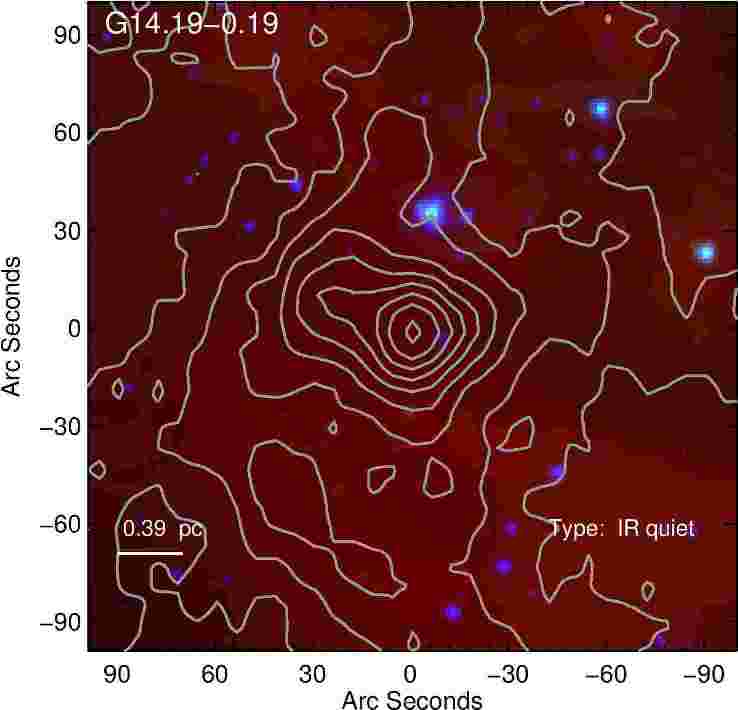}
  \includegraphics[width=6.0cm,angle=90]{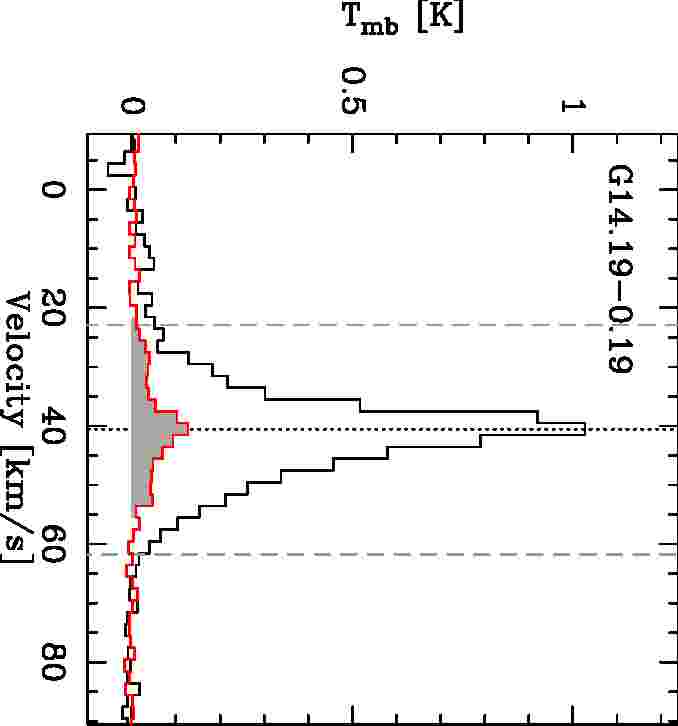}
 \caption{Continued.}
 \end{figure}
\end{landscape}

\begin{landscape}
\begin{figure}
\ContinuedFloat
  \includegraphics[width=6.0cm]{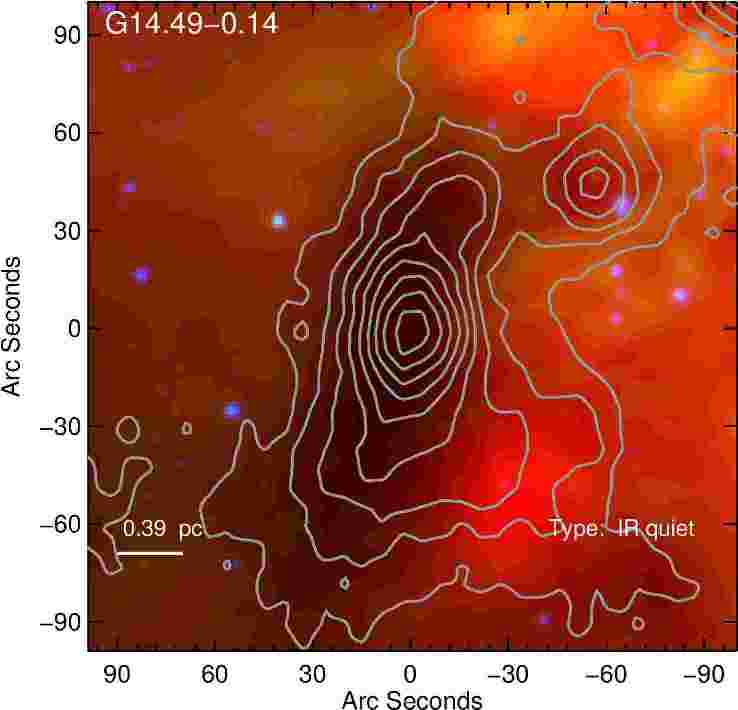}
  \includegraphics[width=6.0cm,angle=90]{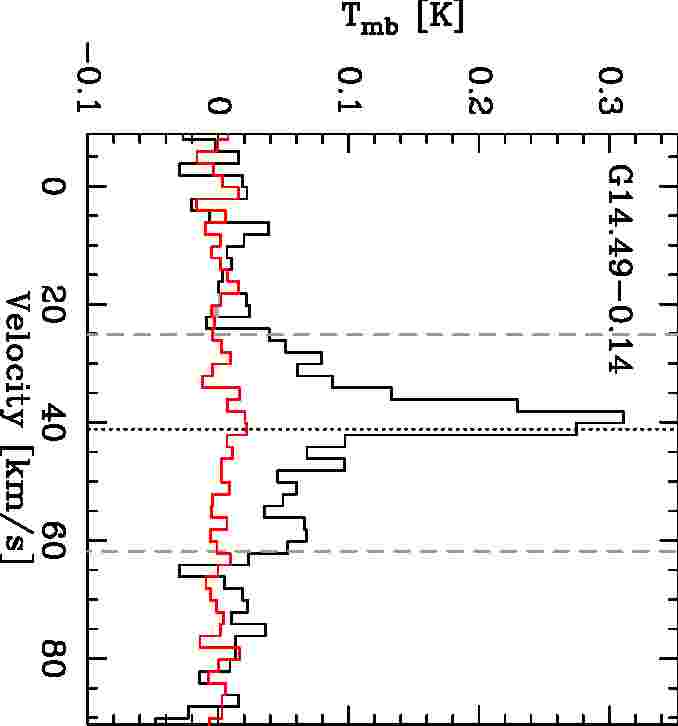}
  \includegraphics[width=6.0cm]{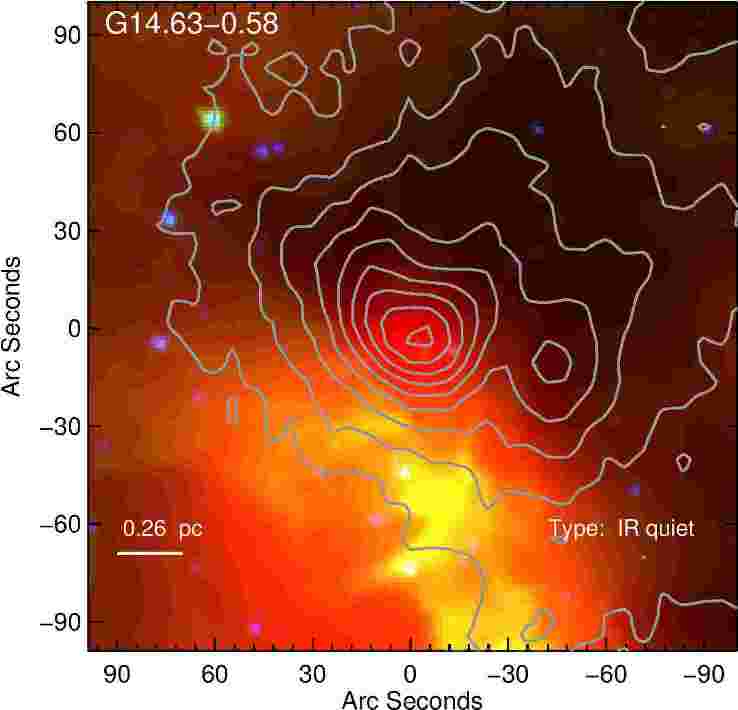}
  \includegraphics[width=6.0cm,angle=90]{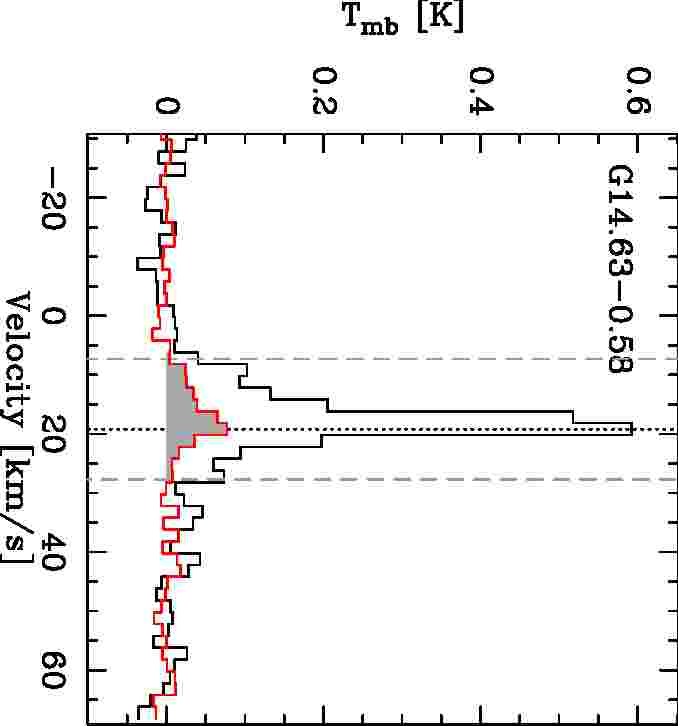}
  \includegraphics[width=6cm,angle=0]{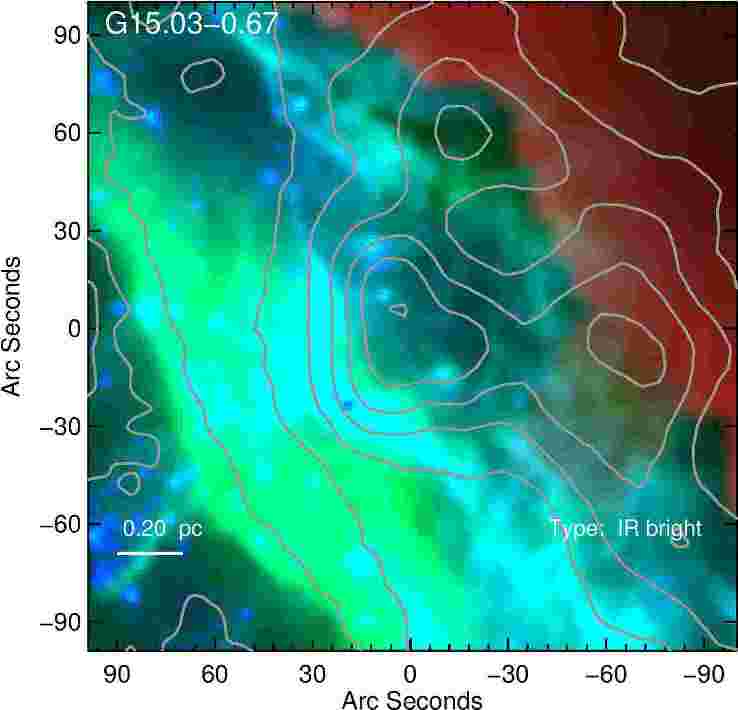}
  \includegraphics[width=6.0cm,angle=90]{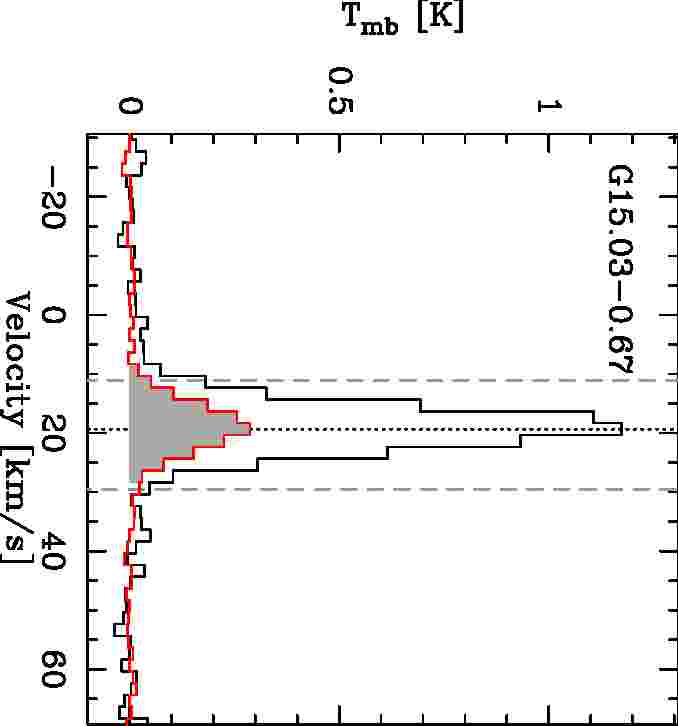}
  \includegraphics[width=6cm,angle=0]{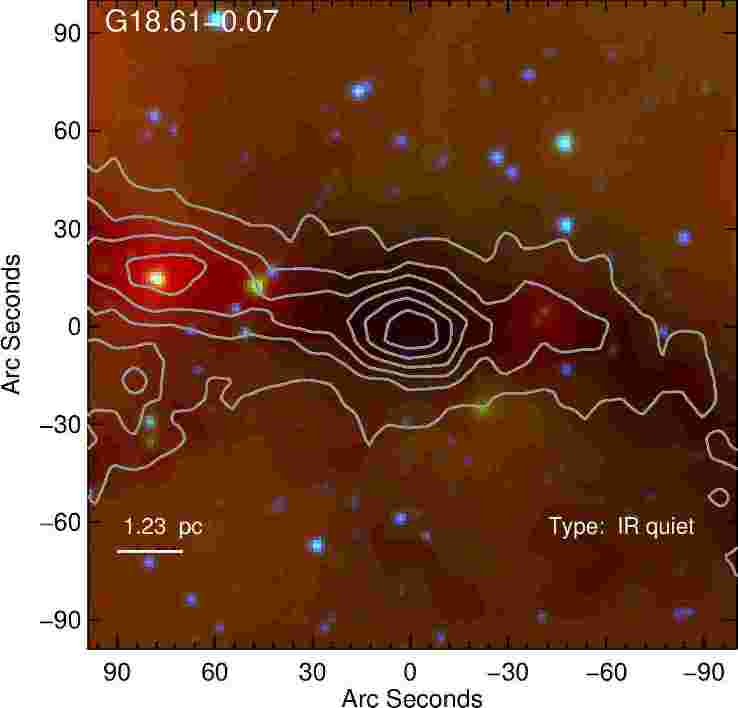}
  \includegraphics[width=6.0cm,angle=90]{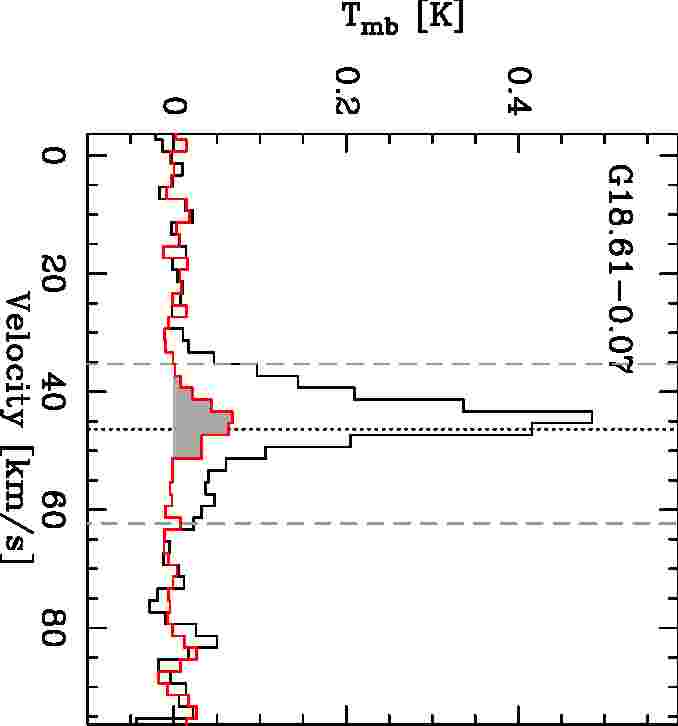}
  \includegraphics[width=6cm,angle=0]{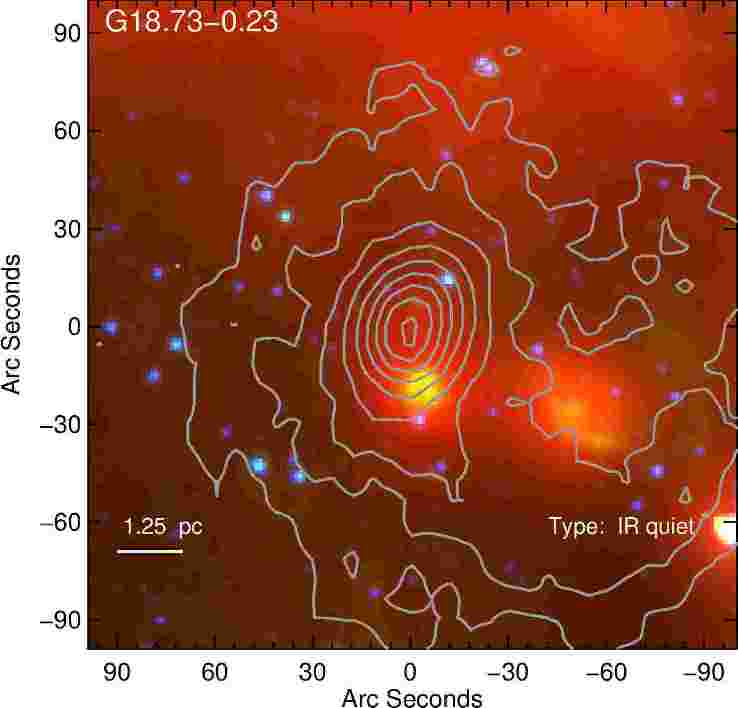}
  \includegraphics[width=6.0cm,angle=90]{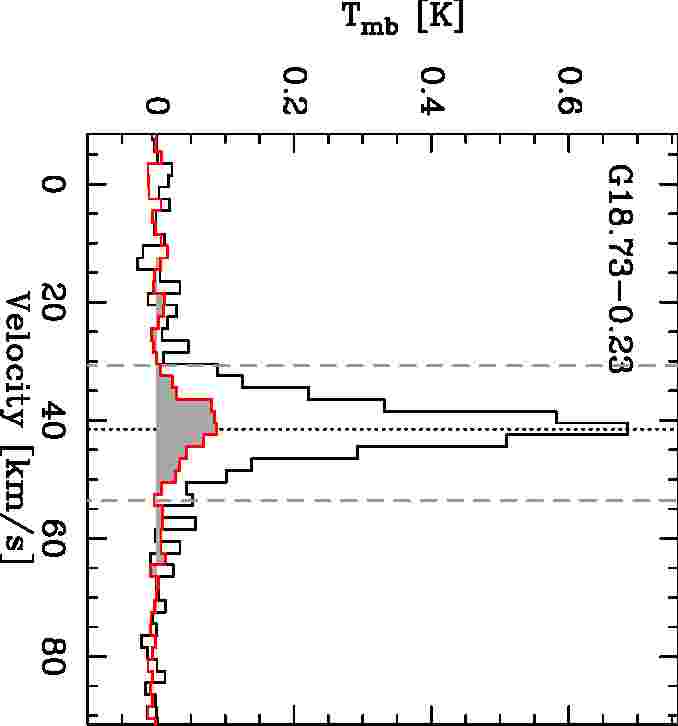}
  \includegraphics[width=6cm,angle=0]{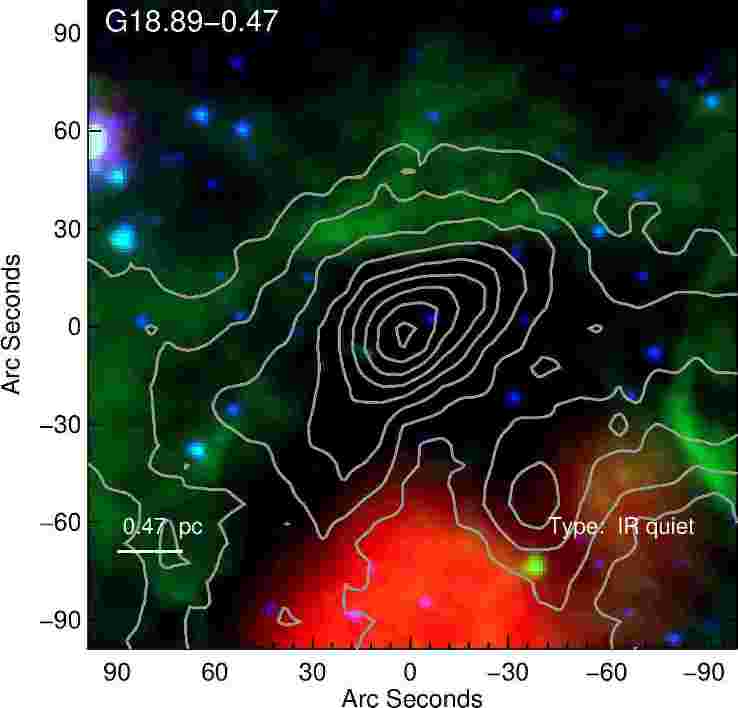}
  \includegraphics[width=6.0cm,angle=90]{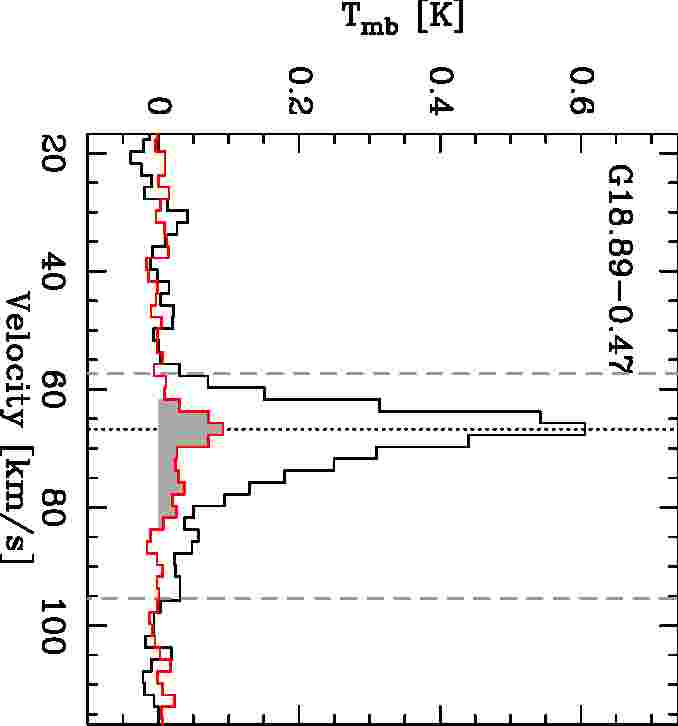}
 \caption{Continued.}
 \end{figure}
 \end{landscape}
 
\begin{landscape}
 \begin{figure}
\ContinuedFloat
  \includegraphics[width=6.0cm,angle=0]{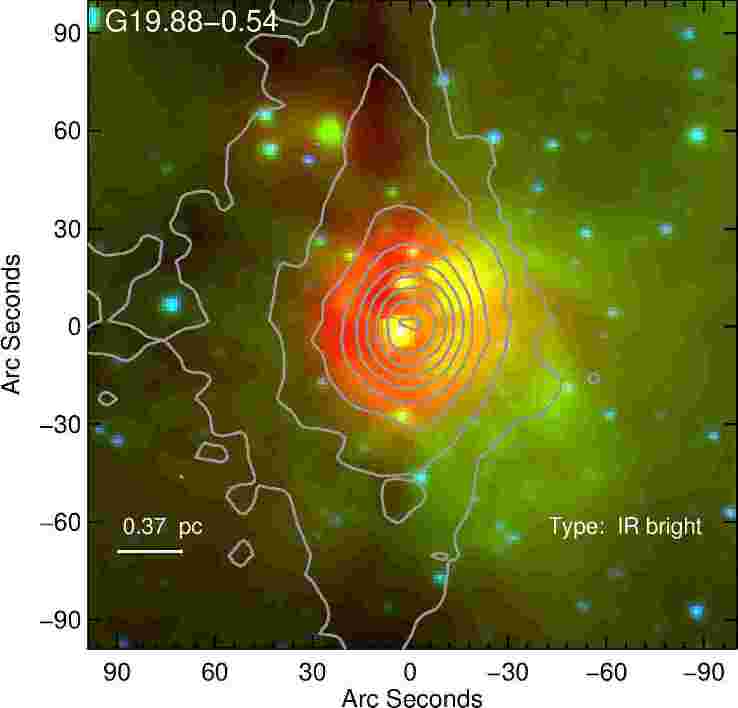}
  \includegraphics[width=6.0cm,angle=90,angle=0]{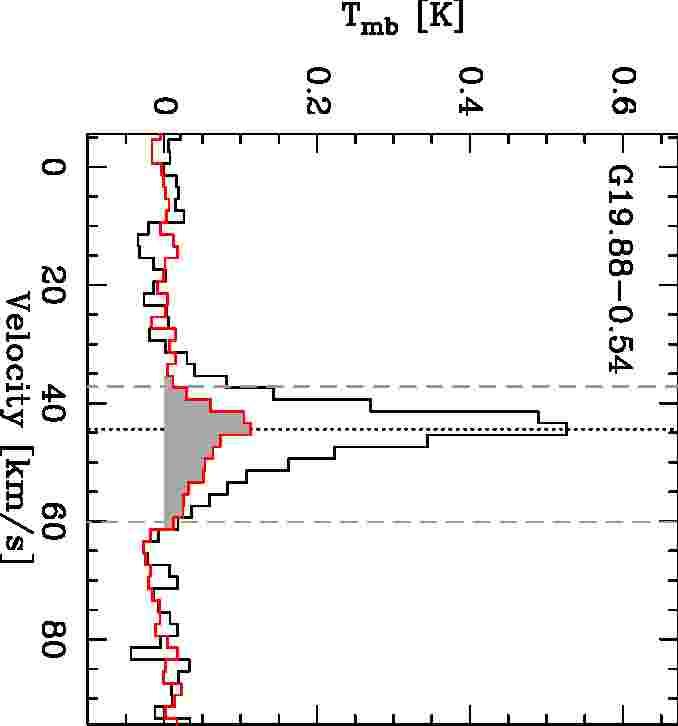}
  \includegraphics[width=6.0cm,angle=0]{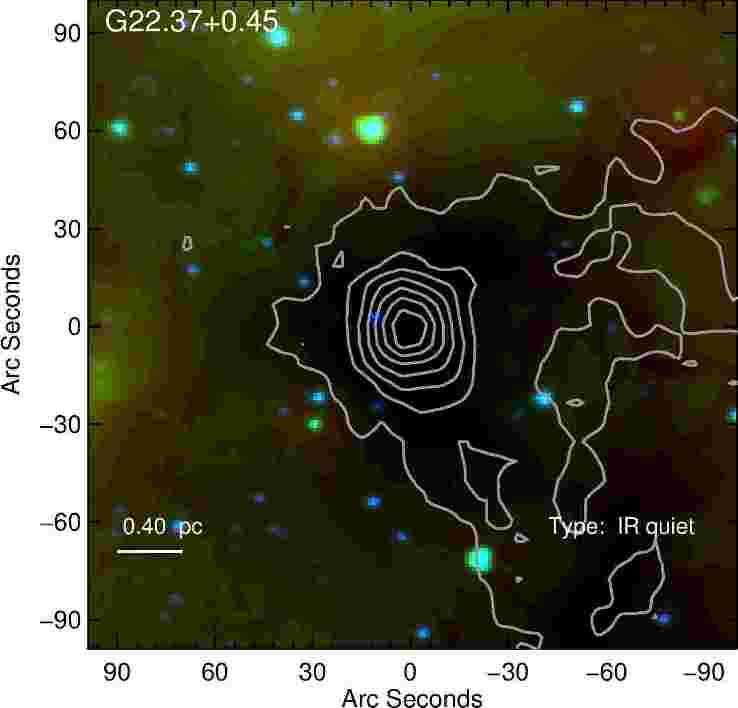}
  \includegraphics[width=6.0cm,angle=90]{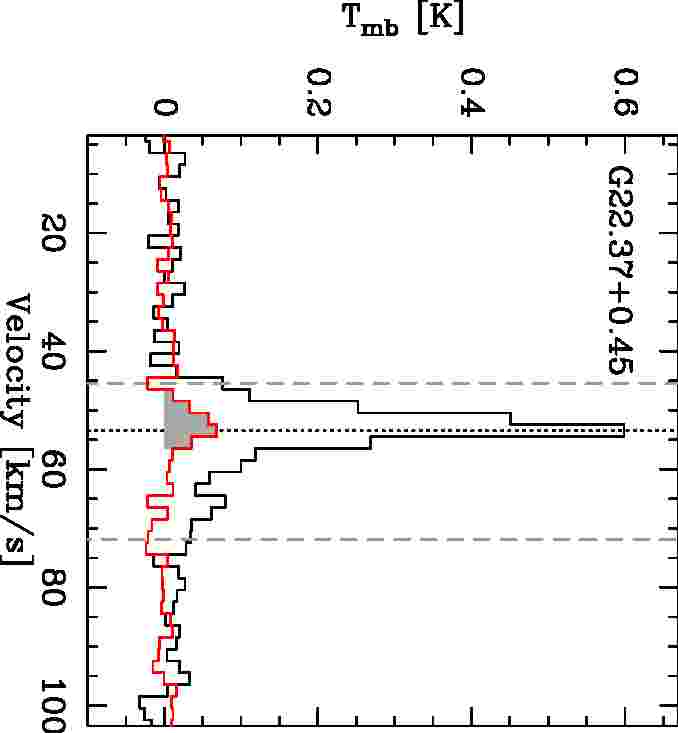}
  \includegraphics[width=6.0cm,angle=0]{FIG_ASTRO_PH/SIO_PILOT_FINAL/G23.21-0.38_3_colour.eps.jpg}
  \includegraphics[width=6.0cm,angle=90]{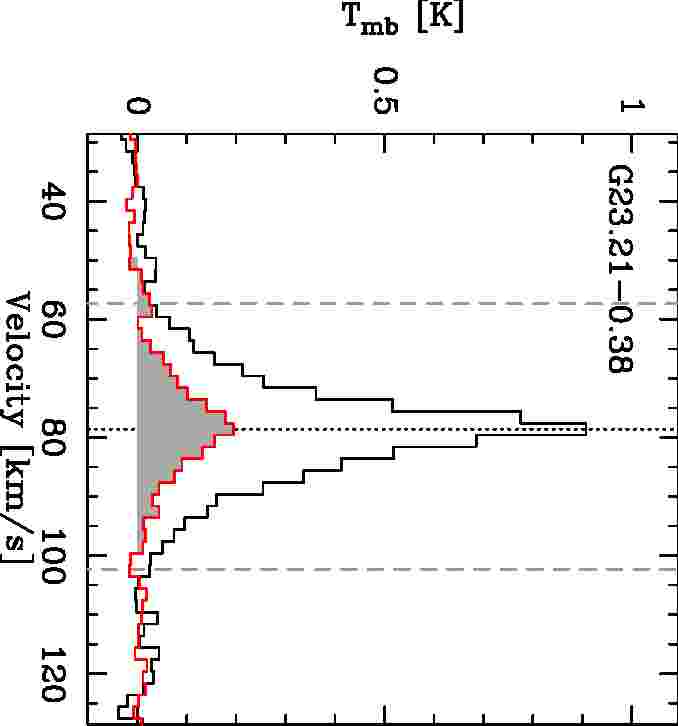}
  \includegraphics[width=6.0cm,angle=0]{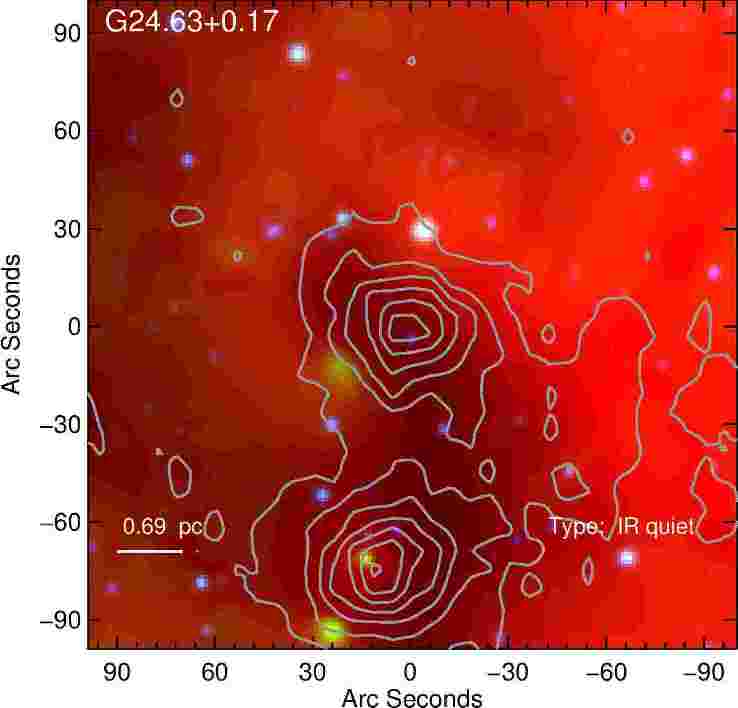}
  \includegraphics[width=5.8cm,angle=90]{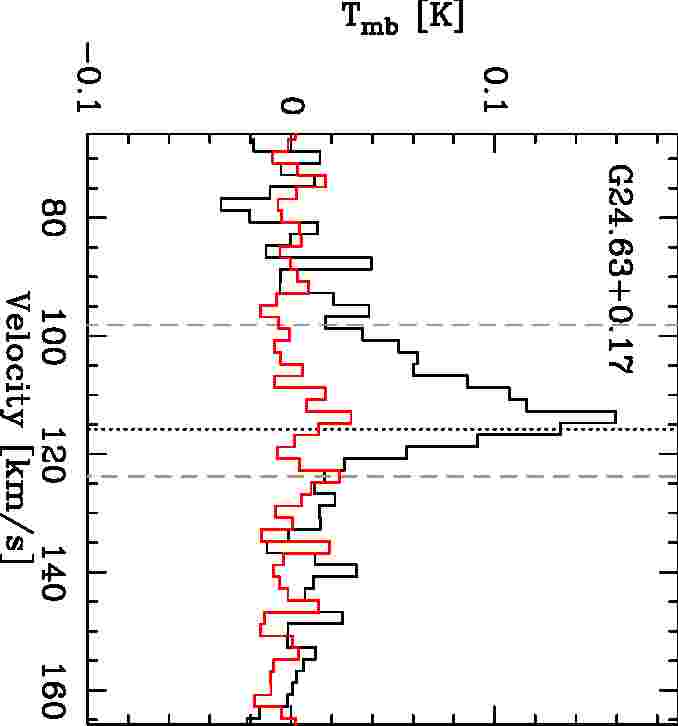}
  \includegraphics[width=6.0cm,angle=0]{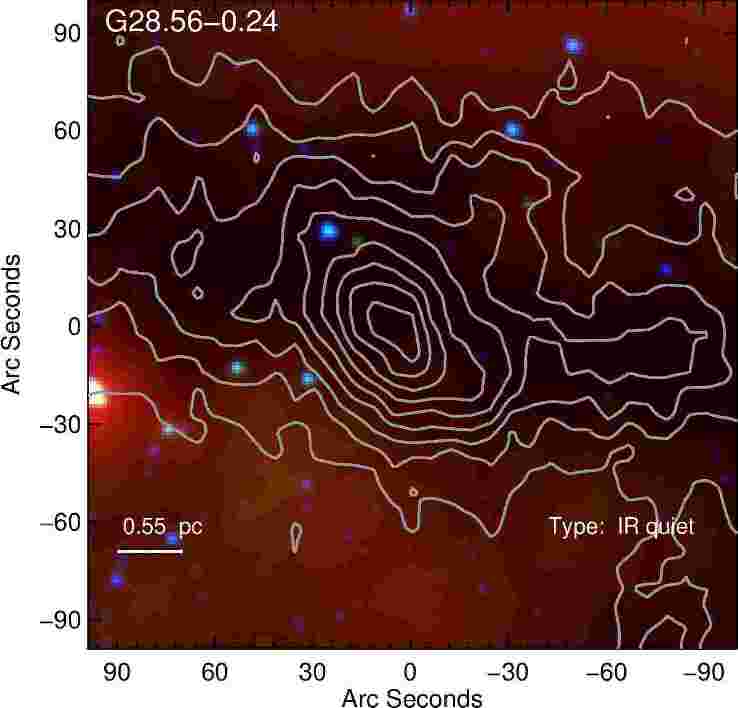}
  \includegraphics[width=6.0cm,angle=90]{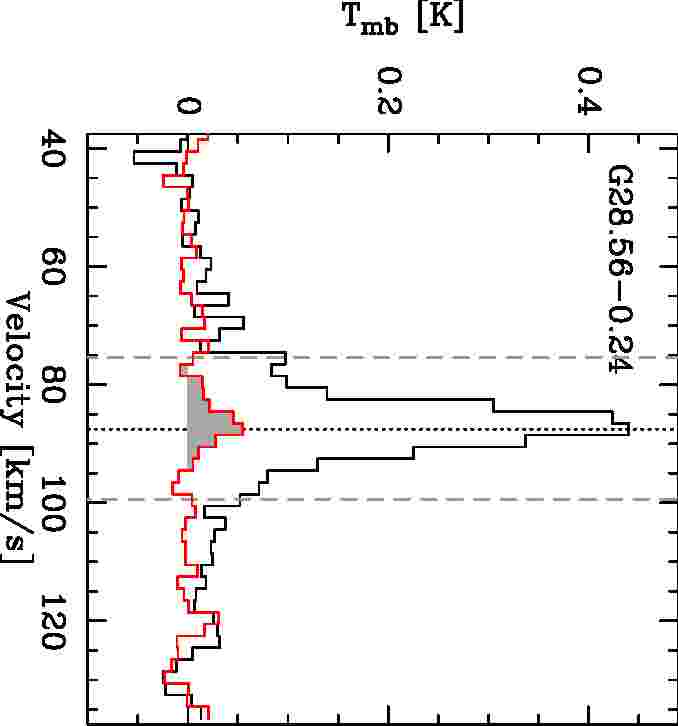}
  \includegraphics[width=6.0cm,angle=0]{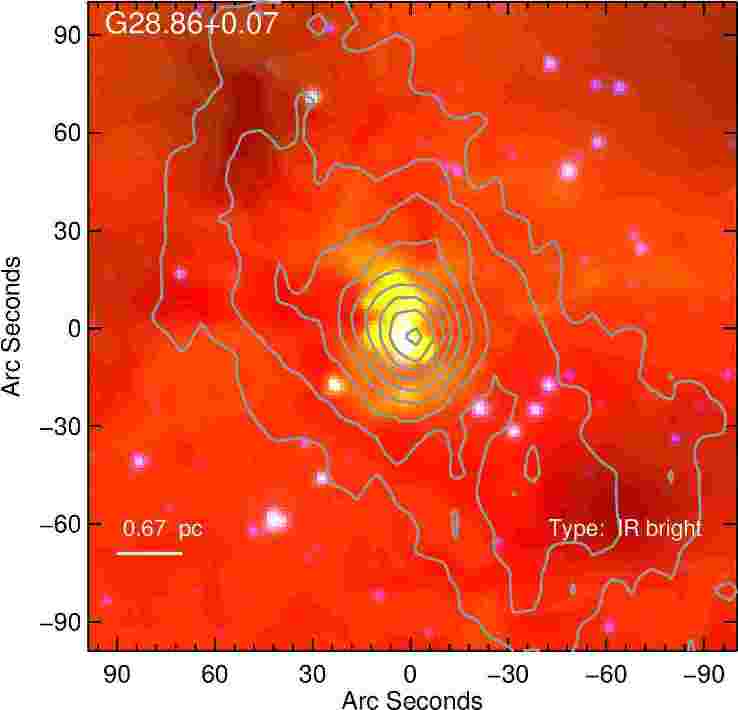}
  \includegraphics[width=6.0cm,angle=90]{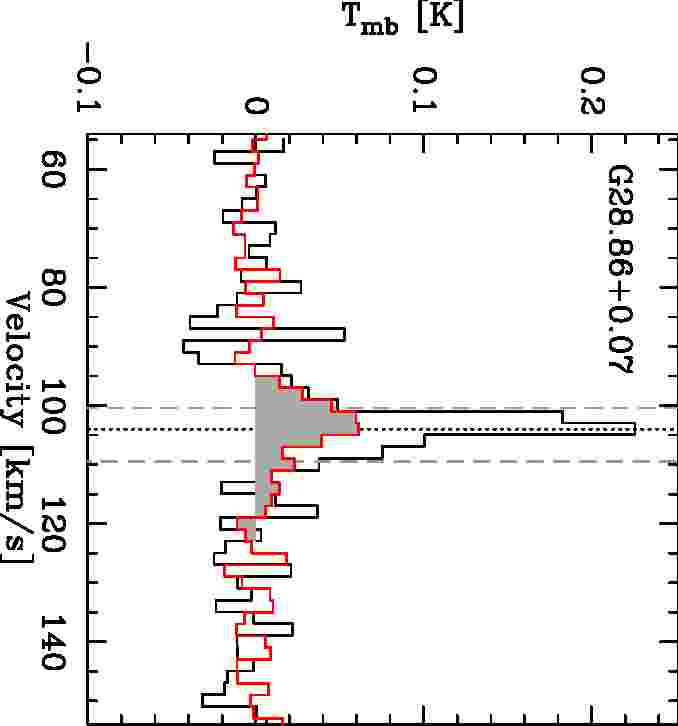}
 \caption{Continued.}
 \end{figure}
 \end{landscape}
 
 \begin{landscape}
 \begin{figure}
\ContinuedFloat
  \includegraphics[width=6.0cm,angle=0]{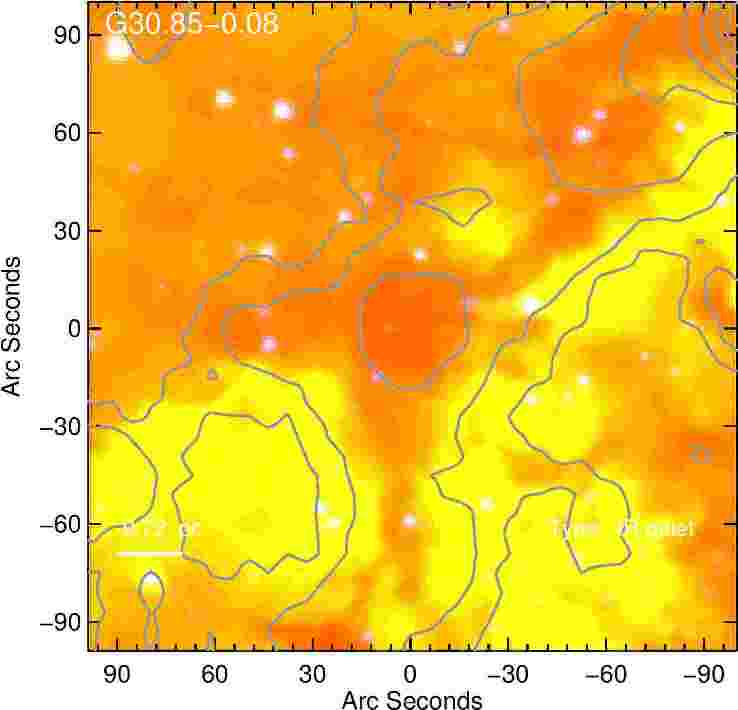}
  \includegraphics[width=6.0cm,angle=90]{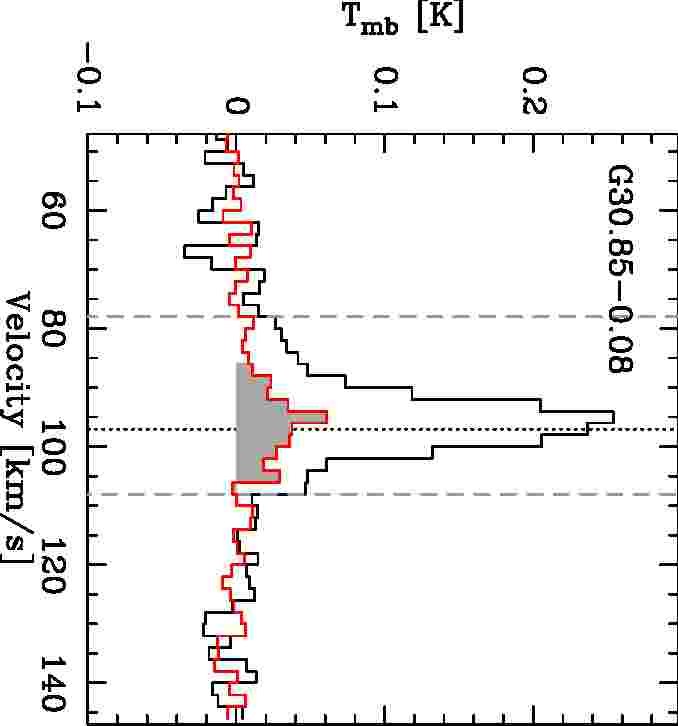}
  \includegraphics[width=6.0cm,angle=0]{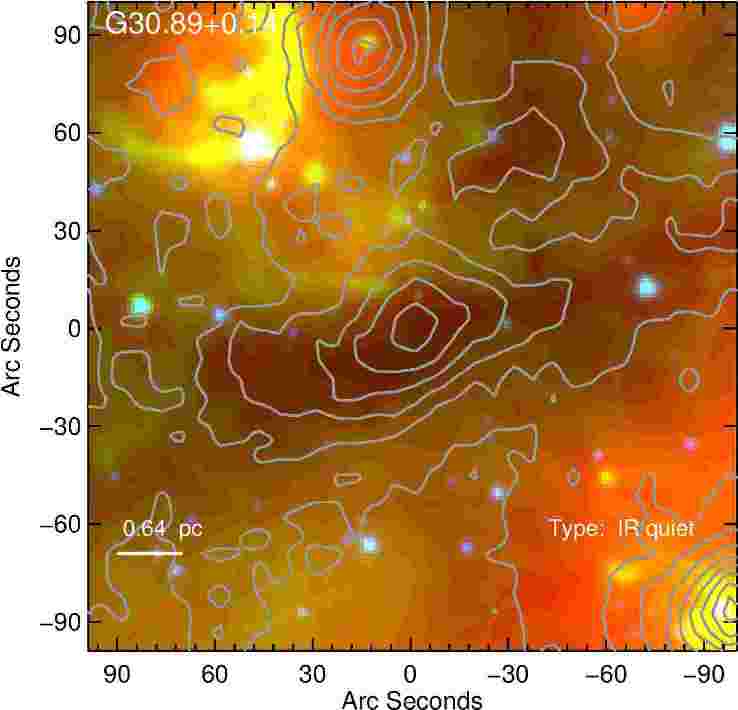}
  \includegraphics[width=6.0cm,angle=90]{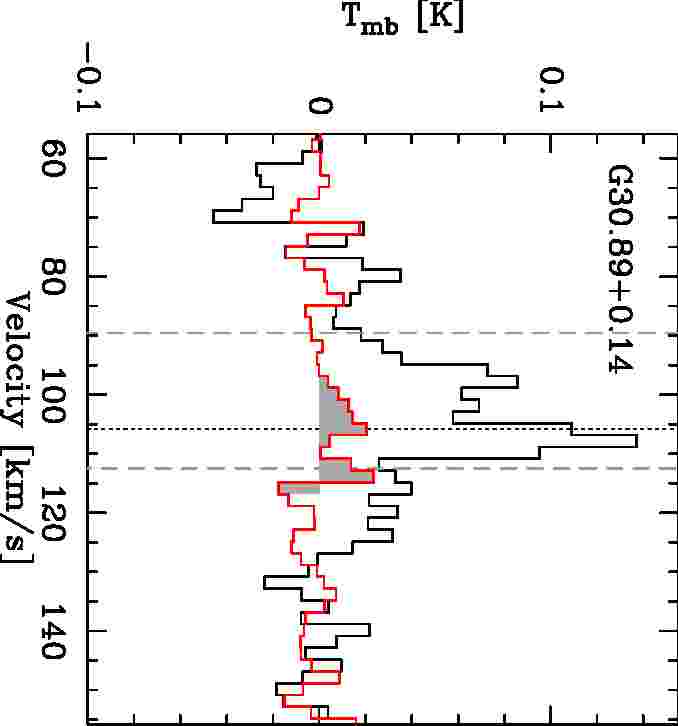}
  \includegraphics[width=6.0cm,angle=0]{FIG_ASTRO_PH/SIO_PILOT_FINAL/G31.41+0.31_3_colour.eps.jpg}
  \includegraphics[width=6.0cm,angle=90]{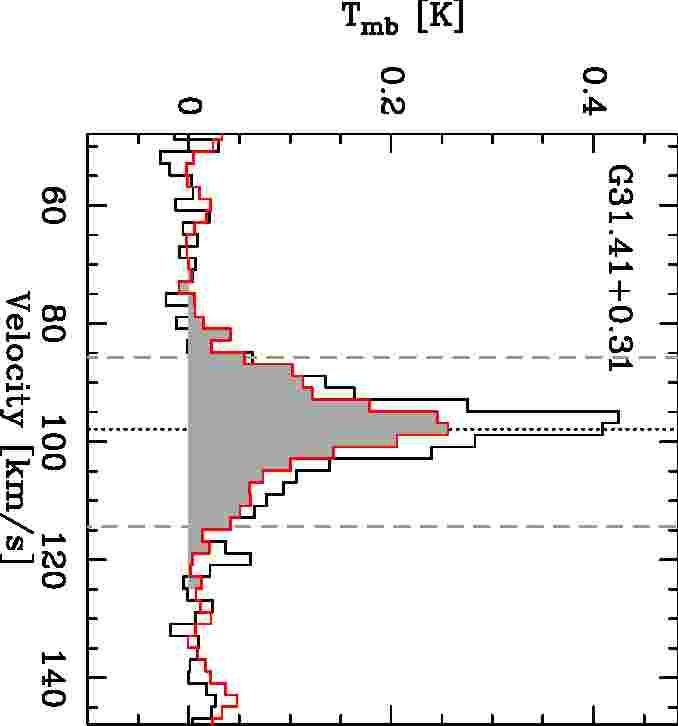}
  \includegraphics[width=6.0cm,angle=0]{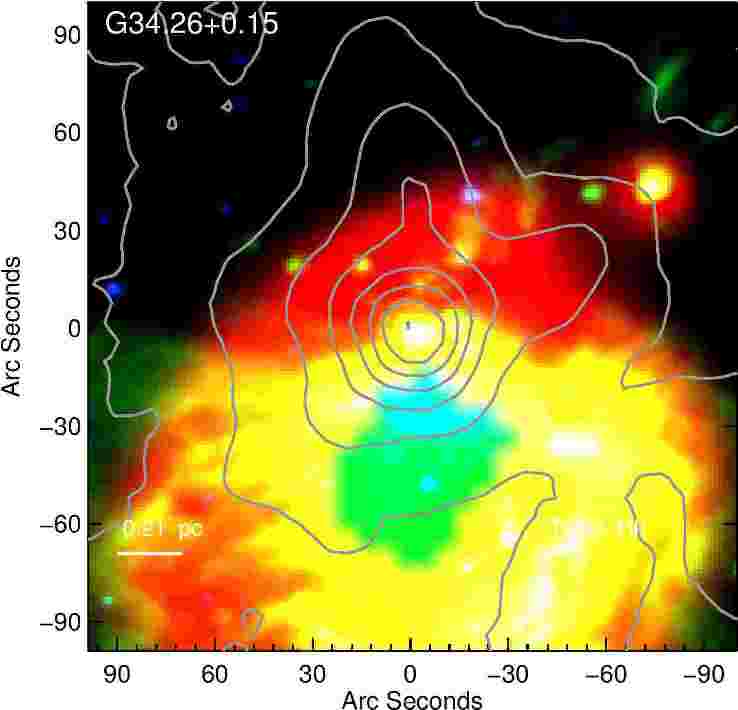}
  \includegraphics[width=6.0cm,angle=90]{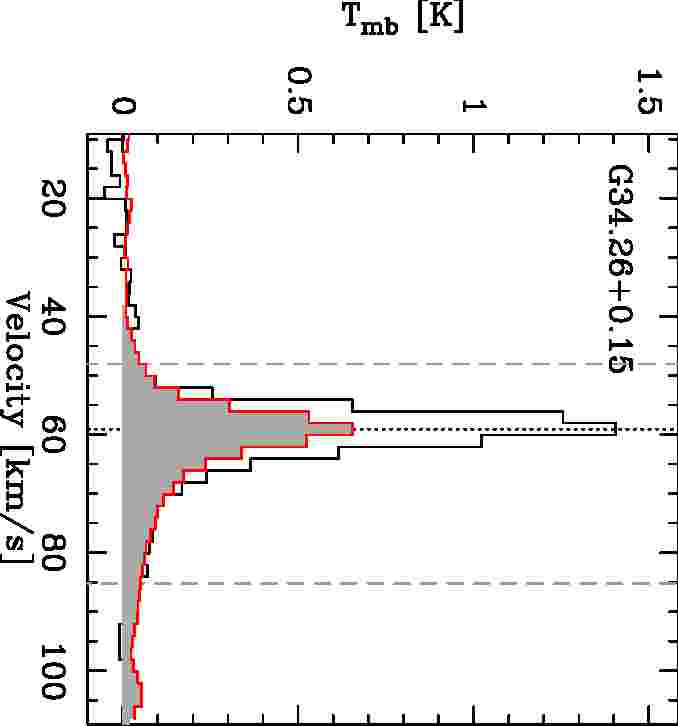}
  \includegraphics[width=6.0cm,angle=0]{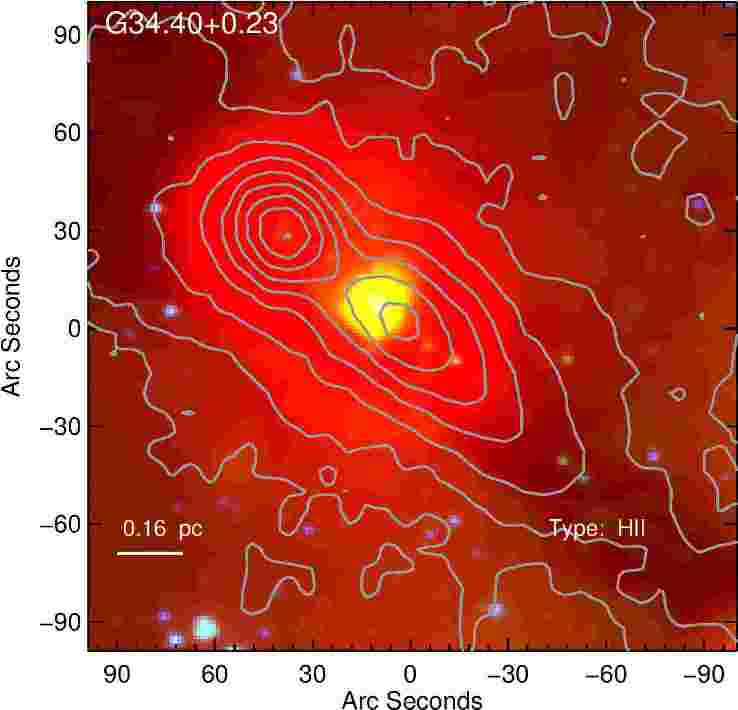}
  \includegraphics[width=6.0cm,angle=90]{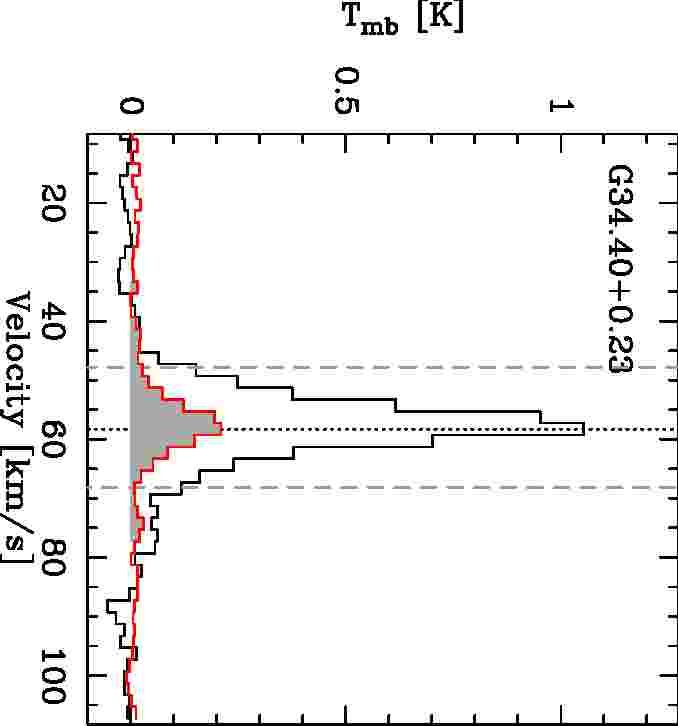}
  \includegraphics[width=6.0cm,angle=0]{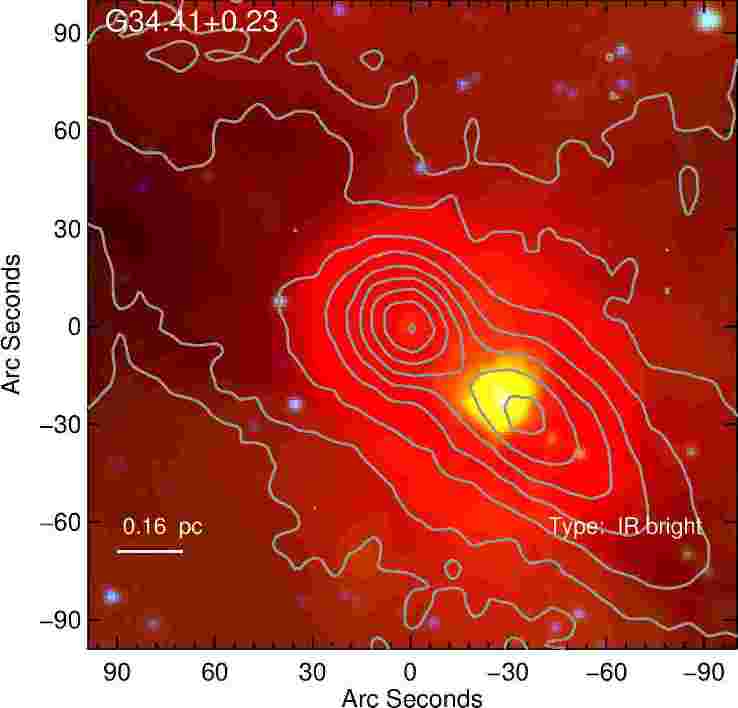}
  \includegraphics[width=6.0cm,angle=90]{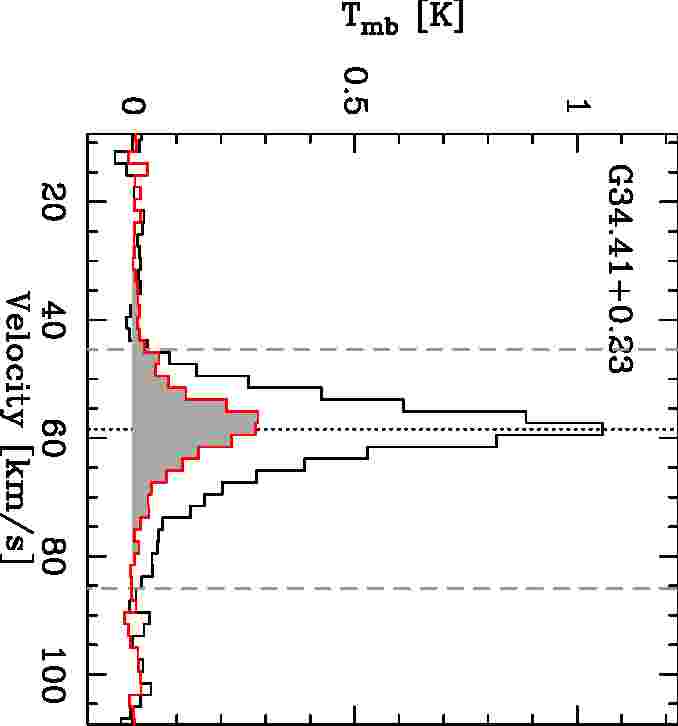}
 \caption{Continued.}
 \end{figure}
 \end{landscape}
 
\begin{landscape} 
\begin{figure}
\ContinuedFloat
  \includegraphics[width=6.0cm,angle=0]{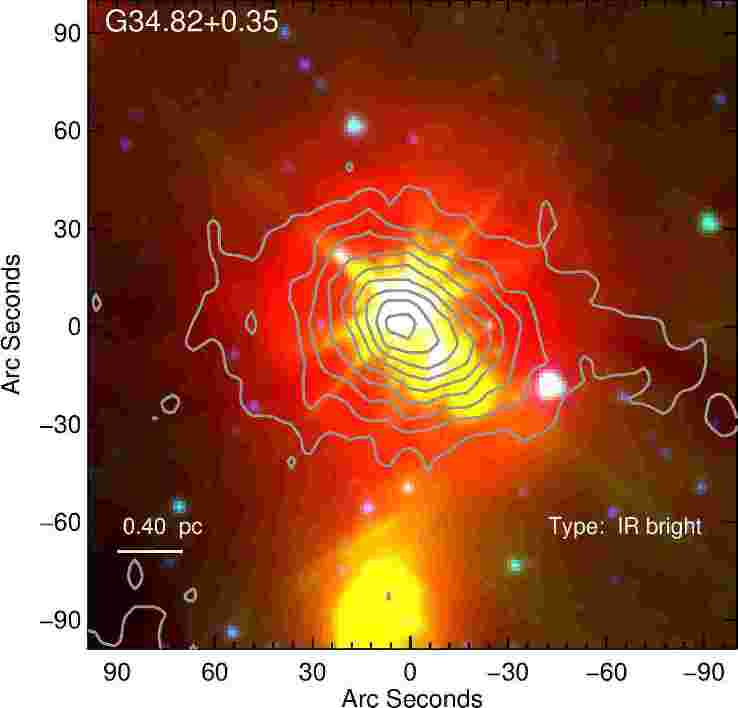}
  \includegraphics[width=6.0cm,angle=90]{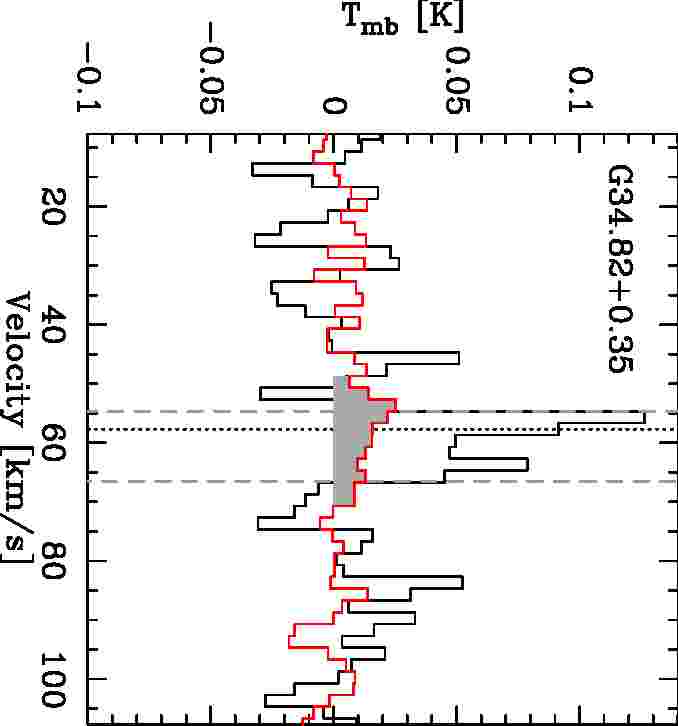}
  \includegraphics[width=6.0cm,angle=0]{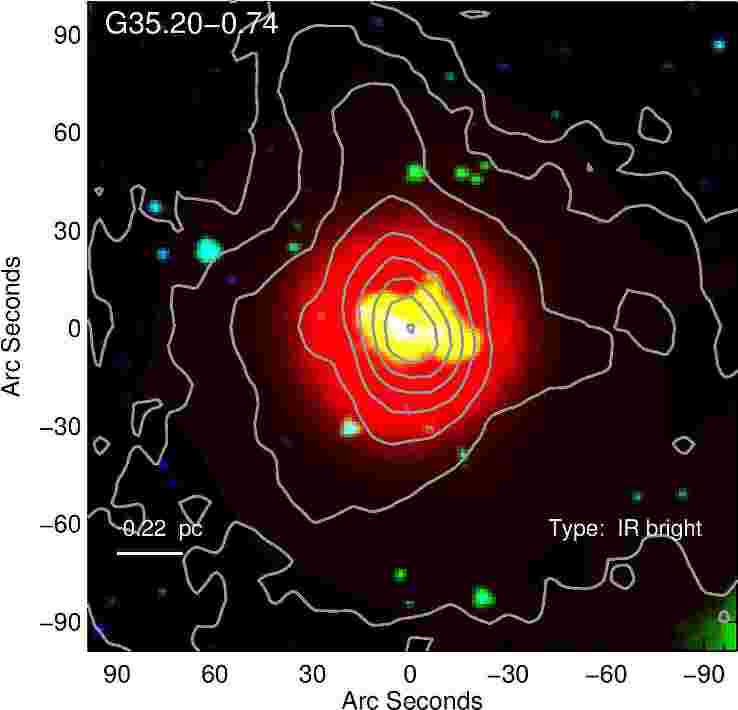}
  \includegraphics[width=6.0cm,angle=90]{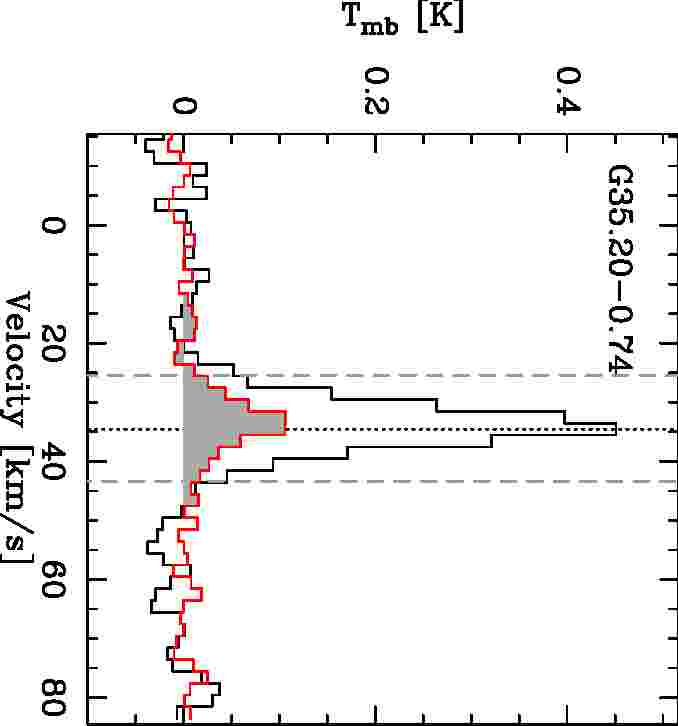}
  \includegraphics[width=6.0cm,angle=0]{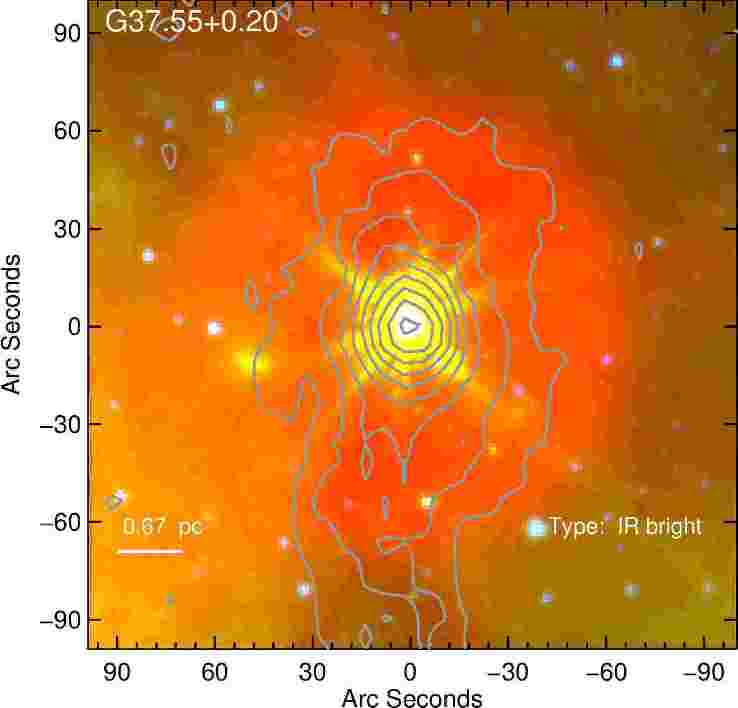}
  \includegraphics[width=5.8cm,angle=90]{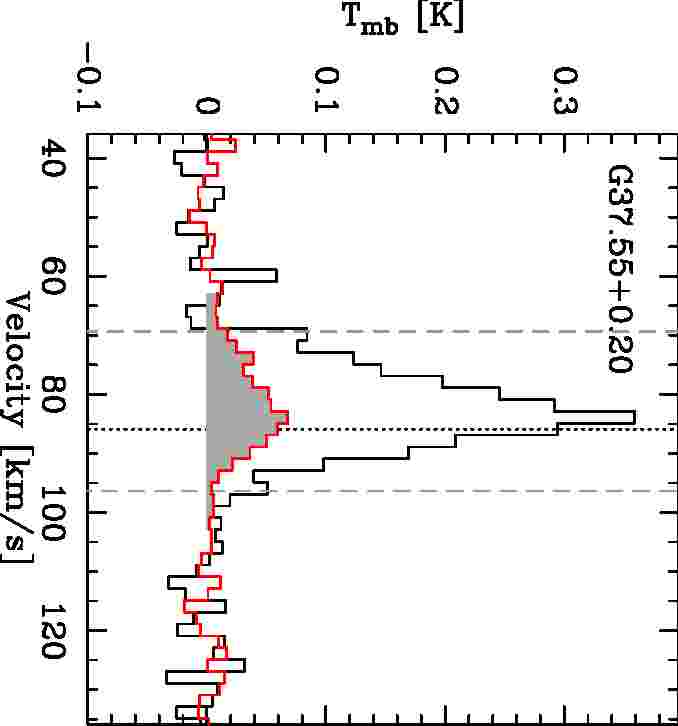}
  \includegraphics[width=6.0cm,angle=0]{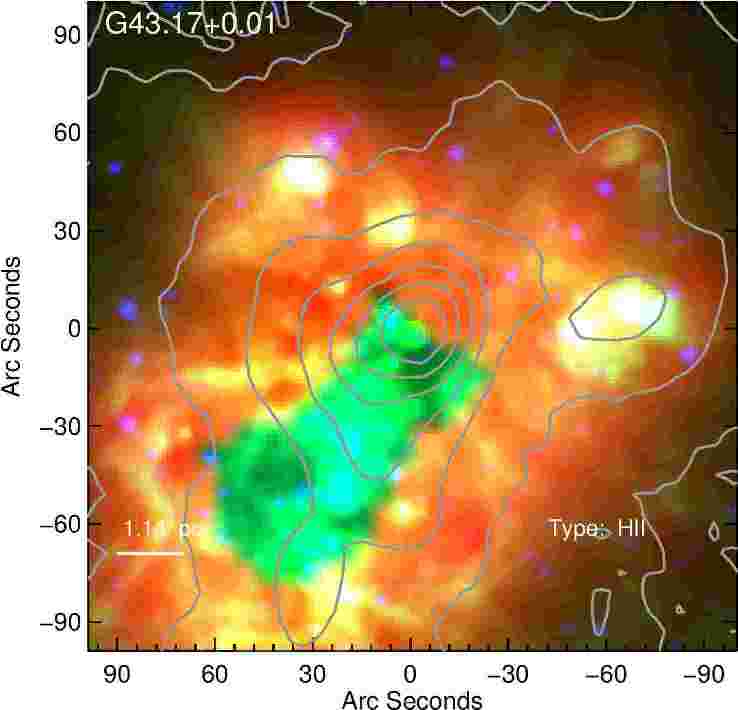}
  \includegraphics[width=6.0cm,angle=90]{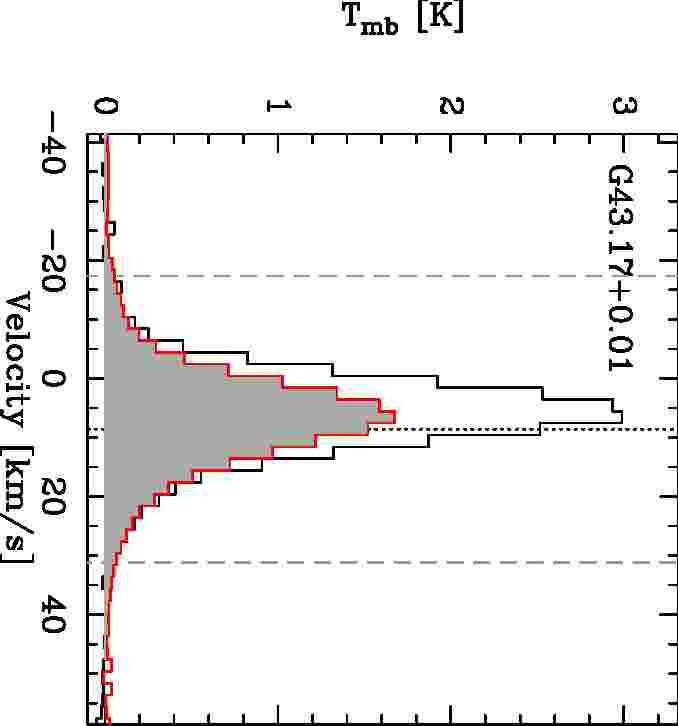}
  \includegraphics[width=6.0cm,angle=0]{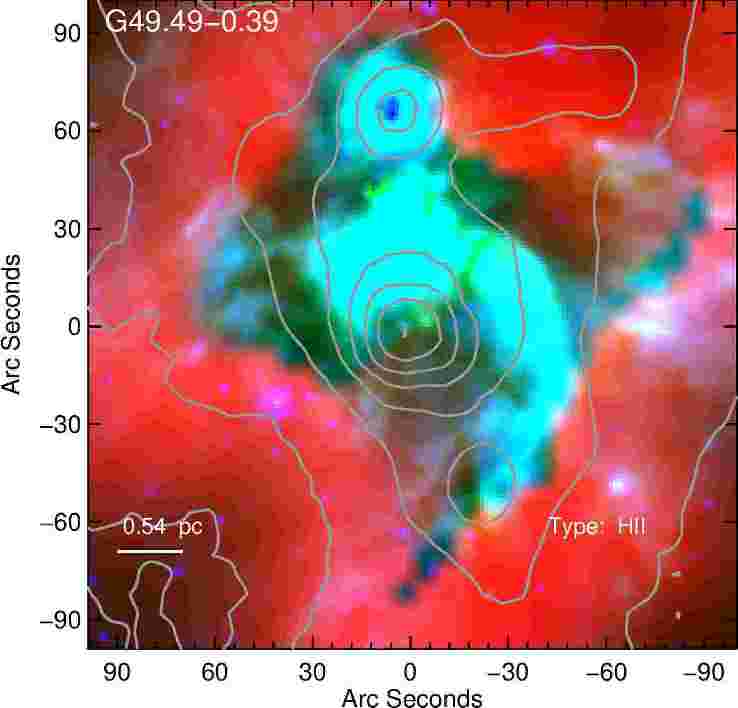}
  \includegraphics[width=6.0cm,angle=90]{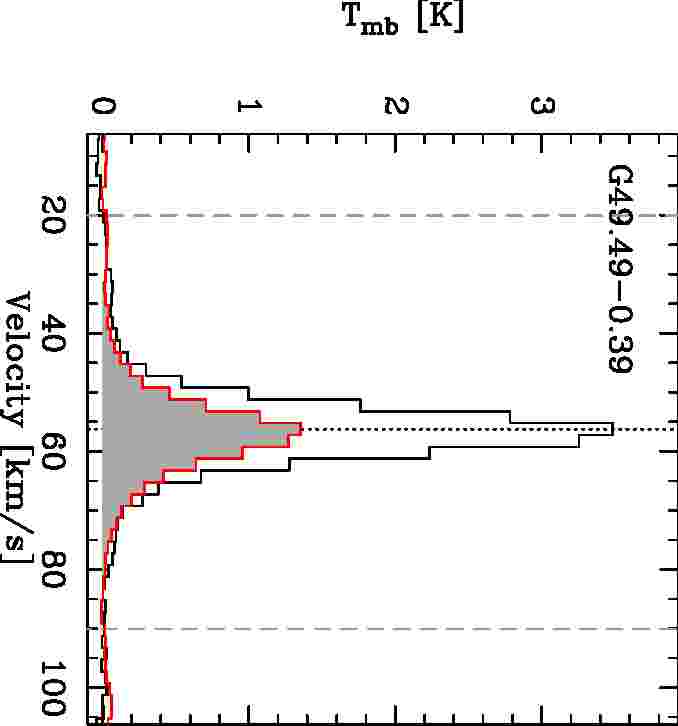}
  \includegraphics[width=6.0cm,angle=0]{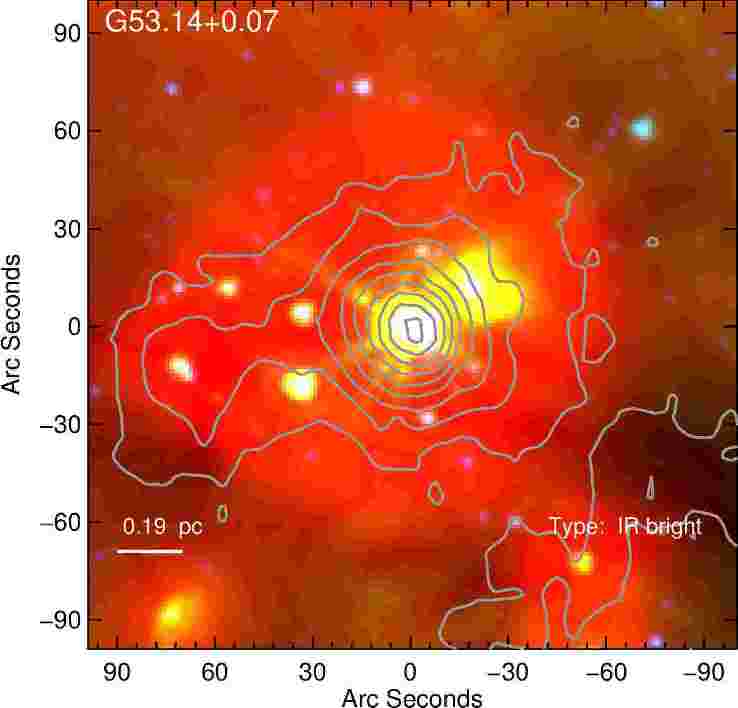}
  \includegraphics[width=6.0cm,angle=90]{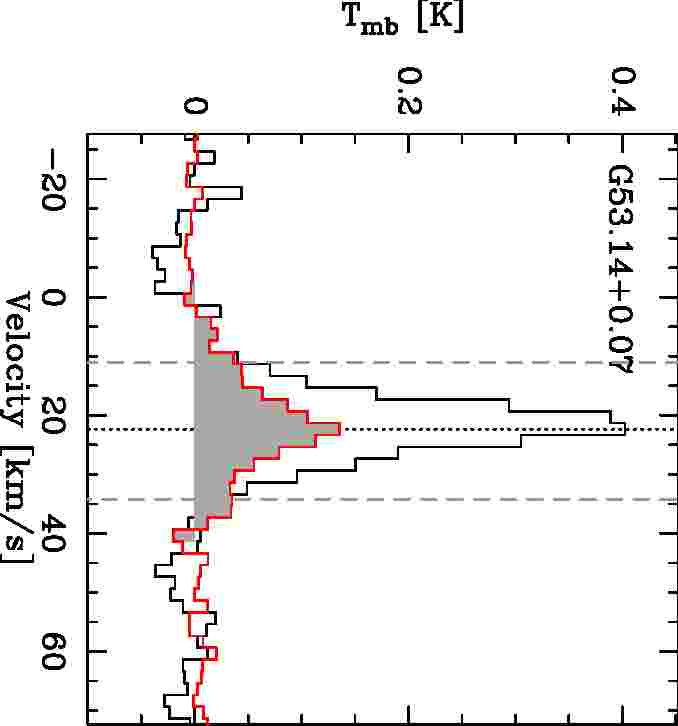}
 \caption{Continued.}
 \end{figure}
 \end{landscape}
 
\begin{landscape} 
\begin{figure}
\ContinuedFloat
  \includegraphics[width=6.0cm,angle=0]{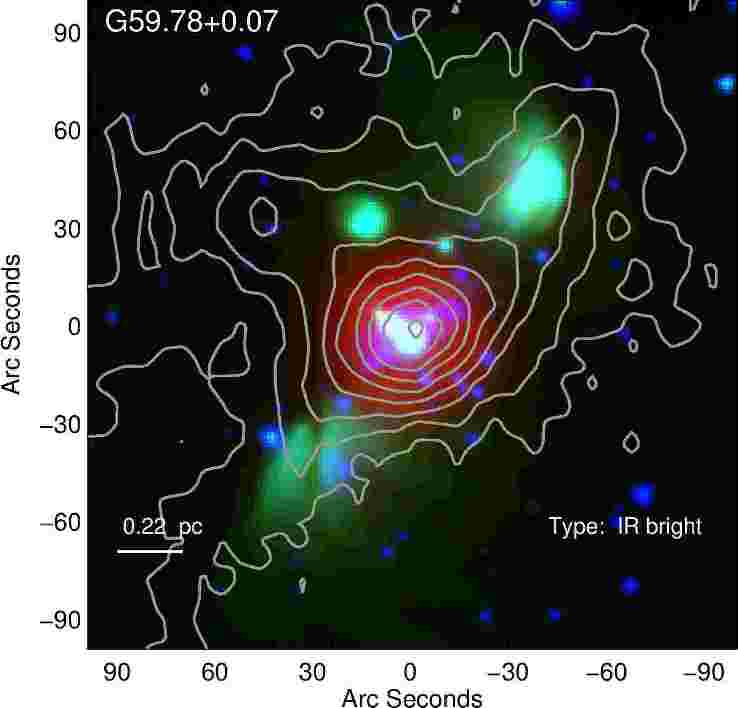}
  \includegraphics[width=6.0cm,angle=90]{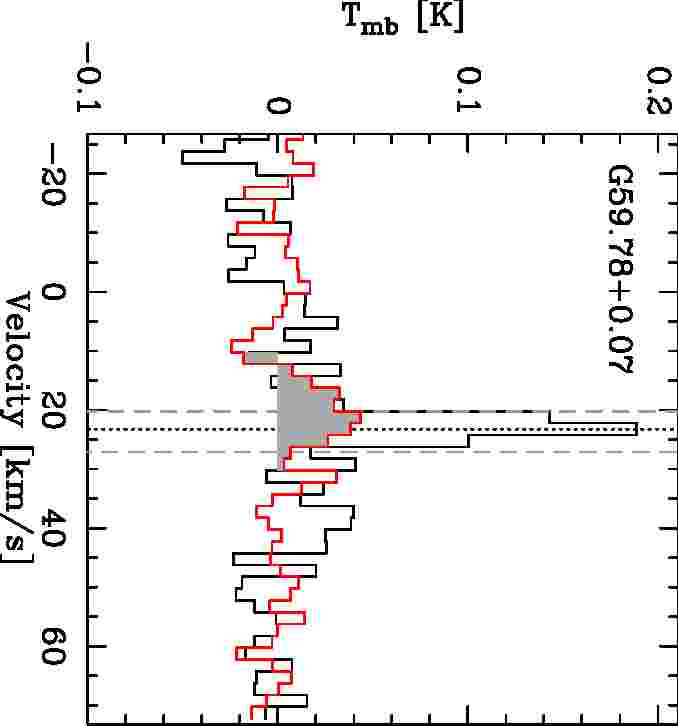}
 \caption{Continued.}
 \end{figure}
 \end{landscape}
 
 \clearpage

\begin{landscape}
\begin{figure}
  \includegraphics[width=6.0cm,angle=0]{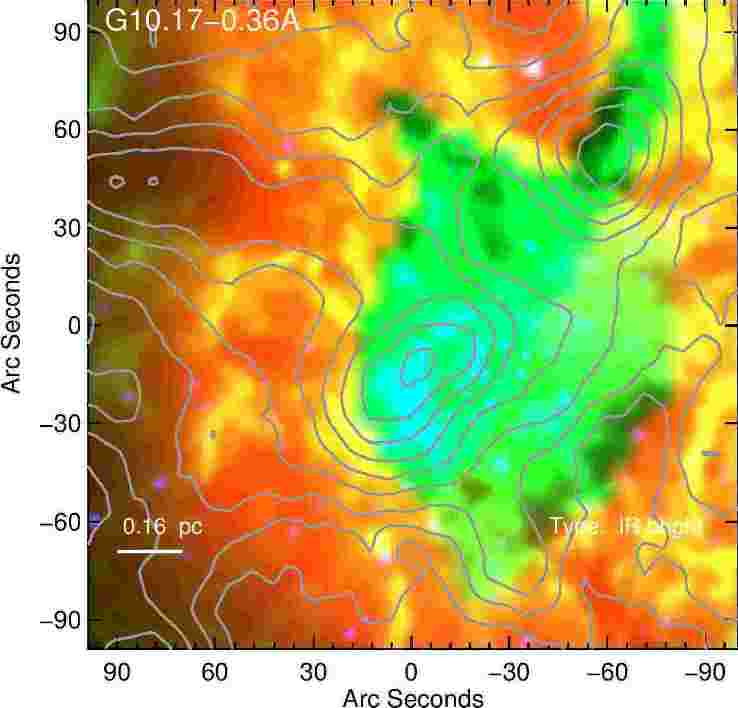}
  \includegraphics[width=6.0cm,angle=90]{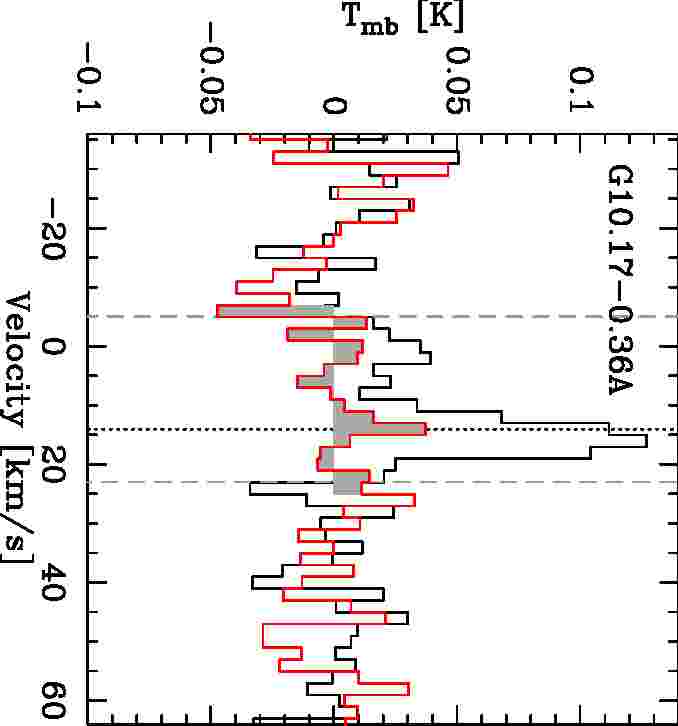}
  \includegraphics[width=6.0cm,angle=0]{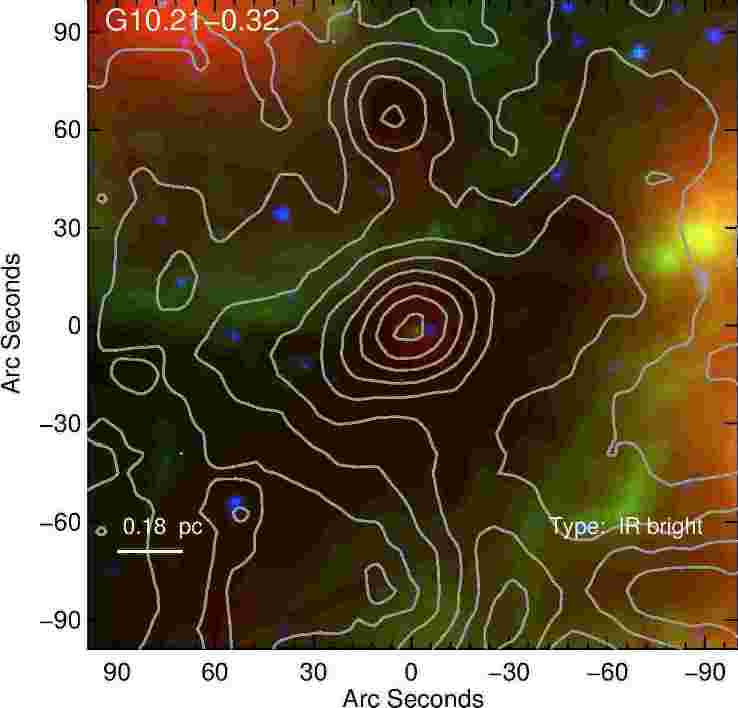}
  \includegraphics[width=6.0cm,angle=90]{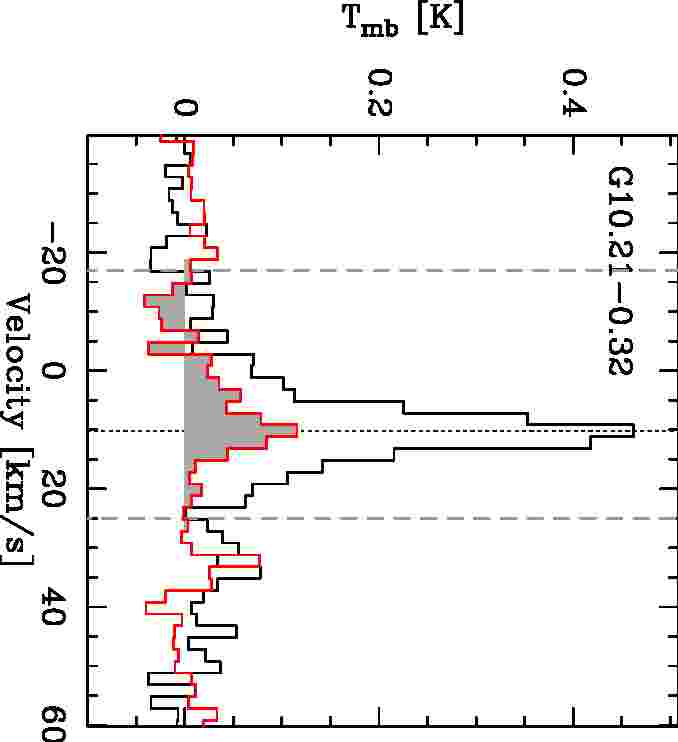}
  \includegraphics[width=6.0cm,angle=0]{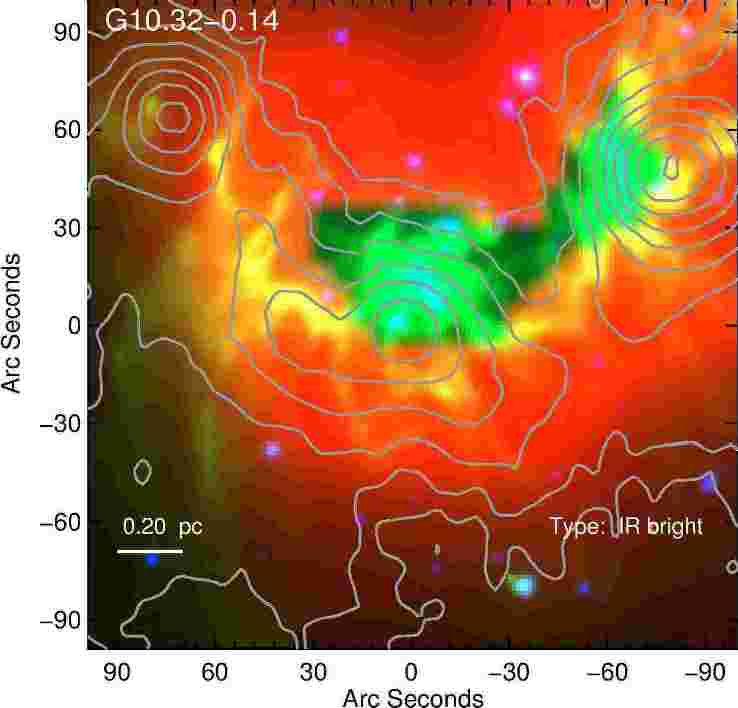}
  \includegraphics[width=5.8cm,angle=90]{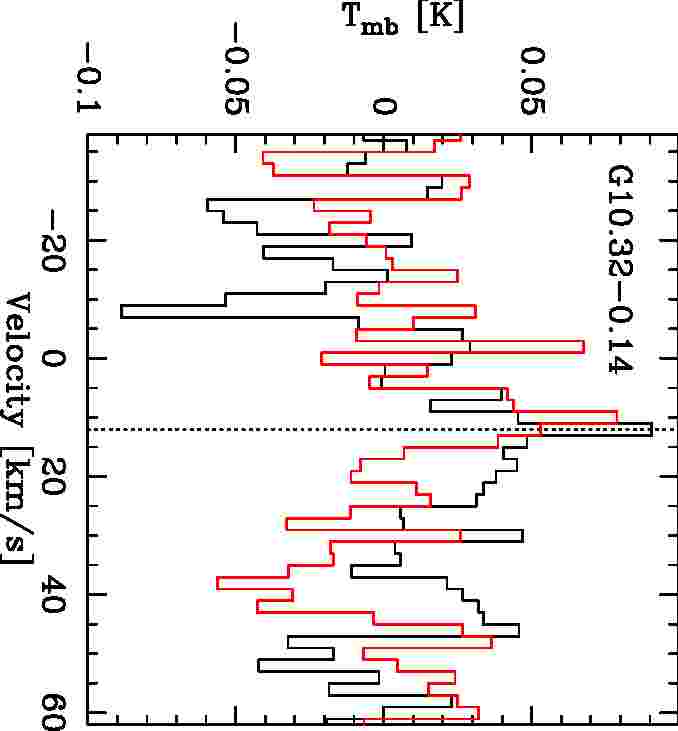} 
  \includegraphics[width=6.0cm,angle=0]{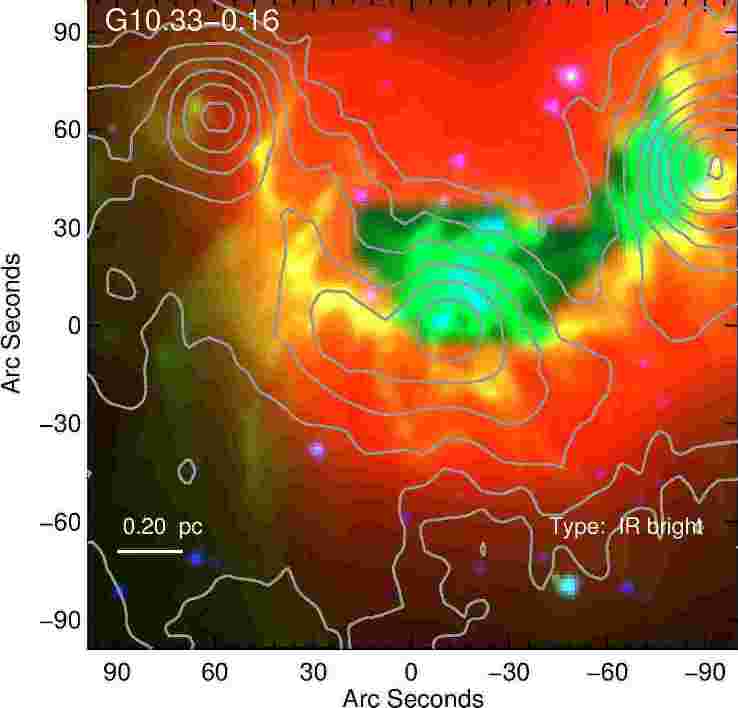}
  \includegraphics[width=6.0cm,angle=90]{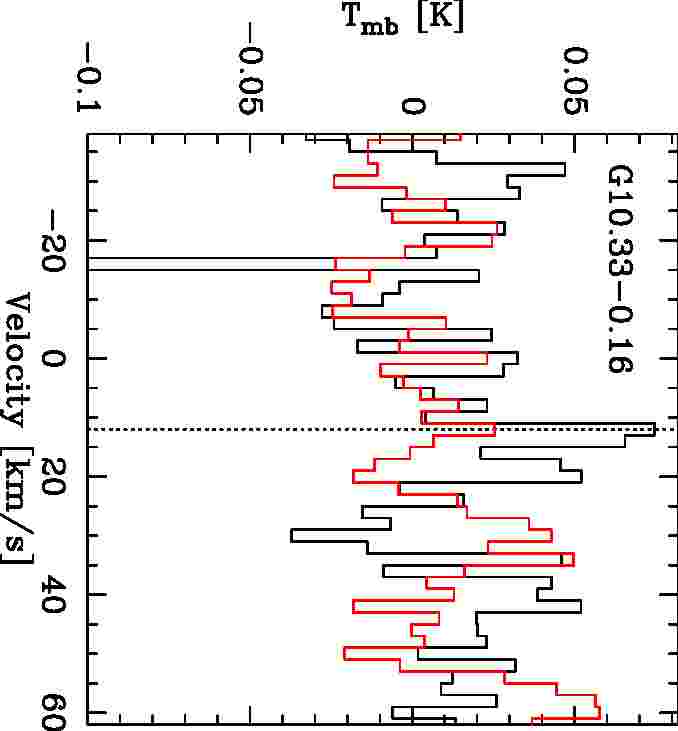}
  \includegraphics[width=6.0cm,angle=0]{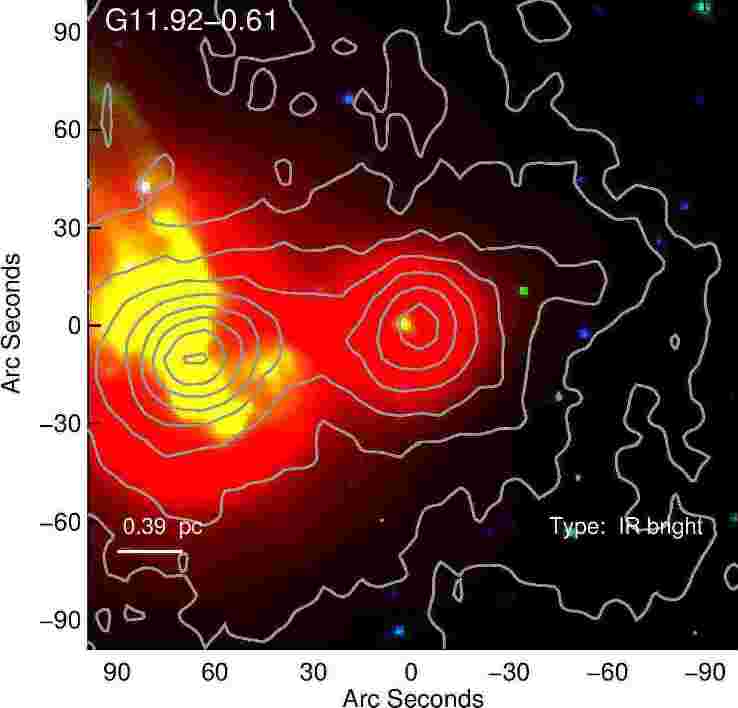}
  \includegraphics[width=6.0cm,angle=90]{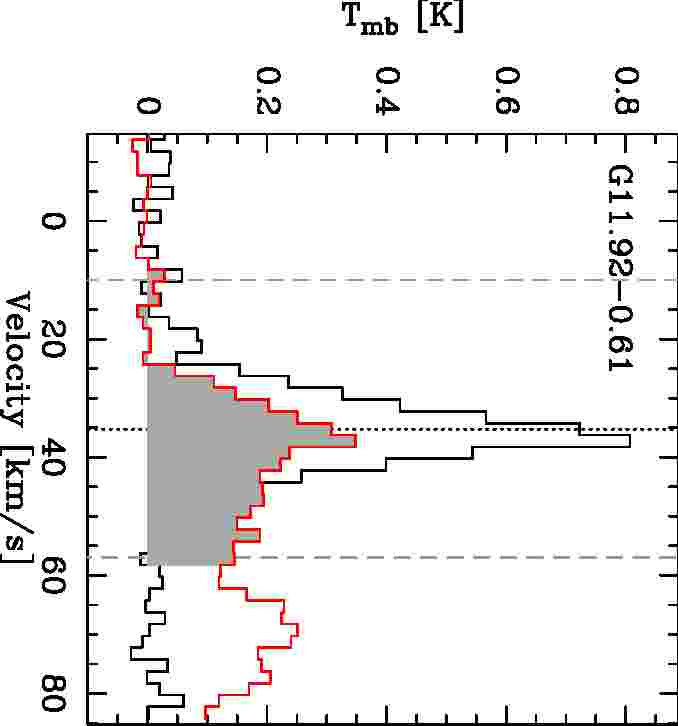}
  \includegraphics[width=6.0cm,angle=0]{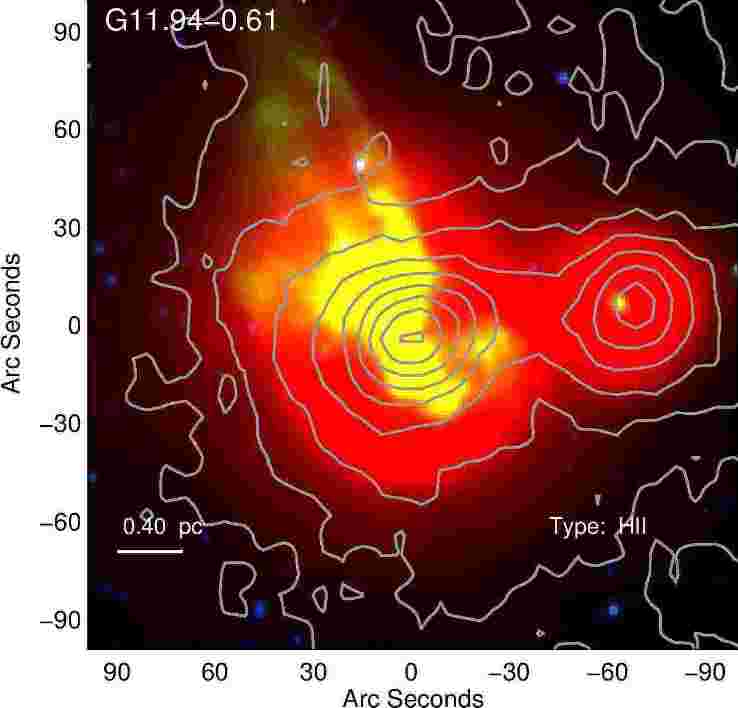}
  \includegraphics[width=6.0cm,angle=90]{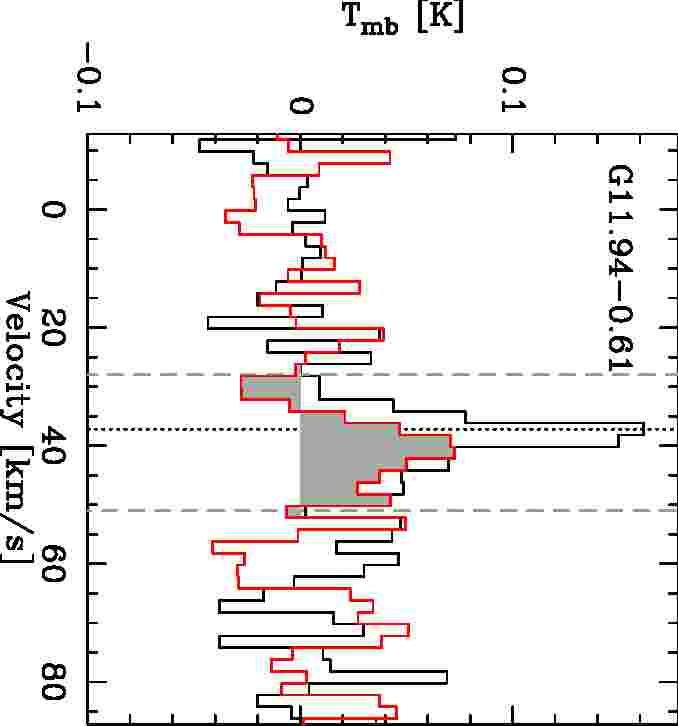}
 \caption{Same as Fig.\,\ref{app:fig1}, for the sources observed in the 
               second IRAM~30m observing campaign. Only those sources
               are shown in colour composite, which have been targeted
               in the SiO ($5-4$) line with APEX.}
 \end{figure}
\end{landscape}

\begin{landscape}
\begin{figure}
\ContinuedFloat 
  \includegraphics[width=6.0cm,angle=0]{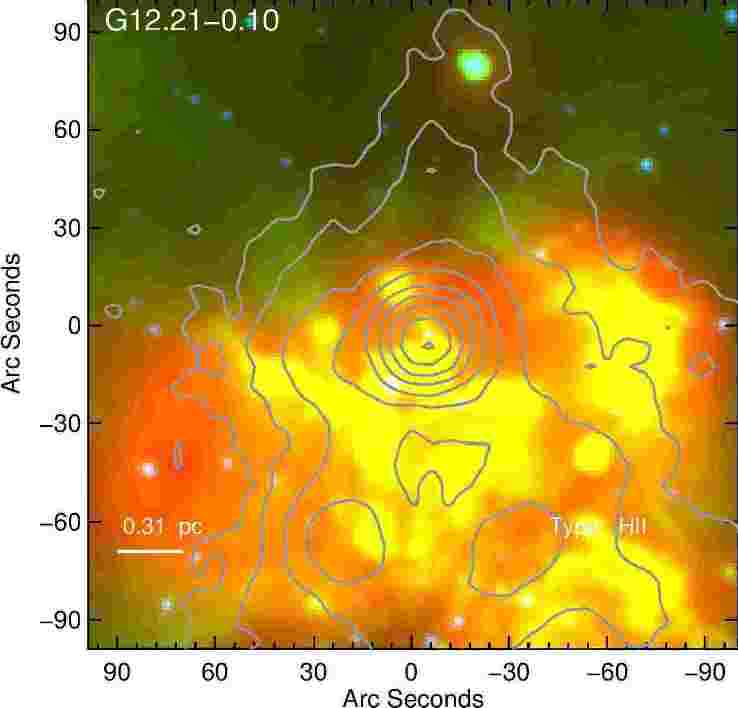}
  \includegraphics[width=6.0cm,angle=90]{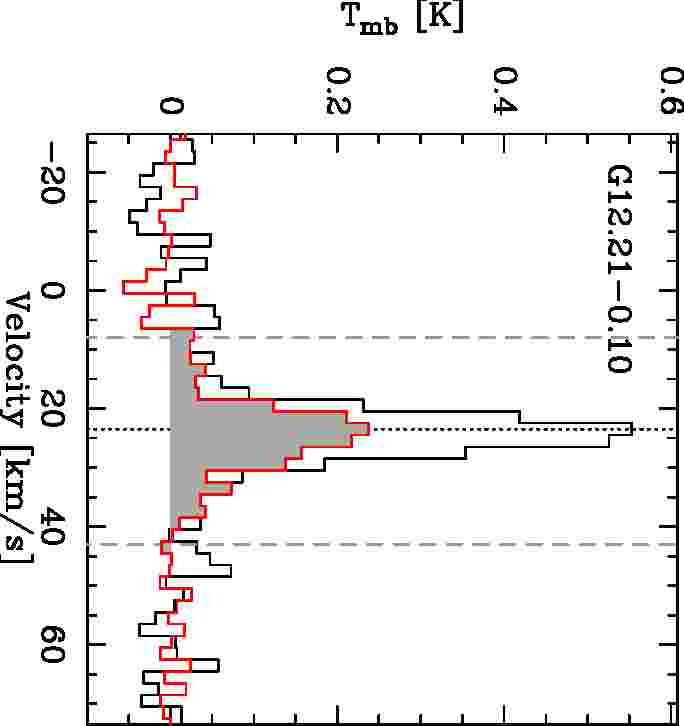}
  \includegraphics[width=6.0cm,angle=0]{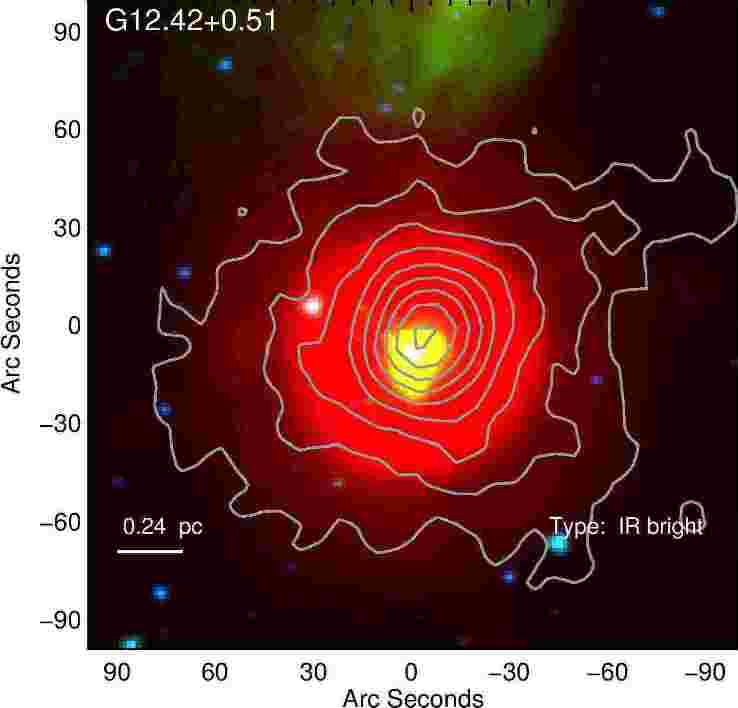}
  \includegraphics[width=6.0cm,angle=90]{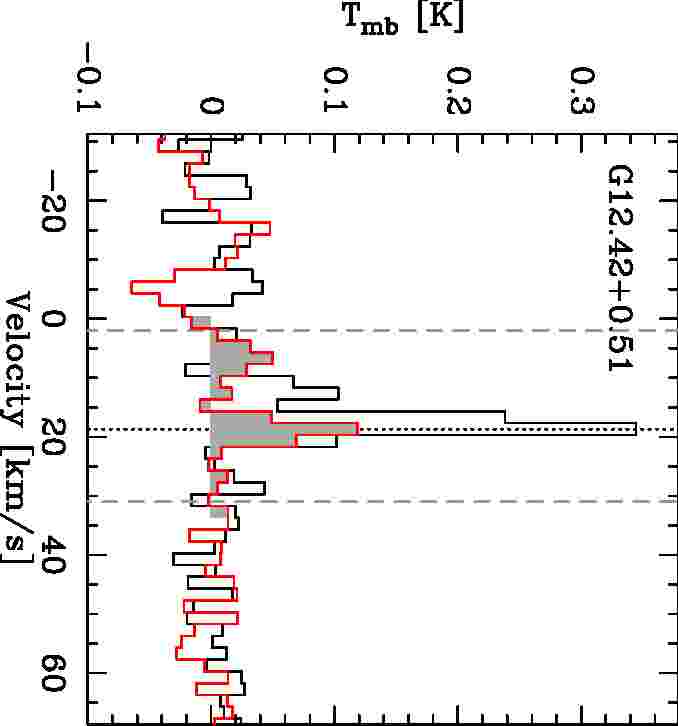}
  \includegraphics[width=6.0cm,angle=0]{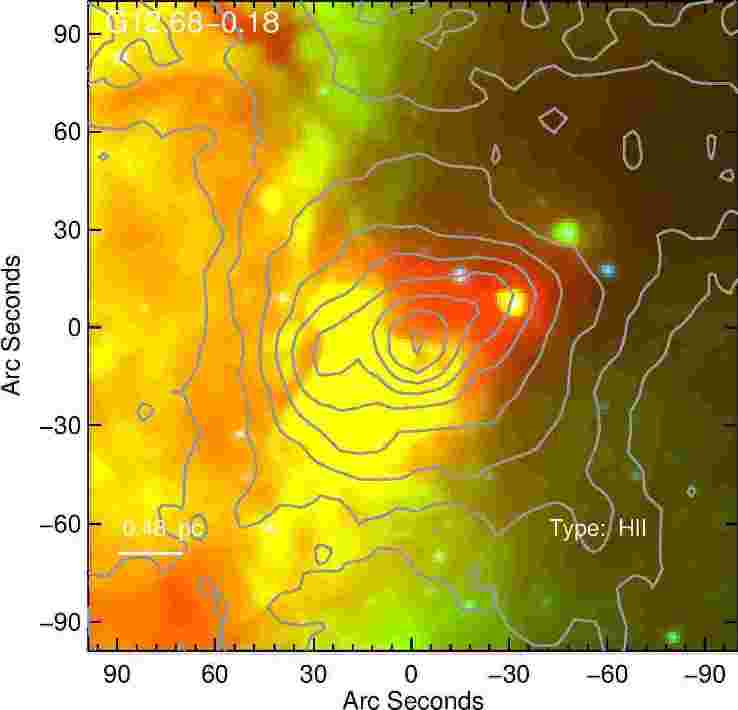}
  \includegraphics[width=6.0cm,angle=90]{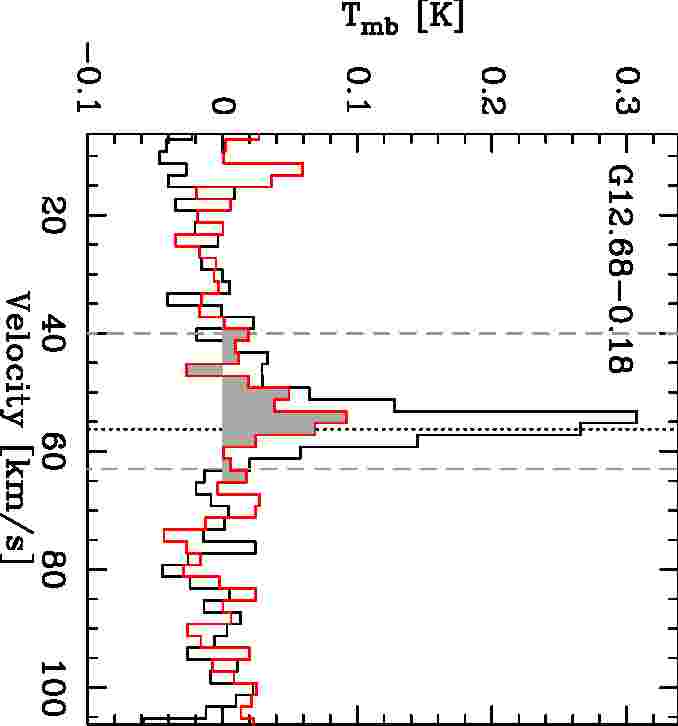}
  \includegraphics[width=6.0cm,angle=0]{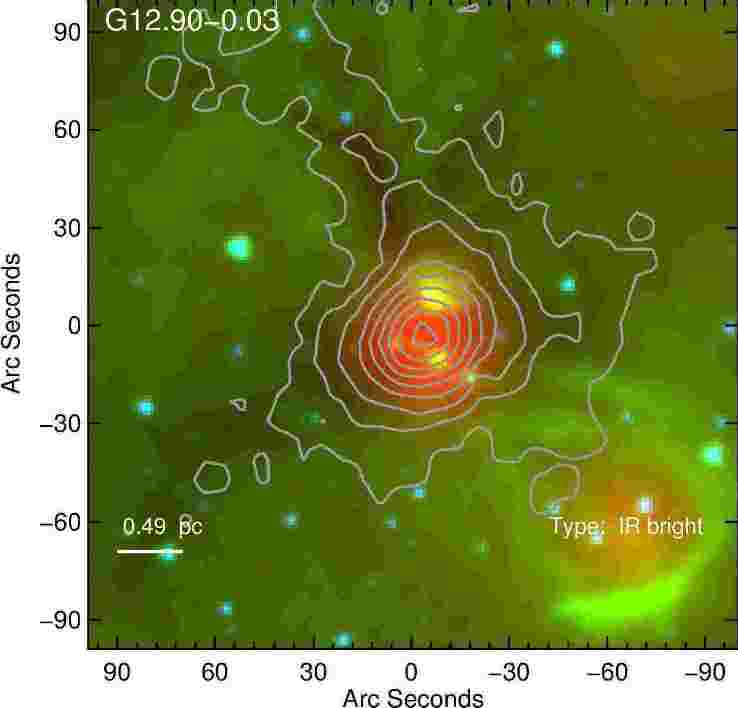}
  \includegraphics[width=6.0cm,angle=90]{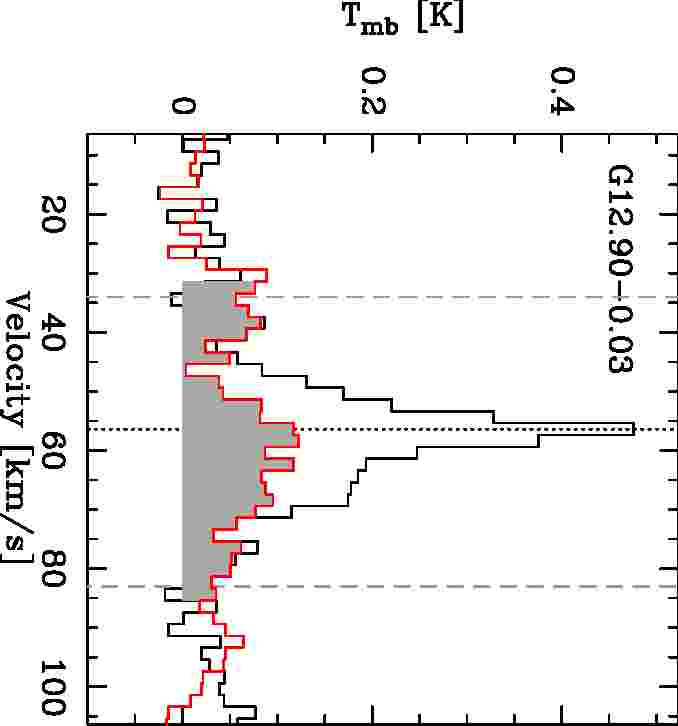}
  \includegraphics[width=6.0cm,angle=0]{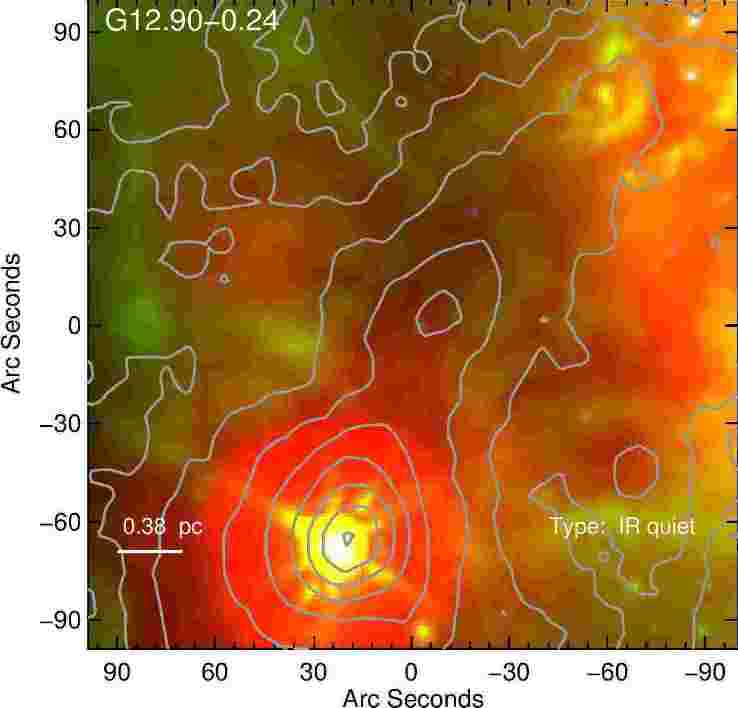}
  \includegraphics[width=6.0cm,angle=90]{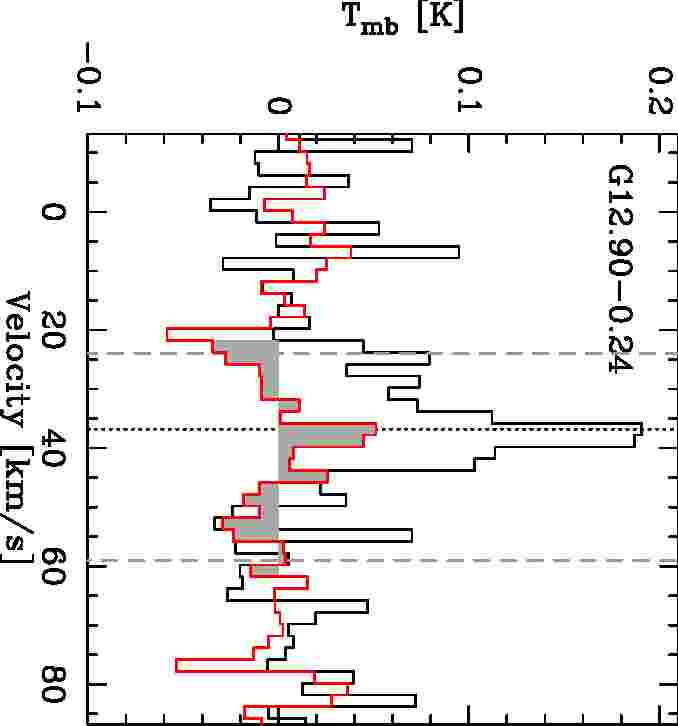}
  \includegraphics[width=6.0cm,angle=0]{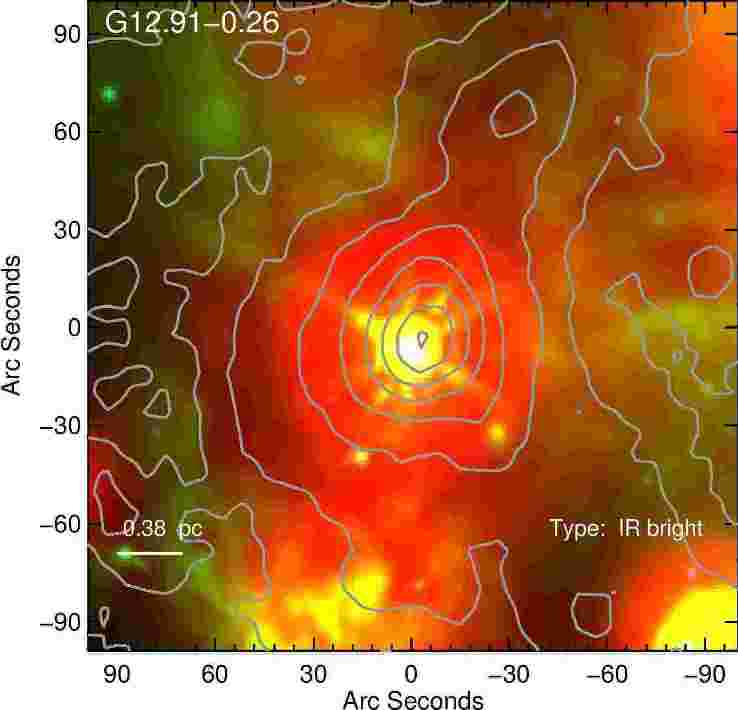}
  \includegraphics[width=6.0cm,angle=90]{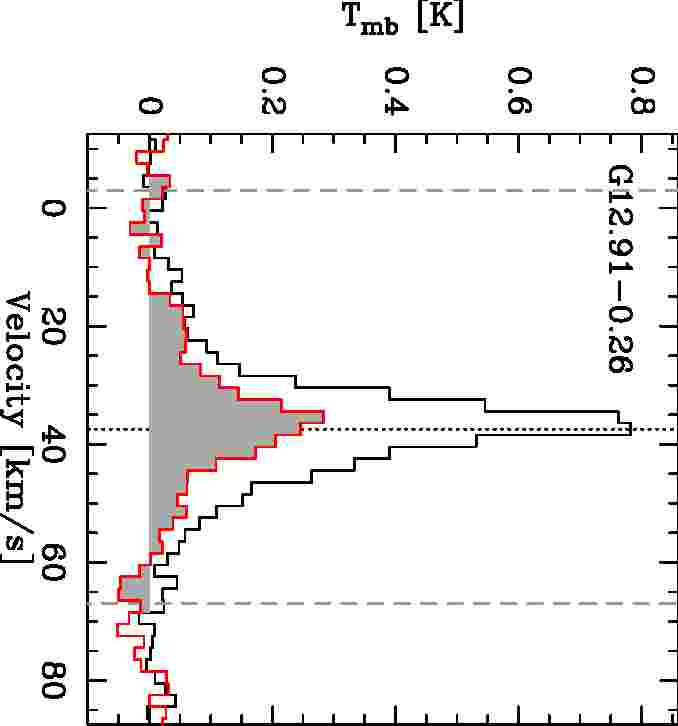}
 \caption{Continued.}
 \end{figure}
 \end{landscape}

\begin{landscape}
\begin{figure}
  \includegraphics[width=6.0cm,angle=0]{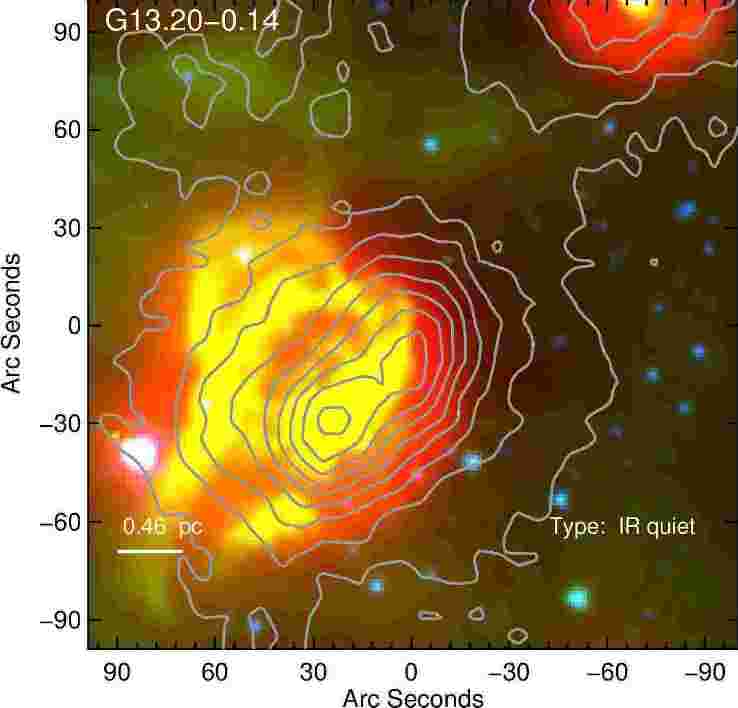}
  \includegraphics[width=6.0cm,angle=90]{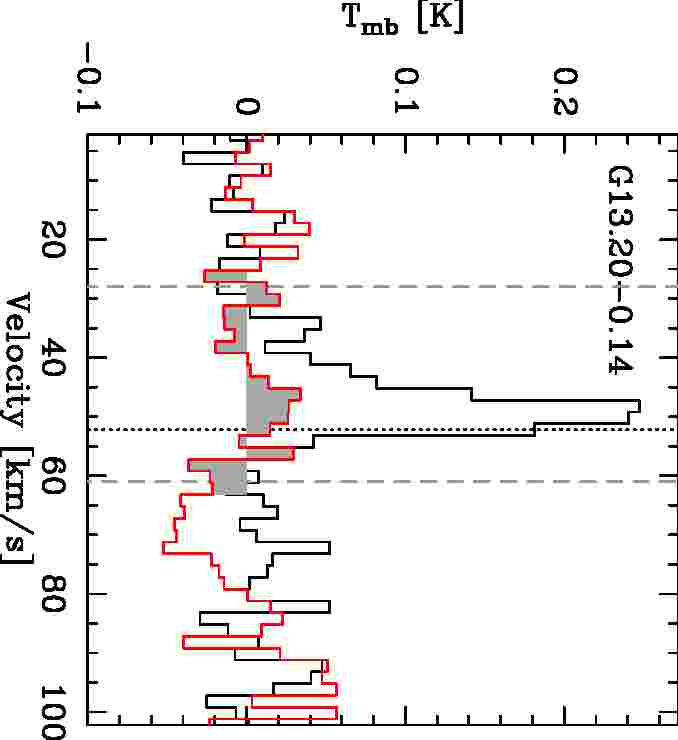}
  \includegraphics[width=6.0cm,angle=0]{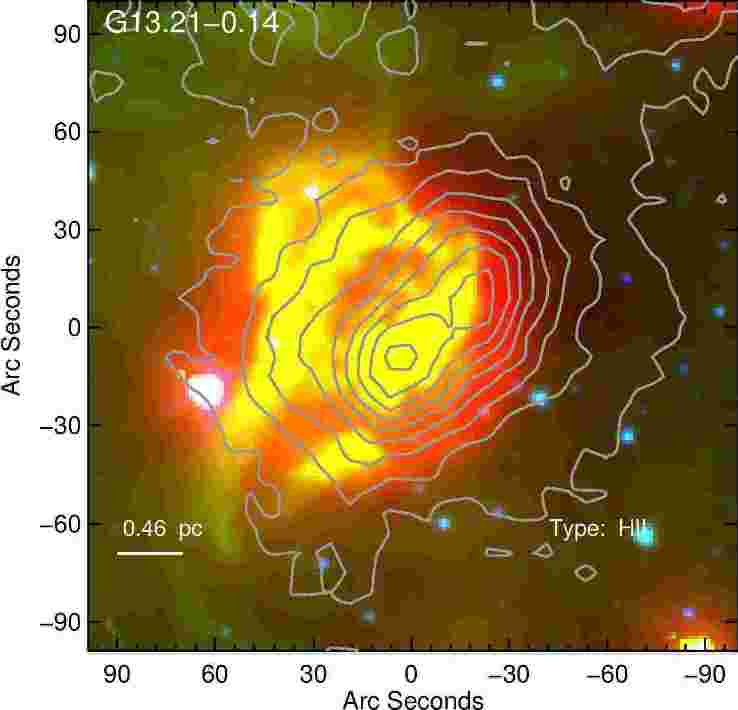}
  \includegraphics[width=6.0cm,angle=90]{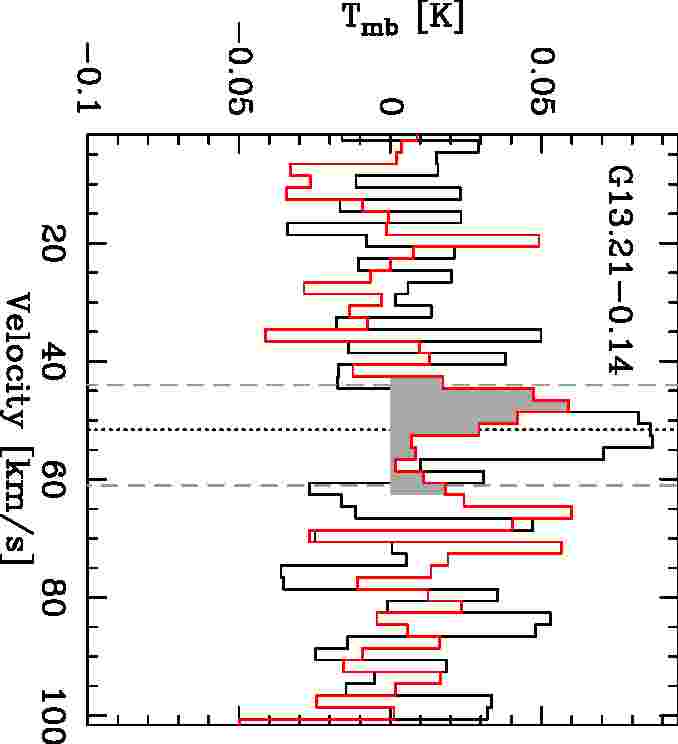}
  \includegraphics[width=6.0cm,angle=0]{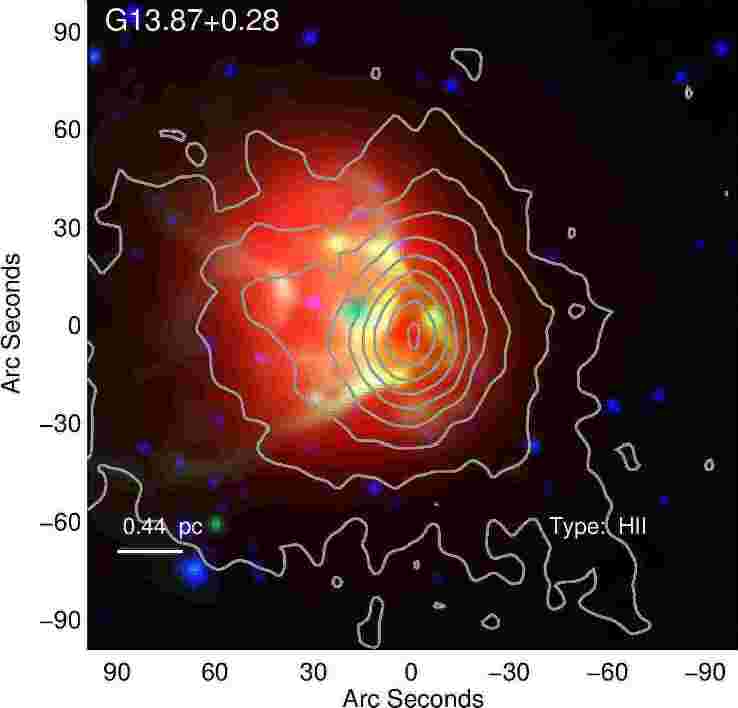}
  \includegraphics[width=5.8cm,angle=90]{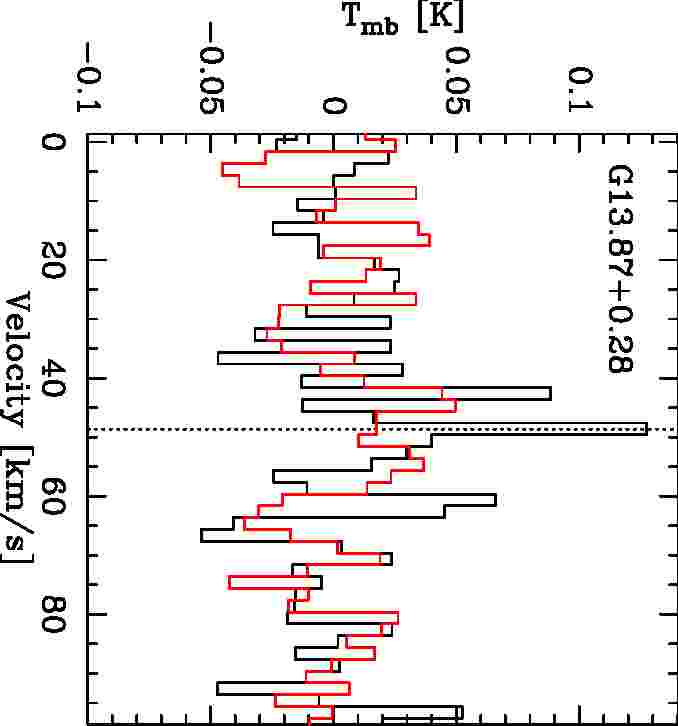} 
  \includegraphics[width=6.0cm,angle=0]{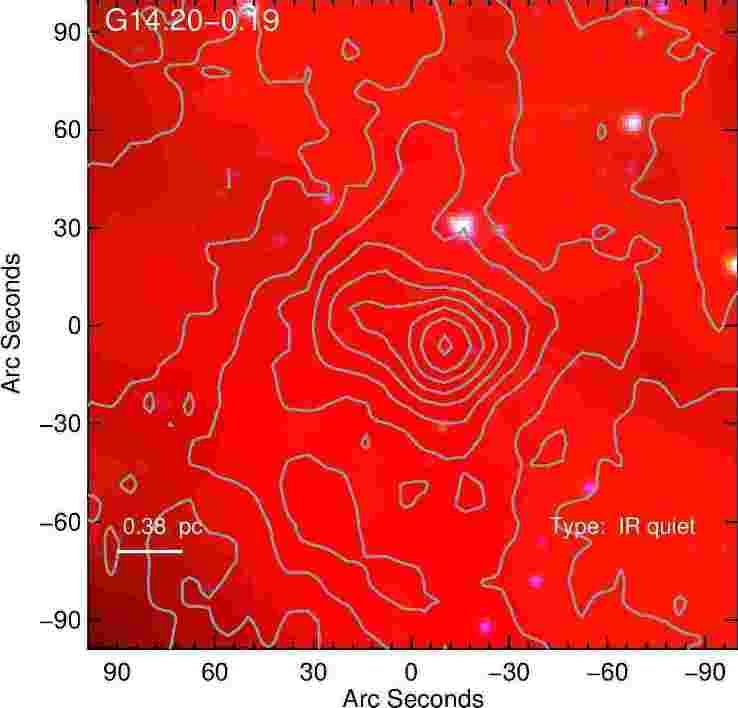}
  \includegraphics[width=6.0cm,angle=90]{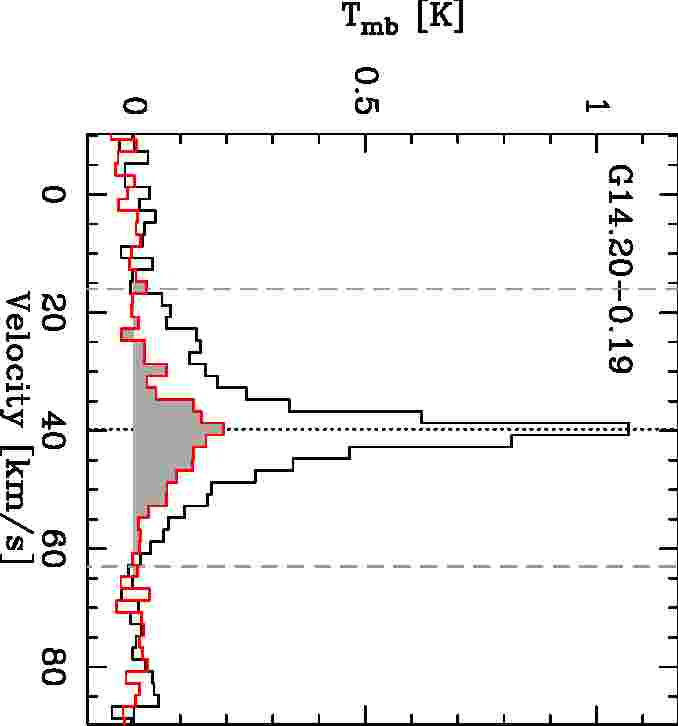}
  \includegraphics[width=6.0cm,angle=0]{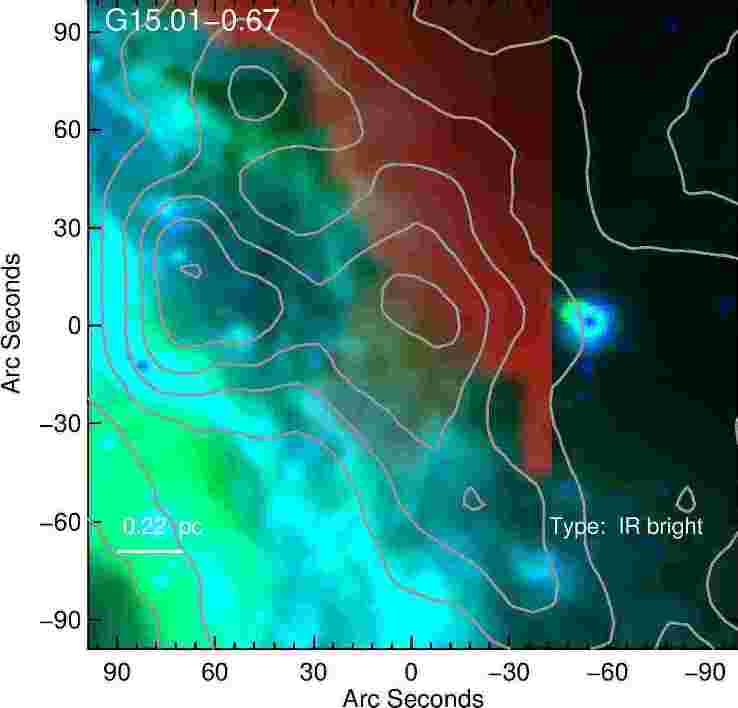}
  \includegraphics[width=6.0cm,angle=90]{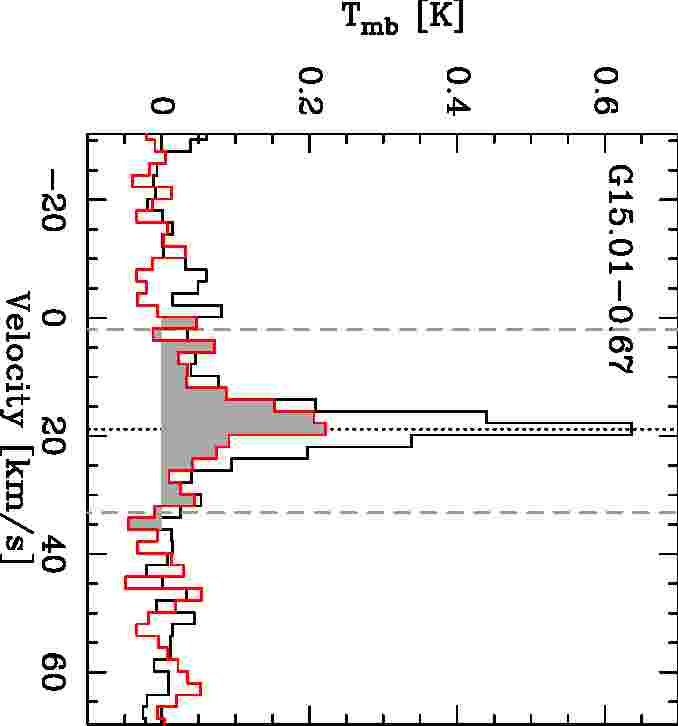}
  \includegraphics[width=6.0cm,angle=0]{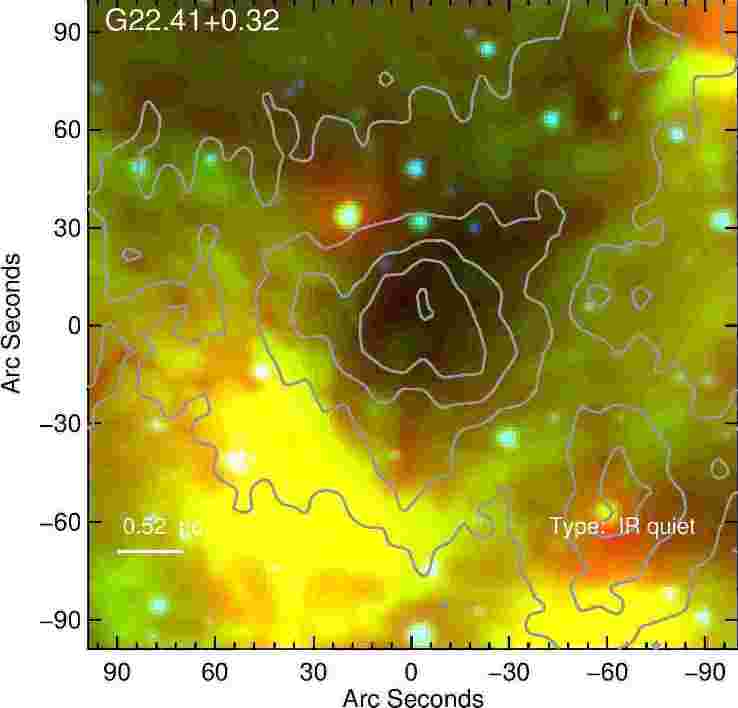}
  \includegraphics[width=6.0cm,angle=90]{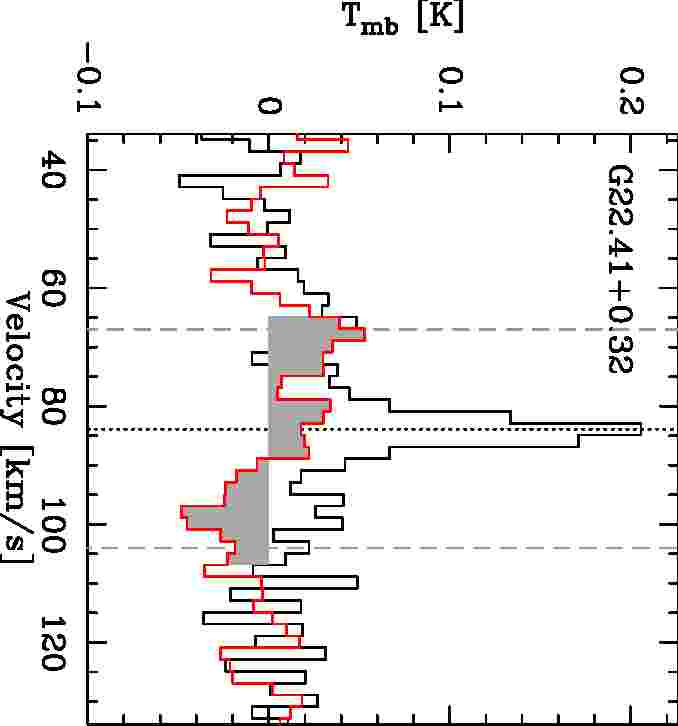}
\ContinuedFloat
 \caption{Continued.}
 \end{figure}
 \end{landscape}

\begin{landscape} 
\begin{figure}
\ContinuedFloat
  \includegraphics[width=6.0cm,angle=0]{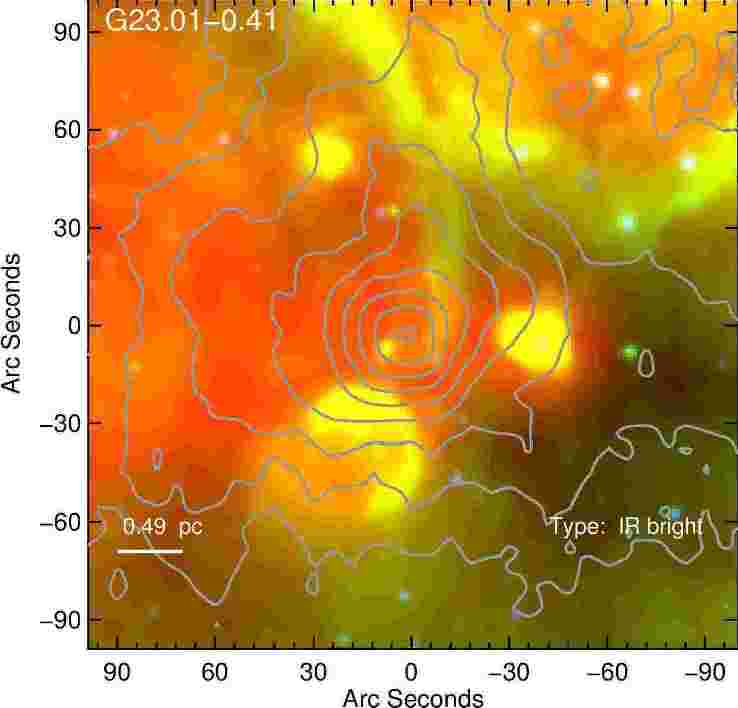}
  \includegraphics[width=6.0cm,angle=90]{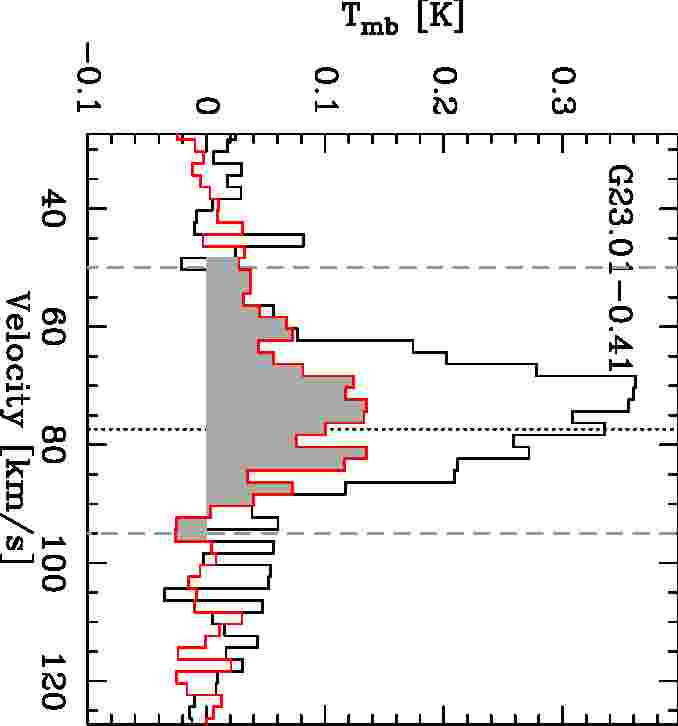}
  \includegraphics[width=6.0cm,angle=0]{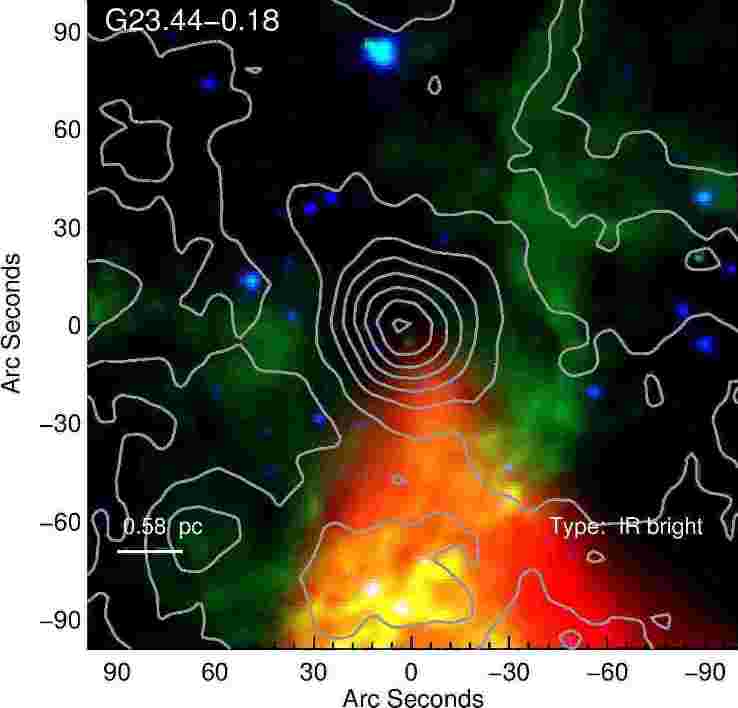}
  \includegraphics[width=6.0cm,angle=90]{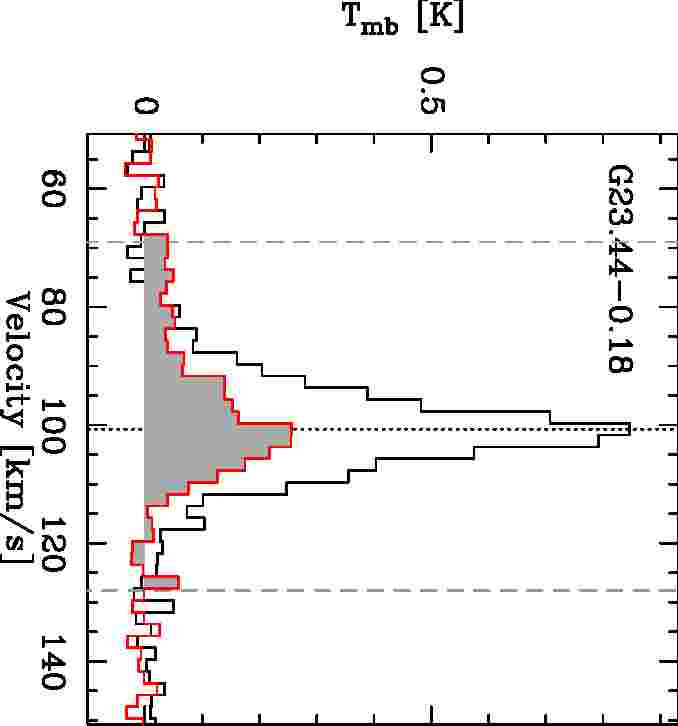}
  \includegraphics[width=6.0cm,angle=0]{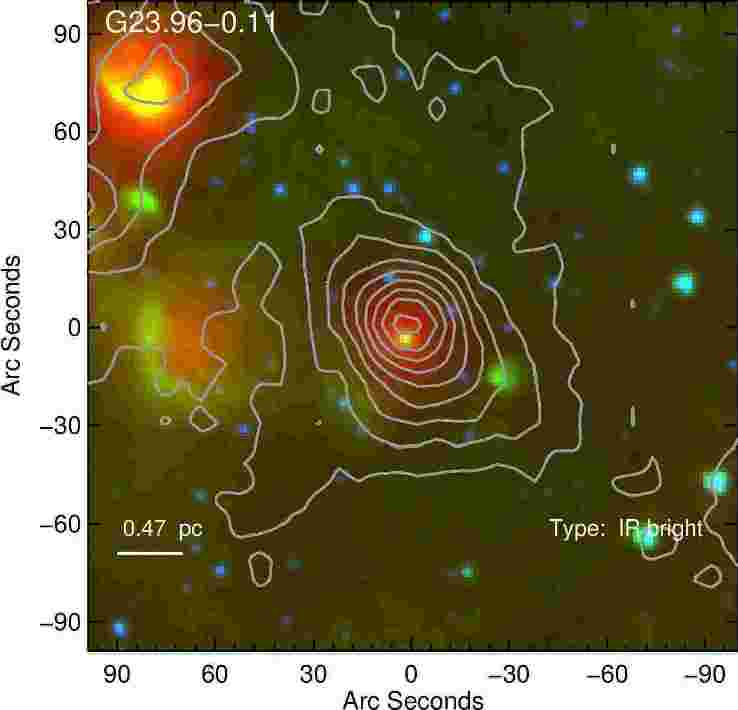}
  \includegraphics[width=6.0cm,angle=90]{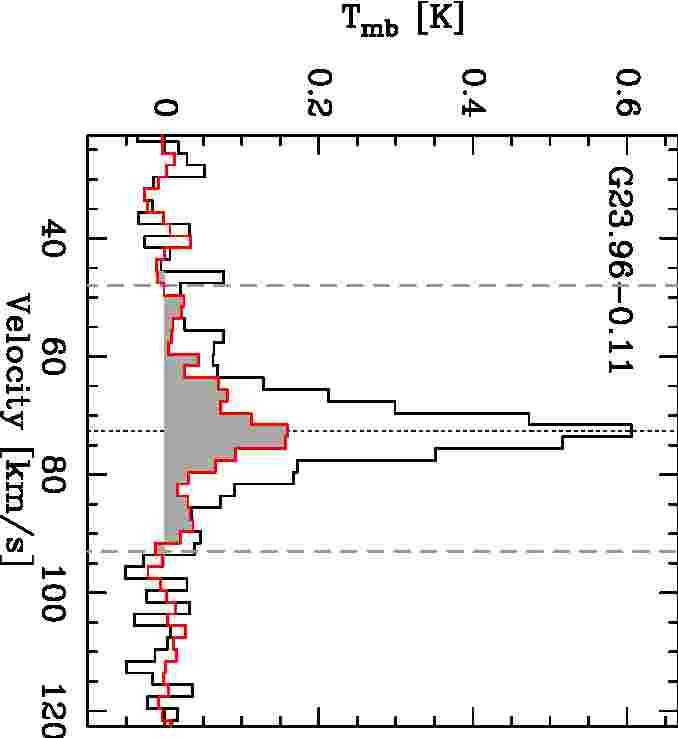}
   \includegraphics[width=6.0cm,angle=0]{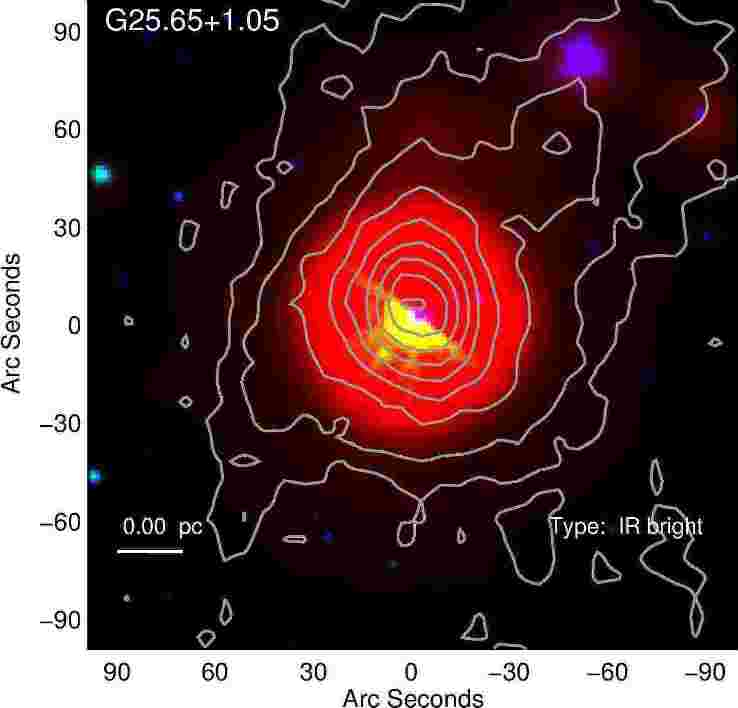}
  \includegraphics[width=6.0cm,angle=90]{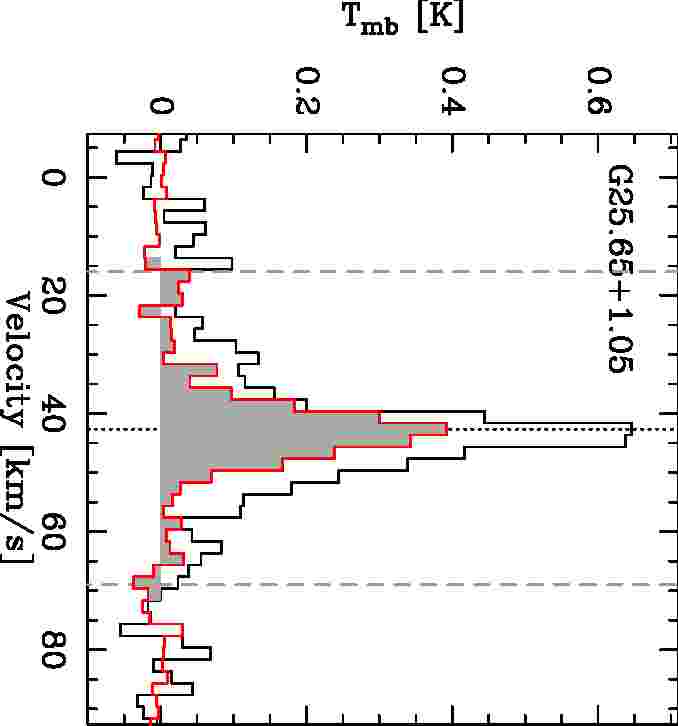}
  \includegraphics[width=6.0cm,angle=0]{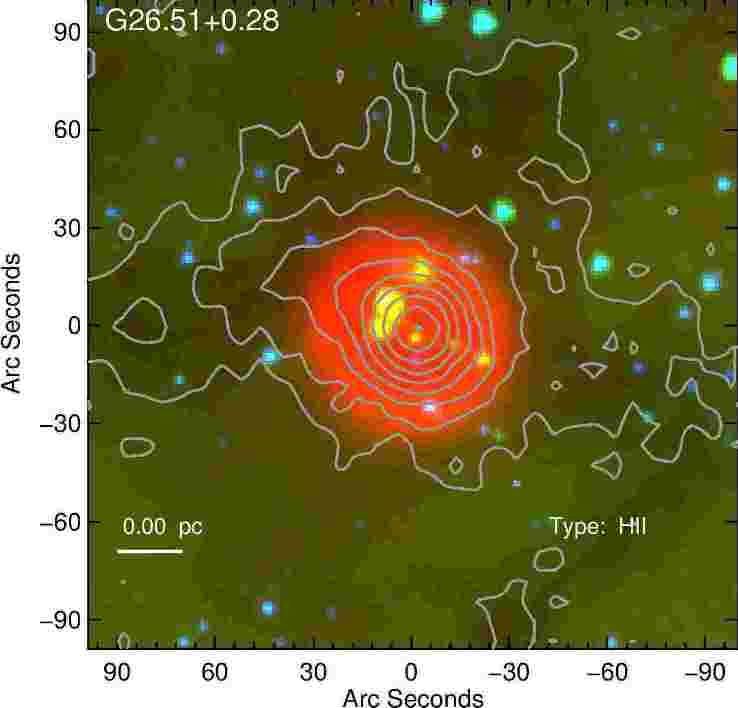}
  \includegraphics[width=6.0cm,angle=90]{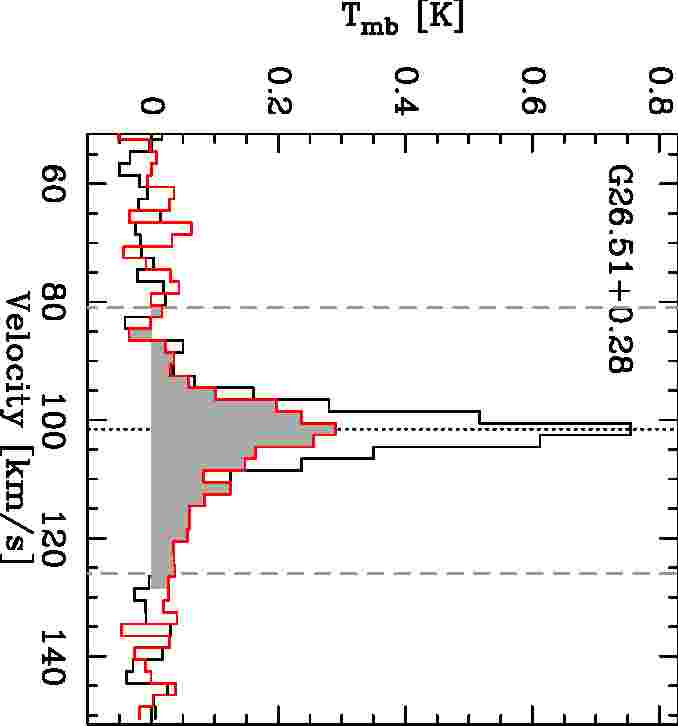}
  \includegraphics[width=6.0cm,angle=0]{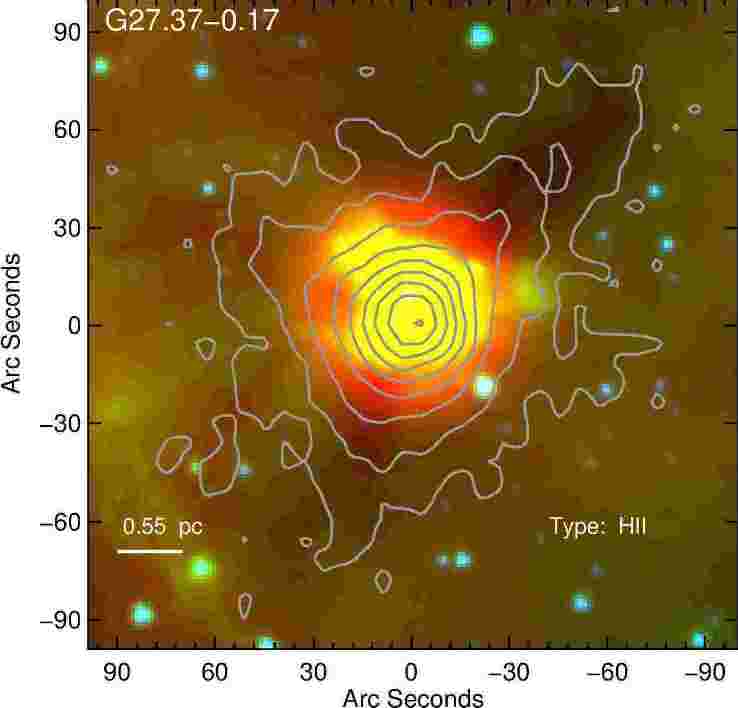}
  \includegraphics[width=6.0cm,angle=90]{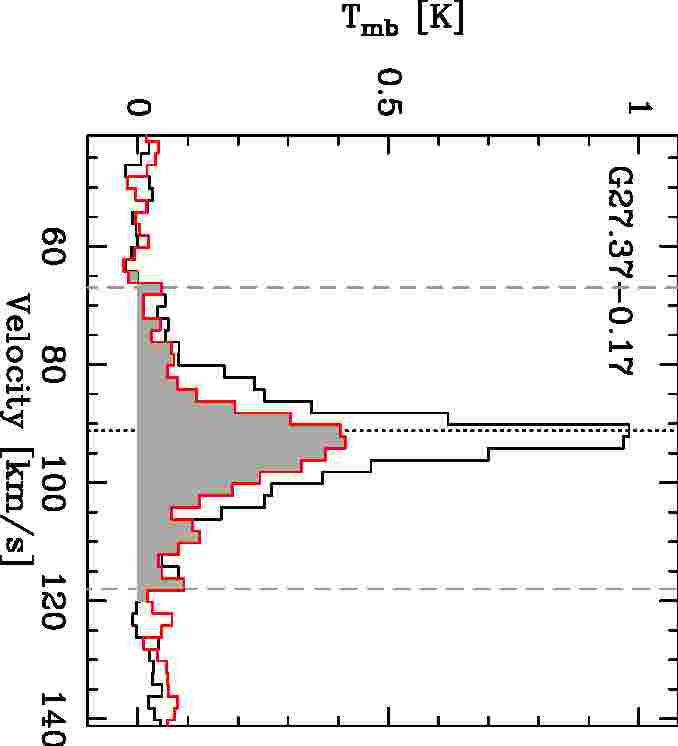}
\caption{Continued.}
 \end{figure}
 \end{landscape}

\begin{landscape} 
\begin{figure}
\ContinuedFloat
  \includegraphics[width=6.0cm,angle=0]{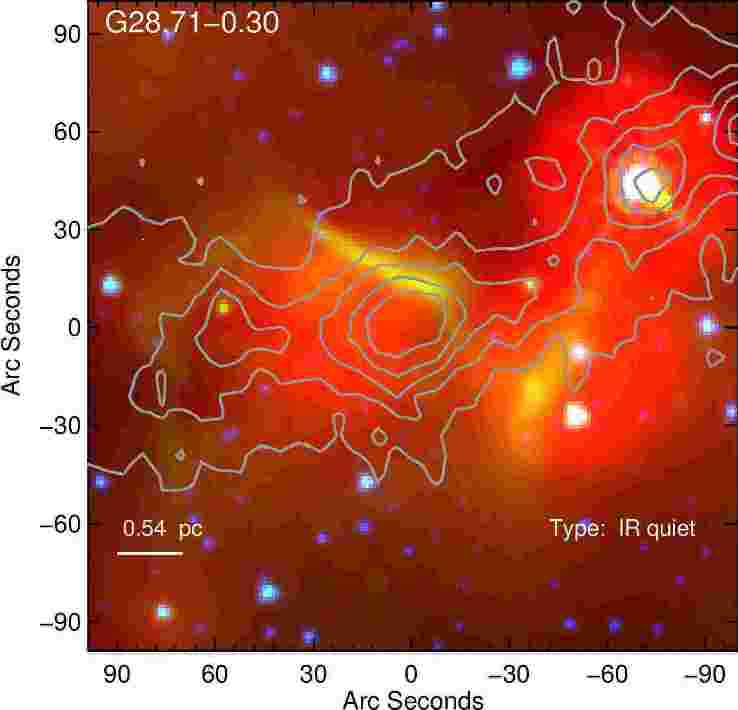}
  \includegraphics[width=6.0cm,angle=90]{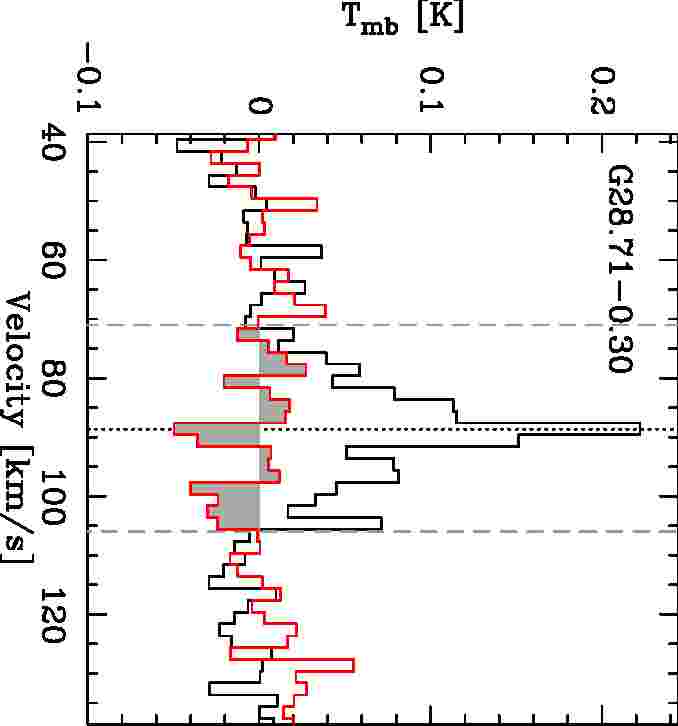}
  \includegraphics[width=6.0cm,angle=0]{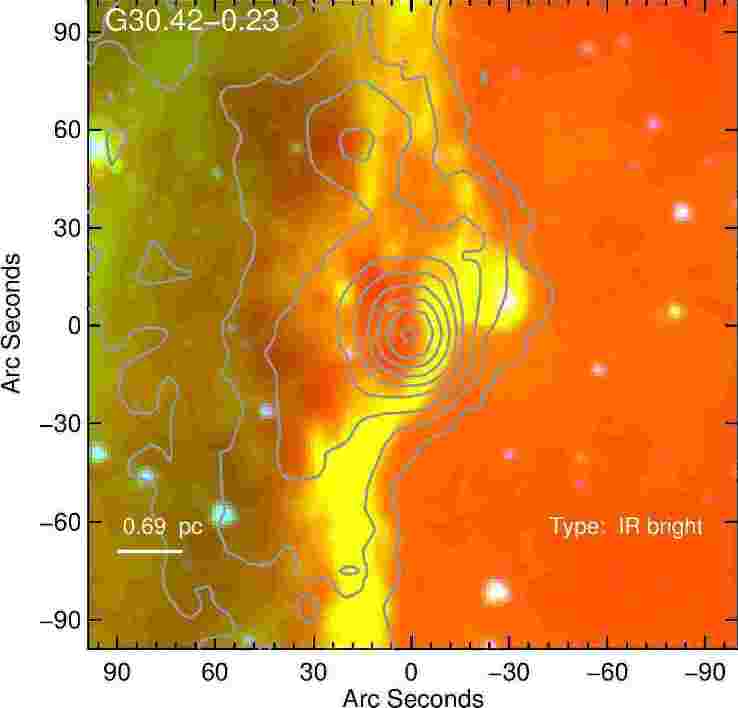}
  \includegraphics[width=6.0cm,angle=90]{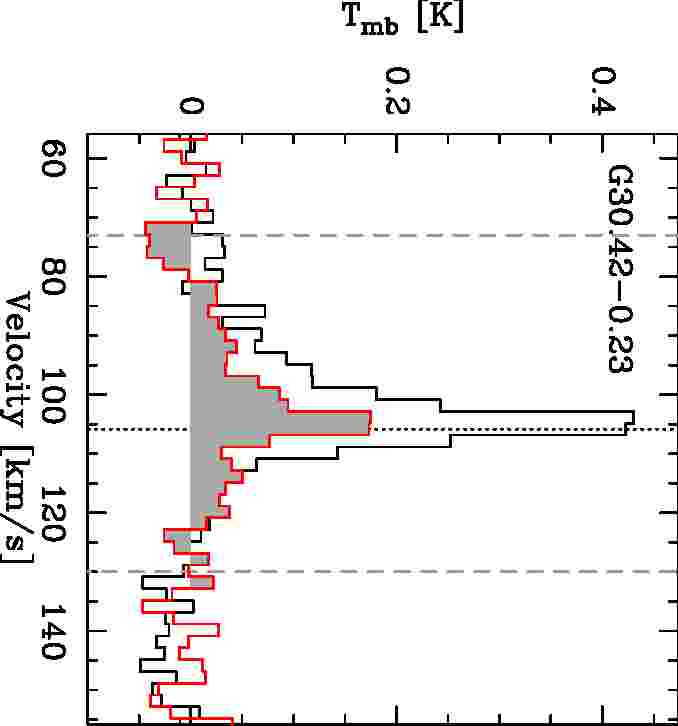}
  \includegraphics[width=6.0cm,angle=0]{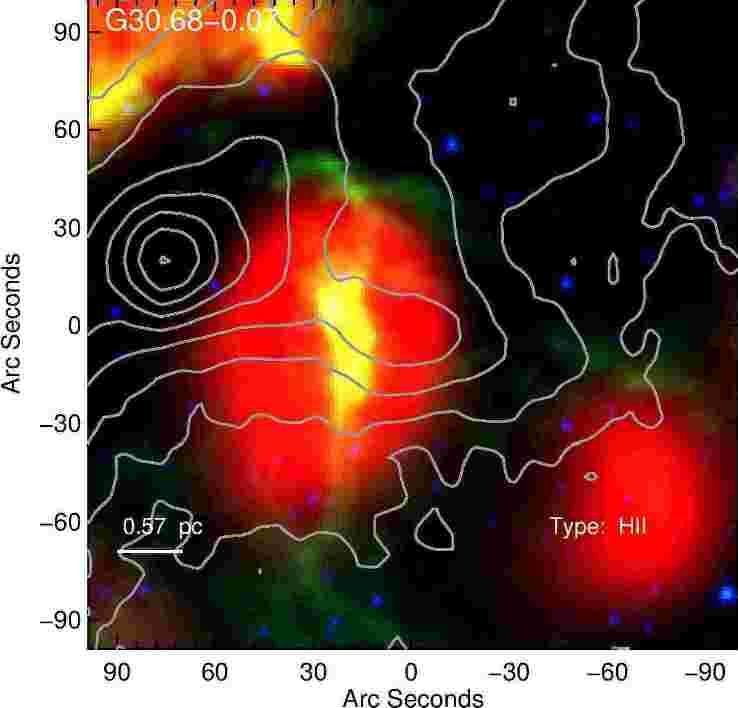}
  \includegraphics[width=6.0cm,angle=90]{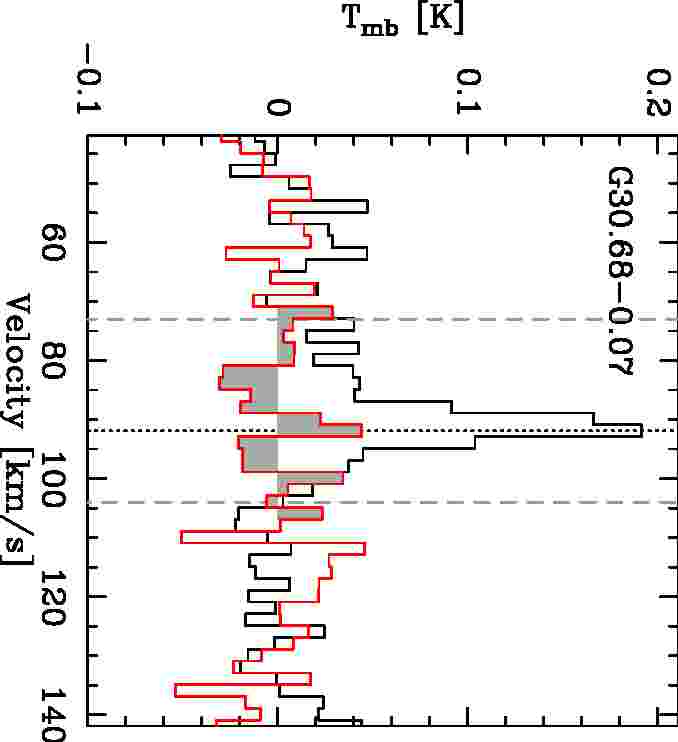}
  \includegraphics[width=6.0cm,angle=0]{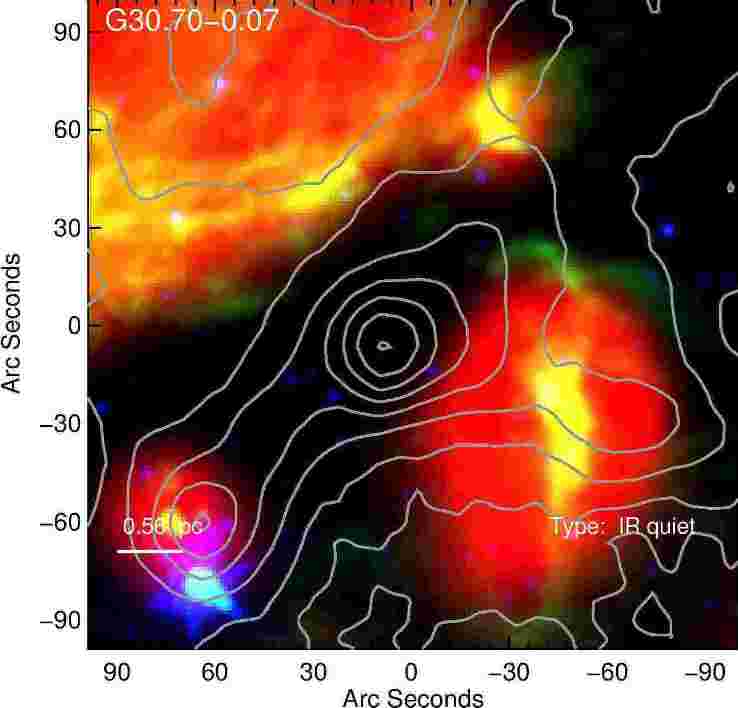}
  \includegraphics[width=6.0cm,angle=90]{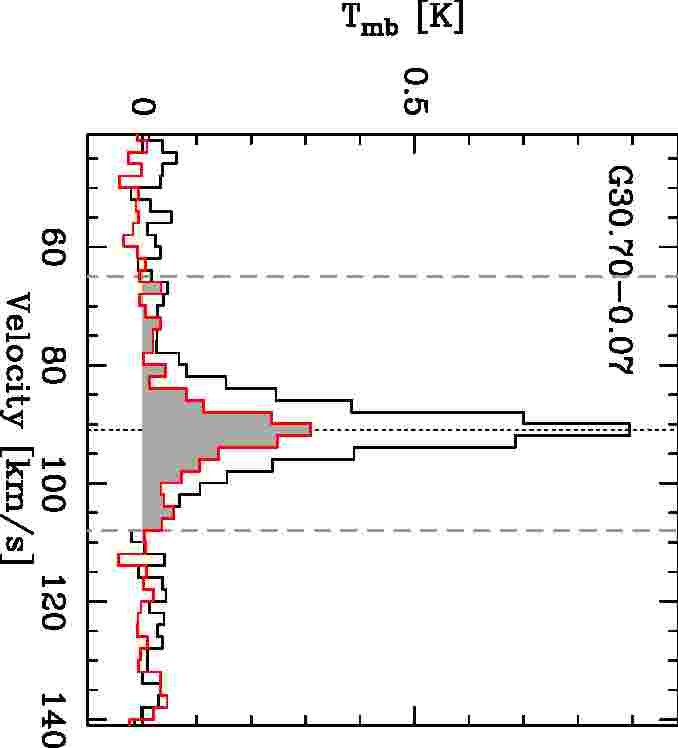}
  \includegraphics[width=6.0cm,angle=0]{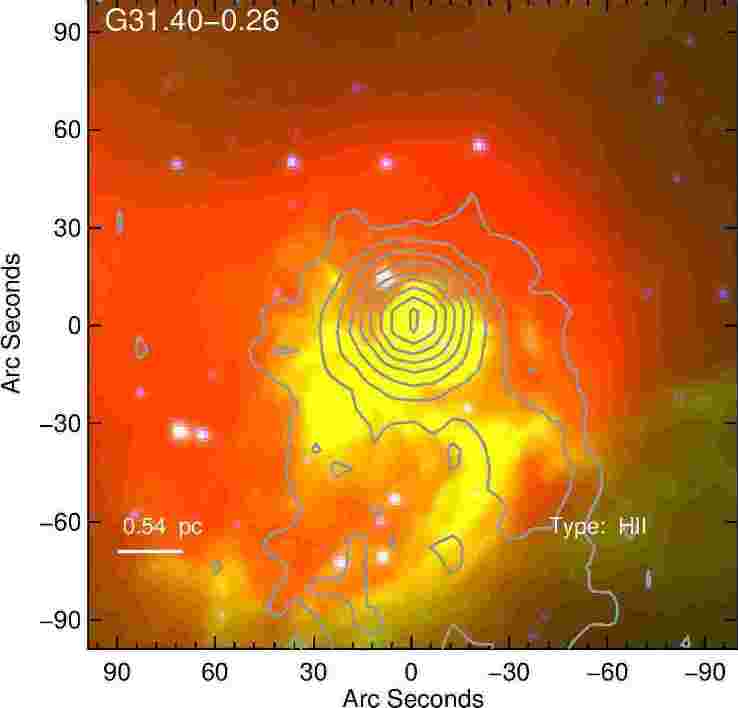}
  \includegraphics[width=6.0cm,angle=90]{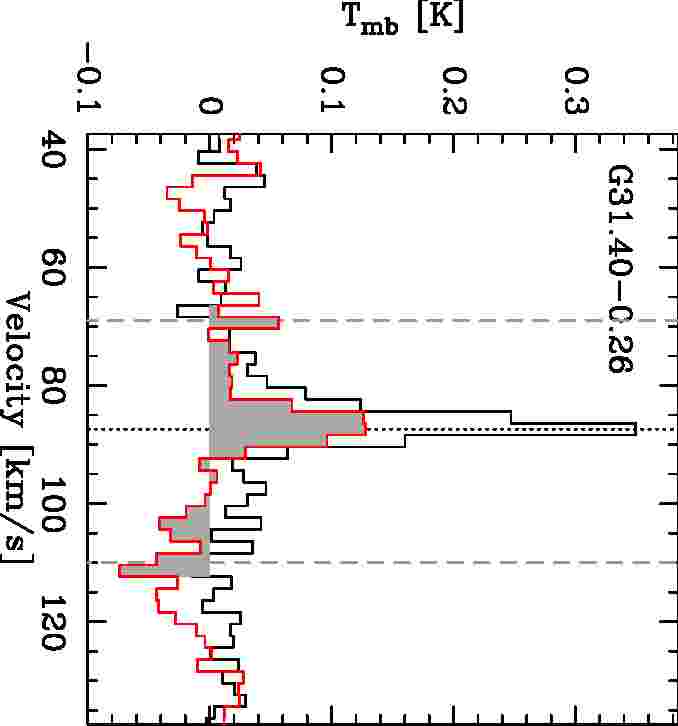}
   \includegraphics[width=6.0cm,angle=0]{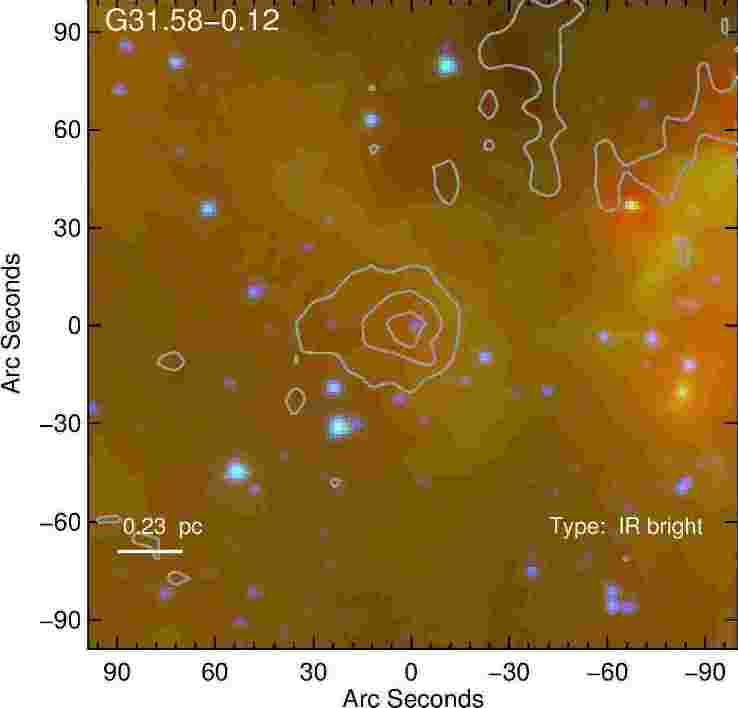}
  \includegraphics[width=5.8cm,angle=90]{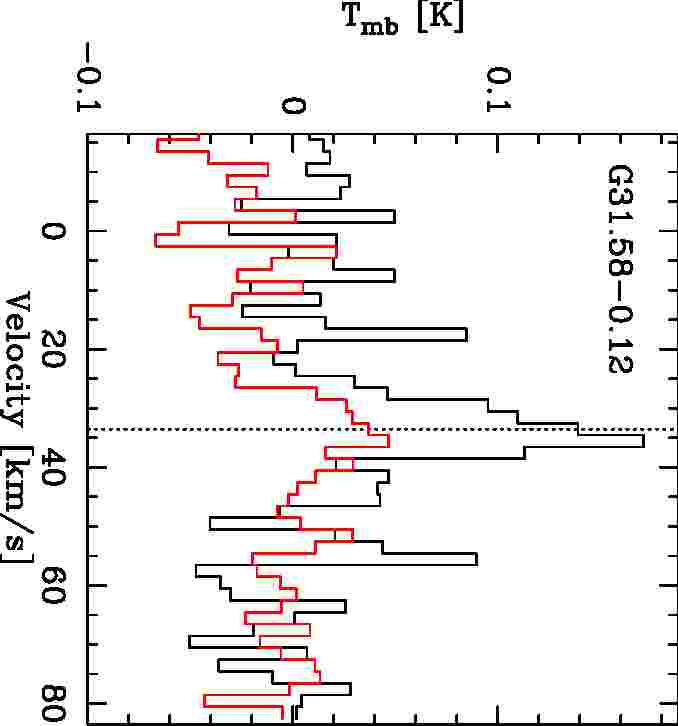} 
\caption{Continued.}
 \end{figure}
 \end{landscape}

\begin{landscape}
\begin{figure}
\ContinuedFloat
  \includegraphics[width=6.0cm,angle=0]{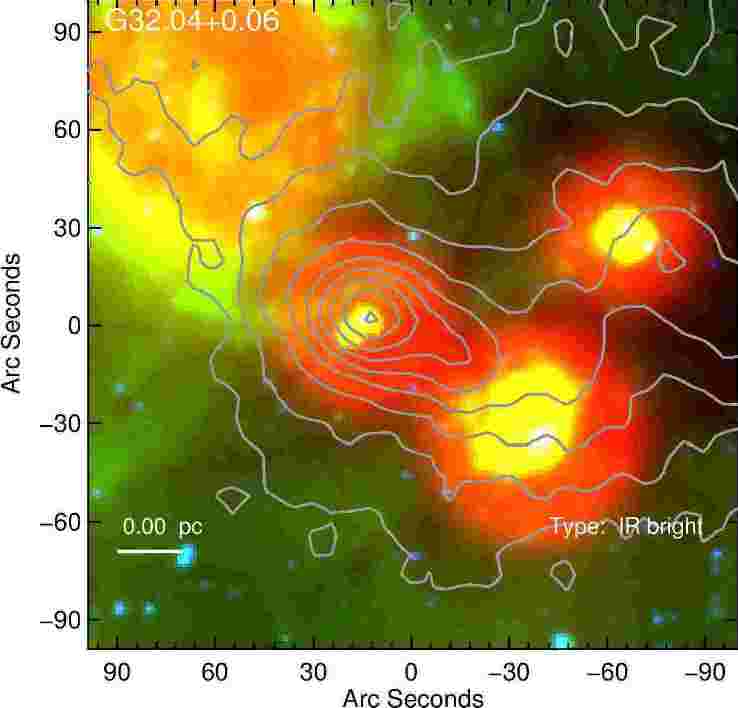}
  \includegraphics[width=6.0cm,angle=90]{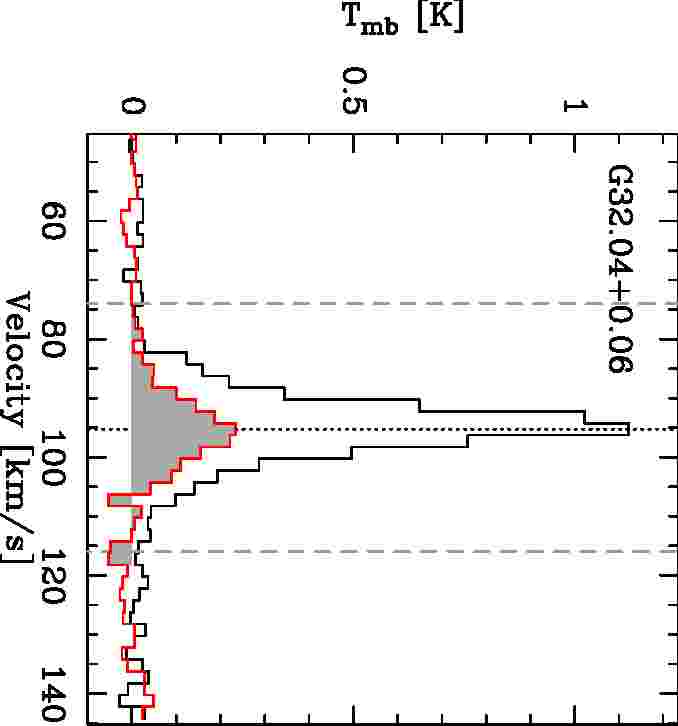}
  \includegraphics[width=6.0cm,angle=0]{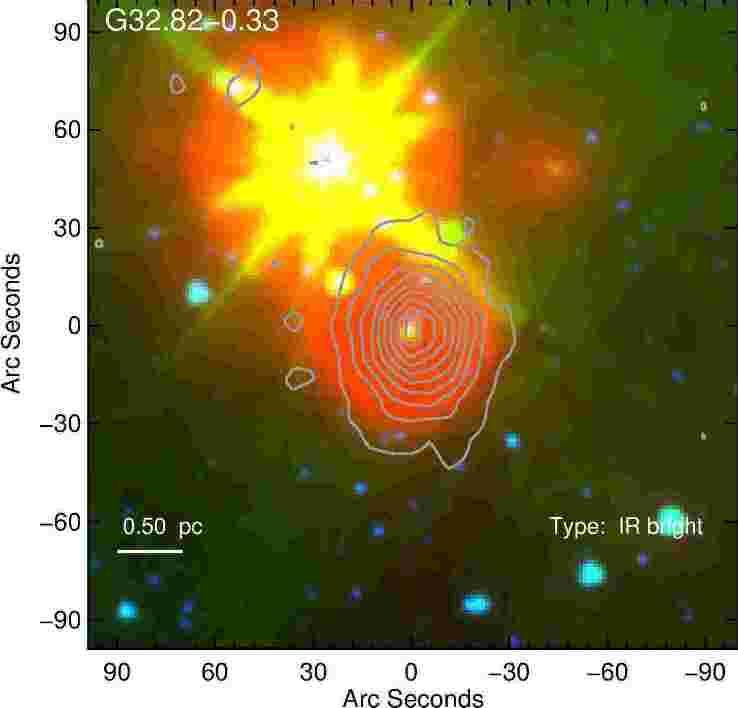}
  \includegraphics[width=6.0cm,angle=90]{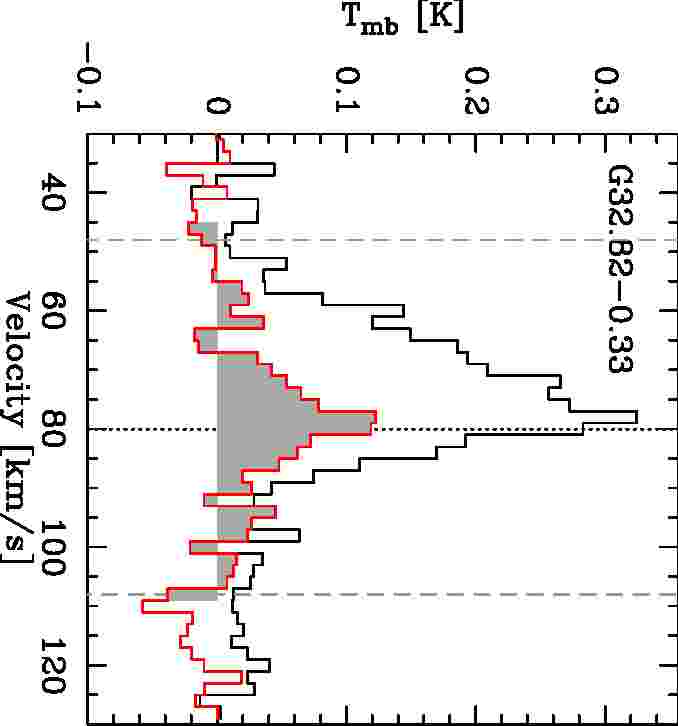}
  \includegraphics[width=6.0cm,angle=0]{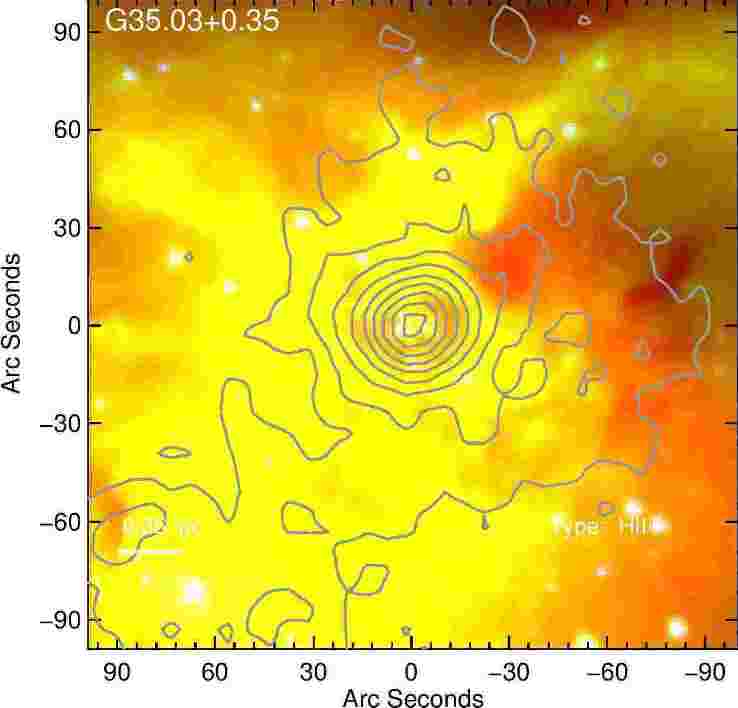}
  \includegraphics[width=6.0cm,angle=90]{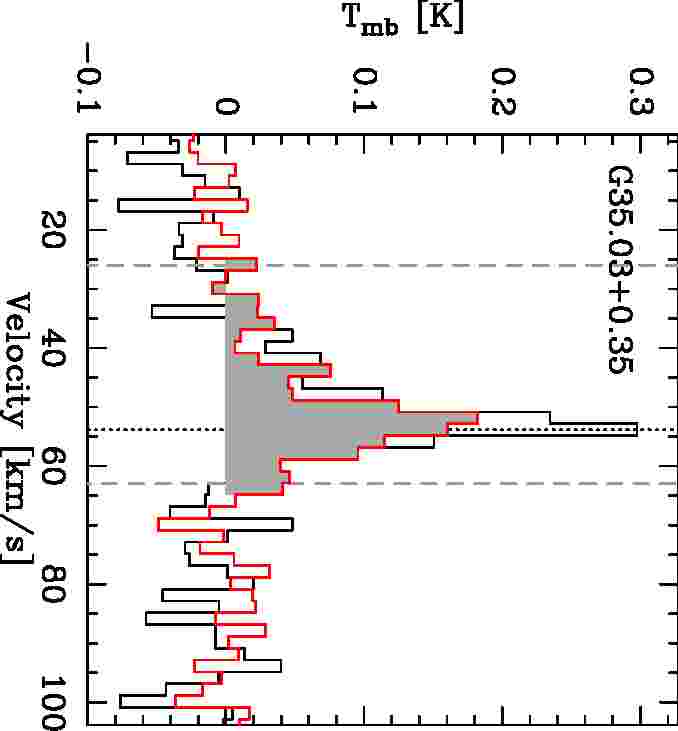}
  \includegraphics[width=6.0cm,angle=0]{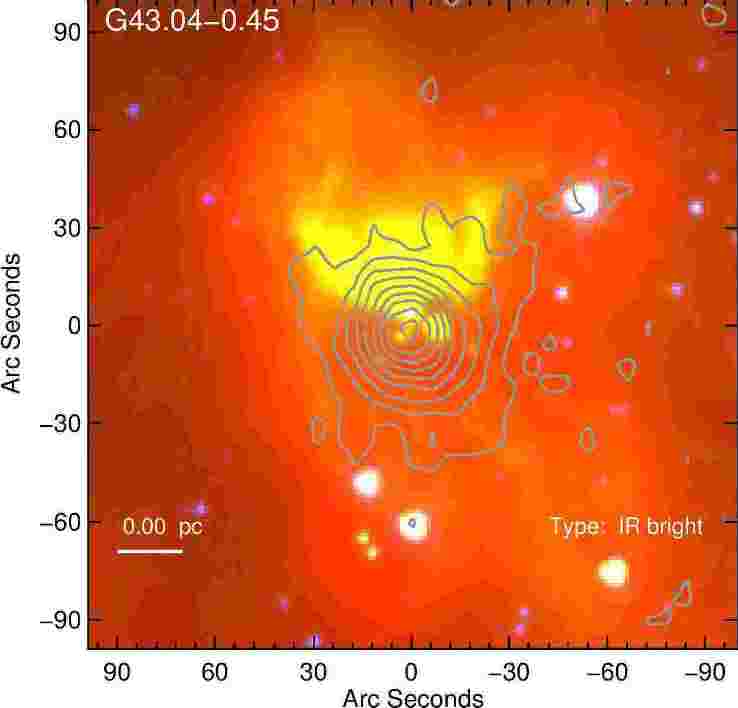}
  \includegraphics[width=6.0cm,angle=90]{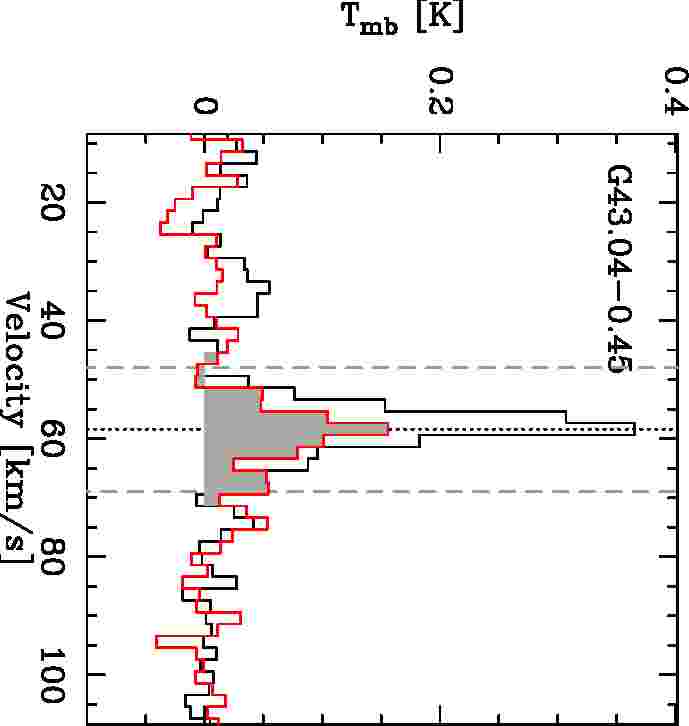}
  \includegraphics[width=6.0cm,angle=0]{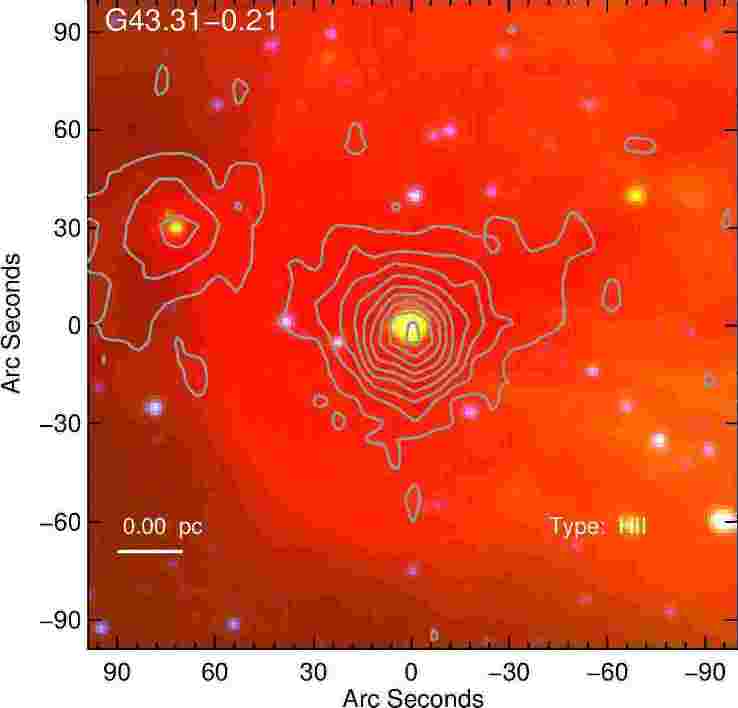}
  \includegraphics[width=6.0cm,angle=90]{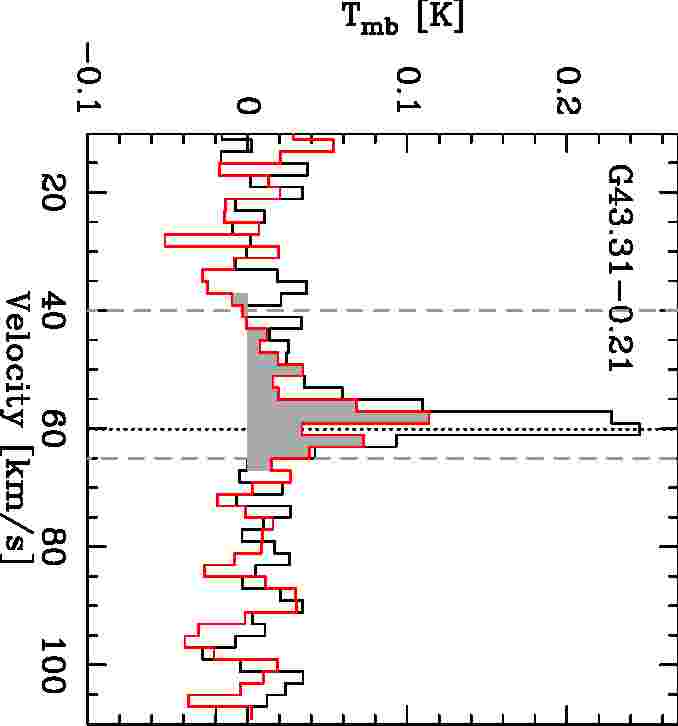}
  \includegraphics[width=6.0cm,angle=0]{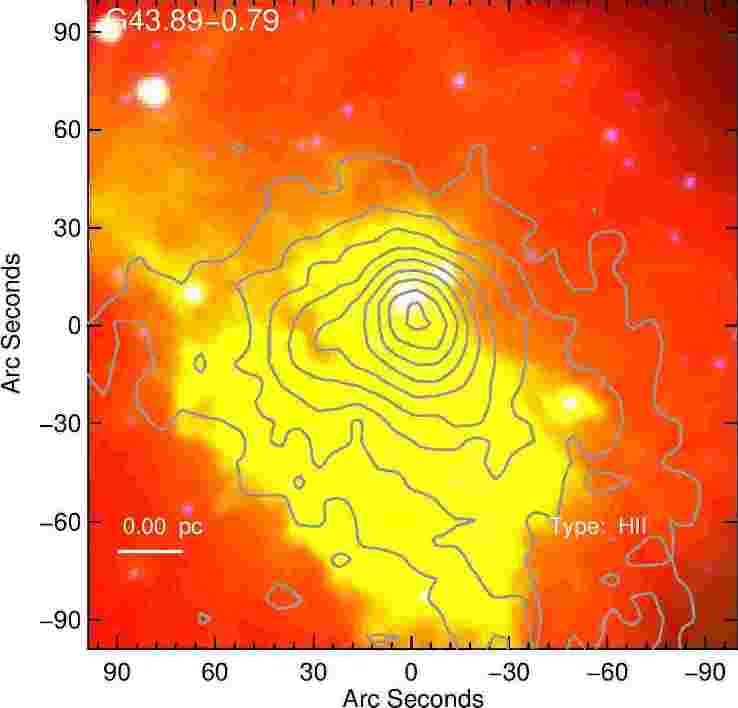}
  \includegraphics[width=6.0cm,angle=90]{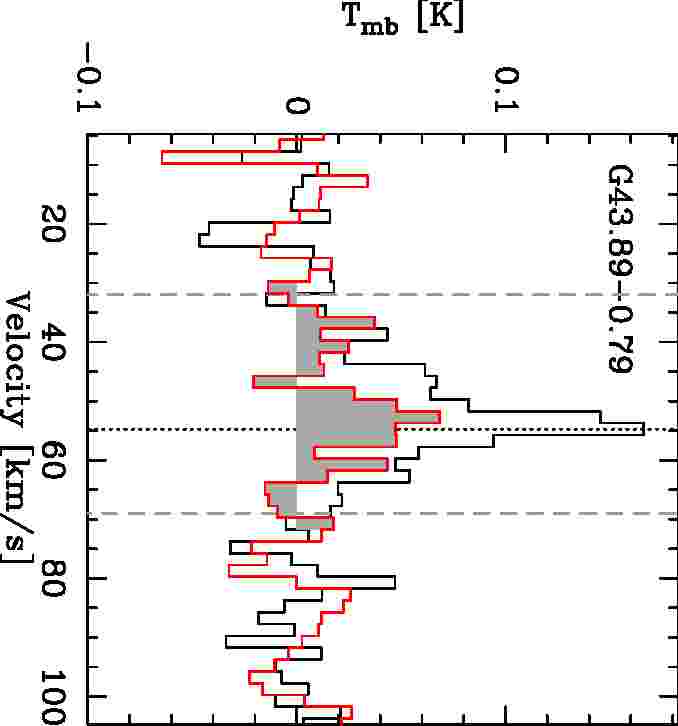}
 \caption{Continued.}
 \end{figure}
 \end{landscape}
\clearpage

\begin{landscape}
\begin{figure}
\ContinuedFloat
  \includegraphics[width=6.0cm,angle=0]{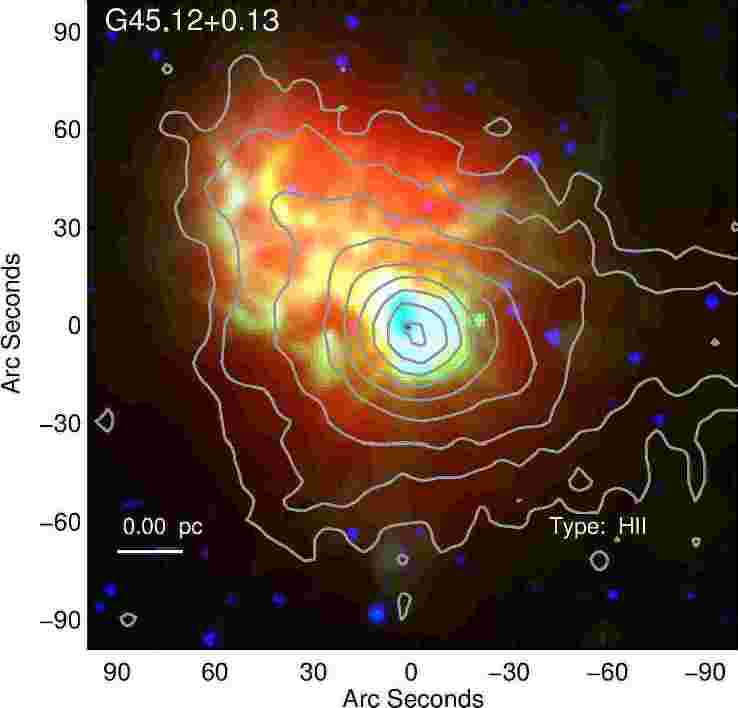}
  \includegraphics[width=5.8cm,angle=90]{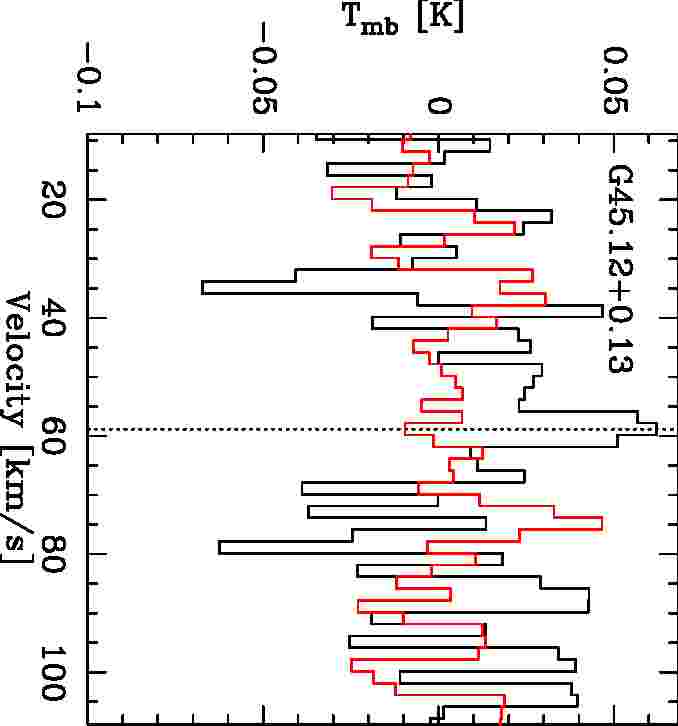} 
  \includegraphics[width=6.0cm,angle=0]{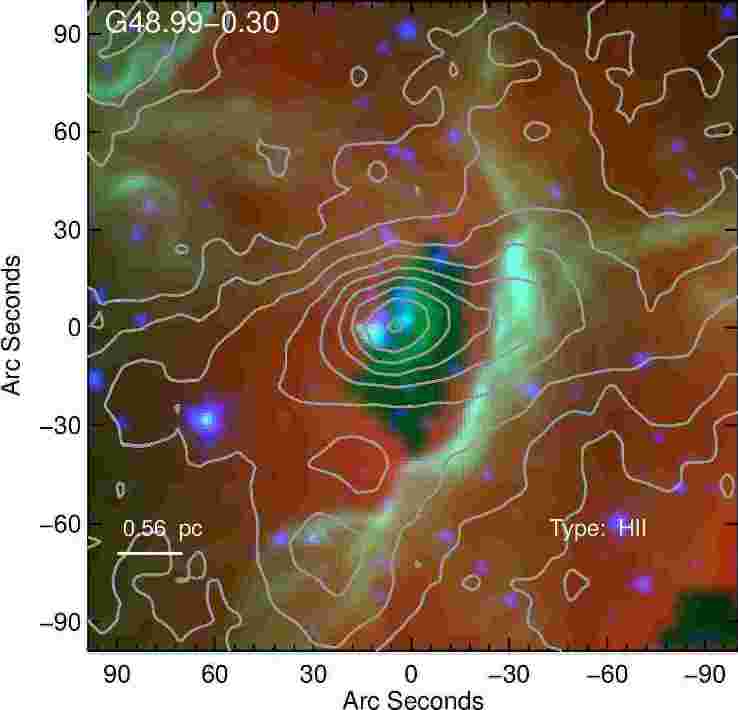}
  \includegraphics[width=6.0cm,angle=90]{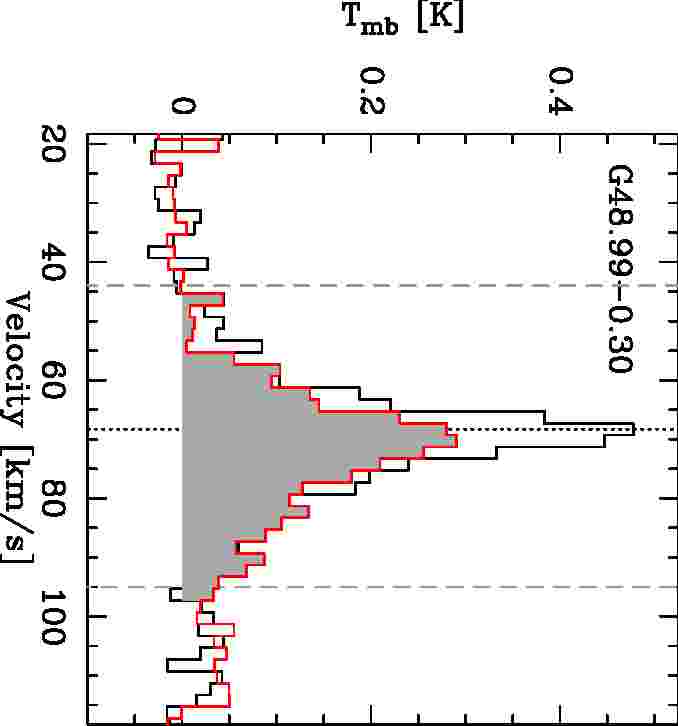}
  \includegraphics[width=6.0cm,angle=0]{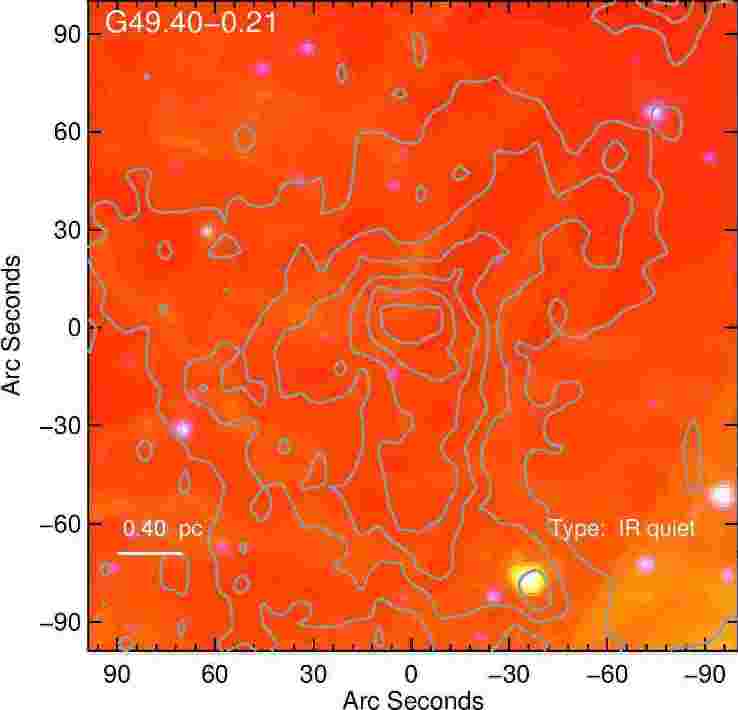}
  \includegraphics[width=6.0cm,angle=90]{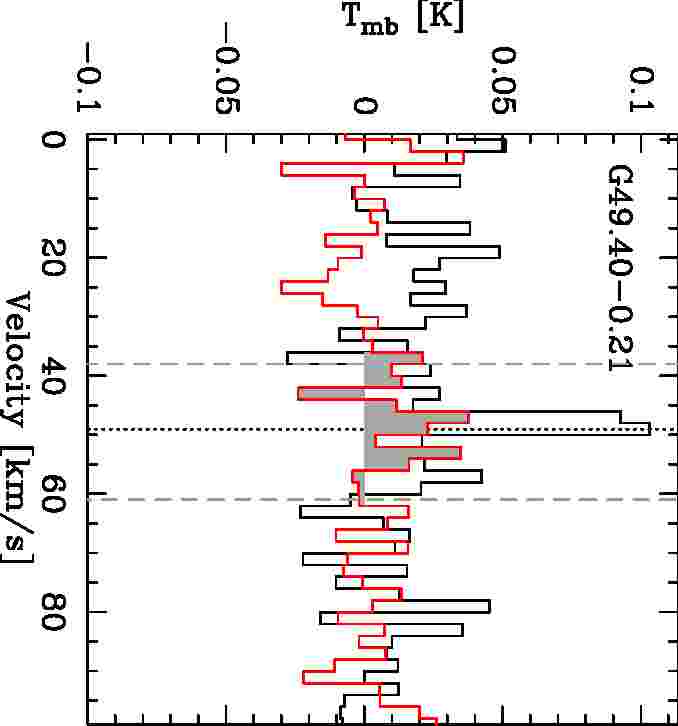}
  \includegraphics[width=6.0cm,angle=0]{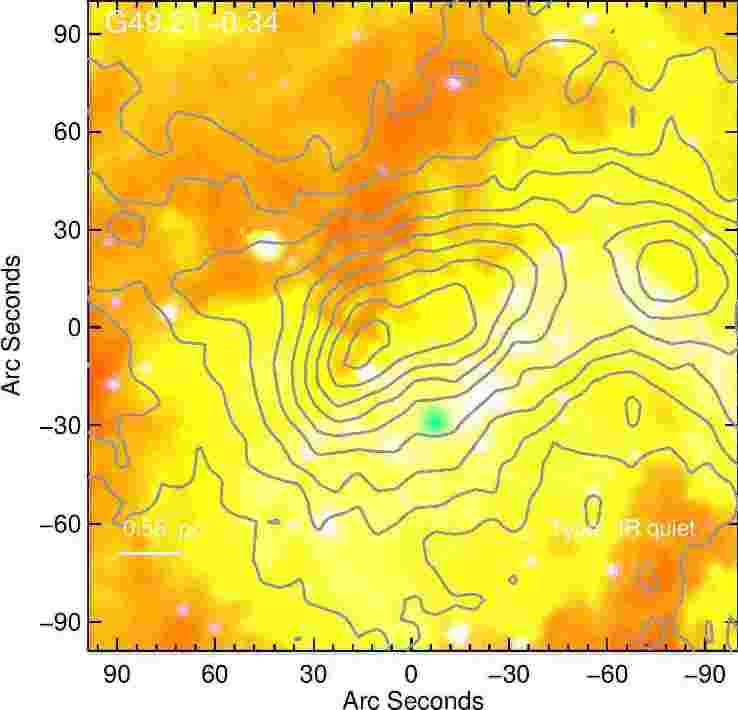}
  \includegraphics[width=5.8cm,angle=90]{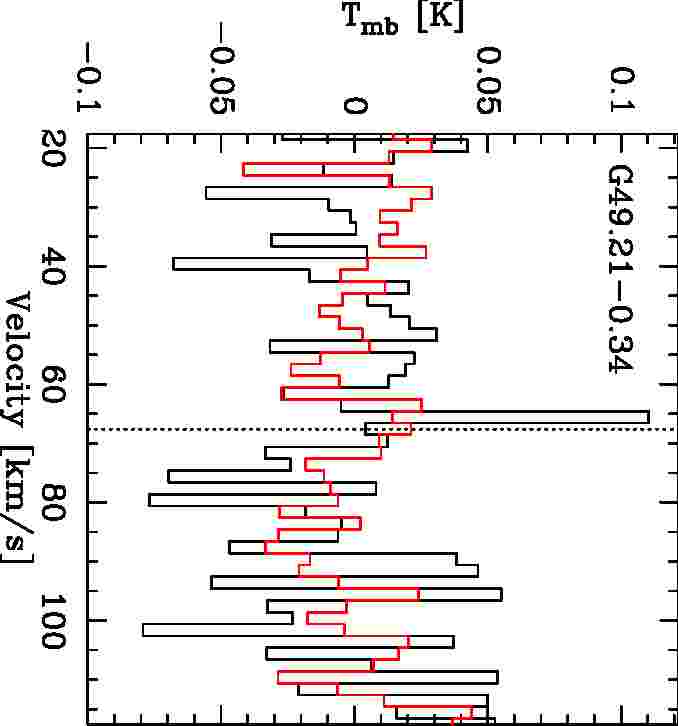} 
  \includegraphics[width=6.0cm,angle=0]{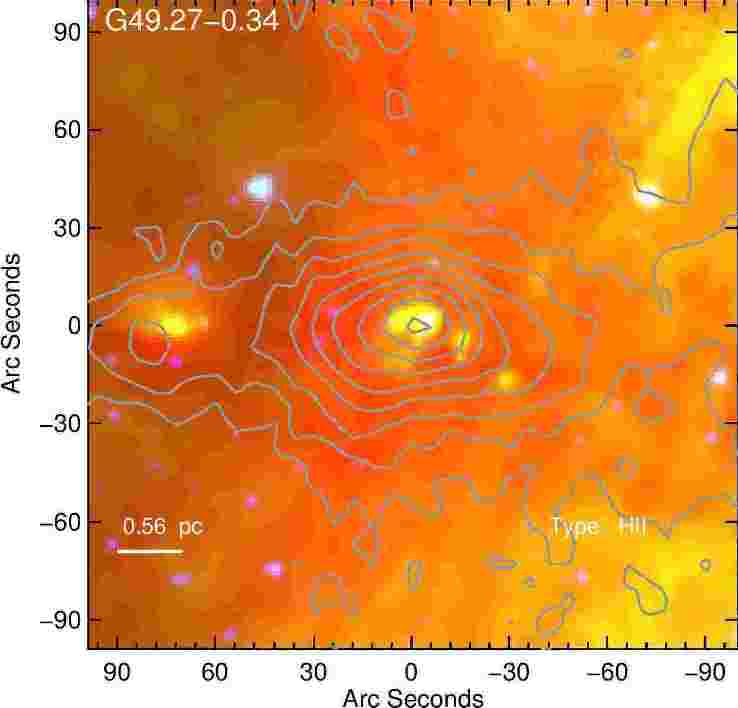}
  \includegraphics[width=6.0cm,angle=90]{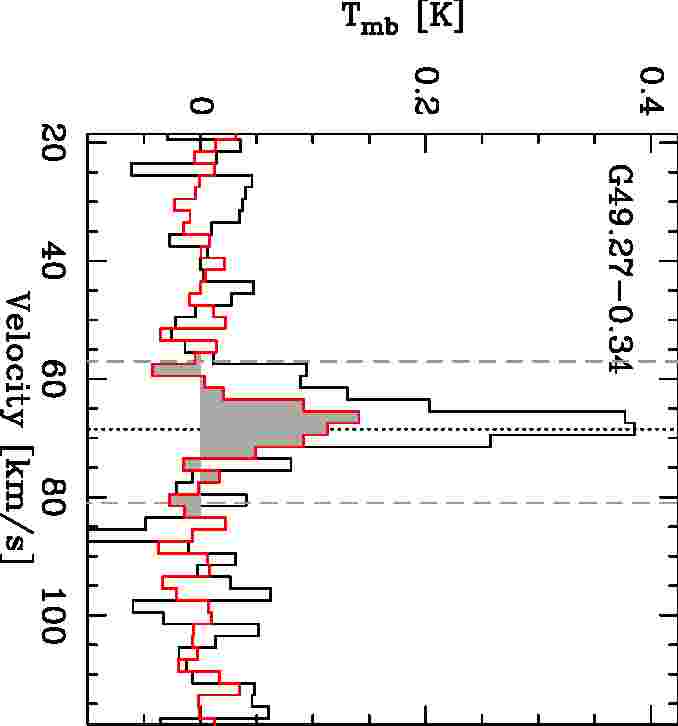}
  \includegraphics[width=6.0cm,angle=0]{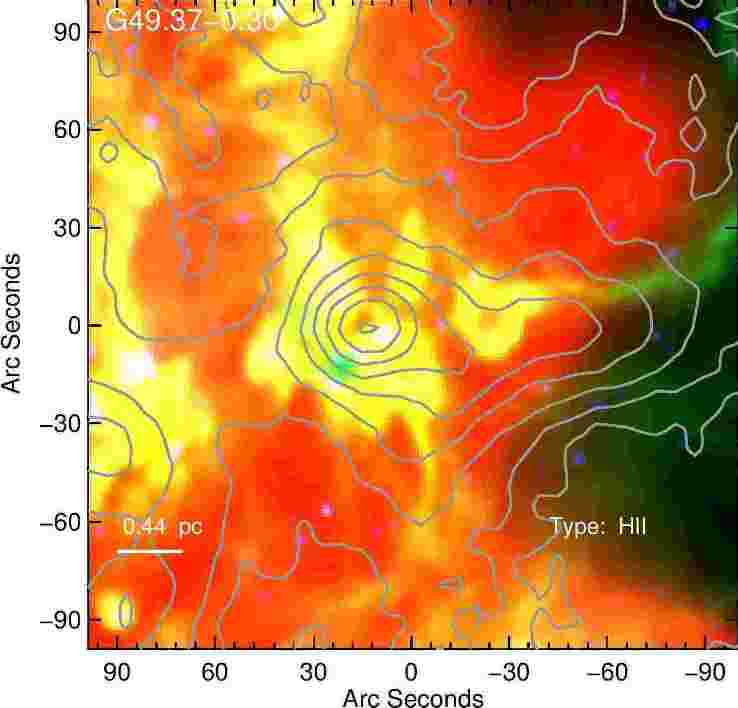}
  \includegraphics[width=6.0cm,angle=90]{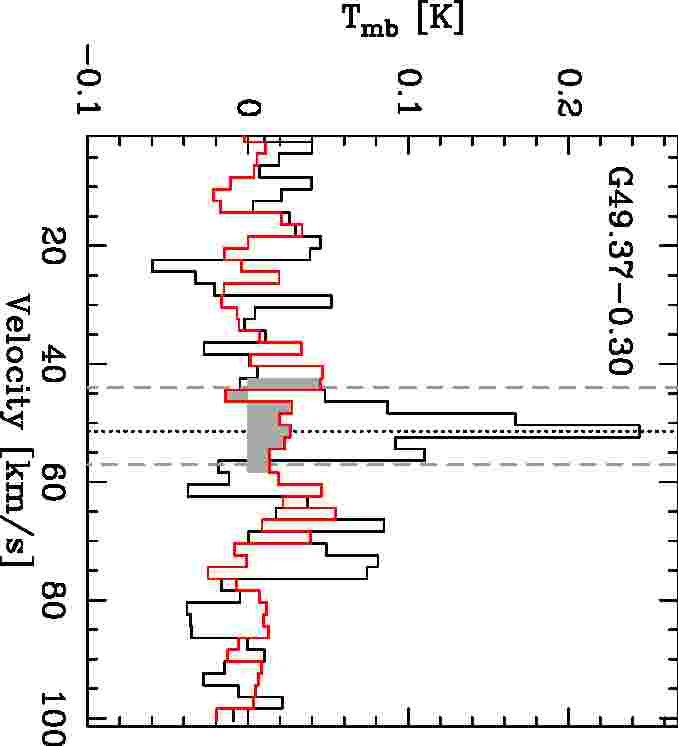}
 \caption{Continued.}
 \end{figure}
 \end{landscape}

\begin{landscape} 
\begin{figure}
\ContinuedFloat
  \includegraphics[width=6.0cm,angle=0]{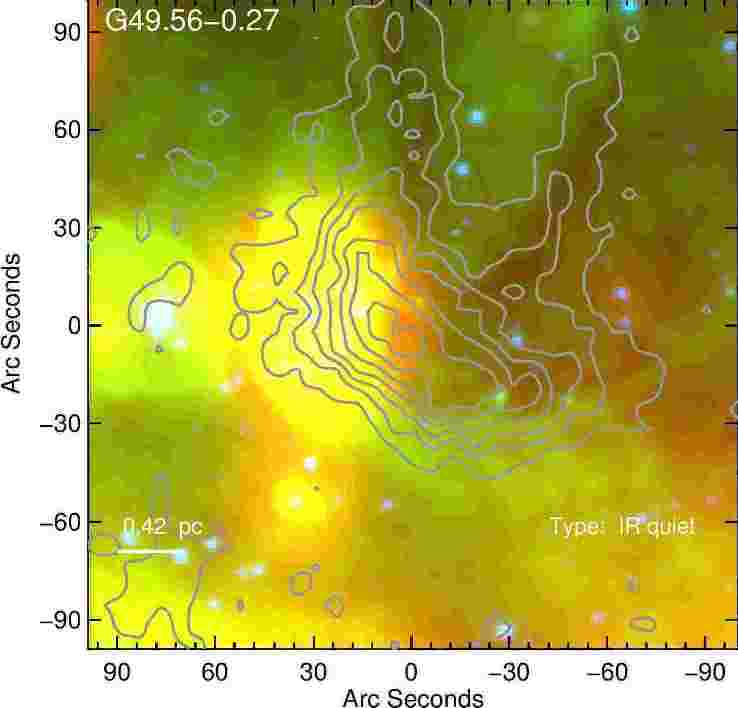}
  \includegraphics[width=6.0cm,angle=90]{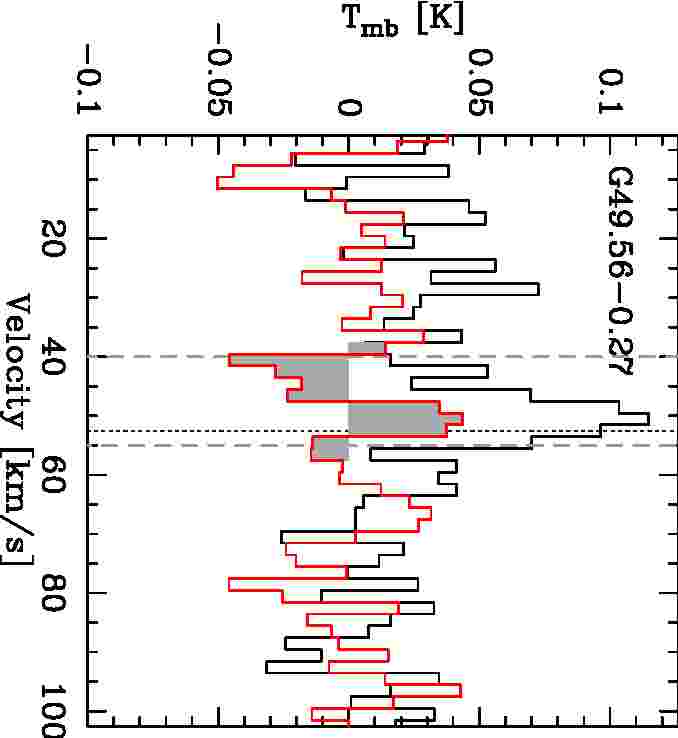}
 \caption{Continued.}
 \end{figure}
 \end{landscape}
\clearpage

\begin{landscape}
\begin{figure}
\ContinuedFloat
  \includegraphics[width=6.0cm,angle=0]{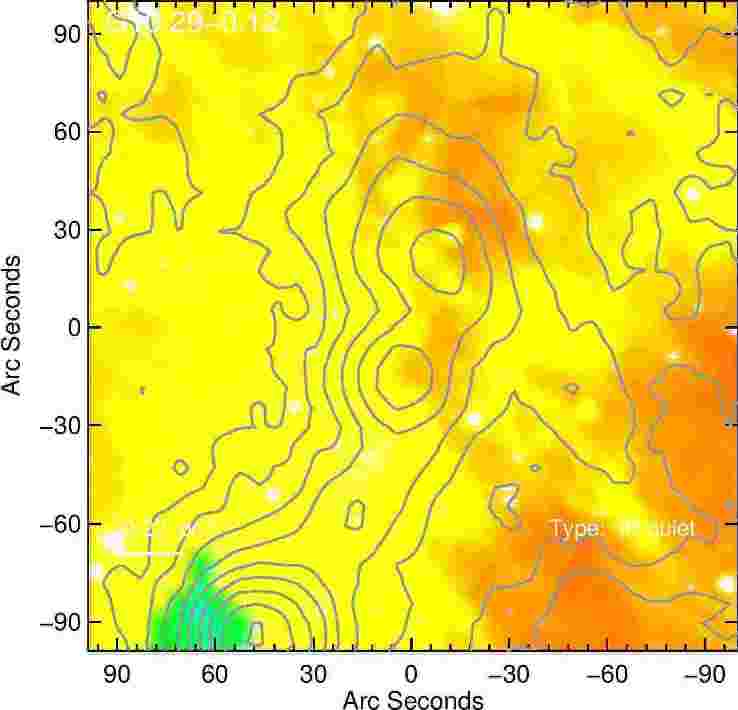}
  \includegraphics[width=5.8cm,angle=90]{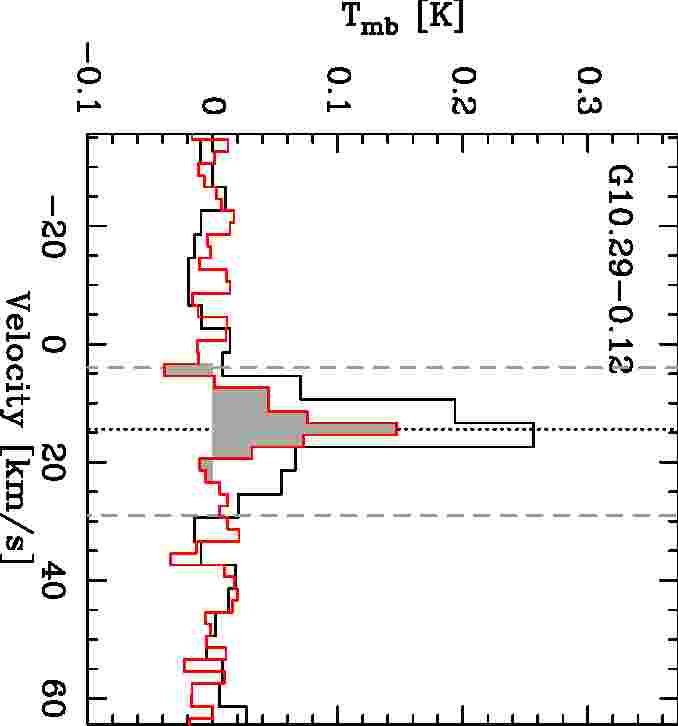} 
  \includegraphics[width=6.0cm,angle=0]{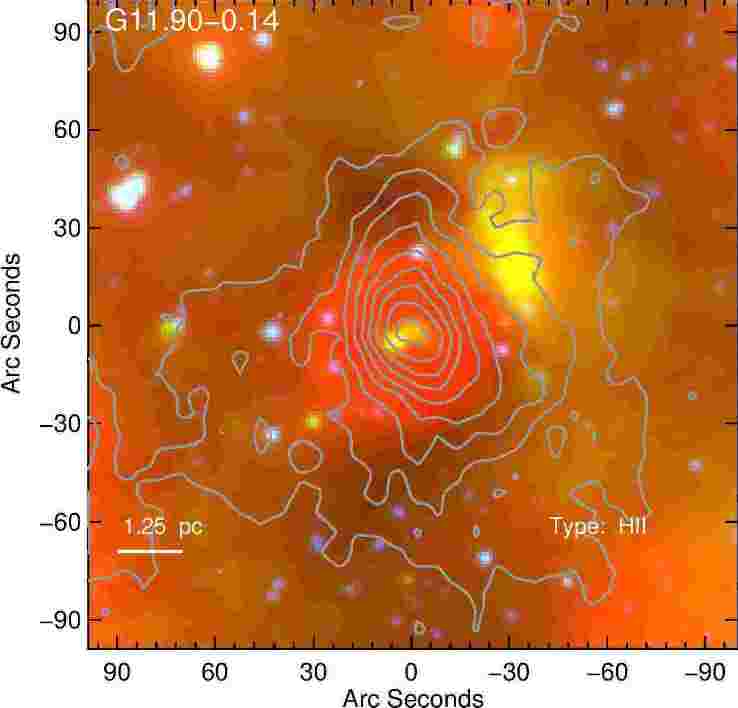}
  \includegraphics[width=6.0cm,angle=90]{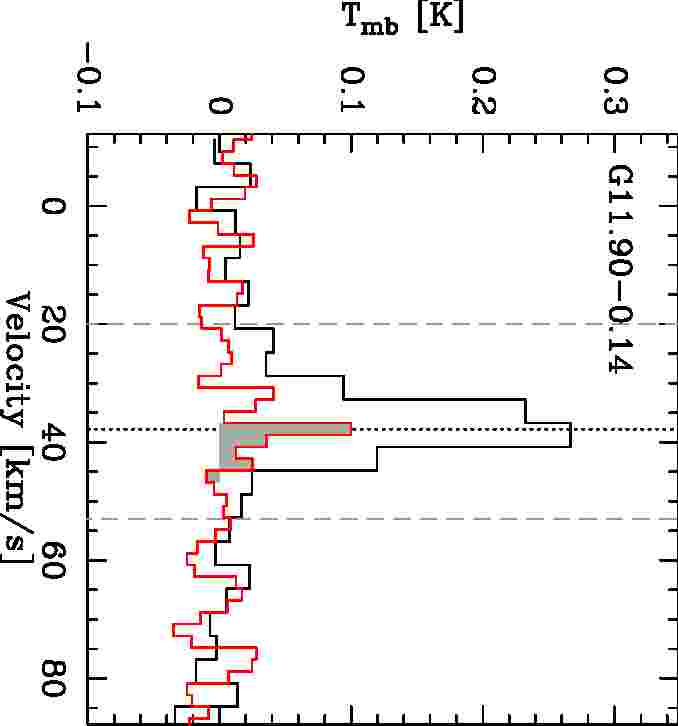}
  \includegraphics[width=6.0cm,angle=0]{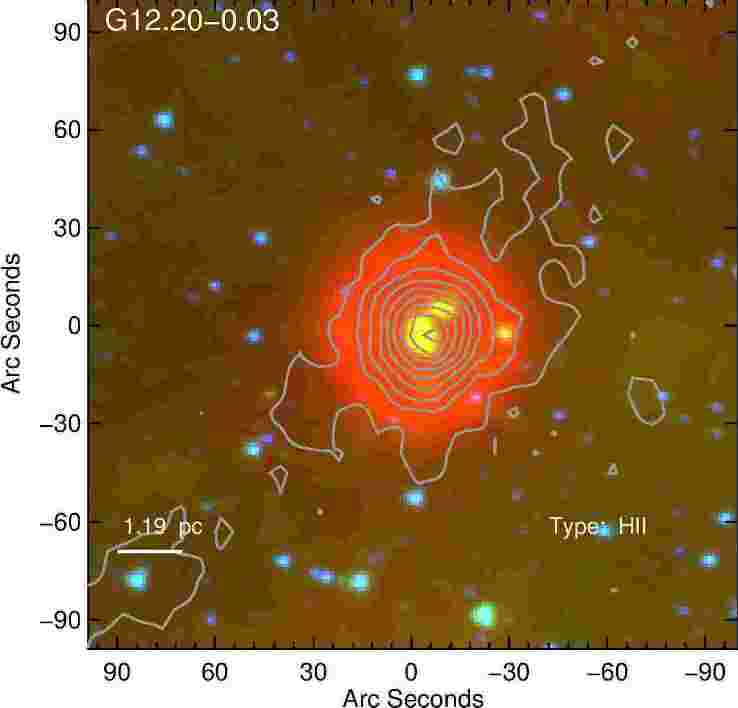}
  \includegraphics[width=5.8cm,angle=90]{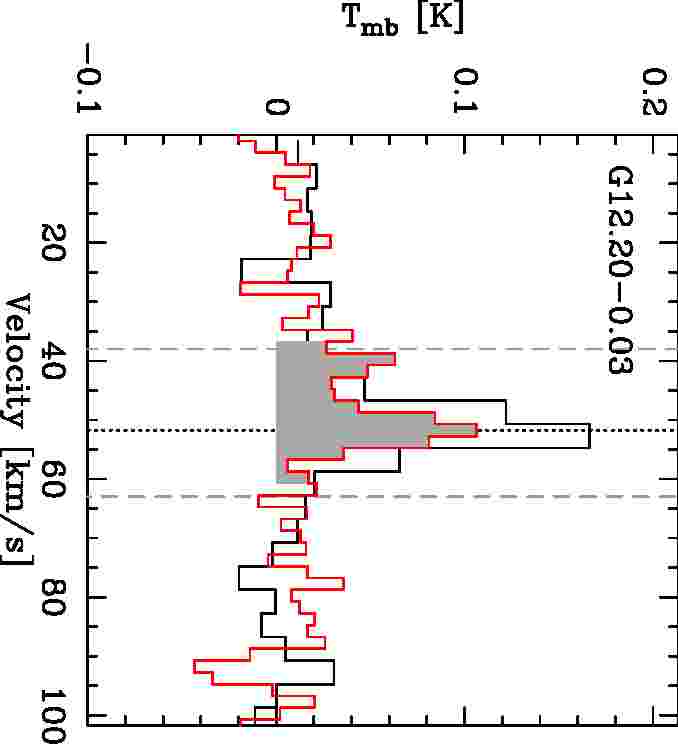} 
  \includegraphics[width=6.0cm,angle=0]{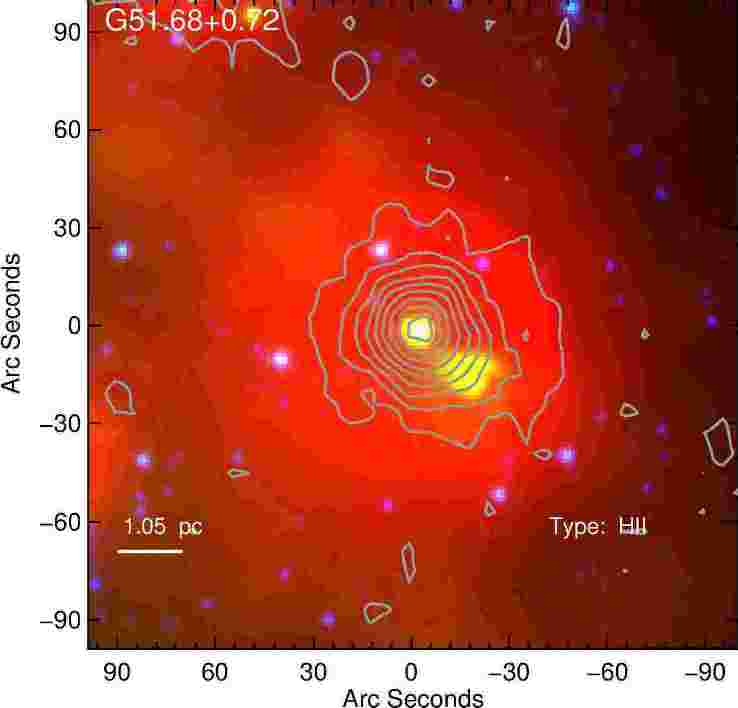}
  \includegraphics[width=6.0cm,angle=90]{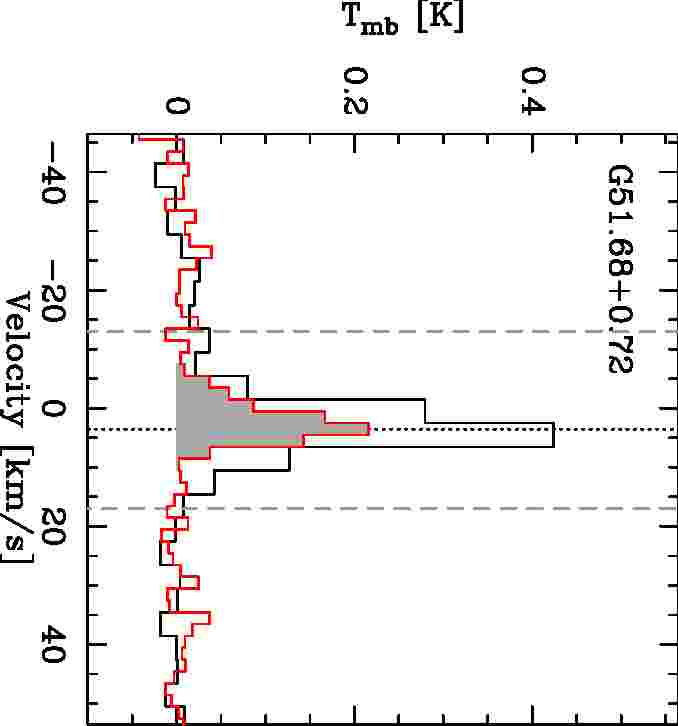}
   \includegraphics[width=6.0cm,angle=0]{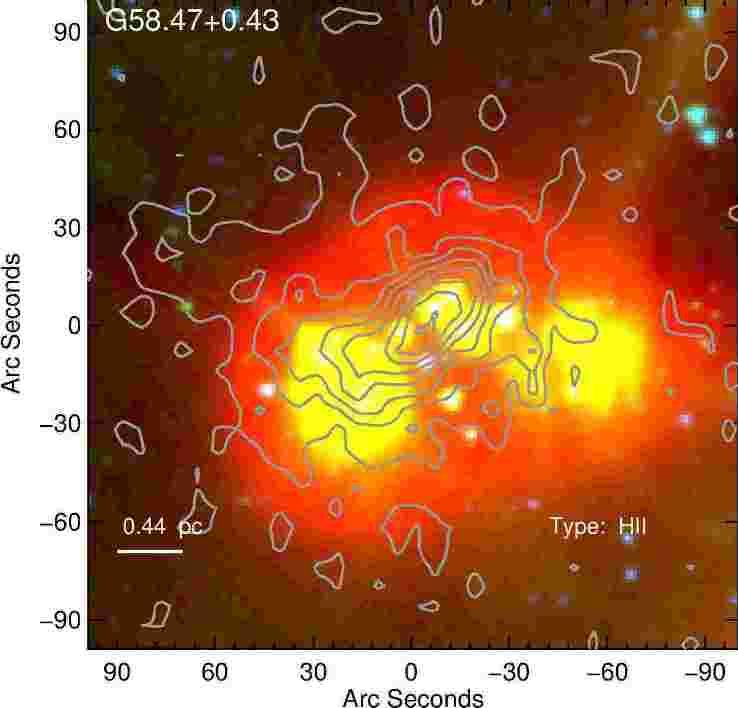}
  \includegraphics[width=6.0cm,angle=90]{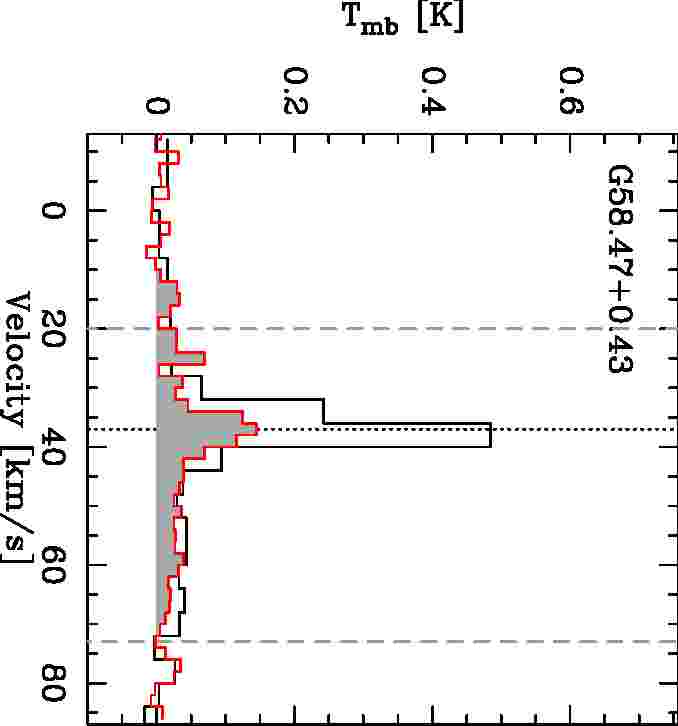}
  \includegraphics[width=6.0cm,angle=0]{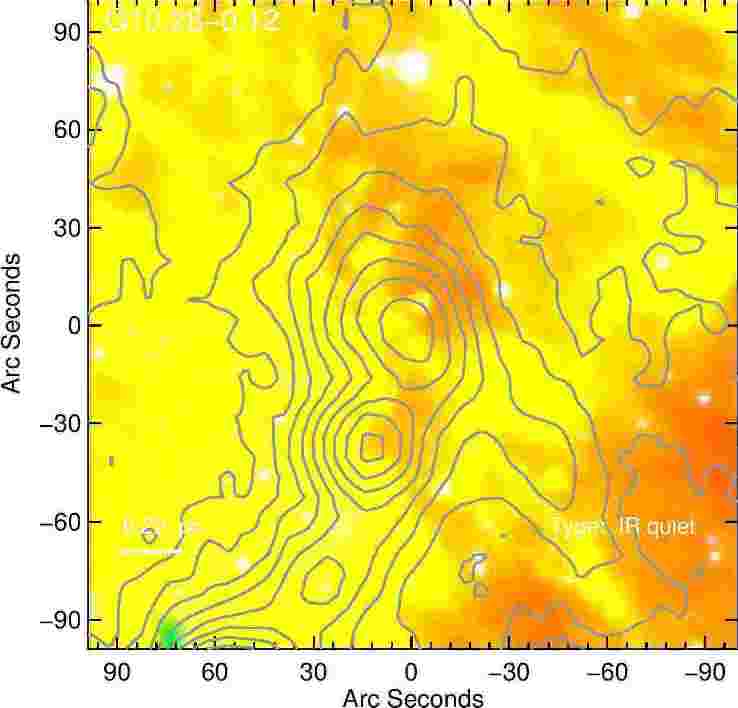}
  \includegraphics[width=6.0cm,angle=90]{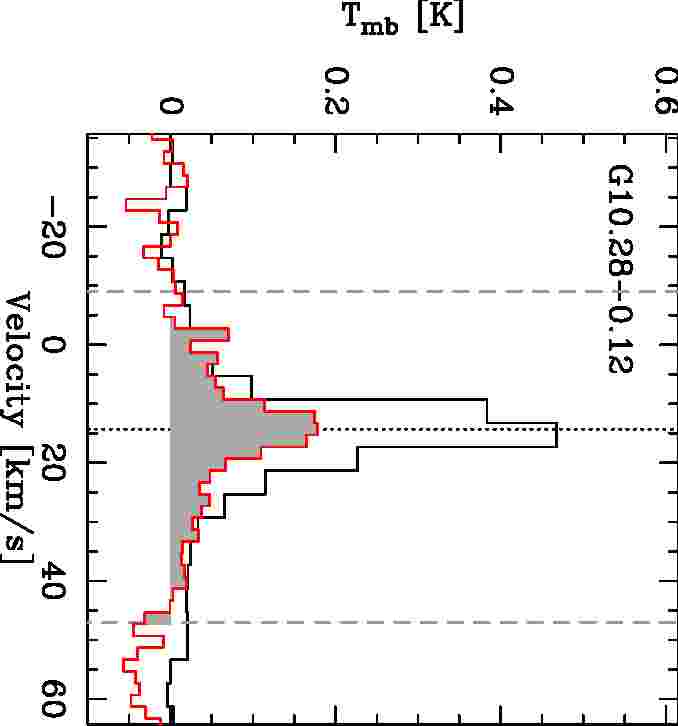}
\caption{Continued.}
\end{figure}
 \end{landscape}

\begin{landscape}
\begin{figure}
\ContinuedFloat
  \includegraphics[width=6.0cm,angle=0]{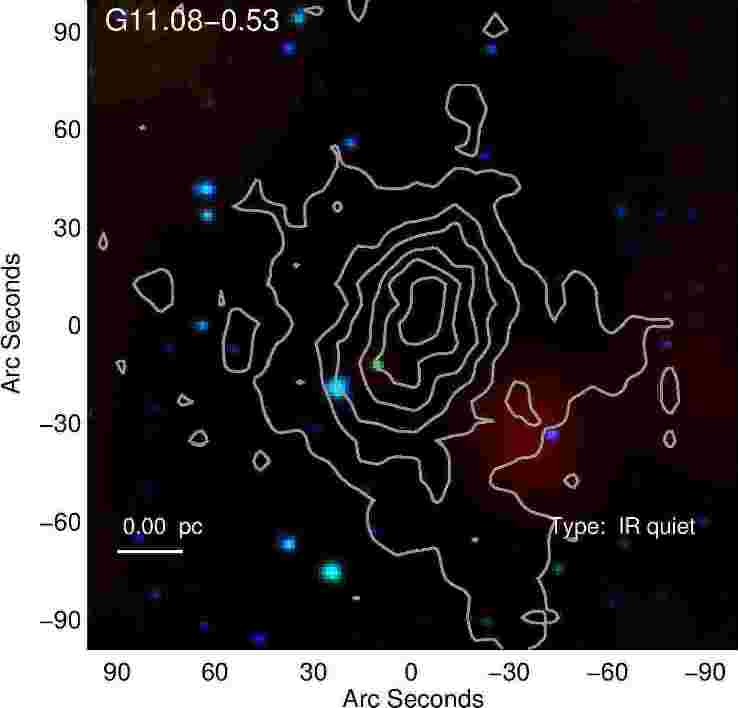}
  \includegraphics[width=6.0cm,angle=90]{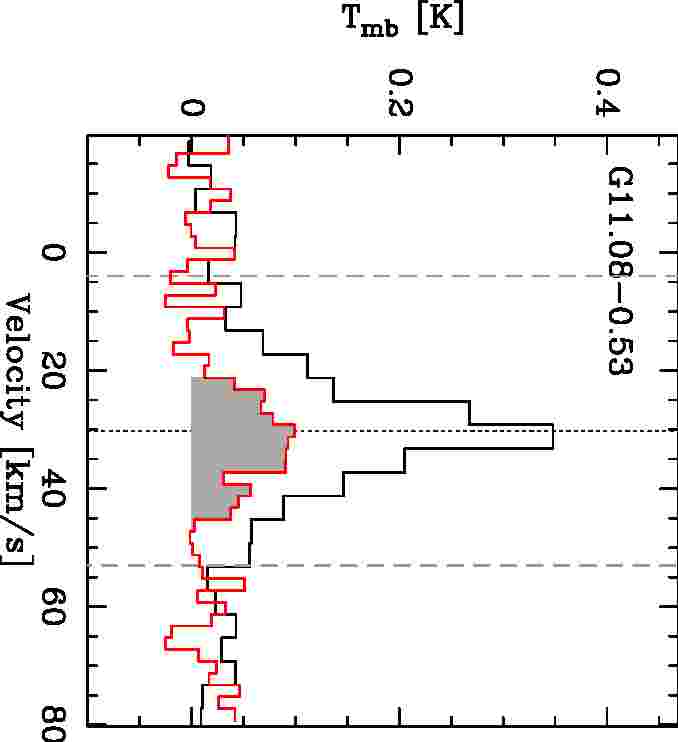}
  \includegraphics[width=6.0cm,angle=0]{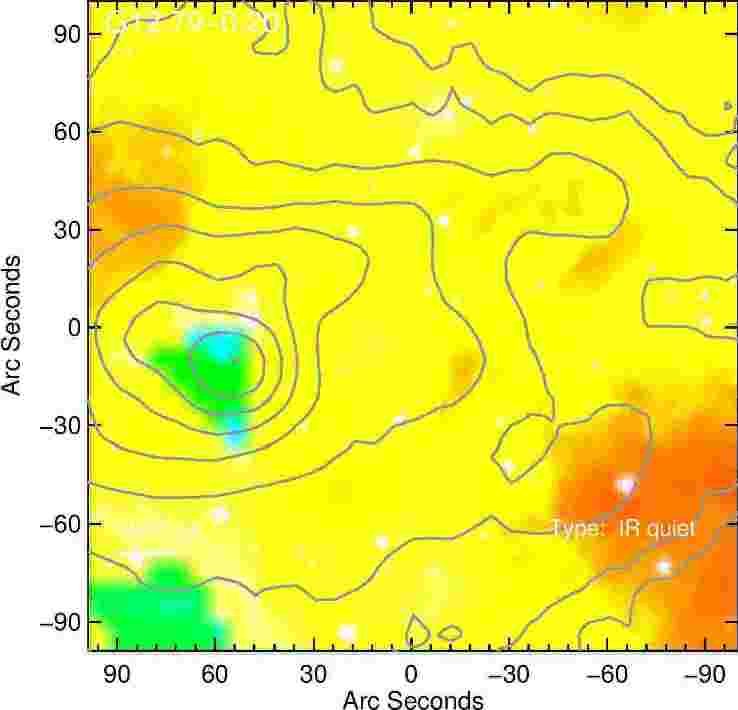}
  \includegraphics[width=6.0cm,angle=90]{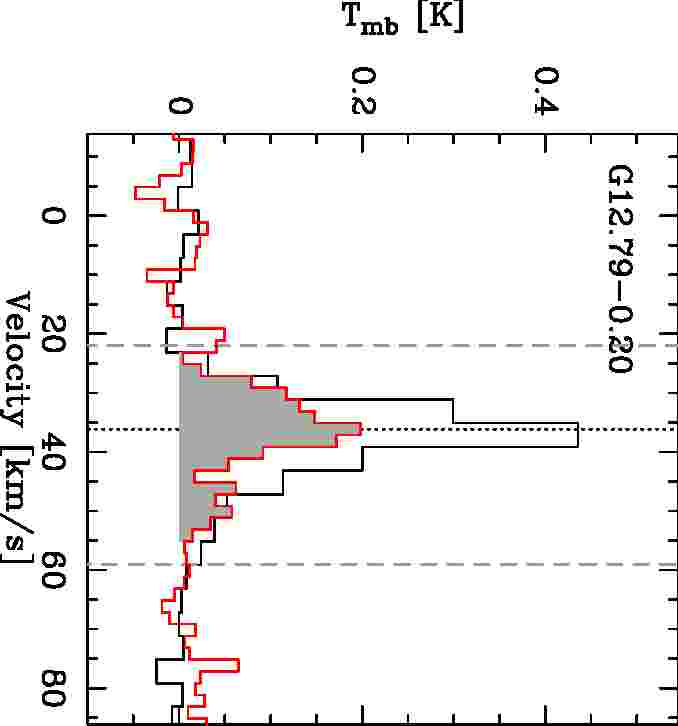}
  \includegraphics[width=6.0cm,angle=0]{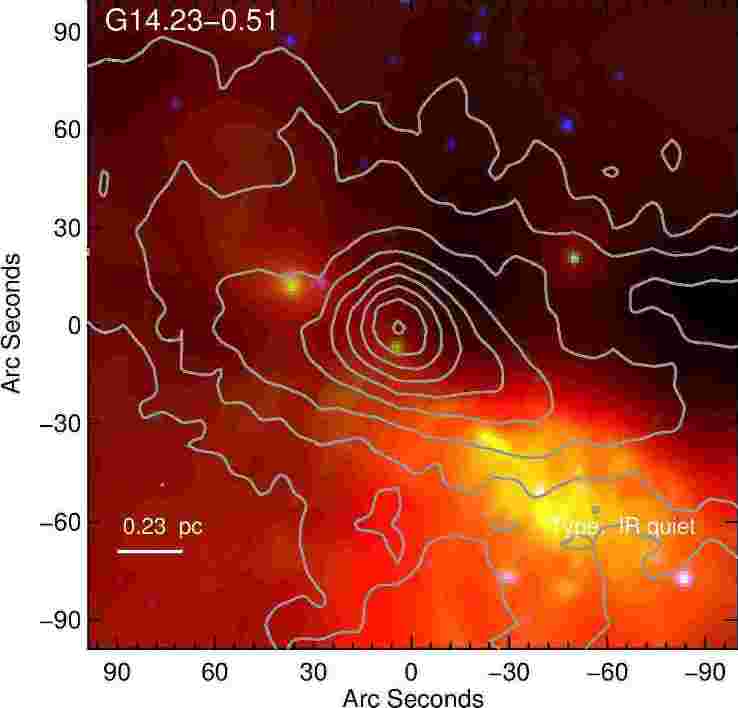}
  \includegraphics[width=6.0cm,angle=90]{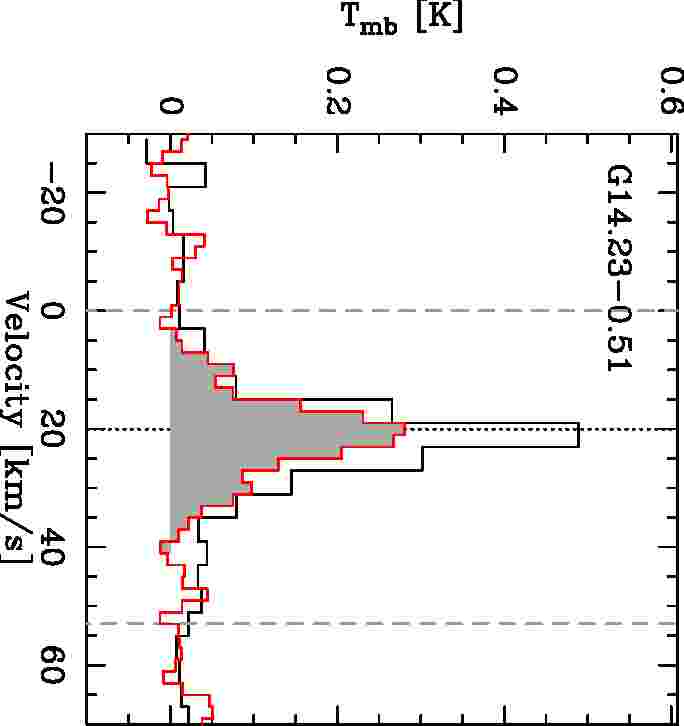}
  \includegraphics[width=6.0cm,angle=0]{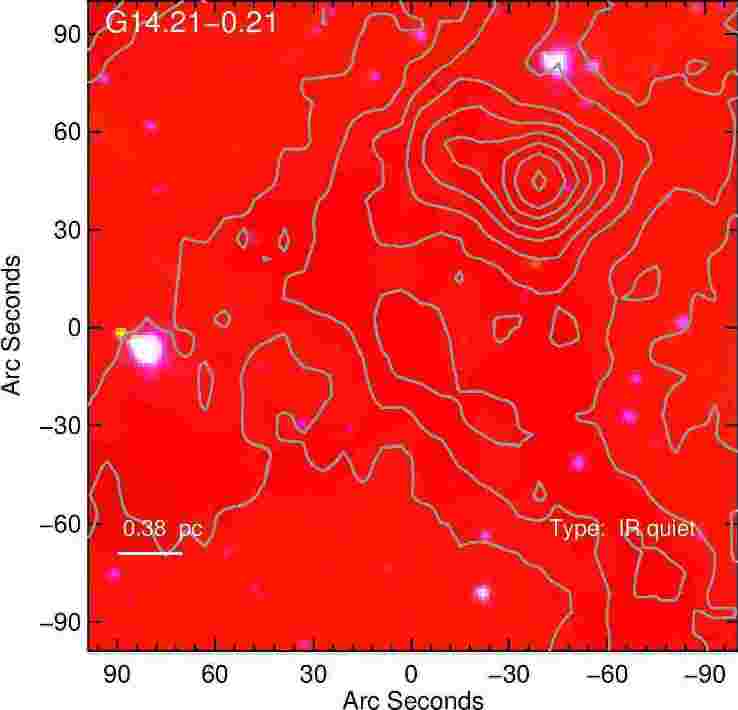}
  \includegraphics[width=6.0cm,angle=90]{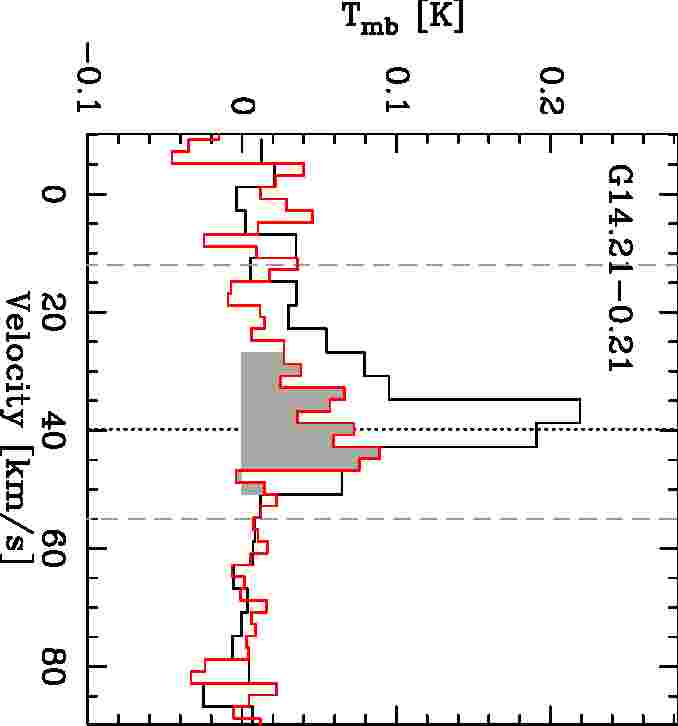}
  \includegraphics[width=6.0cm,angle=0]{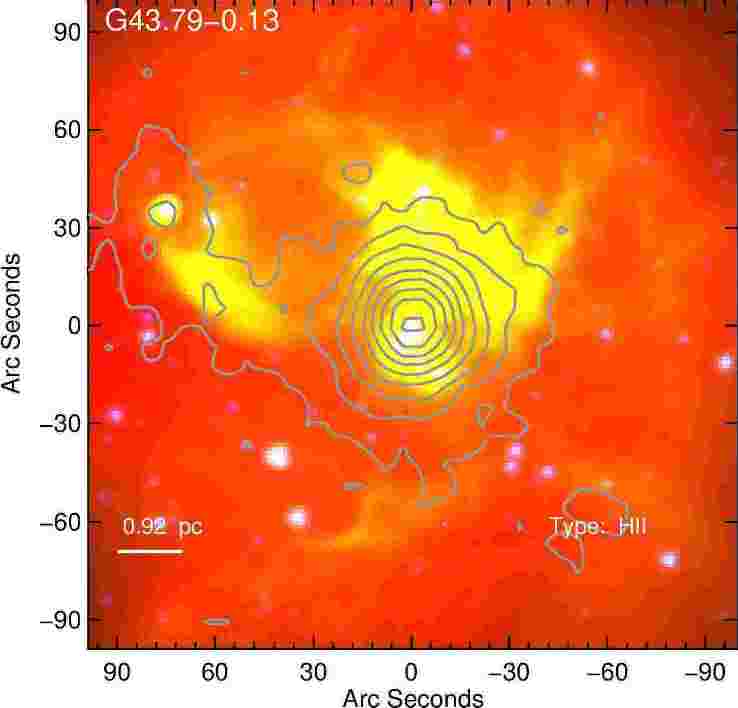}
  \includegraphics[width=6.0cm,angle=90]{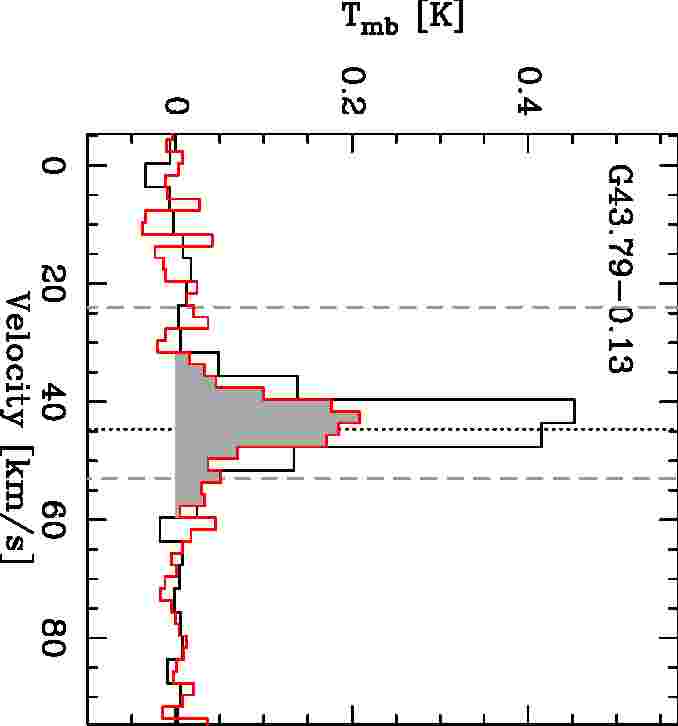}
  \includegraphics[width=6.0cm,angle=0]{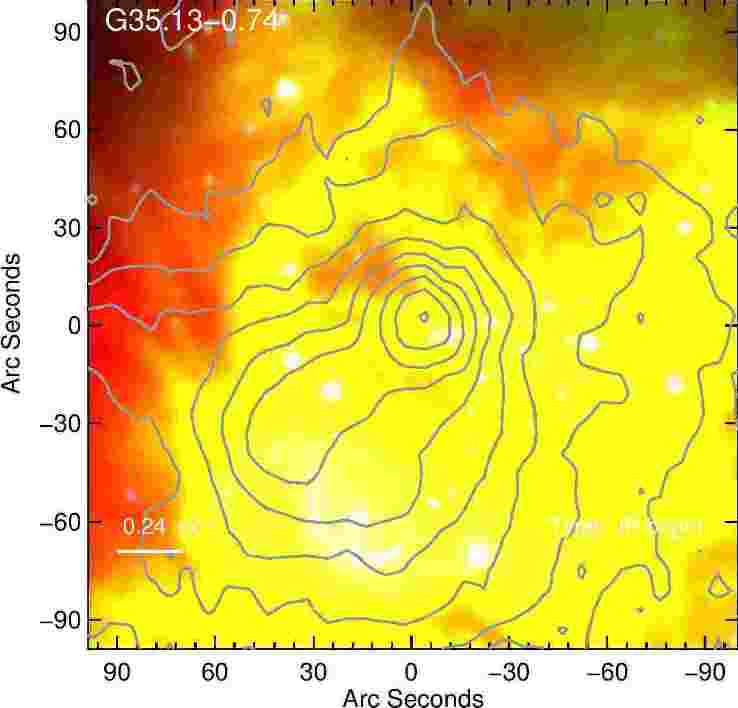}
  \includegraphics[width=6.0cm,angle=90]{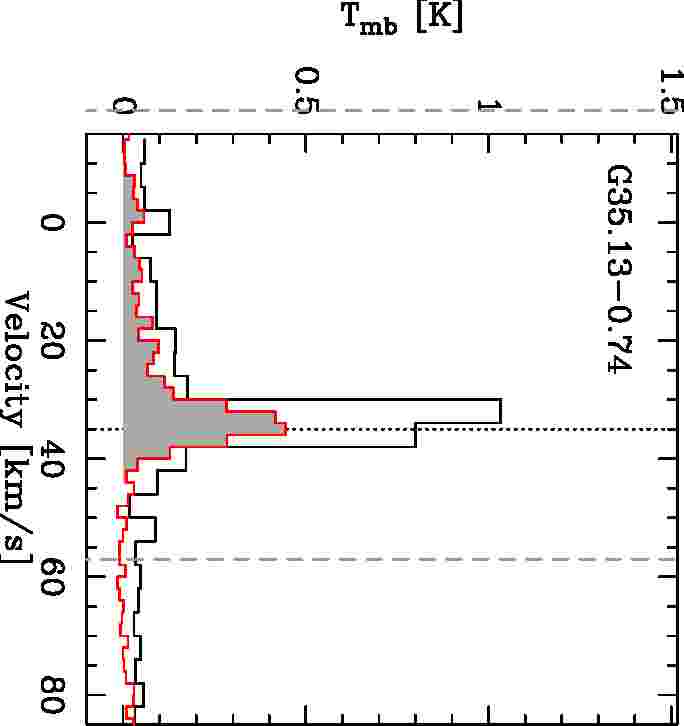}
 \caption{Continued.}
\end{figure}
\end{landscape}

\begin{landscape}
\begin{figure}
\ContinuedFloat
  \includegraphics[width=6.0cm,angle=0]{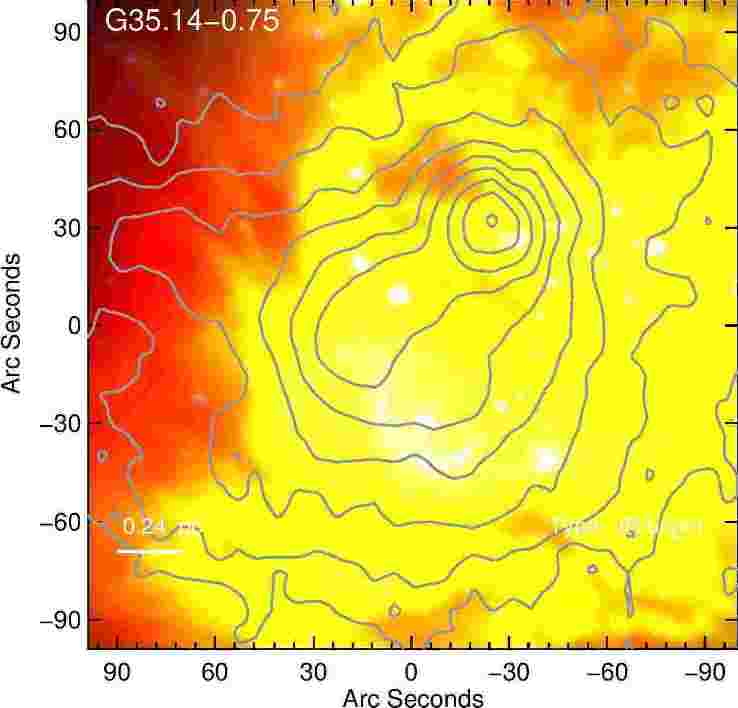}
  \includegraphics[width=6.0cm,angle=90]{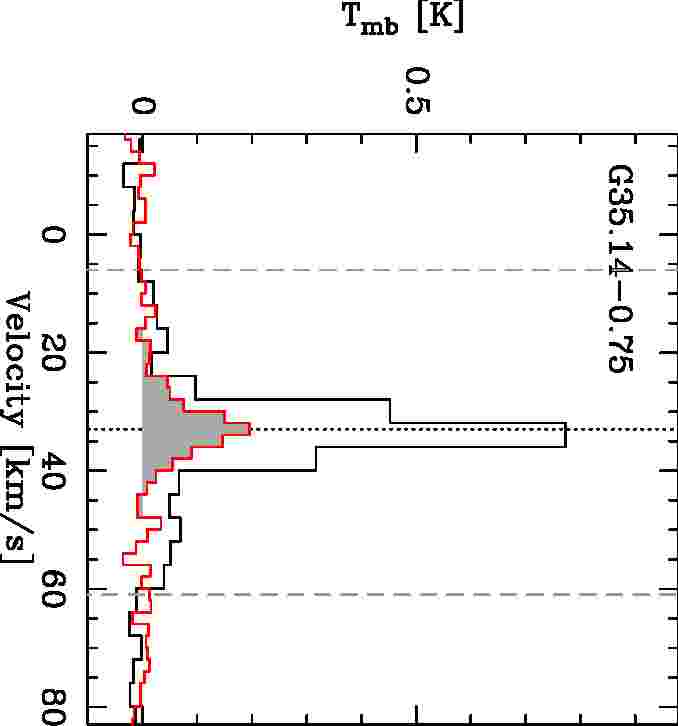}
  \includegraphics[width=6.0cm,angle=0]{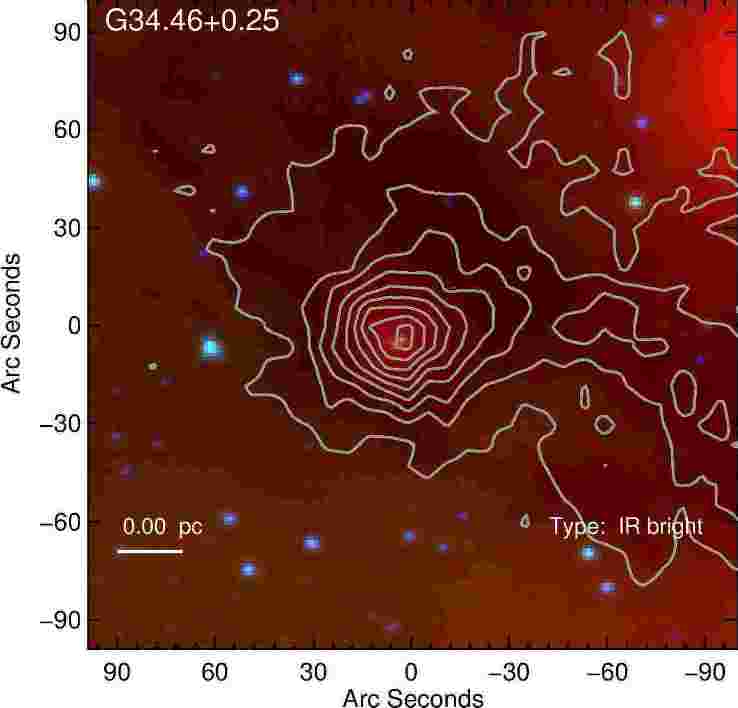}
  \includegraphics[width=6.0cm,angle=90]{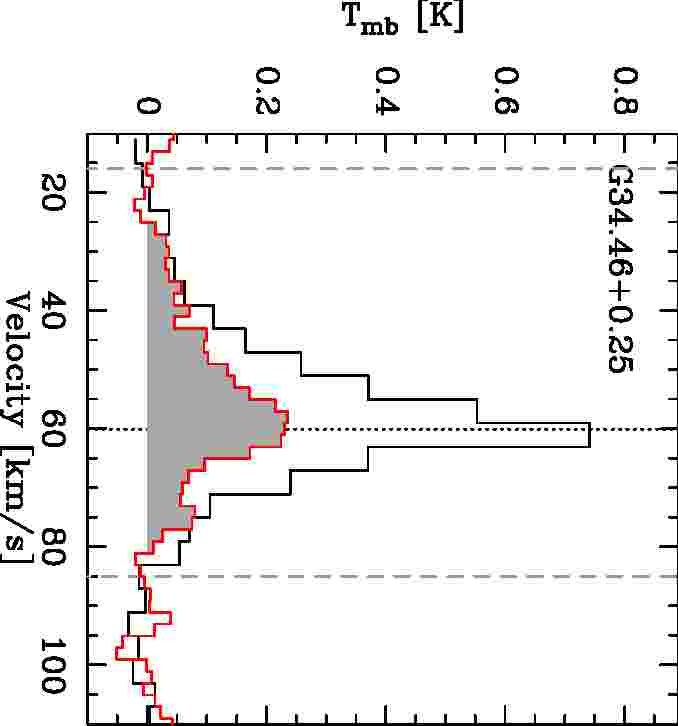}
  \includegraphics[width=6.0cm,angle=0]{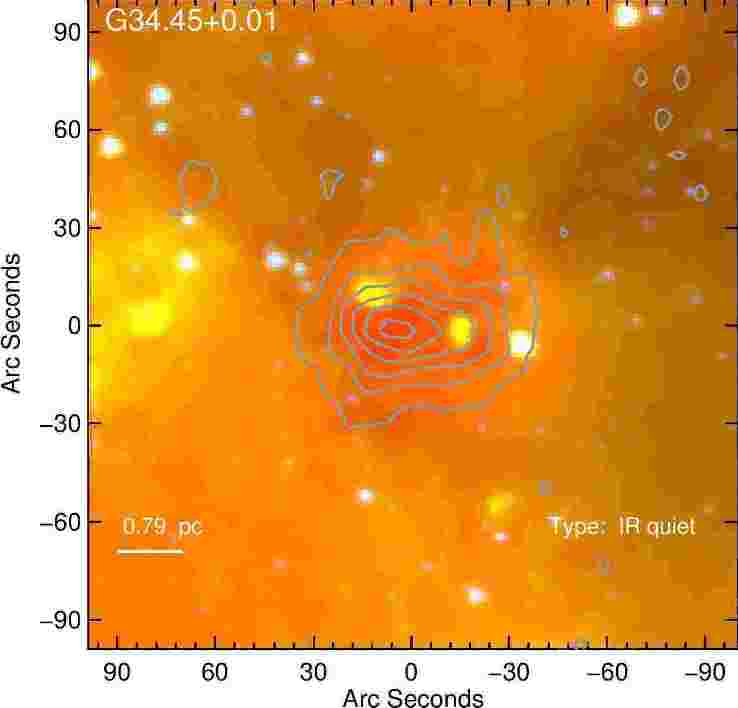}
  \includegraphics[width=6.0cm,angle=90]{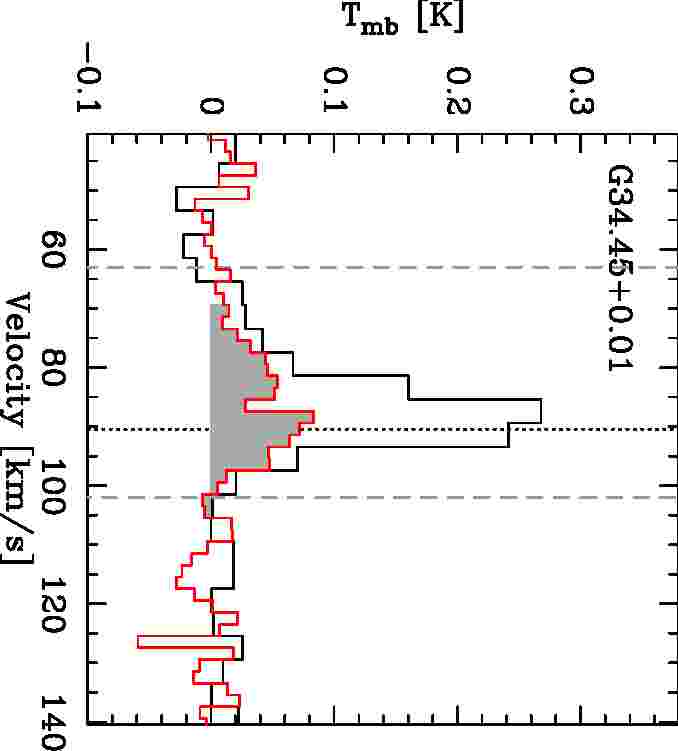}
  \includegraphics[width=6.0cm,angle=0]{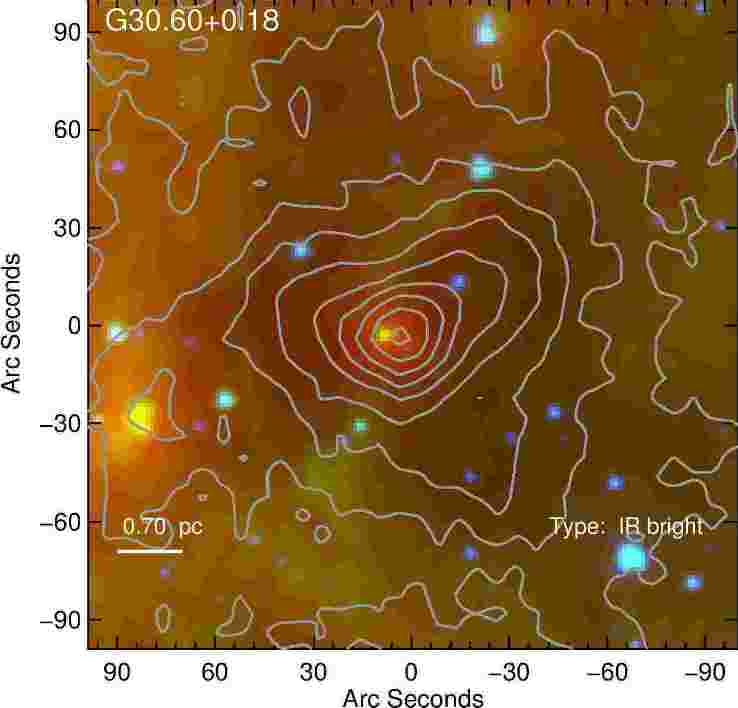}
  \includegraphics[width=6.0cm,angle=90]{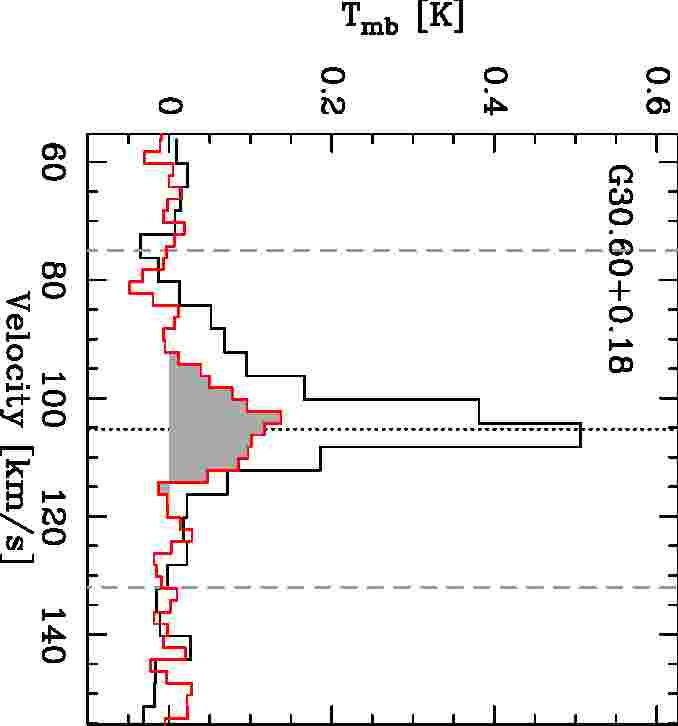}
 \includegraphics[width=6.0cm,angle=0]{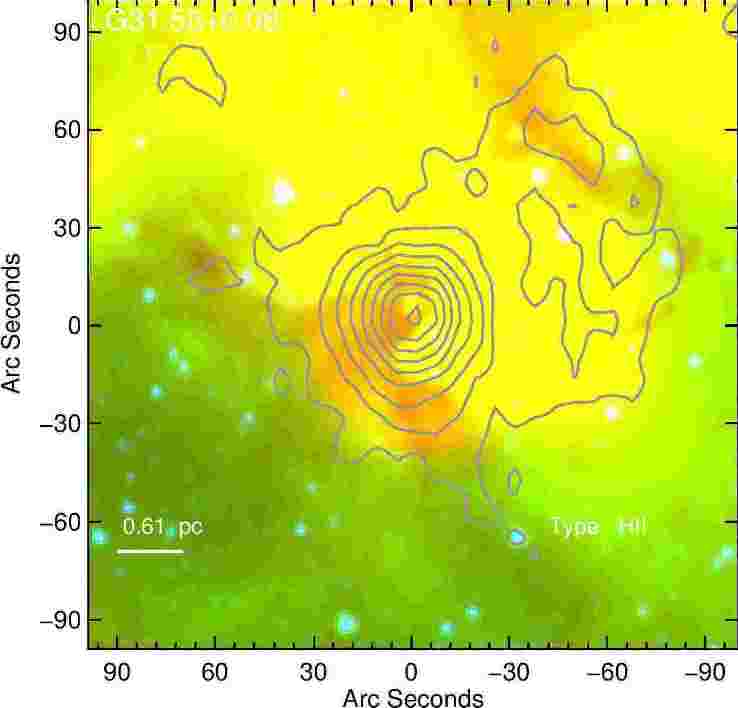}
  \includegraphics[width=6.0cm,angle=90]{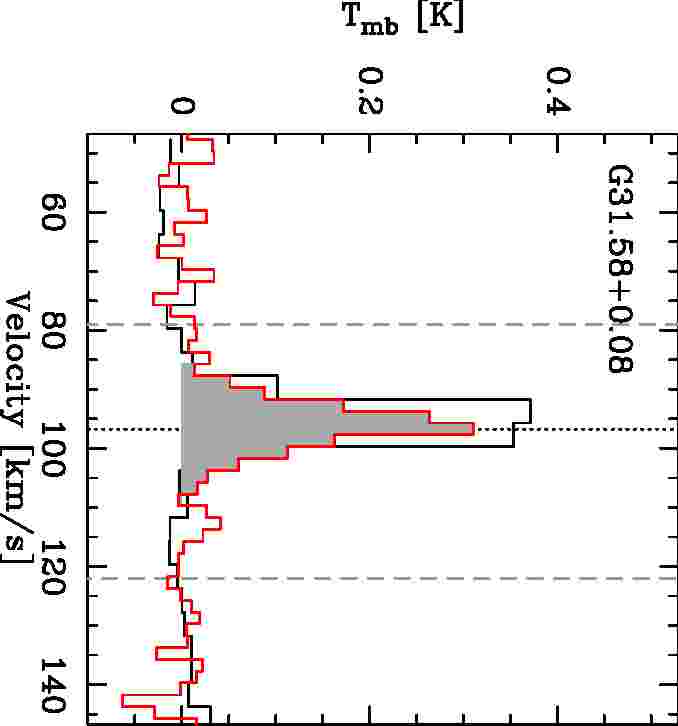}
  \includegraphics[width=6.0cm,angle=0]{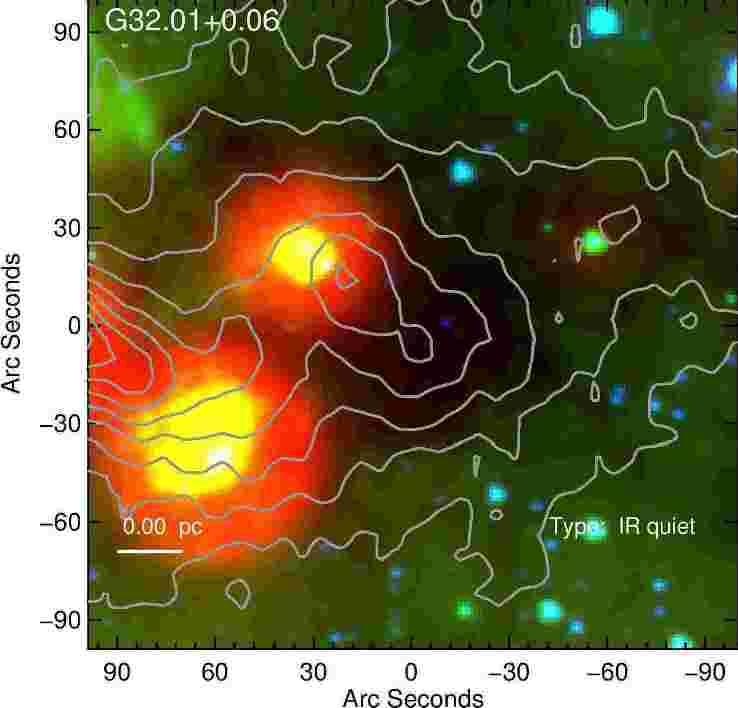}
  \includegraphics[width=6.0cm,angle=90]{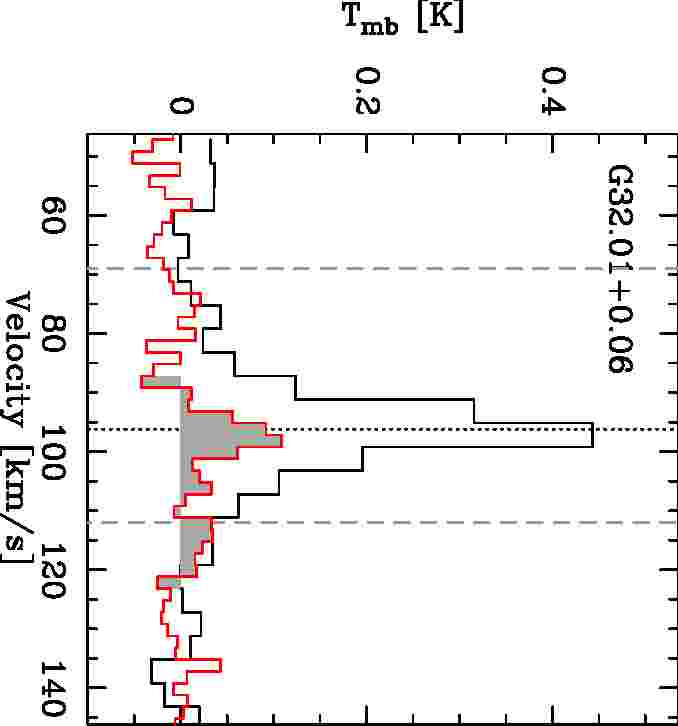}
 \caption{Continued.}
\end{figure}
\end{landscape}

\begin{landscape}
\begin{figure}
\ContinuedFloat
   \includegraphics[width=6.0cm,angle=0]{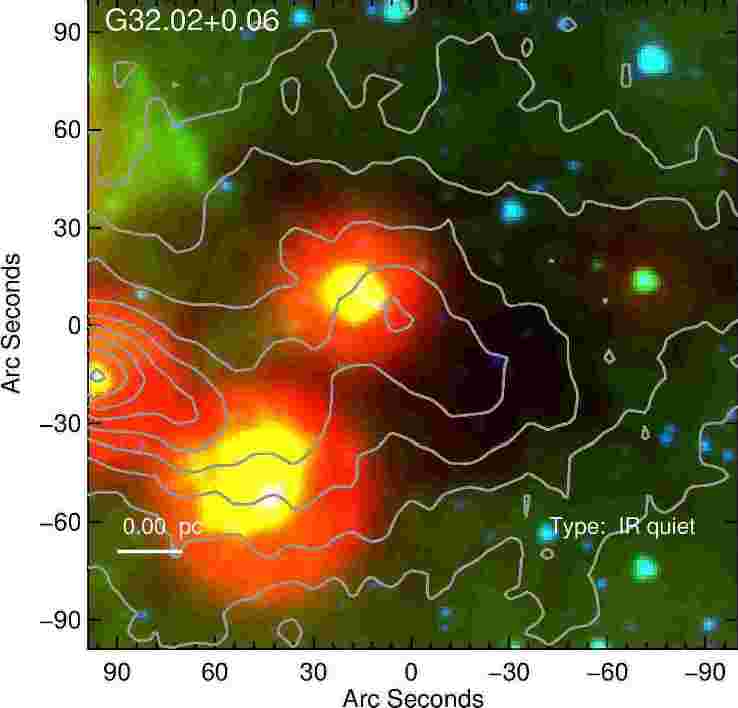}
  \includegraphics[width=6.0cm,angle=90]{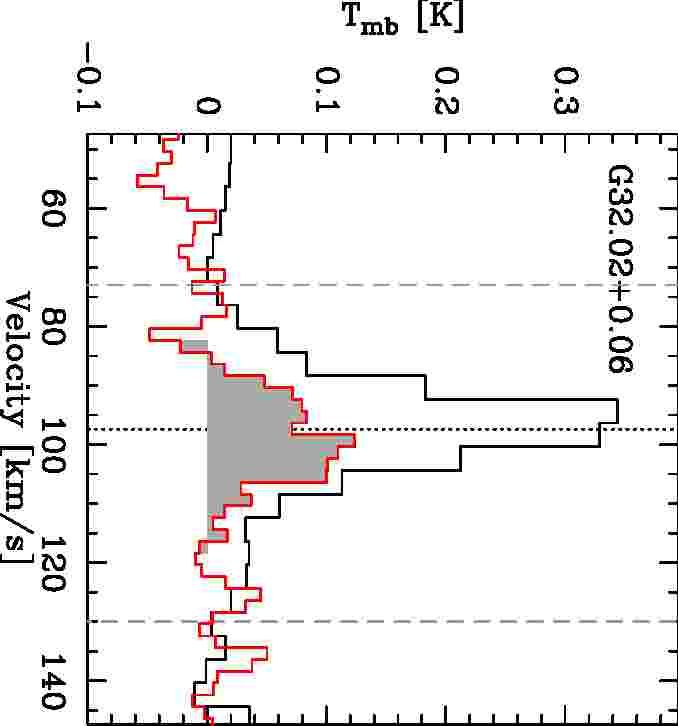}
  \includegraphics[width=6.0cm,angle=0]{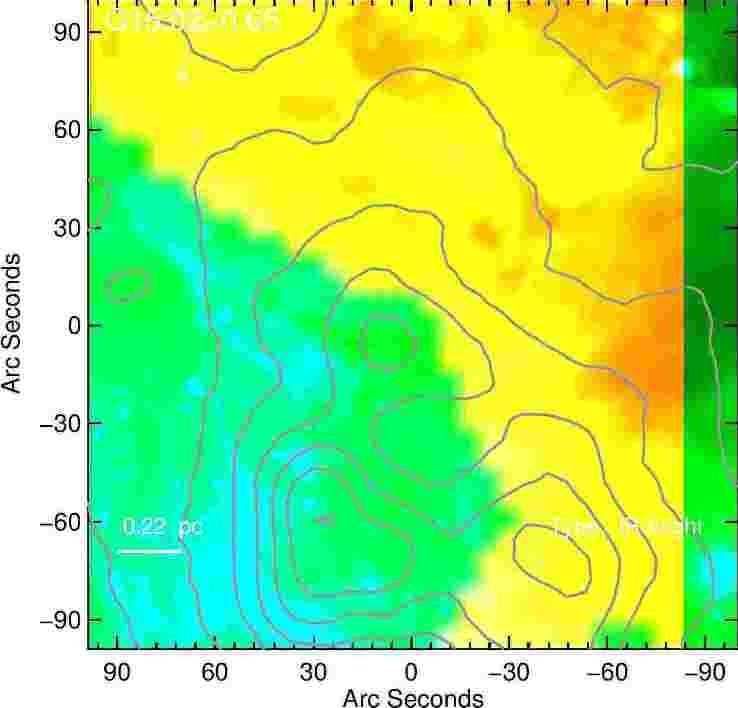}
  \includegraphics[width=6.0cm,angle=90]{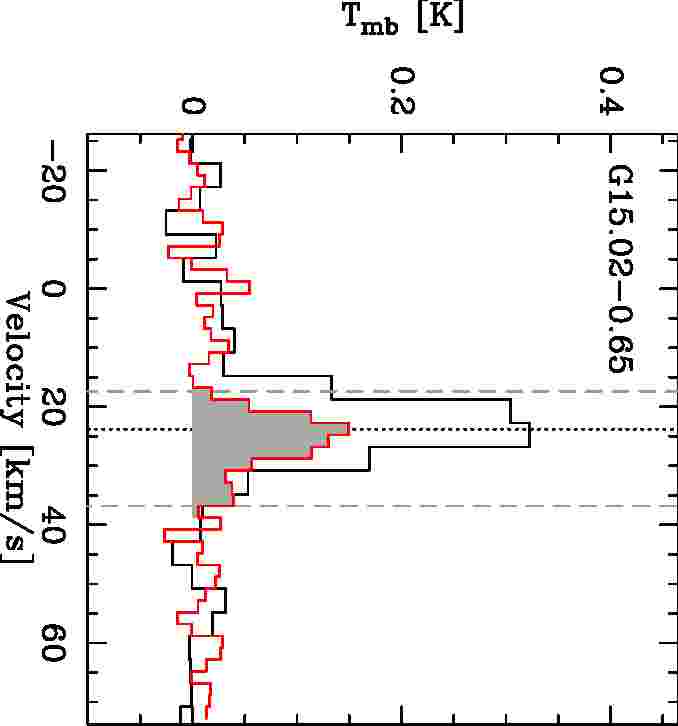}
  \includegraphics[width=6.0cm,angle=0]{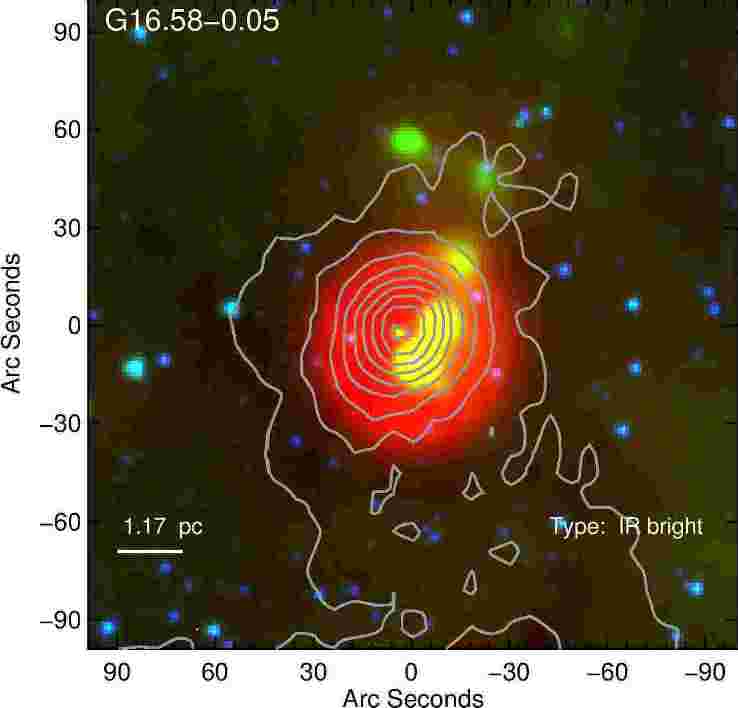}
  \includegraphics[width=6.0cm,angle=90]{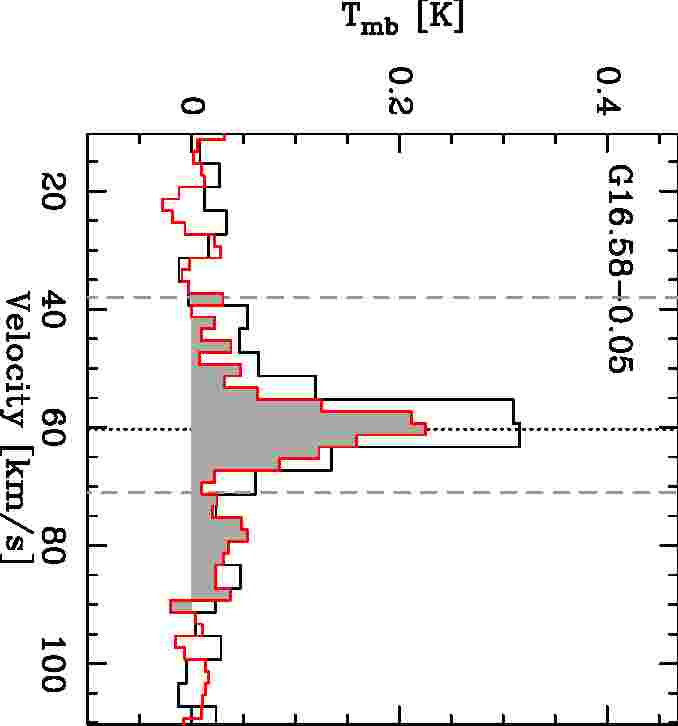}
  \includegraphics[width=6.0cm,angle=0]{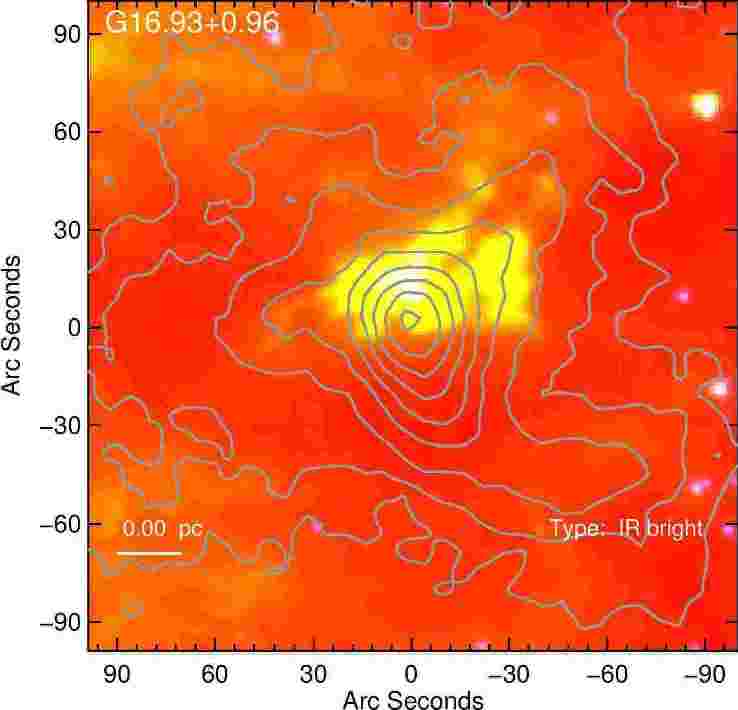}
  \includegraphics[width=6.0cm,angle=90]{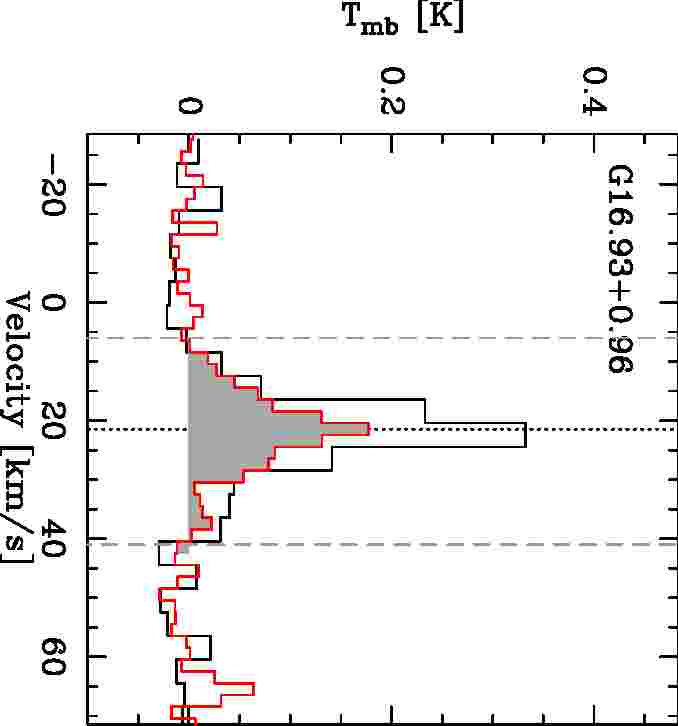}
   \includegraphics[width=6.0cm,angle=0]{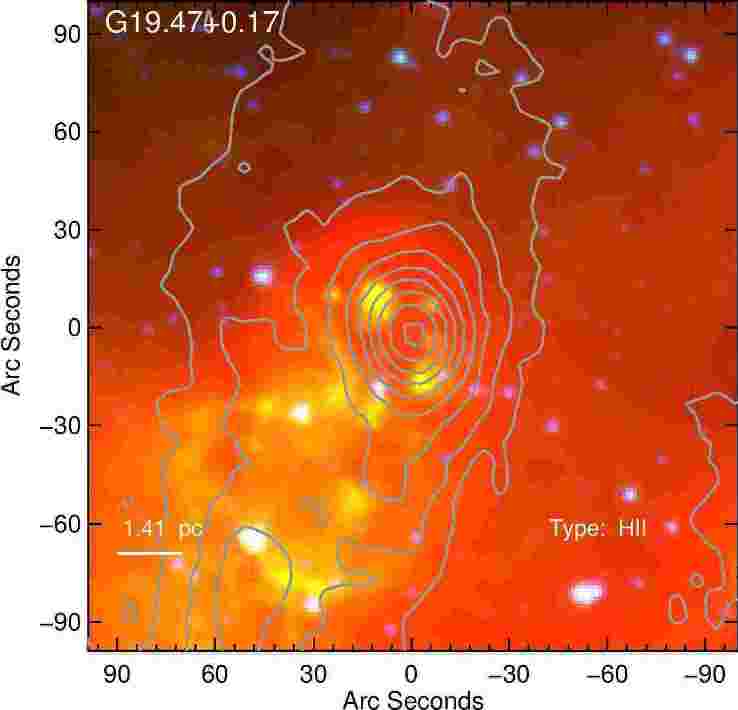}
   \includegraphics[width=6.0cm,angle=90]{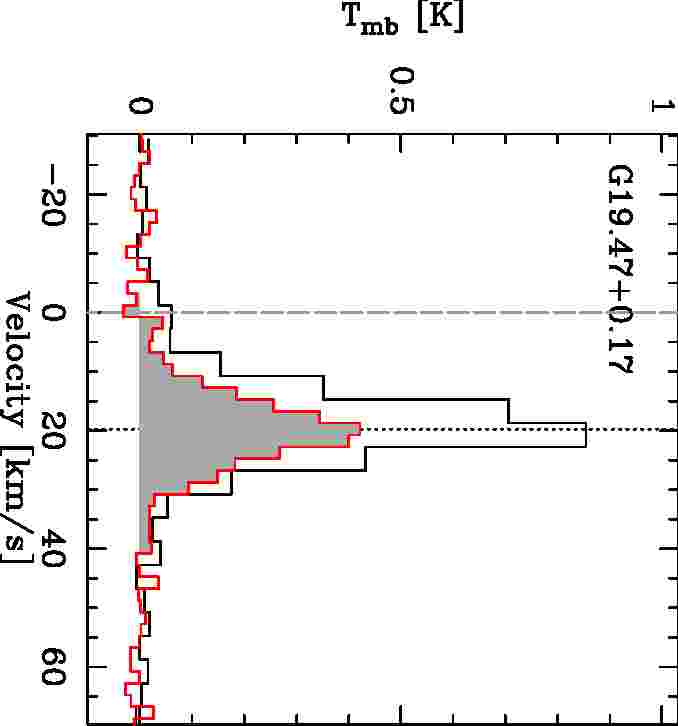}
   \includegraphics[width=6.0cm,angle=0]{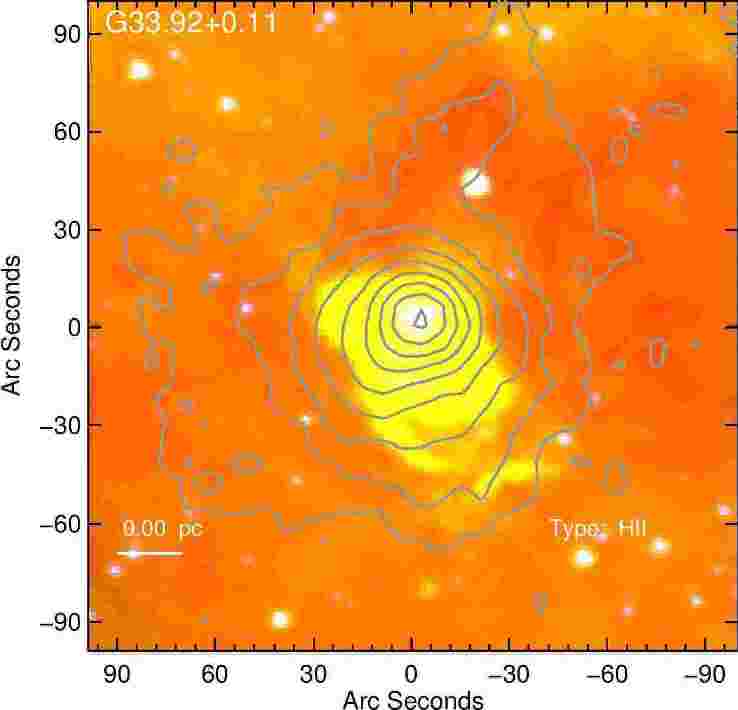}
   \includegraphics[width=6.0cm,angle=90]{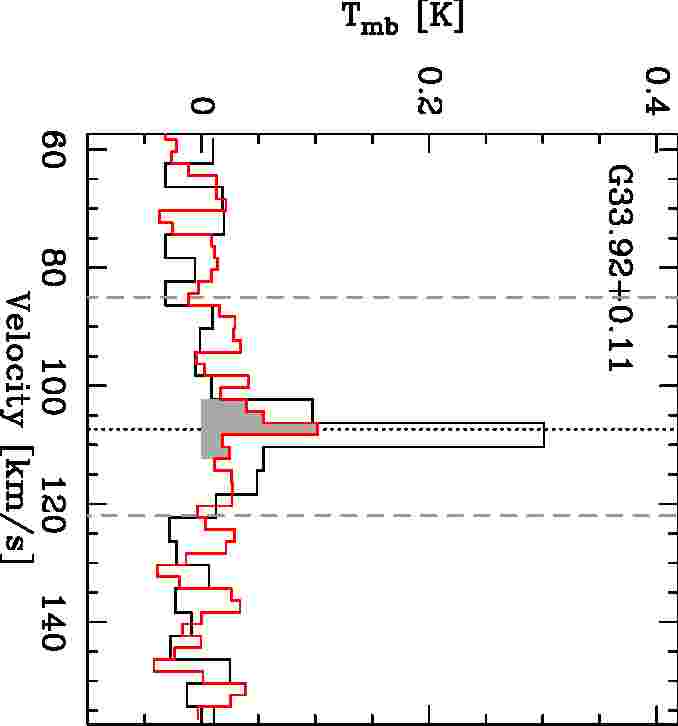}
 \caption{Continued.}
\end{figure}
\end{landscape}

\begin{landscape}
\begin{figure}
\ContinuedFloat
  \includegraphics[width=6.0cm,angle=0]{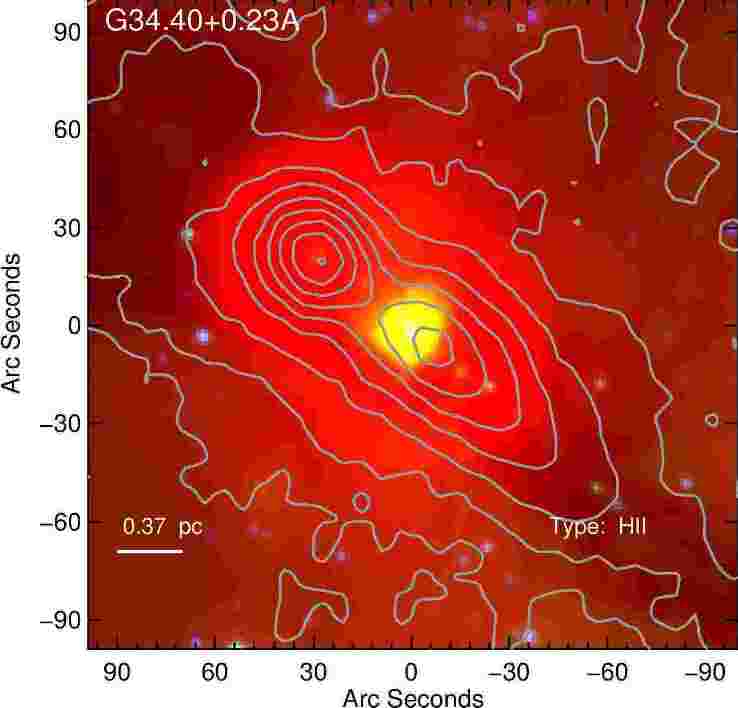}
 \includegraphics[width=6.0cm,angle=90]{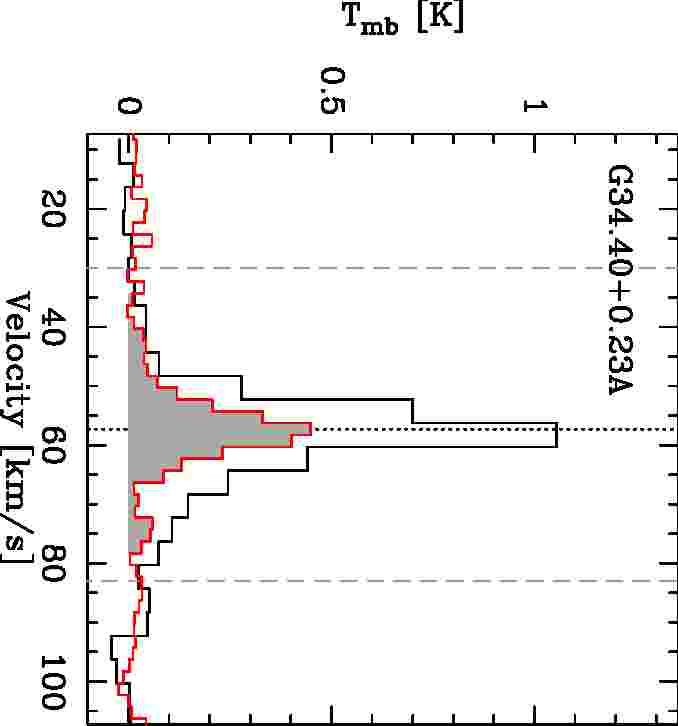}
  \includegraphics[width=6.0cm,angle=0]{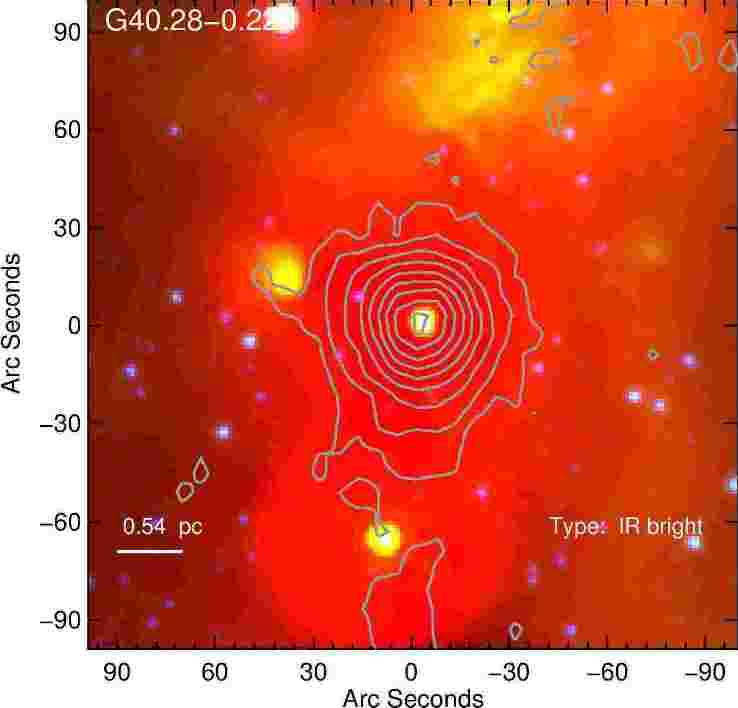}
  \includegraphics[width=6.0cm,angle=90]{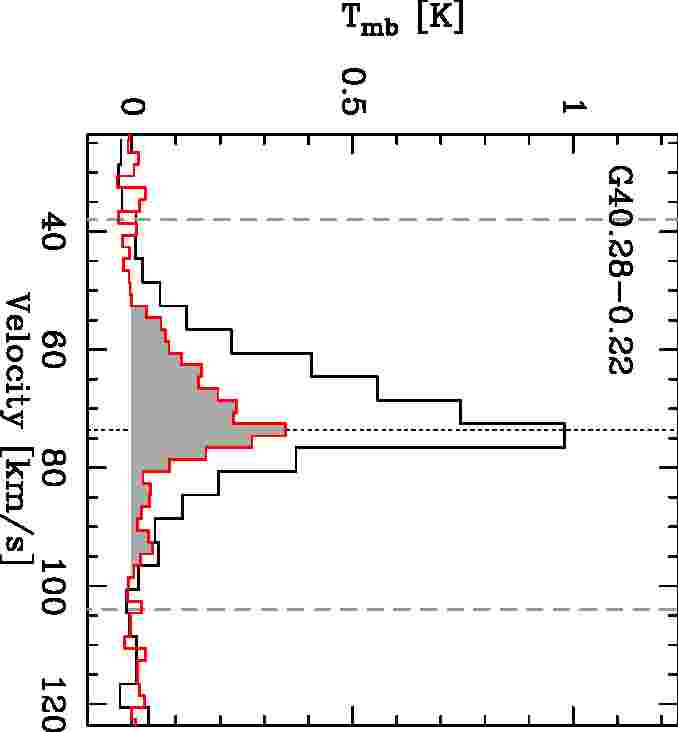}
  \includegraphics[width=6.0cm,angle=0]{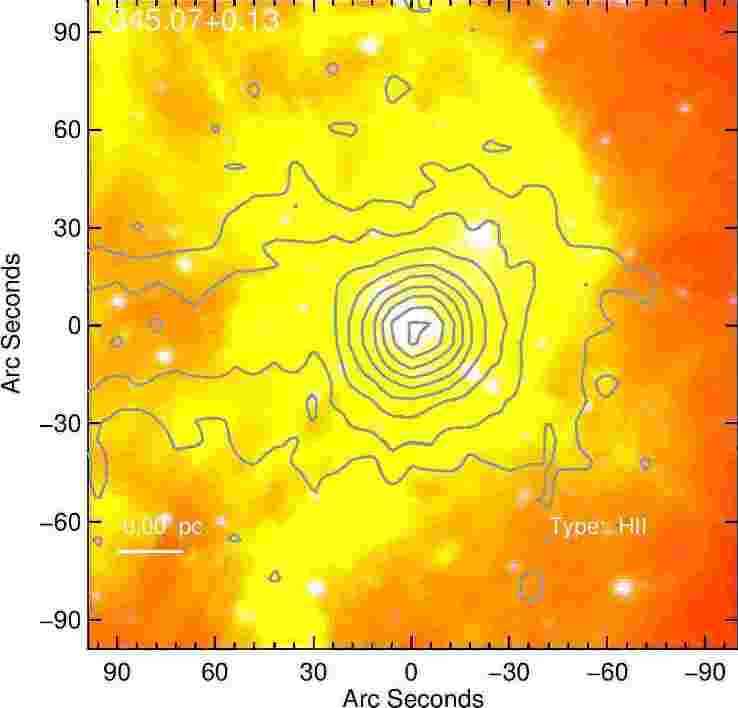}
  \includegraphics[width=6.0cm,angle=90]{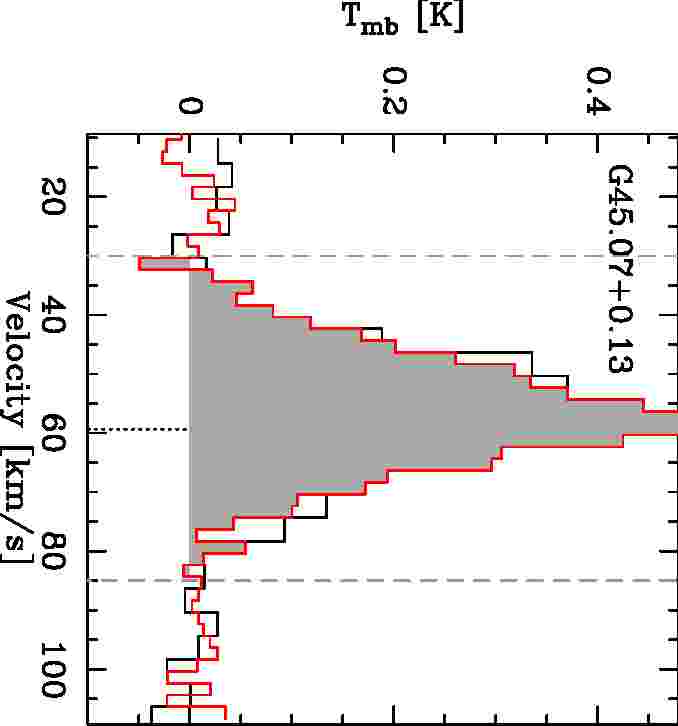}
  \includegraphics[width=6.0cm,angle=0]{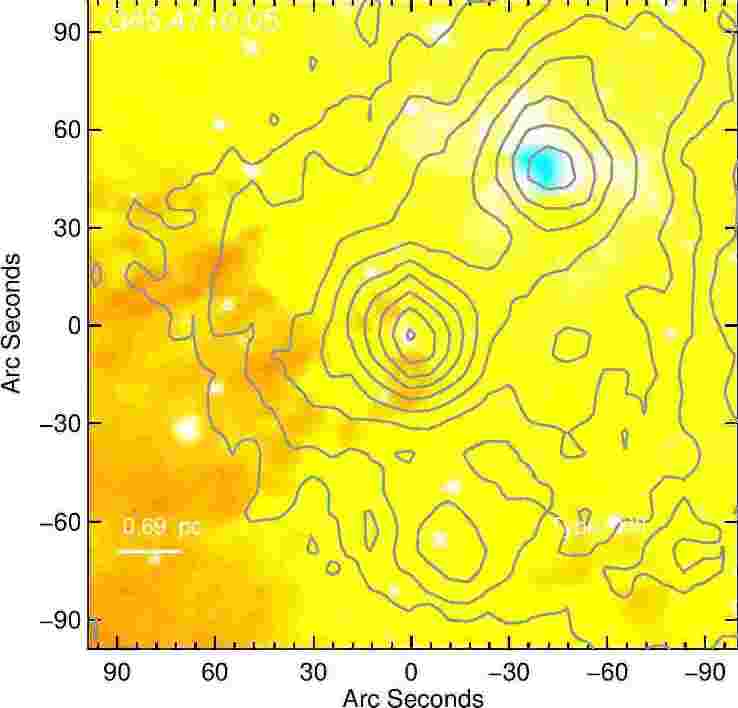}
  \includegraphics[width=6.0cm,angle=90]{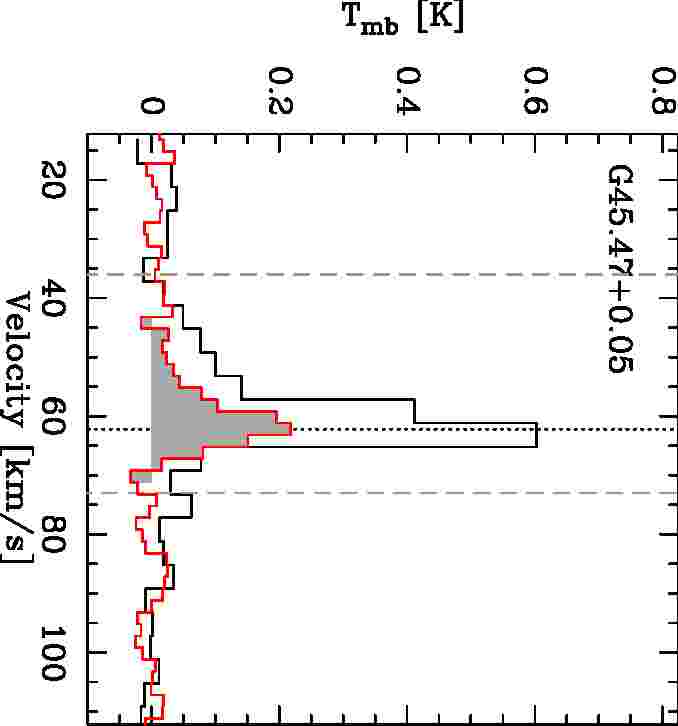}
 \includegraphics[width=6.0cm,angle=0]{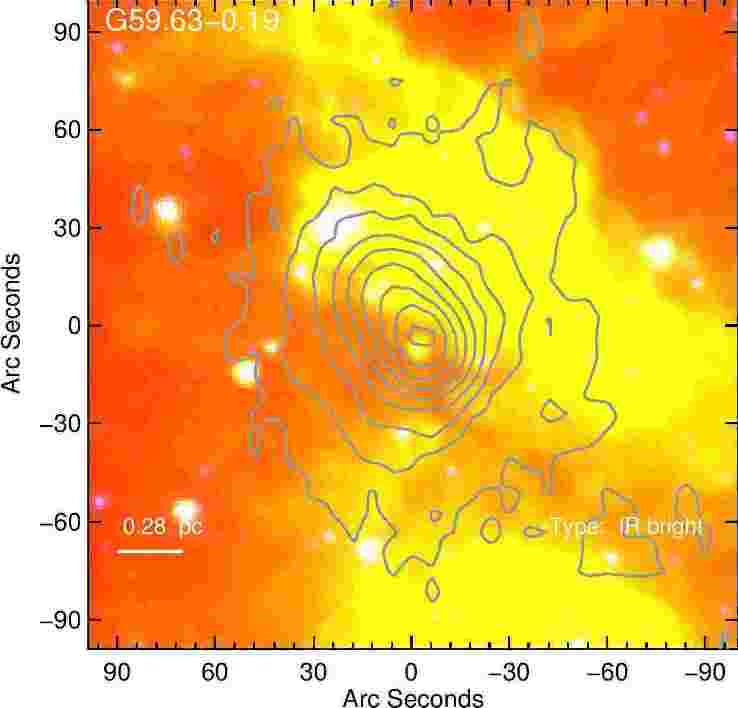}
 \includegraphics[width=6.0cm,angle=90]{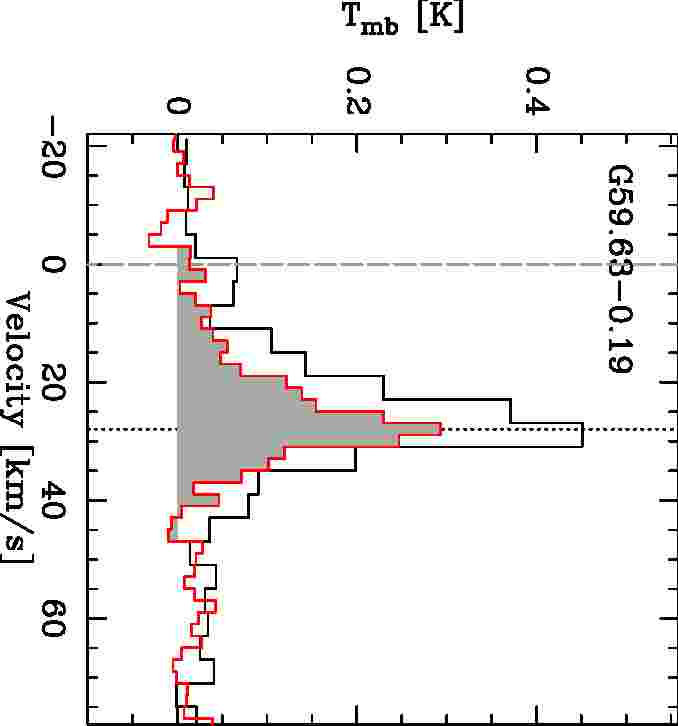}
  \includegraphics[width=6.0cm,angle=0]{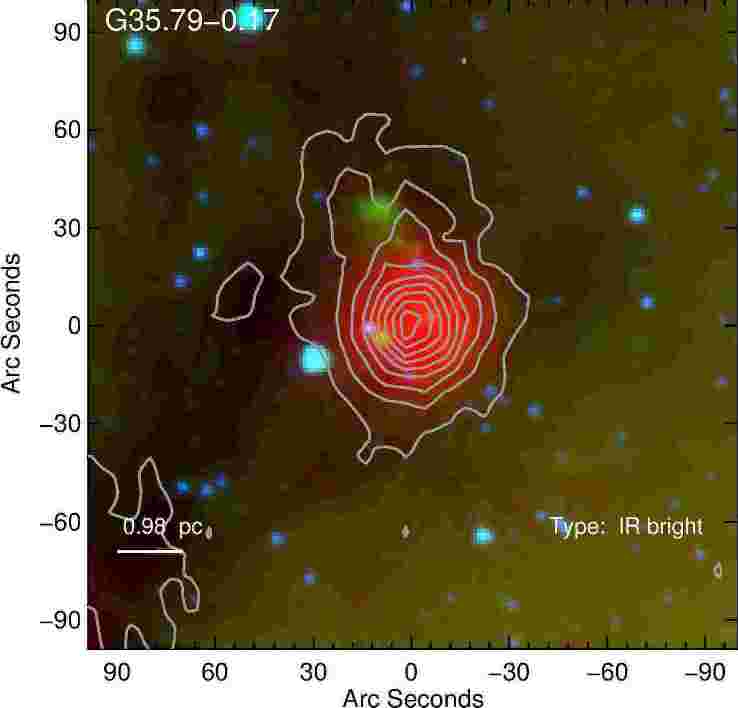}
  \includegraphics[width=6.0cm,angle=90]{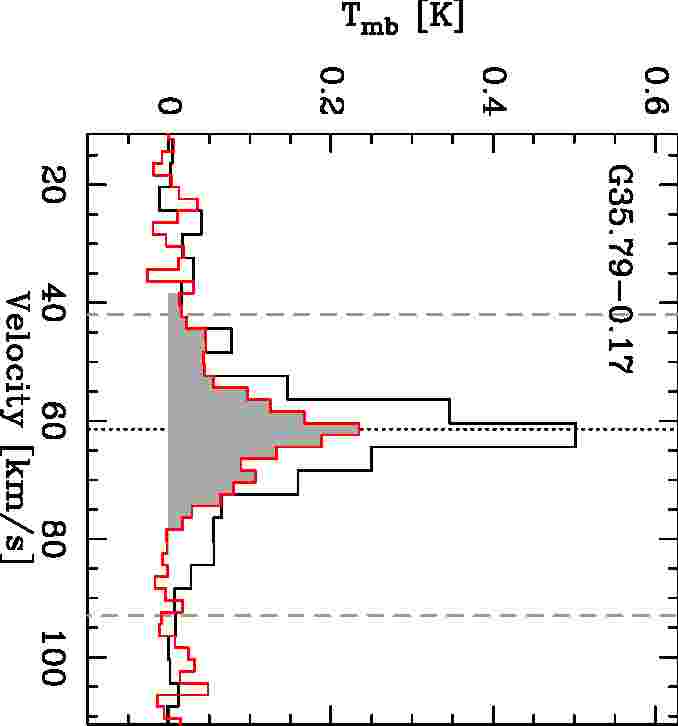}
 \caption{Continued.}
\end{figure}
\end{landscape}

\begin{landscape}
\begin{figure}
\ContinuedFloat
  \includegraphics[width=6.0cm,angle=0]{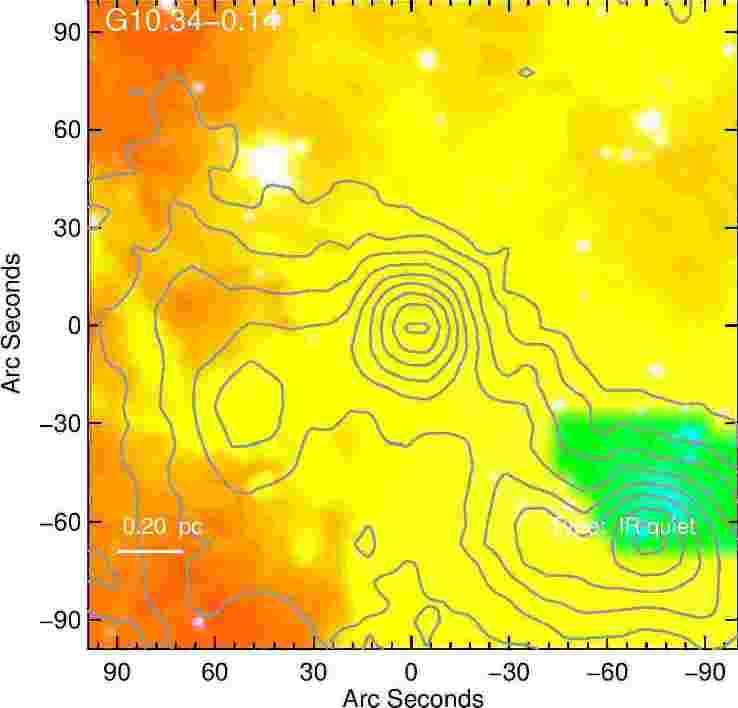}
  \includegraphics[width=6.0cm,angle=90]{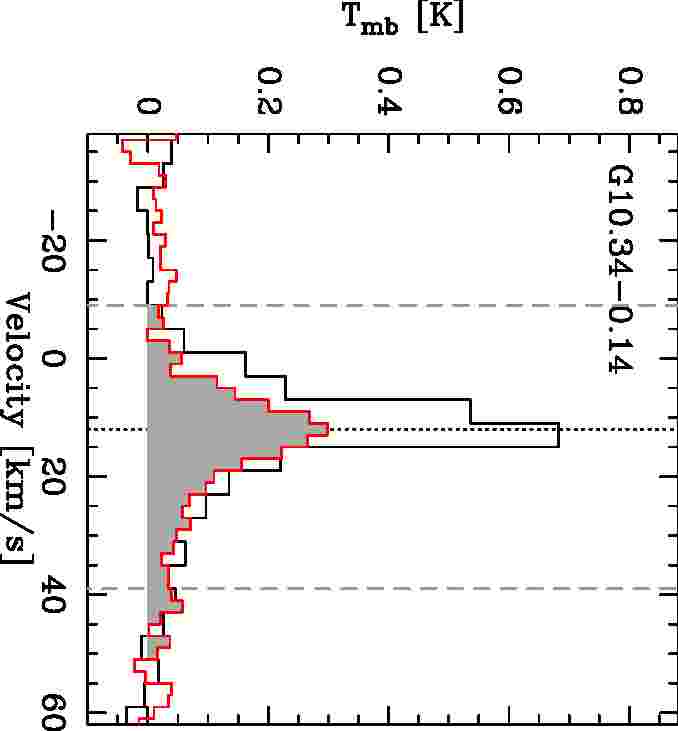}
  \includegraphics[width=6.0cm,angle=0]{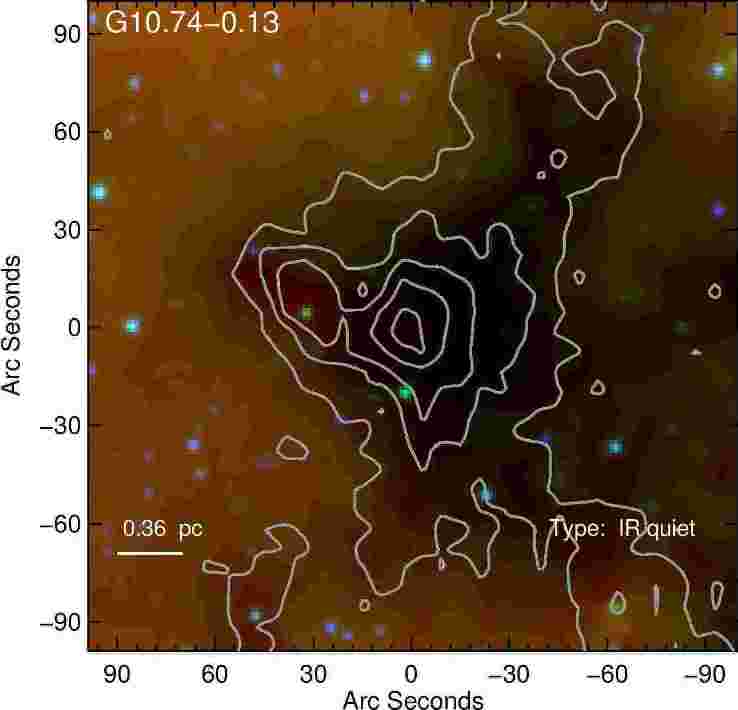}
  \includegraphics[width=6.0cm,angle=90]{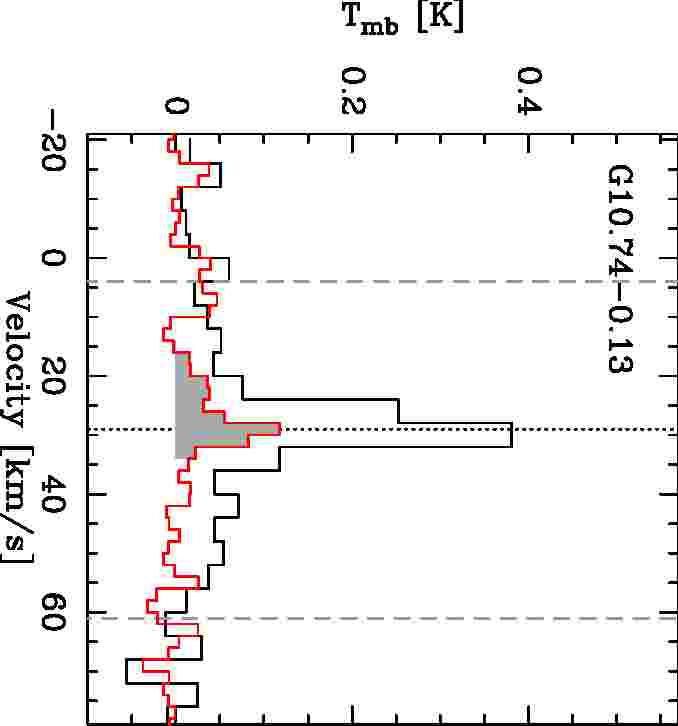}
  \includegraphics[width=6.0cm,angle=0]{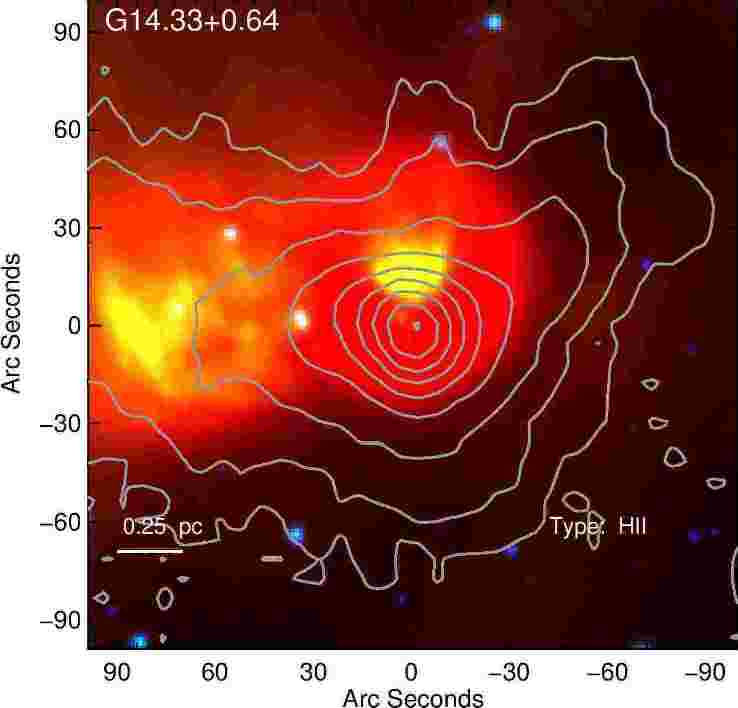}
  \includegraphics[width=6.0cm,angle=90]{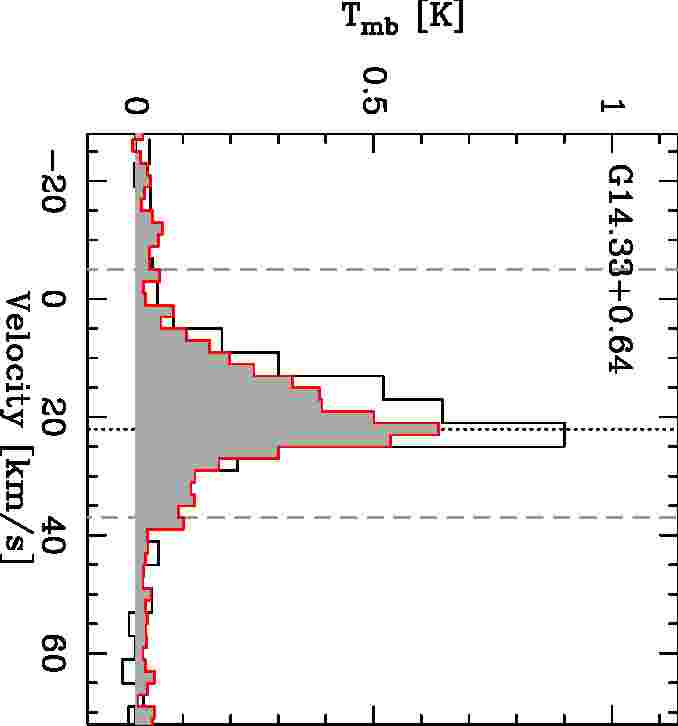}
  \includegraphics[width=6.0cm,angle=0]{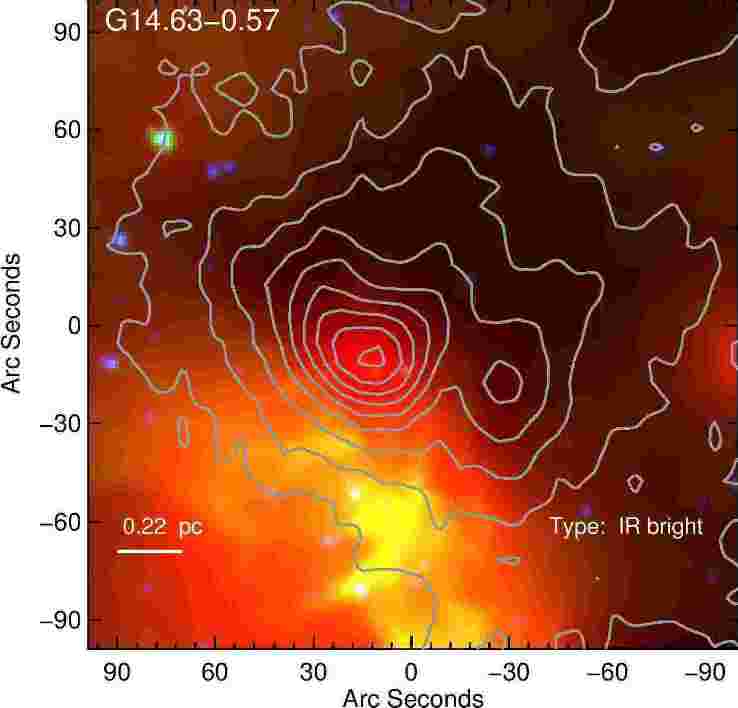}
  \includegraphics[width=6.0cm,angle=90]{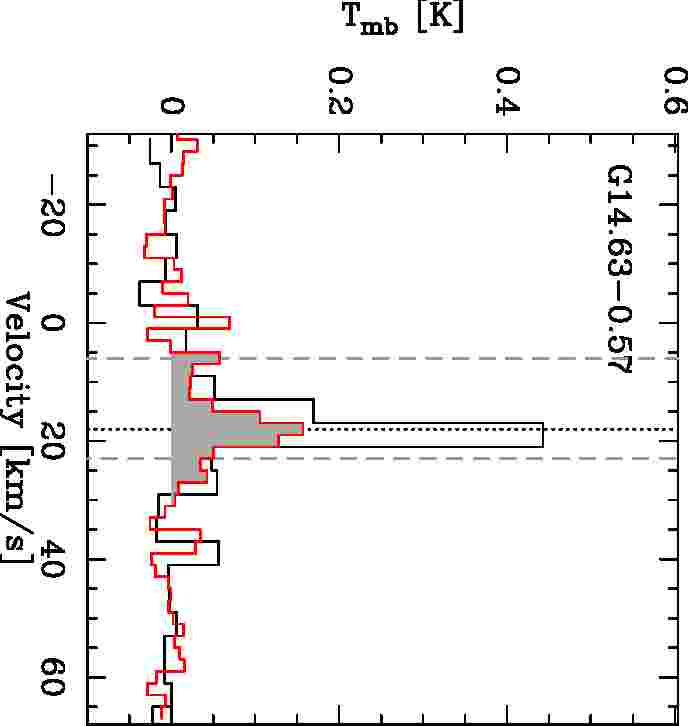}
  \includegraphics[width=6.0cm,angle=0]{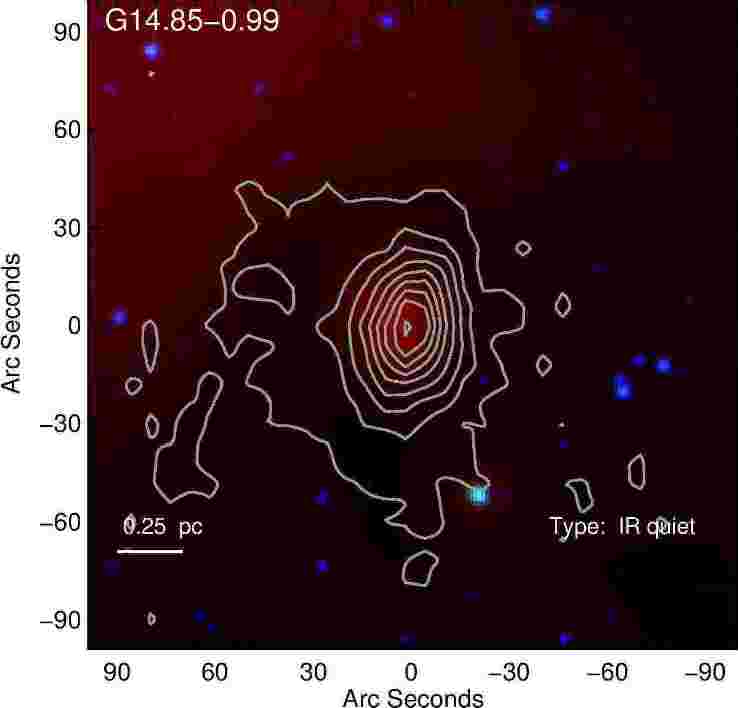}
  \includegraphics[width=6.0cm,angle=90]{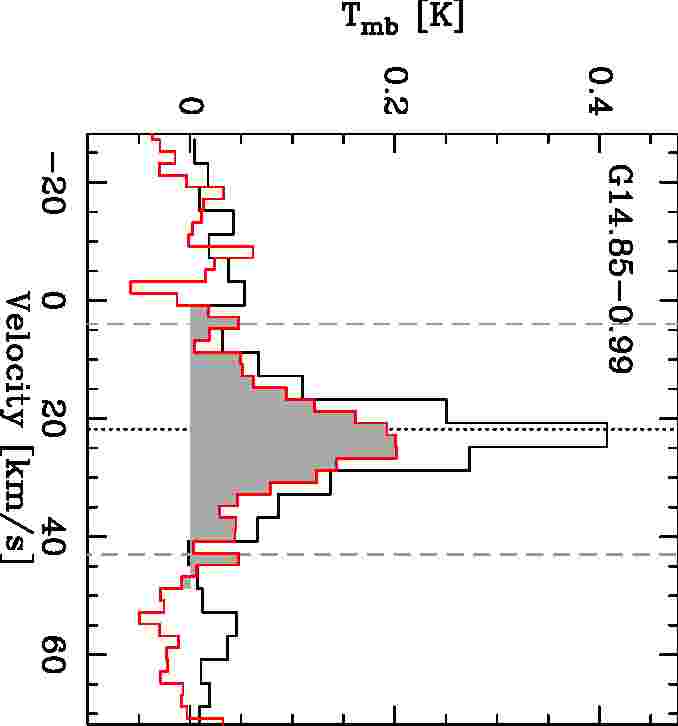}
  \includegraphics[width=6.0cm,angle=0]{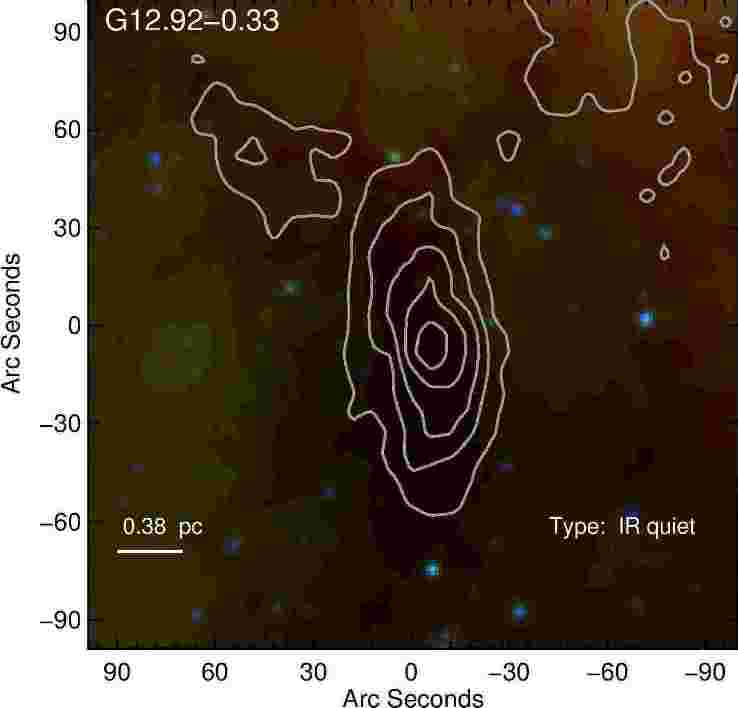}
  \includegraphics[width=6.0cm,angle=90]{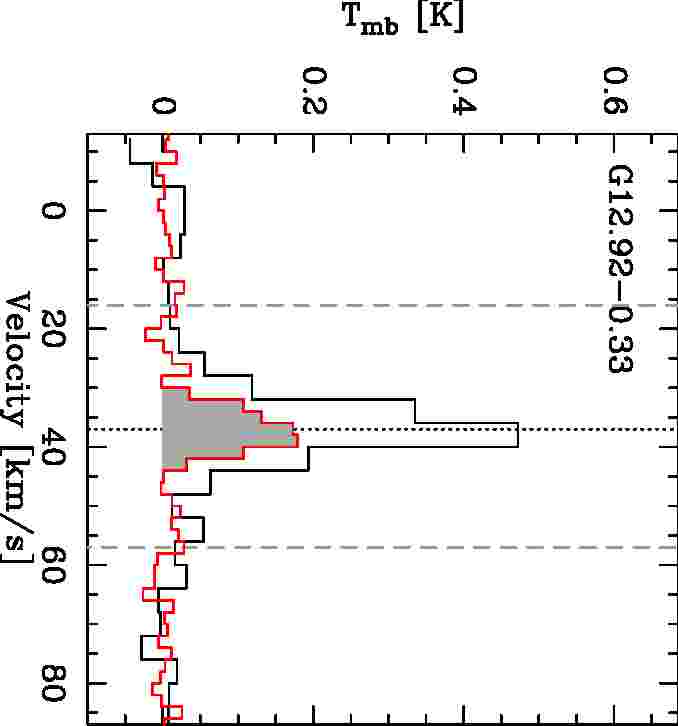}
 \caption{Continued.}
\end{figure}
\end{landscape}

\begin{landscape}
\begin{figure}
\ContinuedFloat
  \includegraphics[width=6.0cm,angle=0]{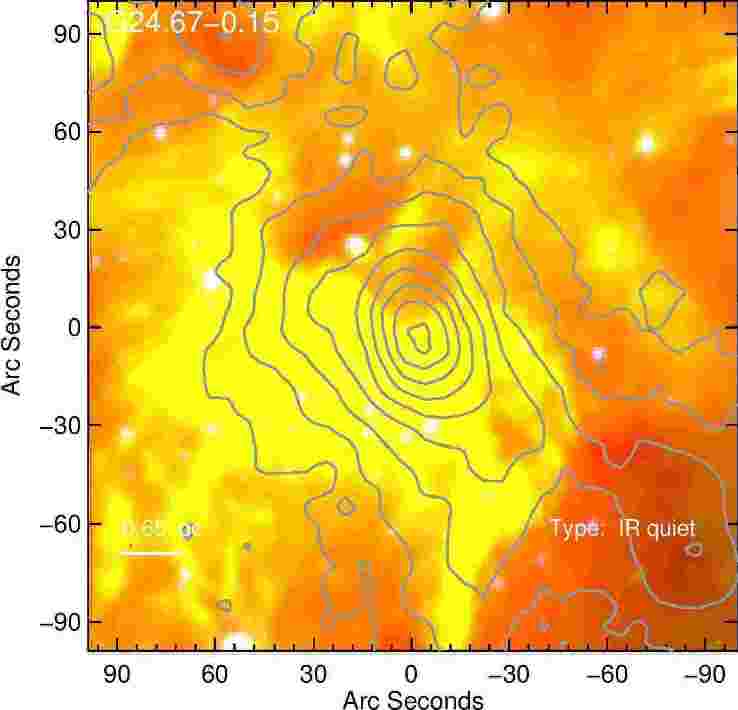}
  \includegraphics[width=6.0cm,angle=90]{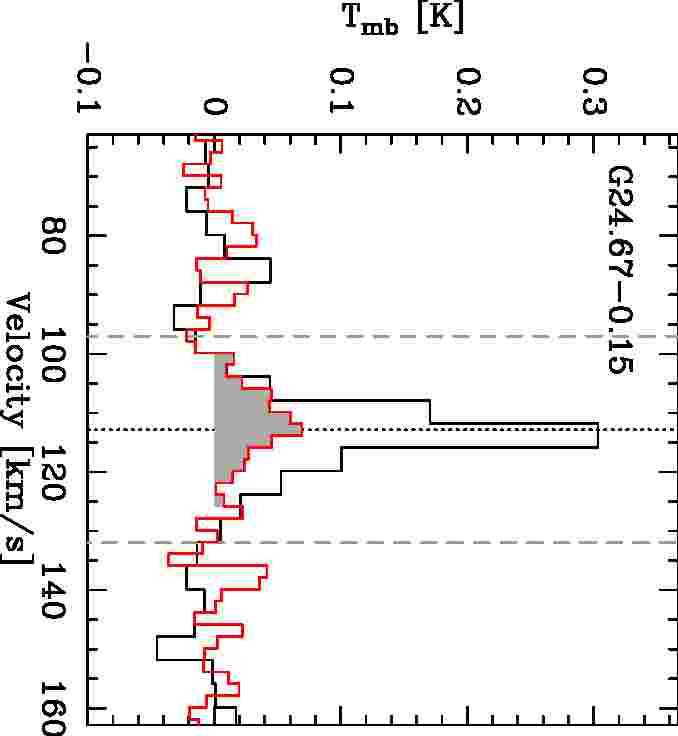}
 \includegraphics[width=6.0cm,angle=0]{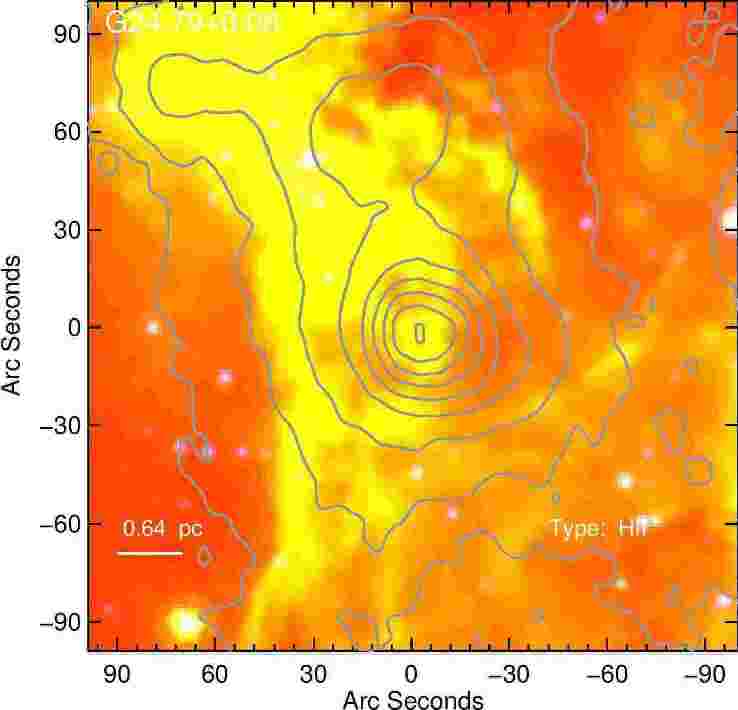}
 \includegraphics[width=6.0cm,angle=90]{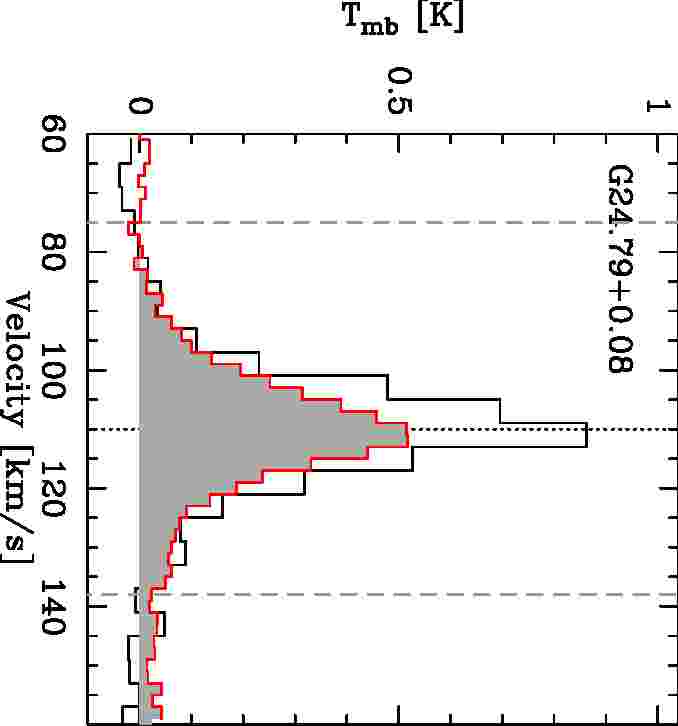}
 \includegraphics[width=6.0cm,angle=0]{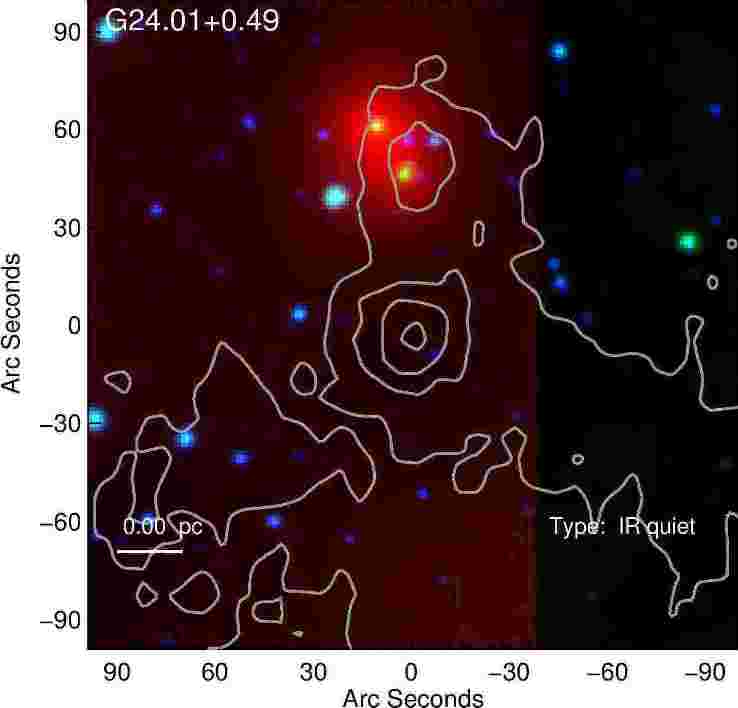}
 \includegraphics[width=6.0cm,angle=90]{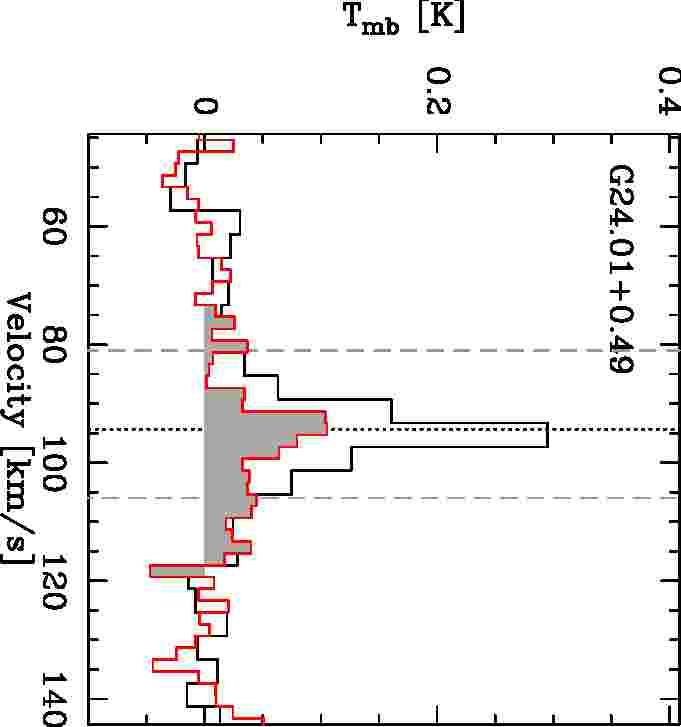}
  \includegraphics[width=6.0cm,angle=0]{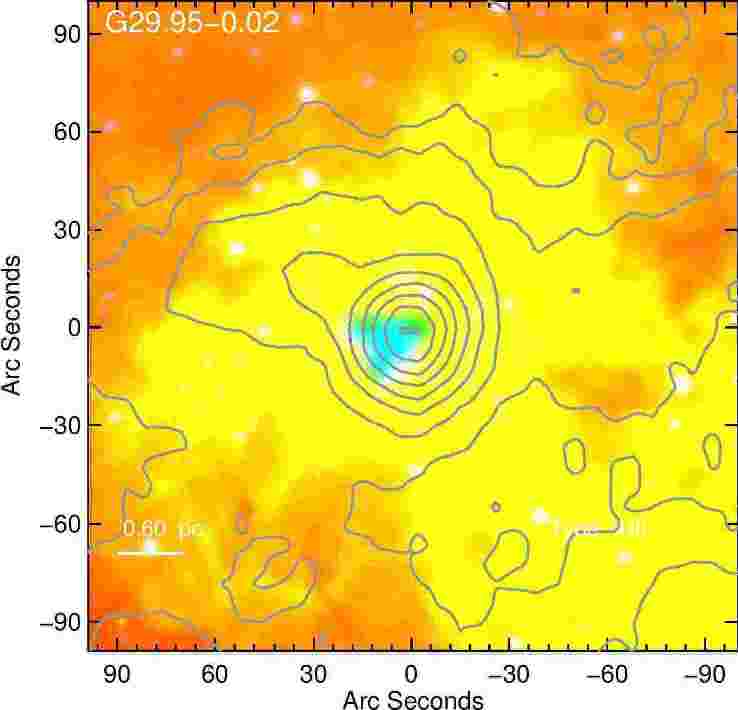}
  \includegraphics[width=6.0cm,angle=90]{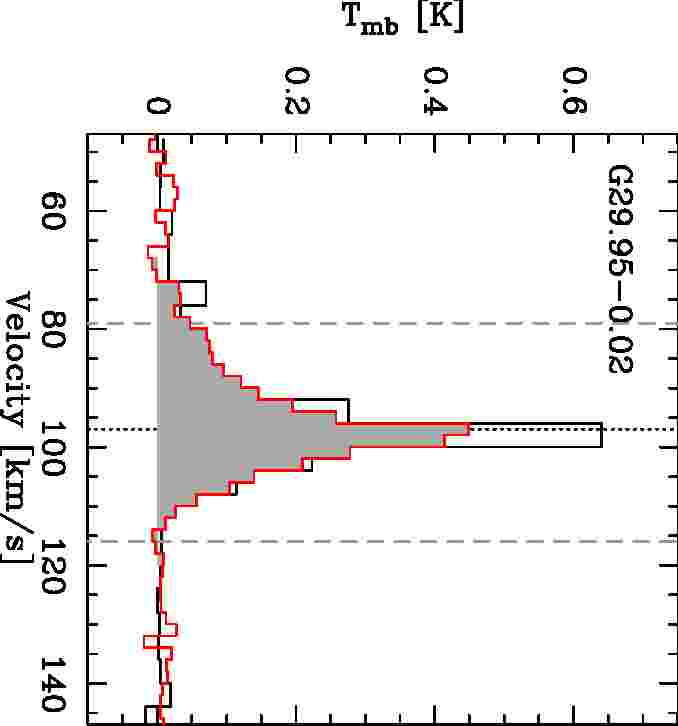}
  \includegraphics[width=6.0cm,angle=0]{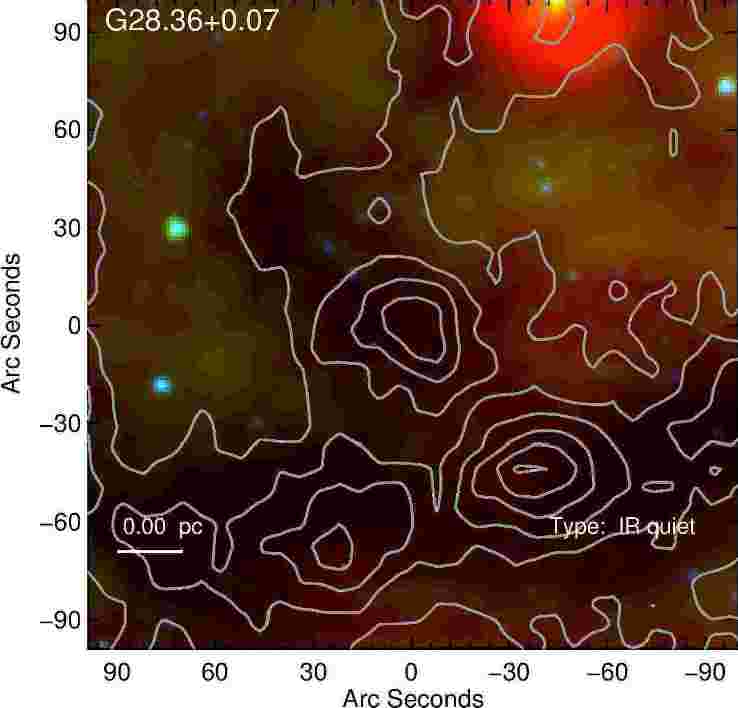}
  \includegraphics[width=6.0cm,angle=90]{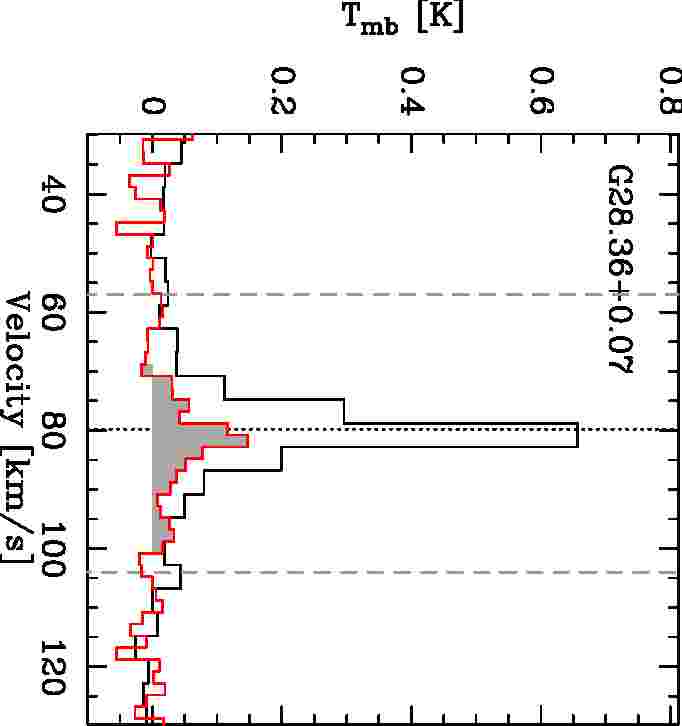}
  \includegraphics[width=6.0cm,angle=0]{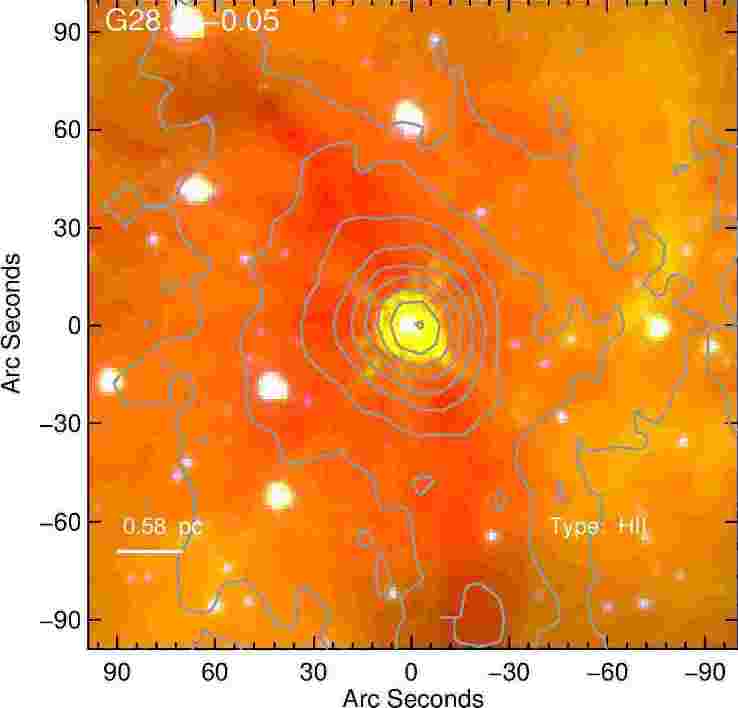}
  \includegraphics[width=6.0cm,angle=90]{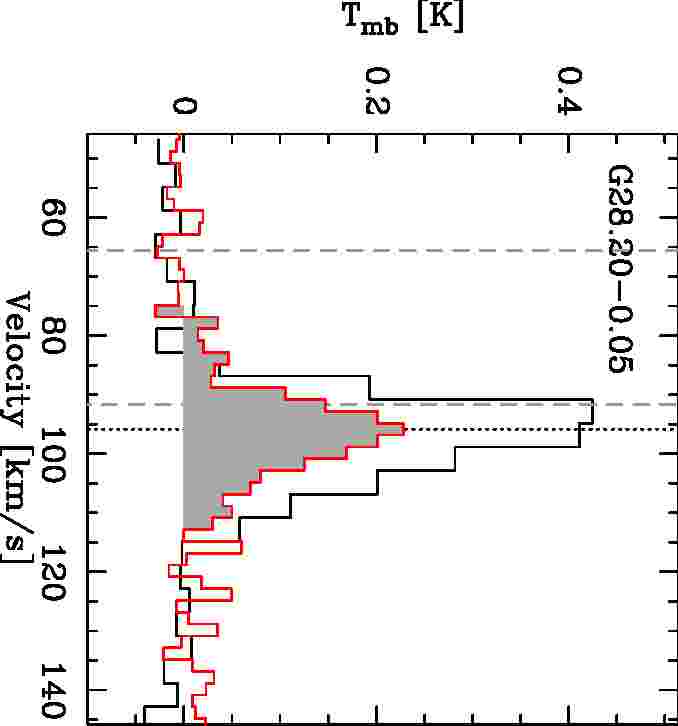}
 \caption{Continued.}
\end{figure}
\end{landscape}

\clearpage

\begin{landscape}
\begin{figure}
\ContinuedFloat
  \includegraphics[width=6.0cm,angle=0]{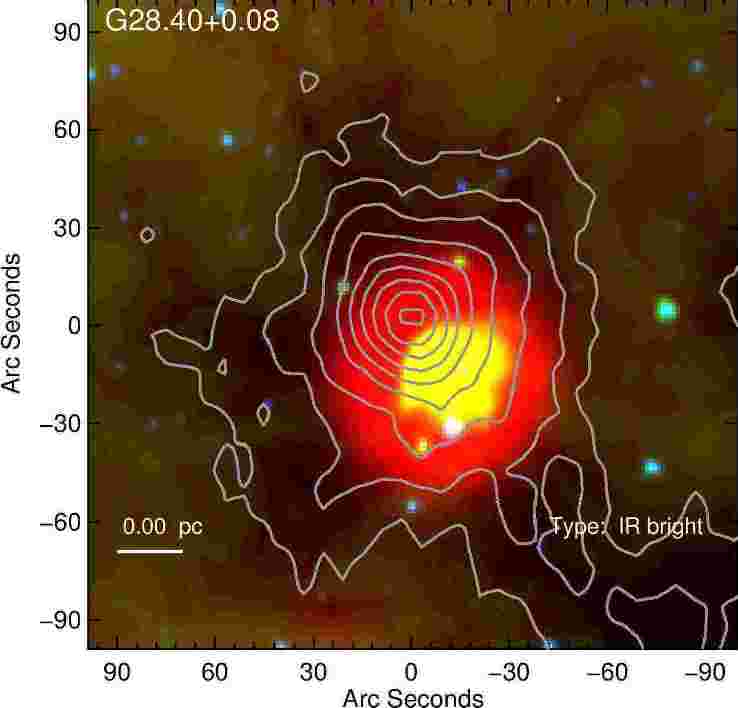}
  \includegraphics[width=6.0cm,angle=90]{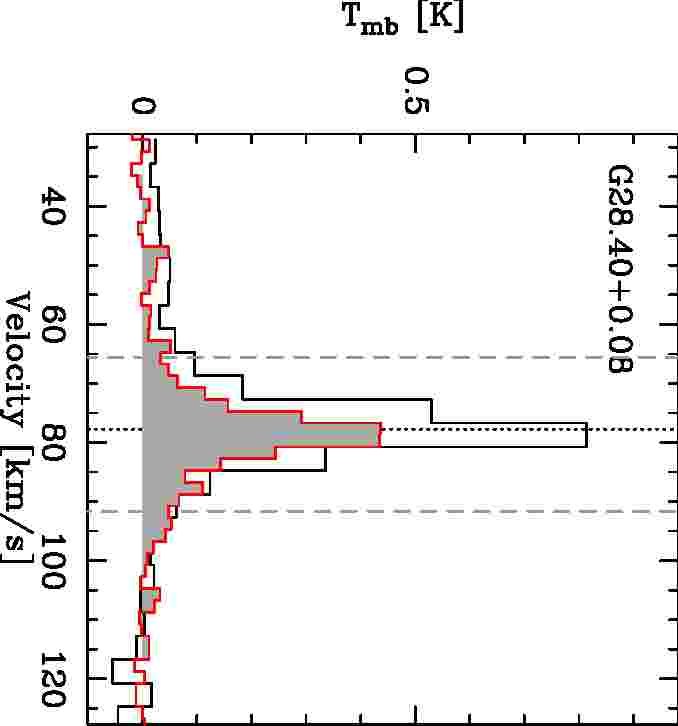}
  \includegraphics[width=6.0cm,angle=0]{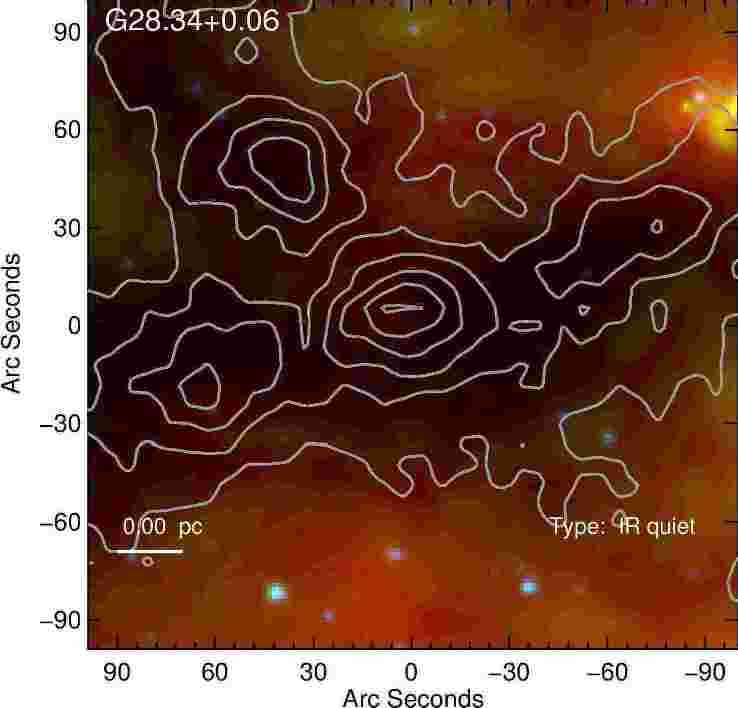}
  \includegraphics[width=6.0cm,angle=90]{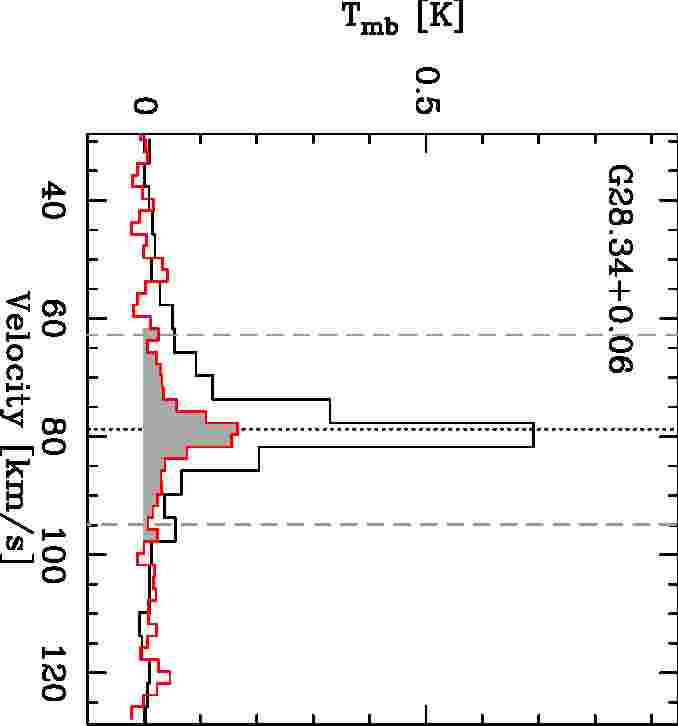}
  \includegraphics[width=6.0cm,angle=0]{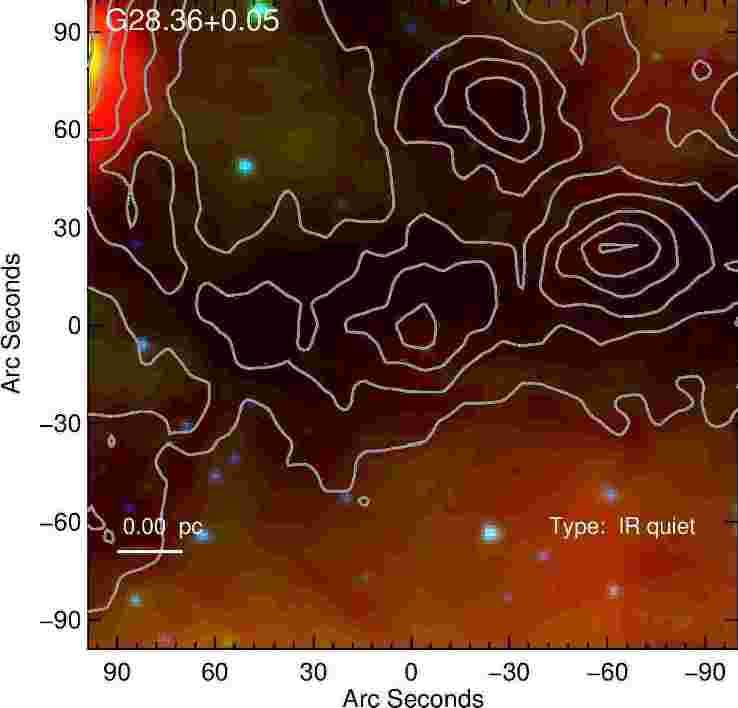}
  \includegraphics[width=6.0cm,angle=90]{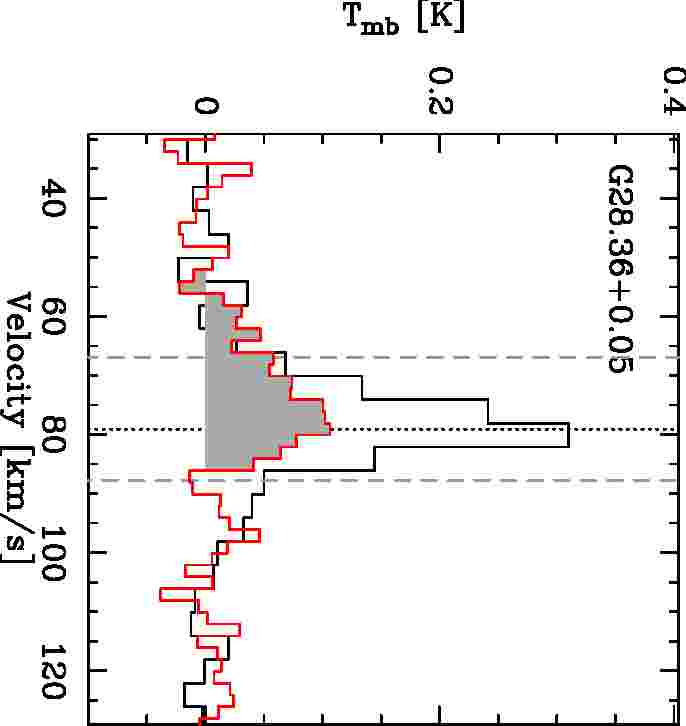}
  \includegraphics[width=6.0cm,angle=0]{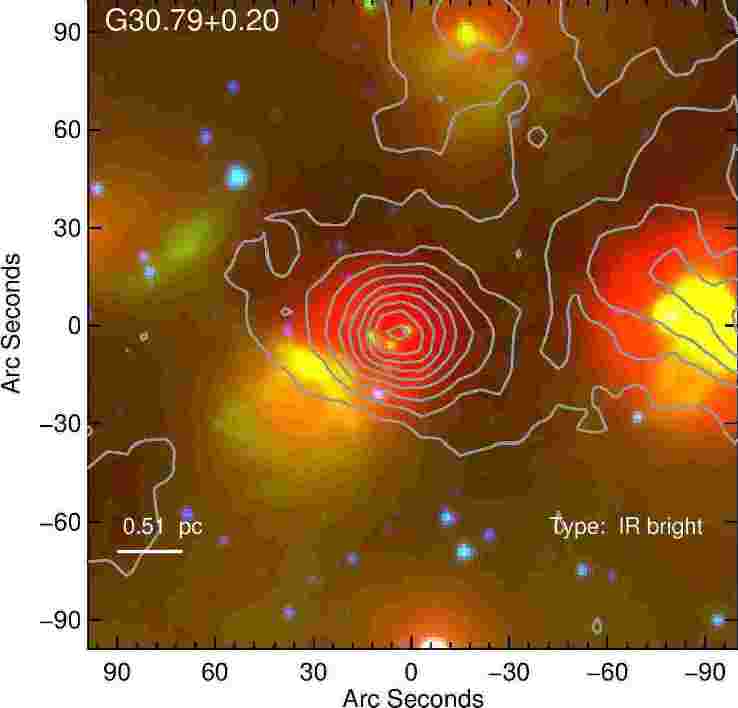}
  \includegraphics[width=6.0cm,angle=90]{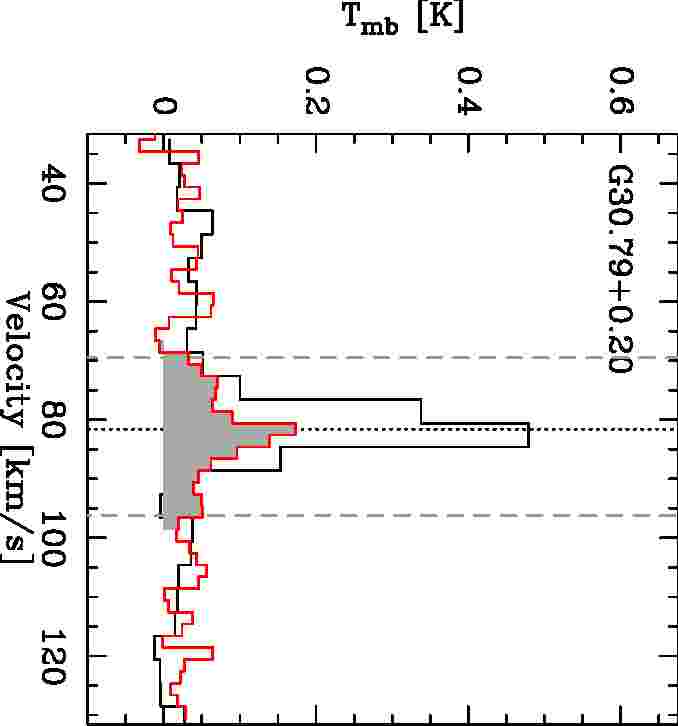}
  \includegraphics[width=6.0cm,angle=0]{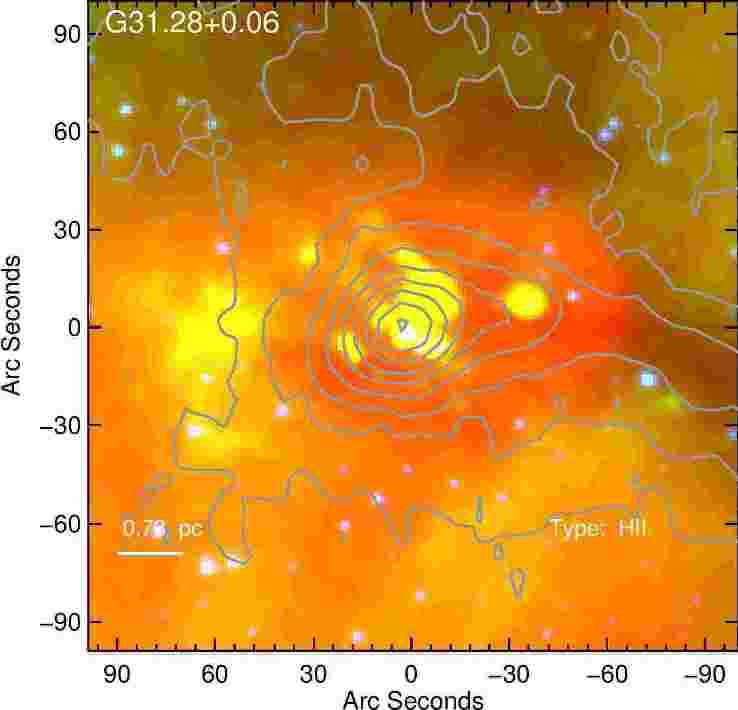}
  \includegraphics[width=6.0cm,angle=90]{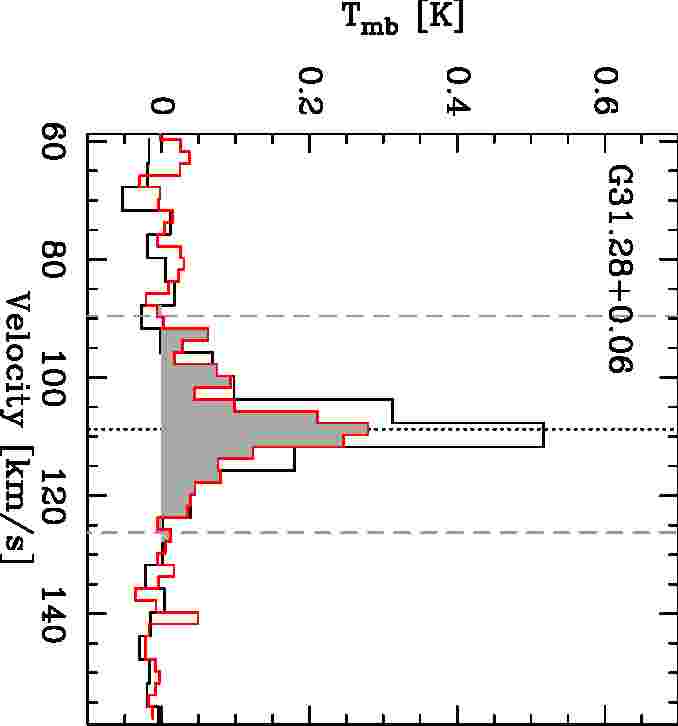}
  \includegraphics[width=6.0cm,angle=0]{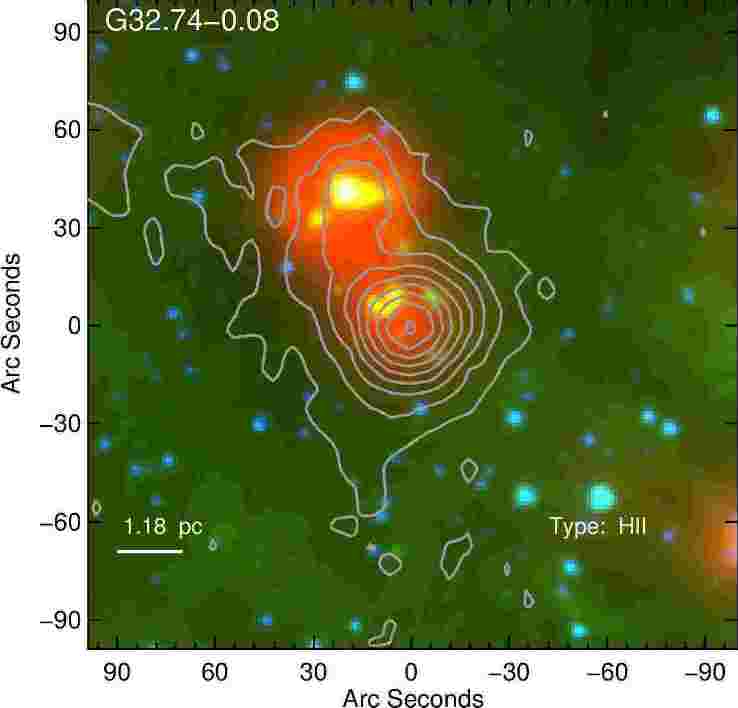}
  \includegraphics[width=6.0cm,angle=90]{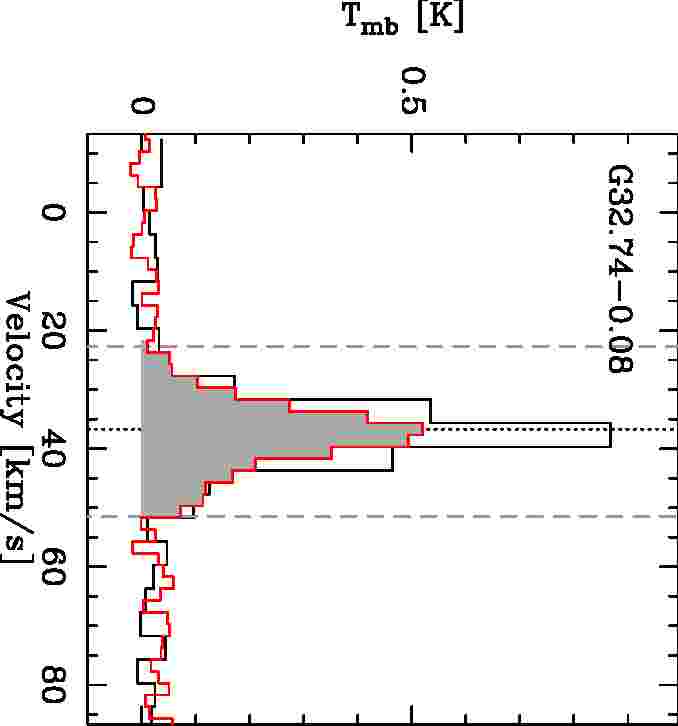}
 \caption{Continued.}
\end{figure}
\end{landscape}
}

\subsection{Spectra of sources observed only in the SiO ($2-1$) transition}\label{app:only21}
\onlfig{
\begin{landscape}
\begin{figure}
\centering
  \includegraphics[width=5.6cm,angle=0]{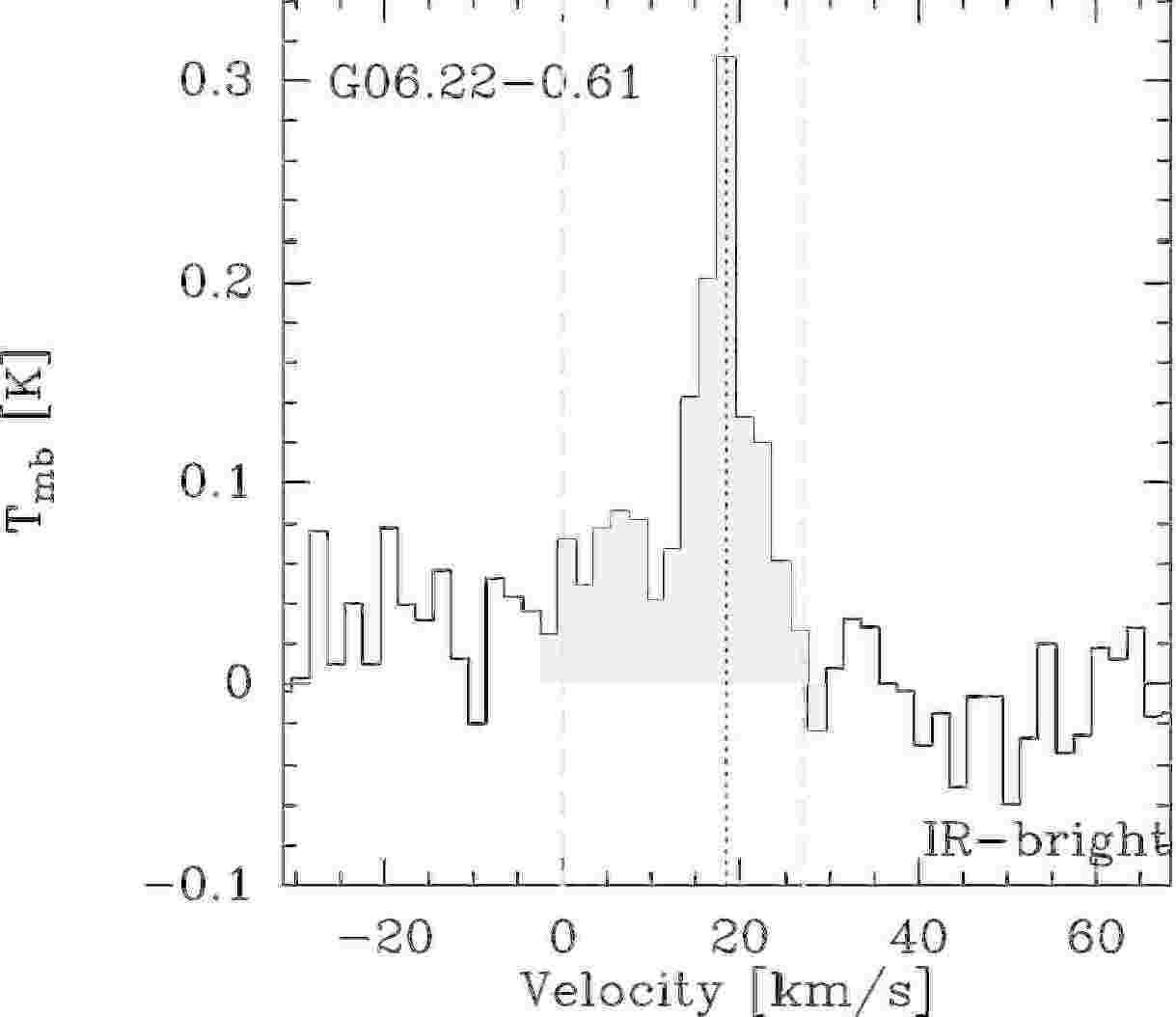} 
  \includegraphics[width=5.6cm,angle=0]{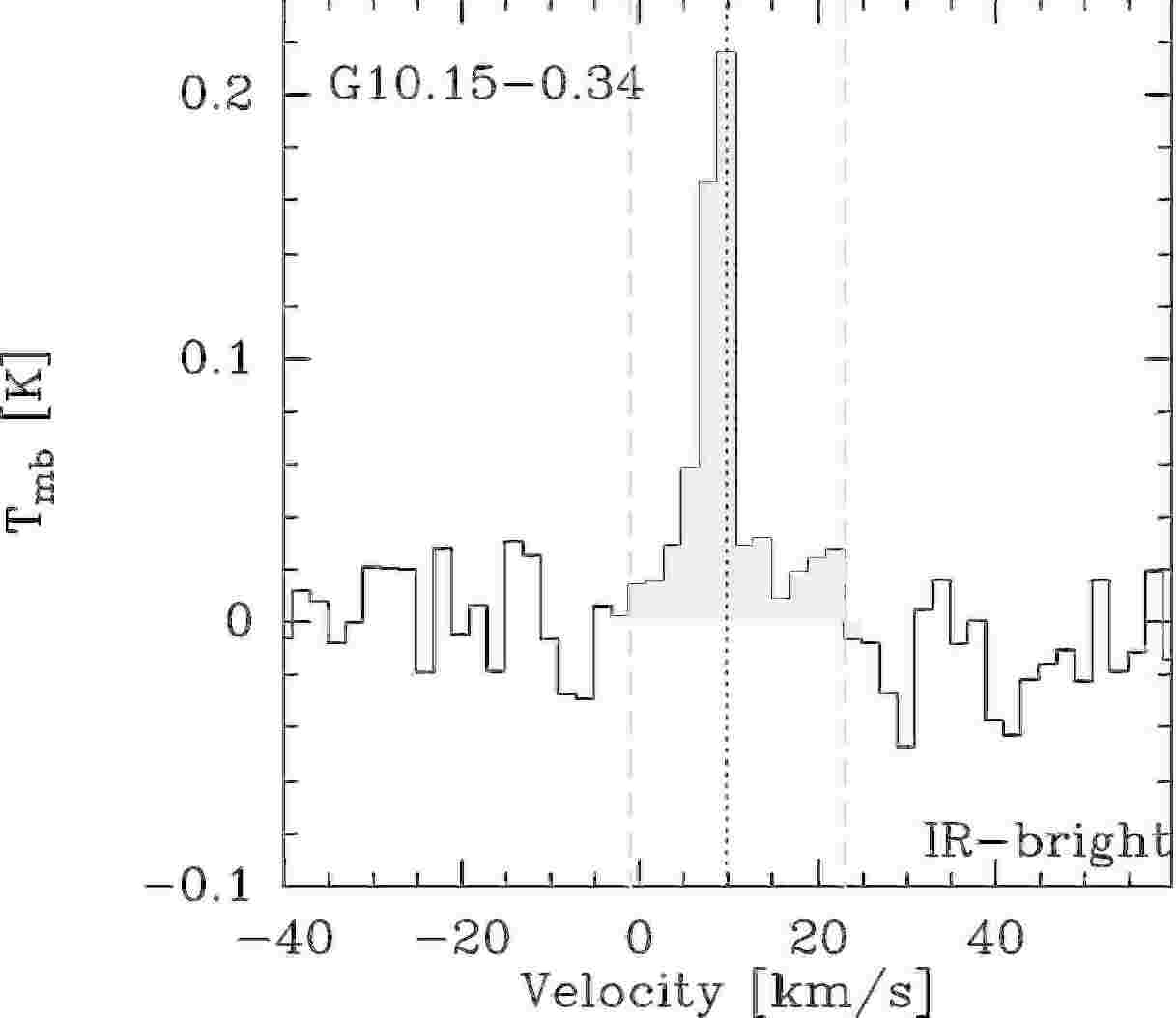} 
  \includegraphics[width=5.6cm,angle=0]{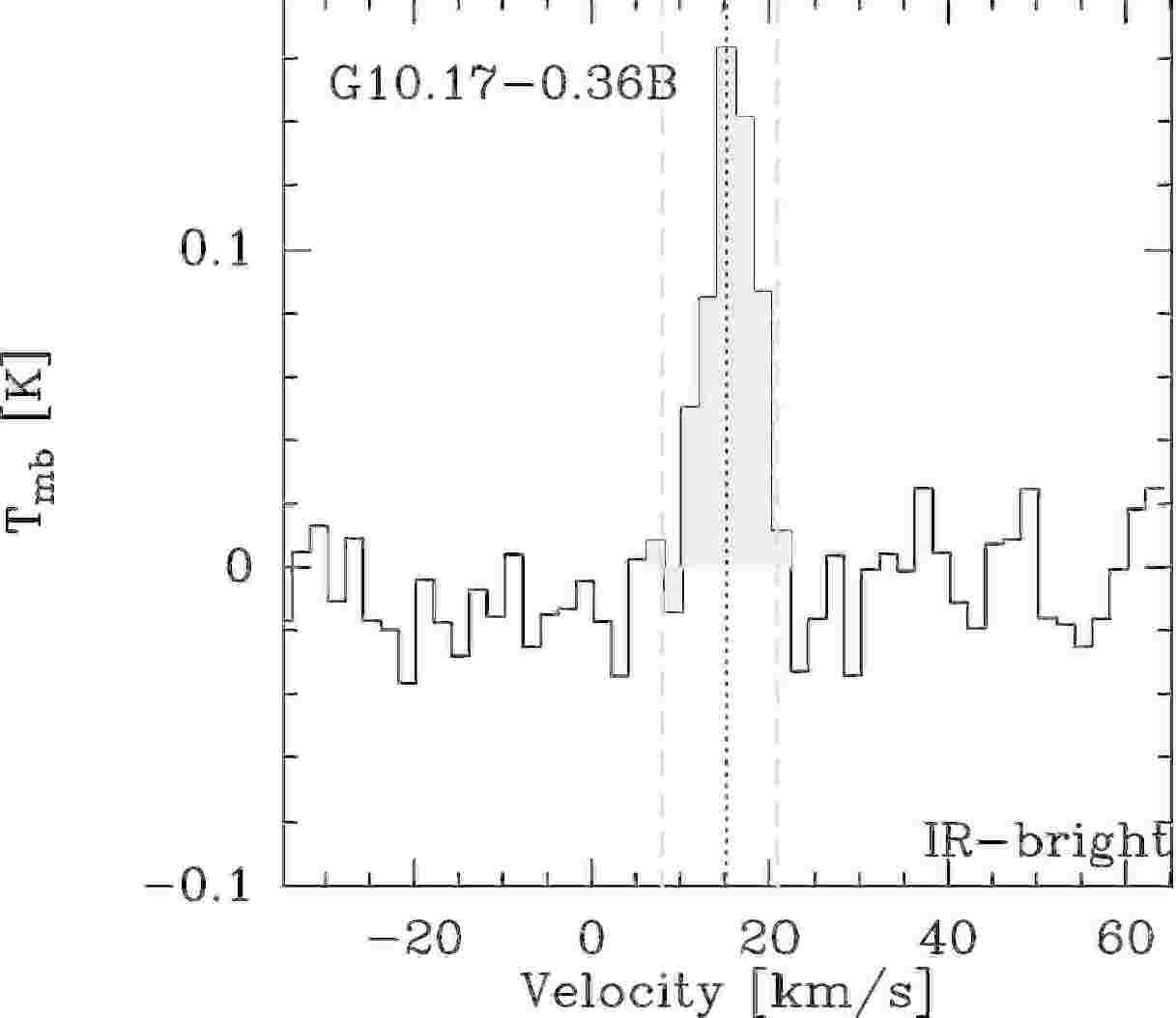} 
  \includegraphics[width=5.6cm,angle=0]{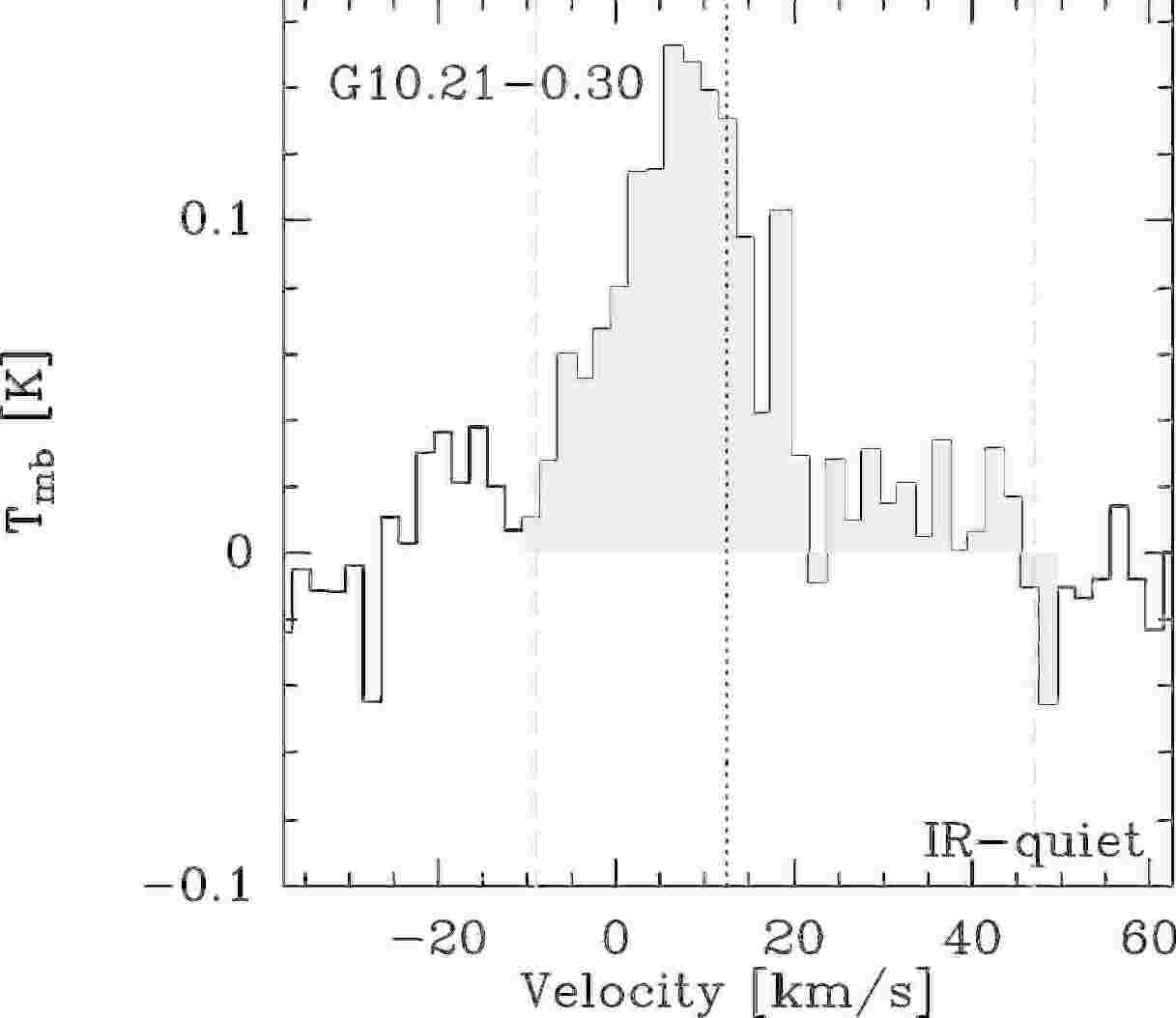} 
  \includegraphics[width=5.6cm,angle=0]{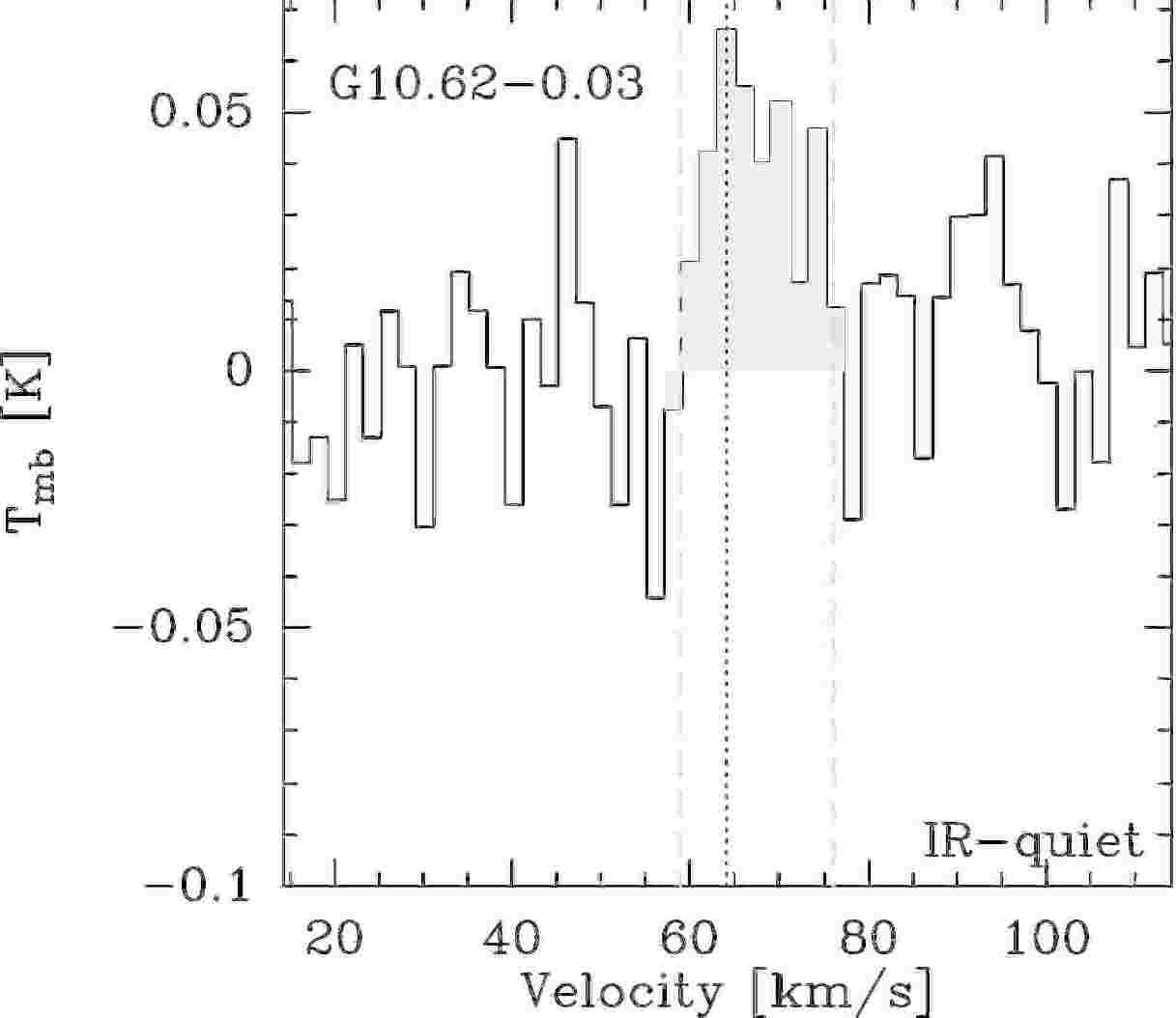} 
  \includegraphics[width=5.6cm,angle=0]{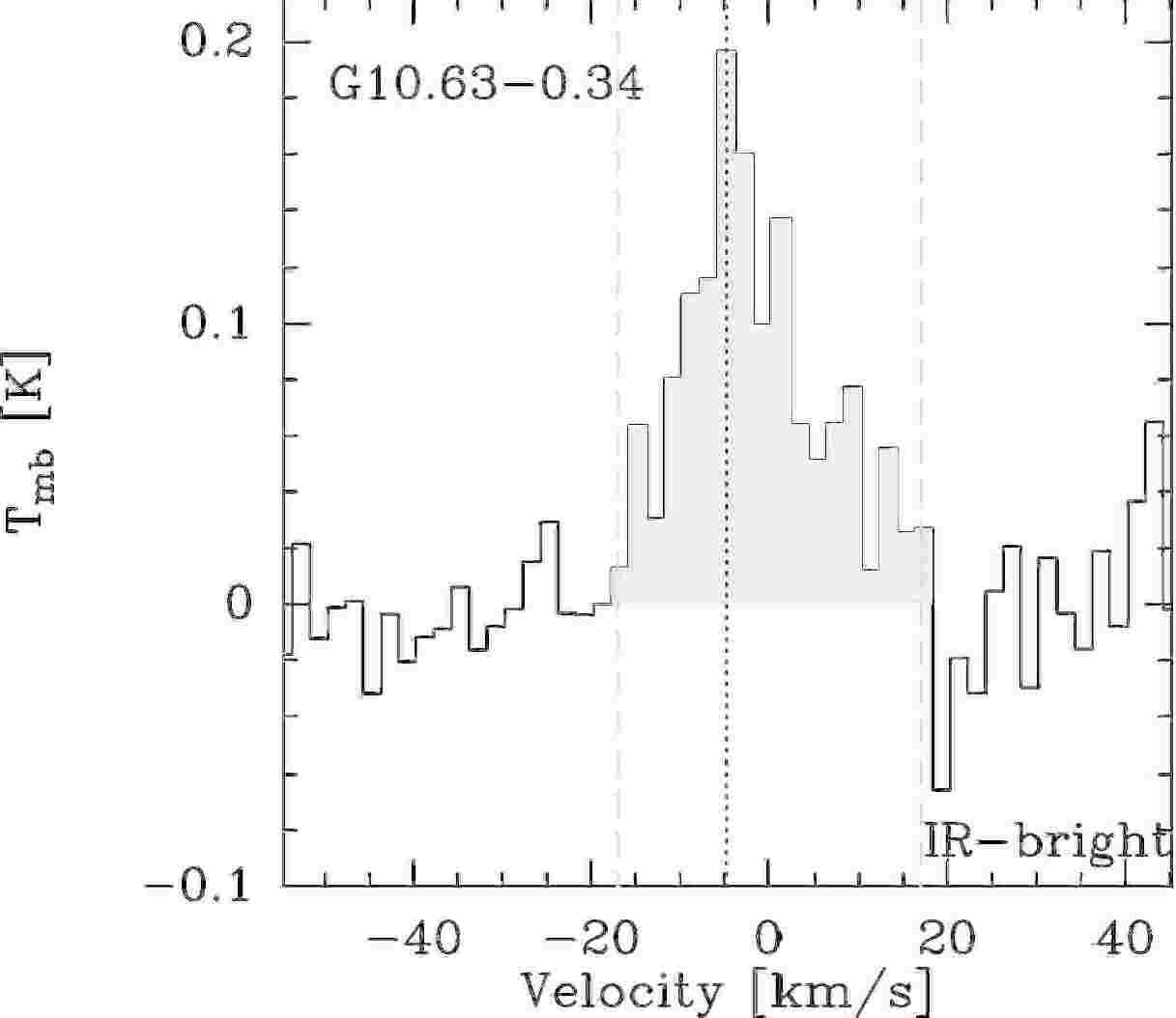} 
  \includegraphics[width=5.6cm,angle=0]{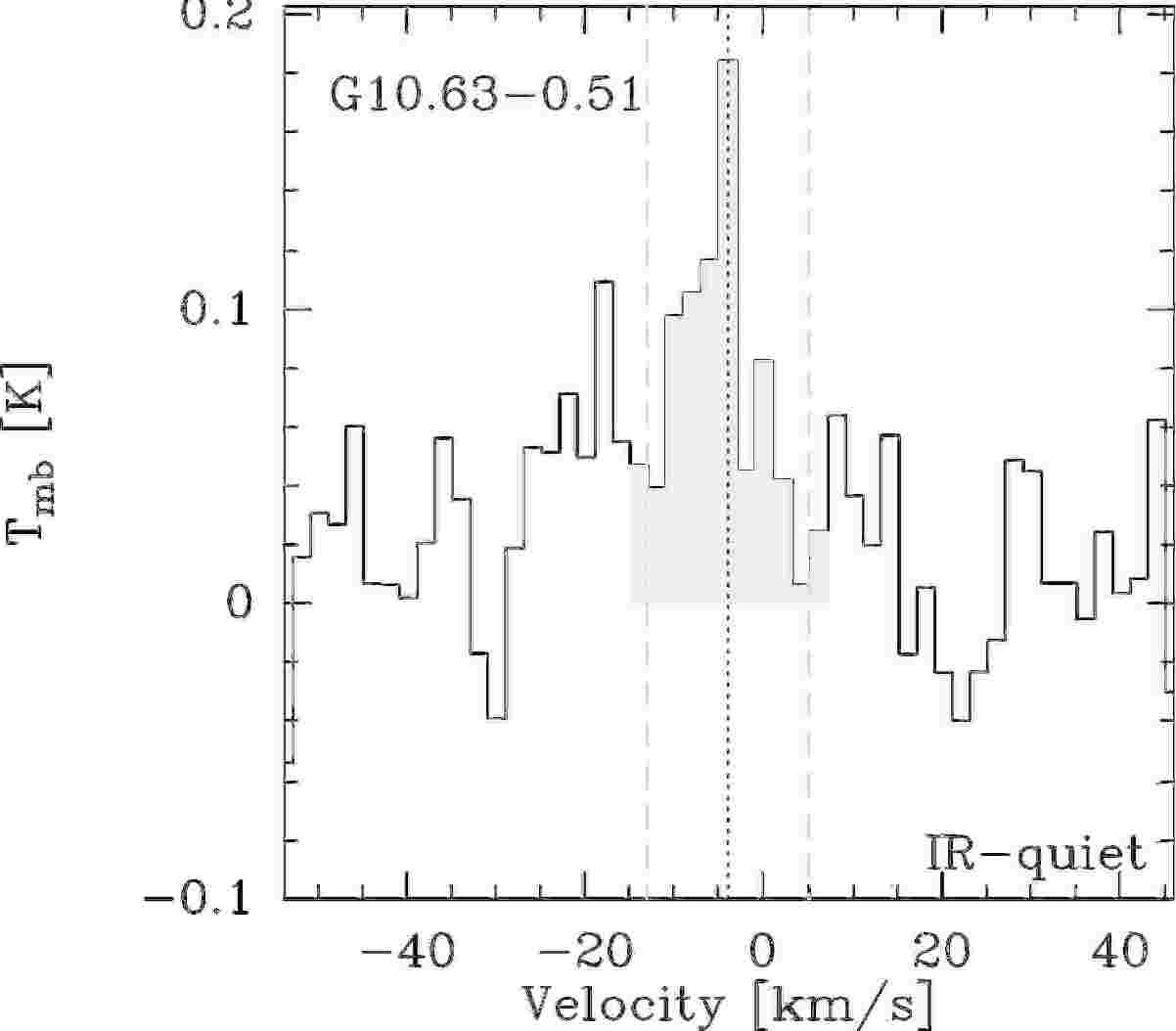} 
  \includegraphics[width=5.6cm,angle=0]{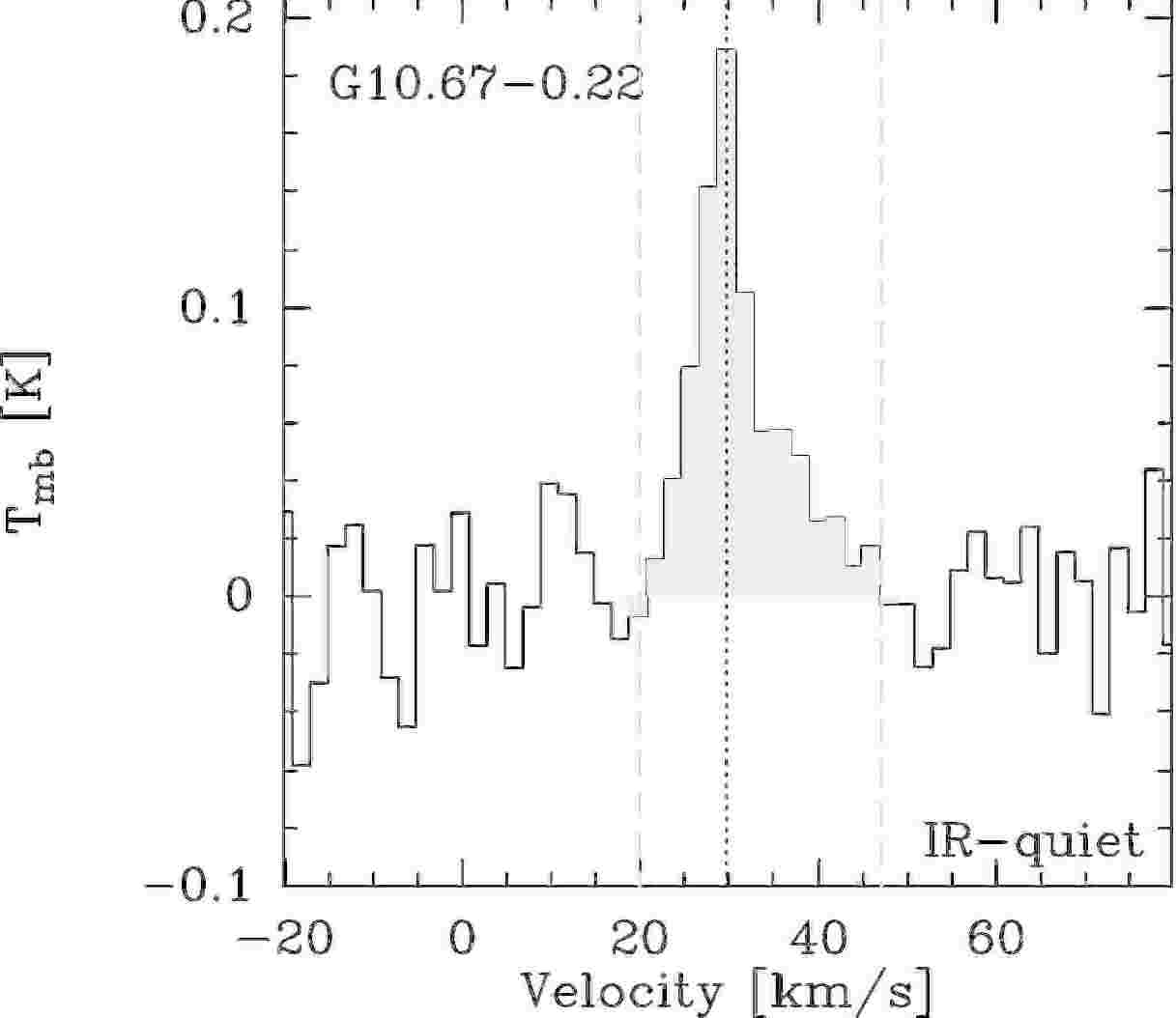} 
  \includegraphics[width=5.6cm,angle=0]{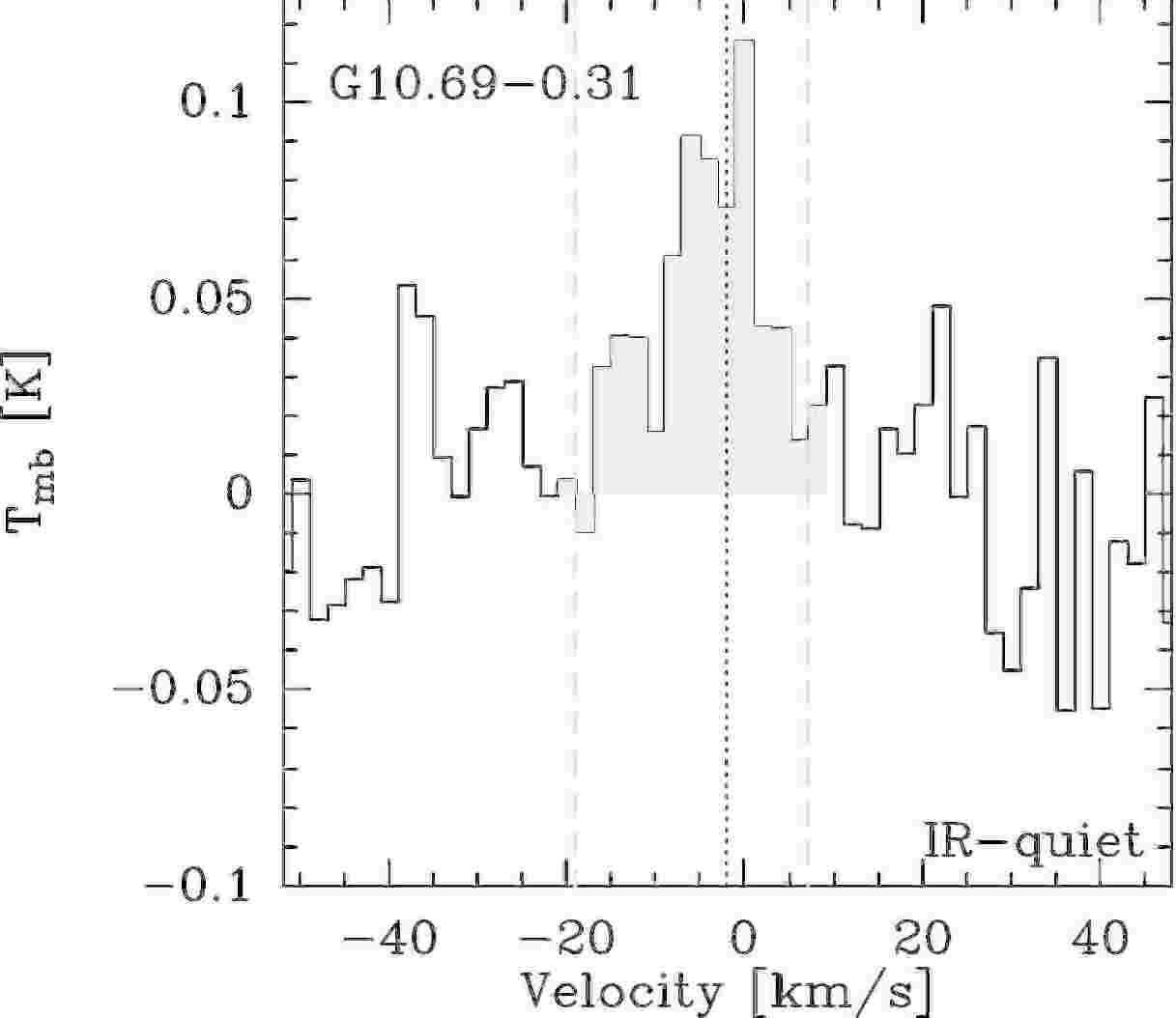} 
  \includegraphics[width=5.6cm,angle=0]{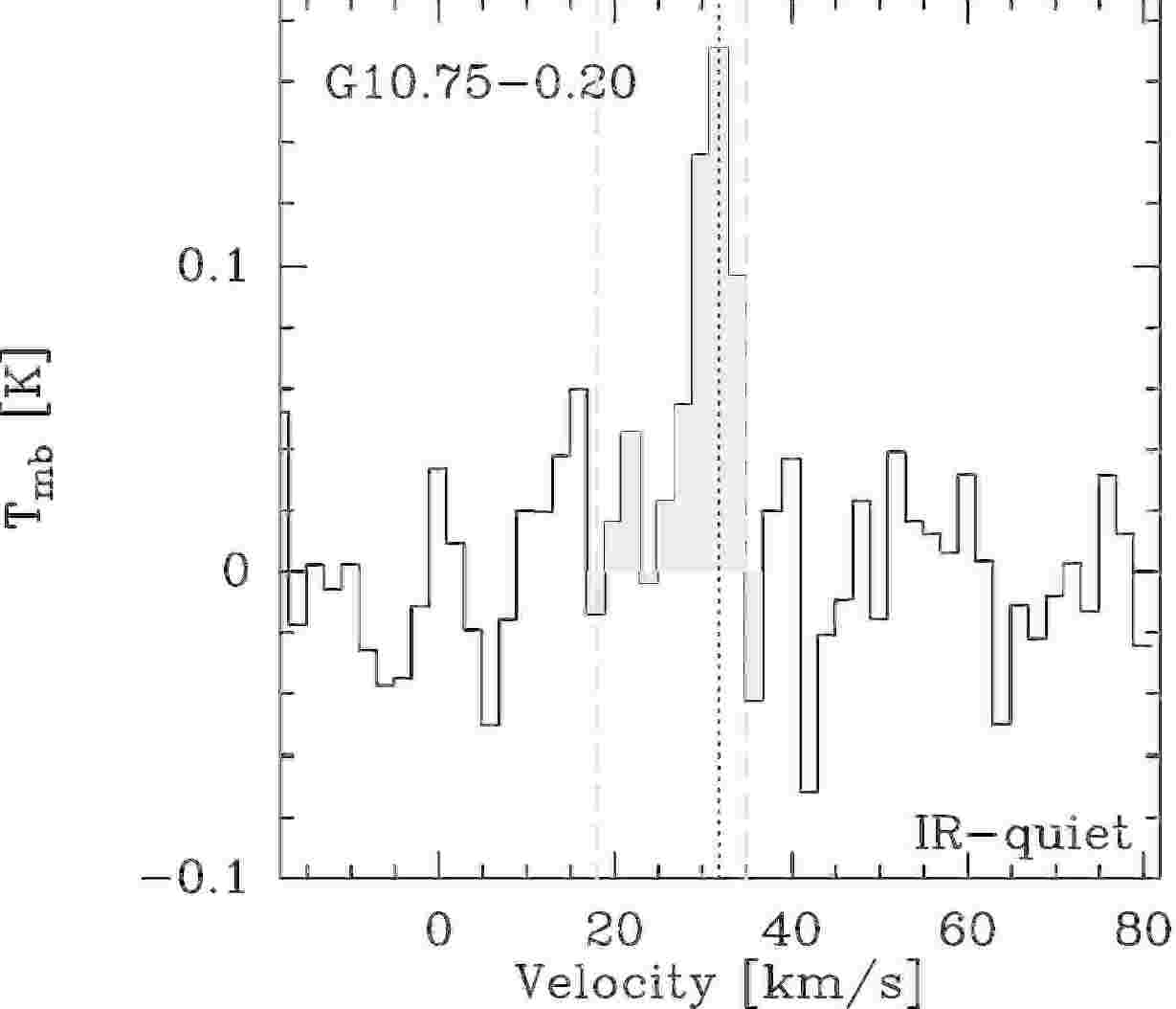} 
  \includegraphics[width=5.6cm,angle=0]{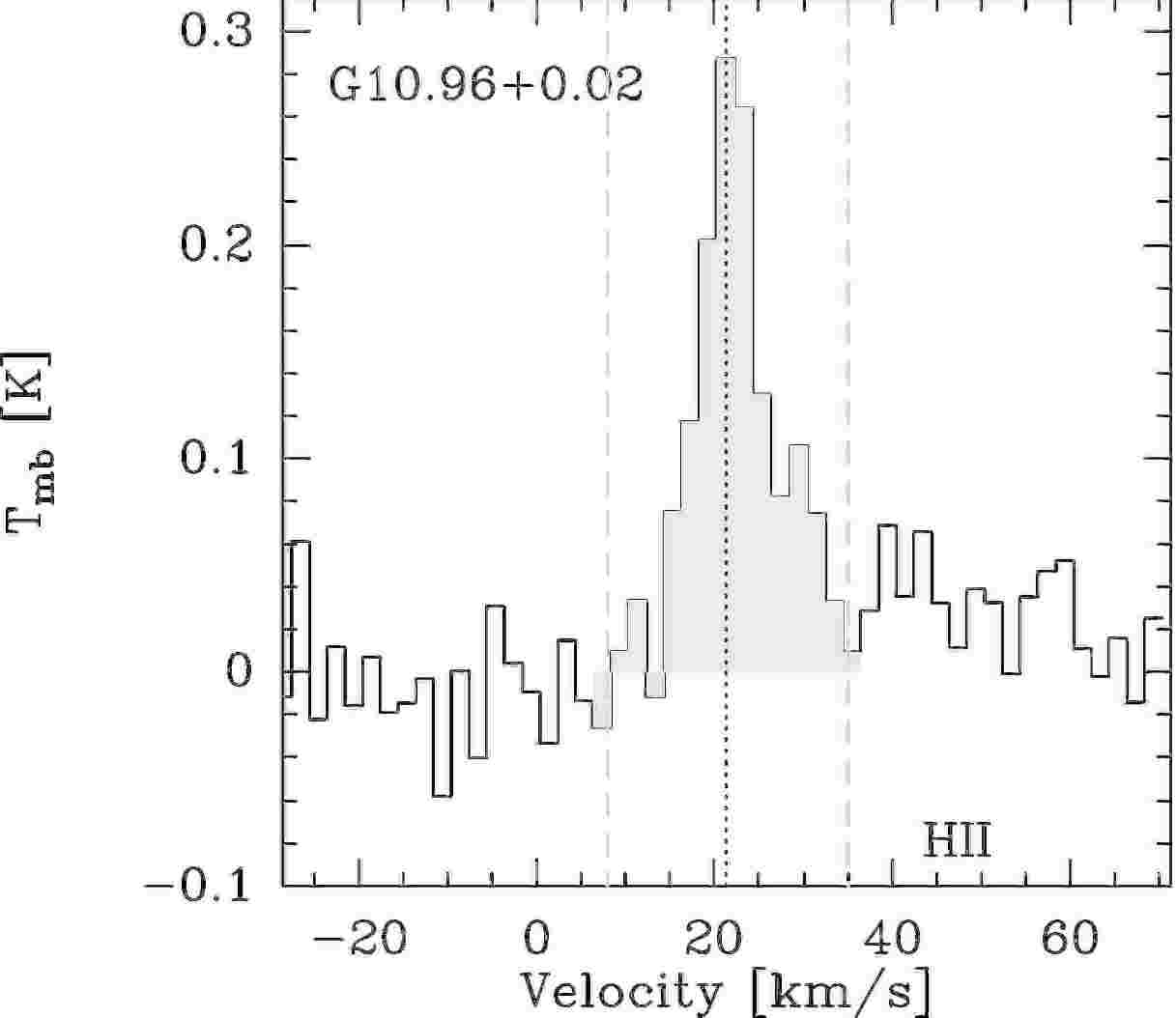} 
  \includegraphics[width=5.6cm,angle=0]{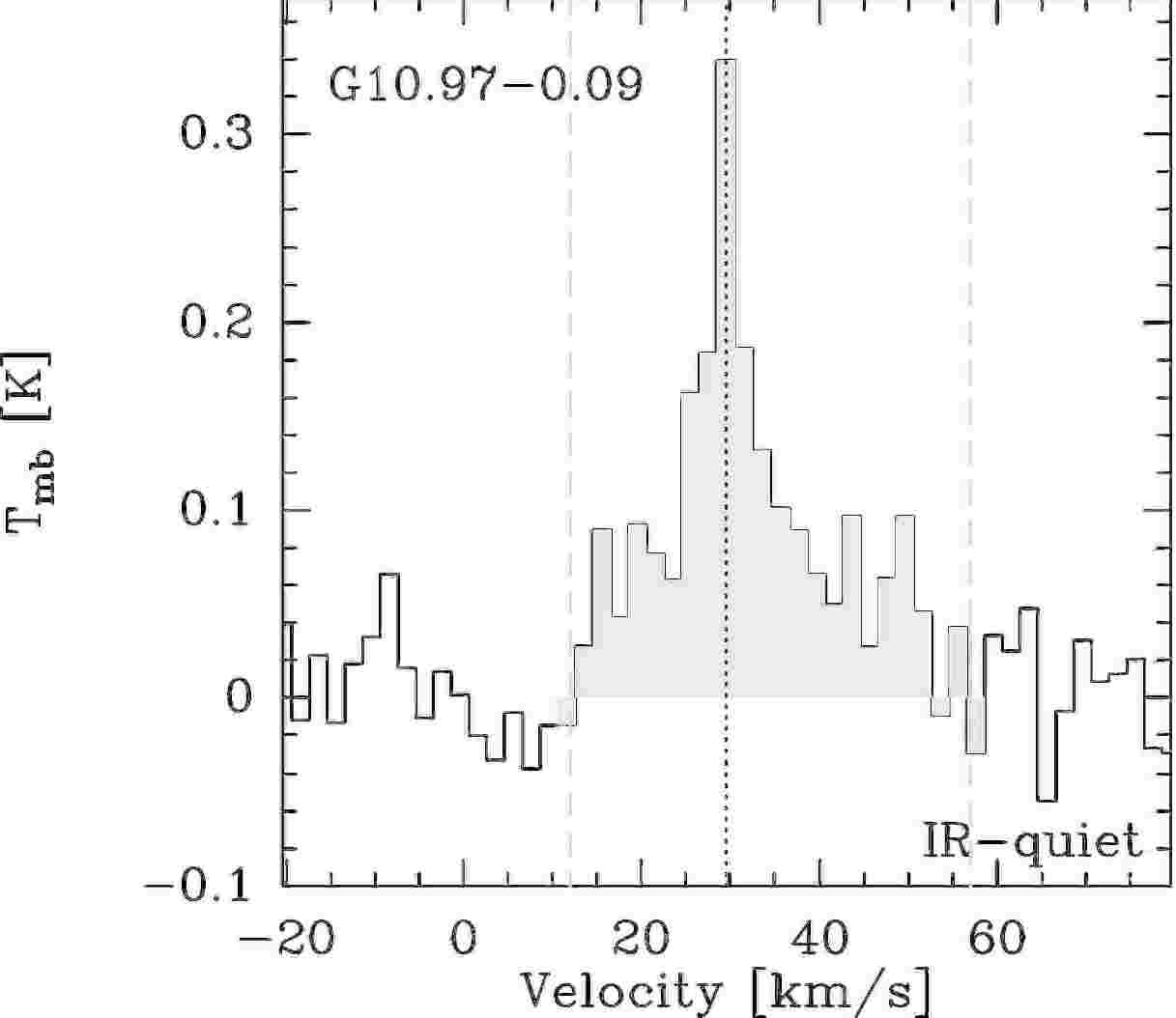} 
 \caption{Spectra of the SiO ($2-1$) transition is shown in black
              with the area of the $FWZP$ shown in grey.
              Dotted line shows
               the systemic velocity of the source 
               (v$_{\rm lsr}$, see also Table\,\ref{tab:table-large-sio21}), dashed lines
               correspond to the velocity range determined from the SiO 
               ($2-1$) transition. }\label{app:fig2}
\end{figure}
\end{landscape}

\begin{landscape}
\begin{figure}
\centering
\ContinuedFloat
  \includegraphics[width=5.6cm,angle=0]{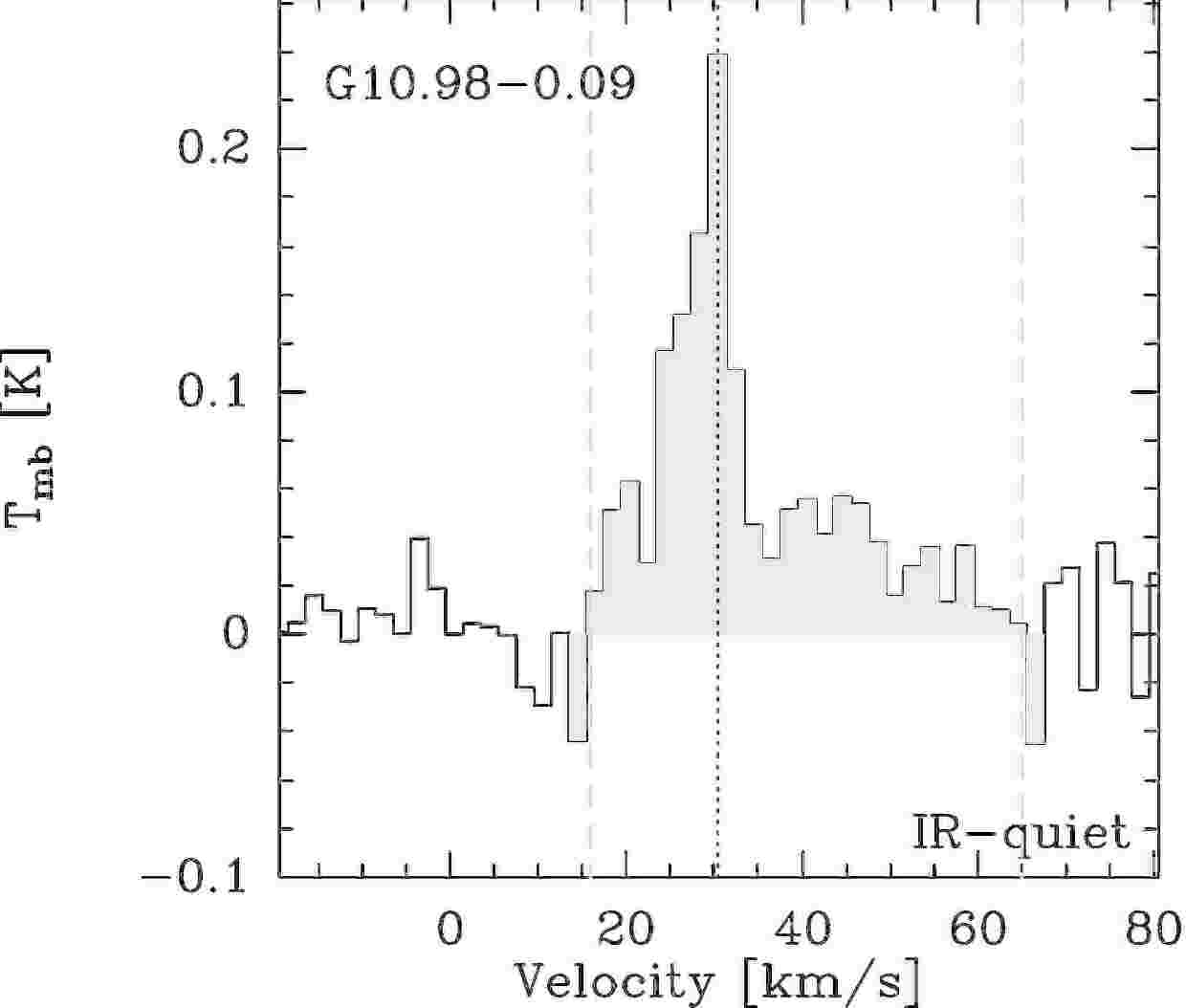} 
  \includegraphics[width=5.6cm,angle=0]{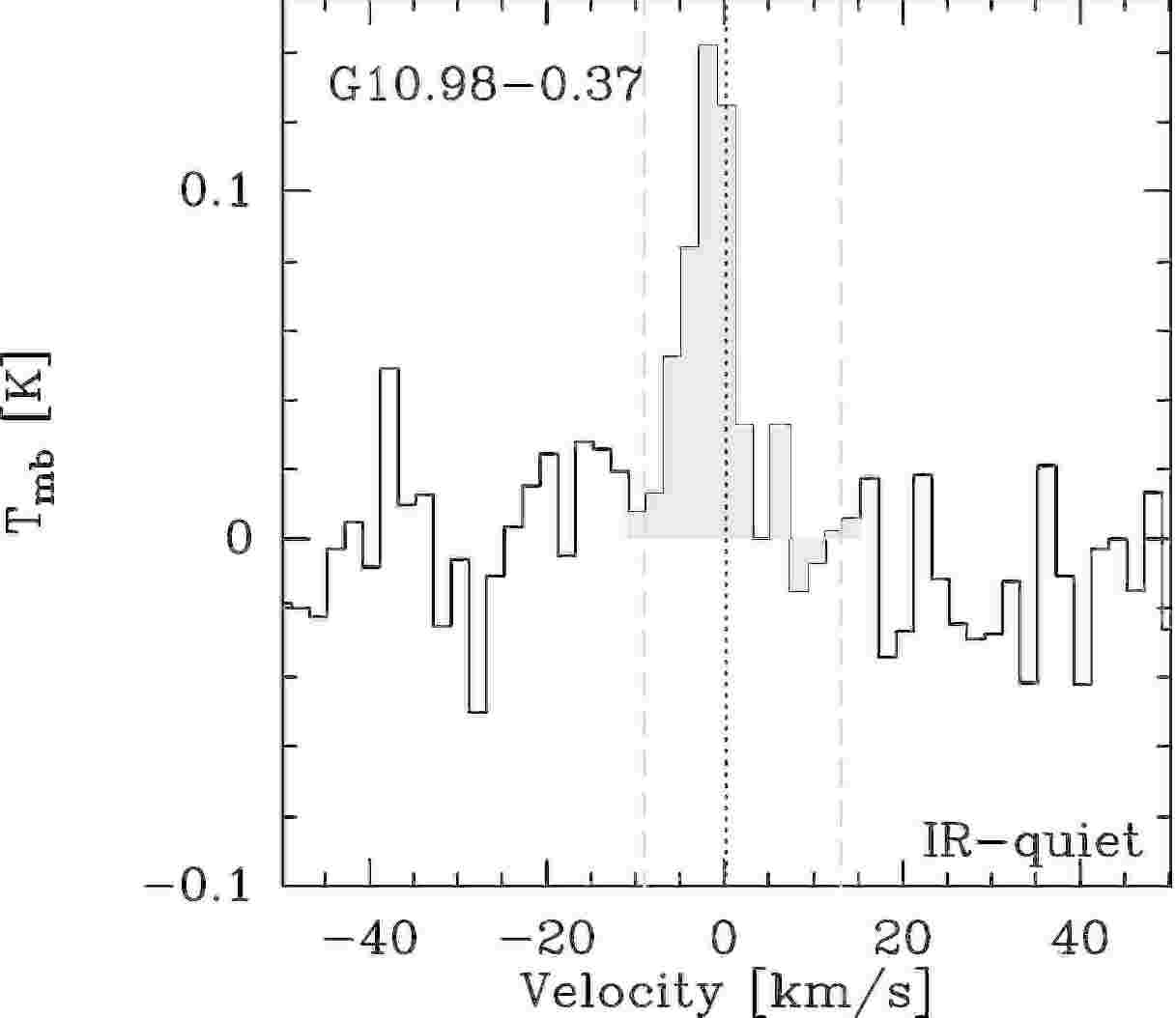} 
  \includegraphics[width=5.6cm,angle=0]{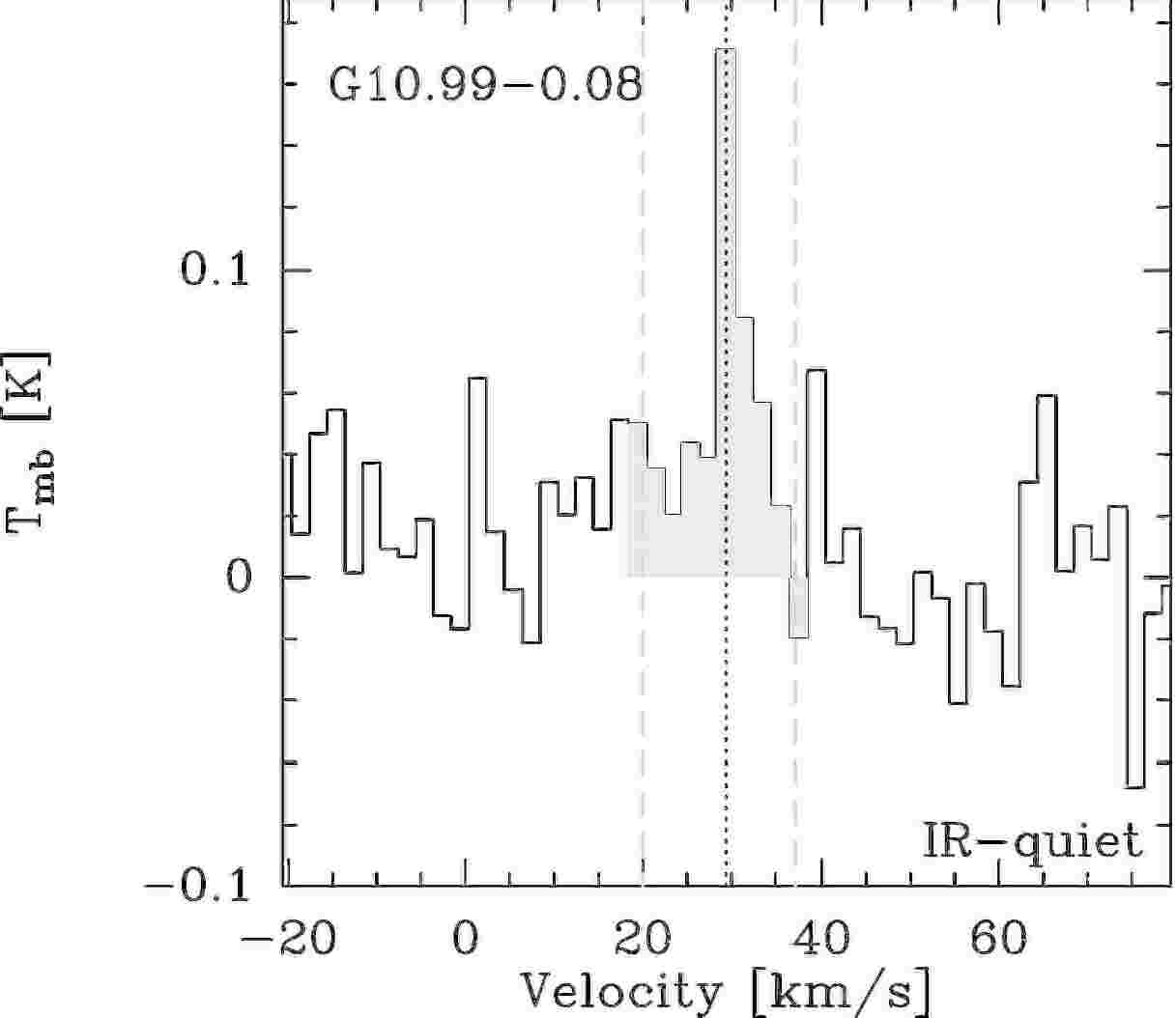} 
  \includegraphics[width=5.6cm,angle=0]{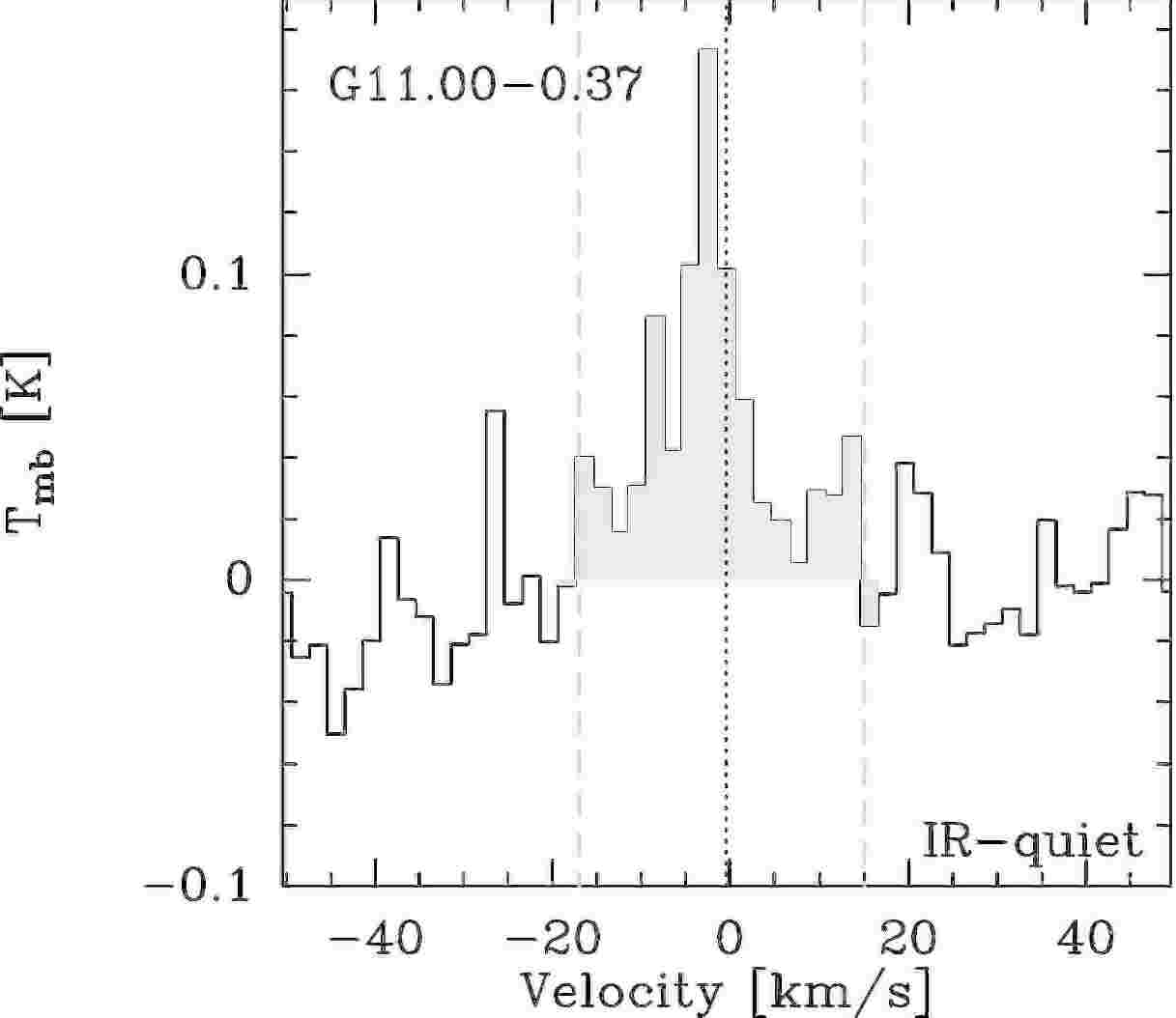} 
  \includegraphics[width=5.6cm,angle=0]{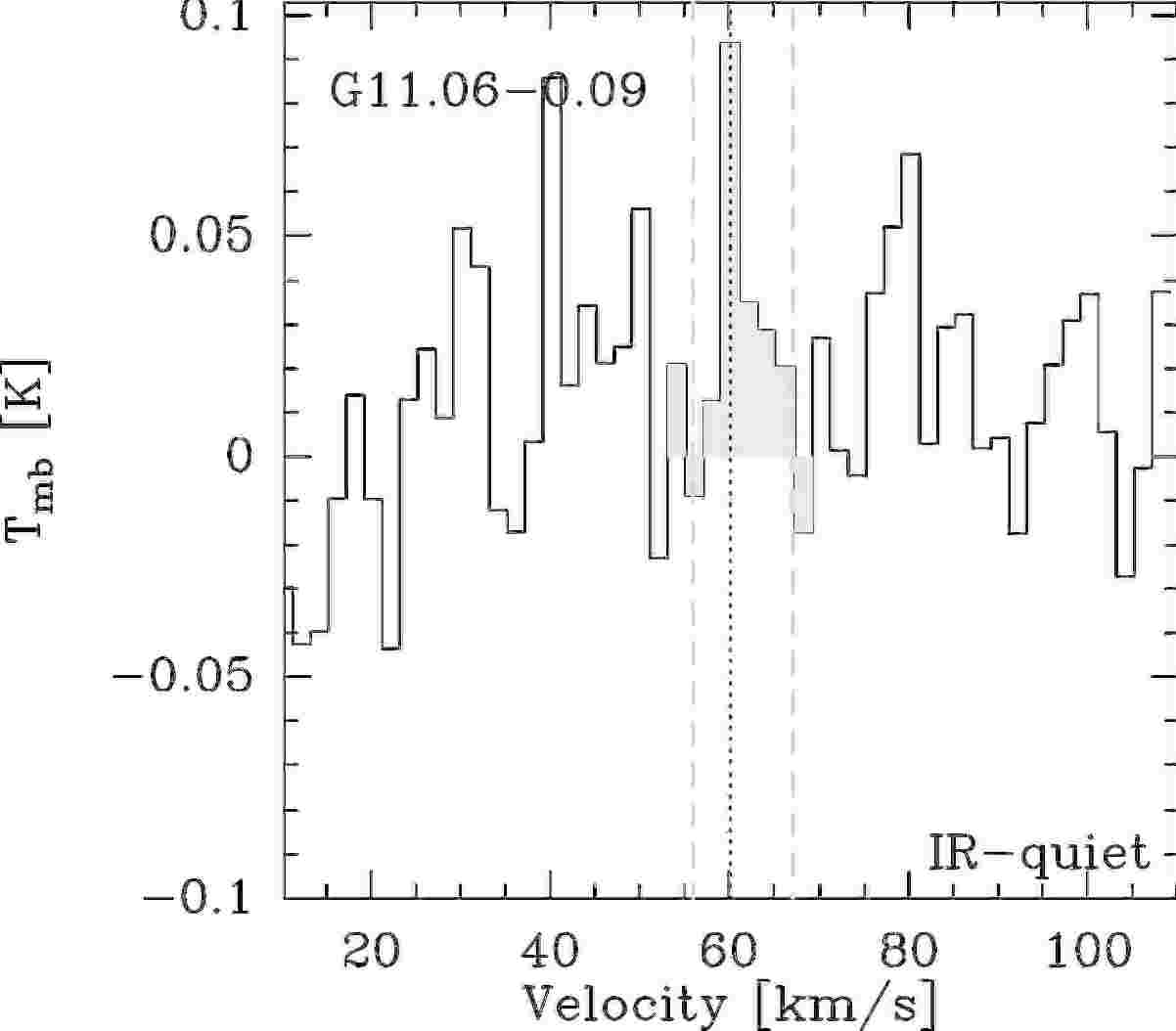} 
  \includegraphics[width=5.6cm,angle=0]{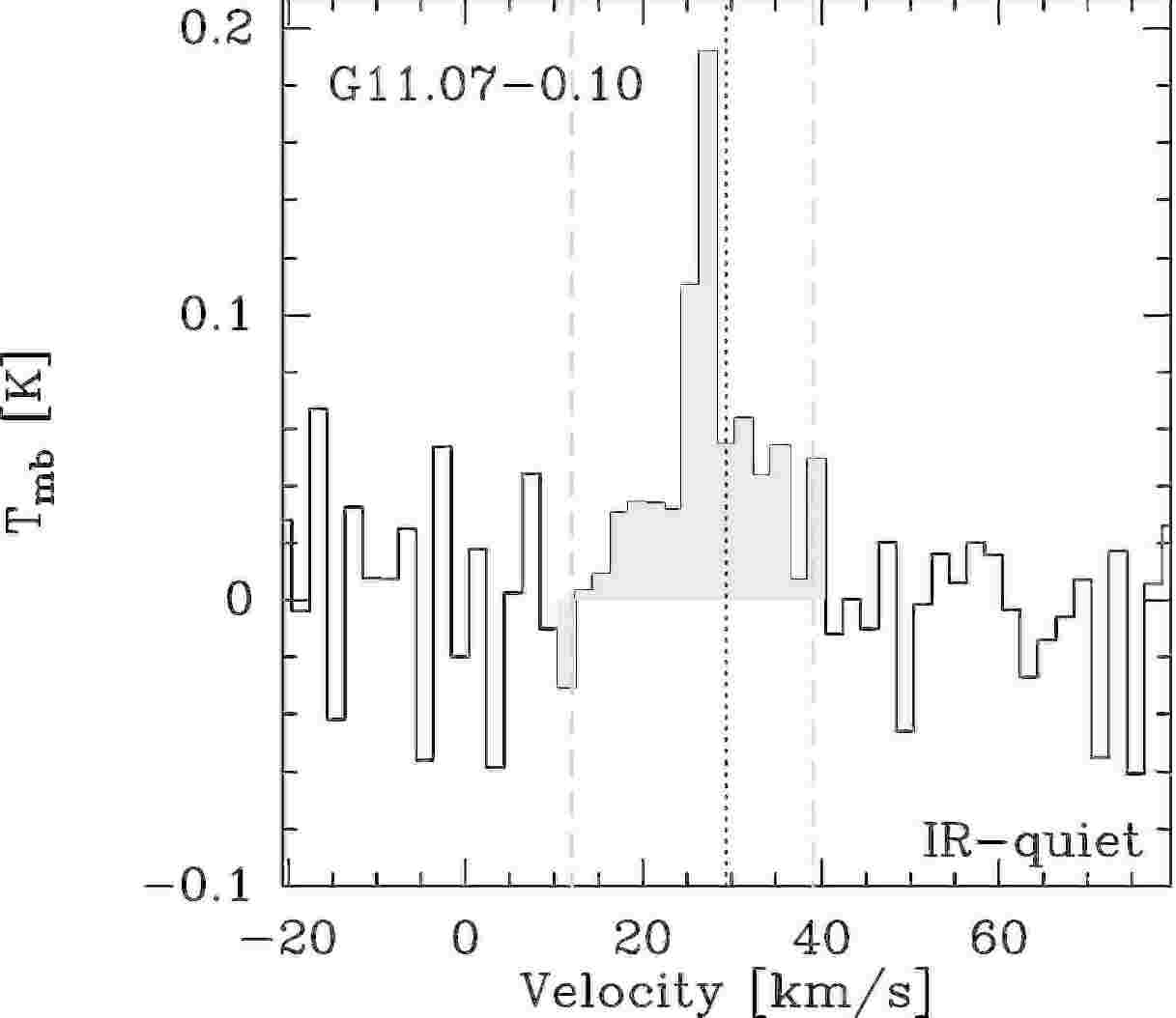} 
  \includegraphics[width=5.6cm,angle=0]{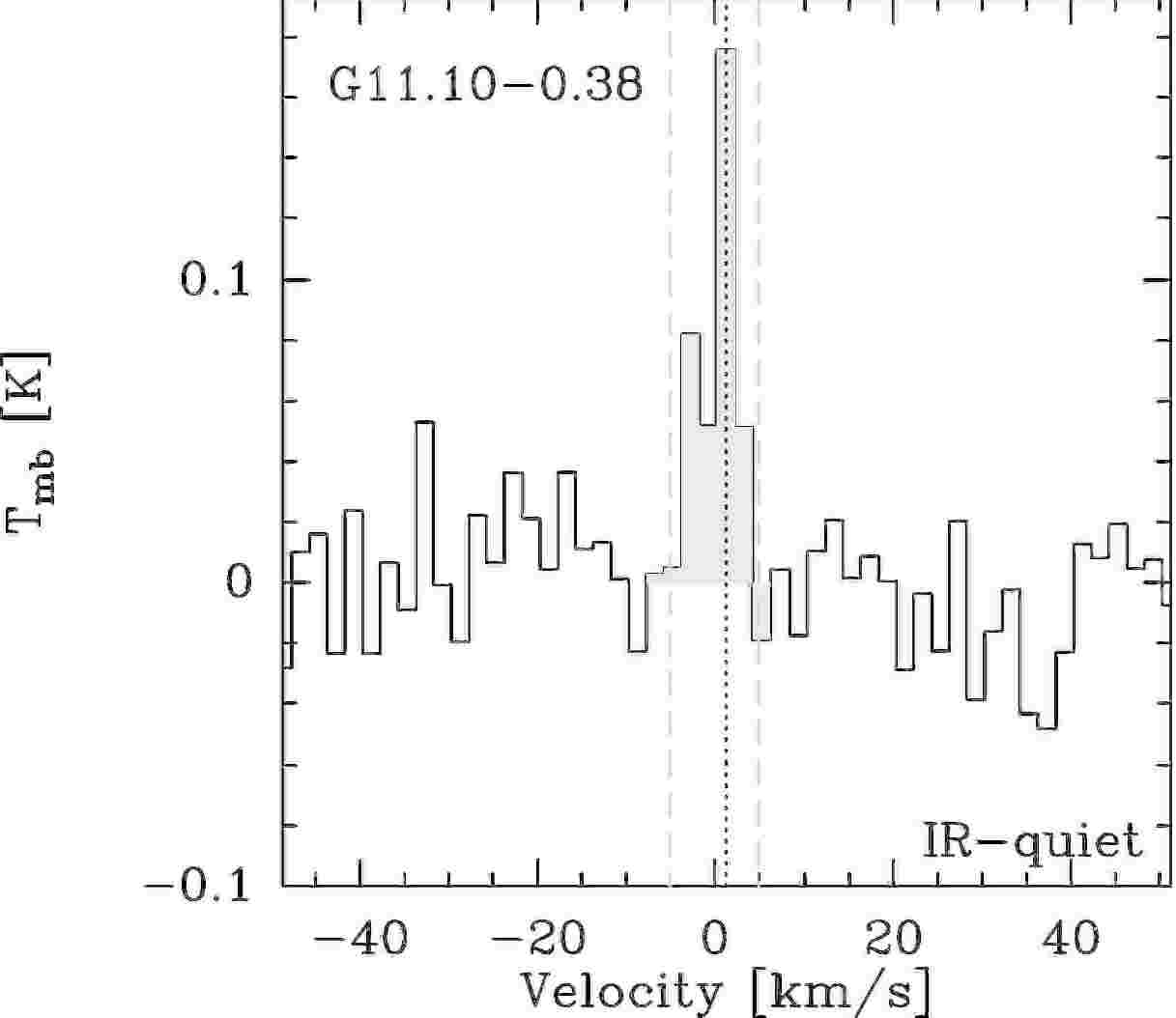} 
  \includegraphics[width=5.6cm,angle=0]{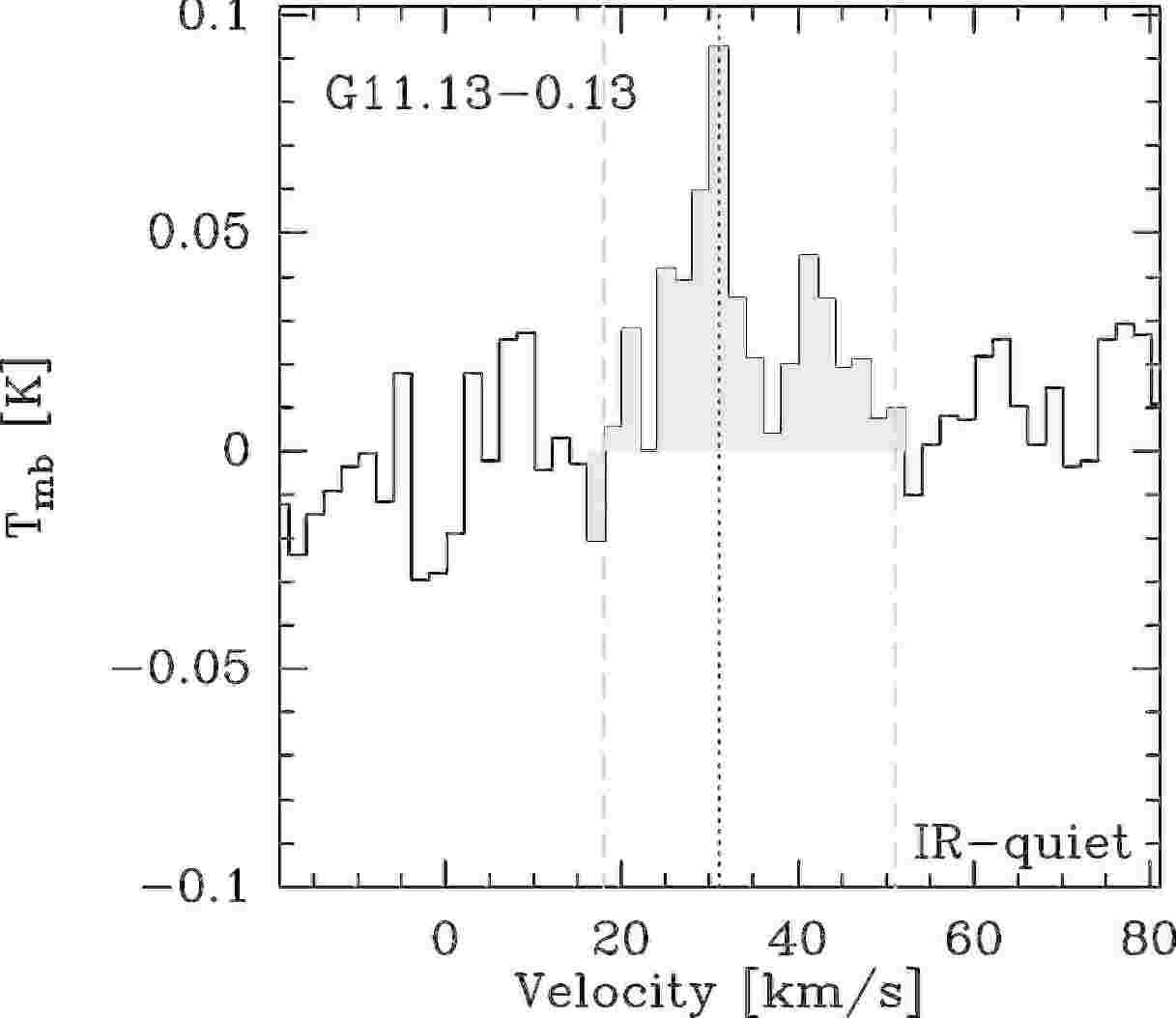} 
  \includegraphics[width=5.6cm,angle=0]{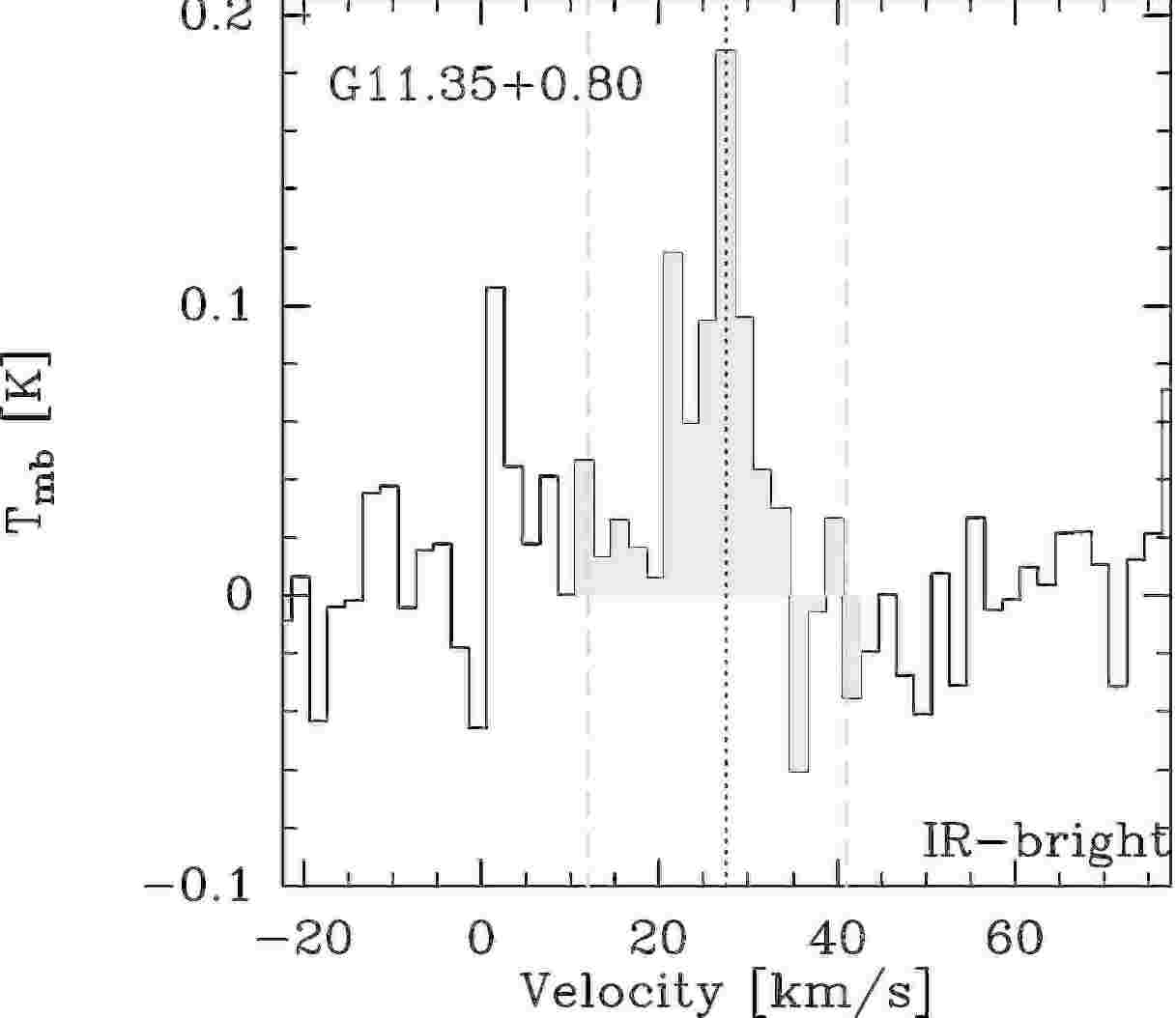} 
  \includegraphics[width=5.6cm,angle=0]{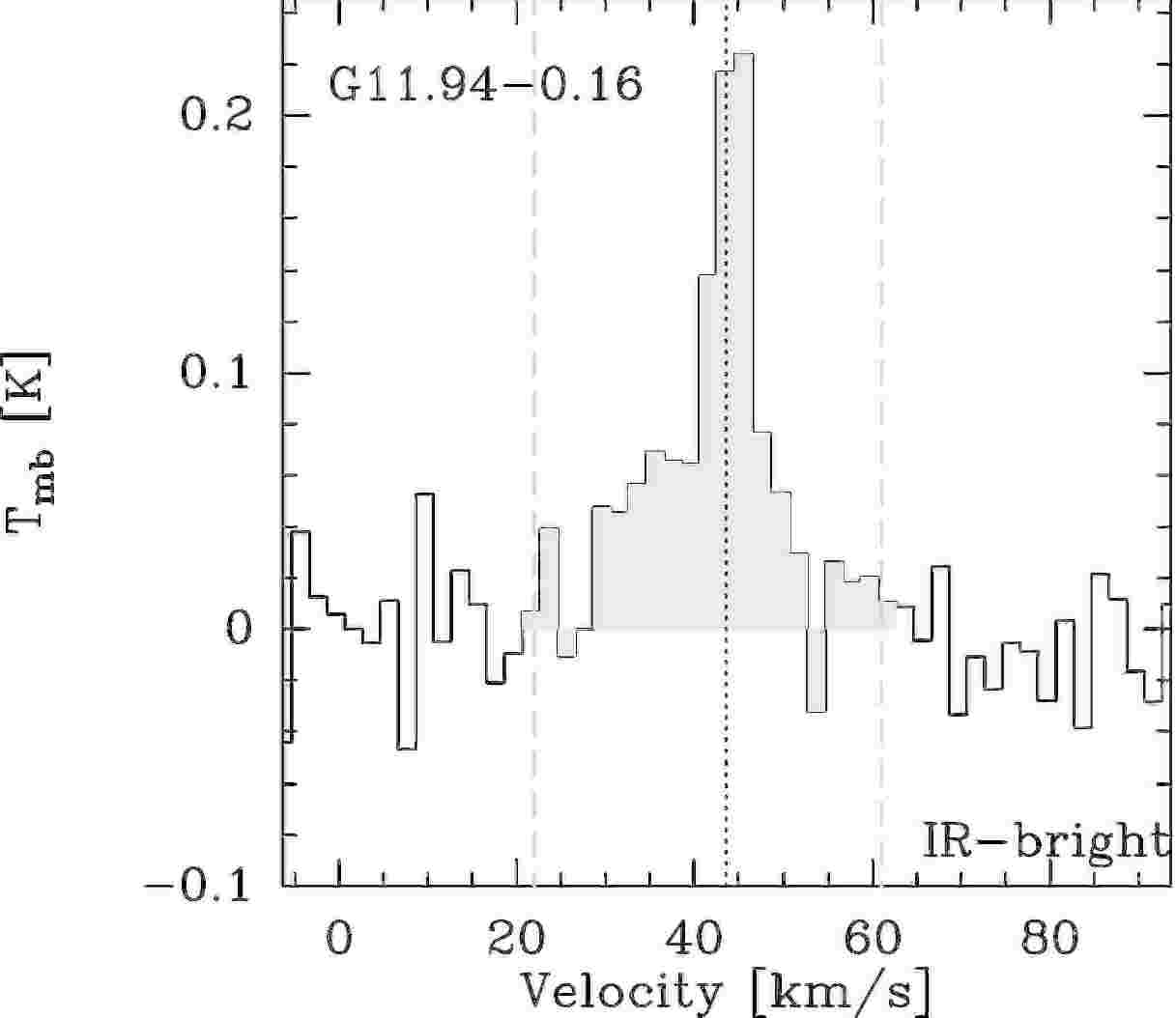} 
  \includegraphics[width=5.6cm,angle=0]{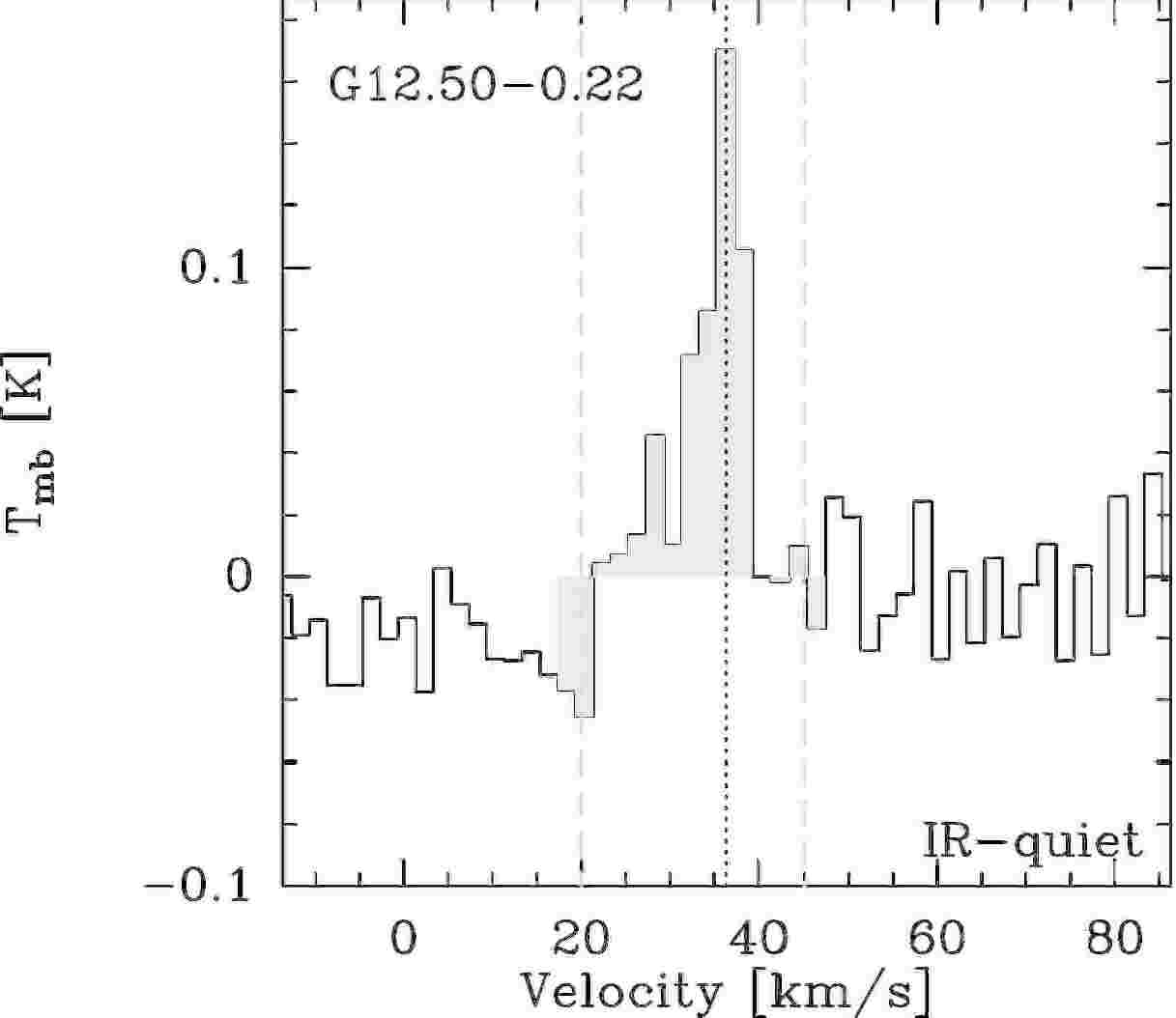} 
  \includegraphics[width=5.6cm,angle=0]{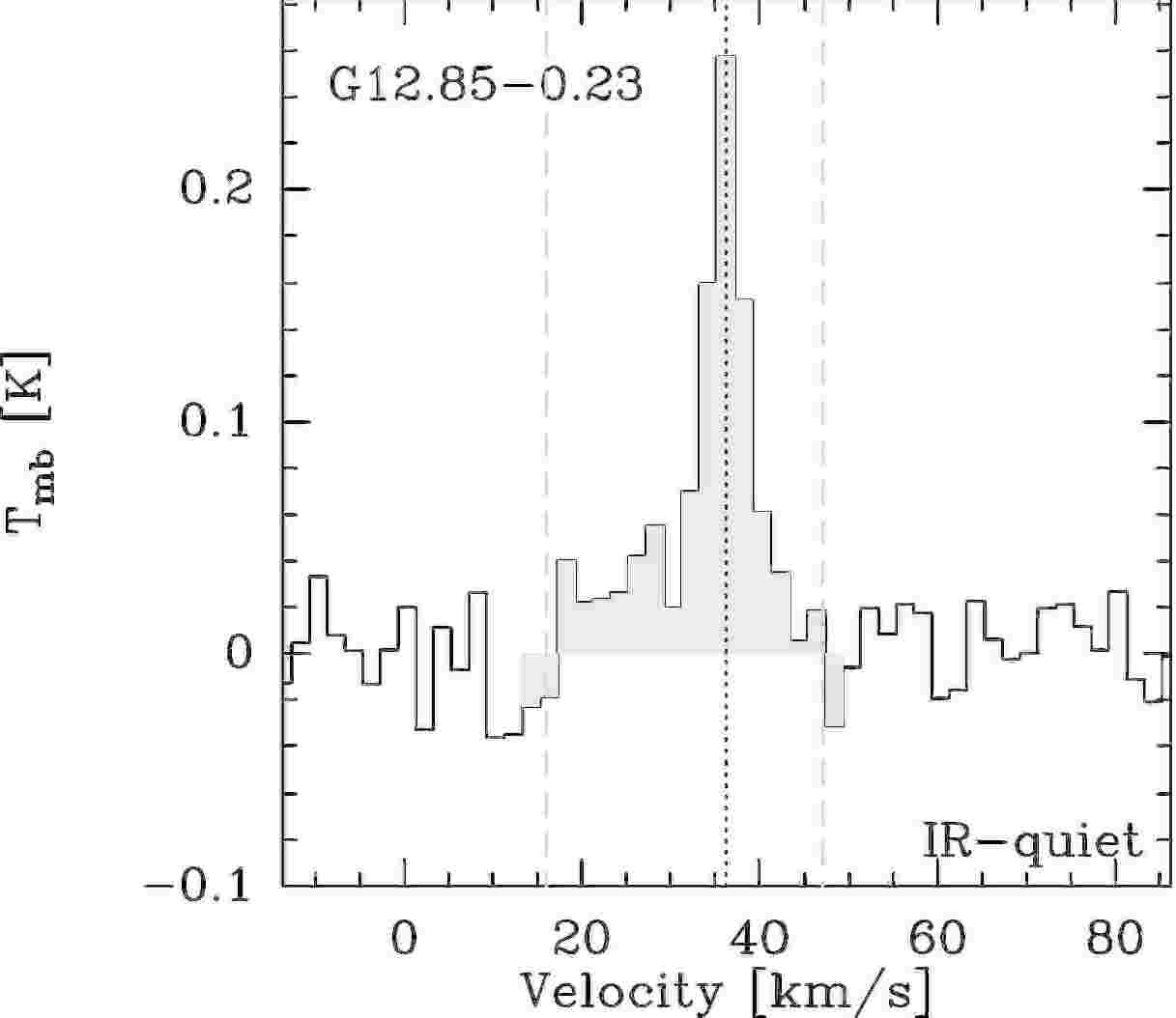} 
 \caption{Continued.}
\end{figure}
\end{landscape}

\begin{landscape}
\begin{figure}
\centering
\ContinuedFloat
  \includegraphics[width=5.6cm,angle=0]{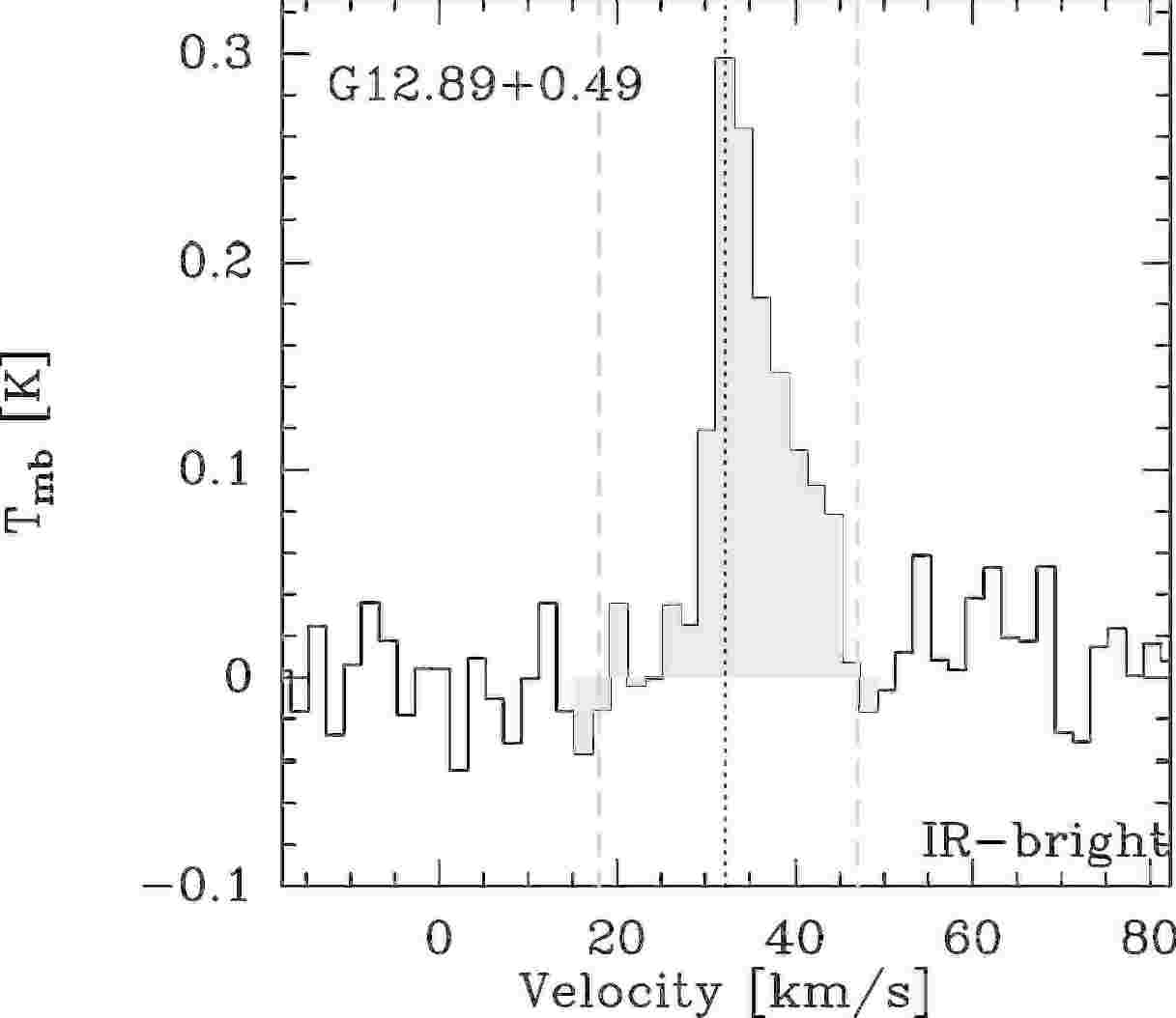} 
  \includegraphics[width=5.6cm,angle=0]{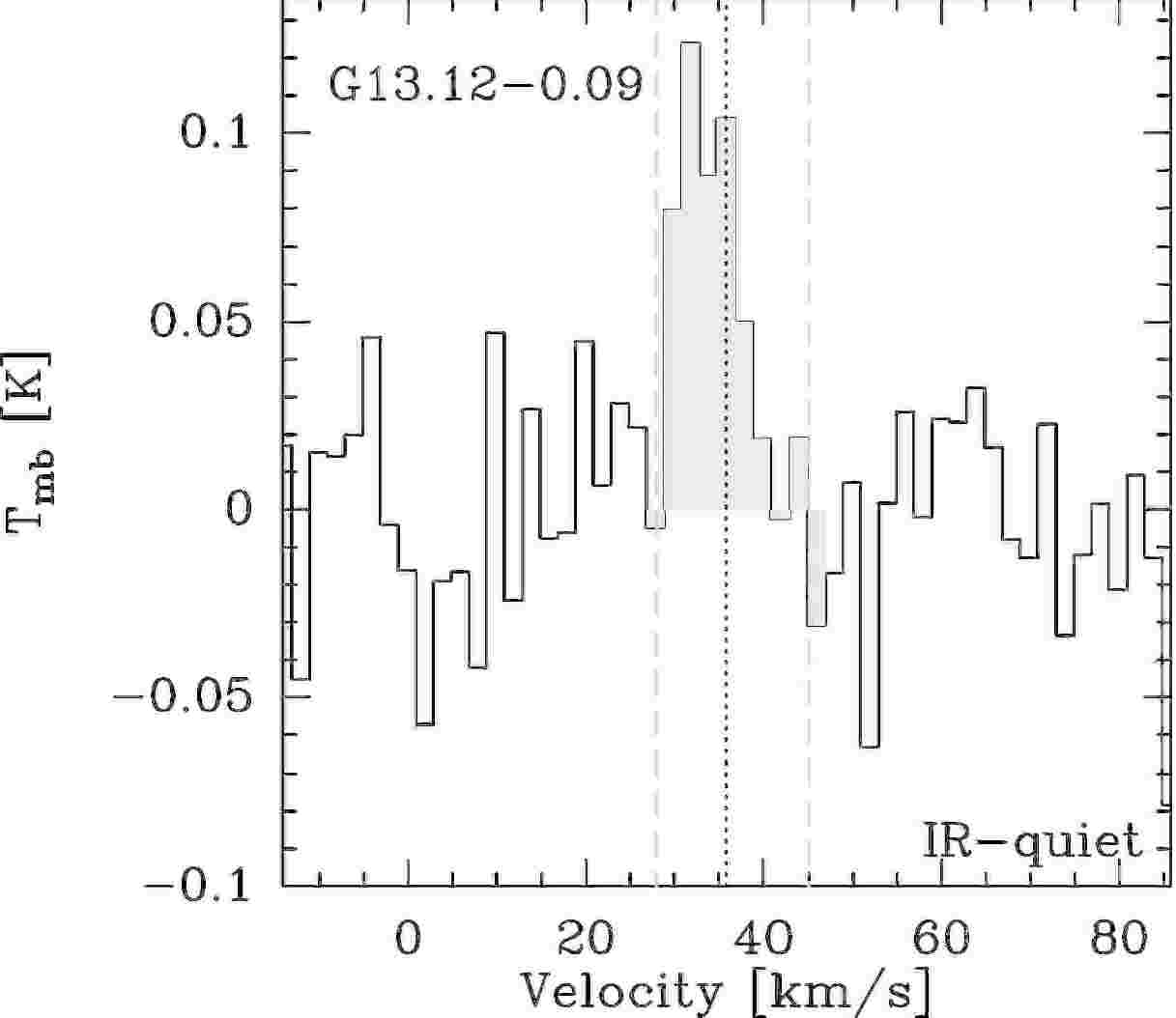} 
  \includegraphics[width=5.6cm,angle=0]{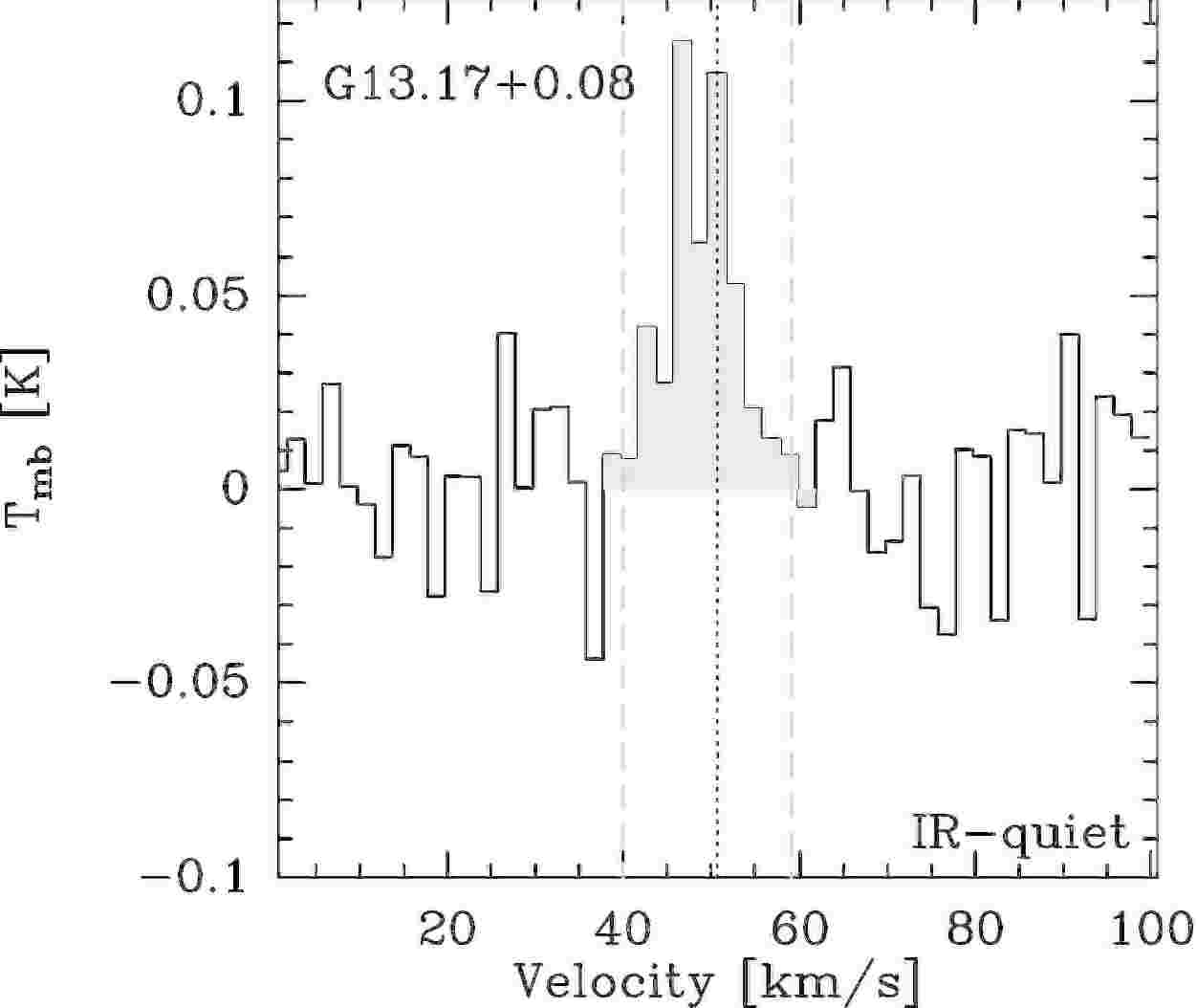} 
  \includegraphics[width=5.6cm,angle=0]{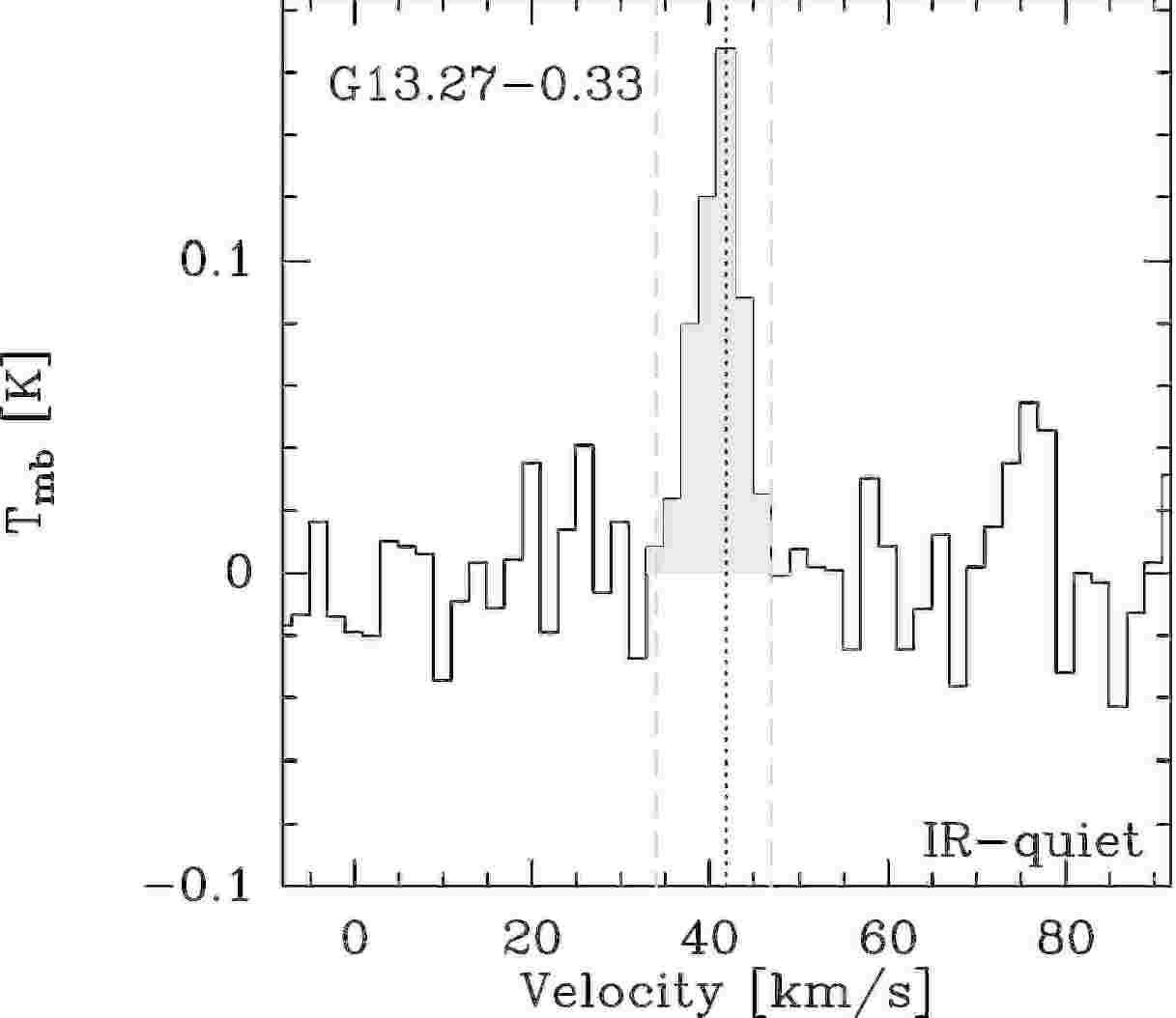} 
  \includegraphics[width=5.6cm,angle=0]{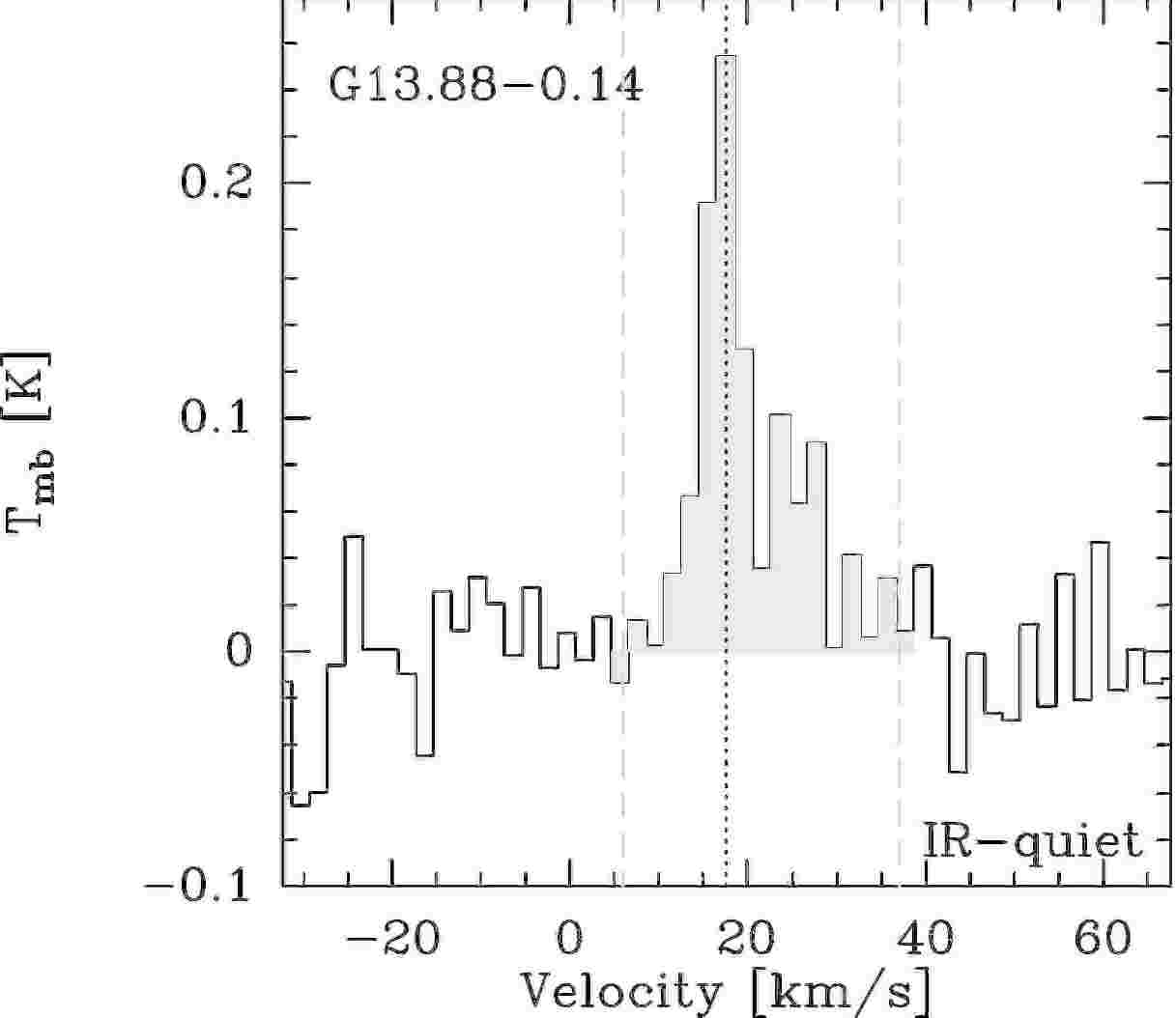} 
  \includegraphics[width=5.6cm,angle=0]{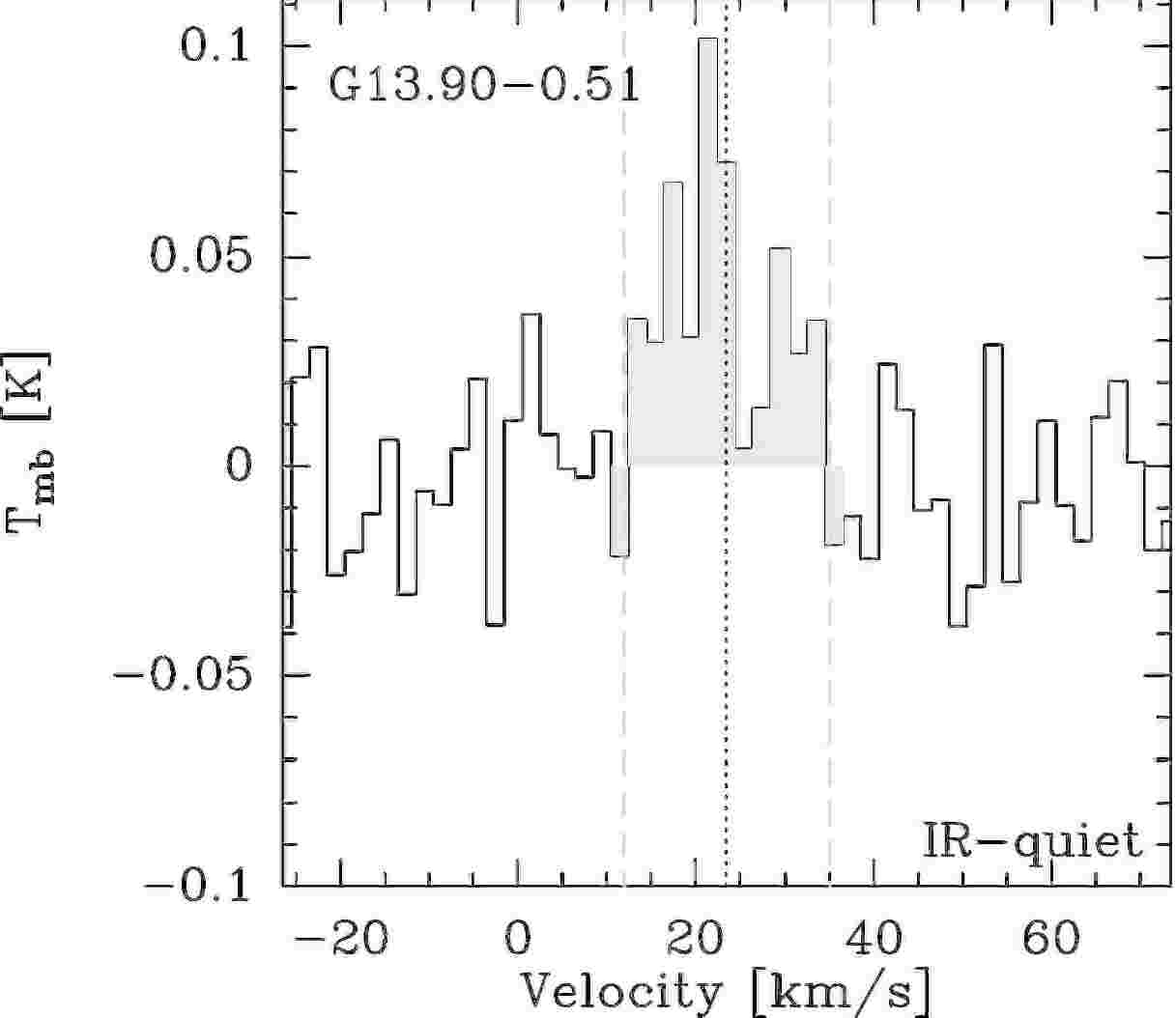} 
  \includegraphics[width=5.6cm,angle=0]{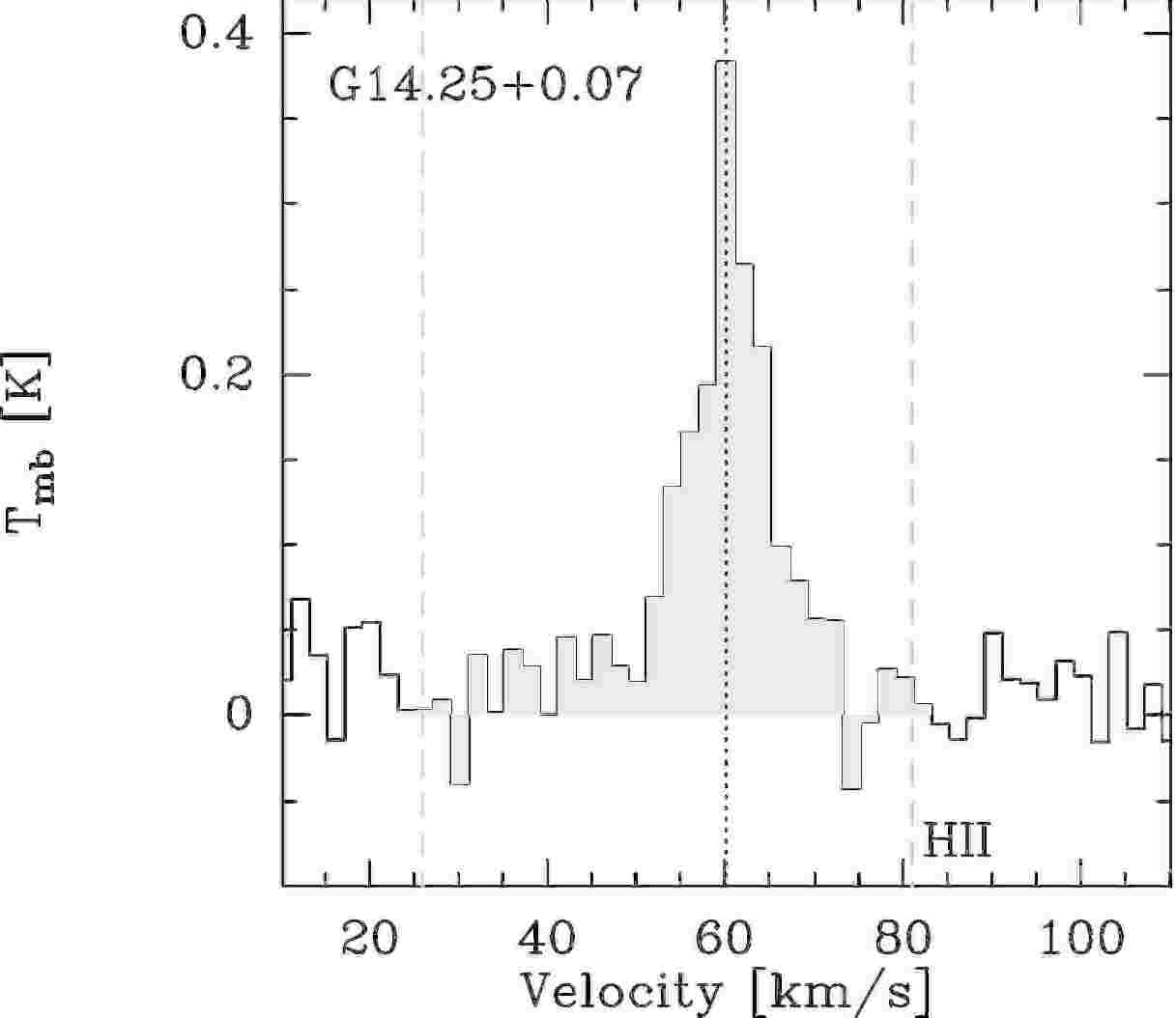} 
  \includegraphics[width=5.6cm,angle=0]{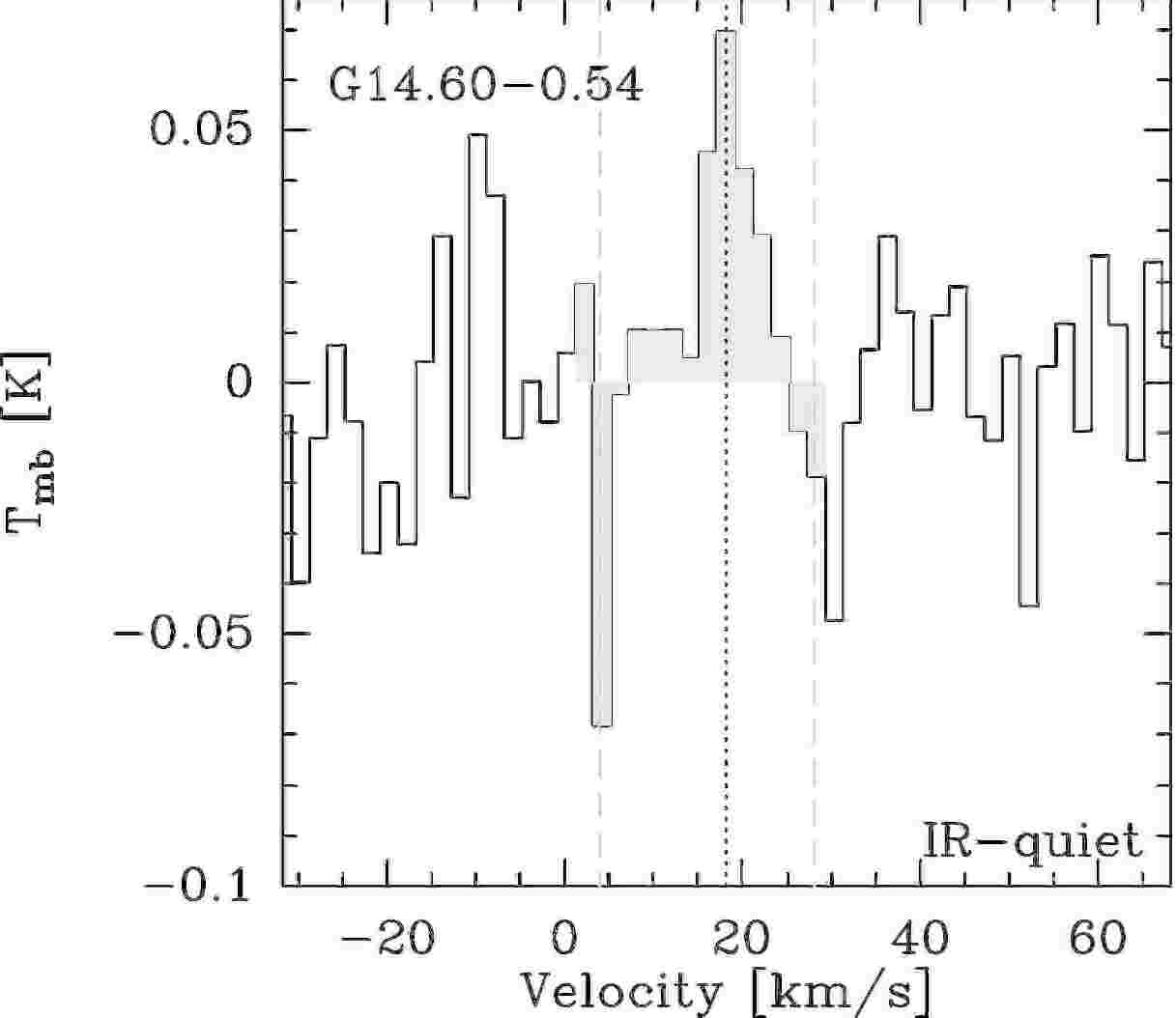} 
  \includegraphics[width=5.6cm,angle=0]{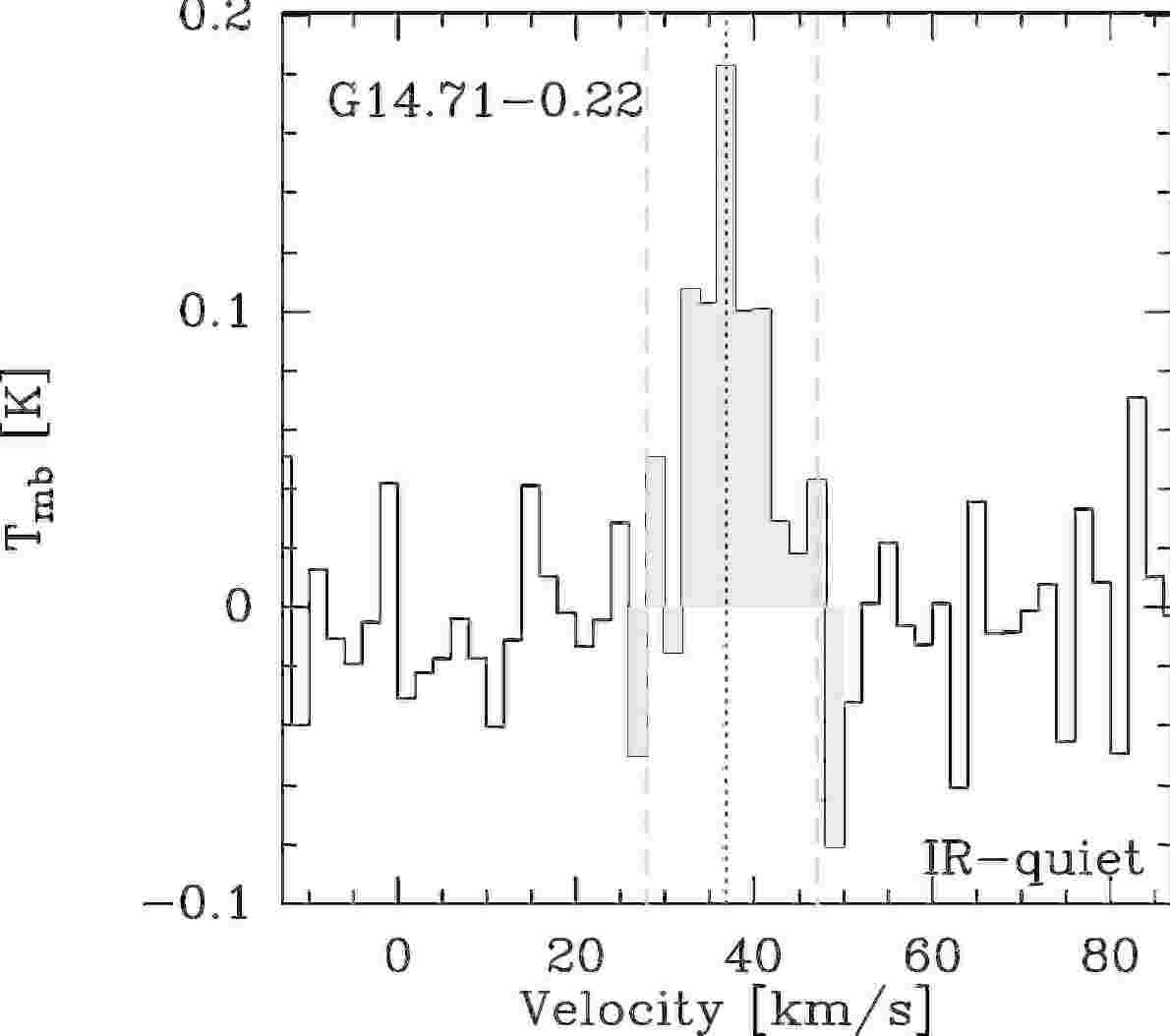} 
  \includegraphics[width=5.6cm,angle=0]{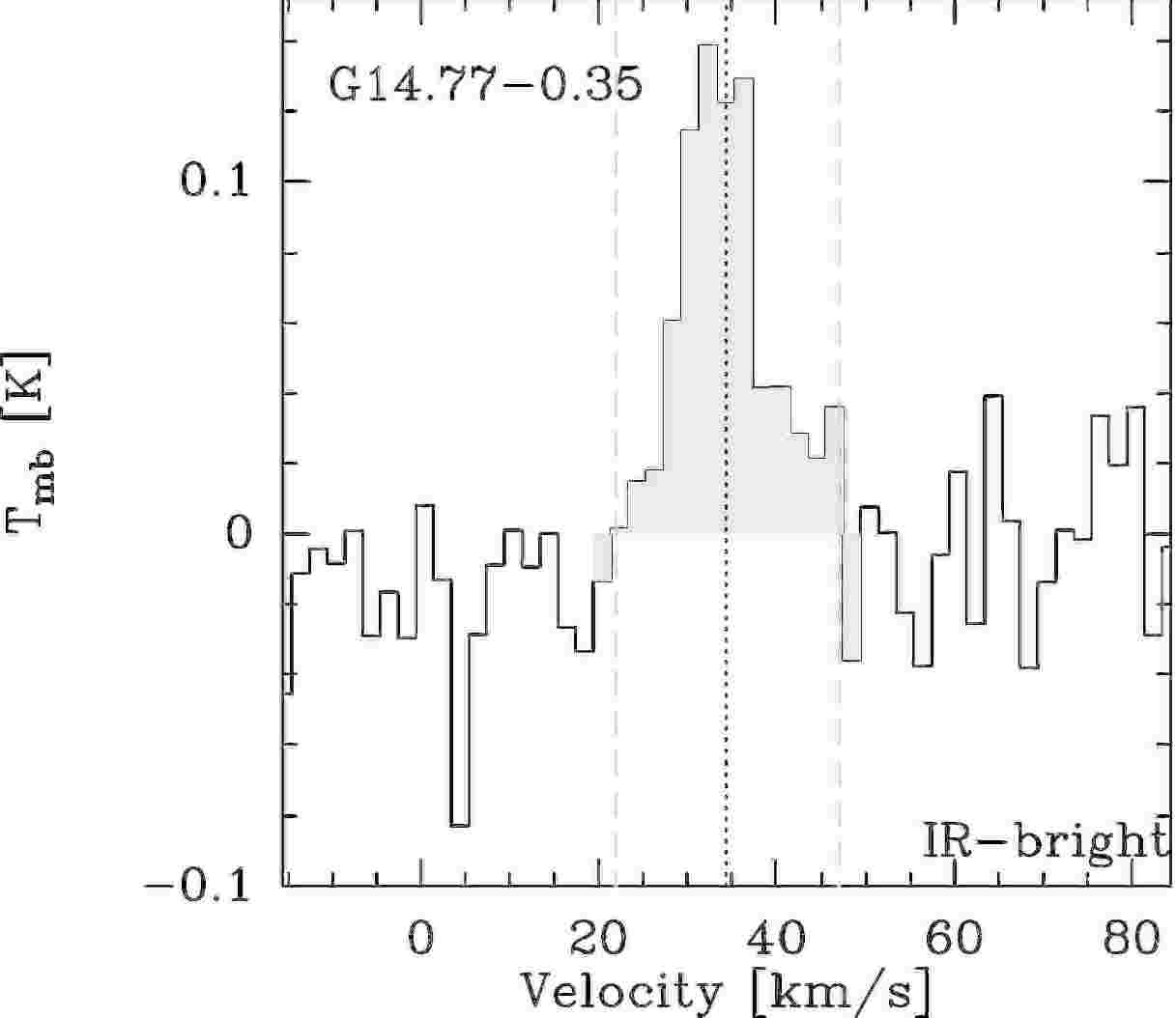} 
  \includegraphics[width=5.6cm,angle=0]{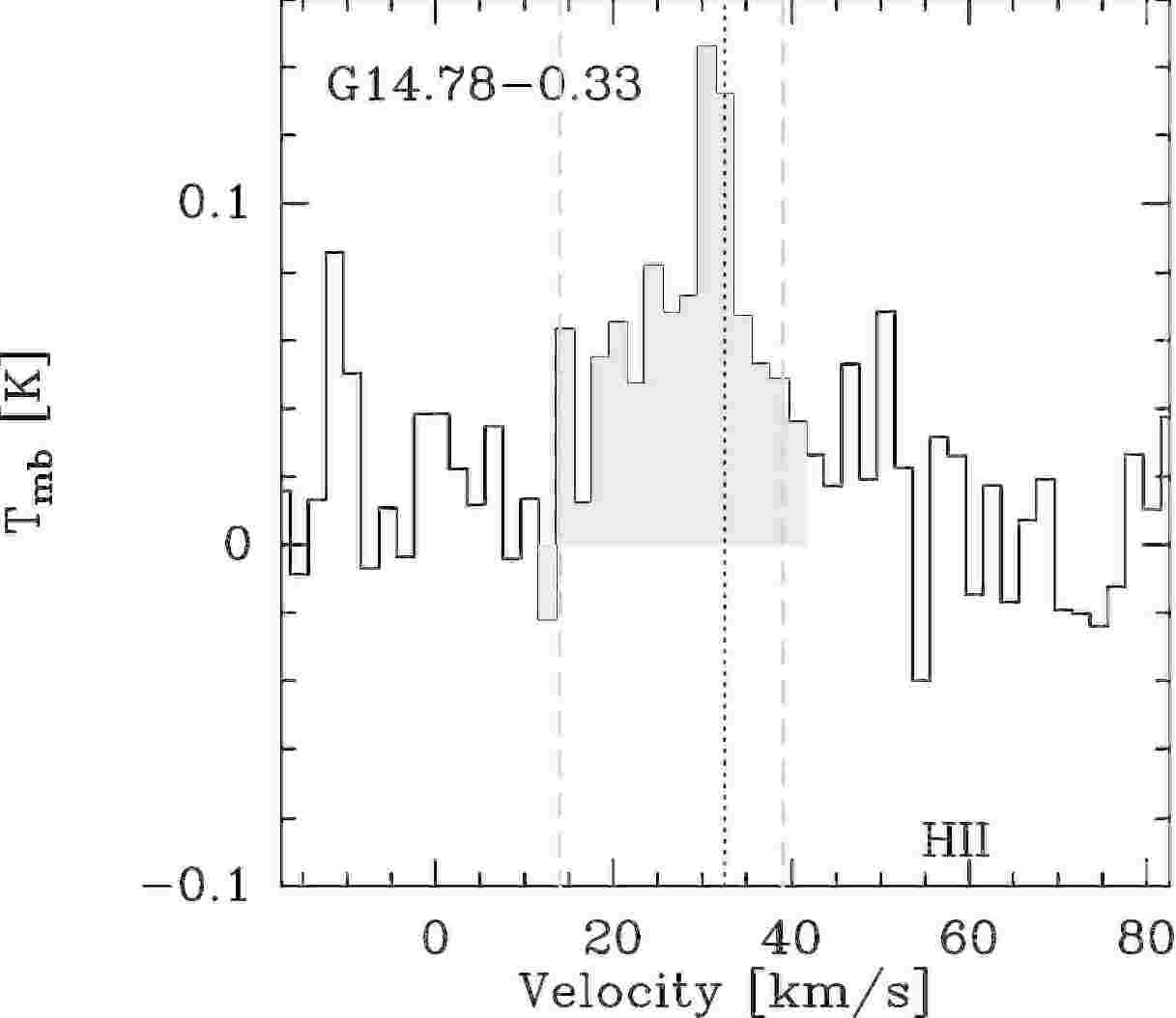} 
  \includegraphics[width=5.6cm,angle=0]{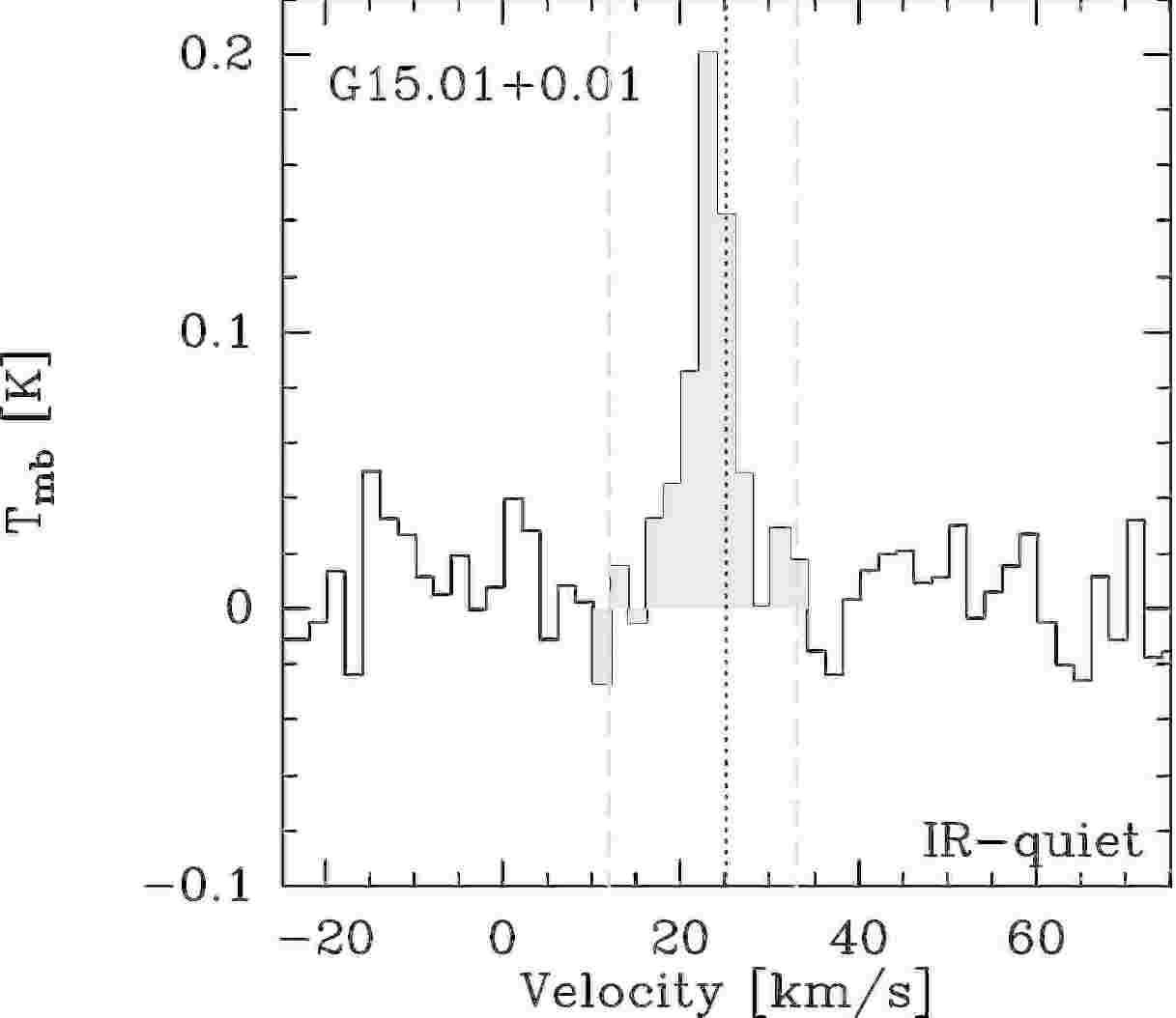} 
 \caption{Continued.}
\end{figure}
\end{landscape}

\begin{landscape}
\begin{figure}
\centering
\ContinuedFloat
  \includegraphics[width=5.6cm,angle=0]{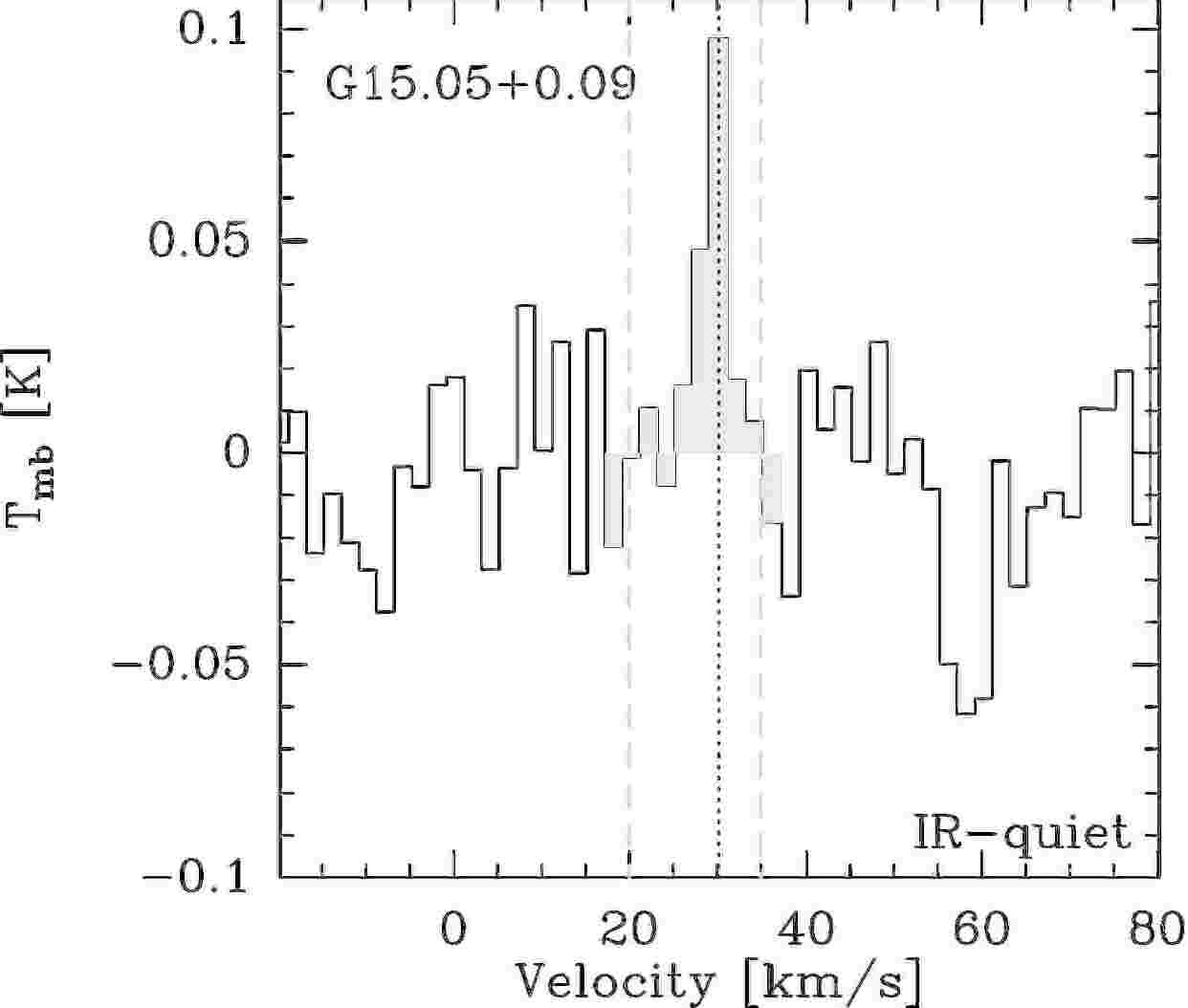} 
  \includegraphics[width=5.6cm,angle=0]{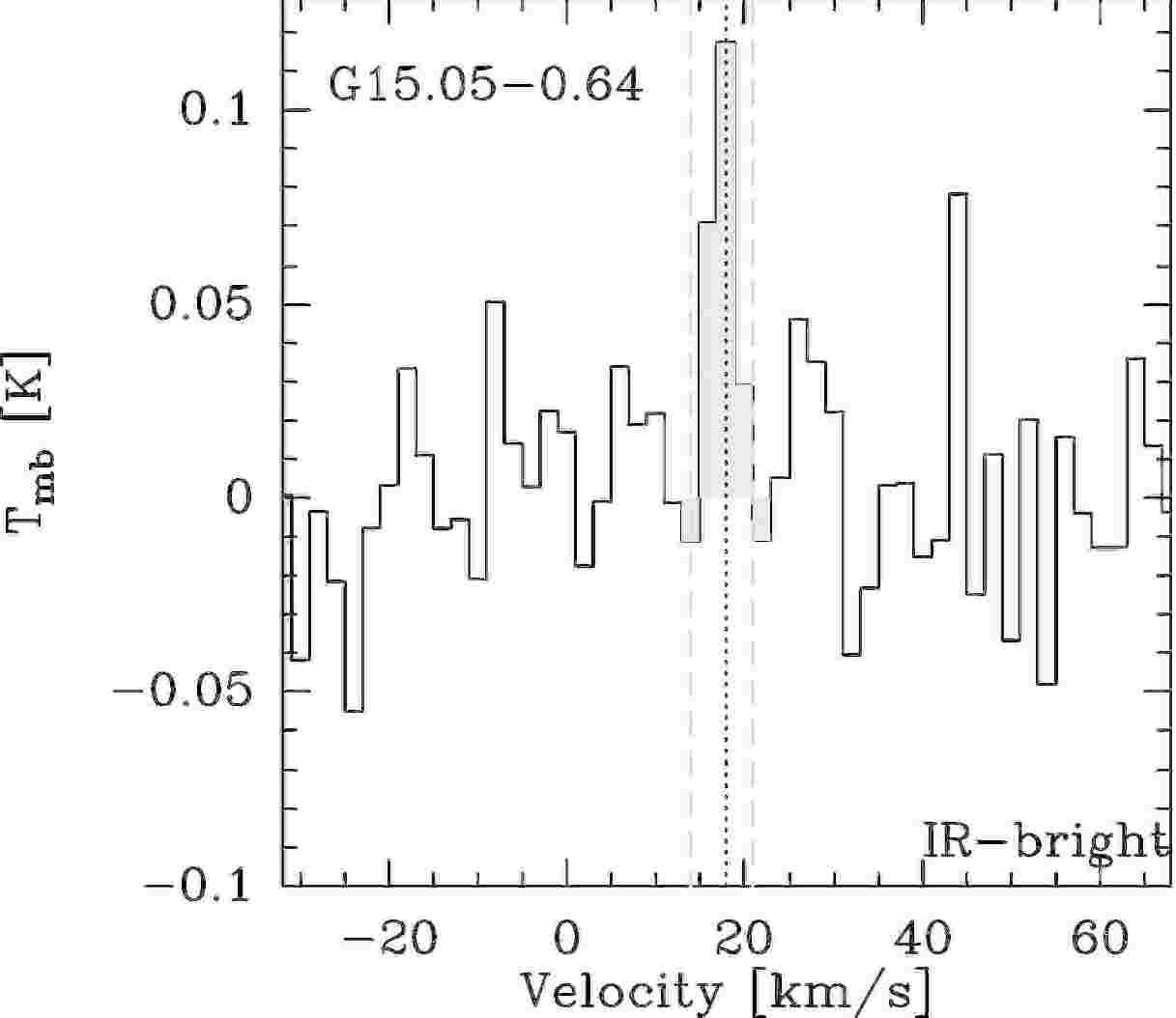} 
  \includegraphics[width=5.6cm,angle=0]{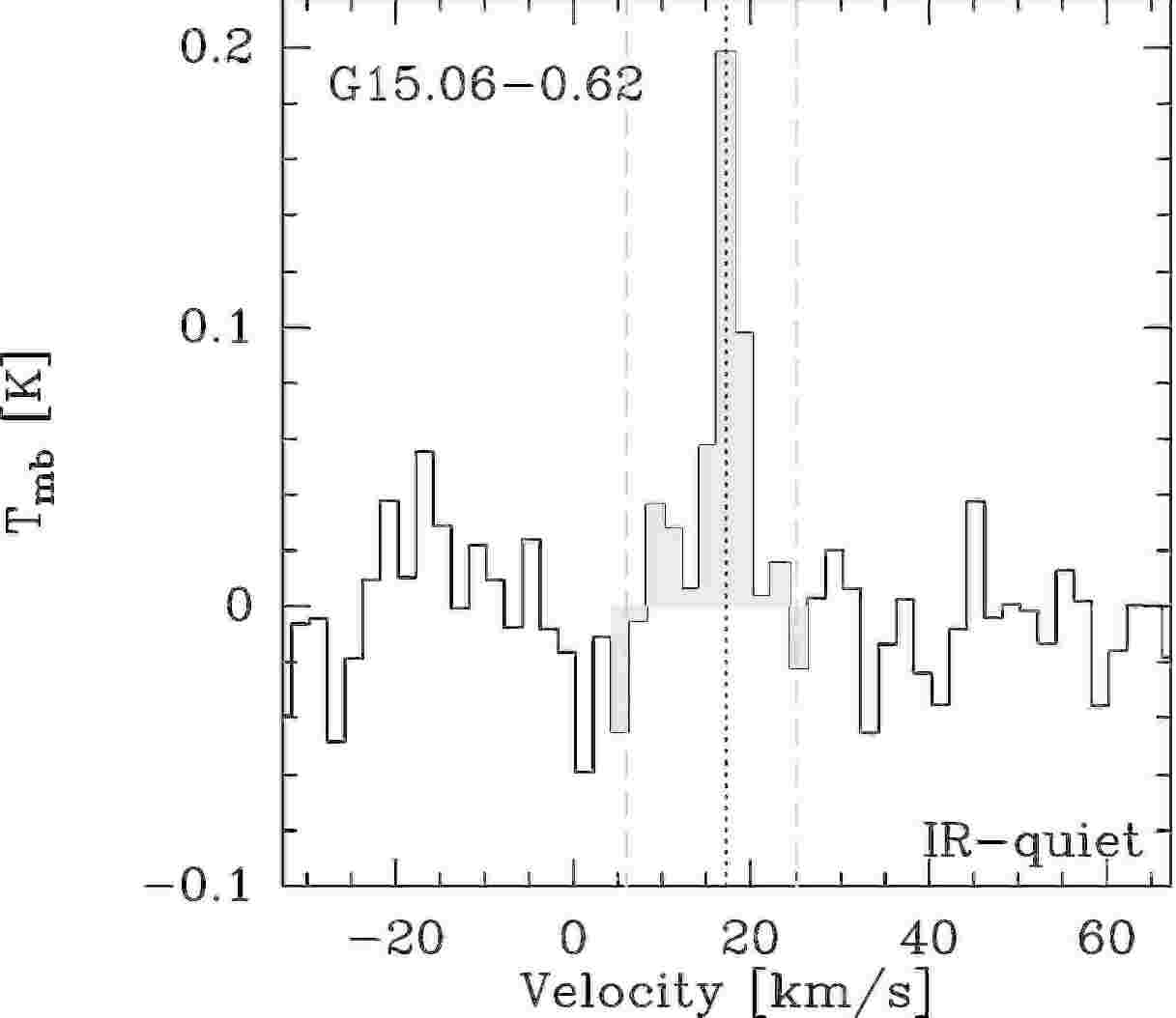} 
  \includegraphics[width=5.6cm,angle=0]{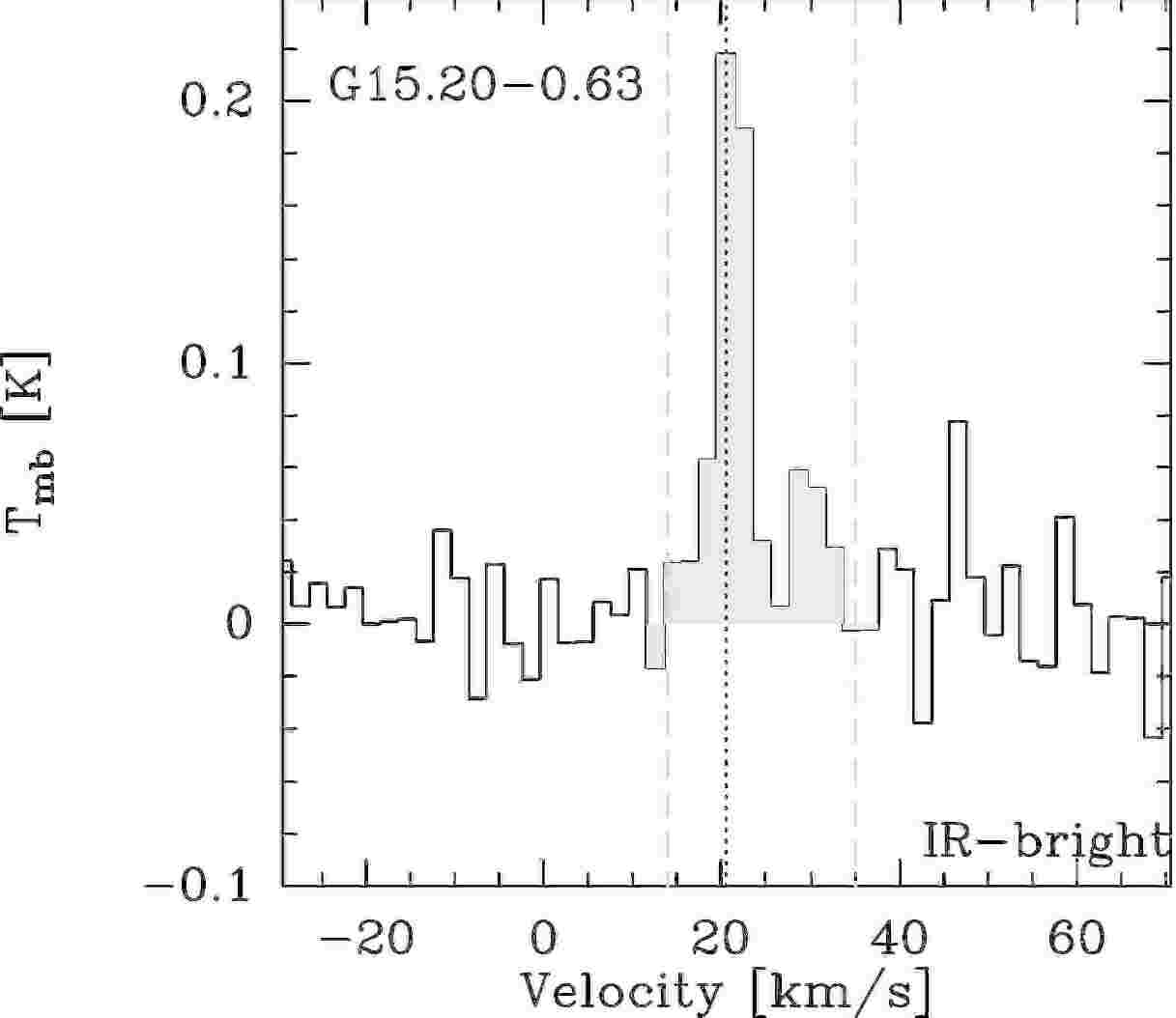} 
  \includegraphics[width=5.6cm,angle=0]{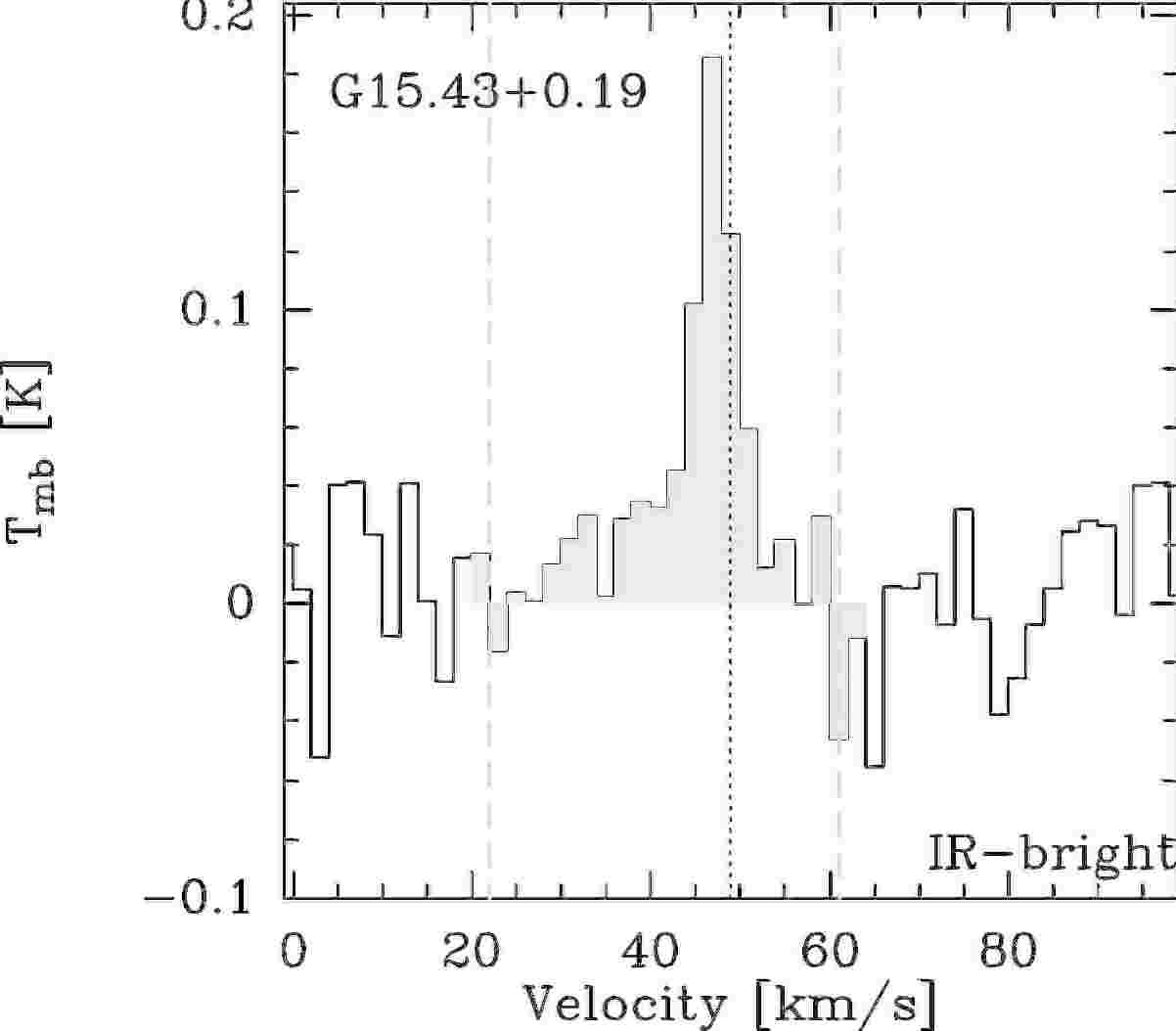} 
  \includegraphics[width=5.6cm,angle=0]{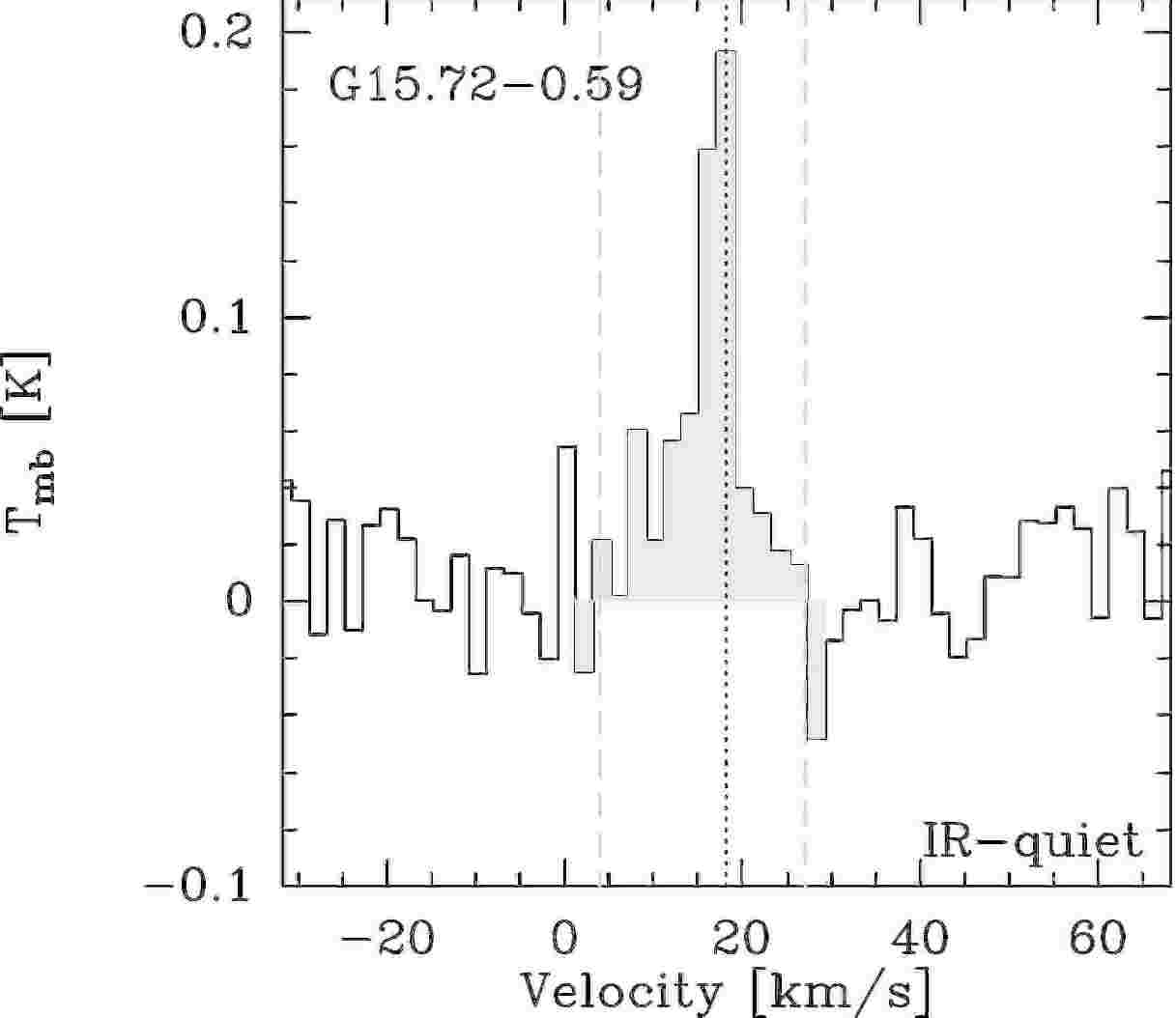} 
  \includegraphics[width=5.6cm,angle=0]{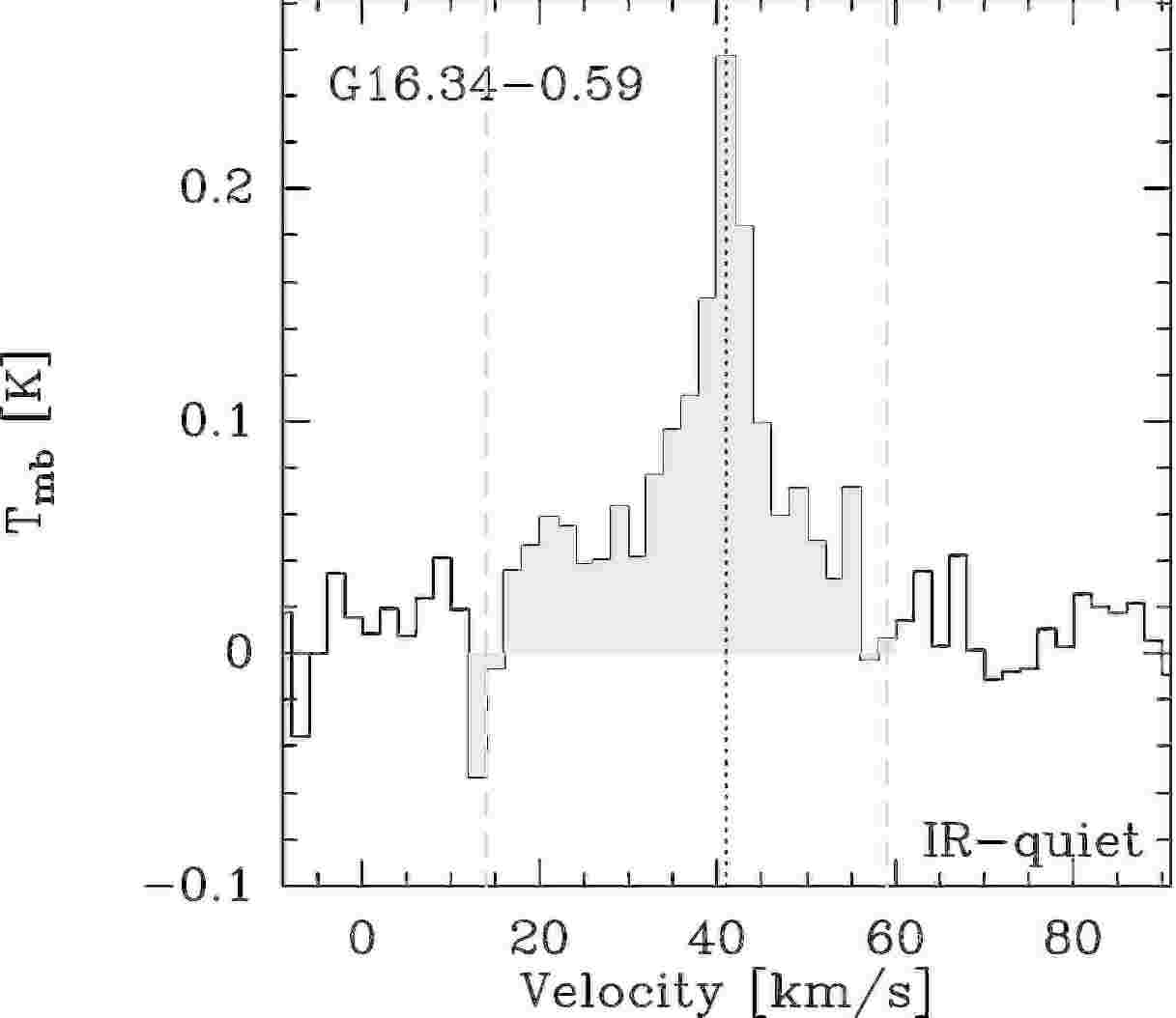} 
  \includegraphics[width=5.6cm,angle=0]{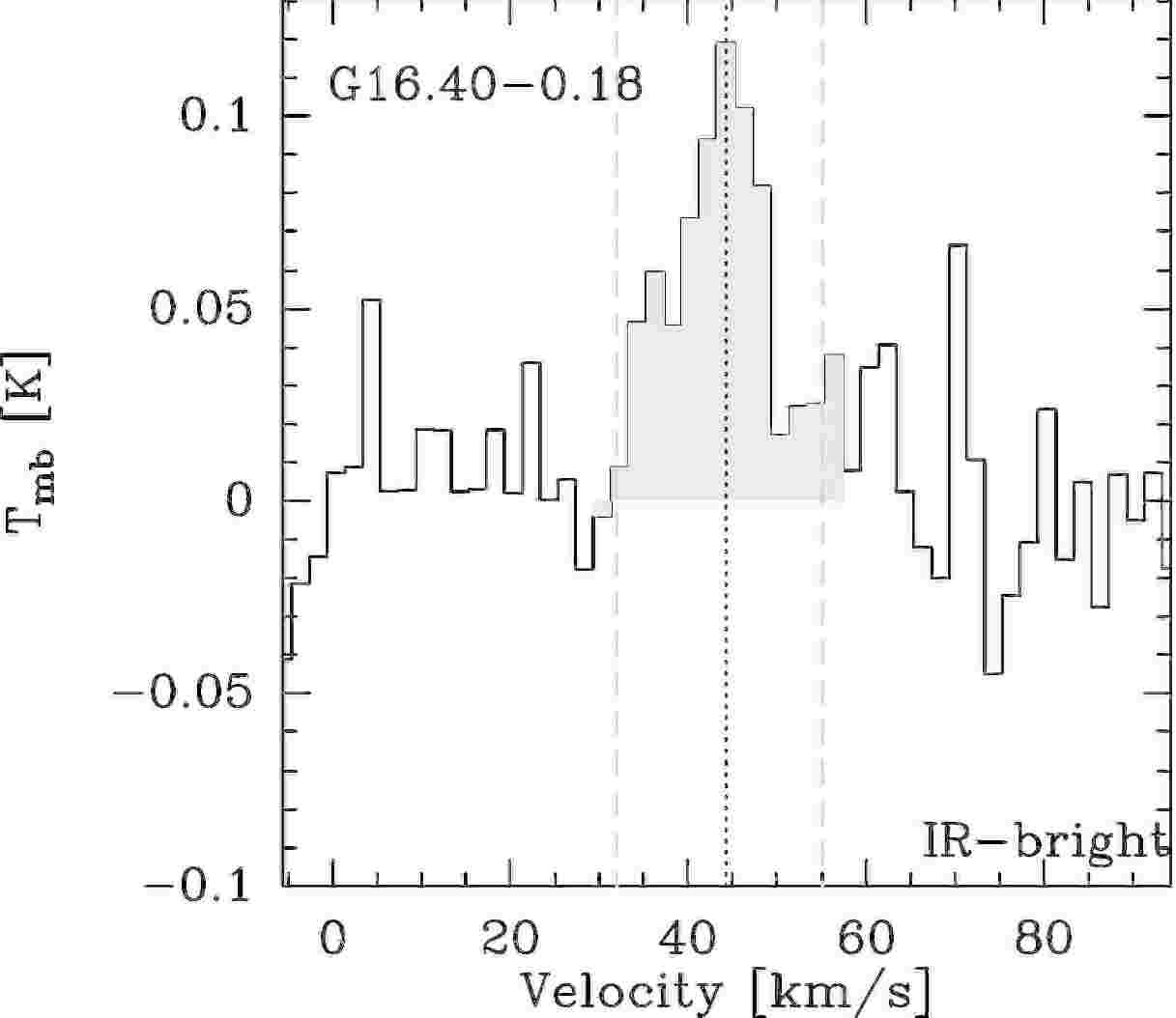} 
  \includegraphics[width=5.6cm,angle=0]{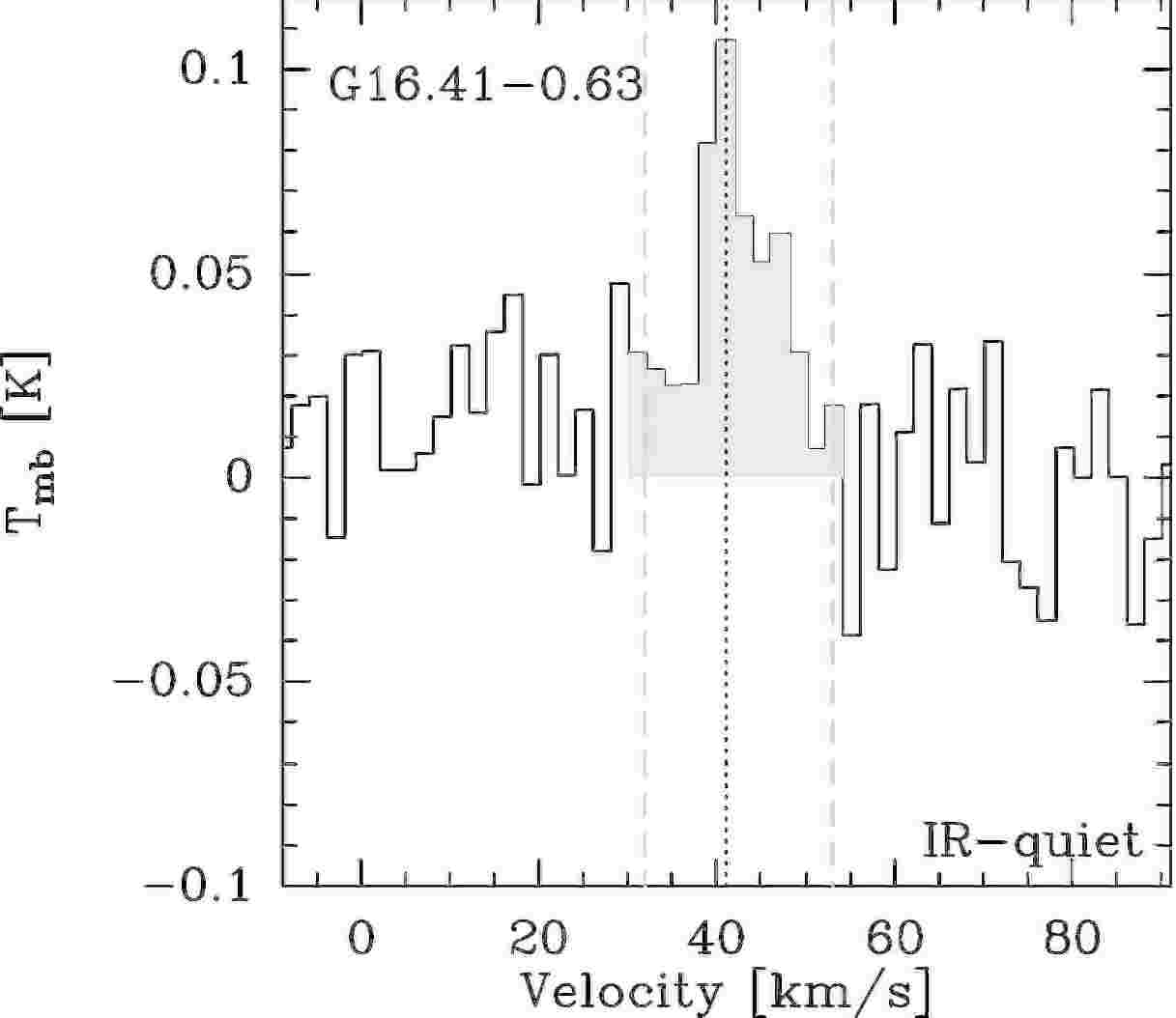} 
  \includegraphics[width=5.6cm,angle=0]{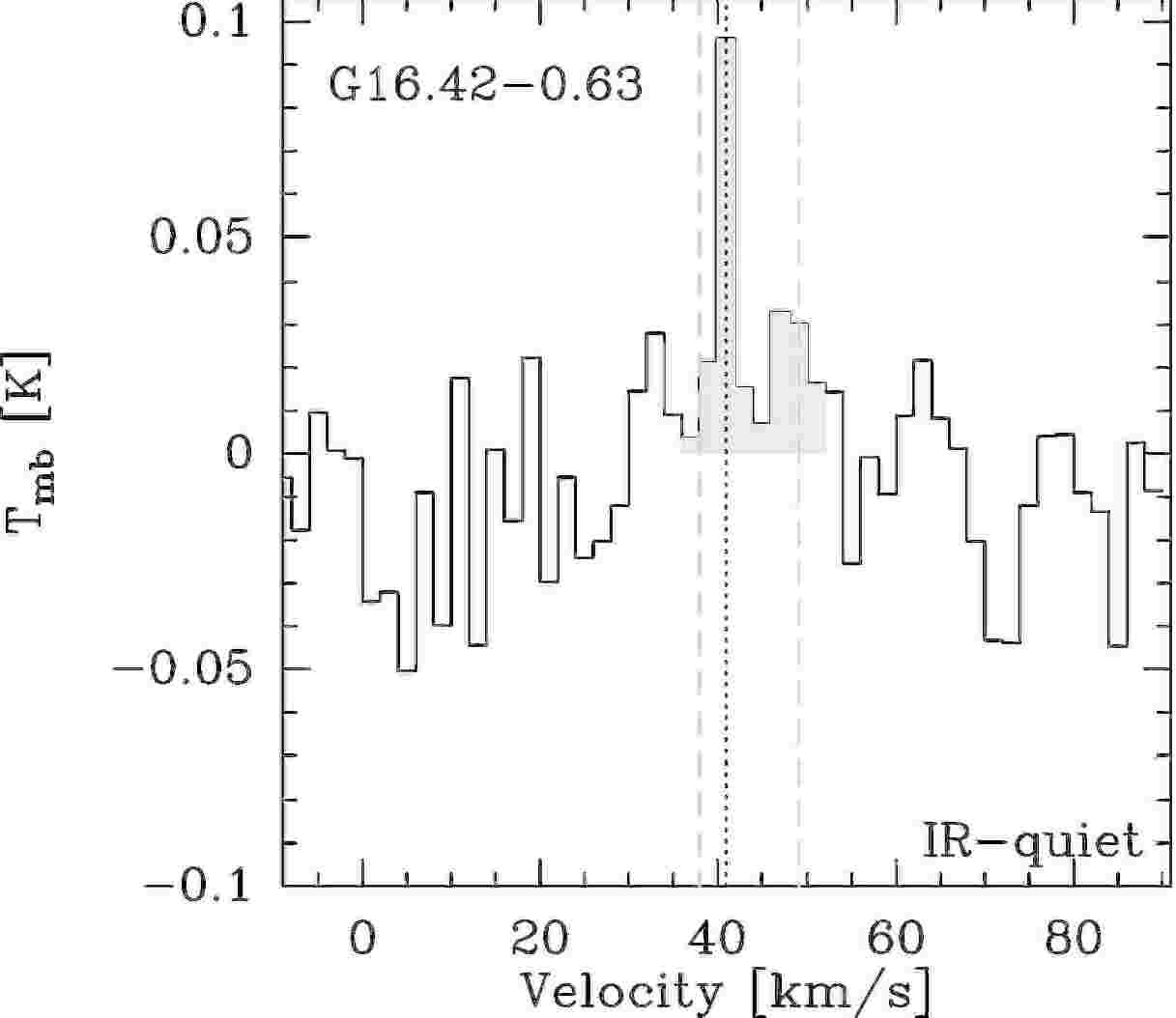} 
  \includegraphics[width=5.6cm,angle=0]{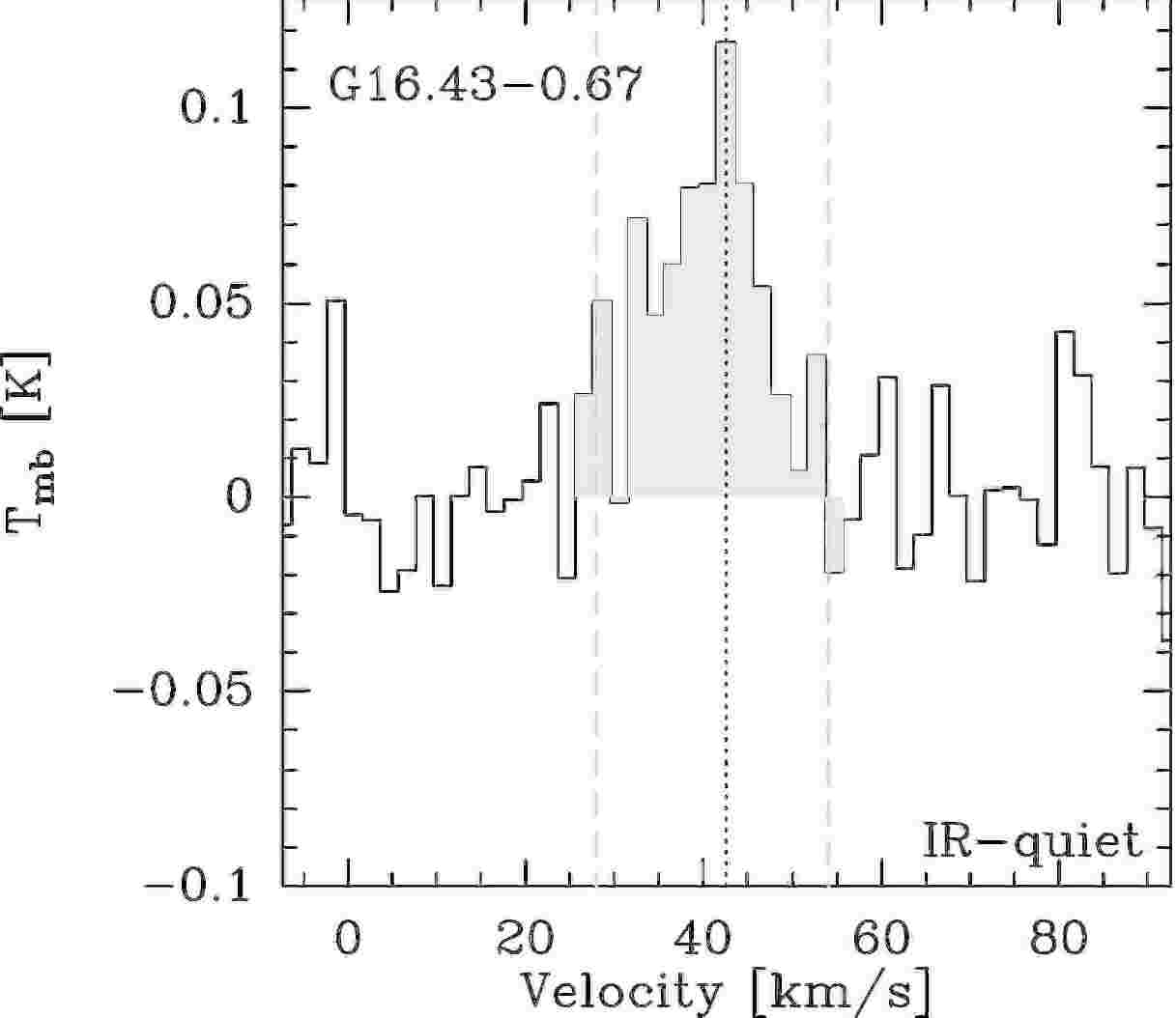} 
  \includegraphics[width=5.6cm,angle=0]{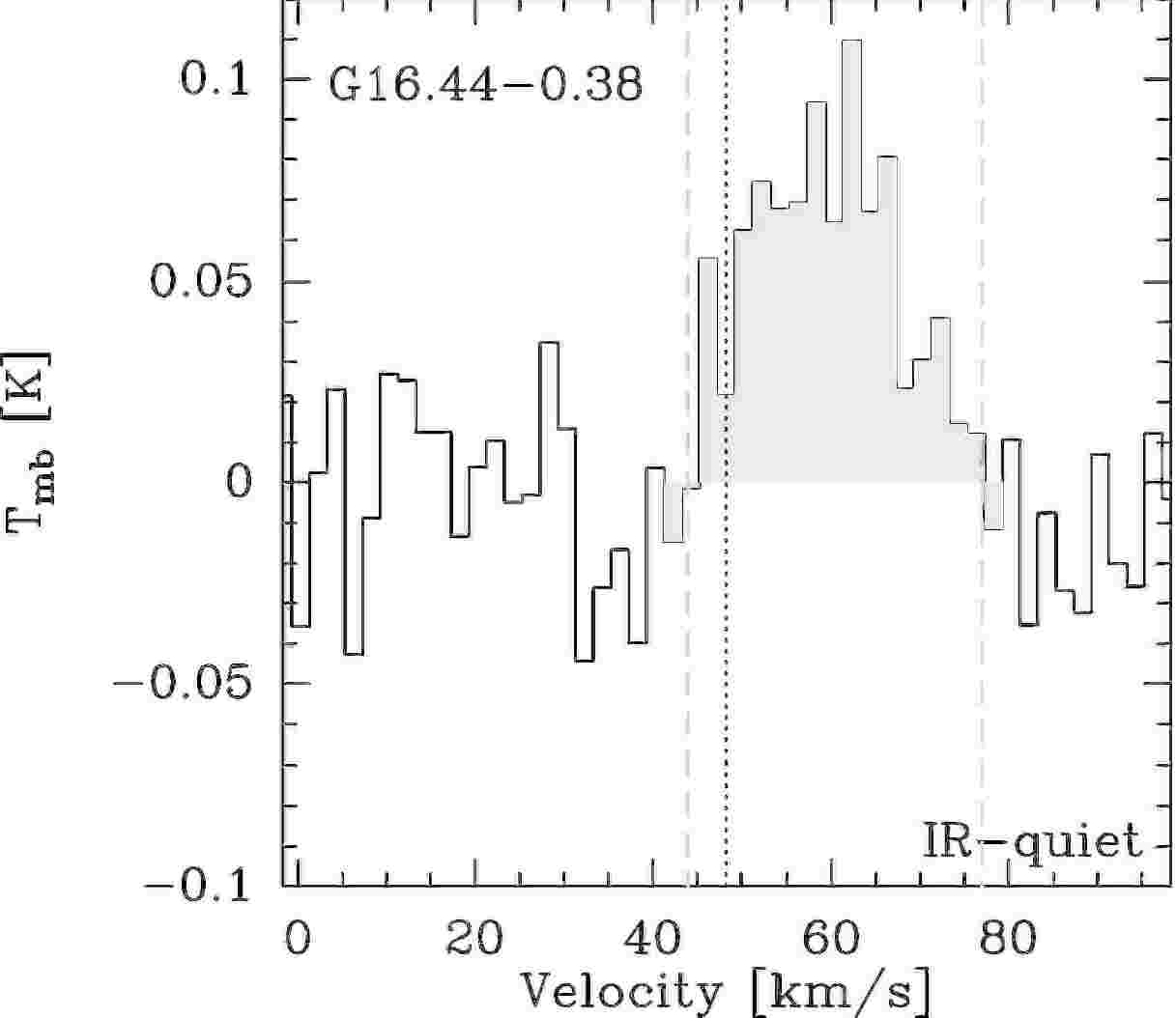} 
 \caption{Continued.}
\end{figure}
\end{landscape}

\begin{landscape}
\begin{figure}
\centering
\ContinuedFloat
  \includegraphics[width=5.6cm,angle=0]{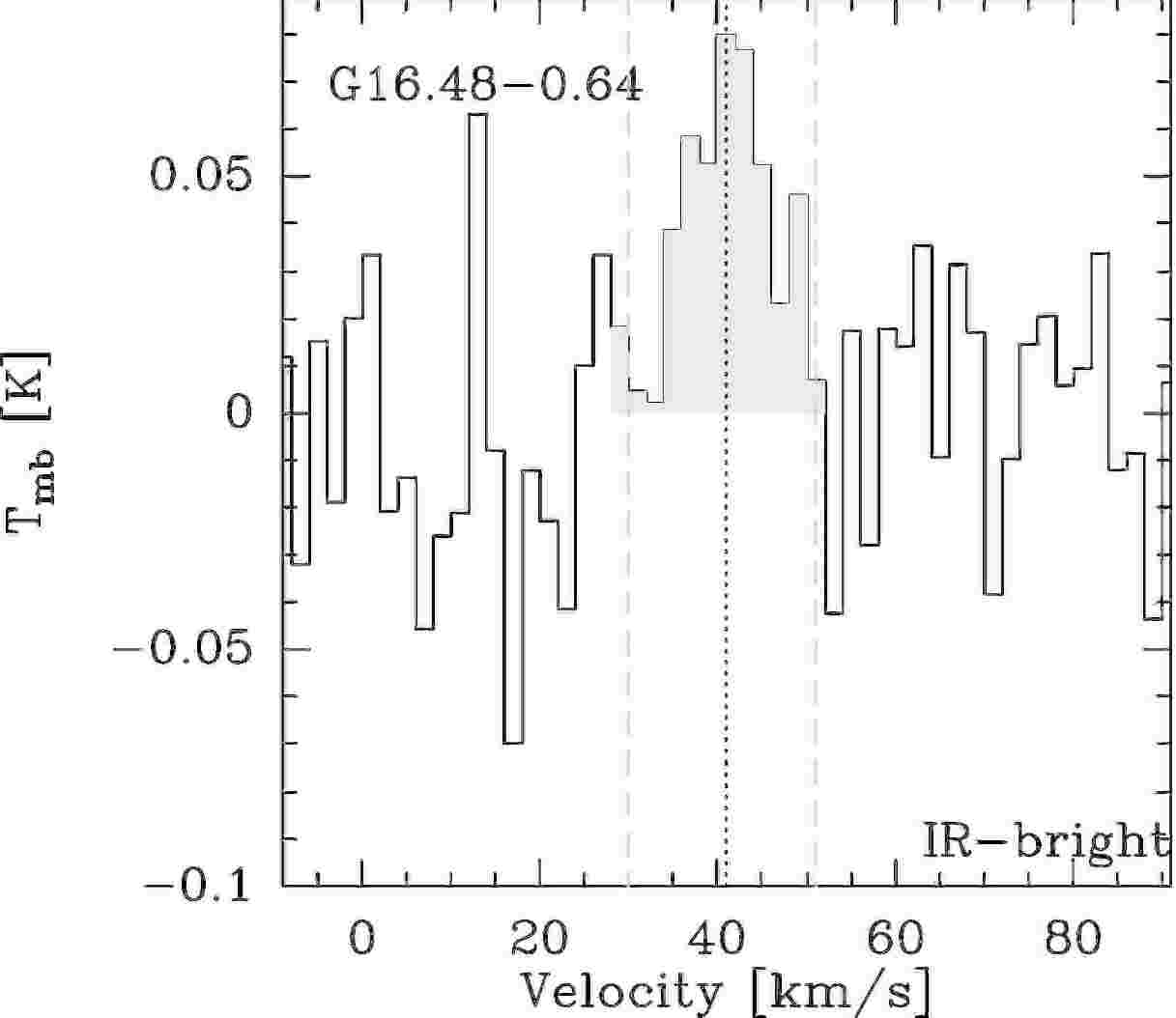} 
  \includegraphics[width=5.6cm,angle=0]{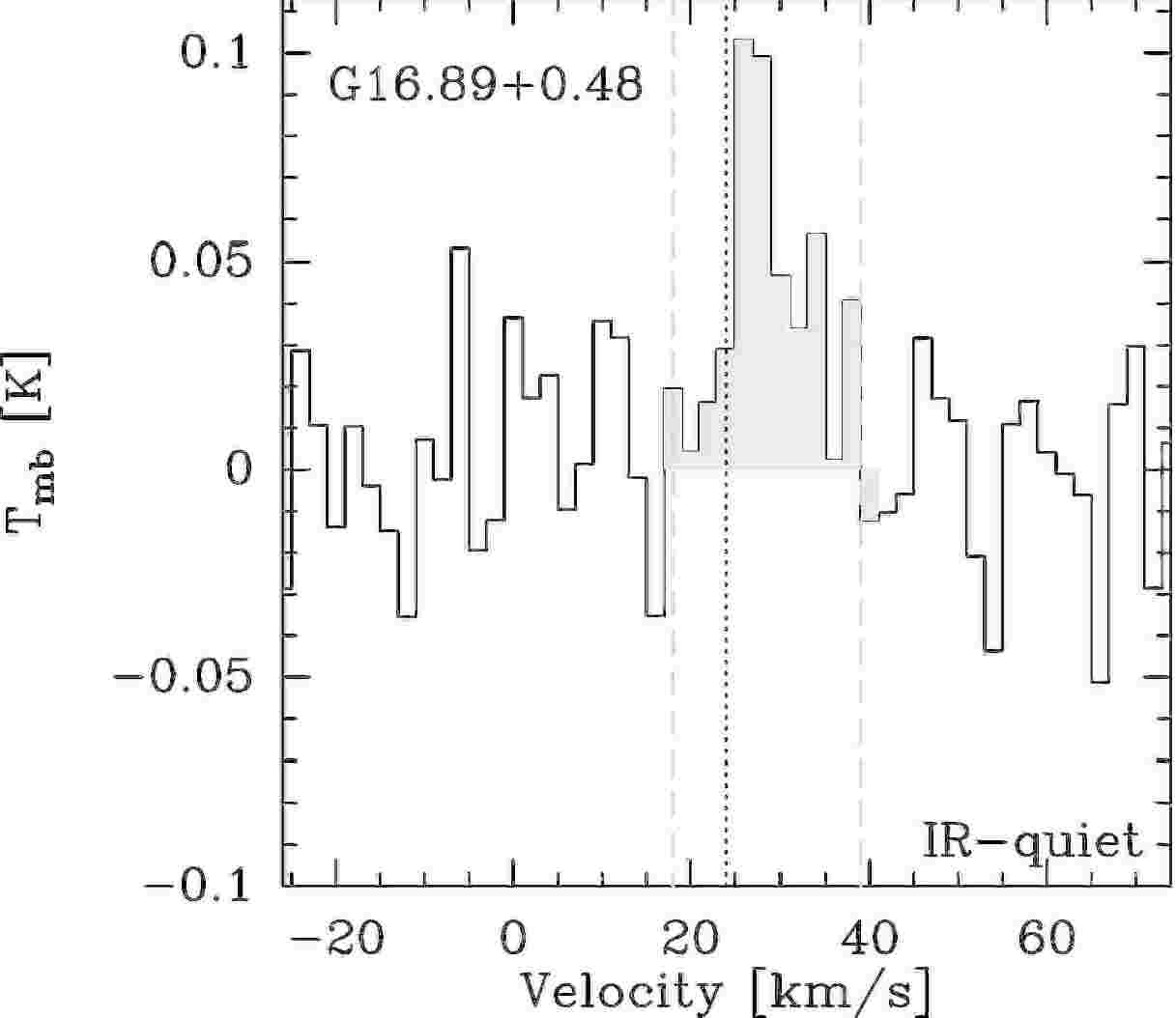} 
  \includegraphics[width=5.6cm,angle=0]{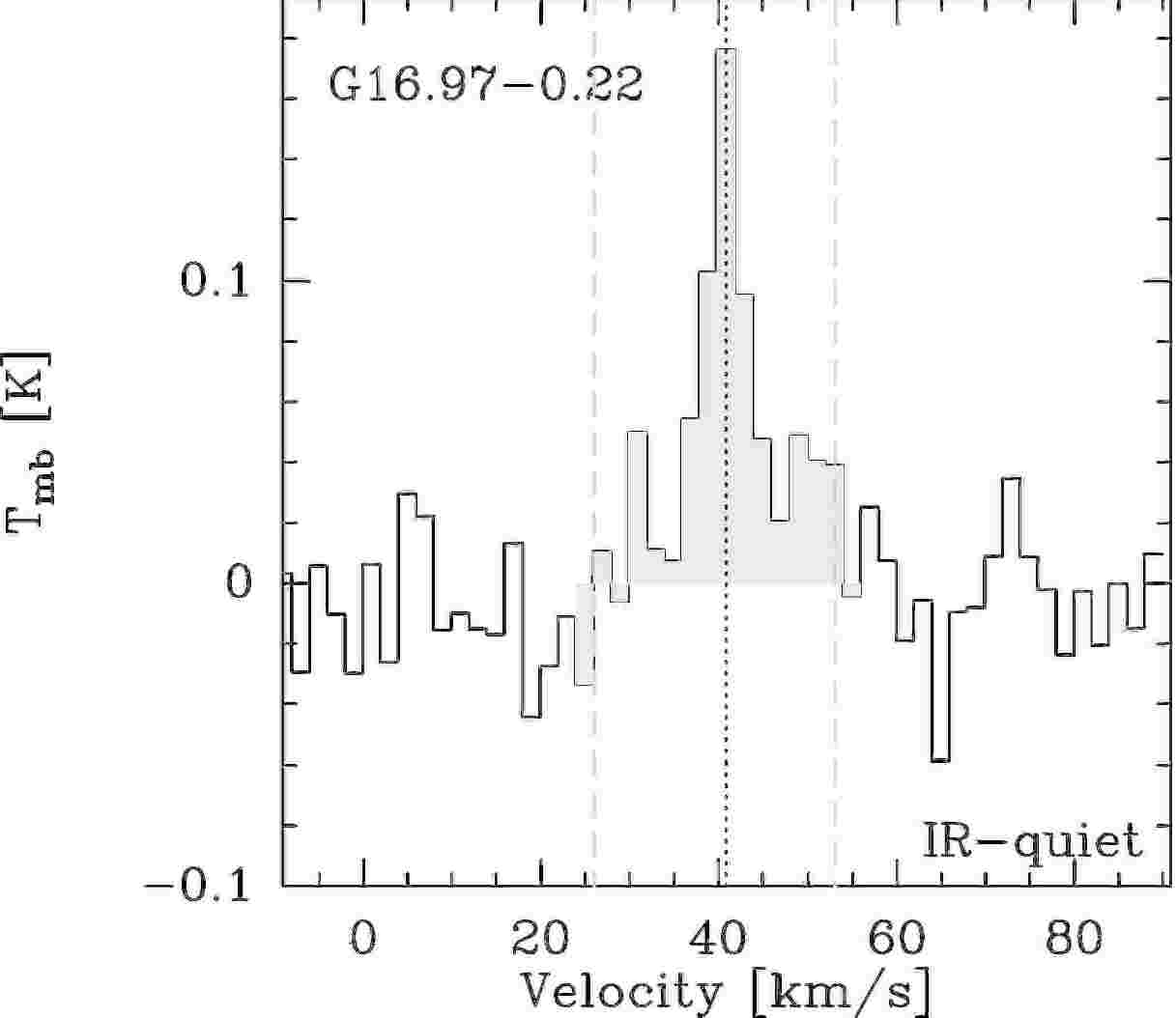} 
  \includegraphics[width=5.6cm,angle=0]{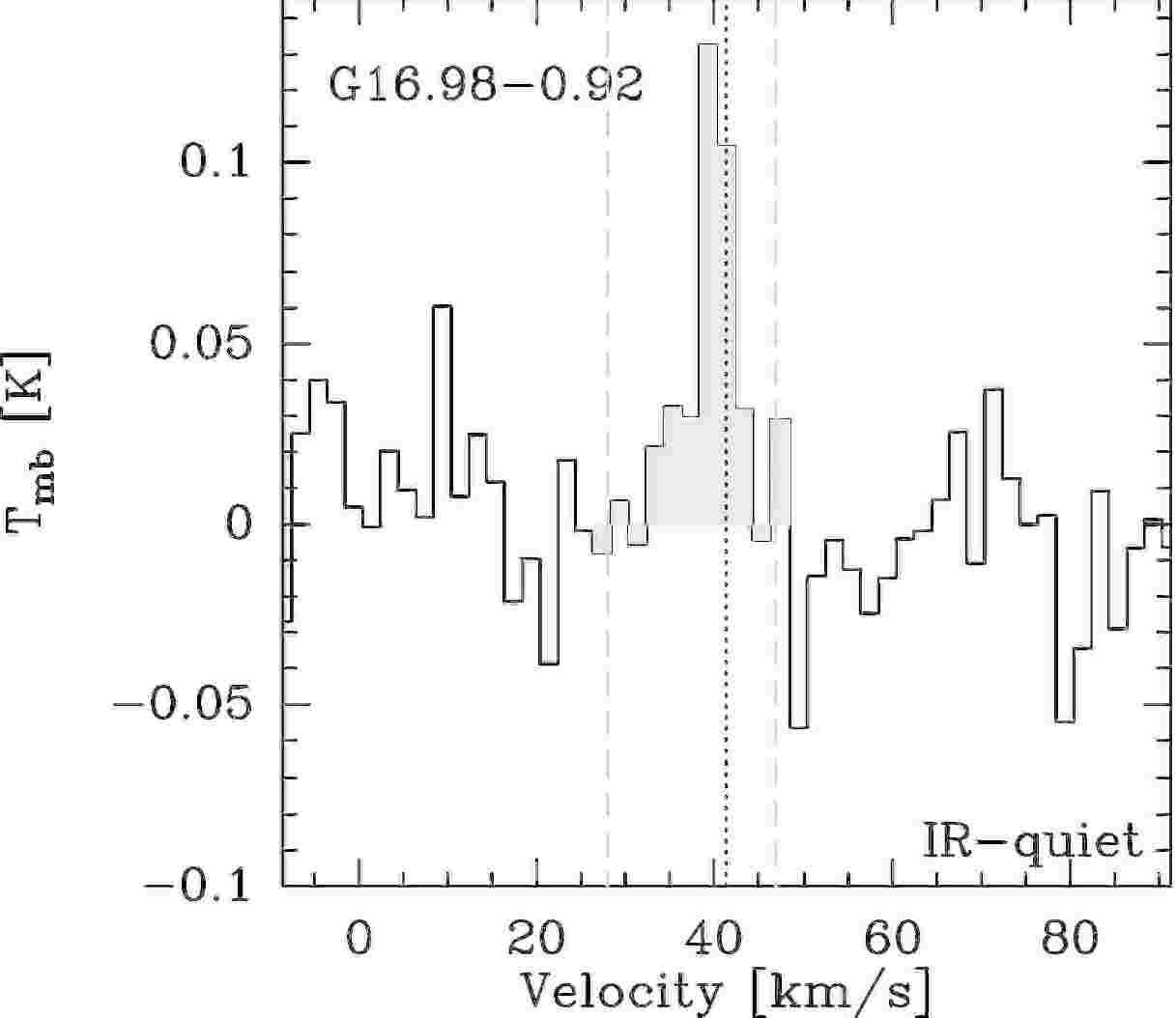} 
  \includegraphics[width=5.6cm,angle=0]{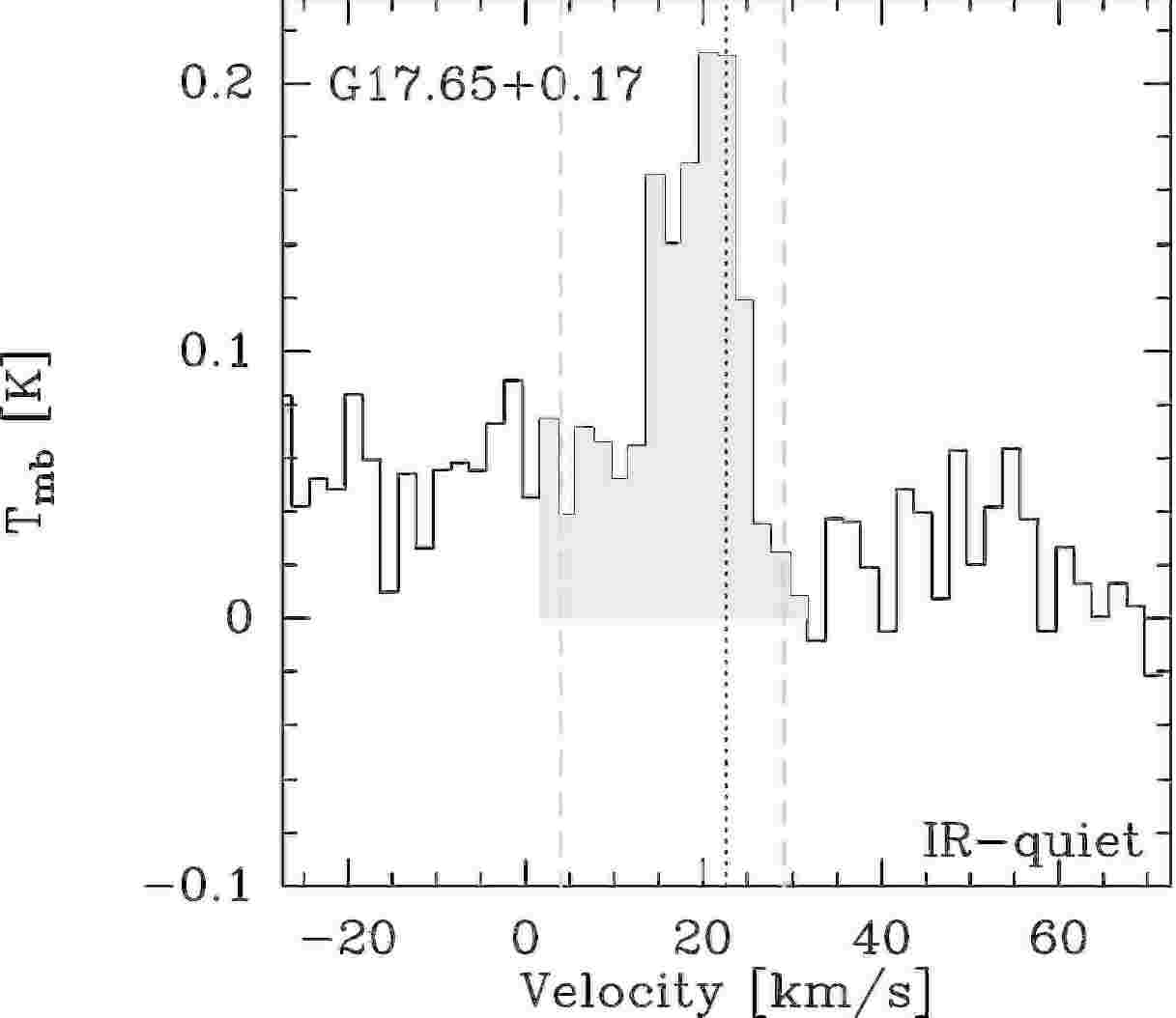} 
  \includegraphics[width=6.0cm,angle=0]{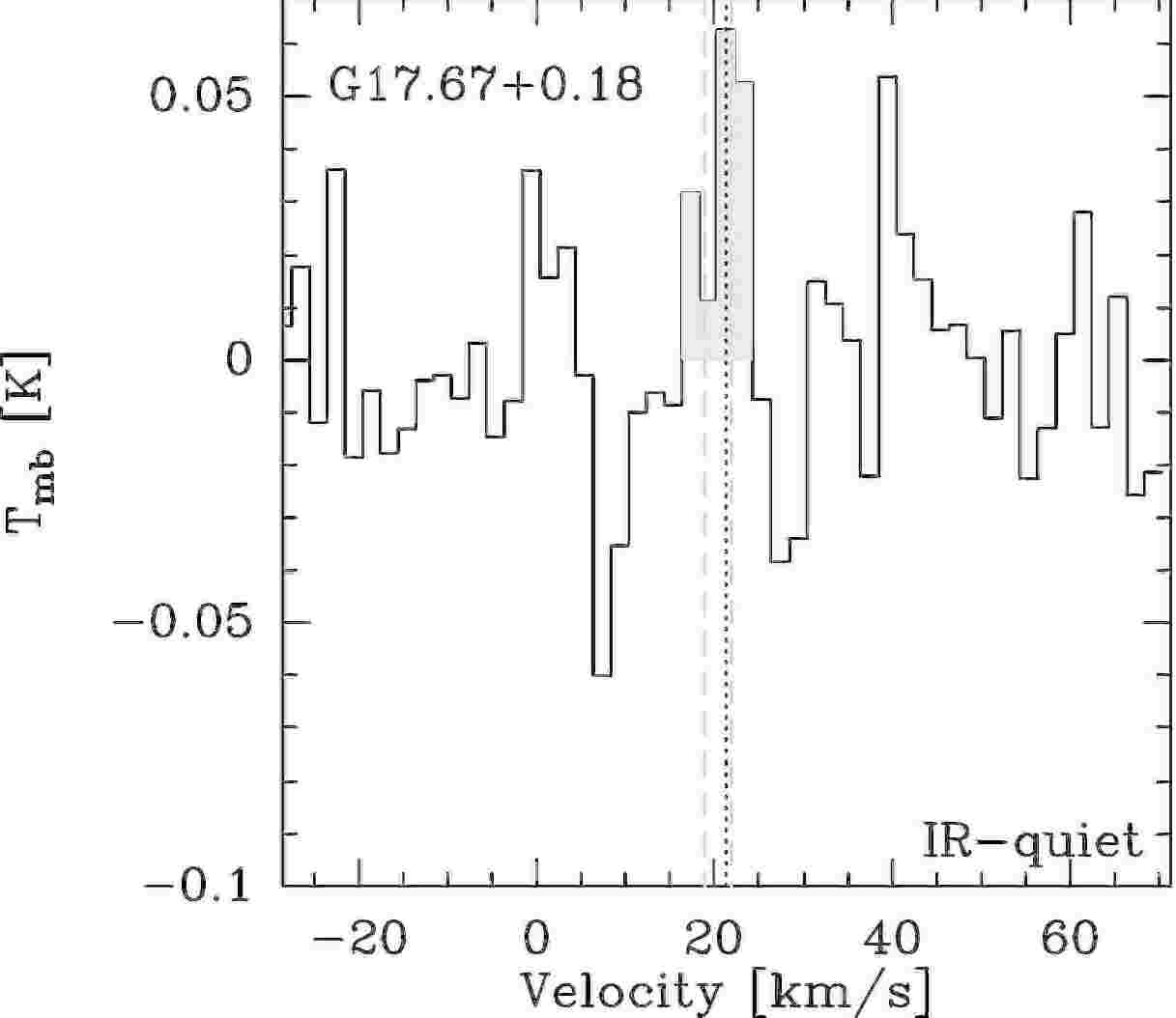} 
  \includegraphics[width=5.6cm,angle=0]{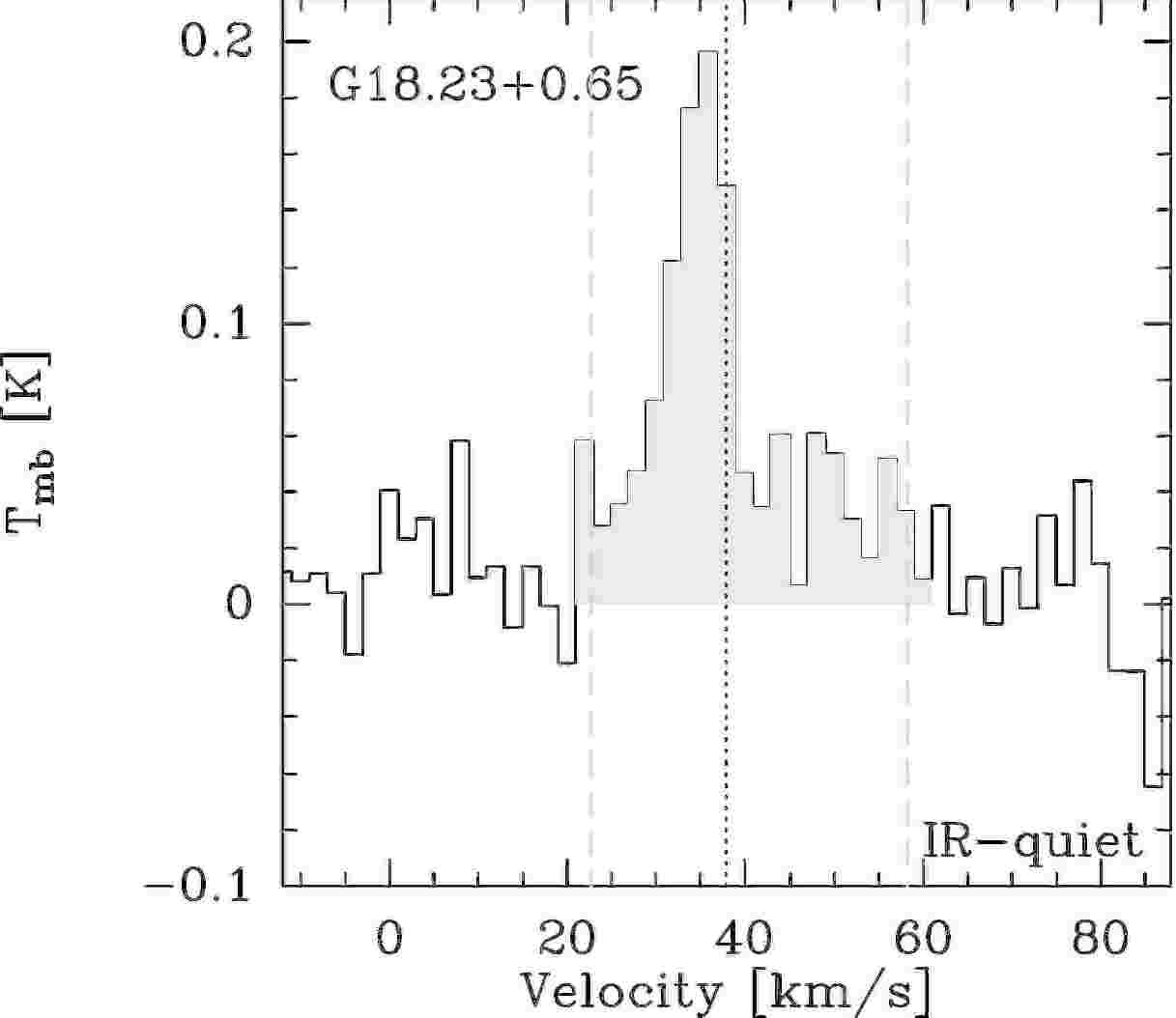} 
  \includegraphics[width=5.6cm,angle=0]{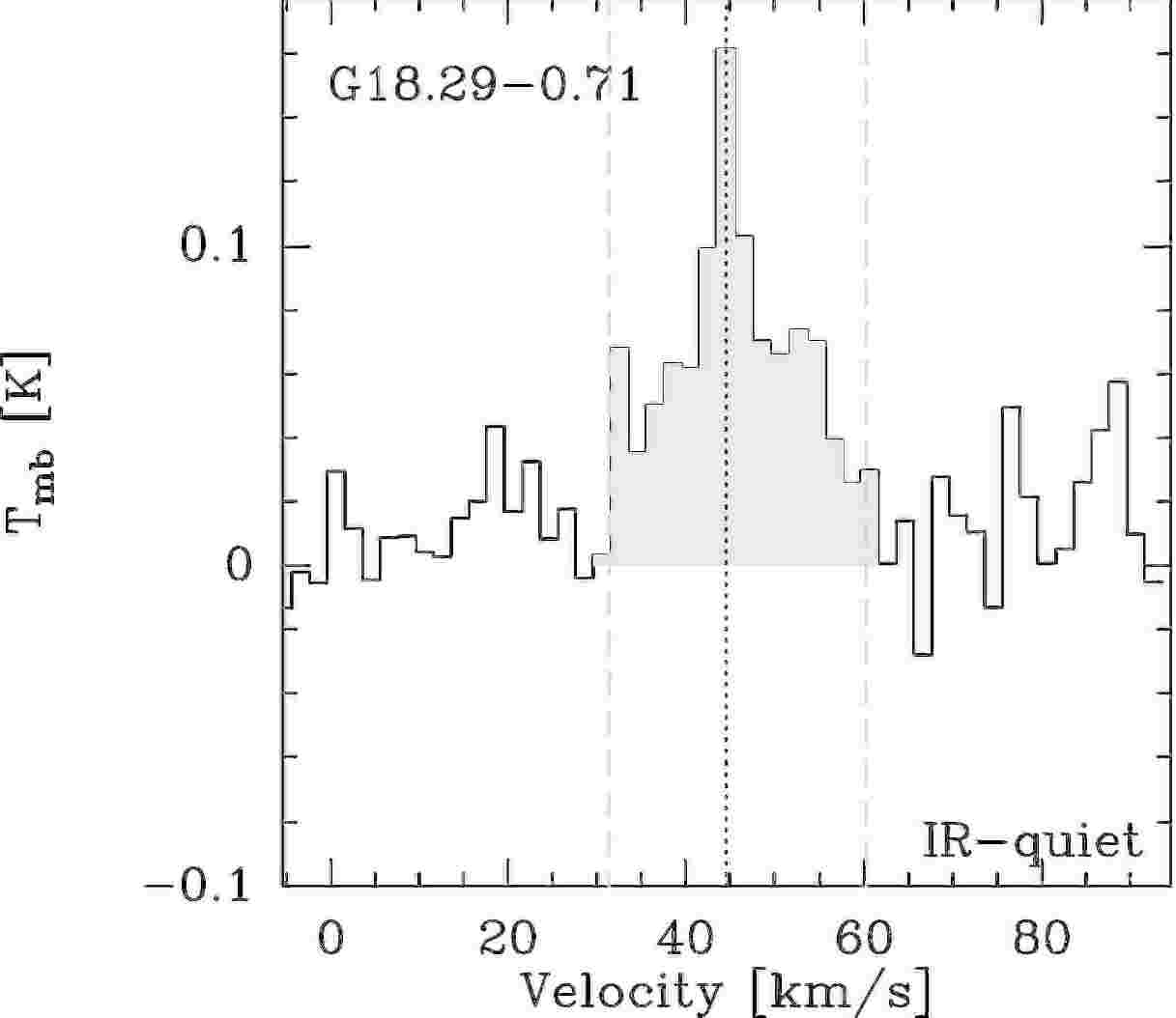} 
  \includegraphics[width=5.6cm,angle=0]{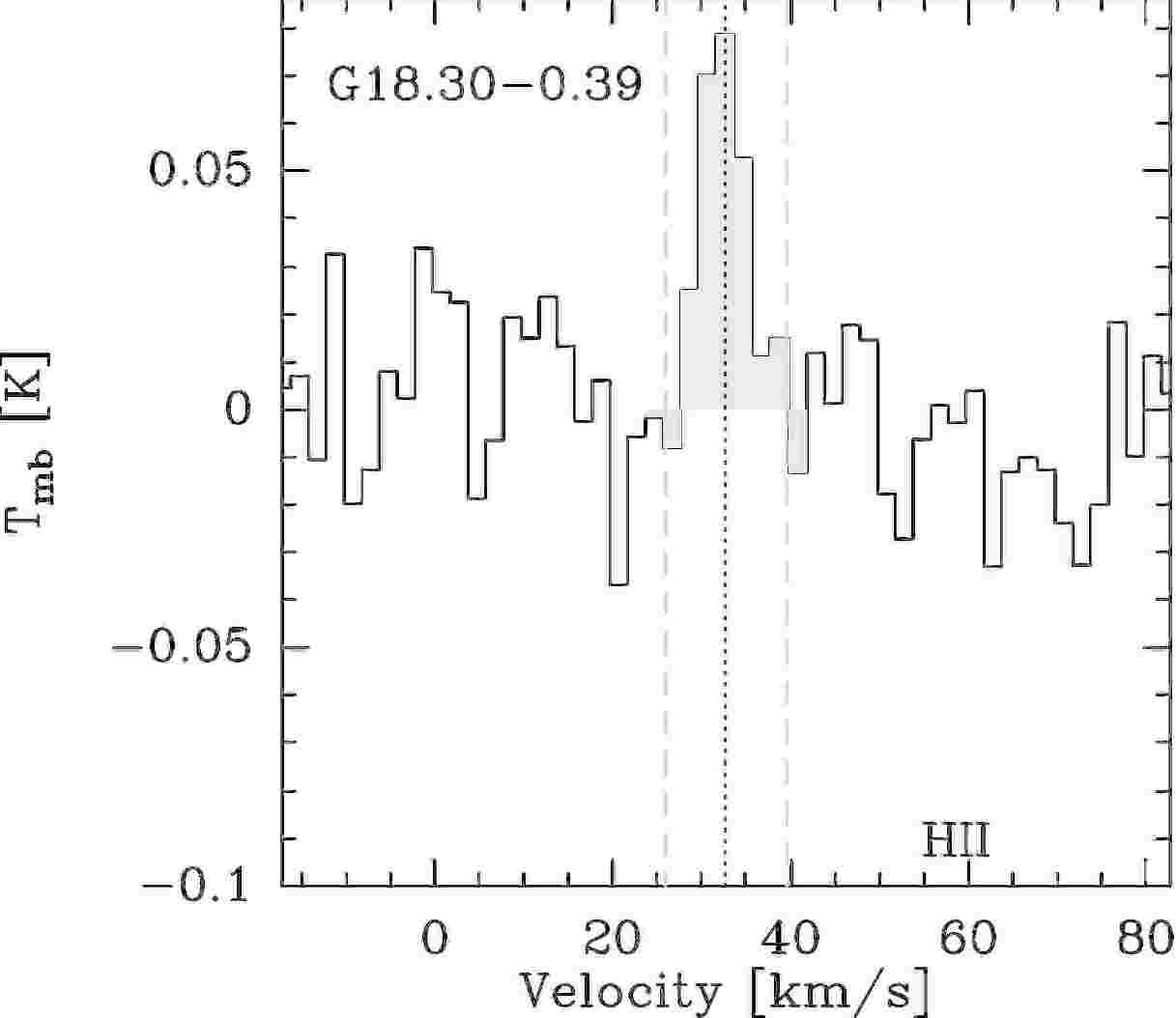} 
  \includegraphics[width=5.6cm,angle=0]{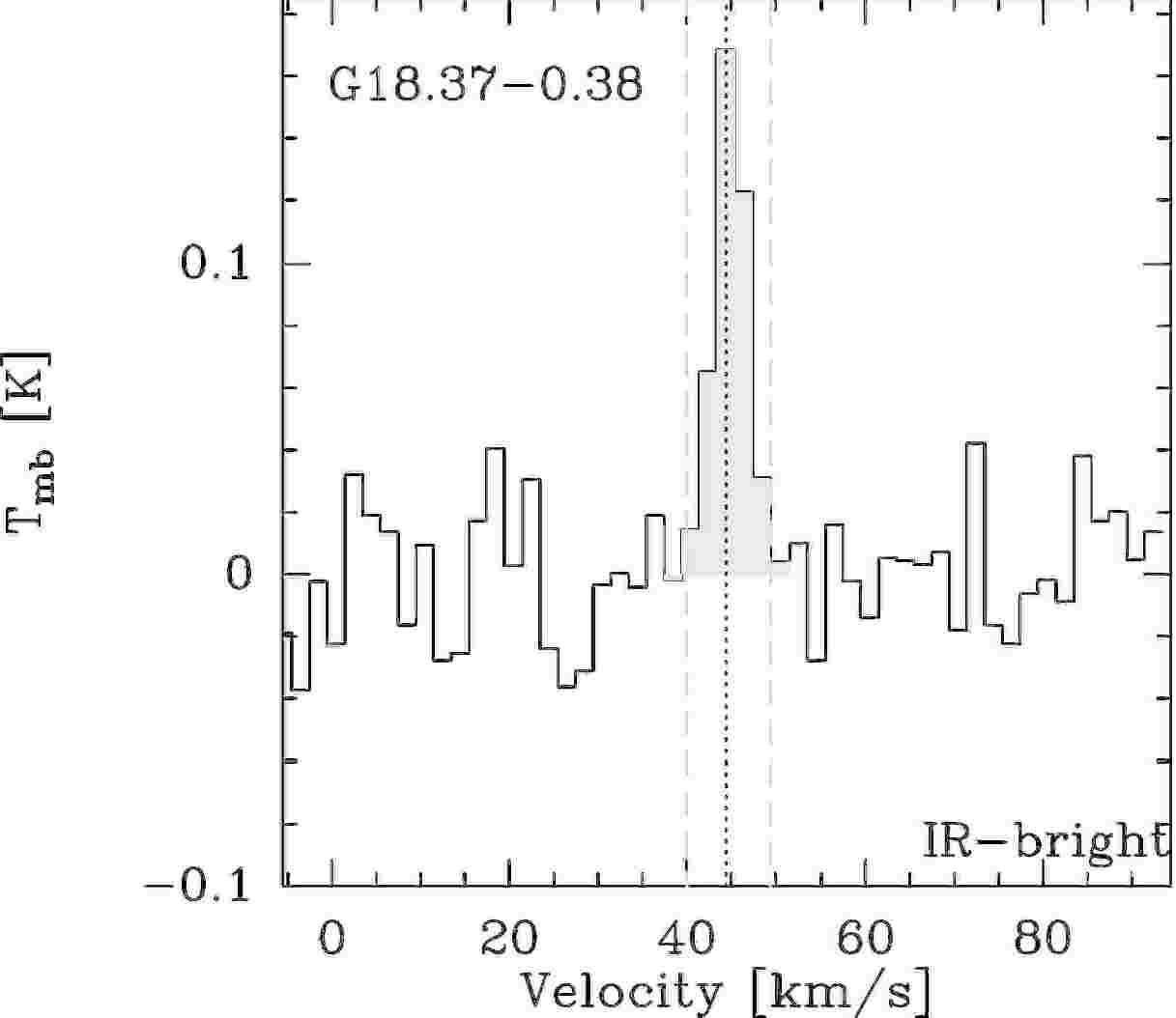} 
  \includegraphics[width=5.6cm,angle=0]{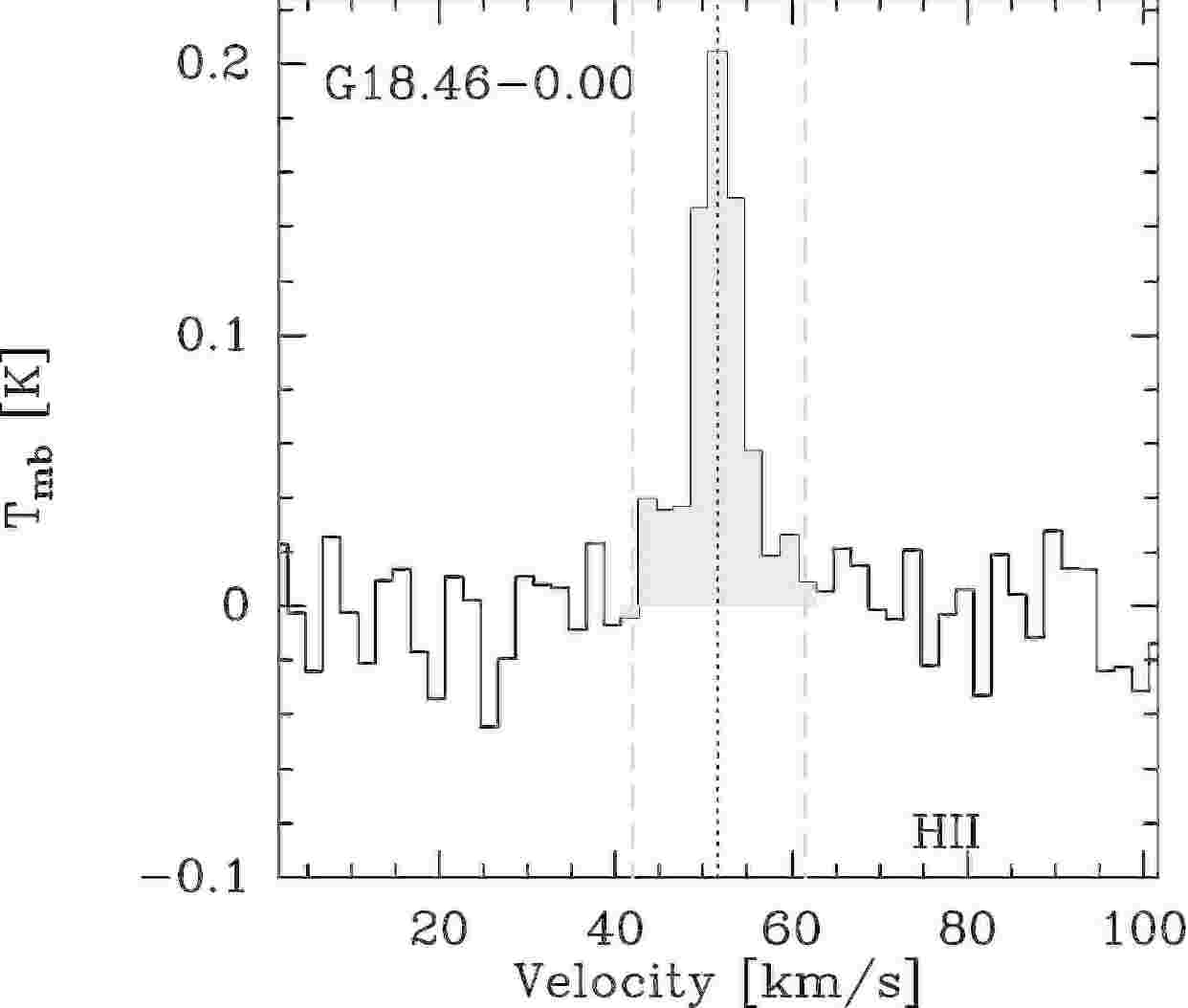} 
  \includegraphics[width=5.6cm,angle=0]{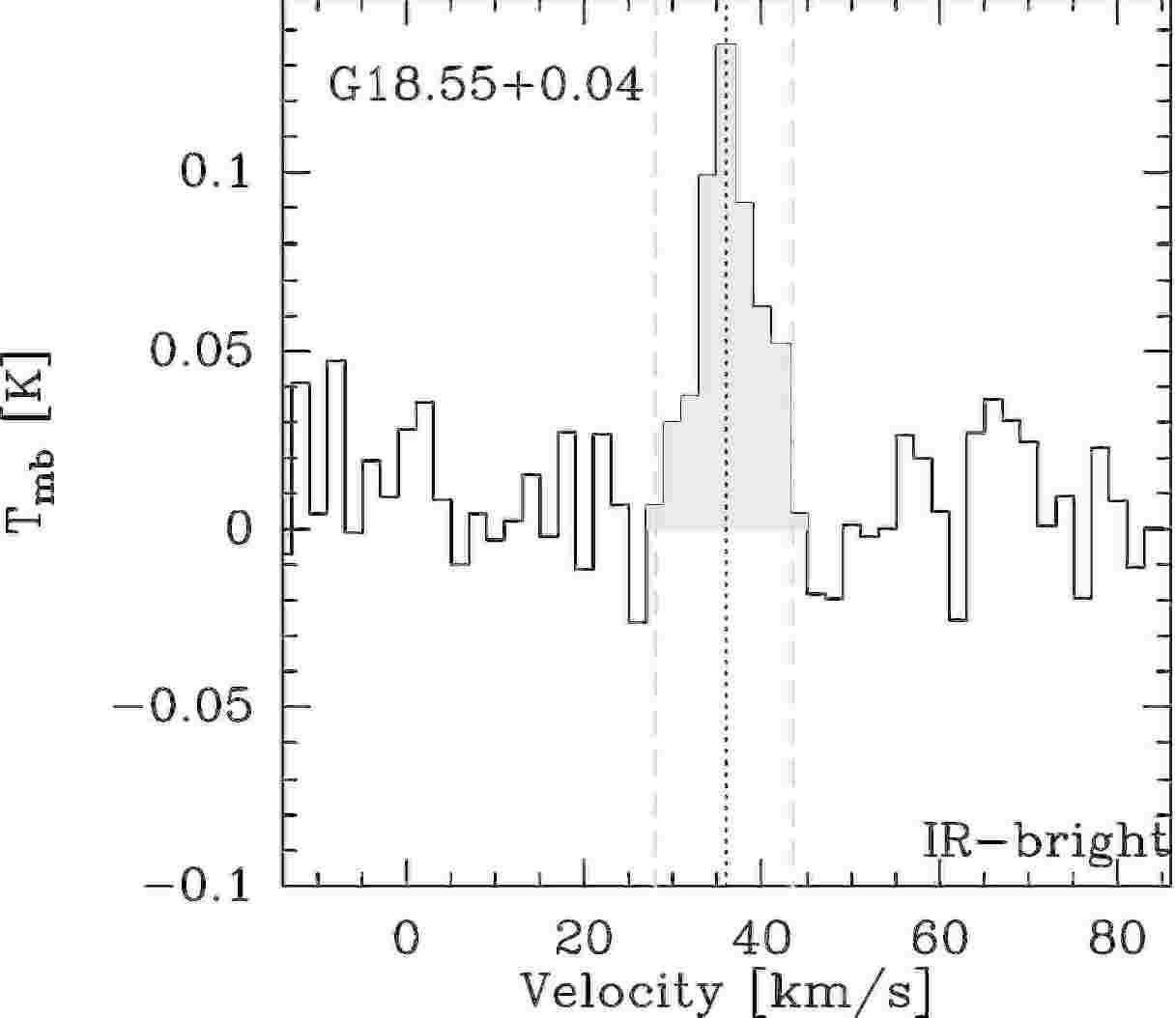} 
 \caption{Continued.}
\end{figure}
\end{landscape}

\begin{landscape}
\begin{figure}
\centering
\ContinuedFloat
 \includegraphics[width=5.6cm,angle=0]{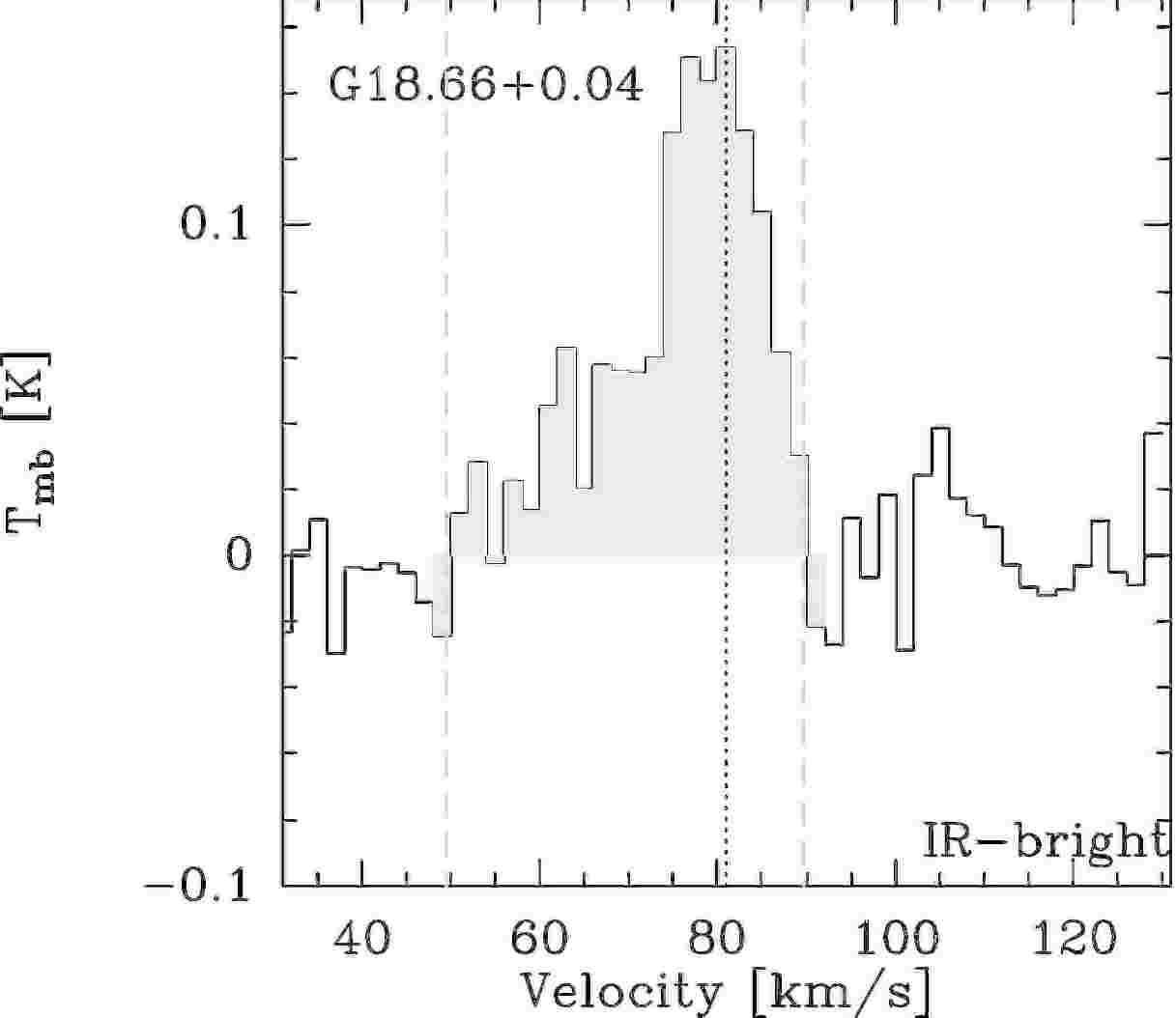} 
 \includegraphics[width=5.6cm,angle=0]{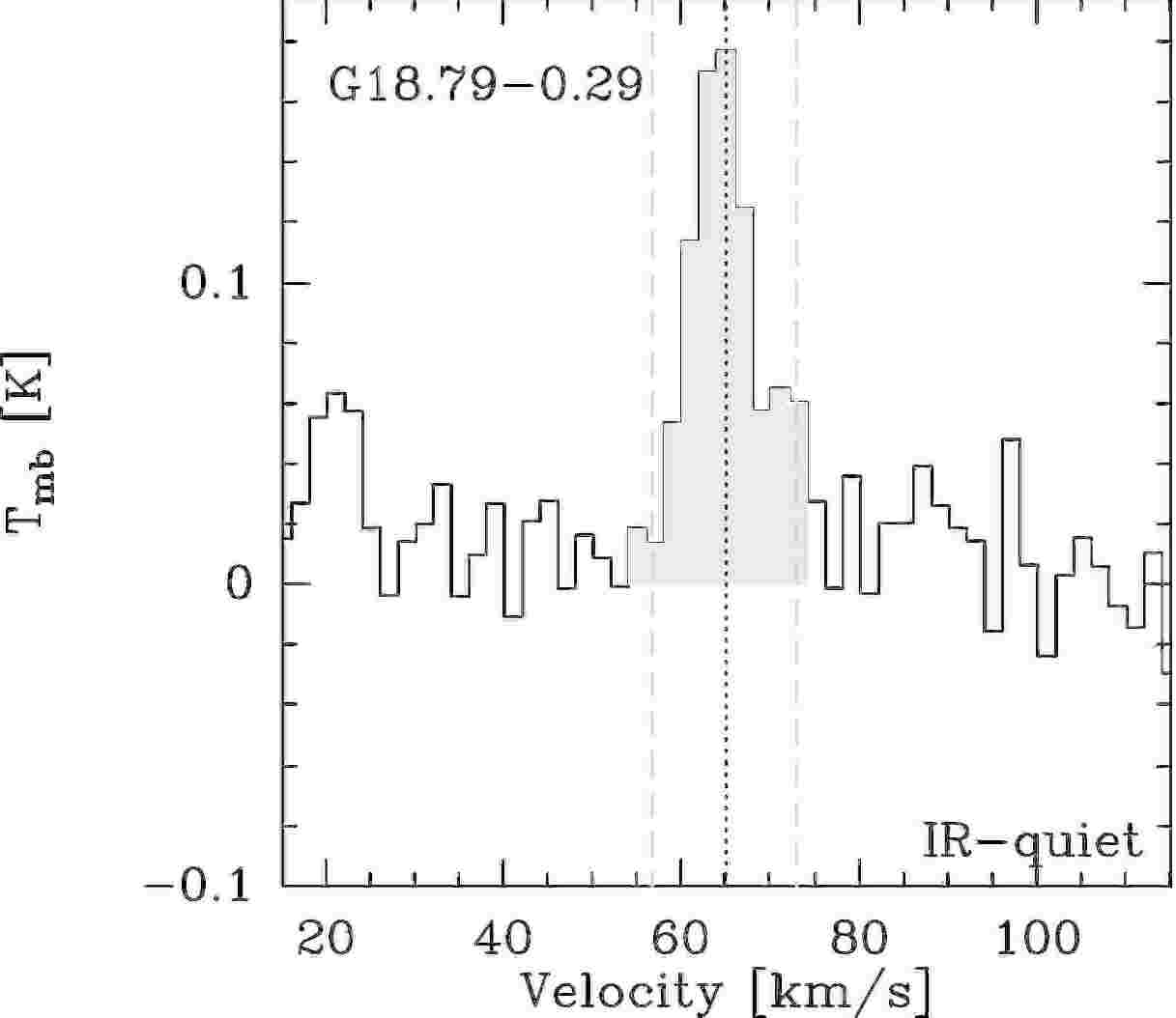} 
  \includegraphics[width=5.6cm,angle=0]{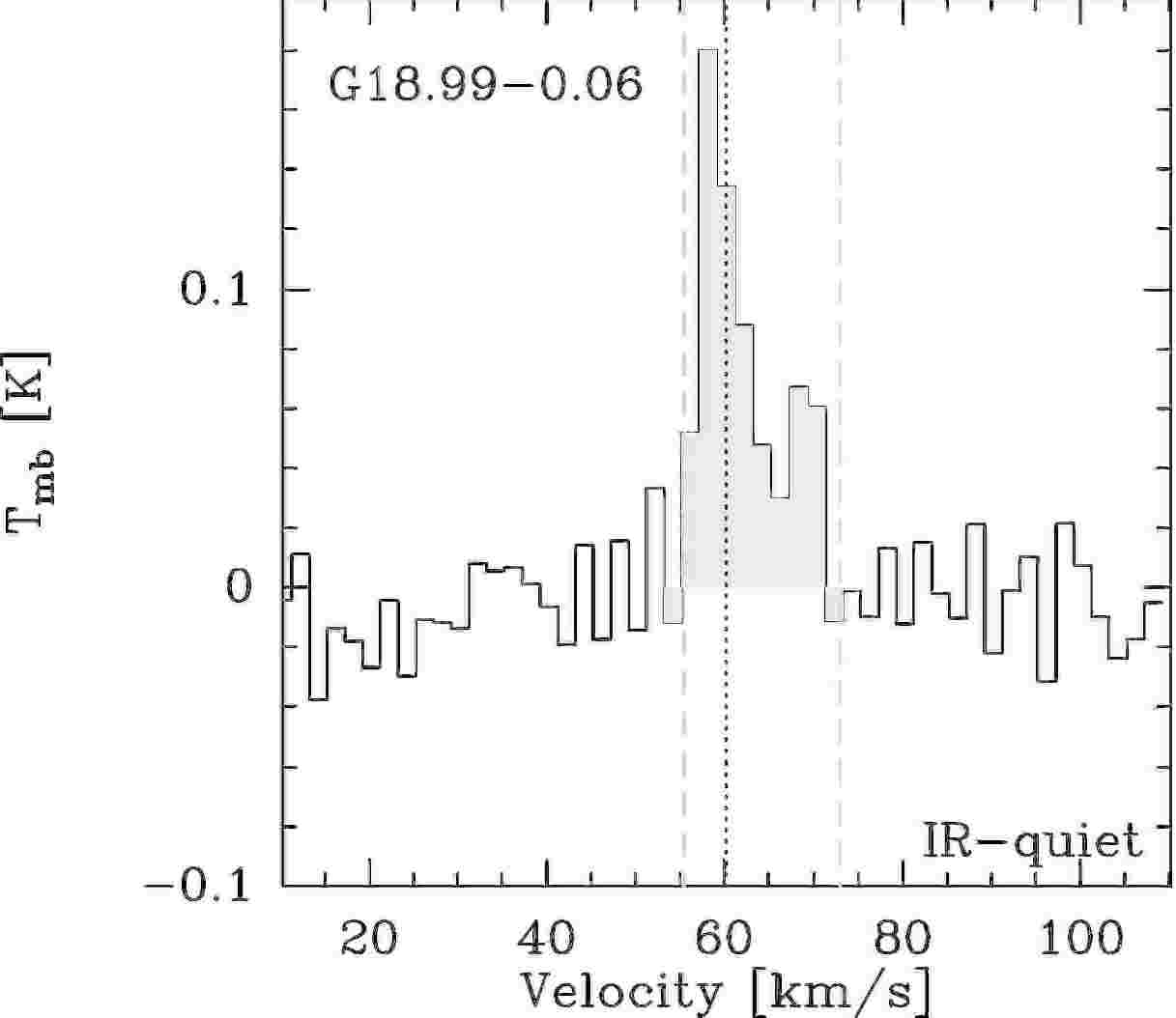} 
  \includegraphics[width=5.6cm,angle=0]{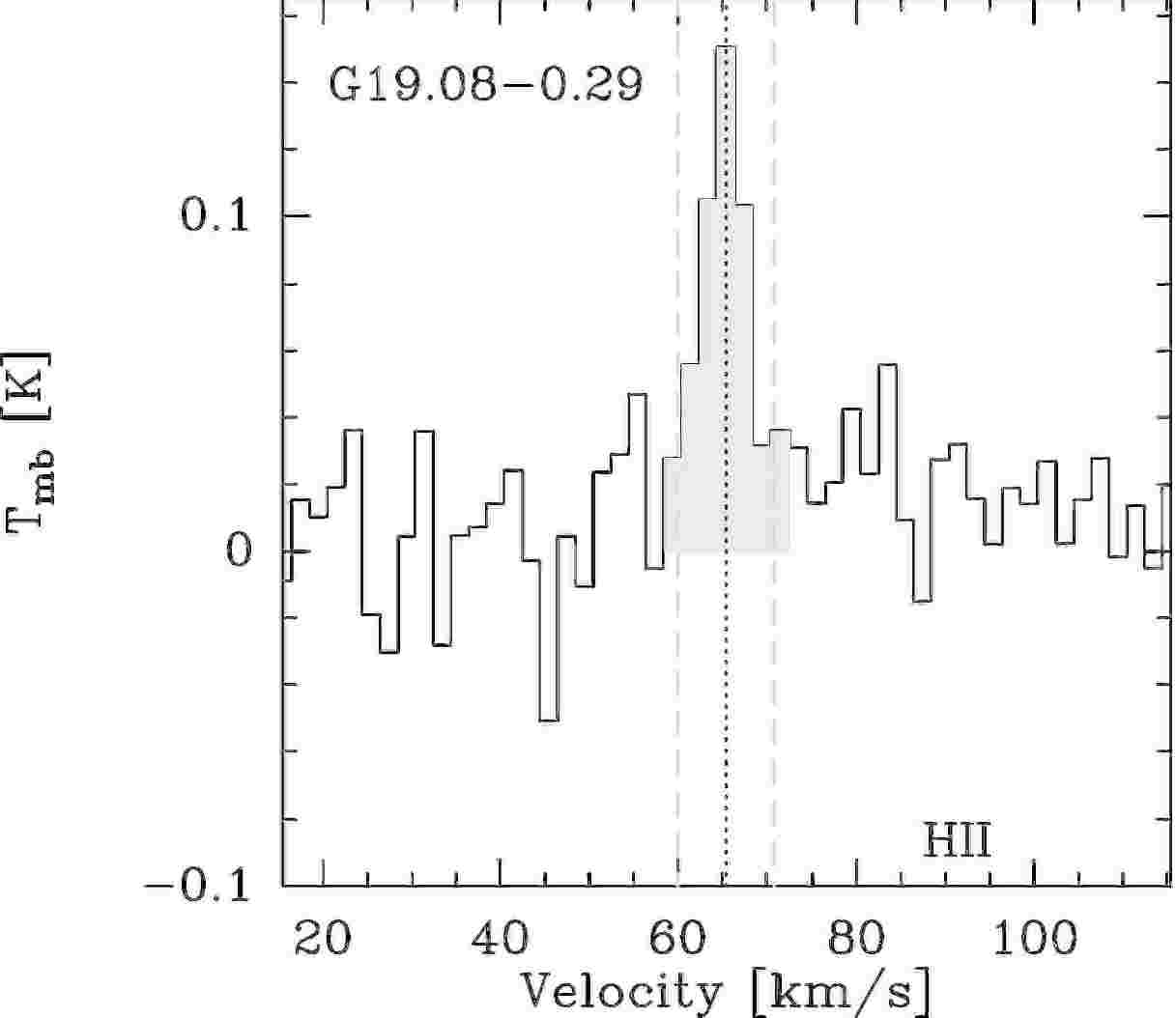} 
  \includegraphics[width=5.6cm,angle=0]{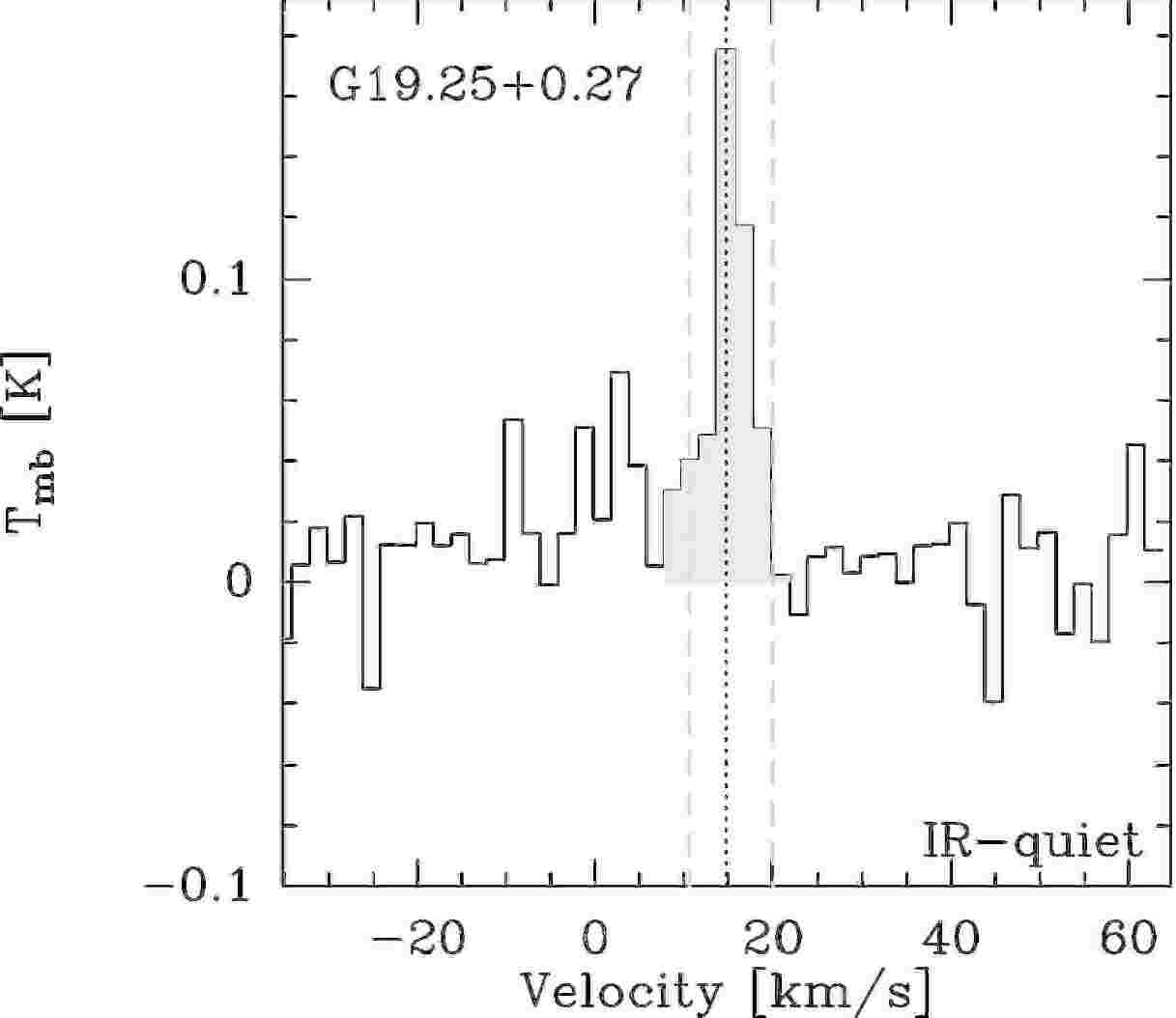} 
  \includegraphics[width=5.6cm,angle=0]{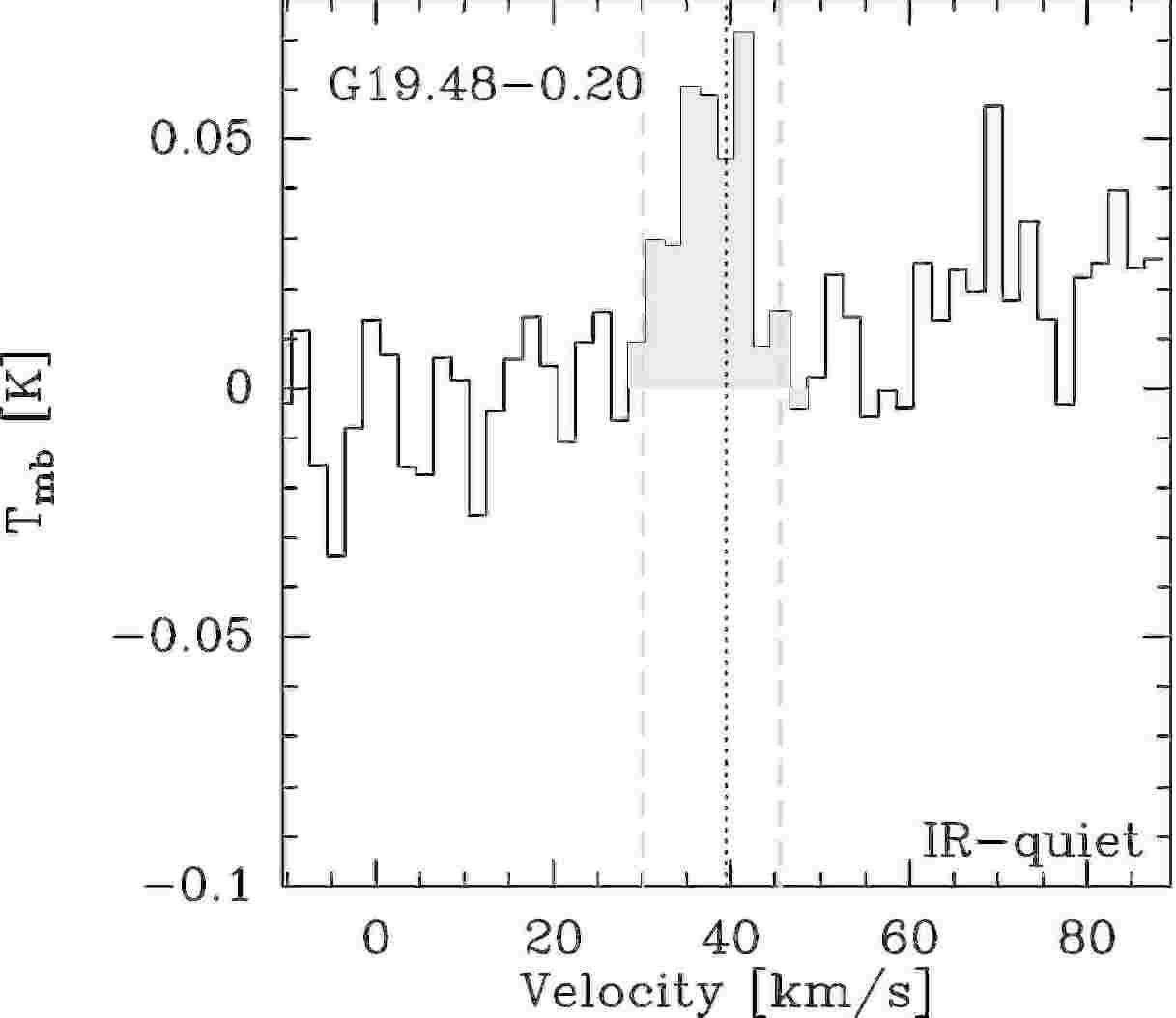} 
  \includegraphics[width=5.6cm,angle=0]{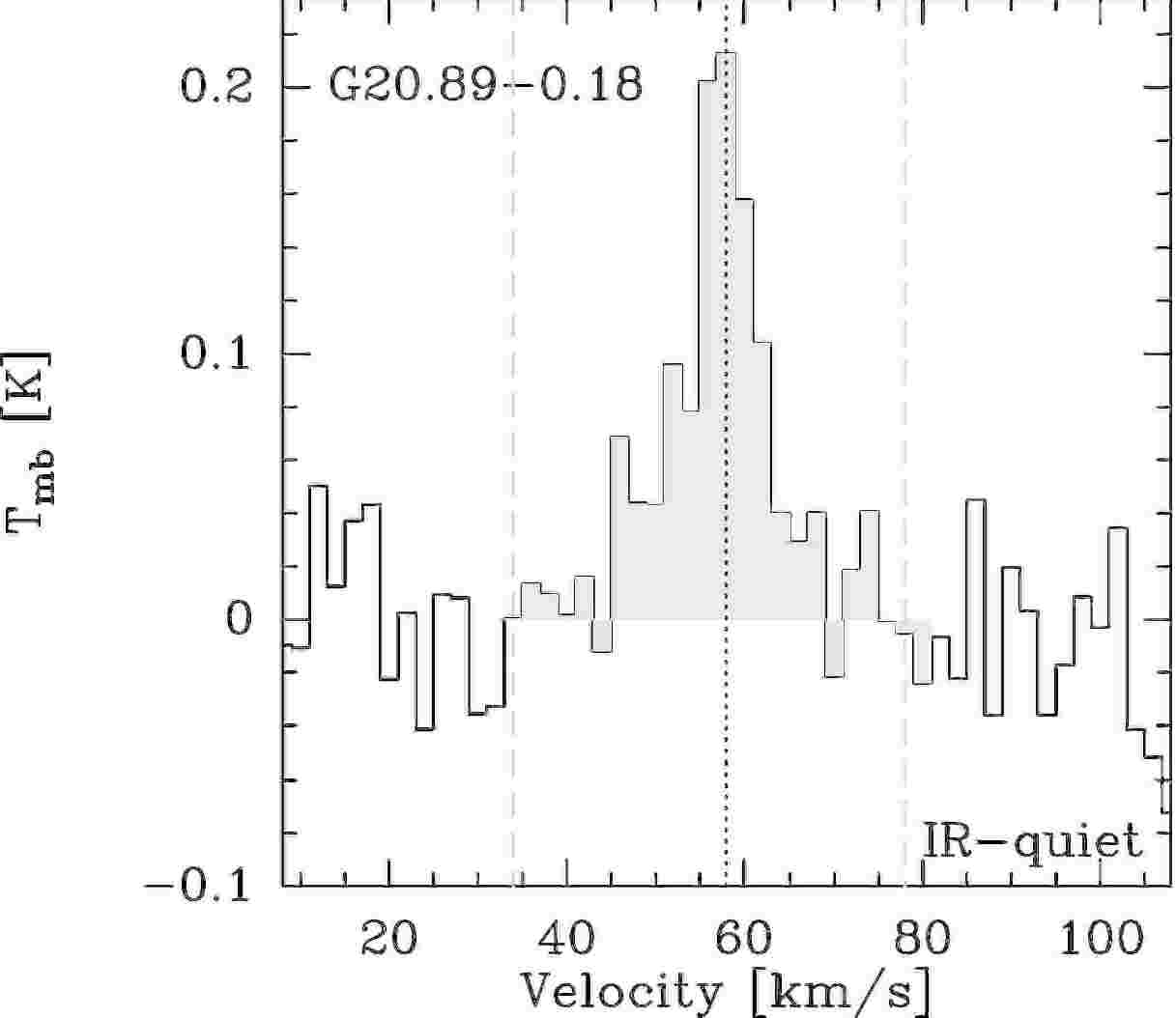} 
  \includegraphics[width=5.6cm,angle=0]{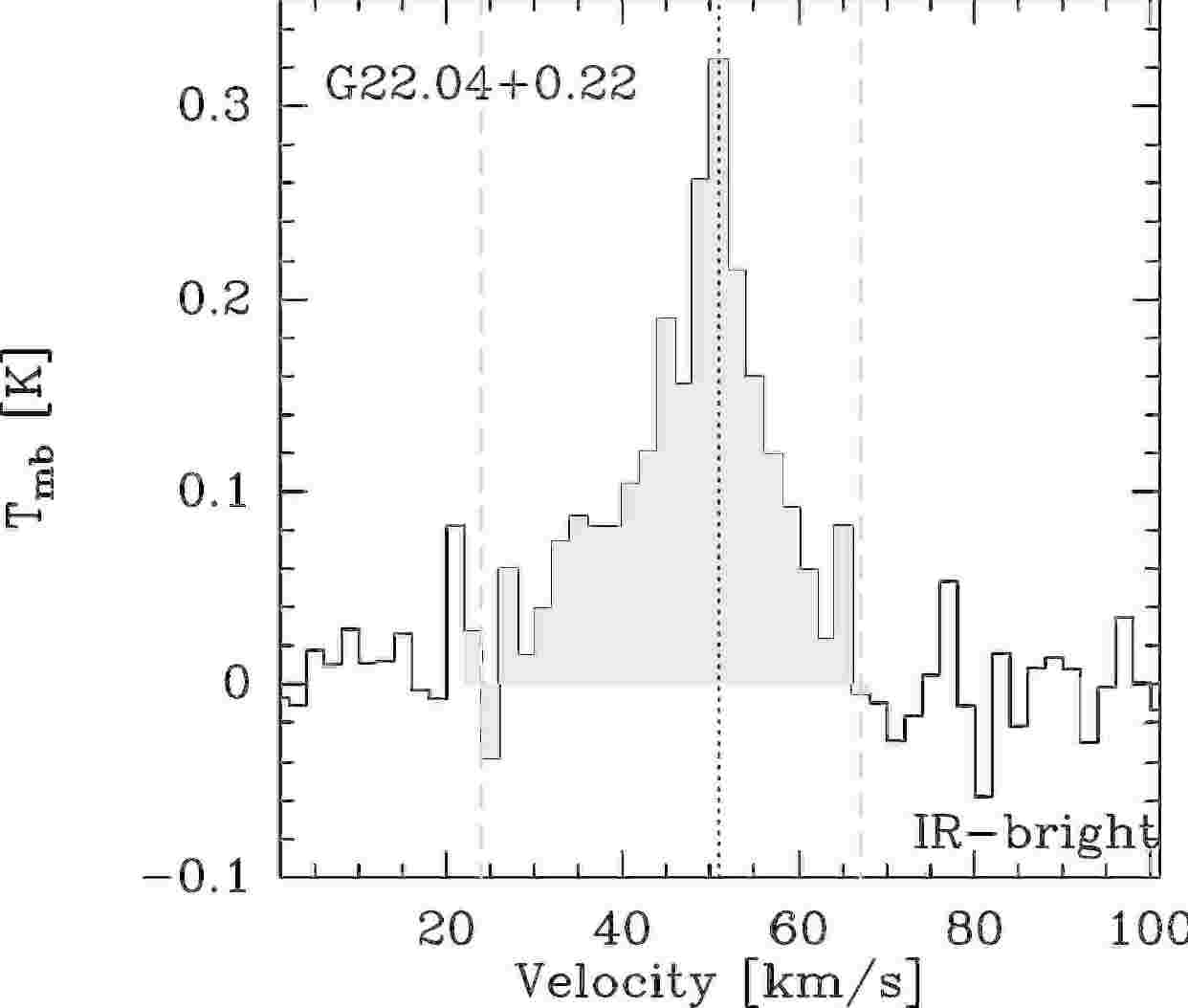} 
  \includegraphics[width=5.6cm,angle=0]{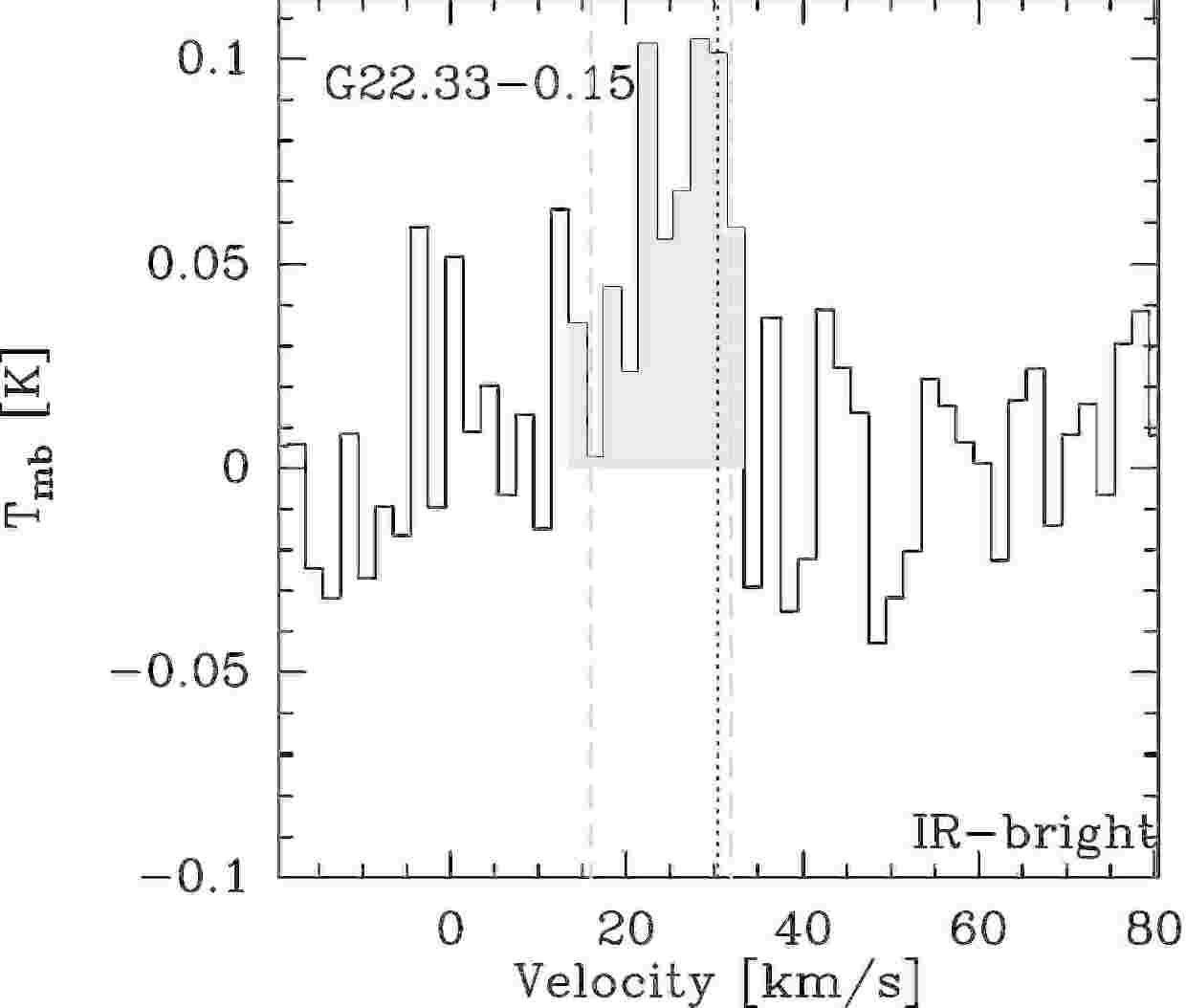} 
  \includegraphics[width=5.6cm,angle=0]{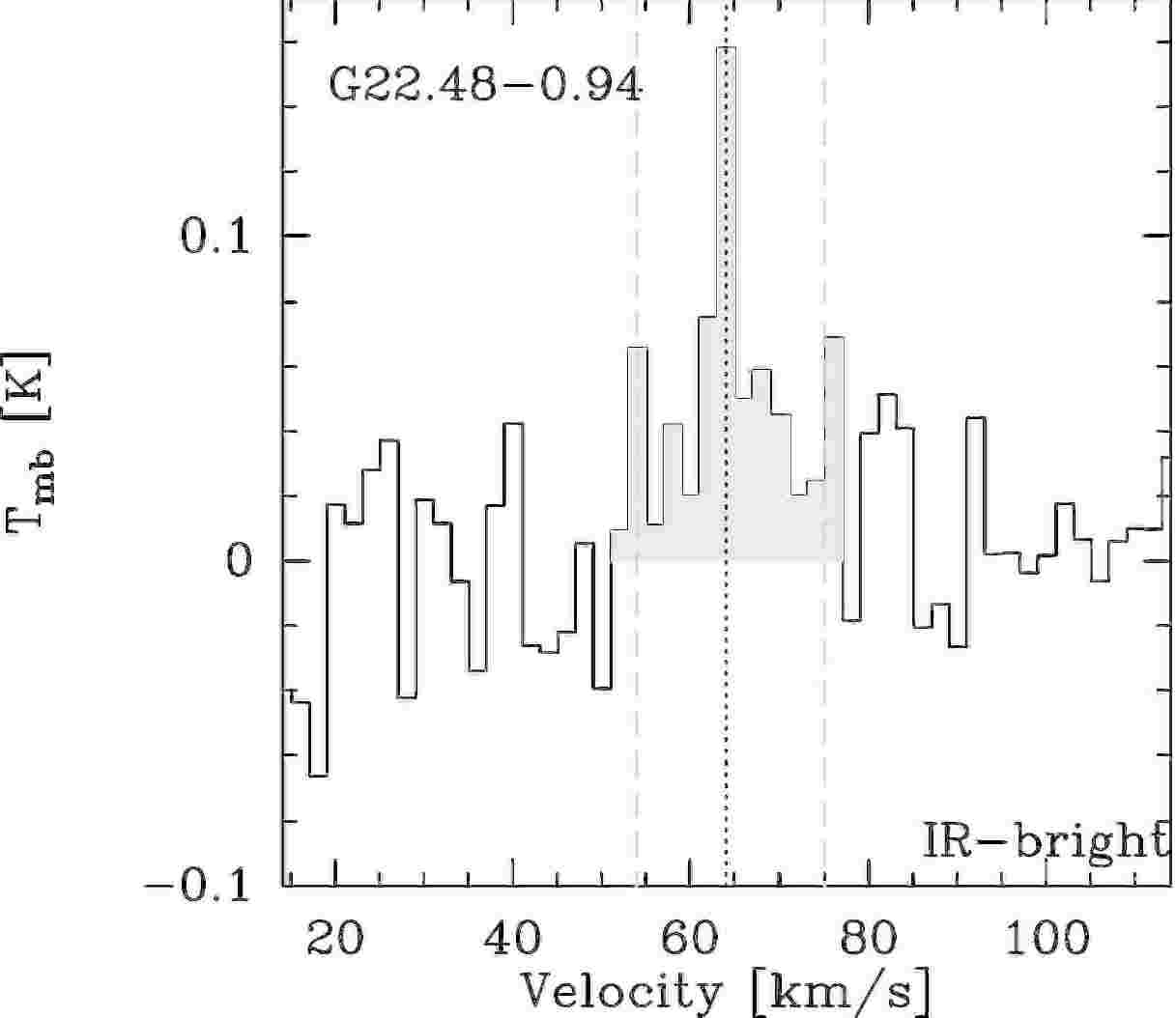} 
  \includegraphics[width=5.6cm,angle=0]{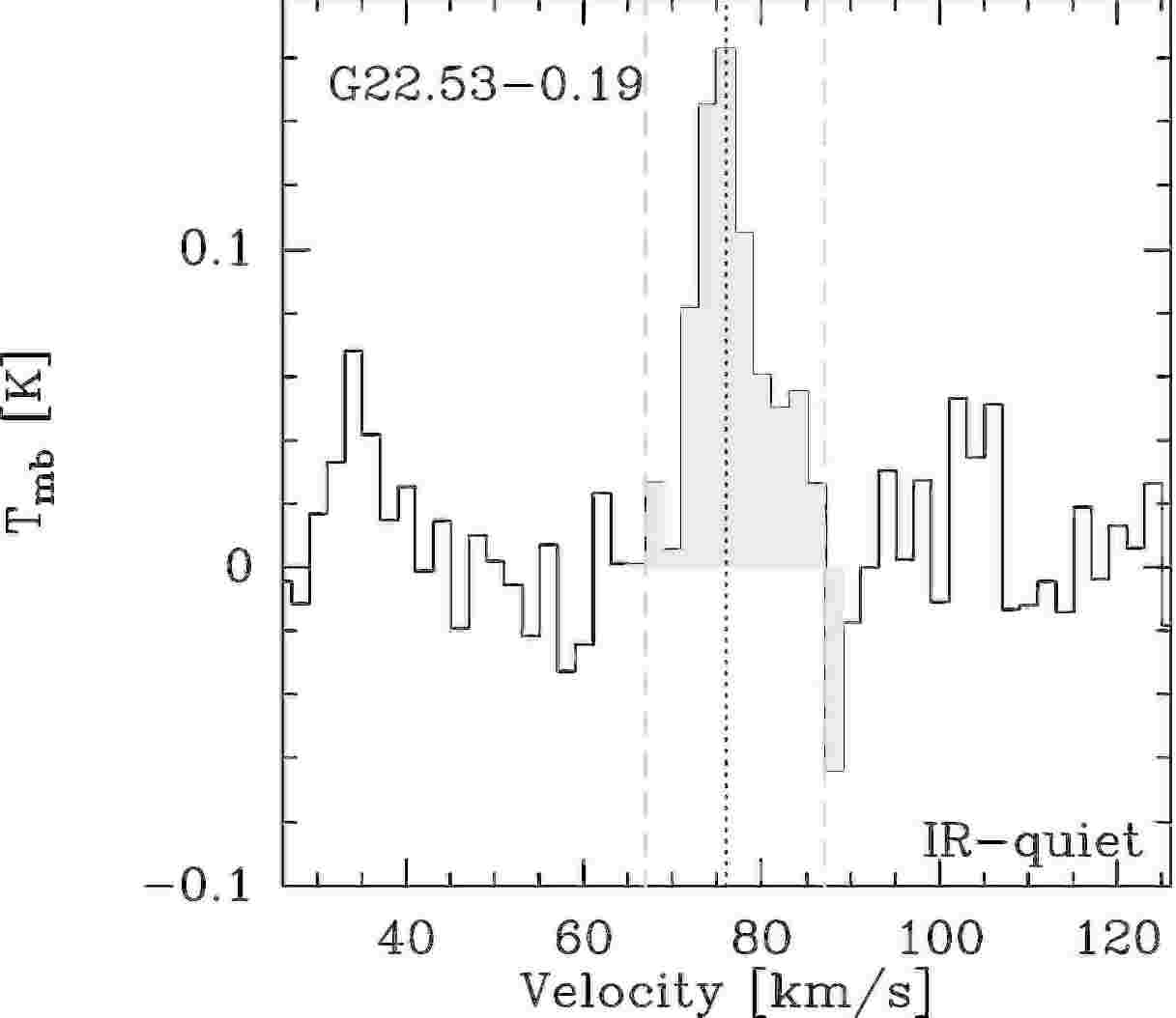} 
  \includegraphics[width=5.6cm,angle=0]{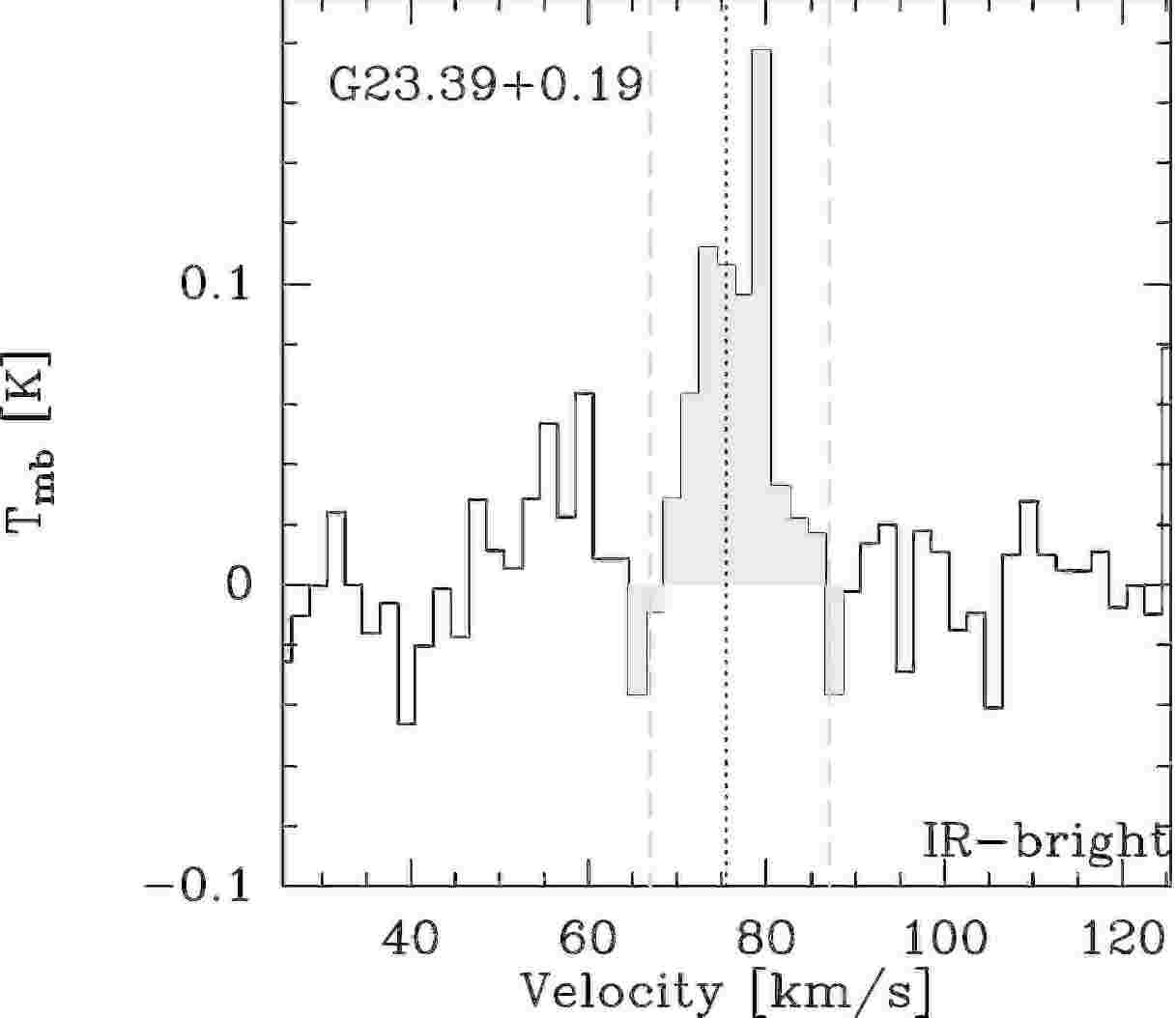} 
 \caption{Continued.}
\end{figure}
\end{landscape}
  
\begin{landscape}
\begin{figure}
\centering
\ContinuedFloat
  \includegraphics[width=5.6cm,angle=0]{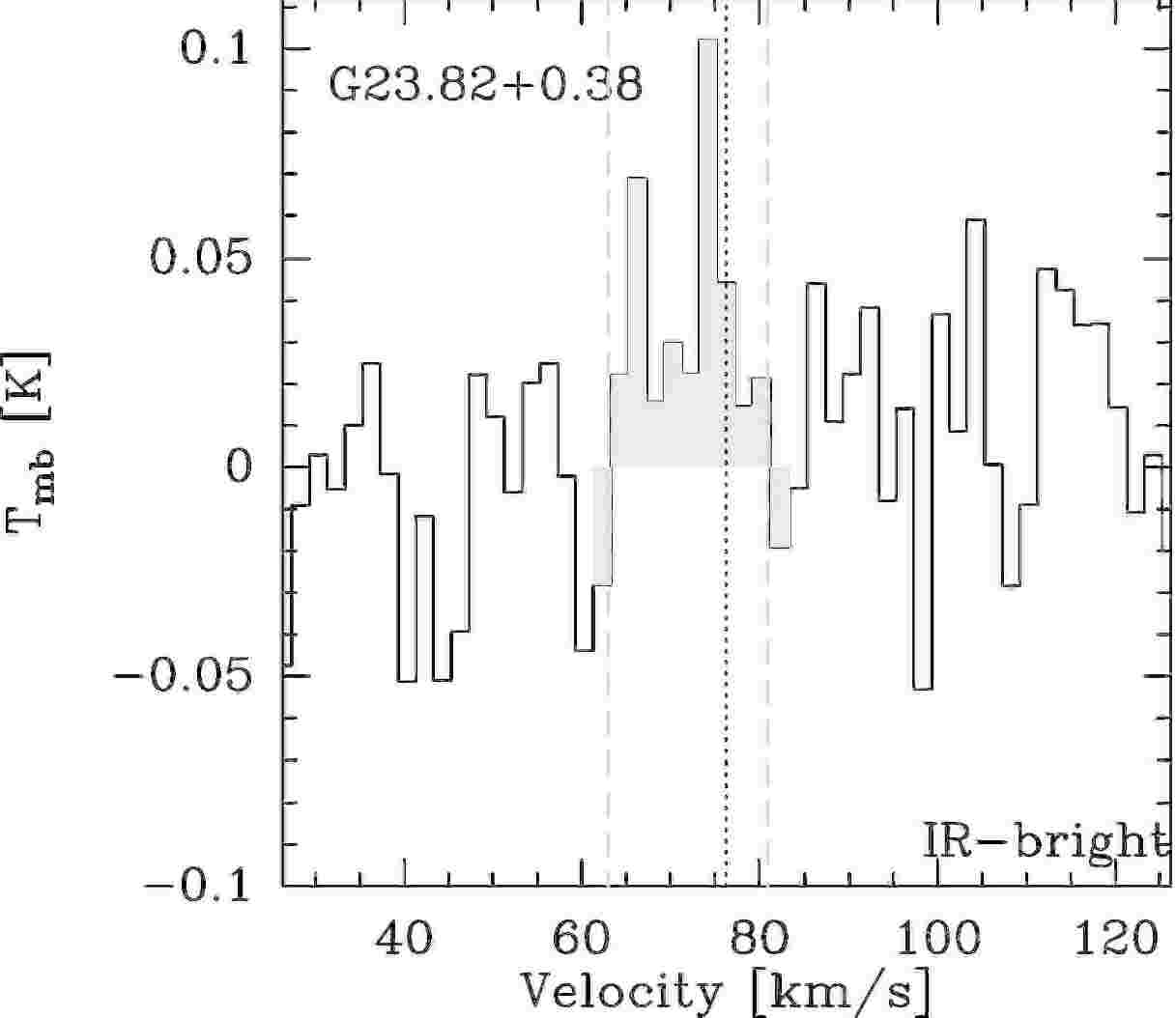} 
  \includegraphics[width=5.6cm,angle=0]{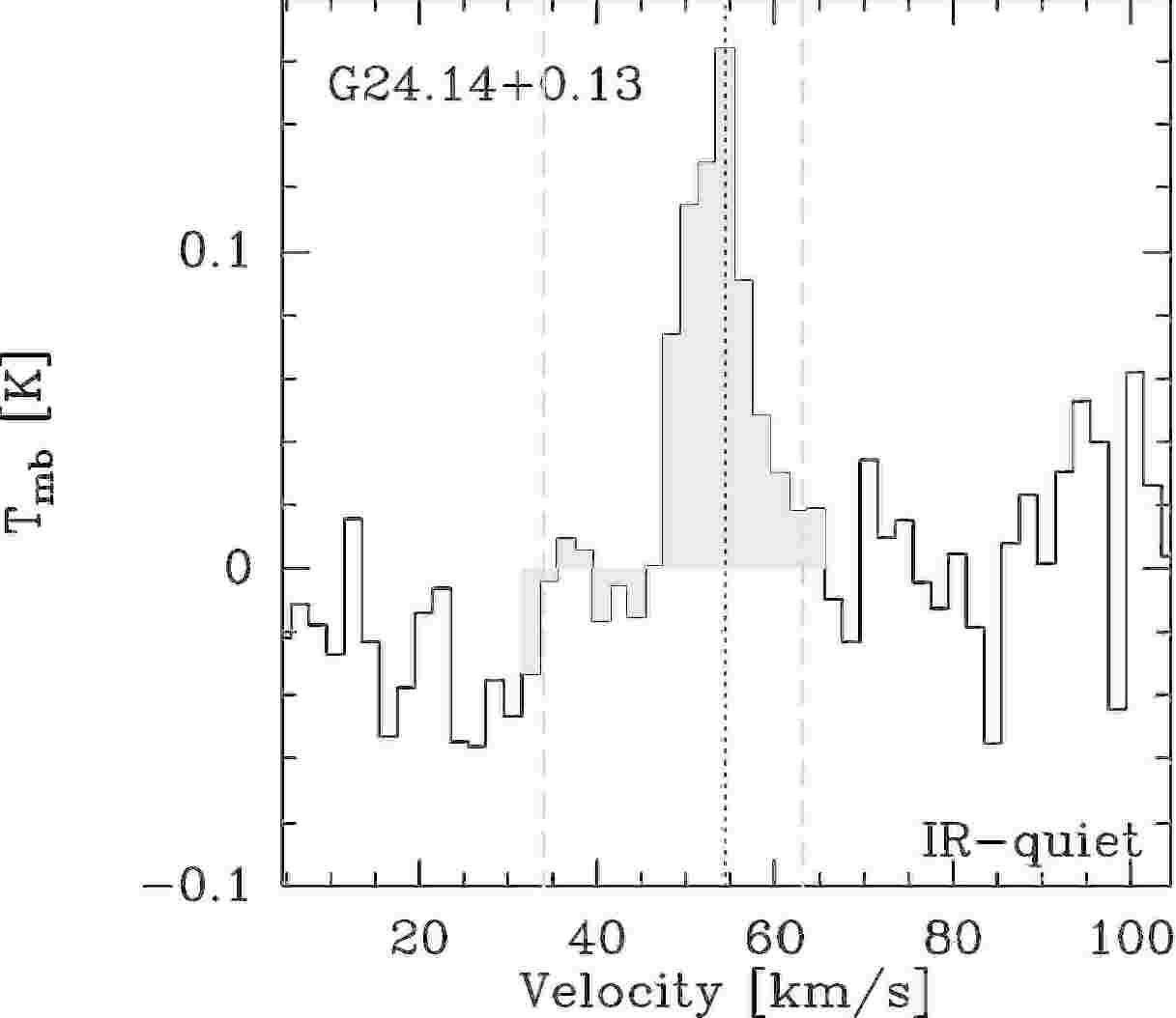} 
  \includegraphics[width=5.6cm,angle=0]{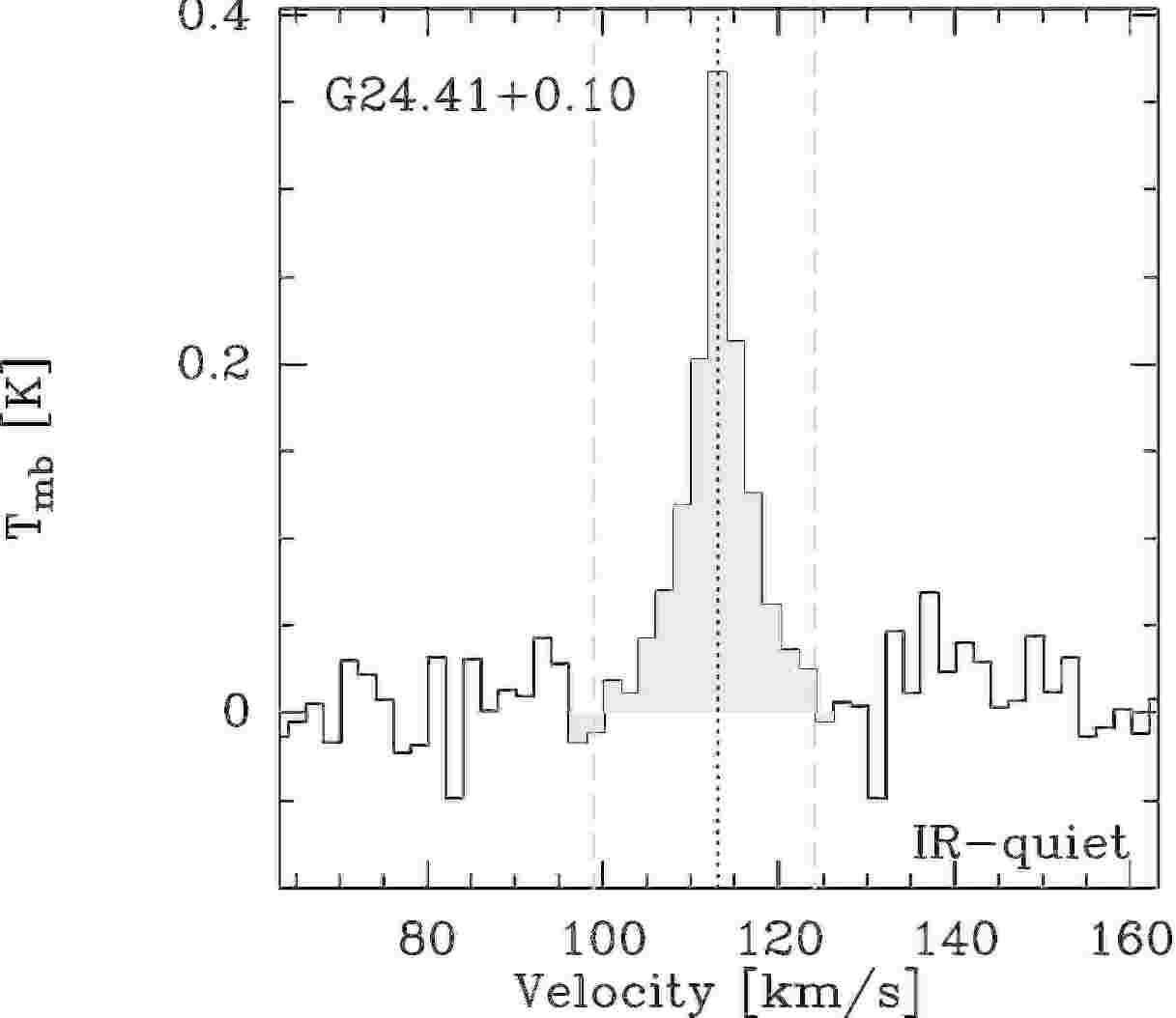} 
  \includegraphics[width=5.6cm,angle=0]{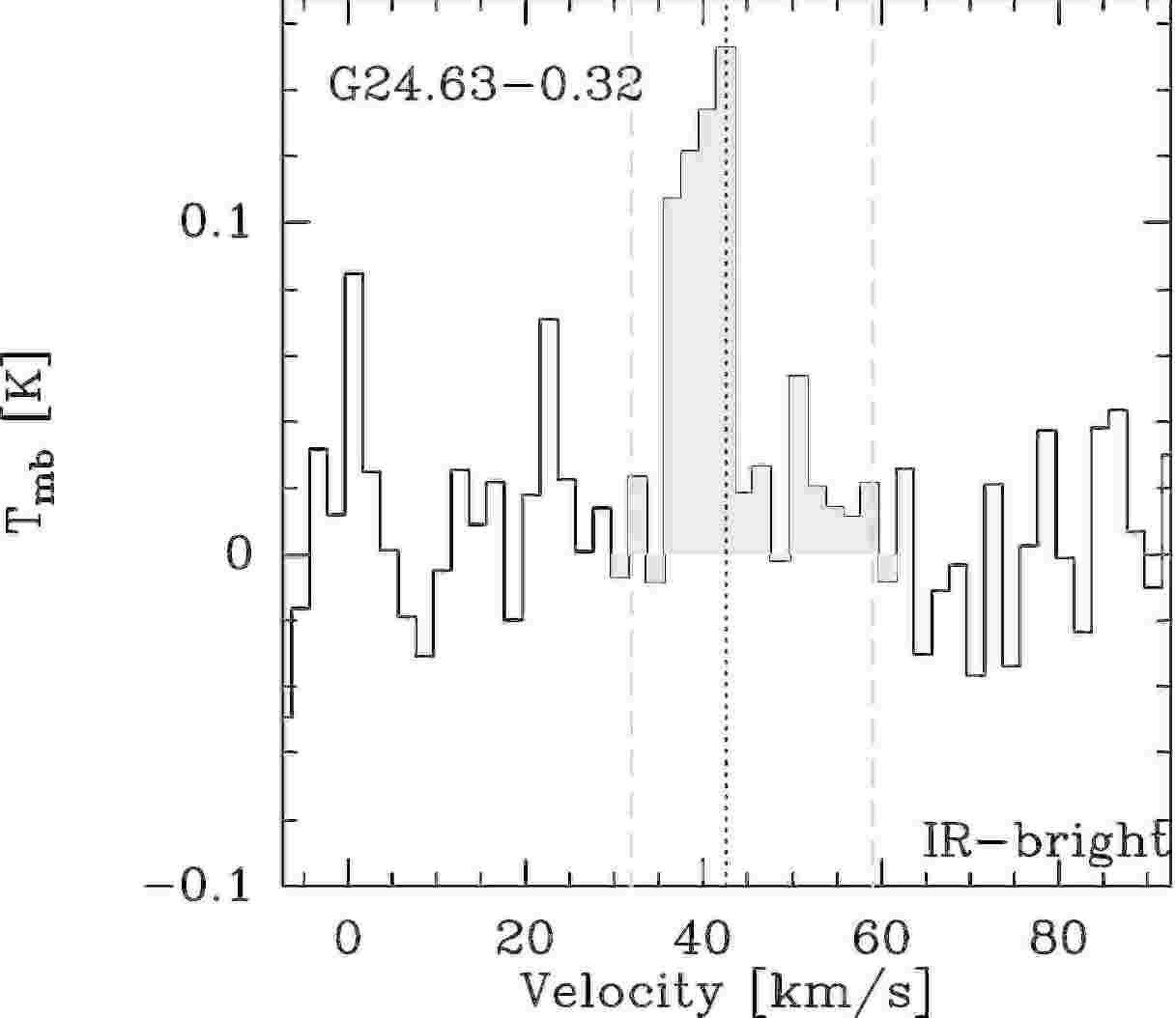} 
  \includegraphics[width=5.6cm,angle=0]{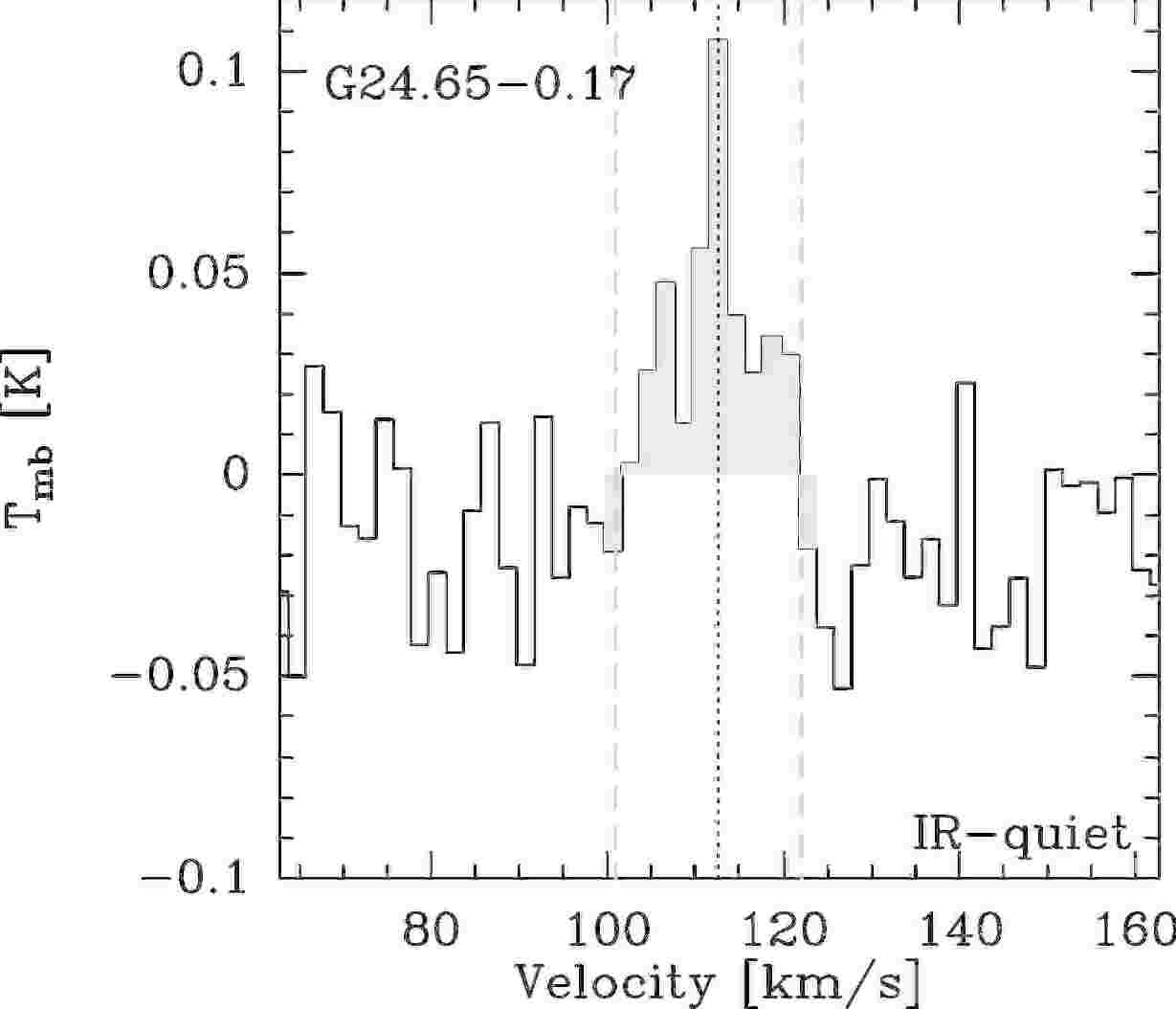} 
  \includegraphics[width=5.6cm,angle=0]{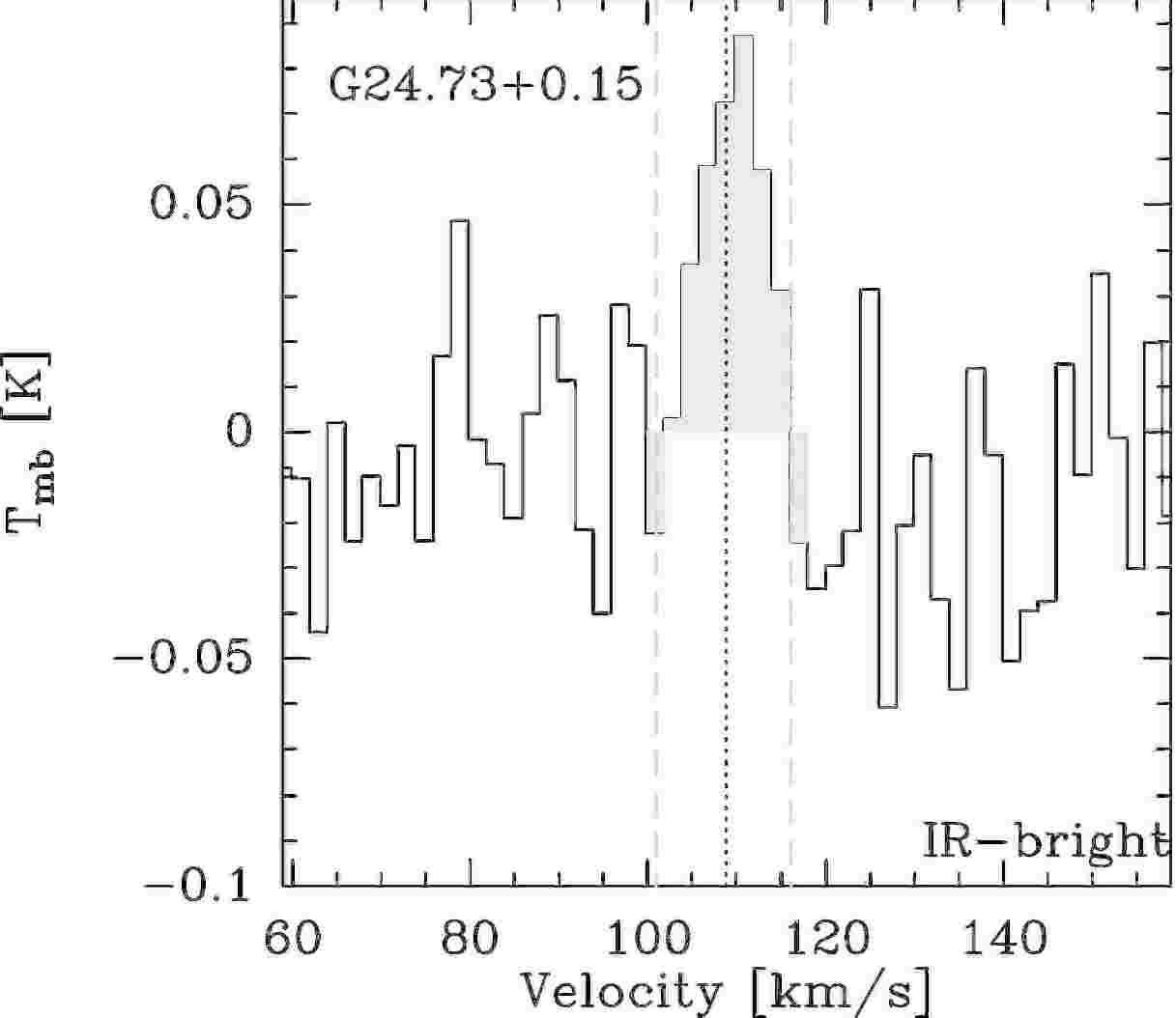} 
  \includegraphics[width=5.6cm,angle=0]{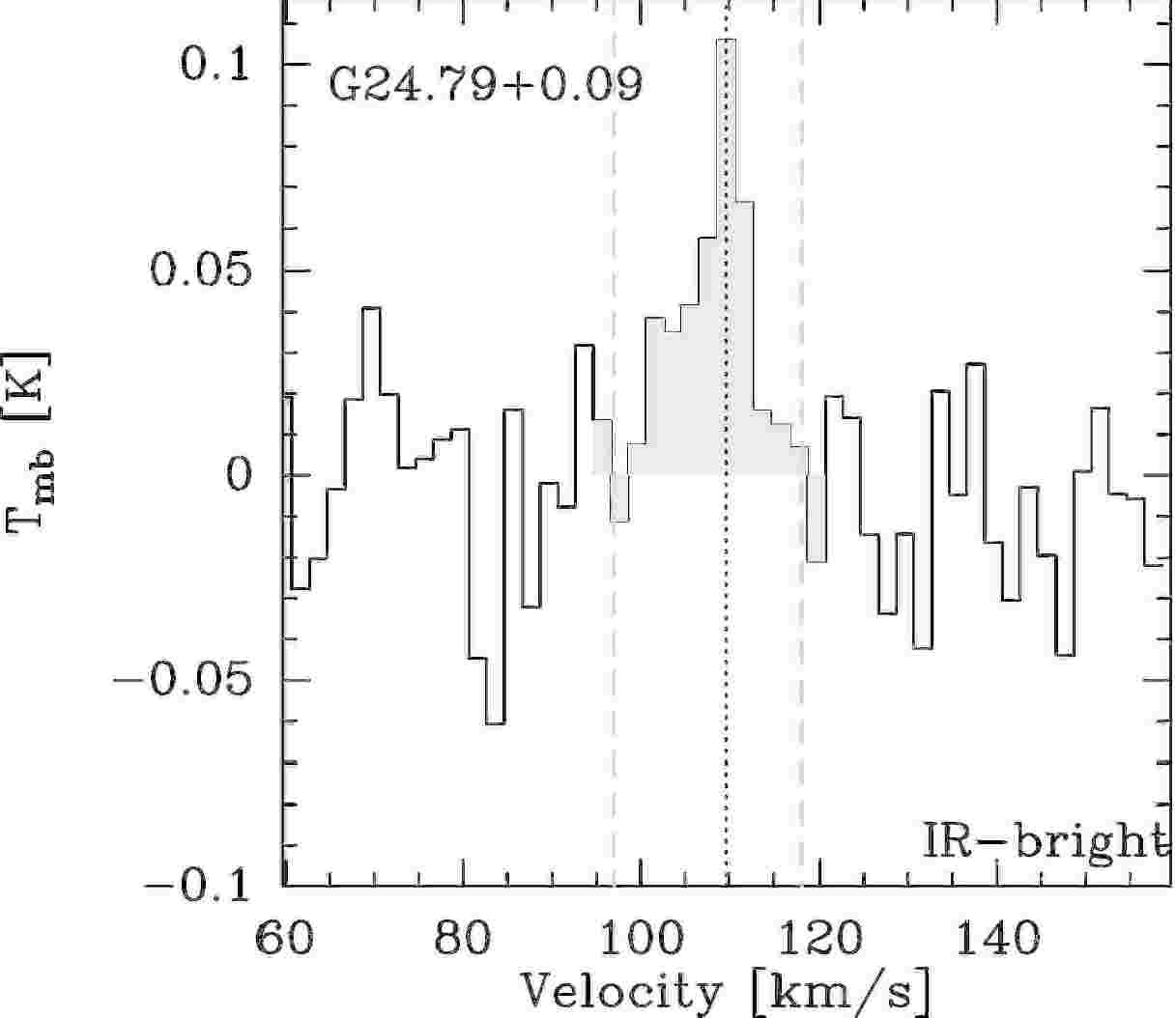} 
  \includegraphics[width=5.6cm,angle=0]{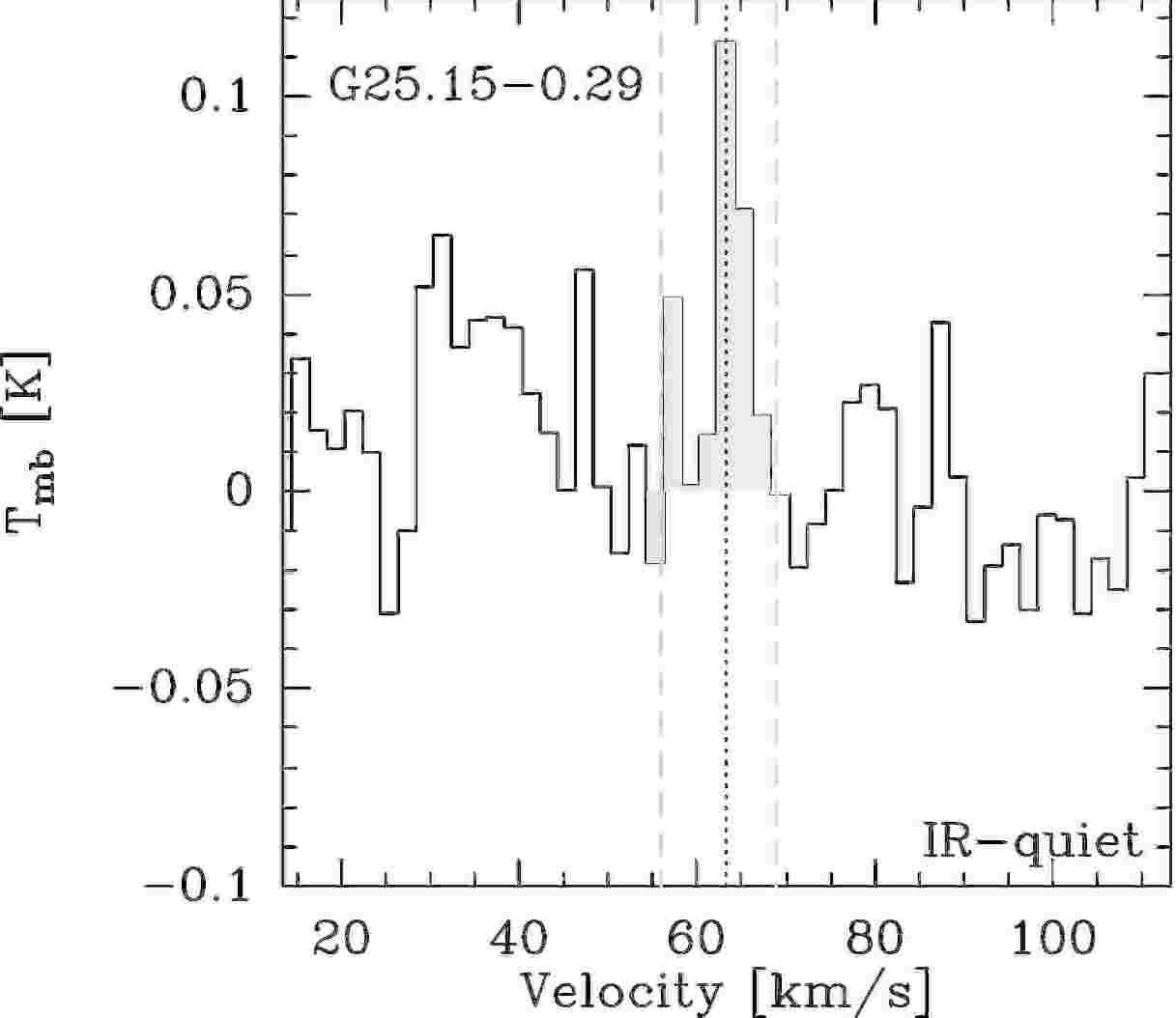} 
  \includegraphics[width=5.6cm,angle=0]{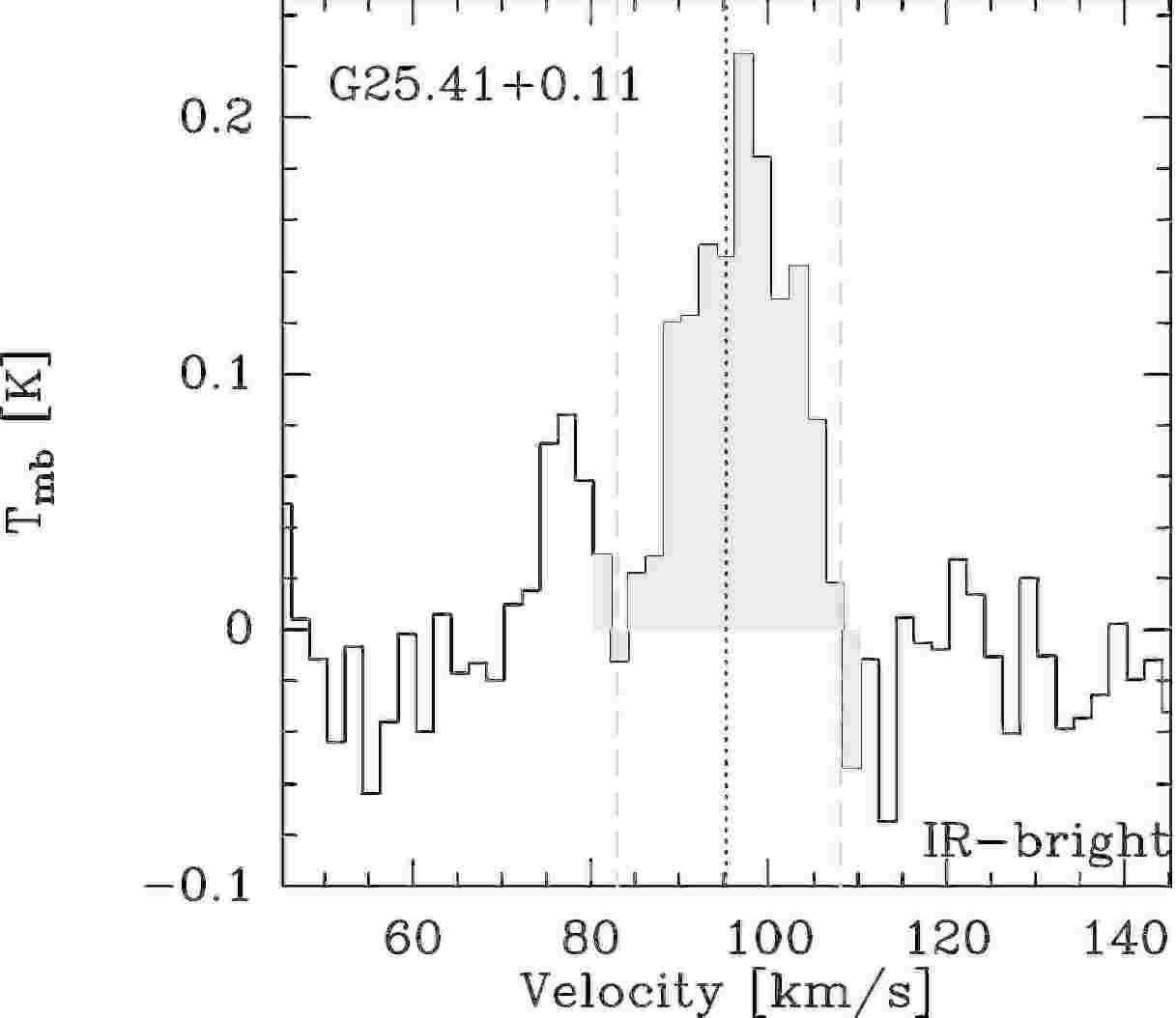} 
  \includegraphics[width=5.6cm,angle=0]{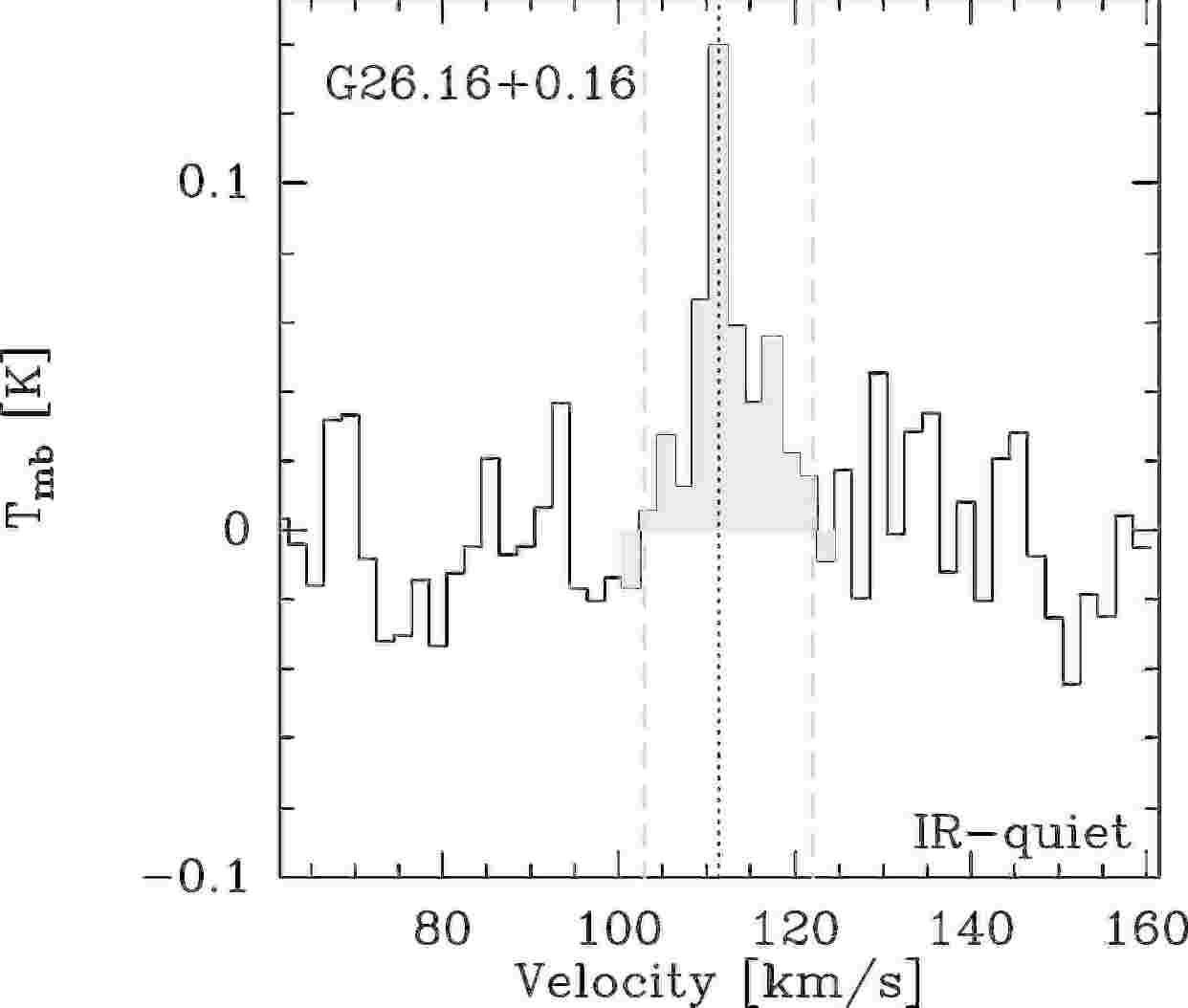} 
 \includegraphics[width=5.6cm,angle=0]{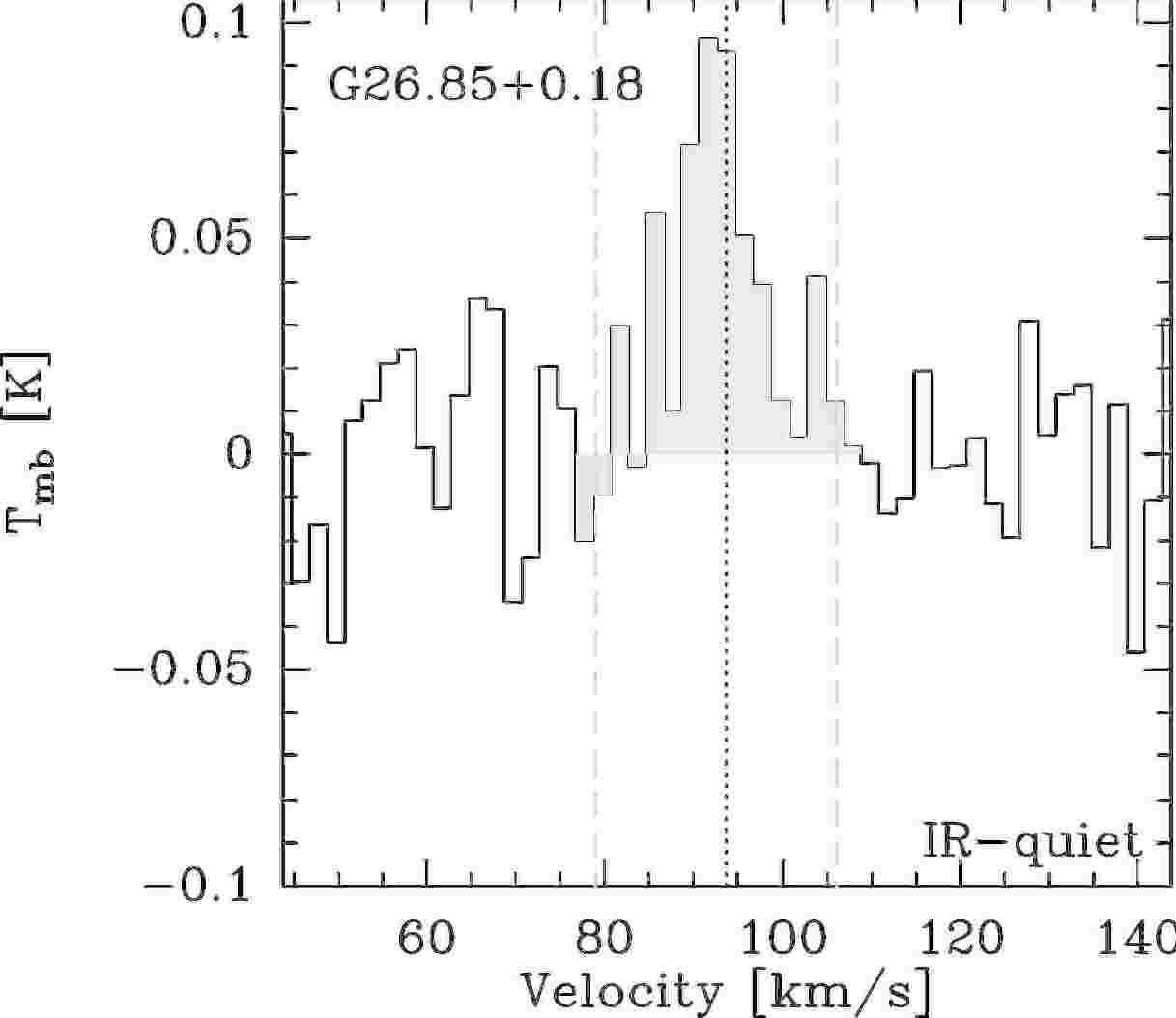} 
  \includegraphics[width=5.6cm,angle=0]{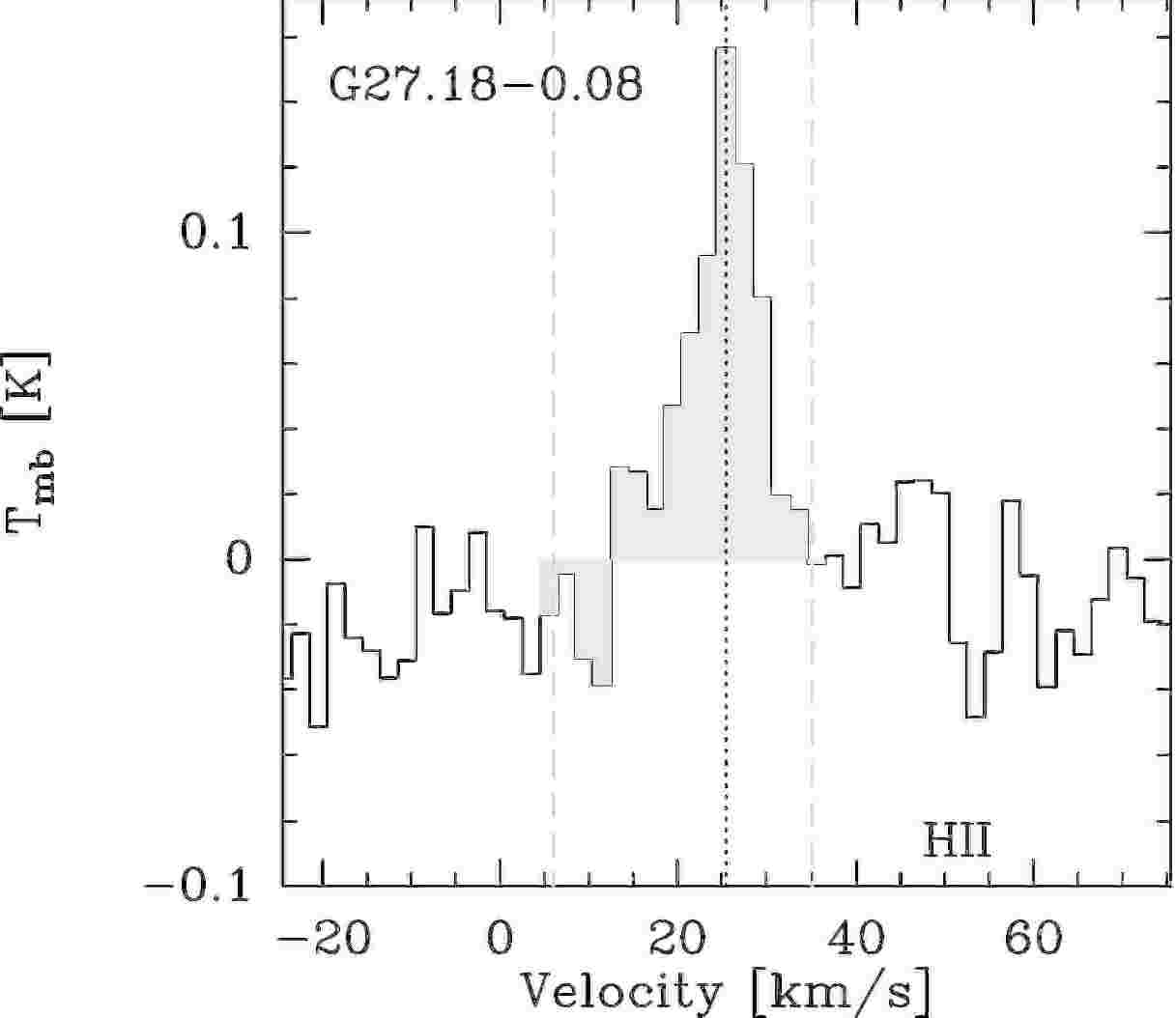} 
 \caption{Continued.}
\end{figure}
\end{landscape}

\begin{landscape}
\begin{figure}
\centering
\ContinuedFloat
  \includegraphics[width=5.6cm,angle=0]{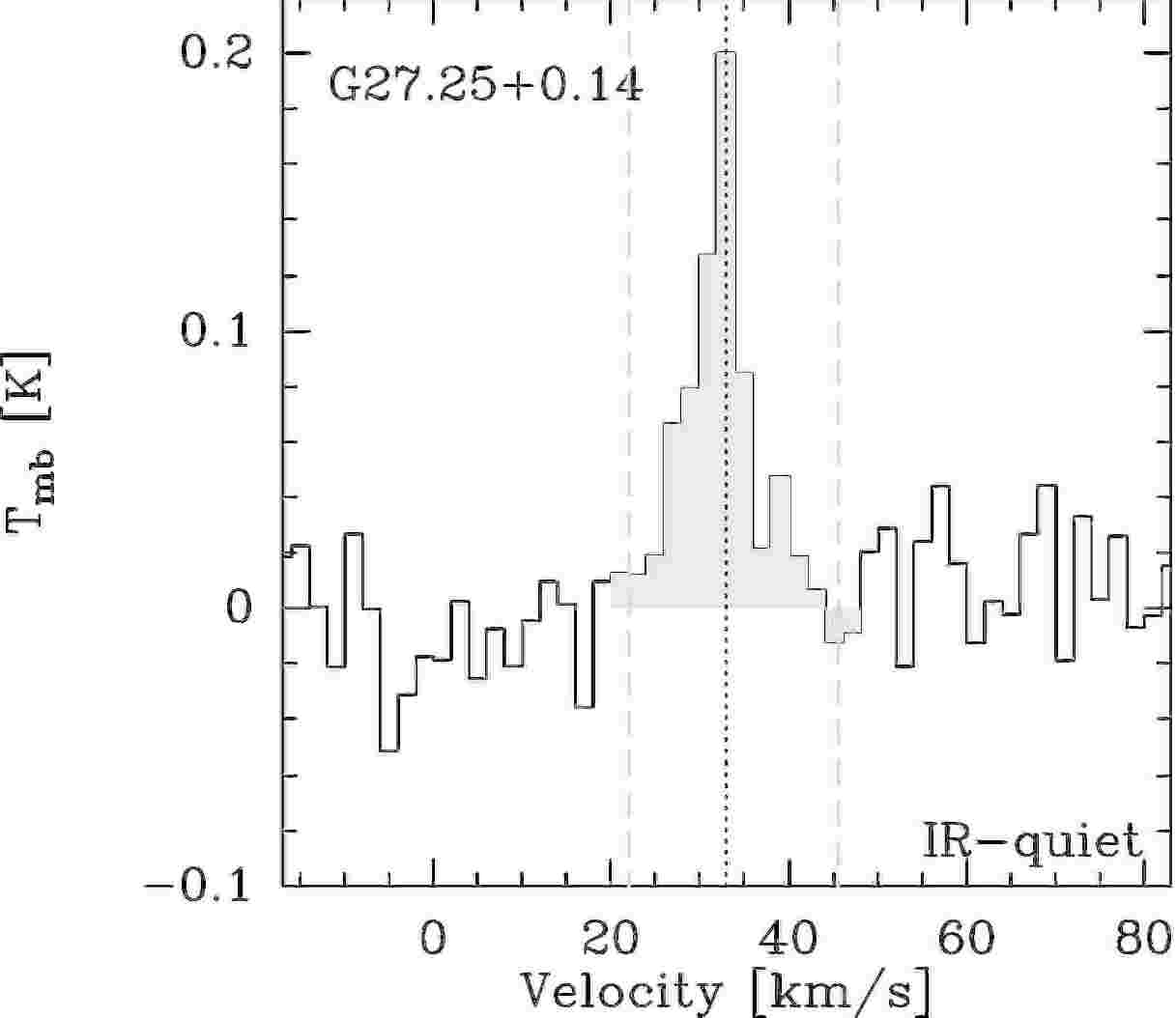} 
  \includegraphics[width=5.6cm,angle=0]{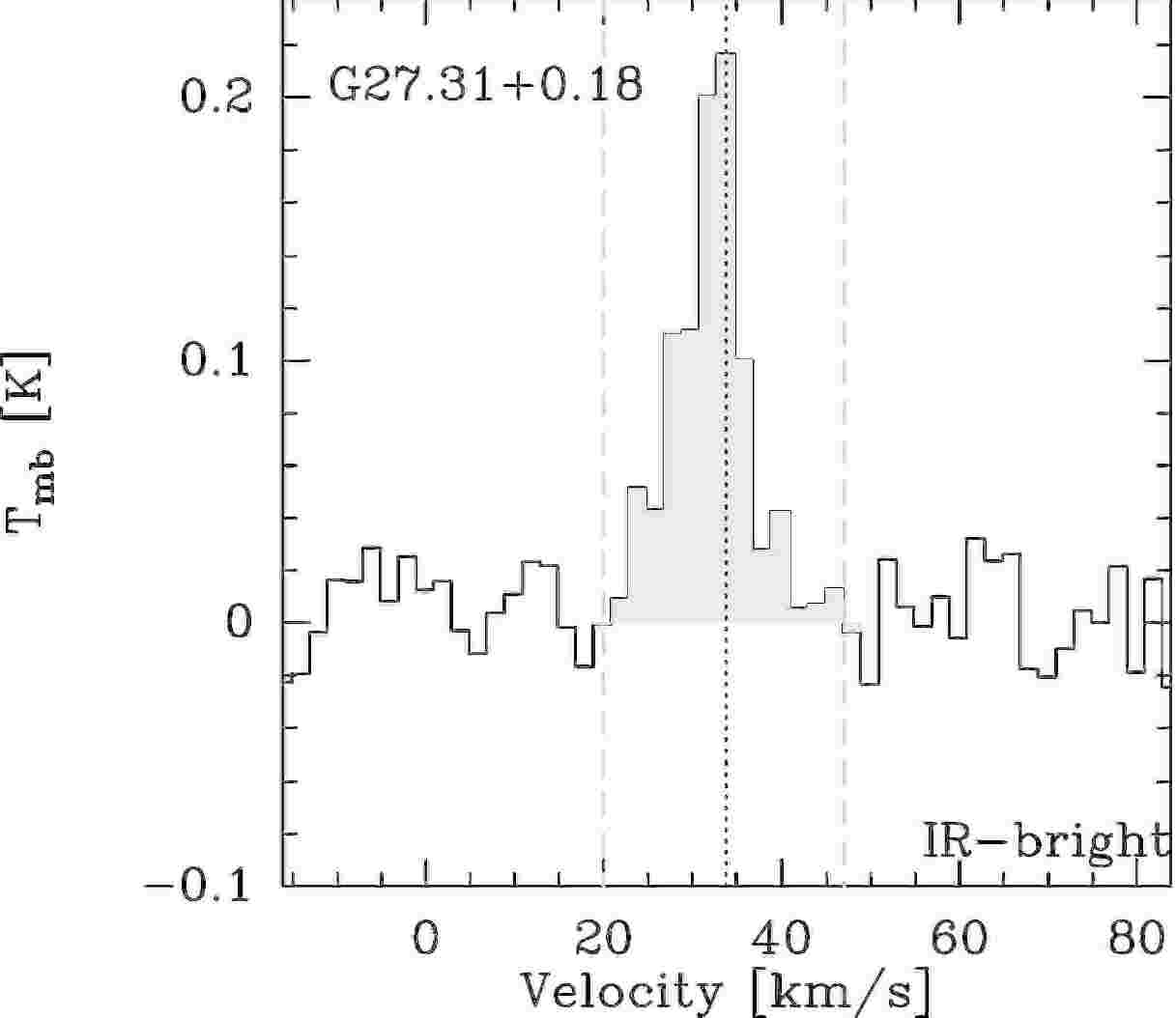} 
  \includegraphics[width=5.6cm,angle=0]{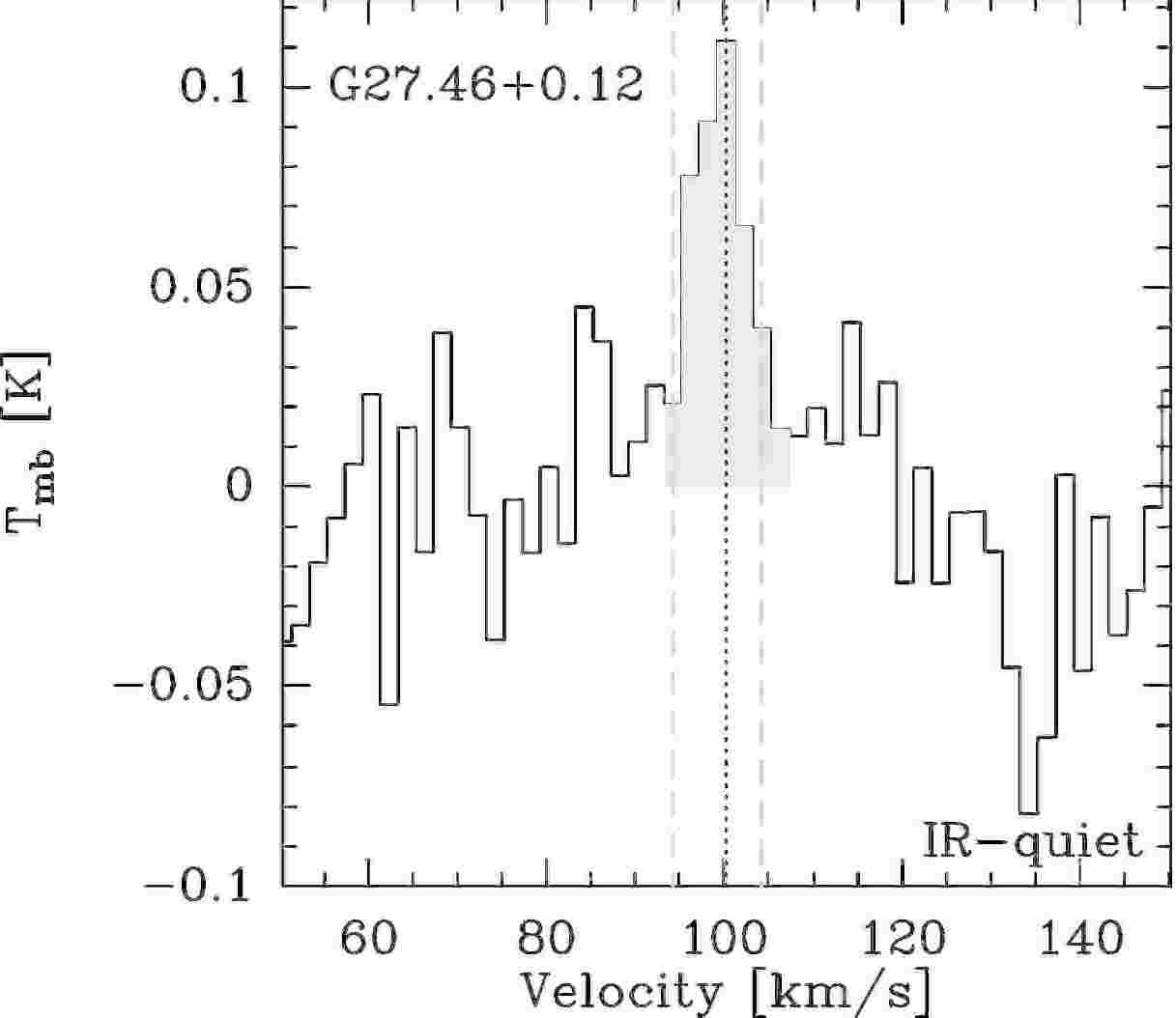} 
  \includegraphics[width=5.6cm,angle=0]{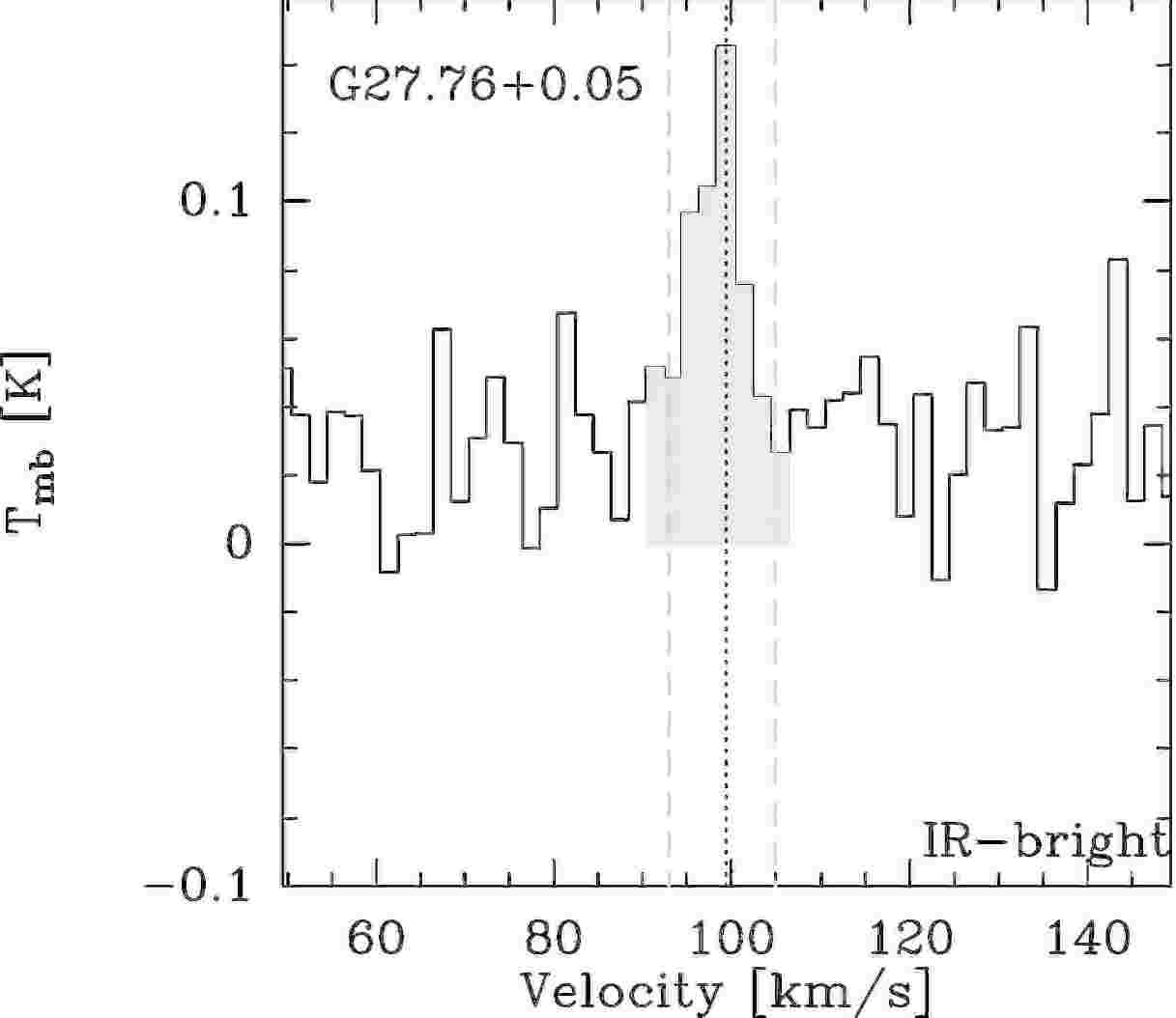} 
  \includegraphics[width=5.6cm,angle=0]{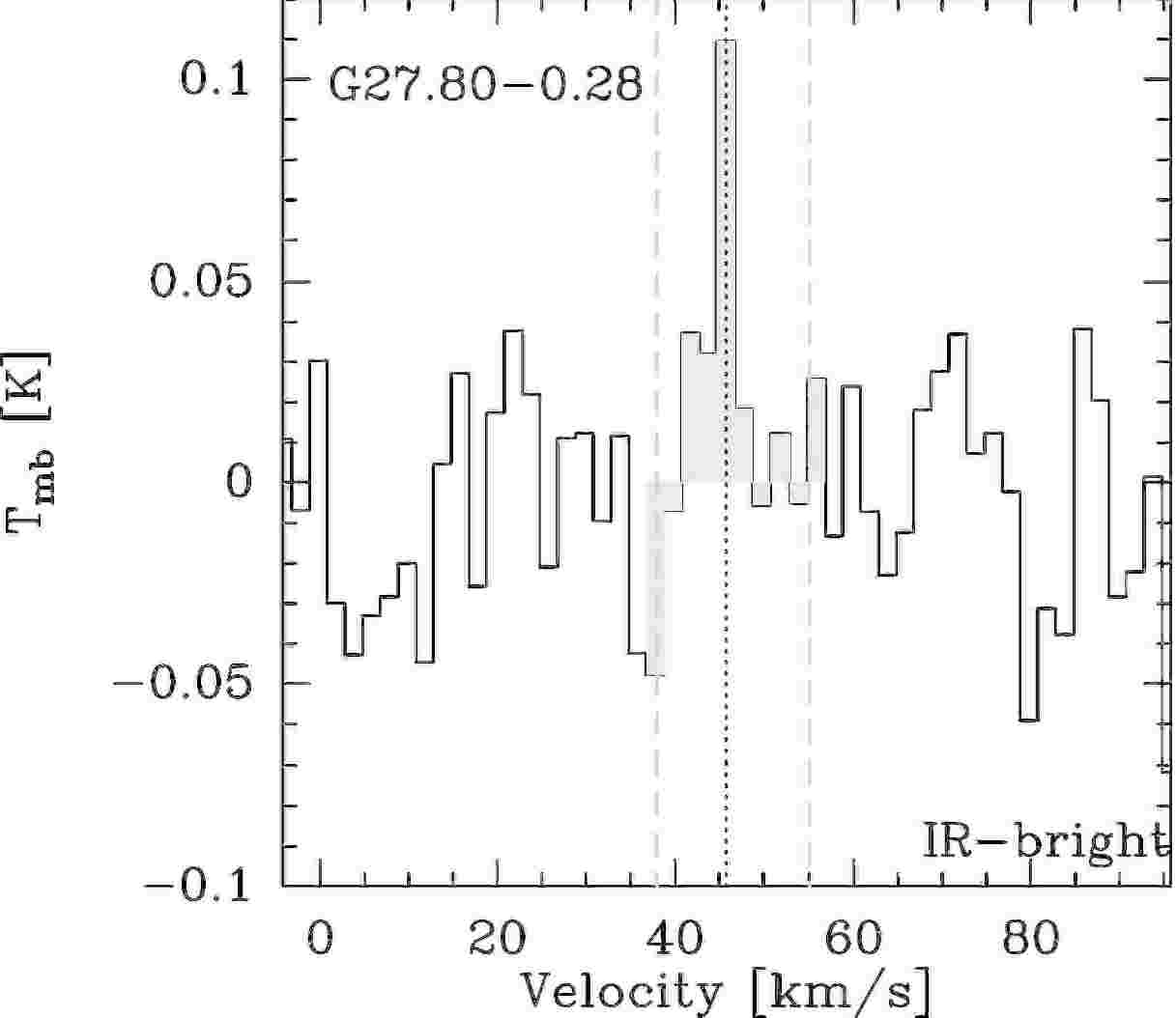} 
  \includegraphics[width=5.6cm,angle=0]{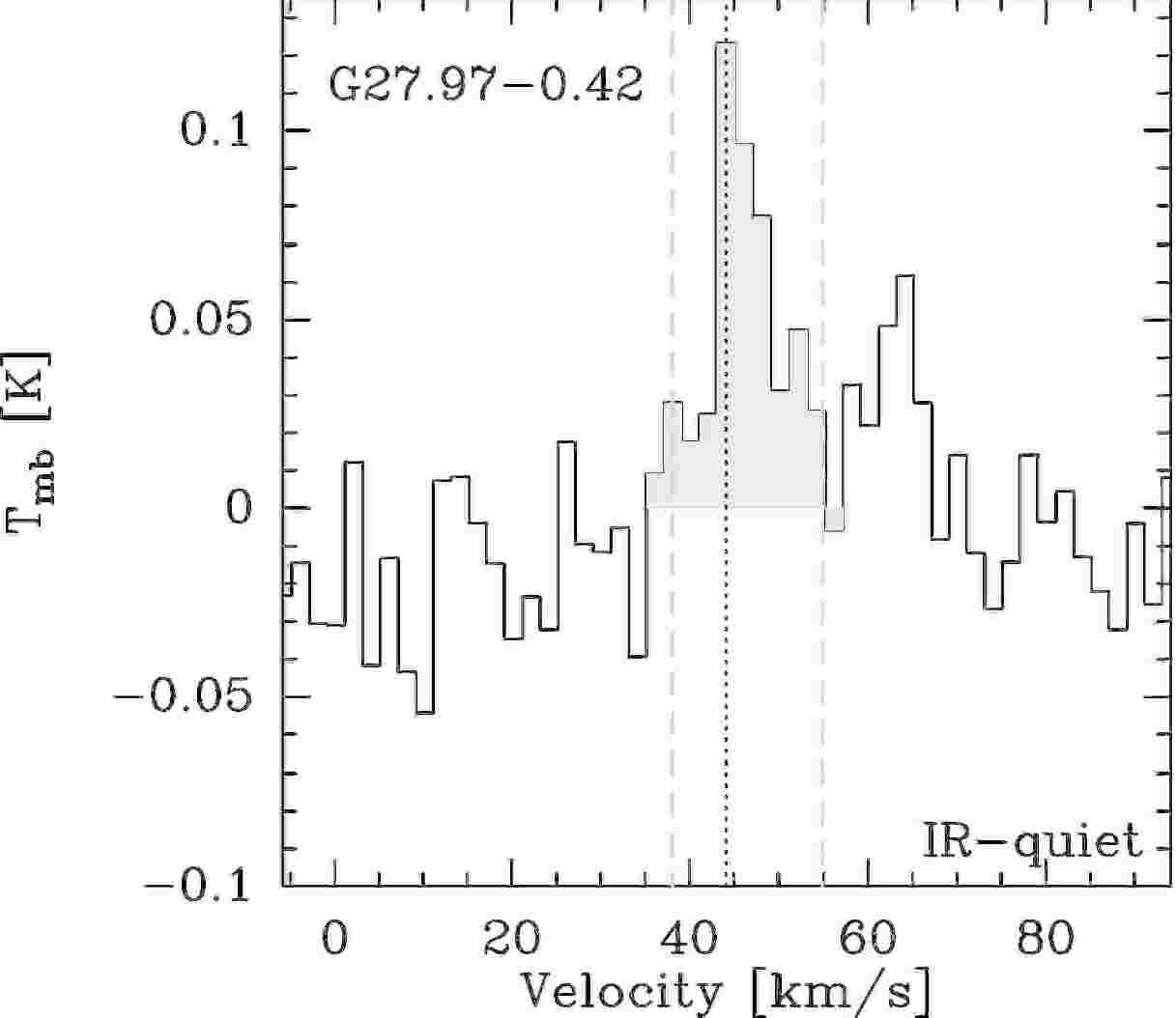} 
  \includegraphics[width=5.6cm,angle=0]{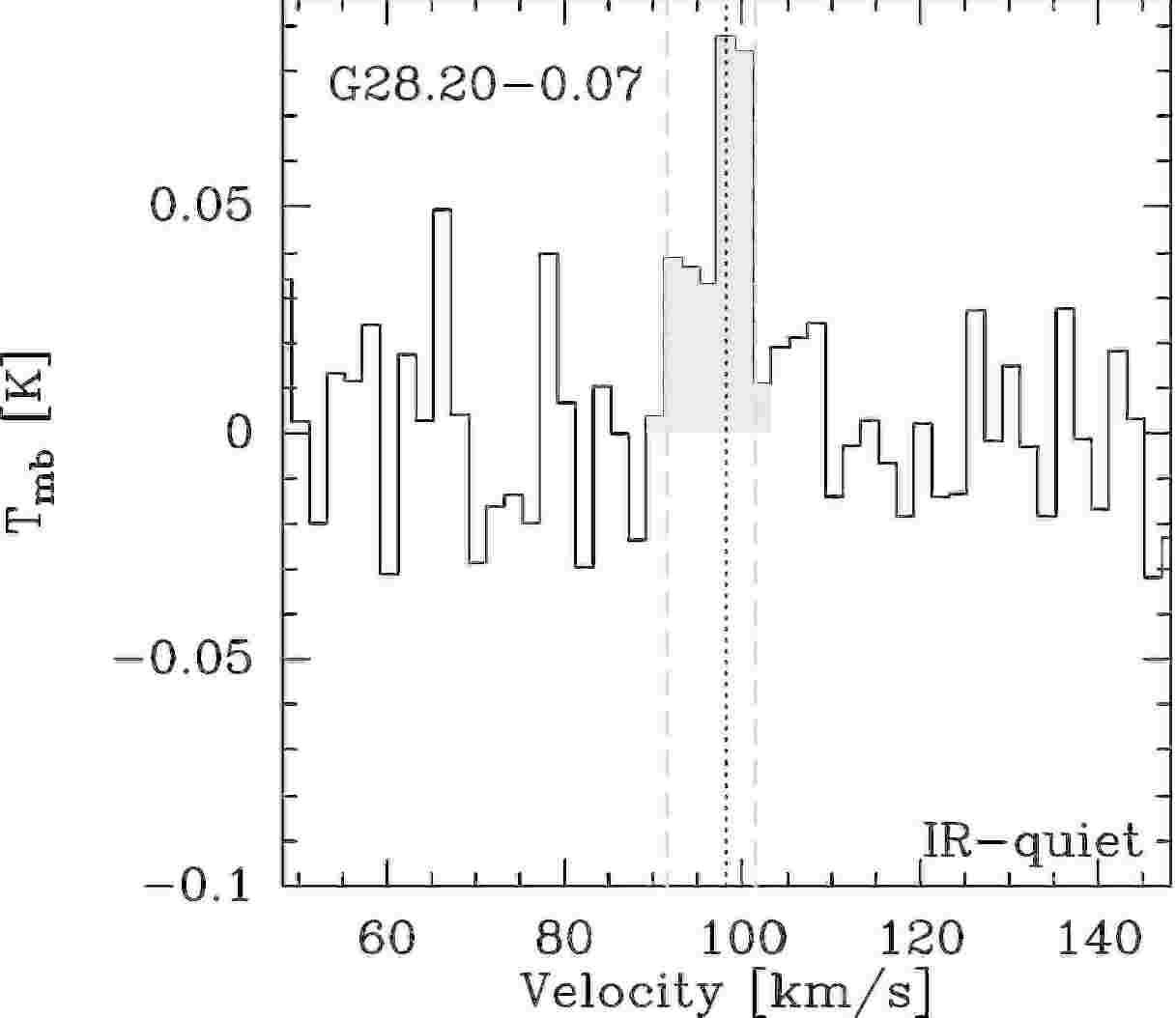} 
  \includegraphics[width=5.6cm,angle=0]{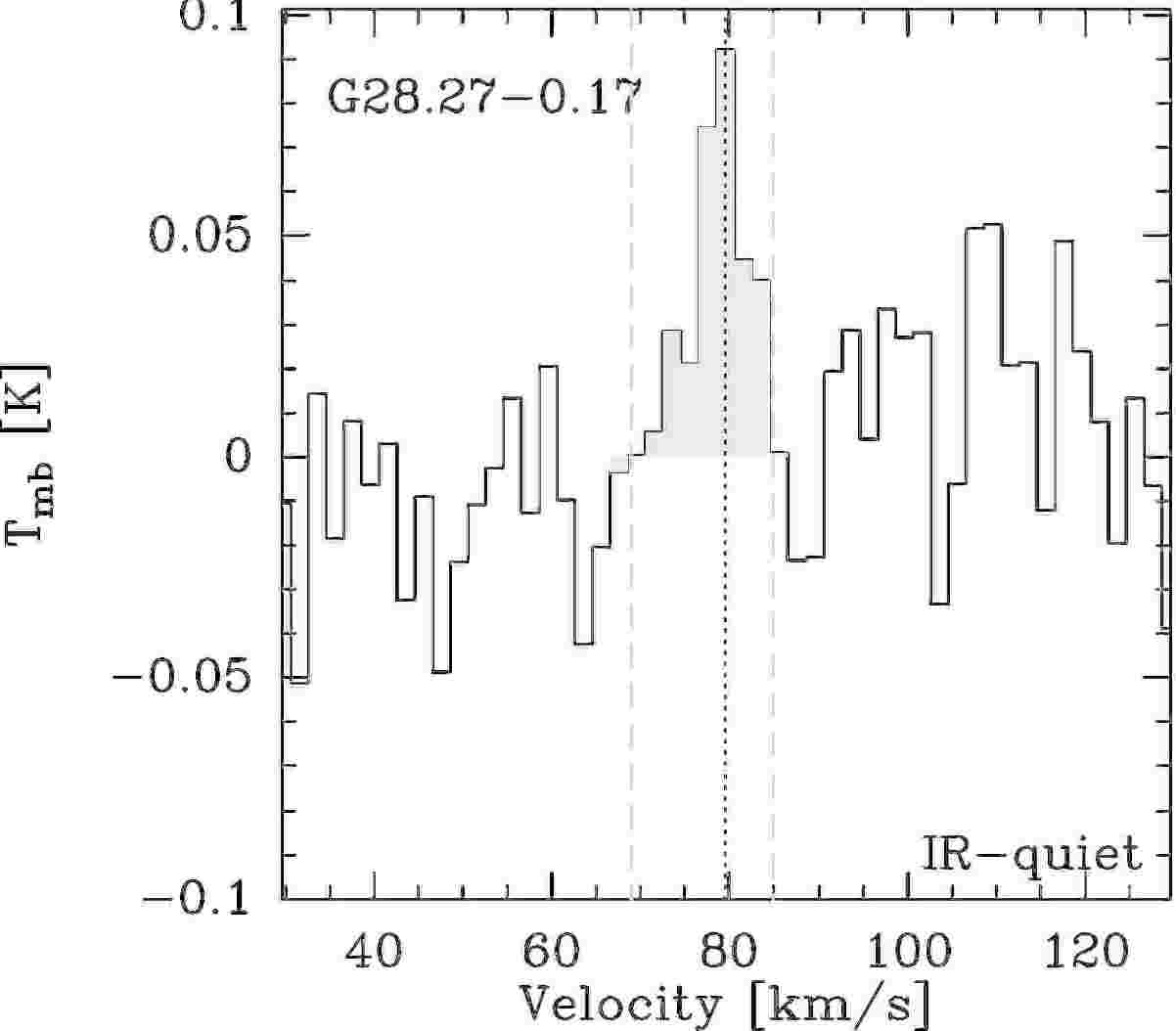} 
  \includegraphics[width=5.6cm,angle=0]{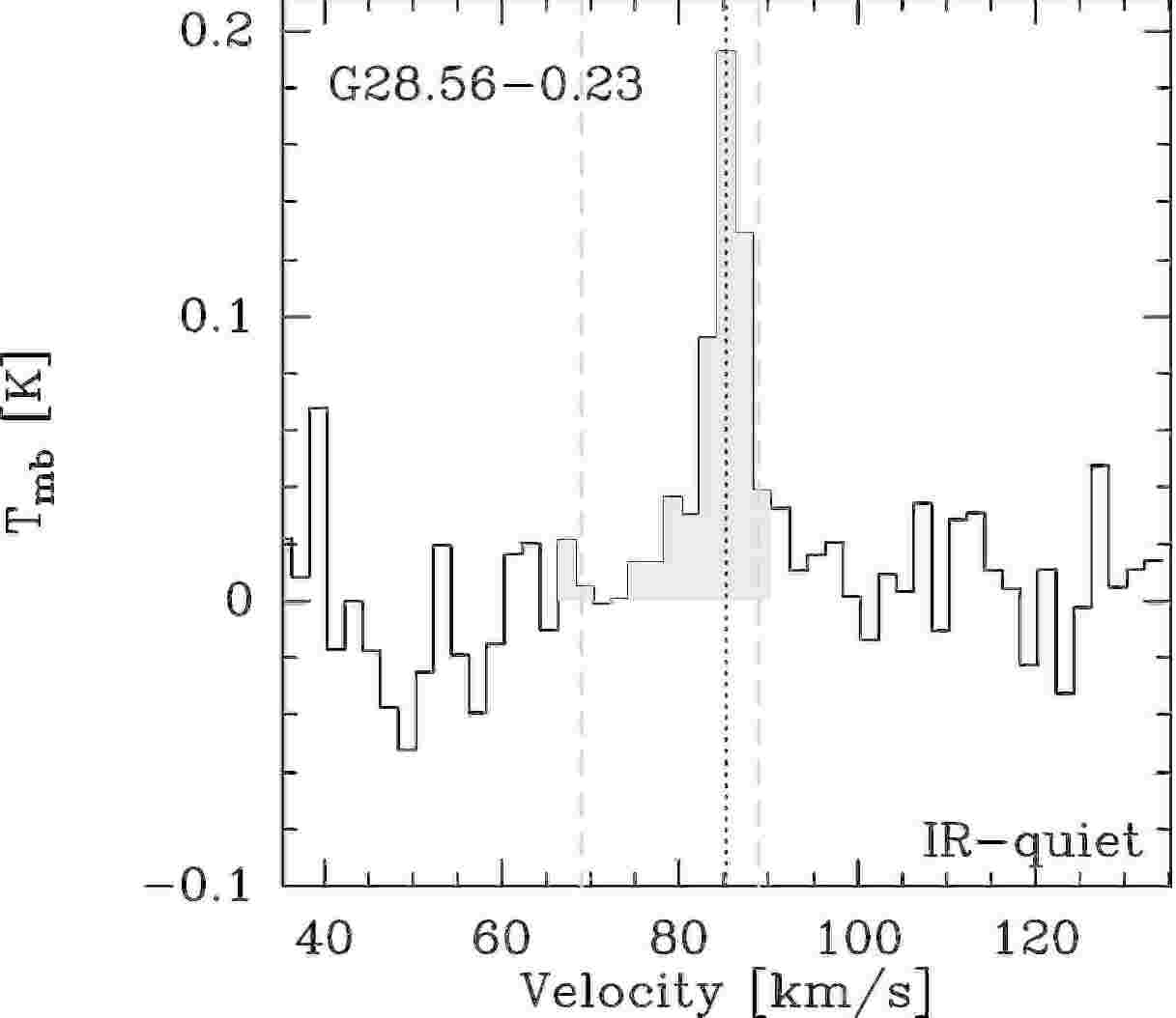} 
   \includegraphics[width=6.0cm,angle=0]{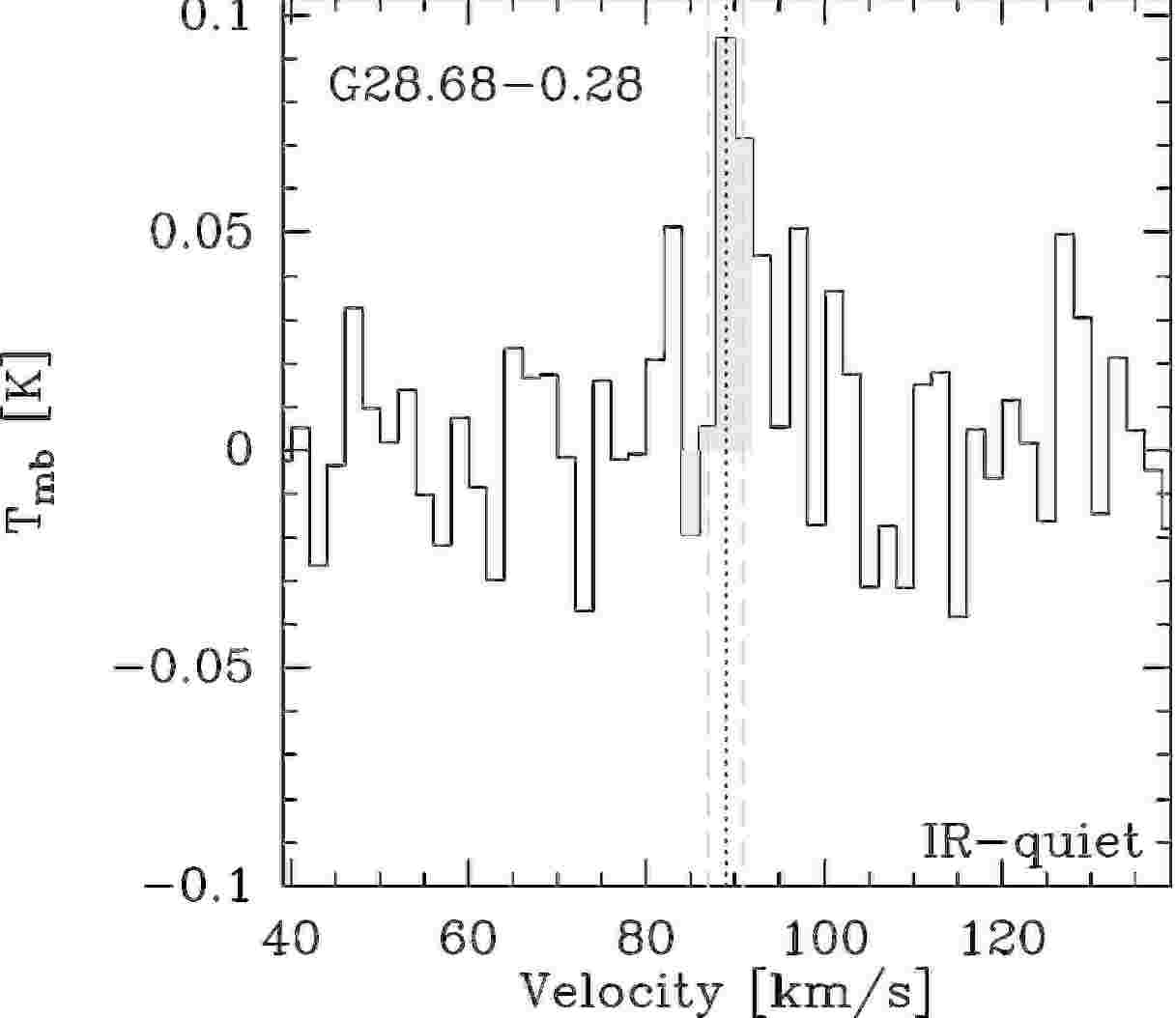} 
 \includegraphics[width=5.6cm,angle=0]{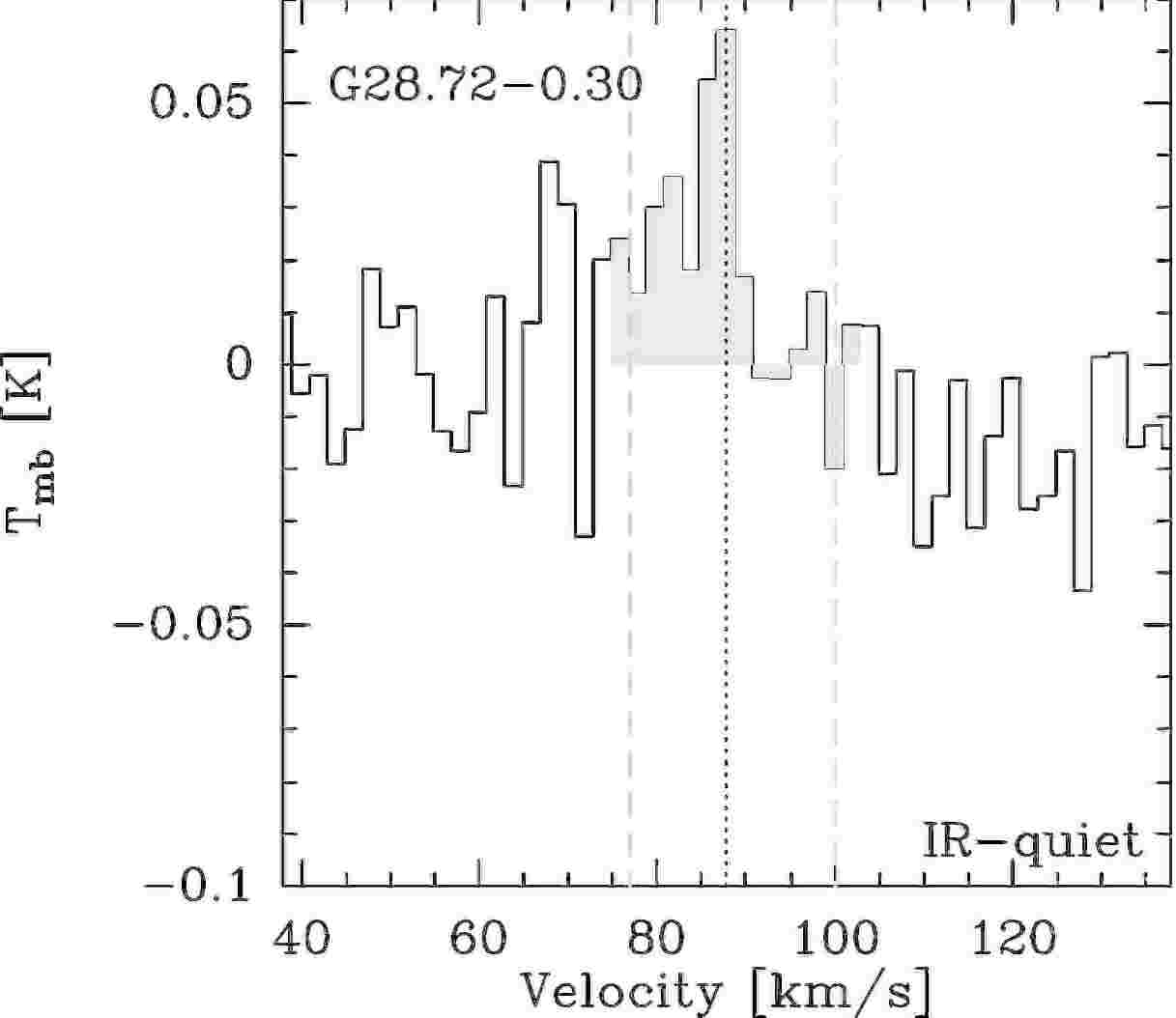} 
  \includegraphics[width=5.6cm,angle=0]{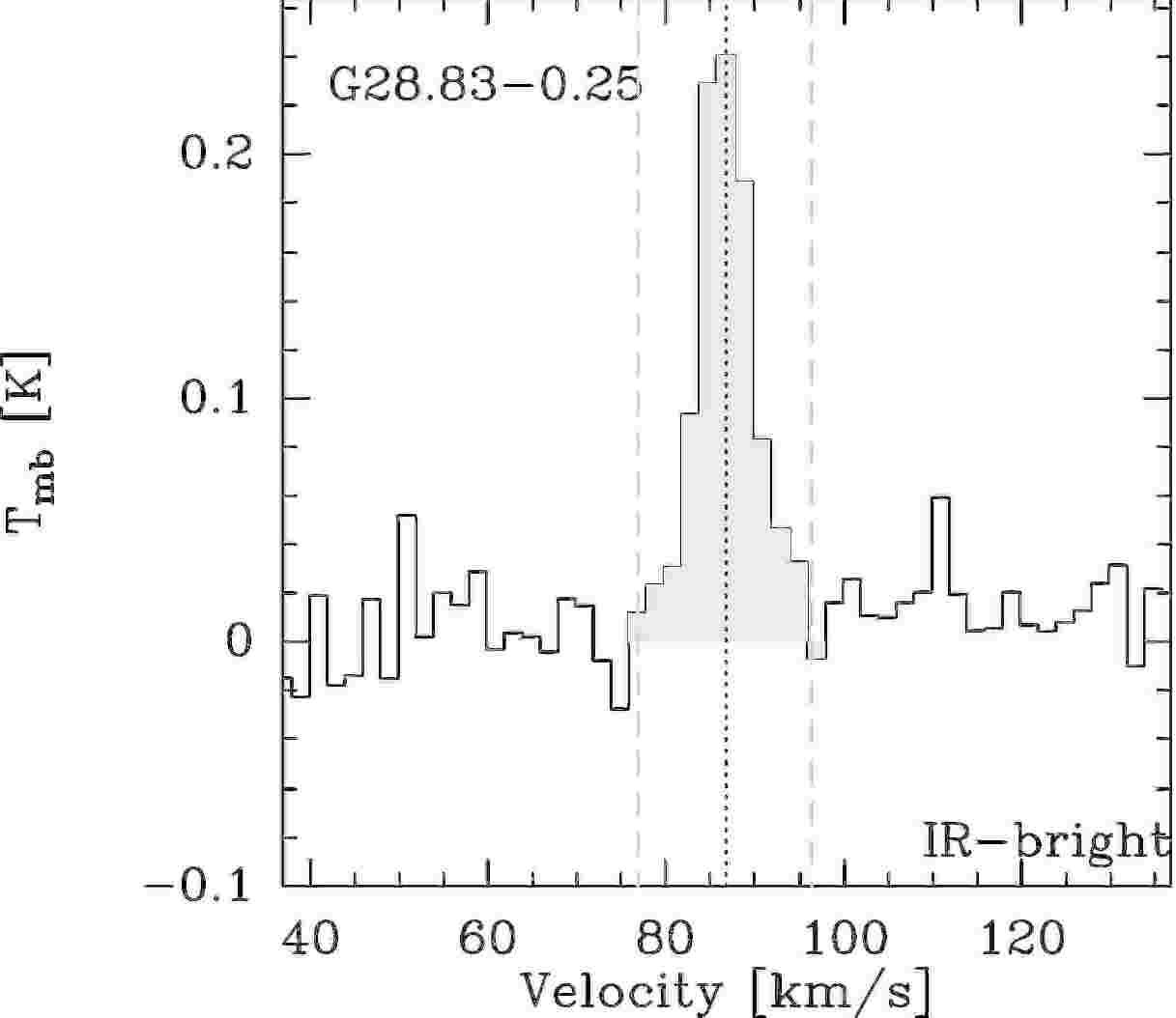} 
 \caption{Continued.}
\end{figure}
\end{landscape}

\begin{landscape}
\begin{figure}
\centering
\ContinuedFloat
  \includegraphics[width=5.6cm,angle=0]{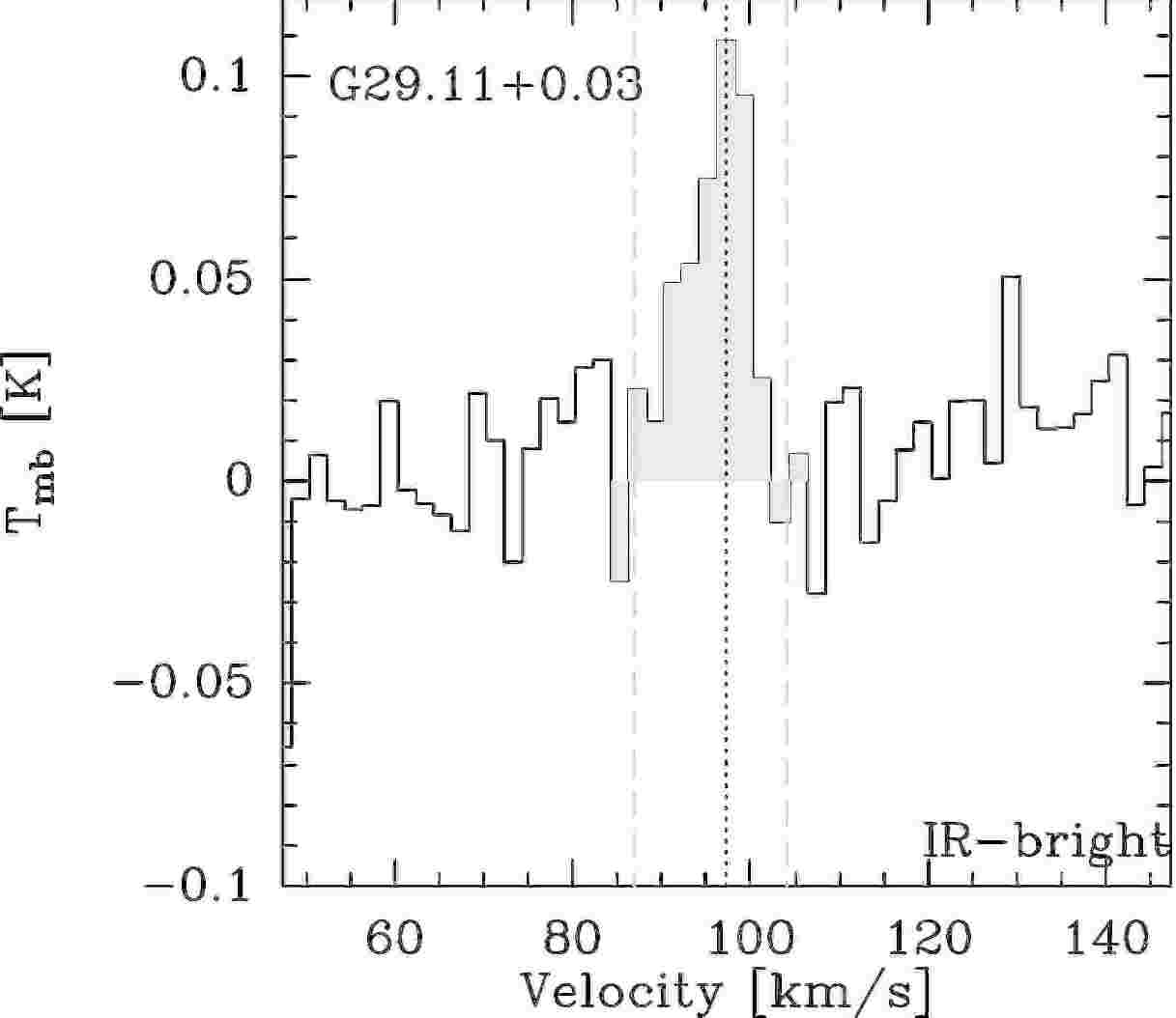} 
  \includegraphics[width=5.6cm,angle=0]{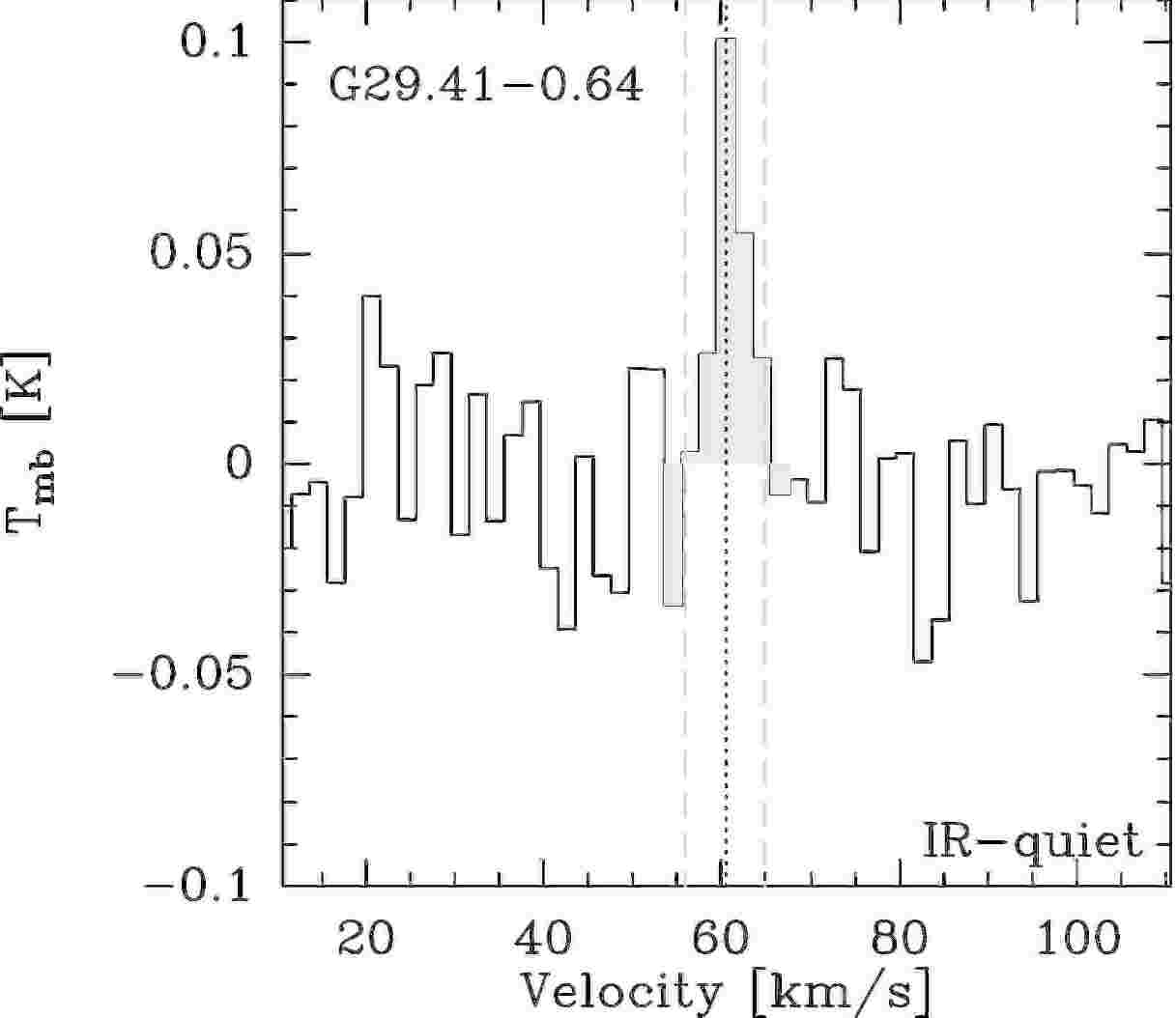} 
  \includegraphics[width=5.6cm,angle=0]{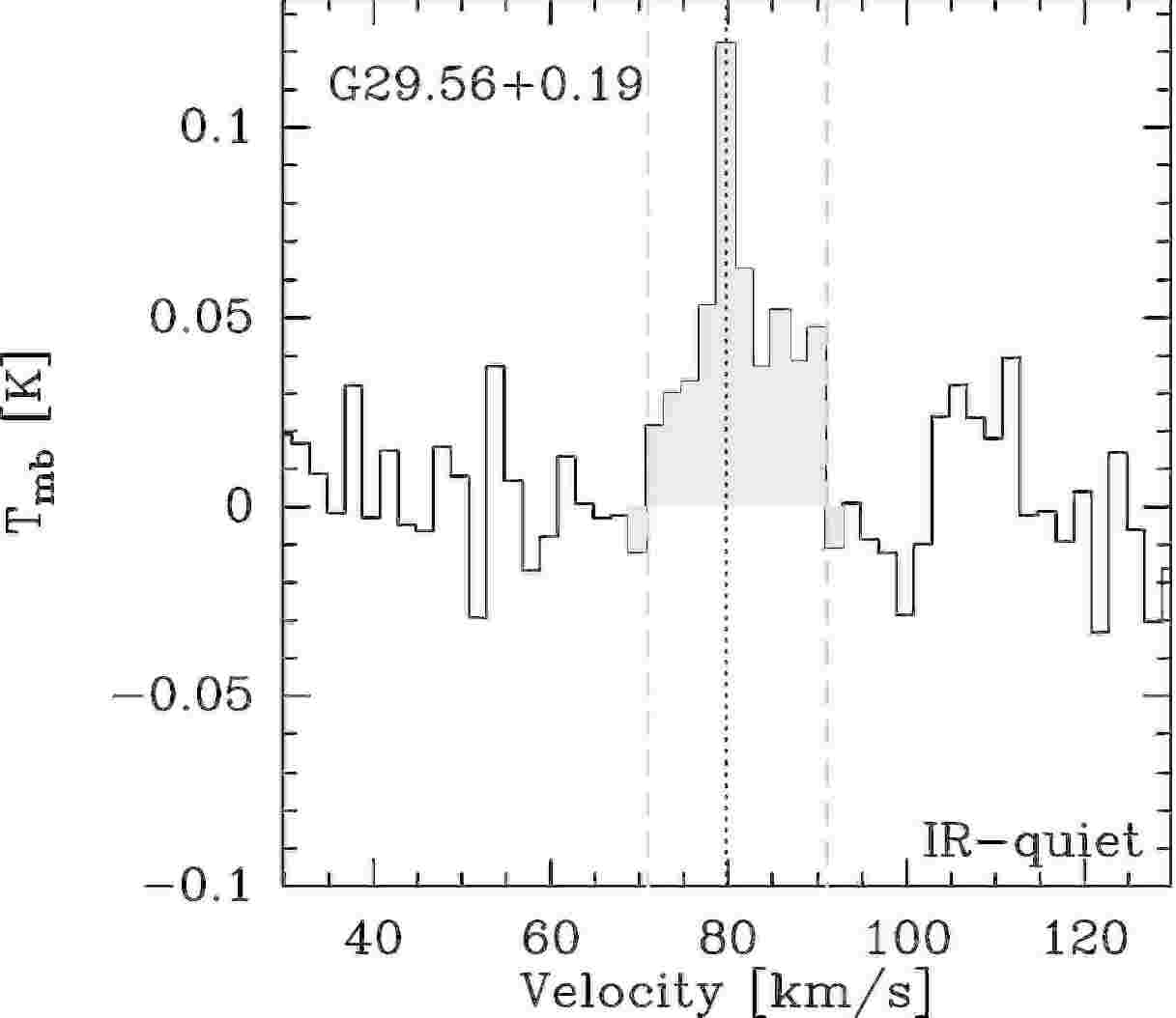} 
  \includegraphics[width=5.6cm,angle=0]{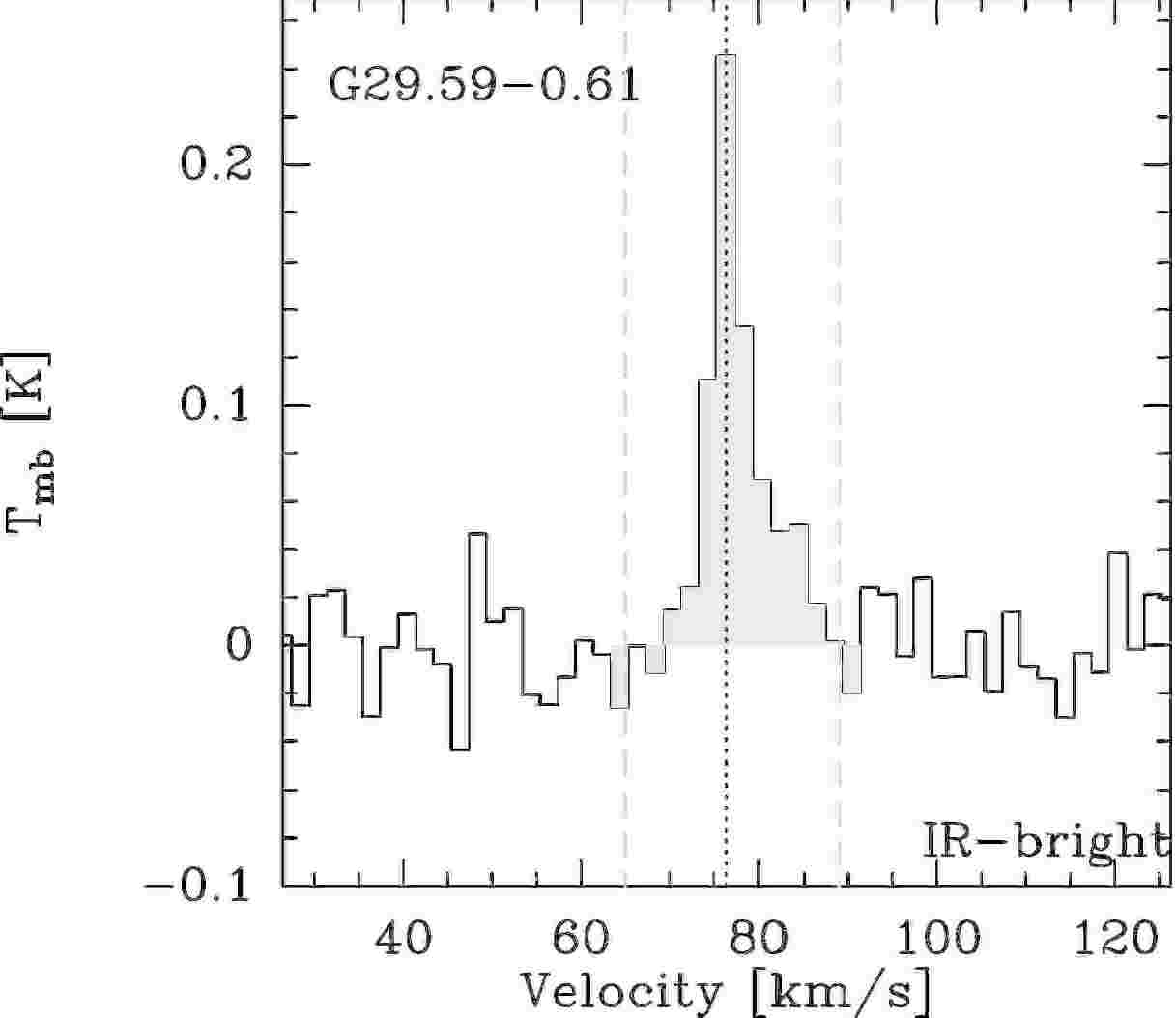} 
  \includegraphics[width=5.6cm,angle=0]{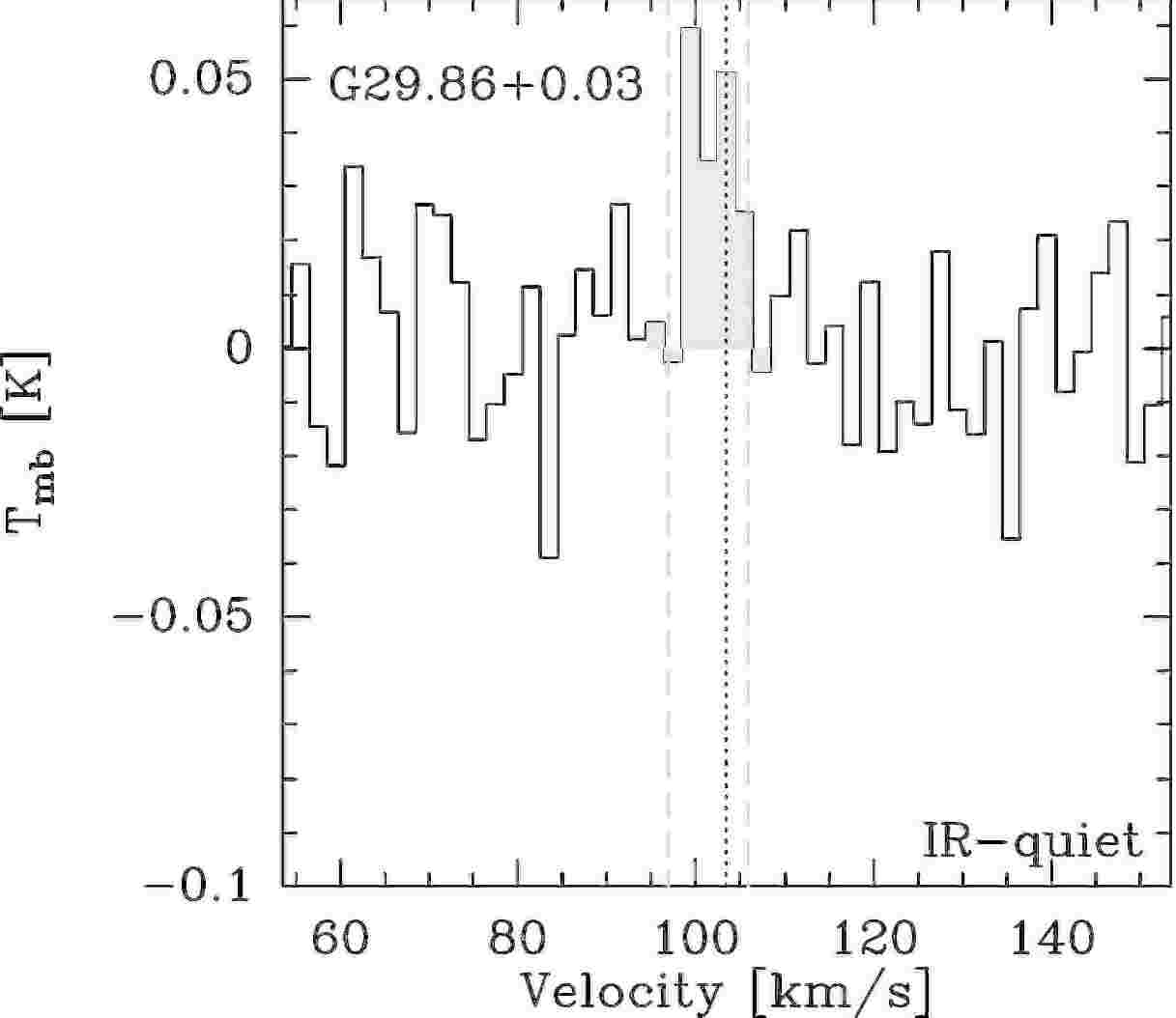} 
  \includegraphics[width=5.6cm,angle=0]{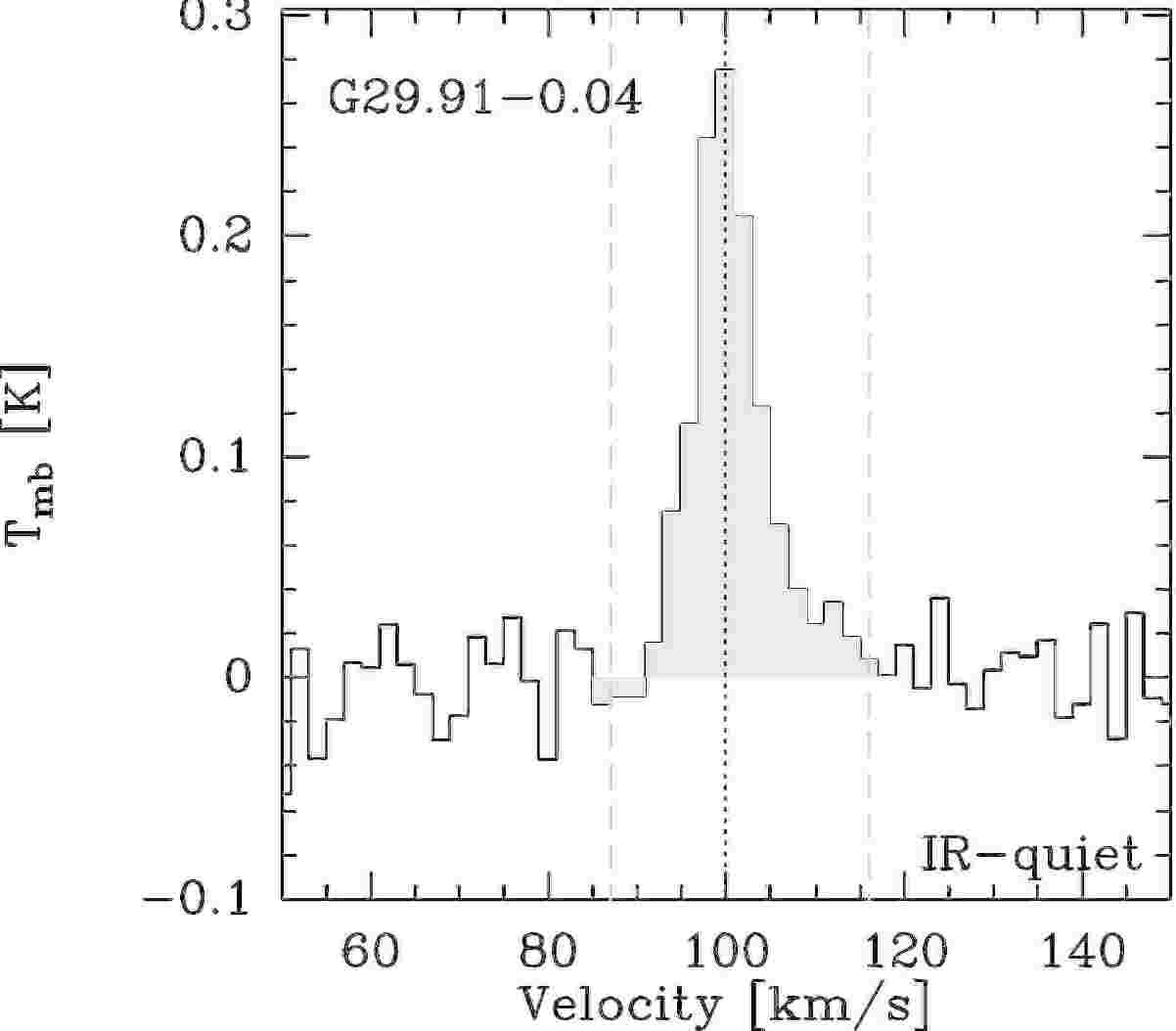} 
  \includegraphics[width=5.6cm,angle=0]{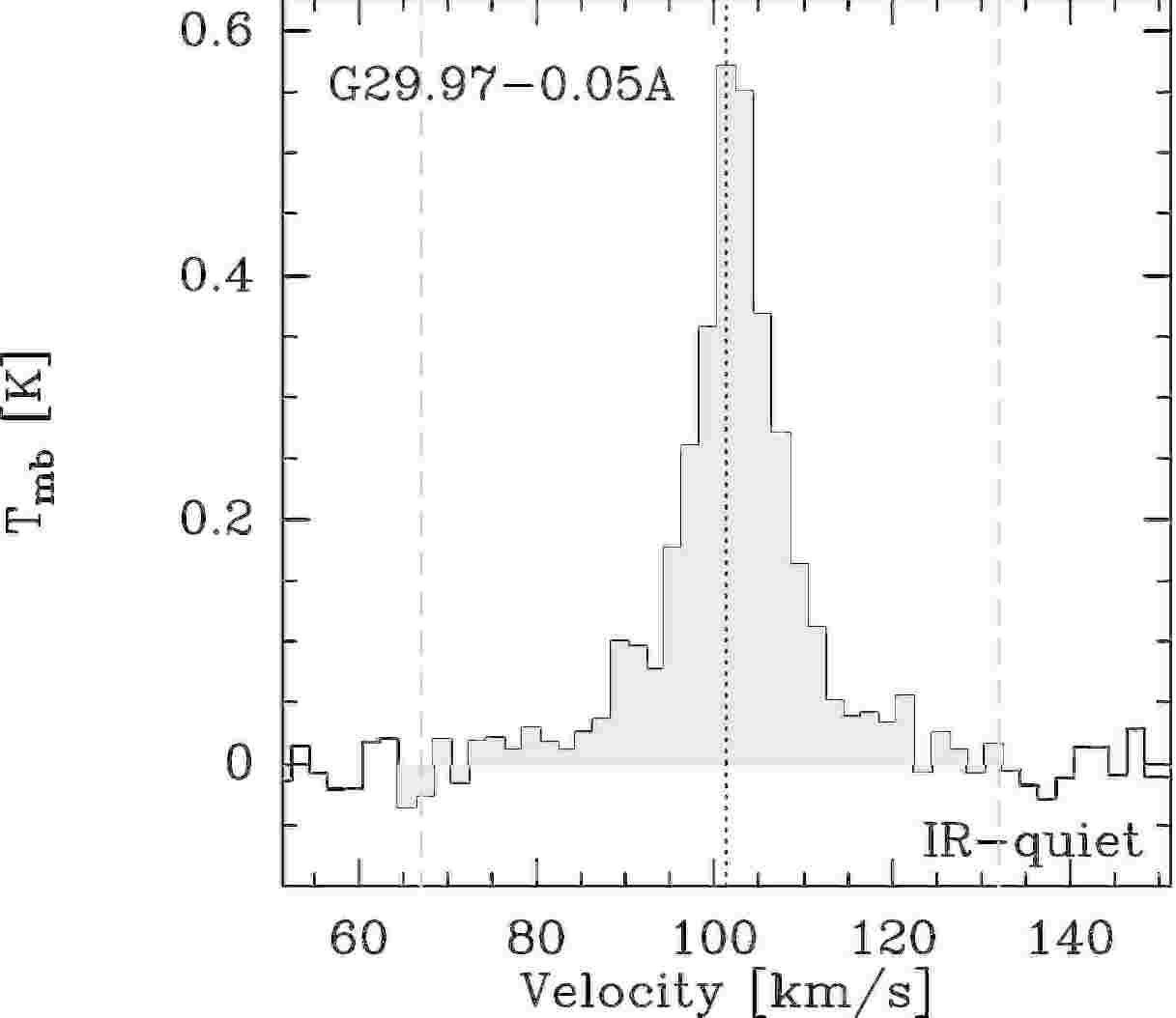} 
  \includegraphics[width=5.6cm,angle=0]{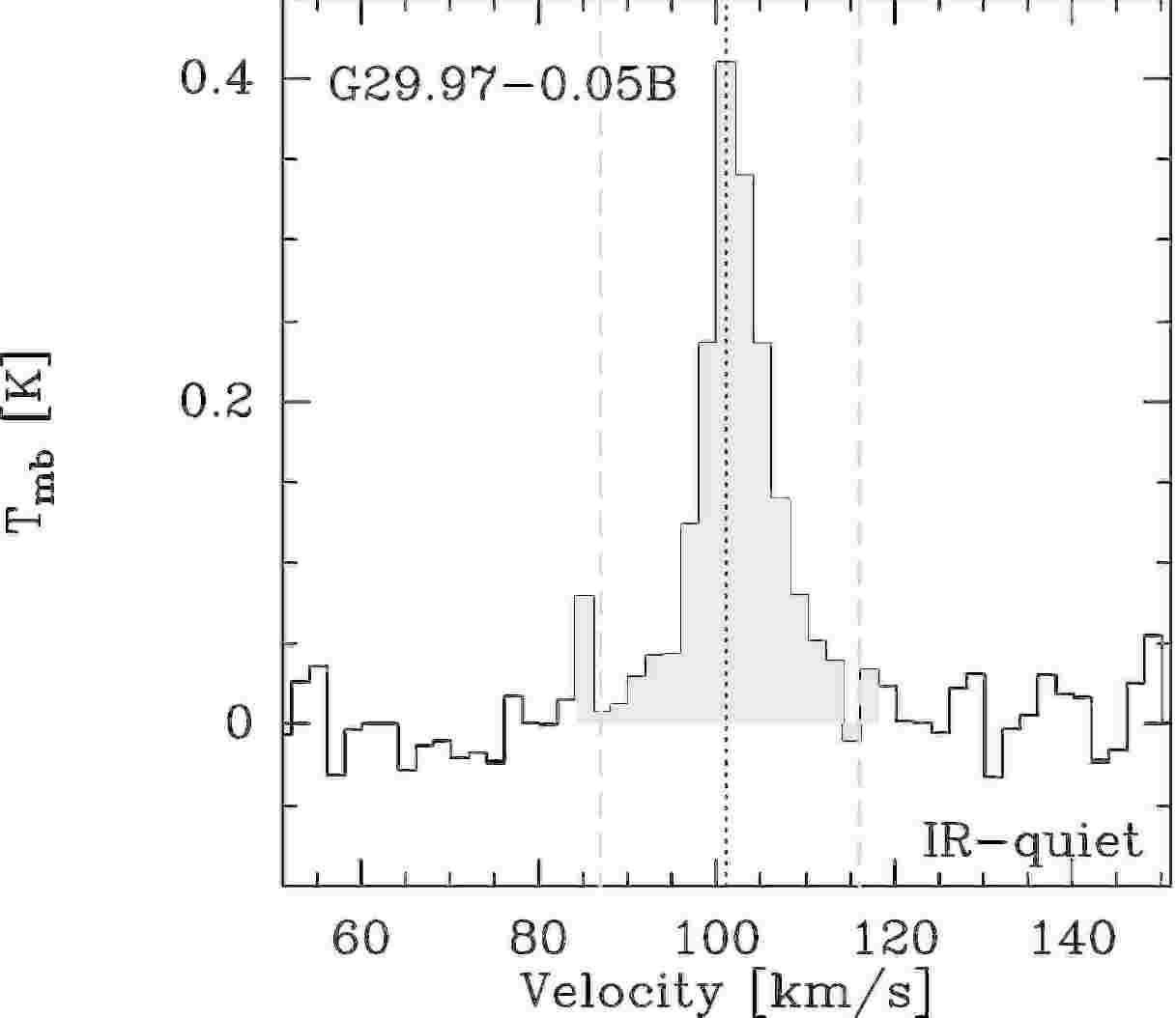} 
  \includegraphics[width=5.6cm,angle=0]{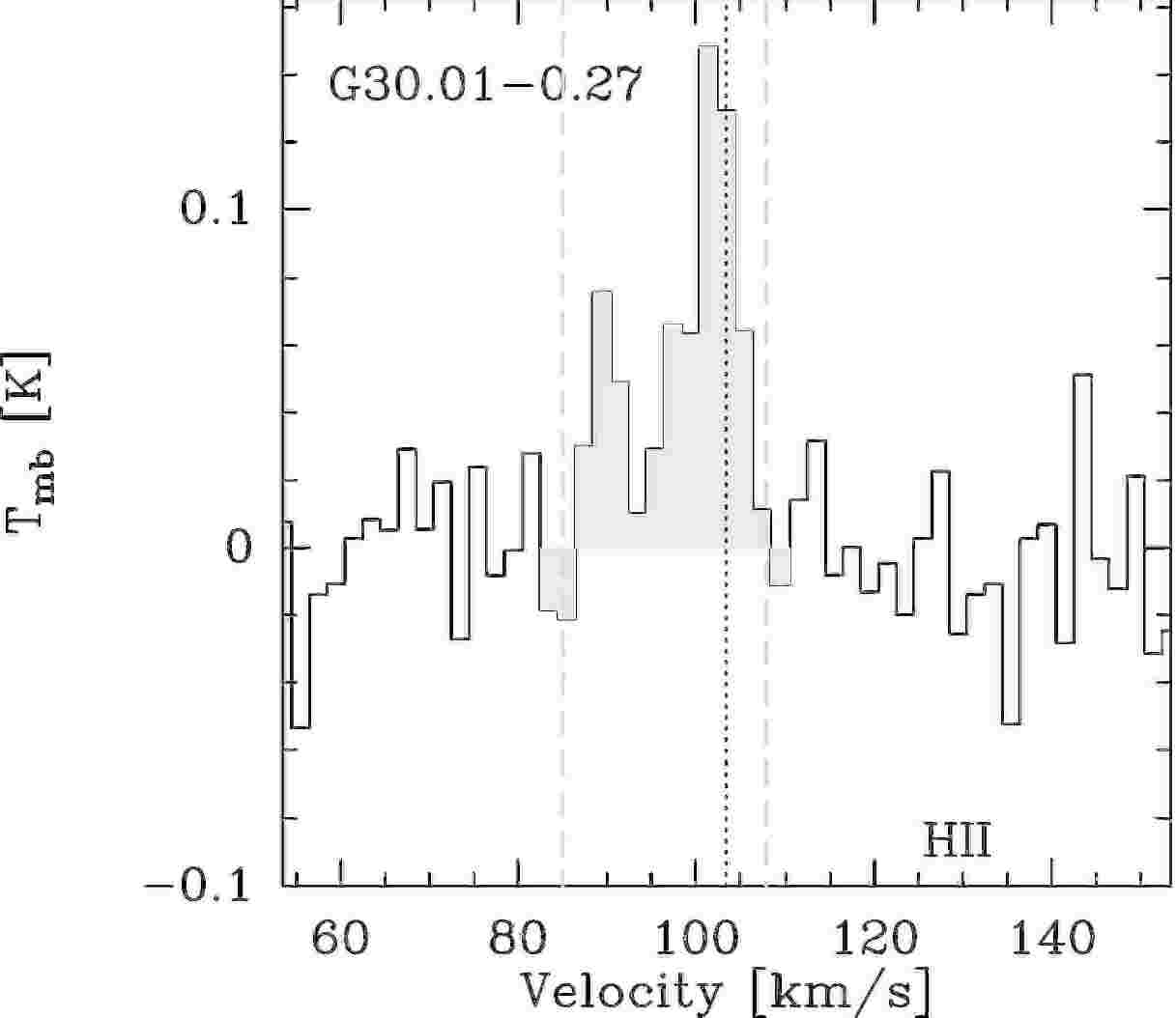} 
  \includegraphics[width=5.6cm,angle=0]{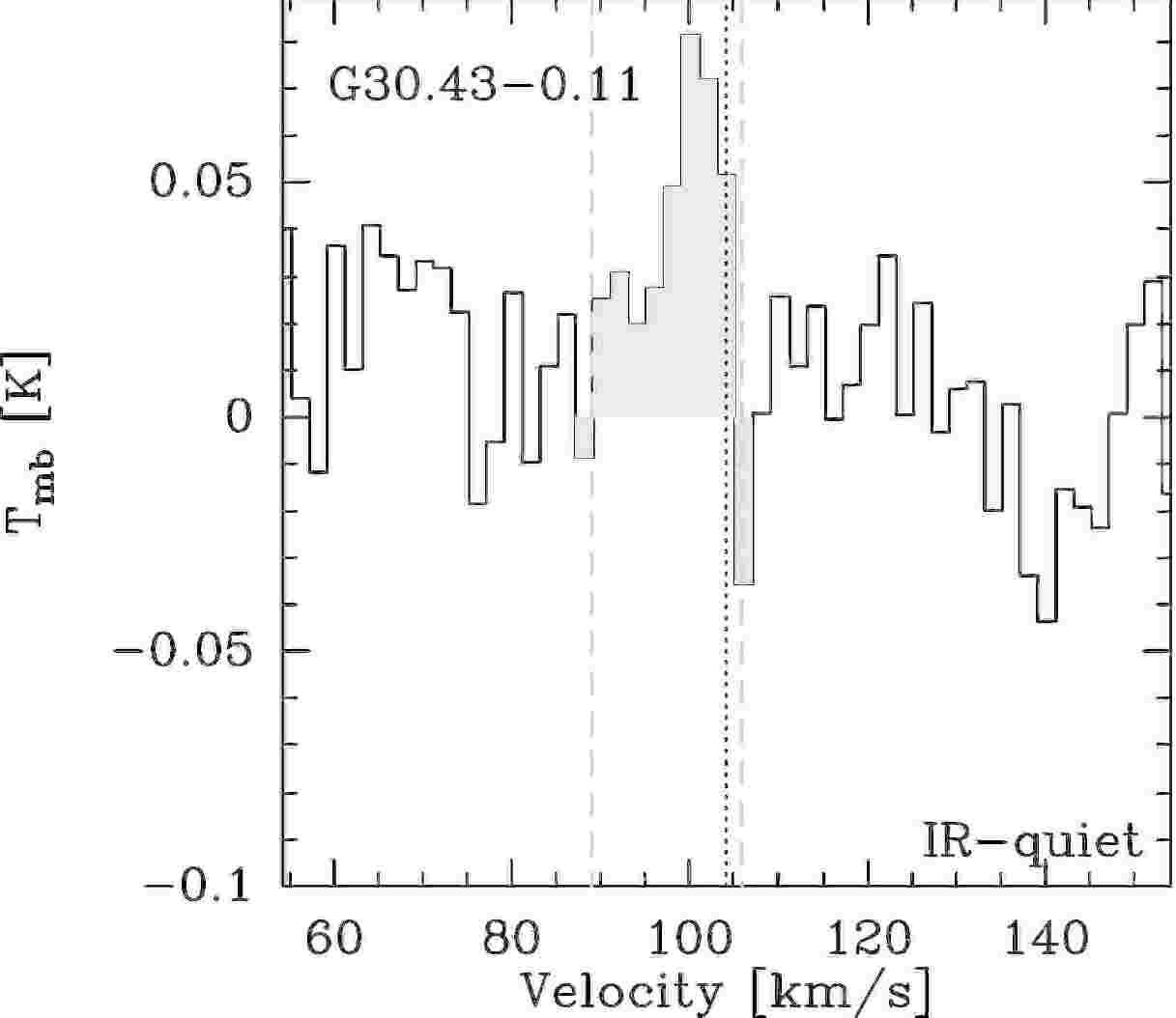} 
  \includegraphics[width=5.6cm,angle=0]{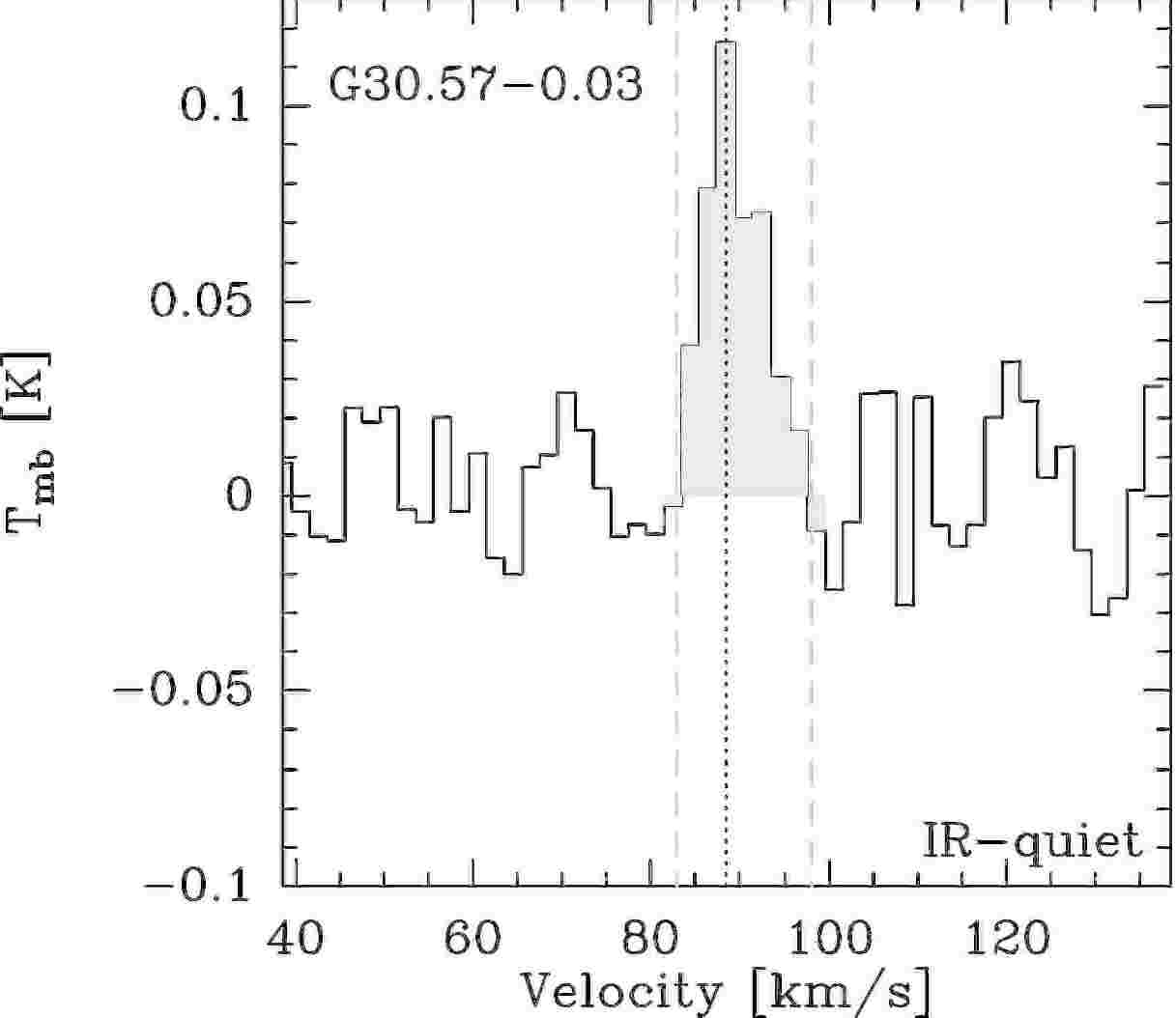} 
  \includegraphics[width=5.6cm,angle=0]{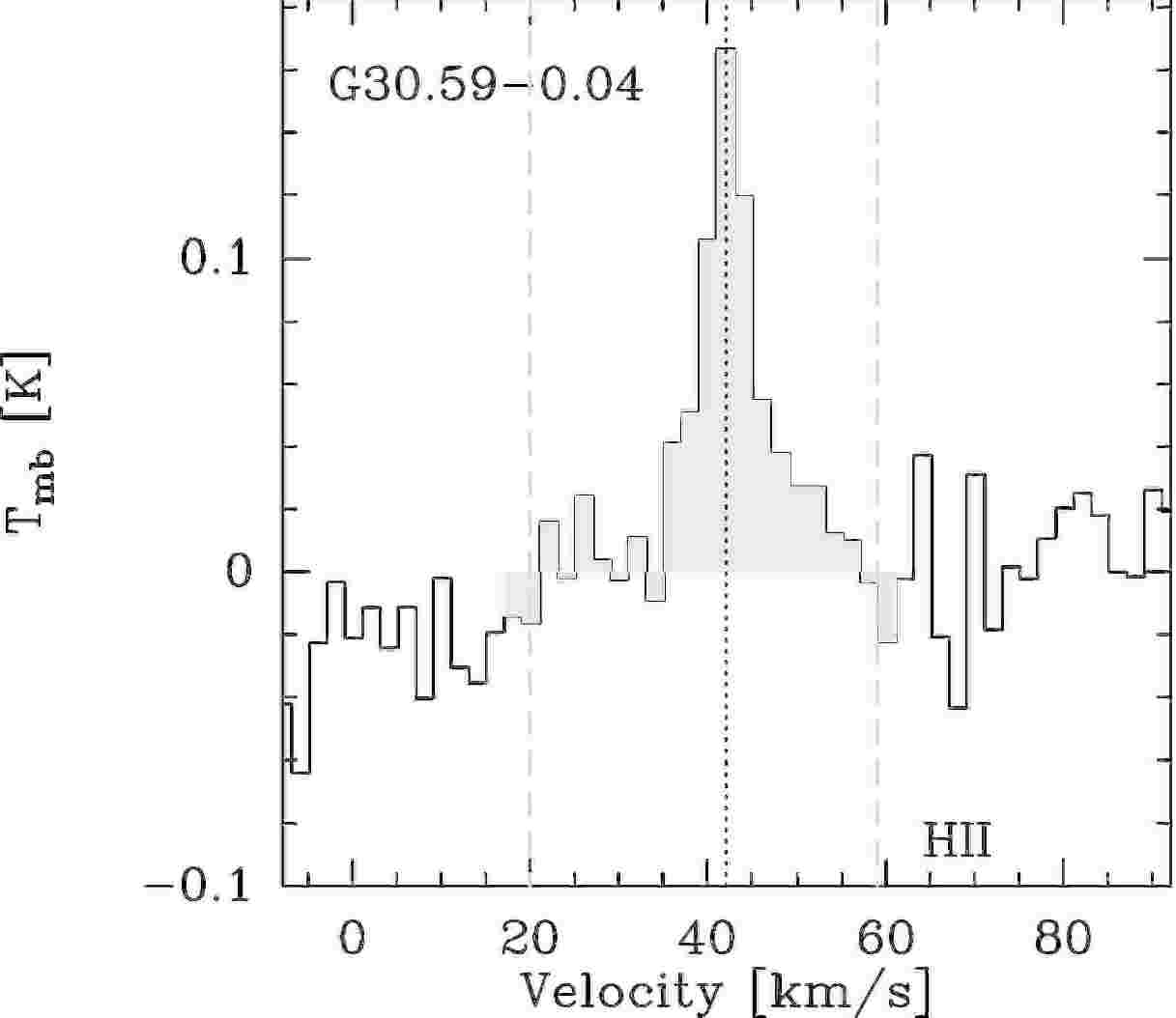} 
 \caption{Continued.}
\end{figure}
\end{landscape}

\begin{landscape}
\begin{figure}
\centering
\ContinuedFloat
  \includegraphics[width=5.6cm,angle=0]{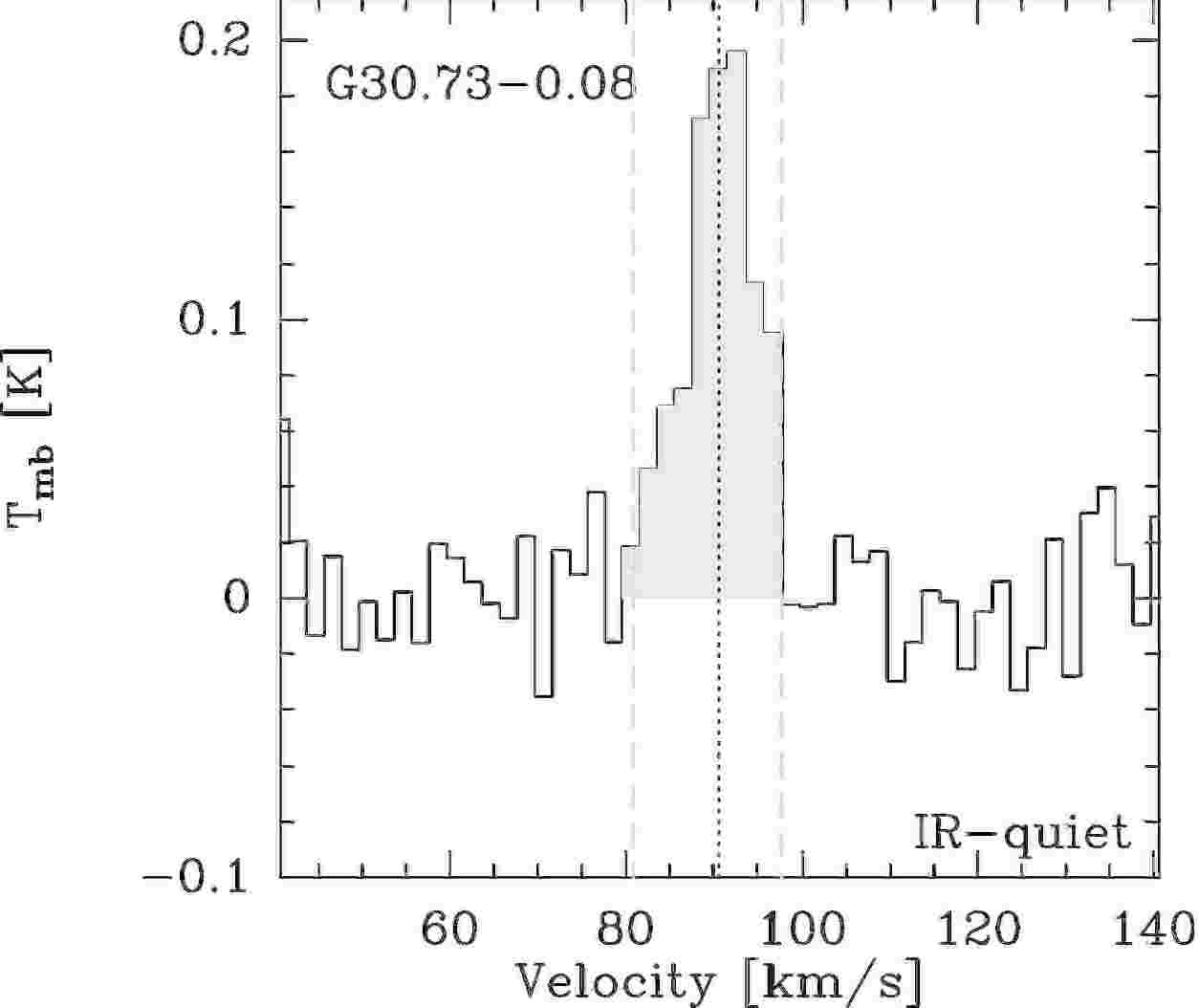} 
  \includegraphics[width=5.6cm,angle=0]{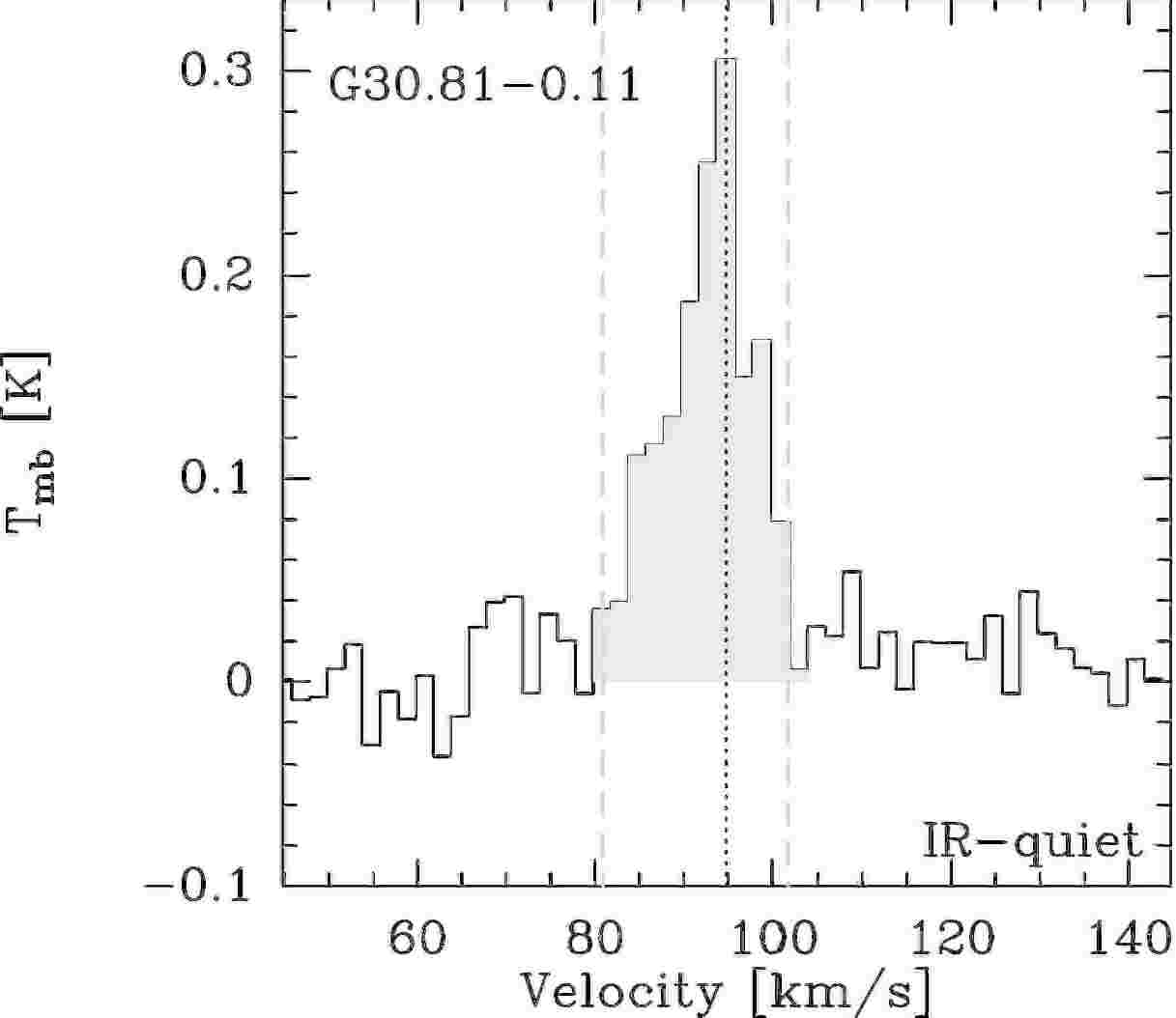} 
 \includegraphics[width=5.6cm,angle=0]{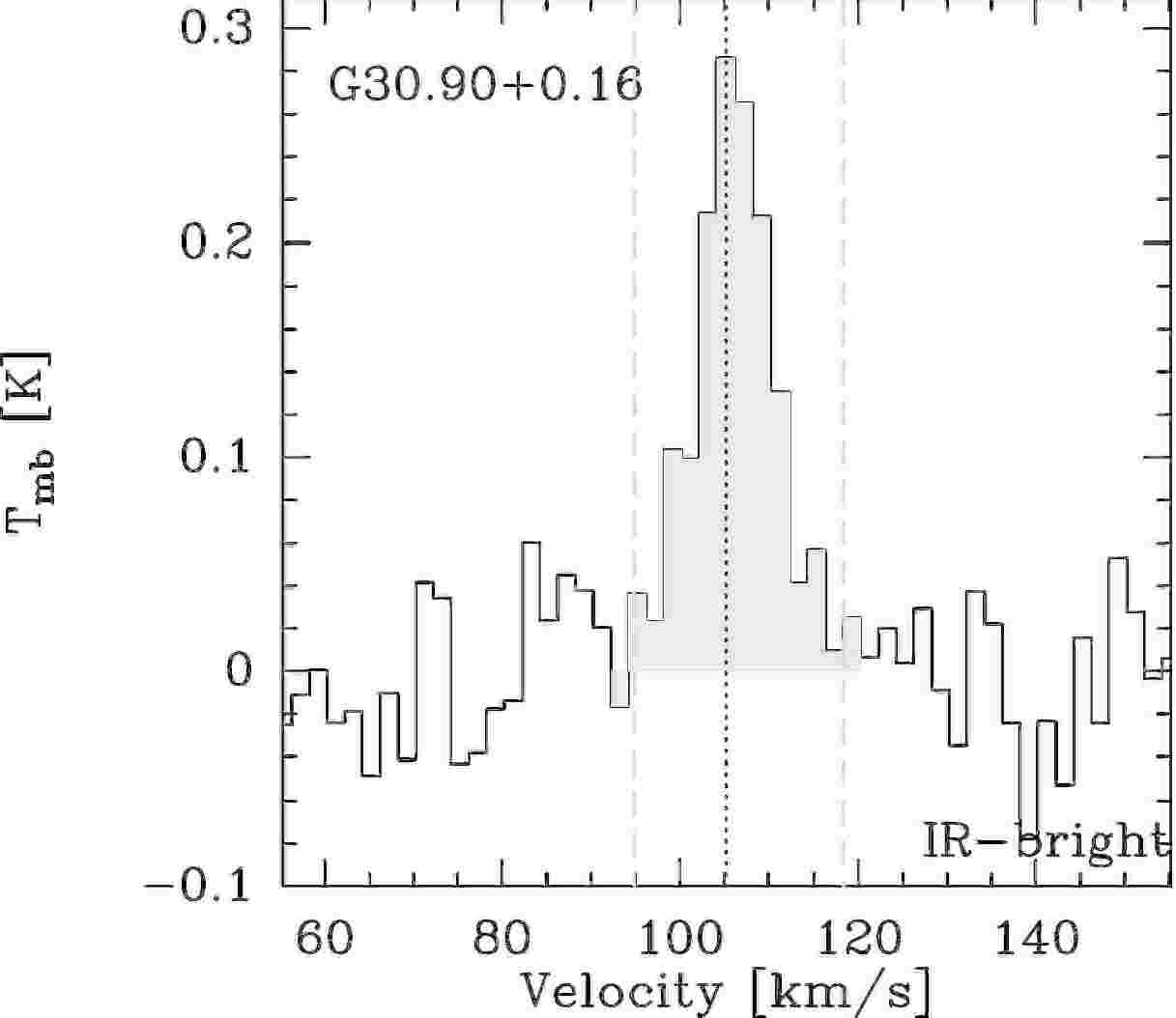} 
  \includegraphics[width=5.6cm,angle=0]{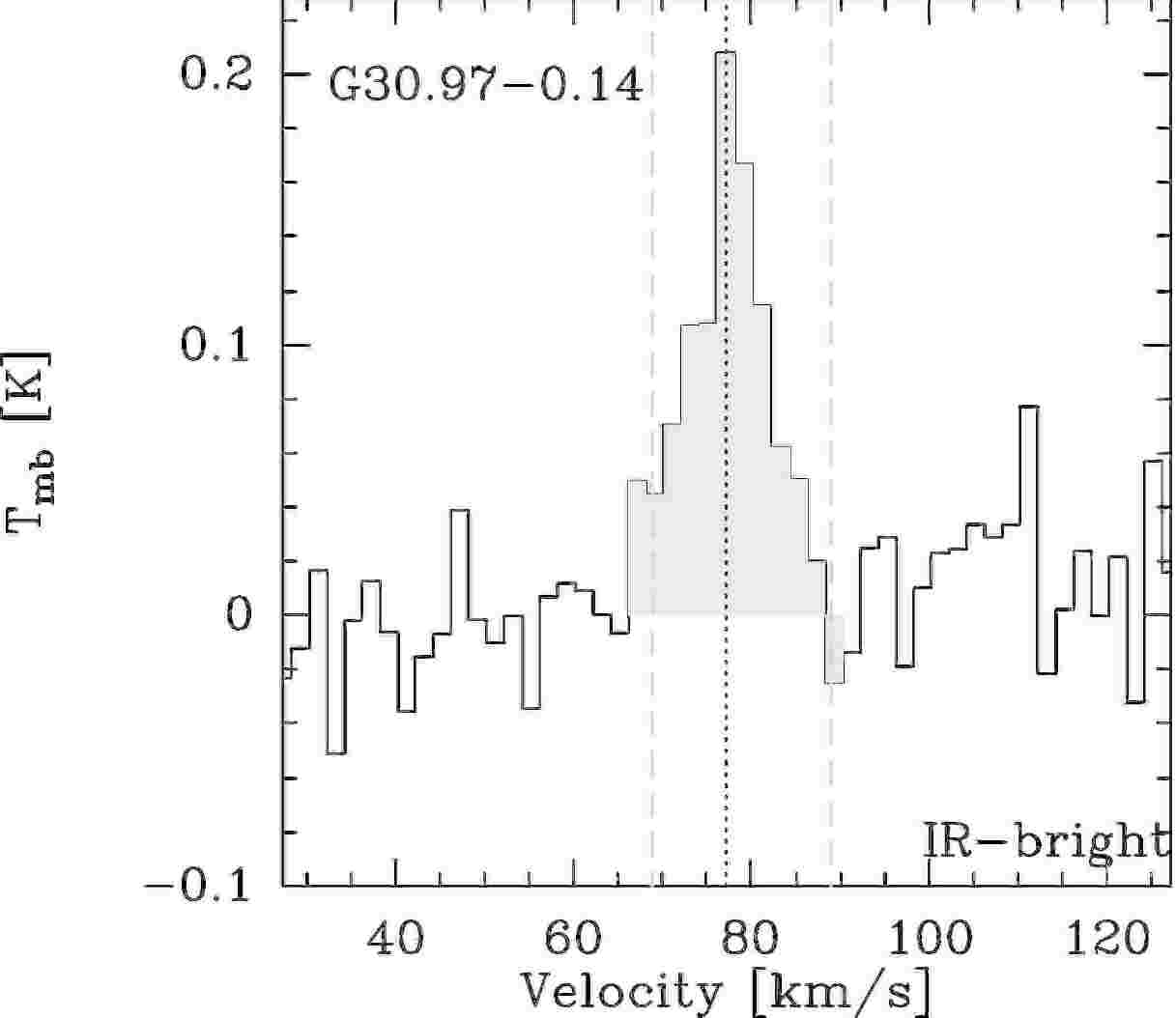} 
  \includegraphics[width=5.6cm,angle=0]{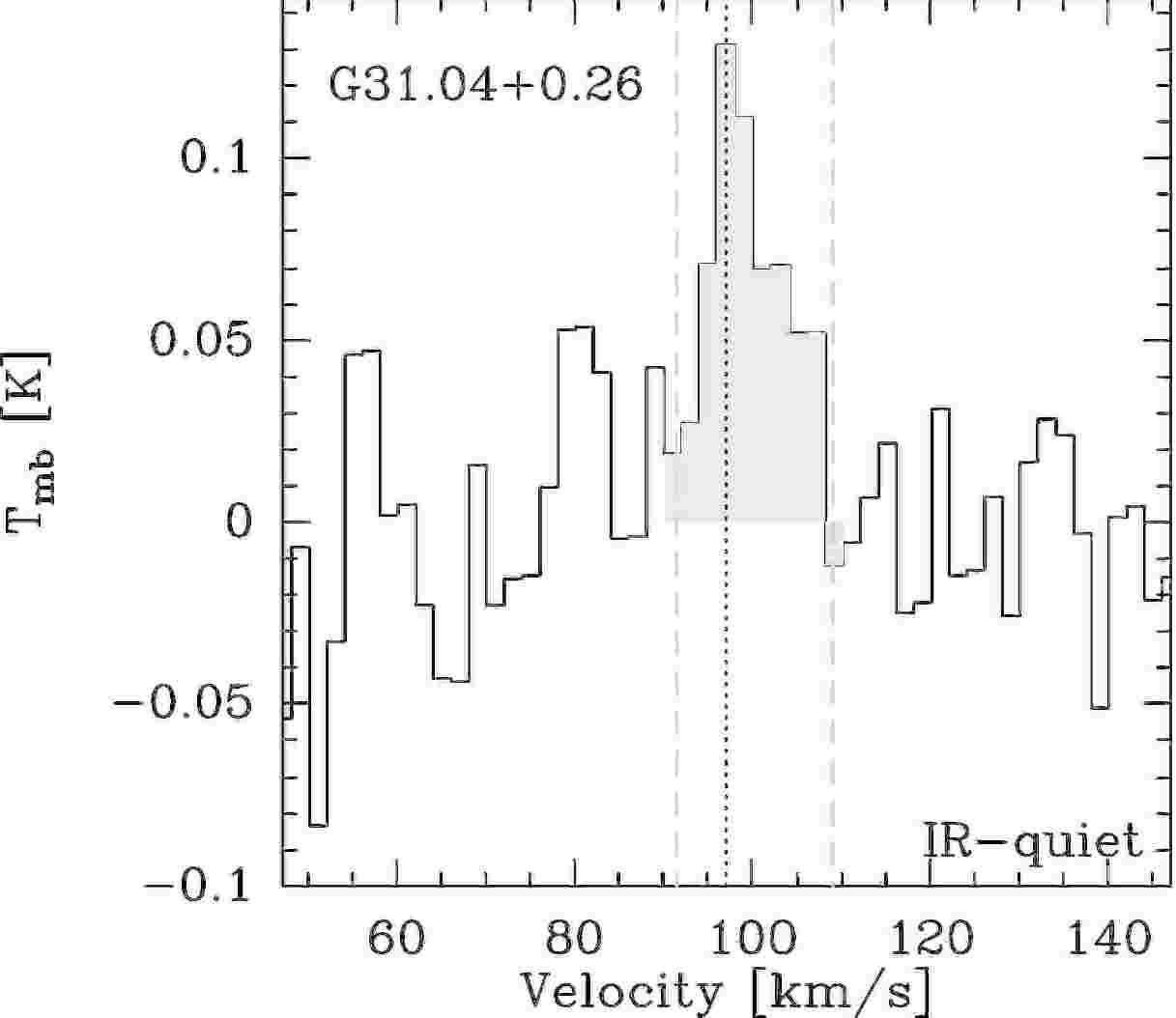} 
  \includegraphics[width=5.6cm,angle=0]{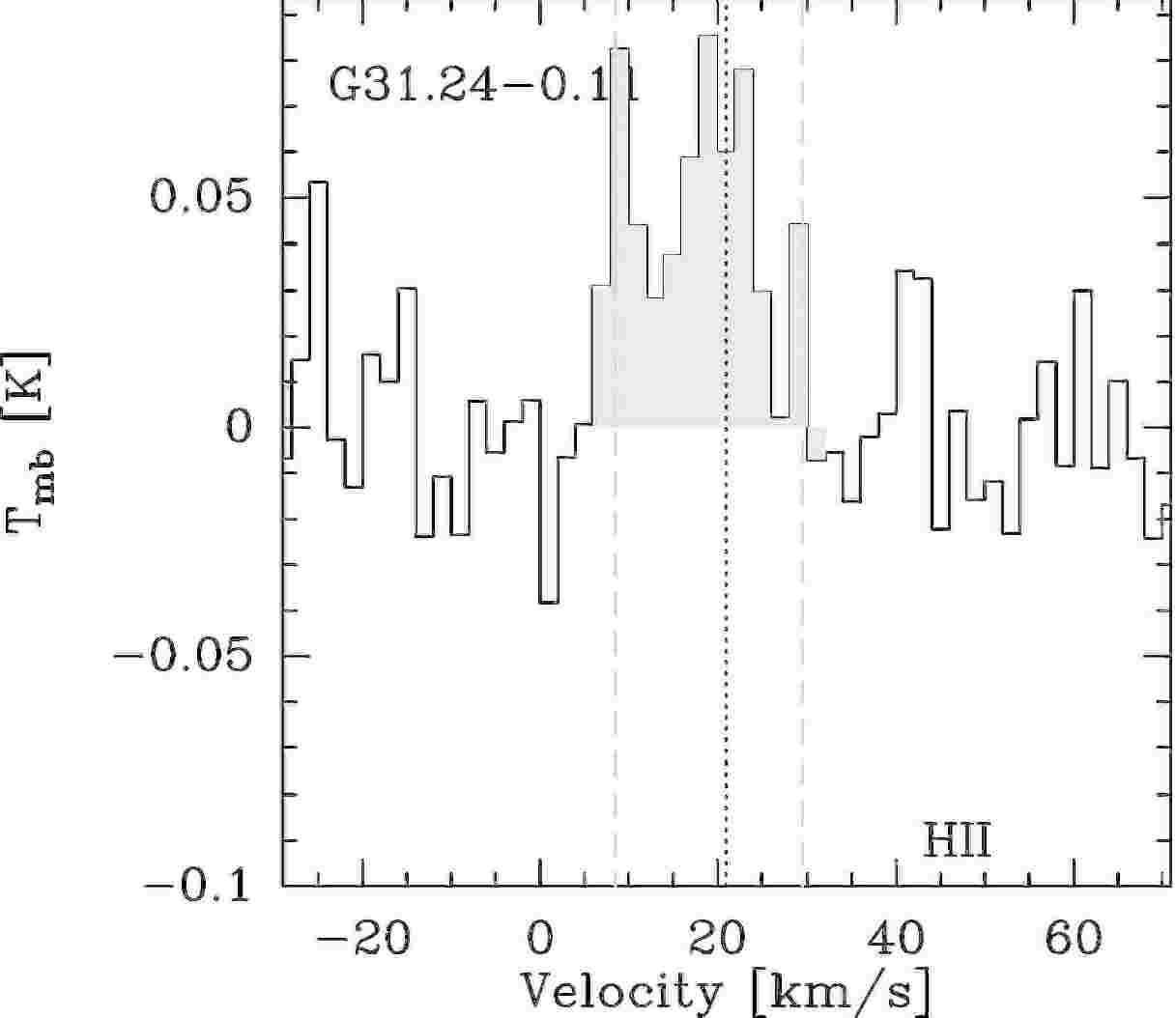} 
  \includegraphics[width=5.6cm,angle=0]{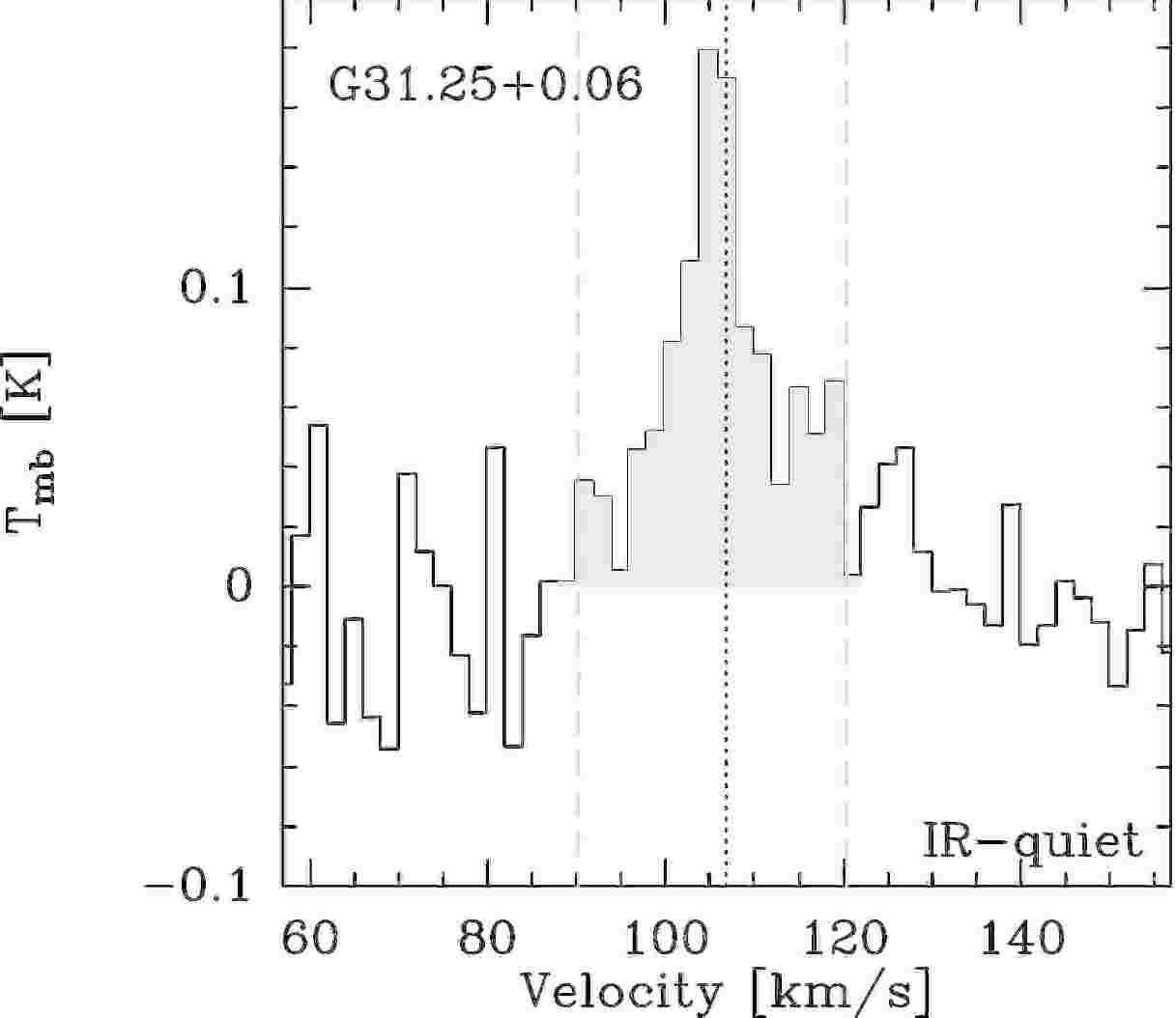} 
 \includegraphics[width=5.6cm,angle=0]{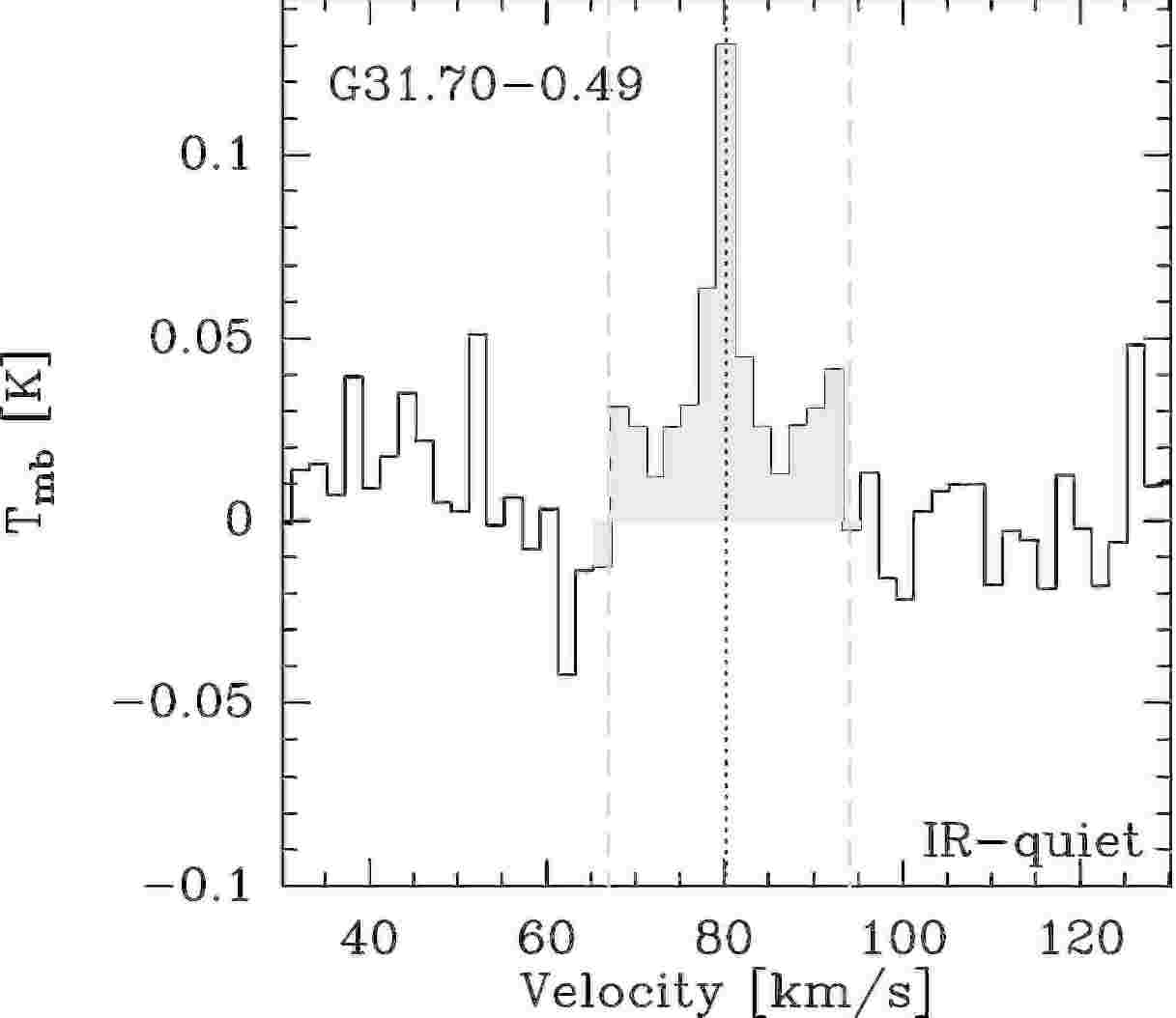} 
  \includegraphics[width=5.6cm,angle=0]{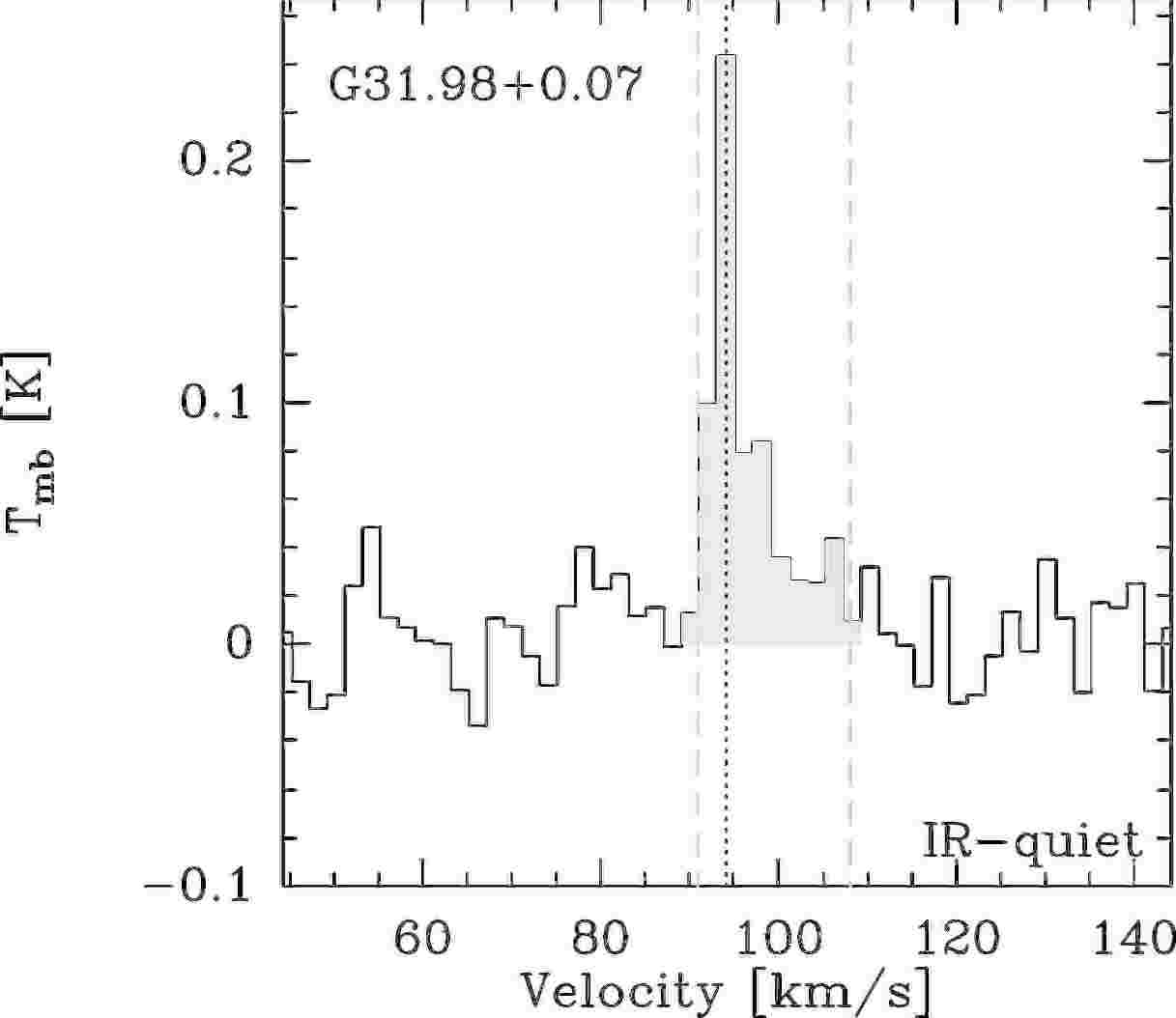} 
  \includegraphics[width=5.6cm,angle=0]{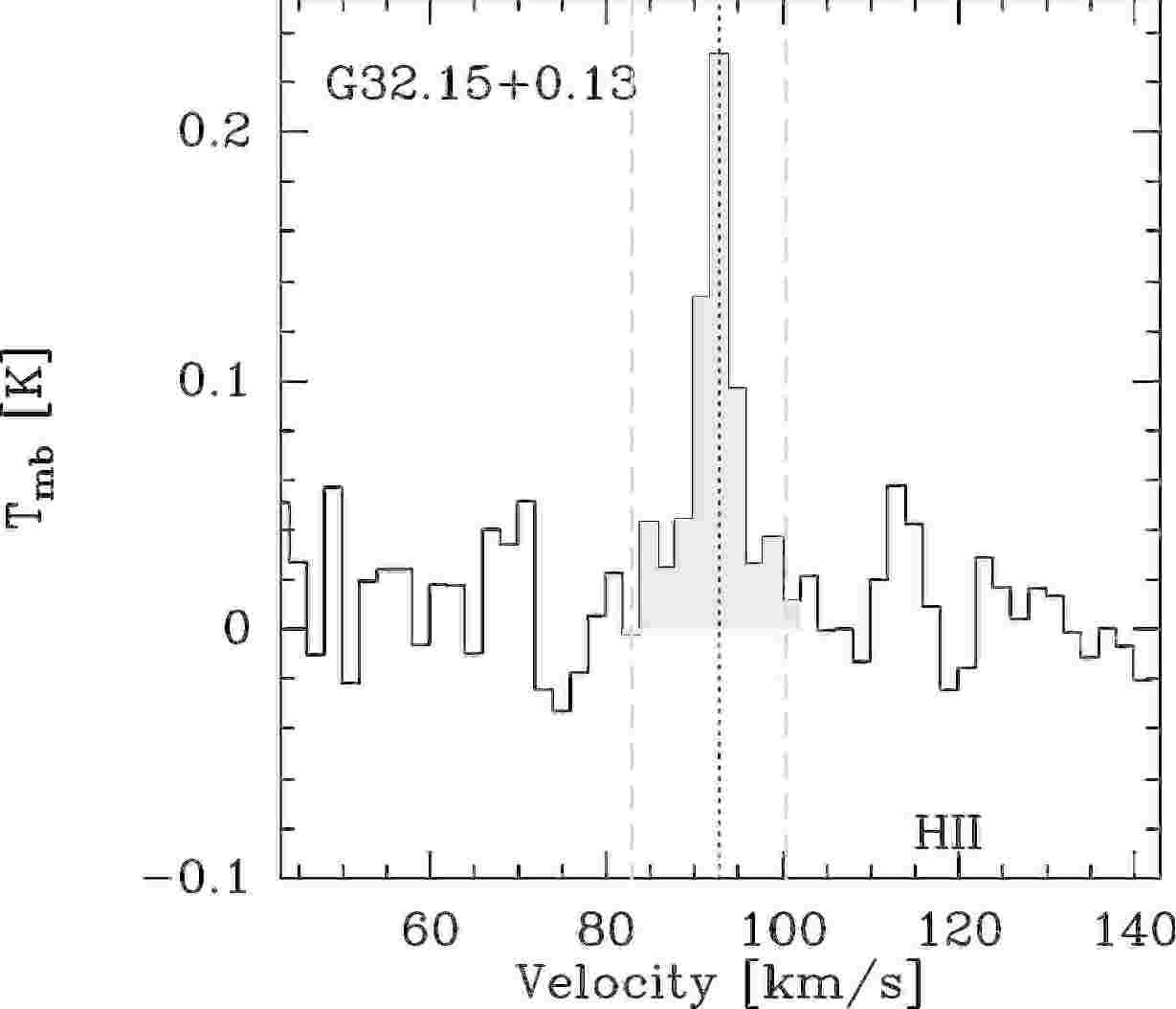} 
  \includegraphics[width=5.6cm,angle=0]{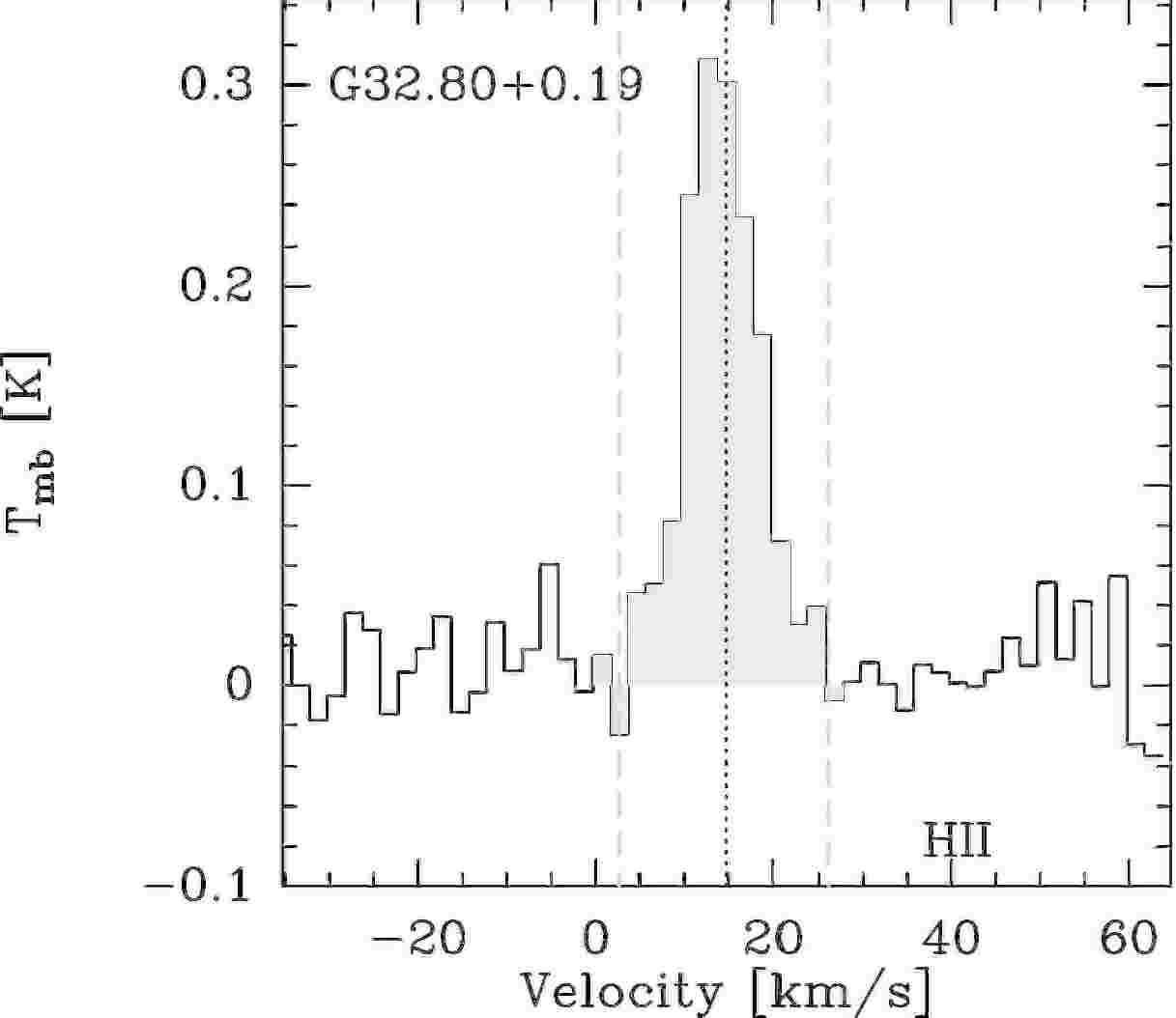} 
  \includegraphics[width=5.6cm,angle=0]{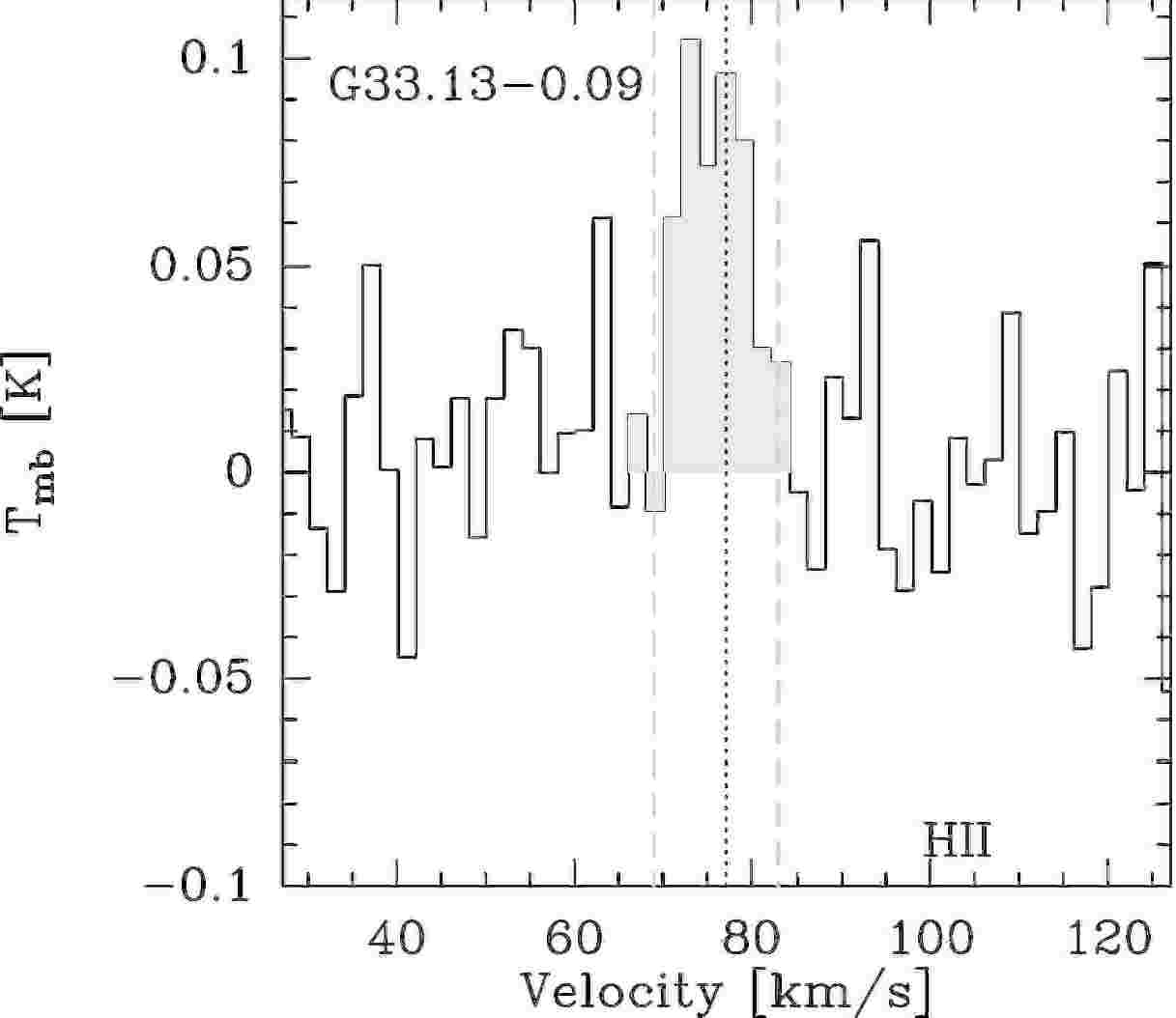} 
 \caption{Continued.}
\end{figure}
\end{landscape}

\begin{landscape}
\begin{figure}
\centering
\ContinuedFloat
  \includegraphics[width=5.6cm,angle=0]{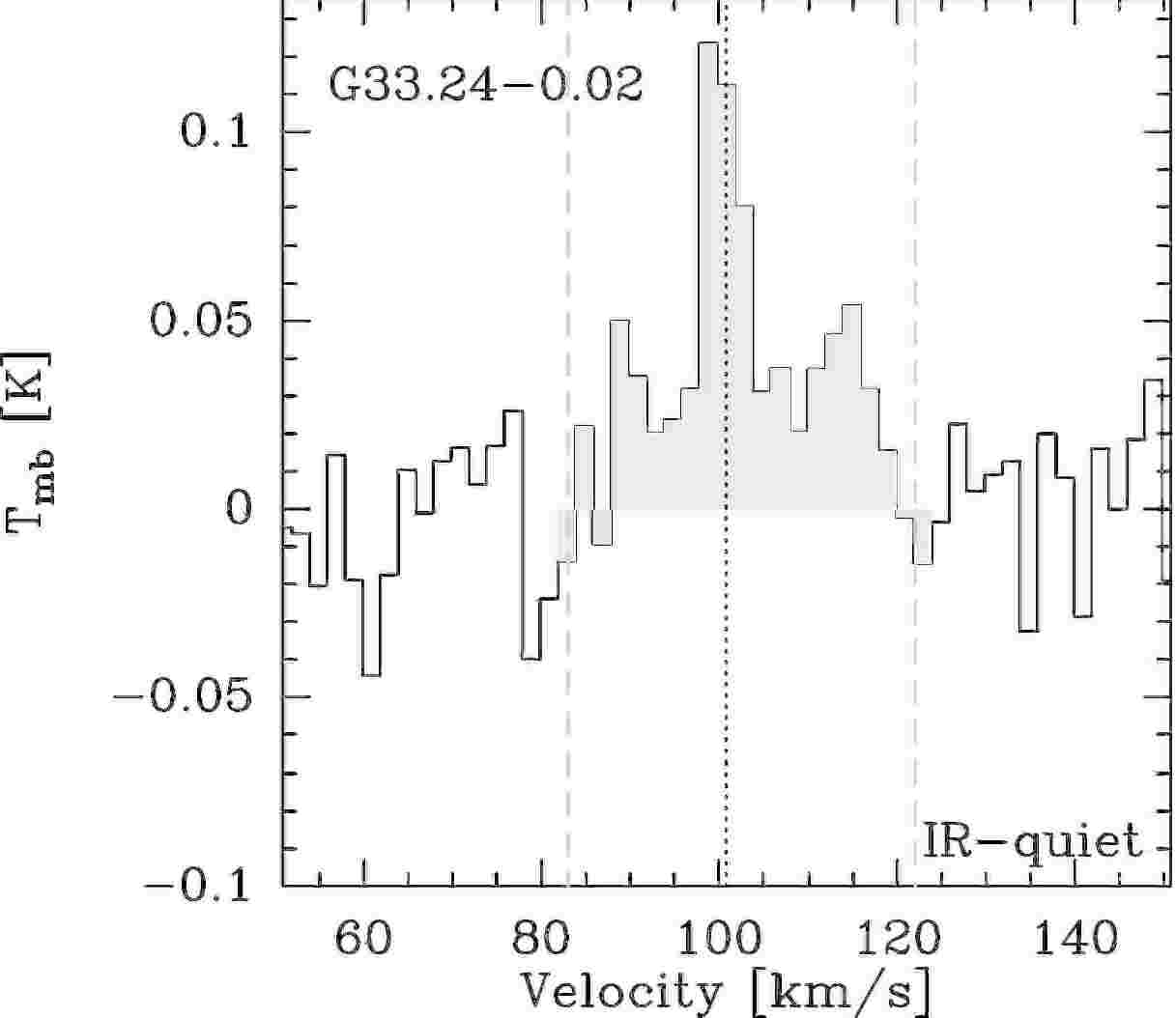} 
  \includegraphics[width=5.6cm,angle=0]{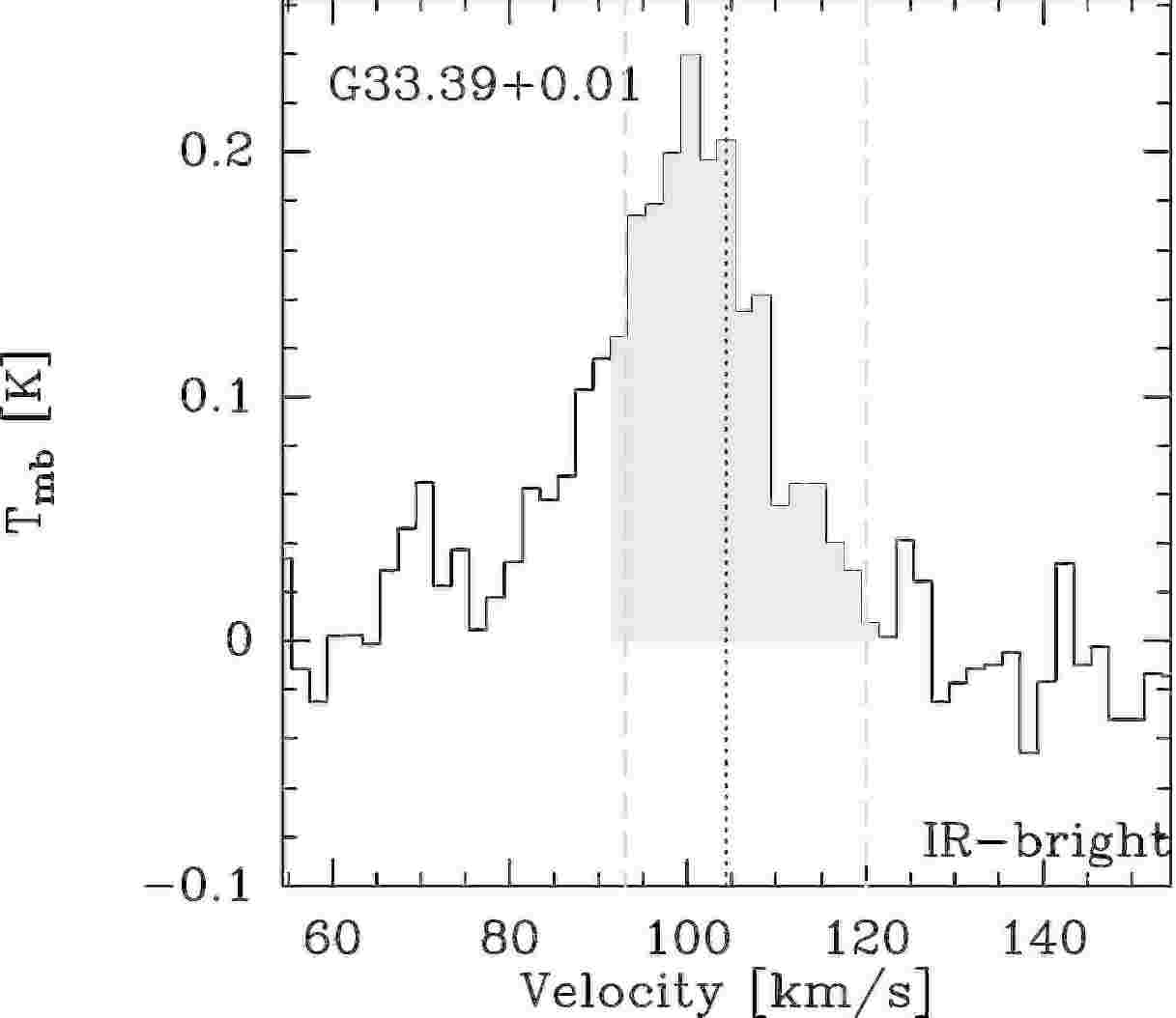} 
  \includegraphics[width=5.6cm,angle=0]{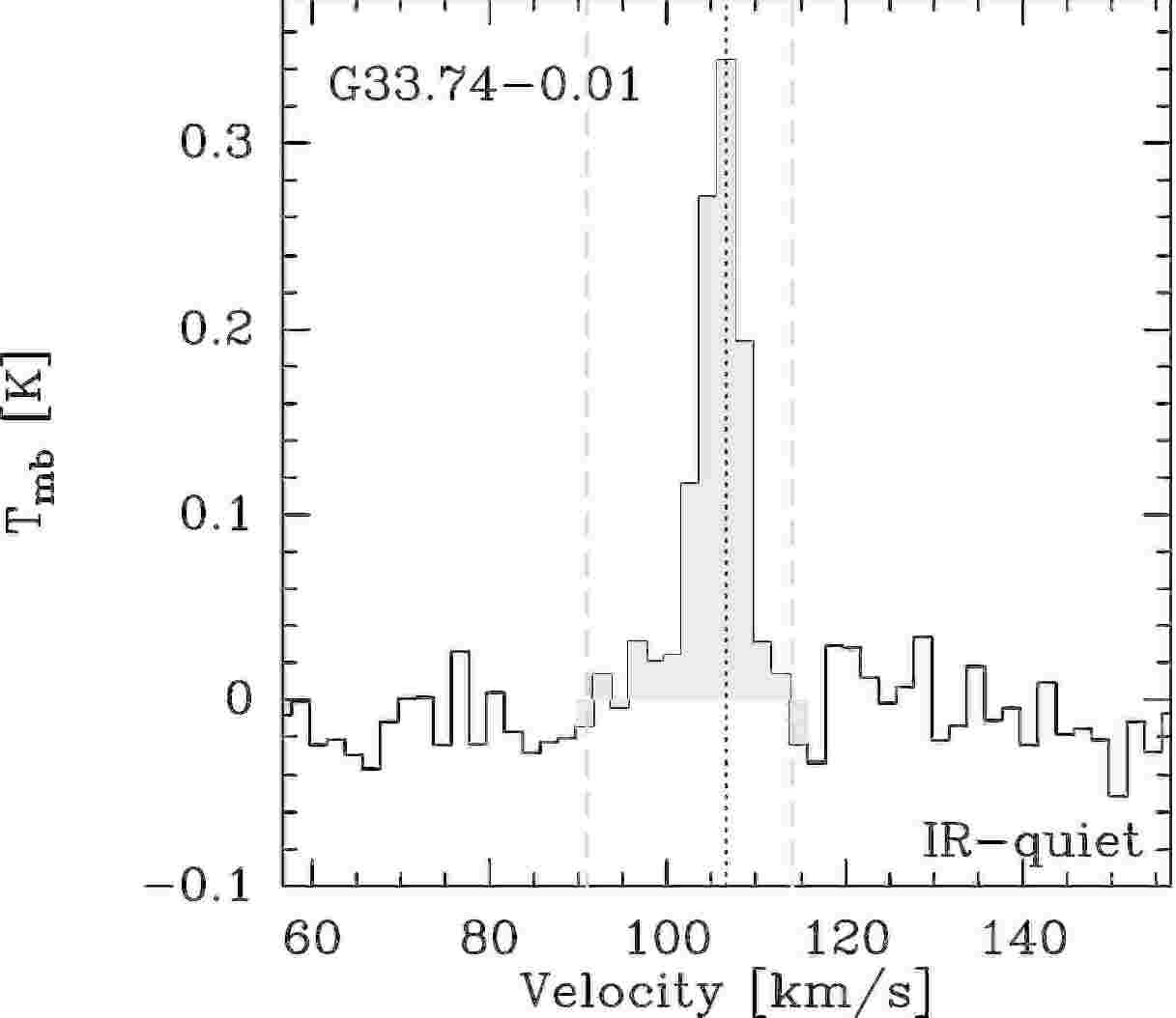} 
 \includegraphics[width=5.6cm,angle=0]{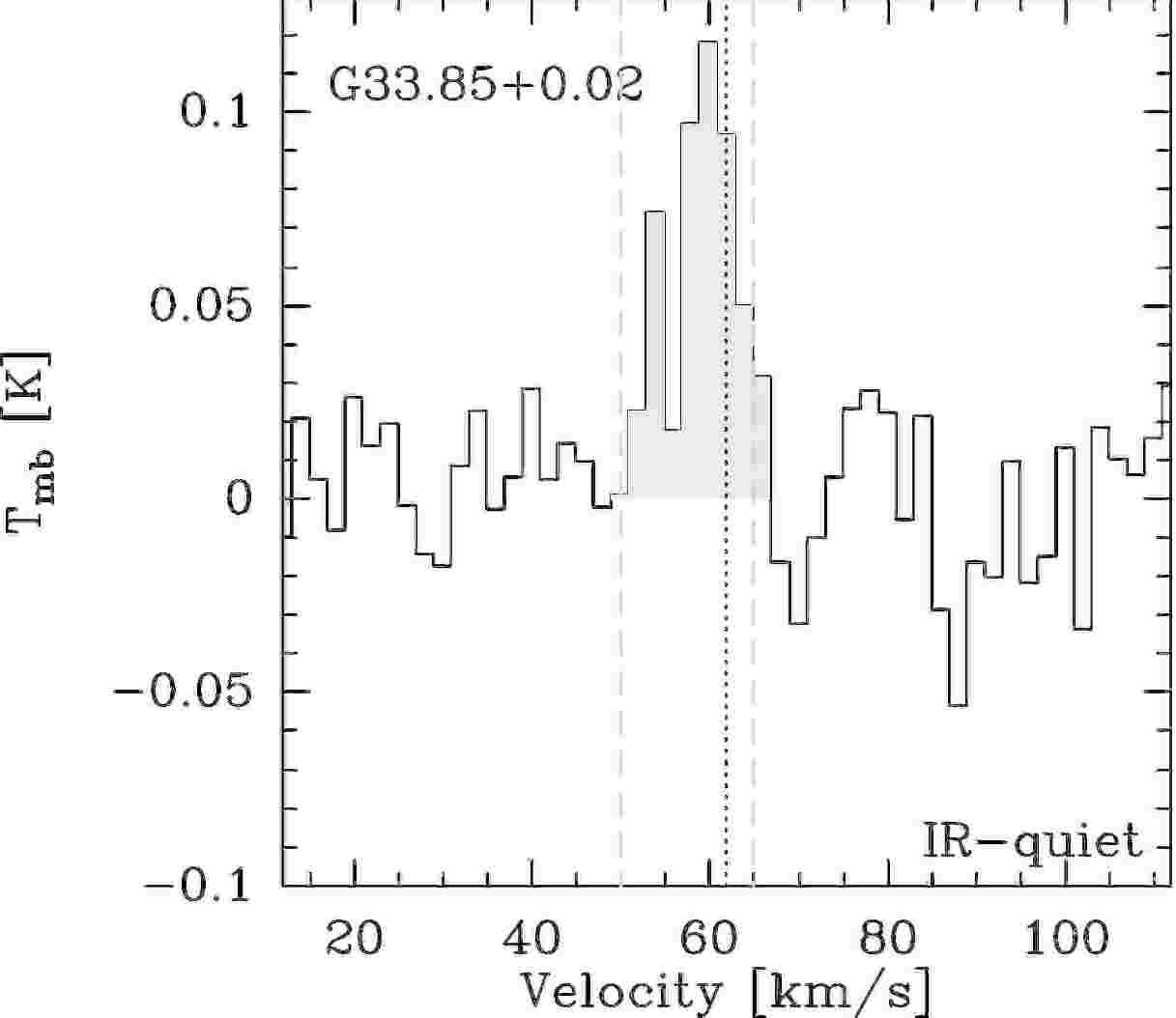} 
  \includegraphics[width=5.6cm,angle=0]{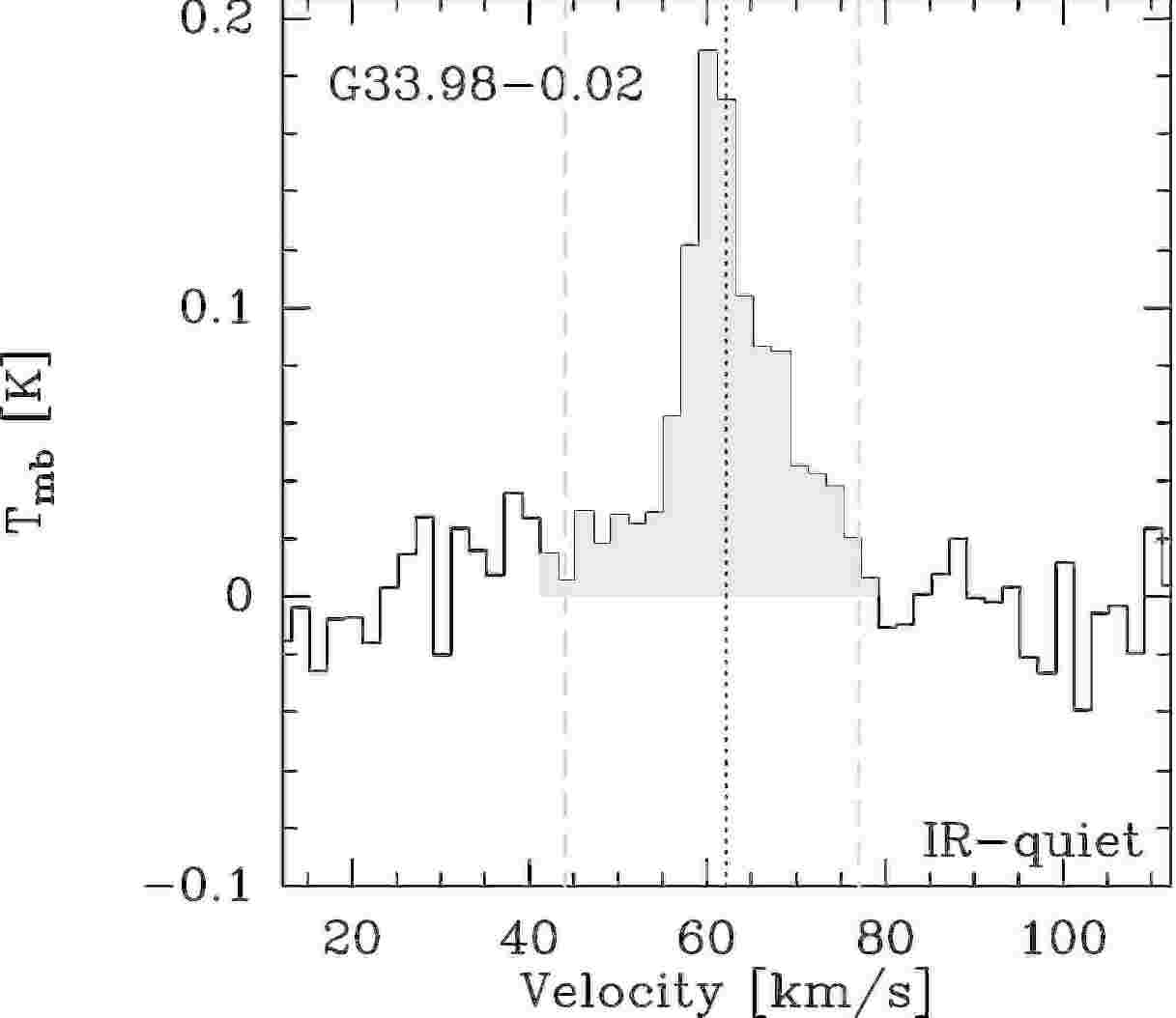} 
  \includegraphics[width=5.6cm,angle=0]{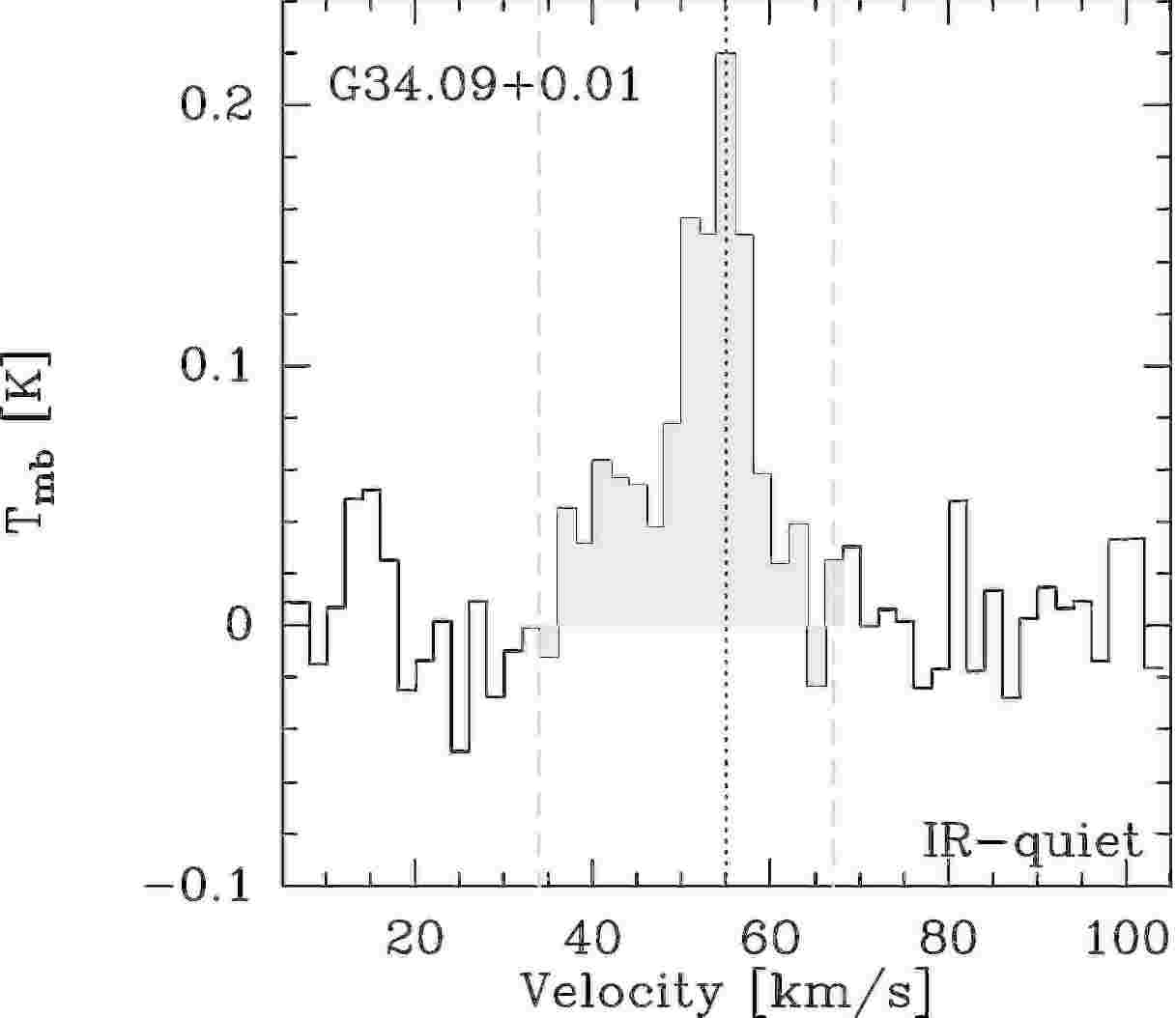} 
   \includegraphics[width=5.6cm,angle=0]{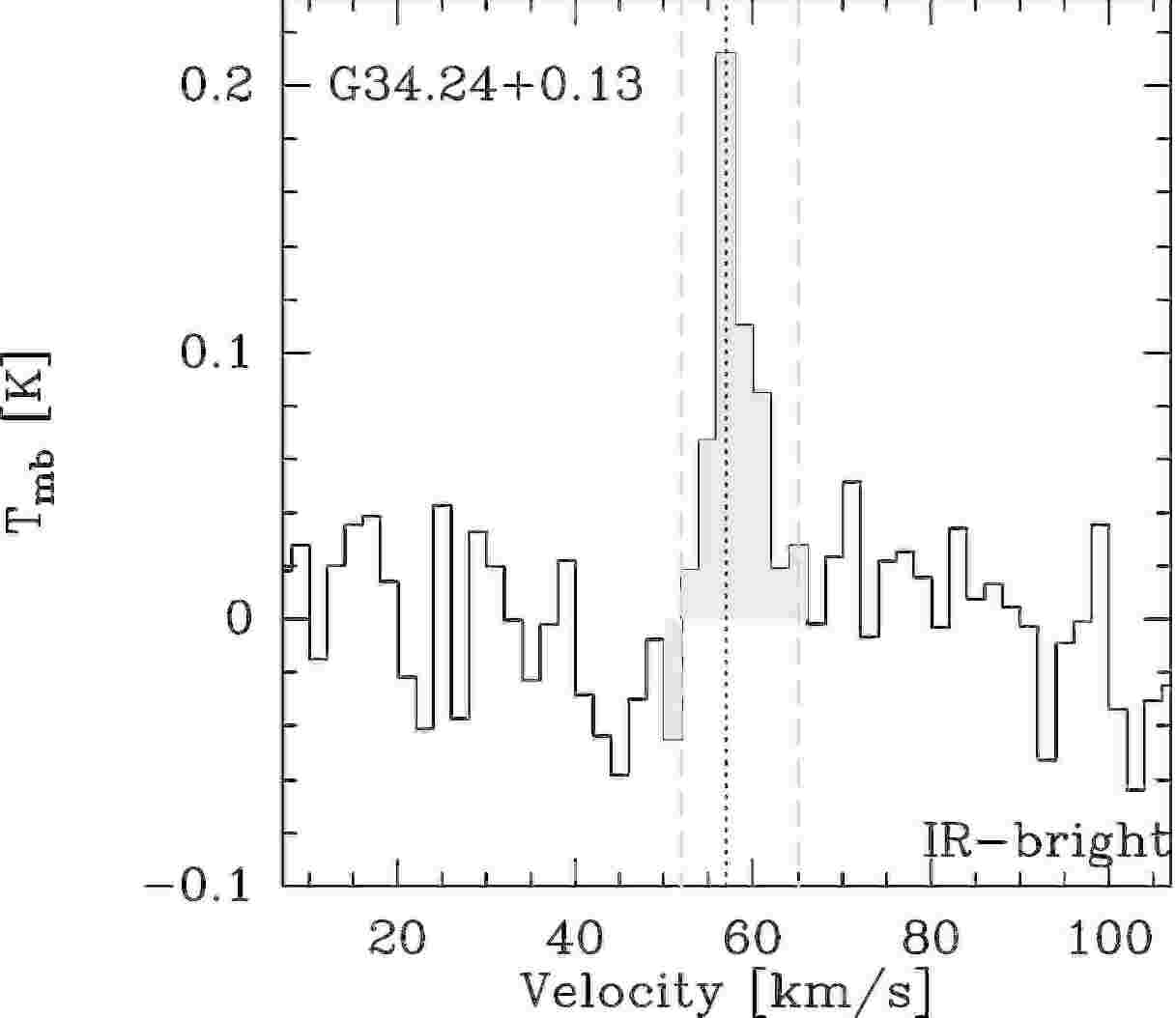} 
  \includegraphics[width=5.6cm,angle=0]{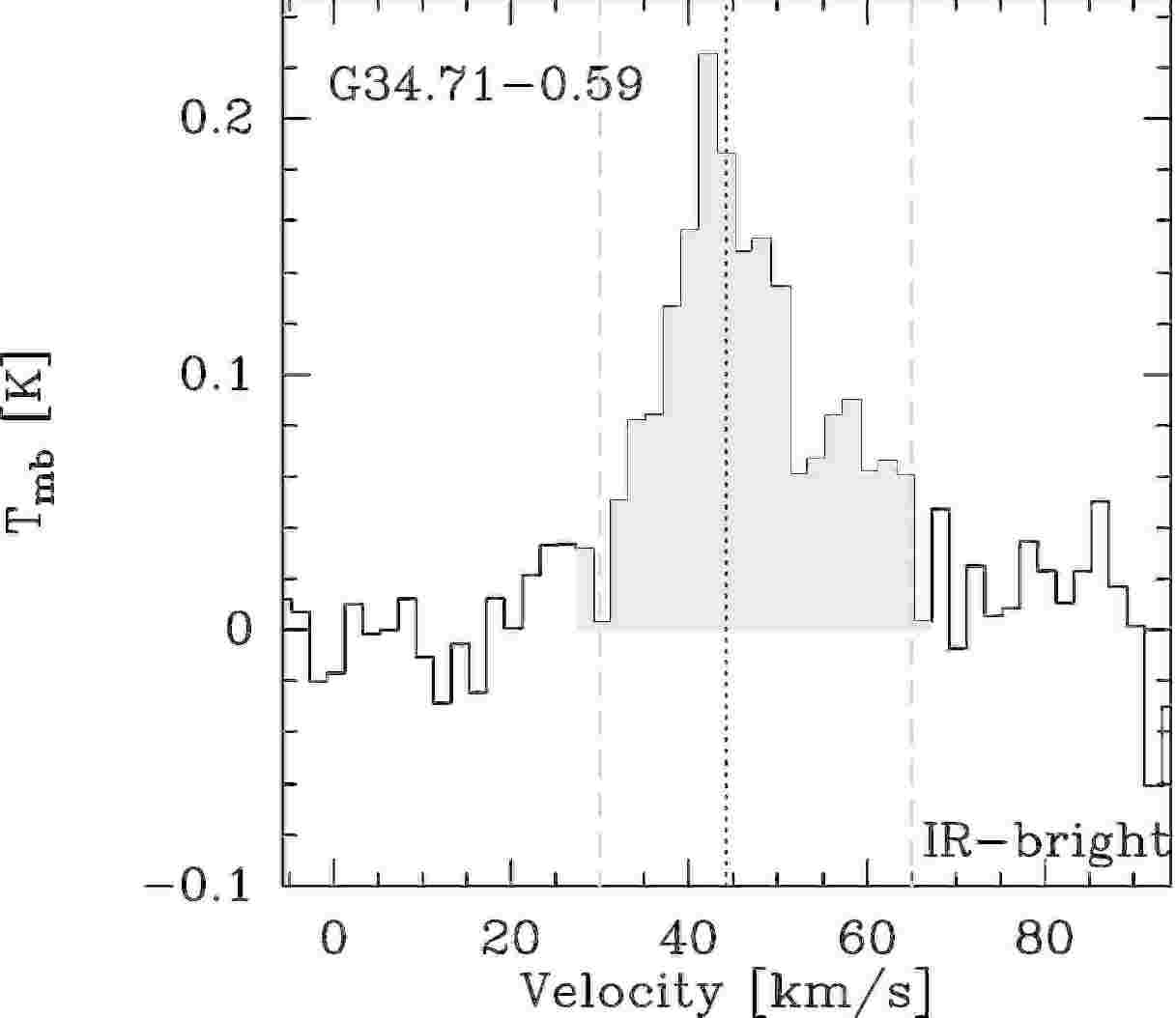} 
  \includegraphics[width=5.6cm,angle=0]{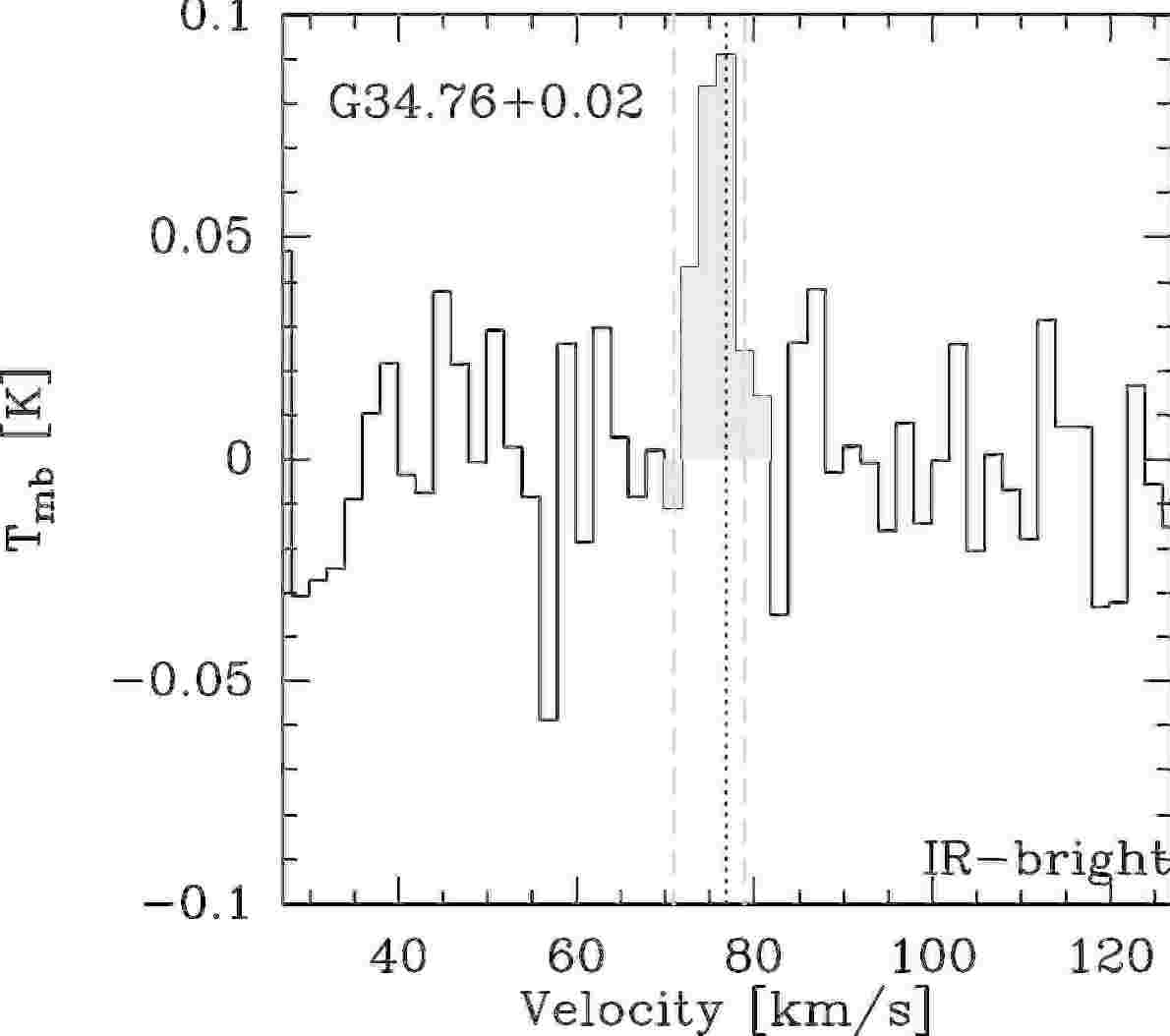} 
  \includegraphics[width=5.6cm,angle=0]{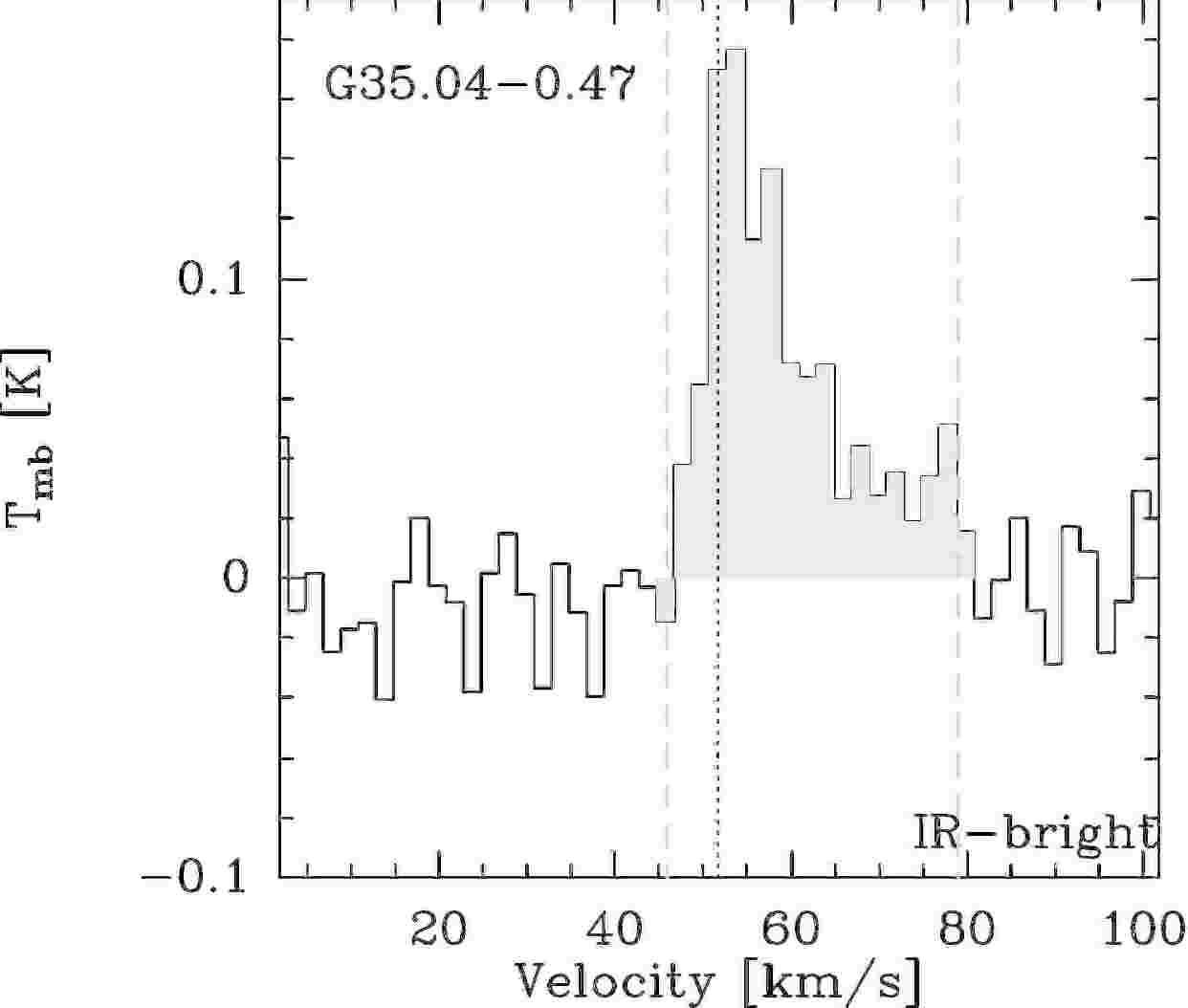} 
  \includegraphics[width=5.6cm,angle=0]{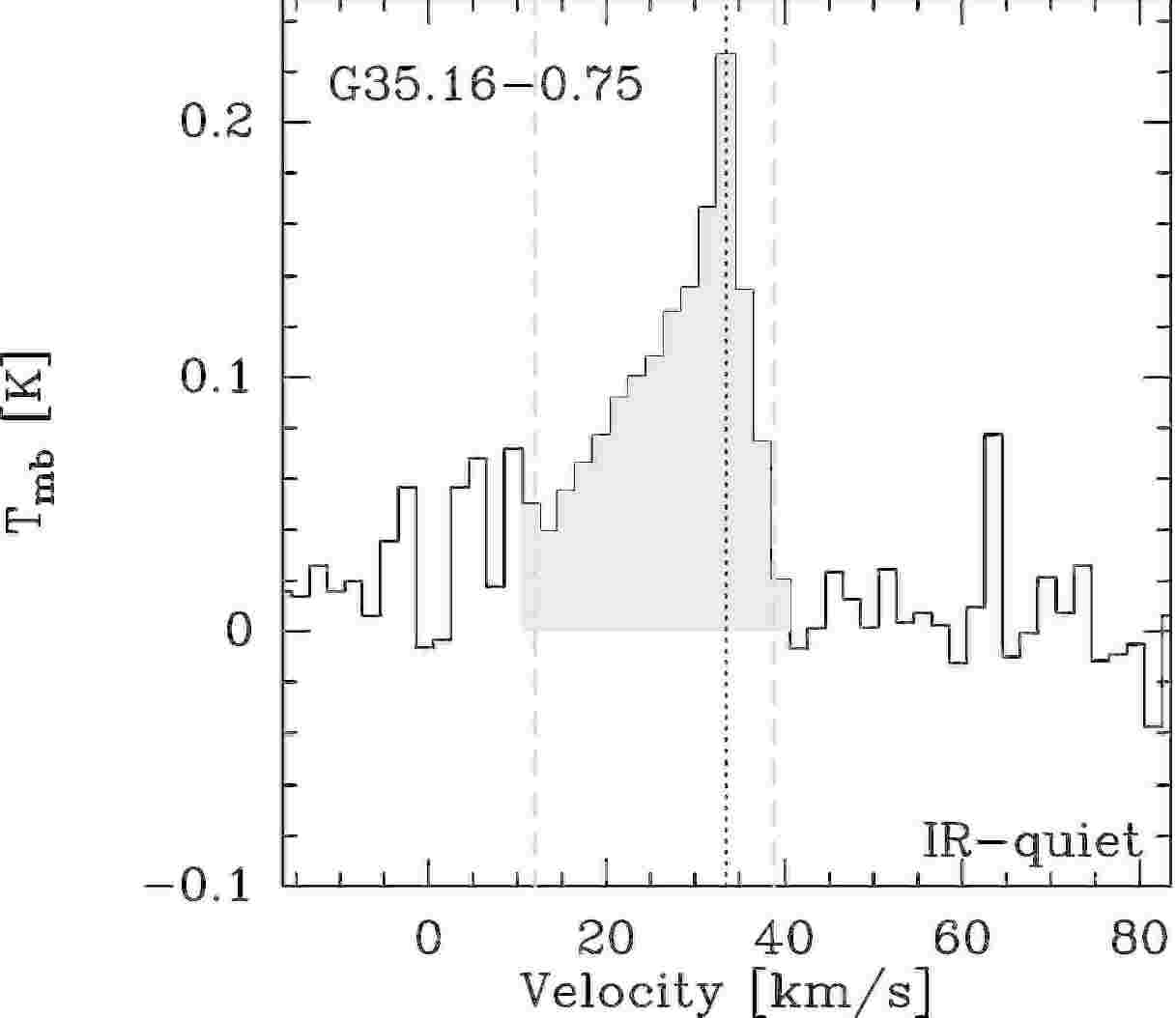} 
  \includegraphics[width=5.6cm,angle=0]{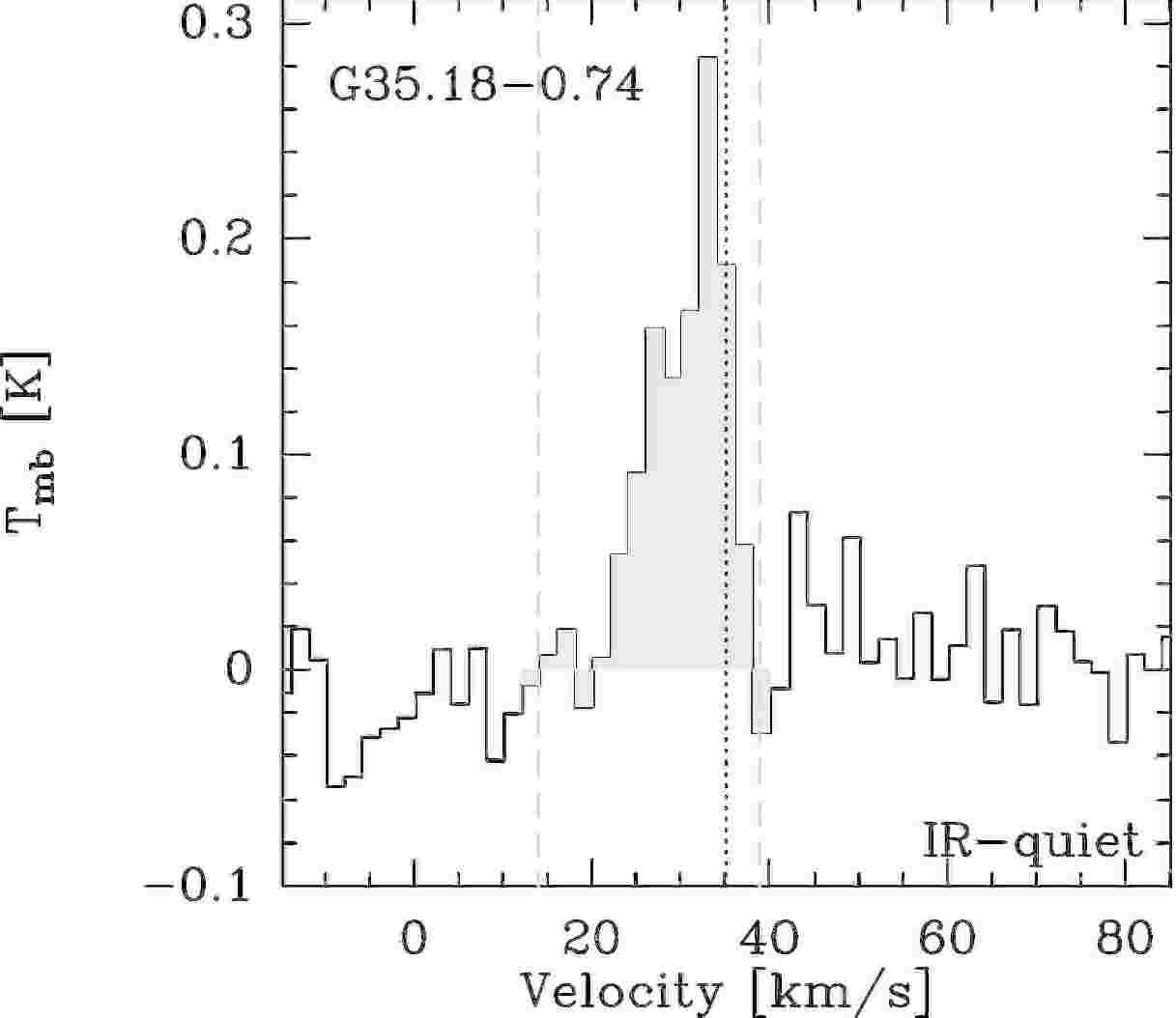} 
 \caption{Continued.}
\end{figure}
\end{landscape}

\begin{landscape}
\begin{figure}
\centering
\ContinuedFloat
  \includegraphics[width=5.6cm,angle=0]{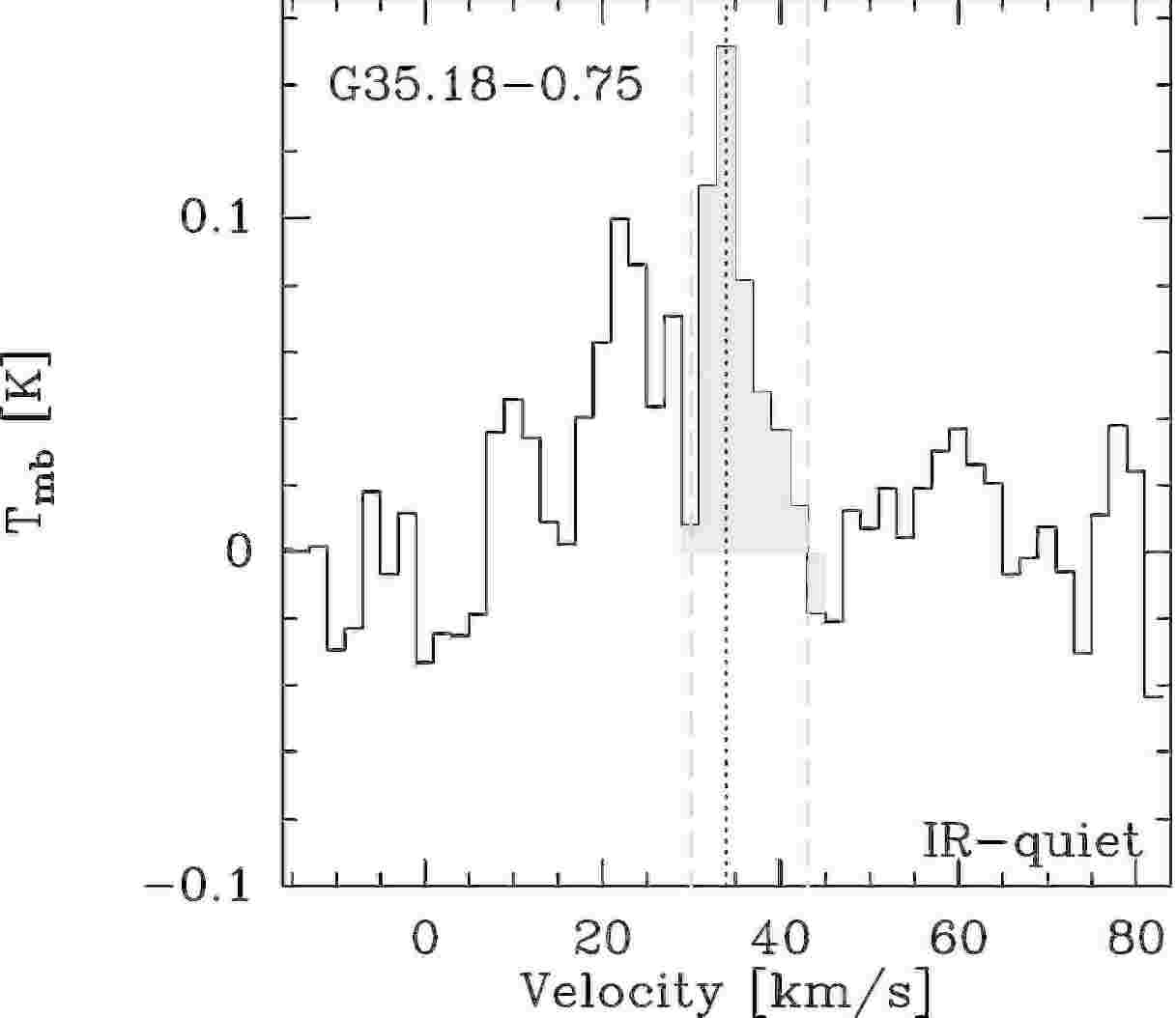} 
  \includegraphics[width=5.6cm,angle=0]{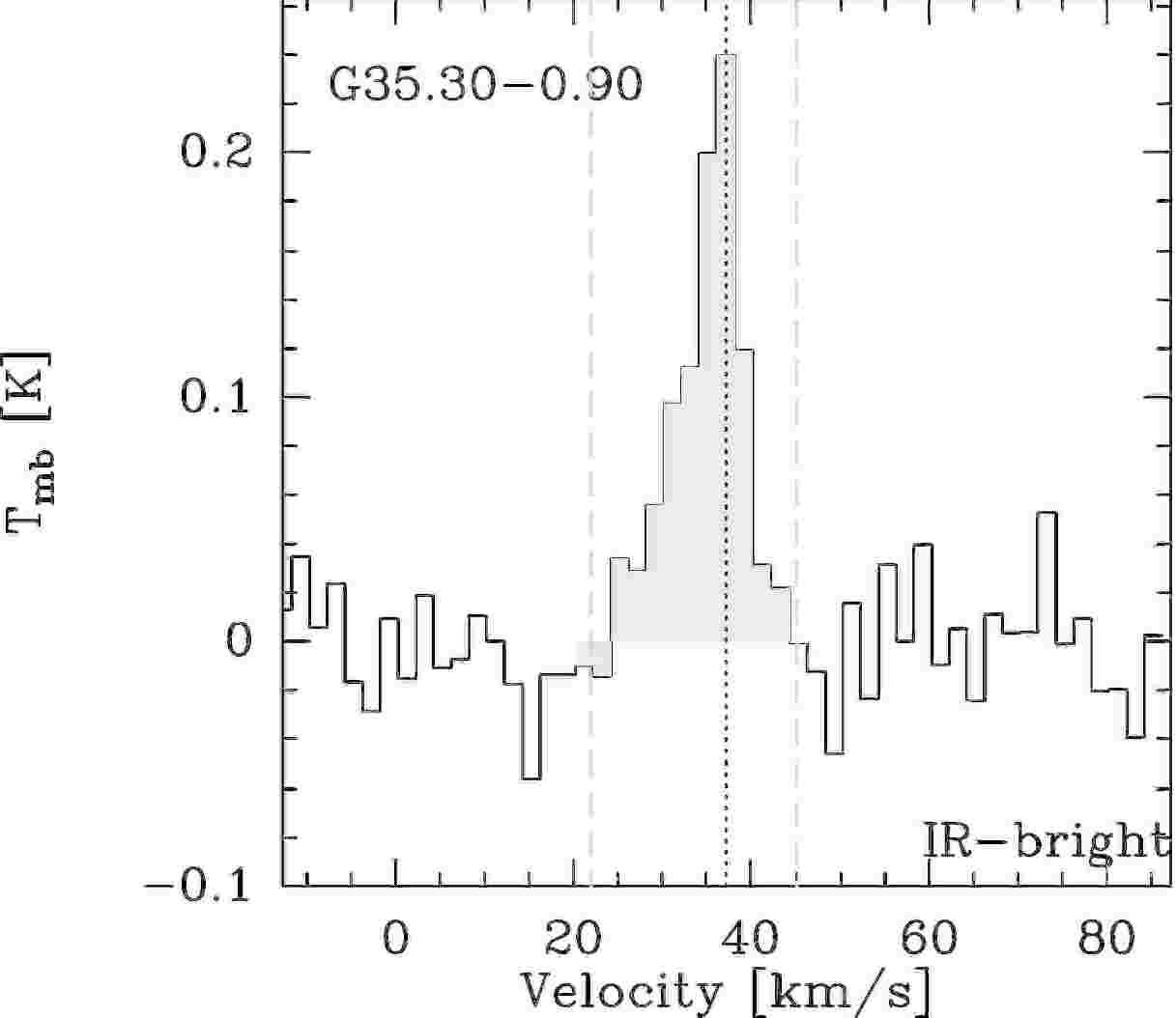} 
  \includegraphics[width=5.6cm,angle=0]{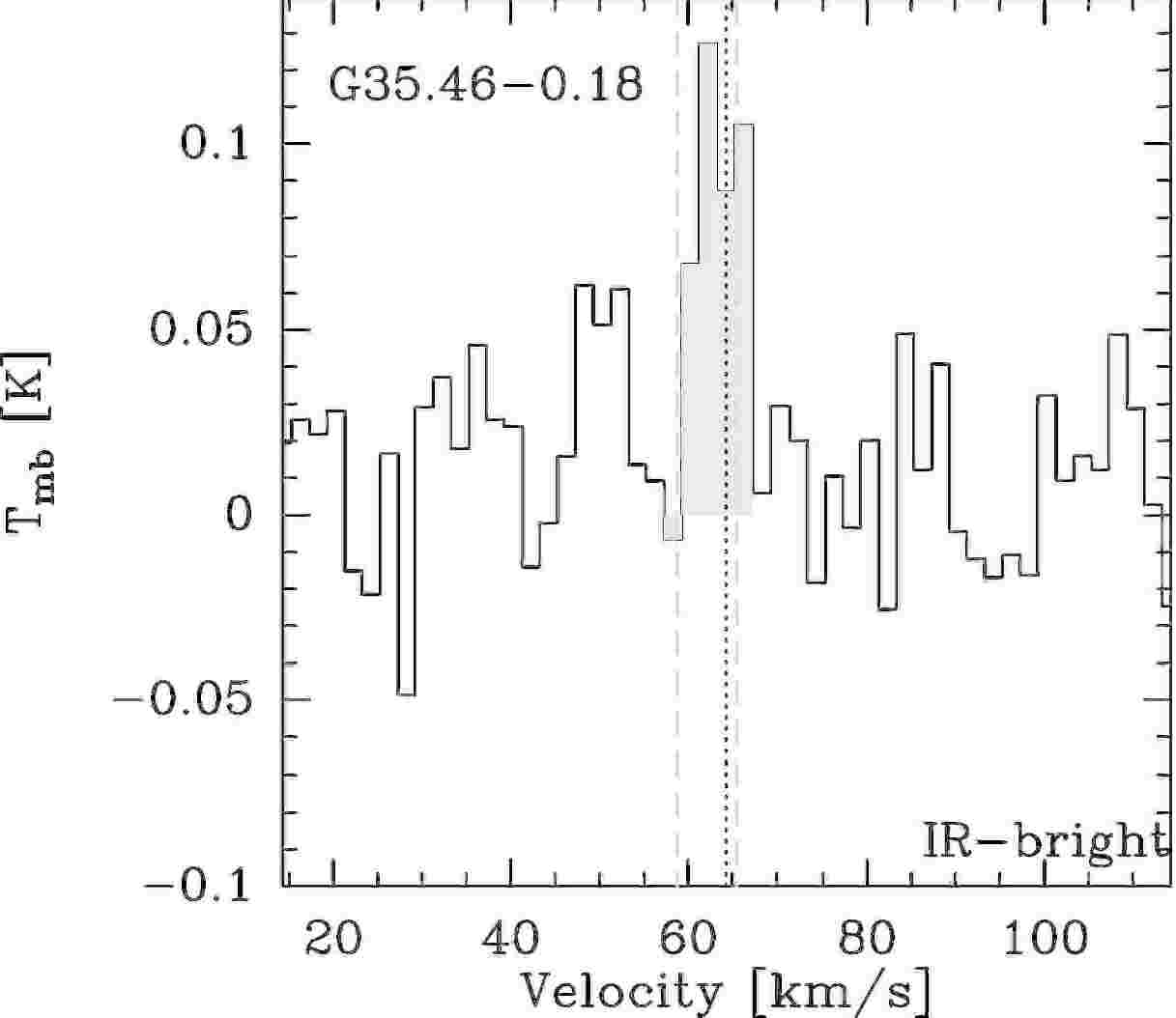} 
  \includegraphics[width=5.6cm,angle=0]{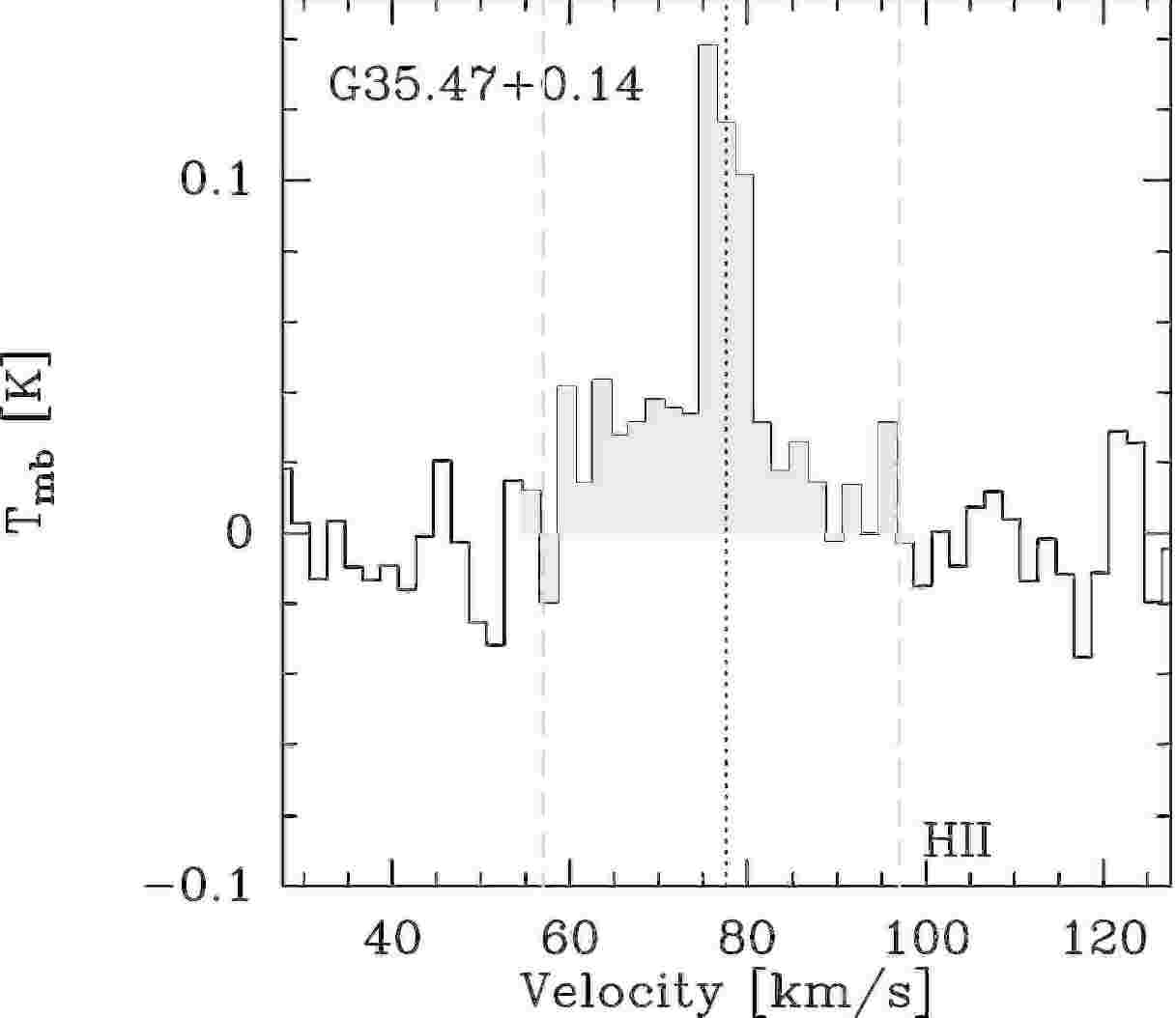} 
  \includegraphics[width=5.6cm,angle=0]{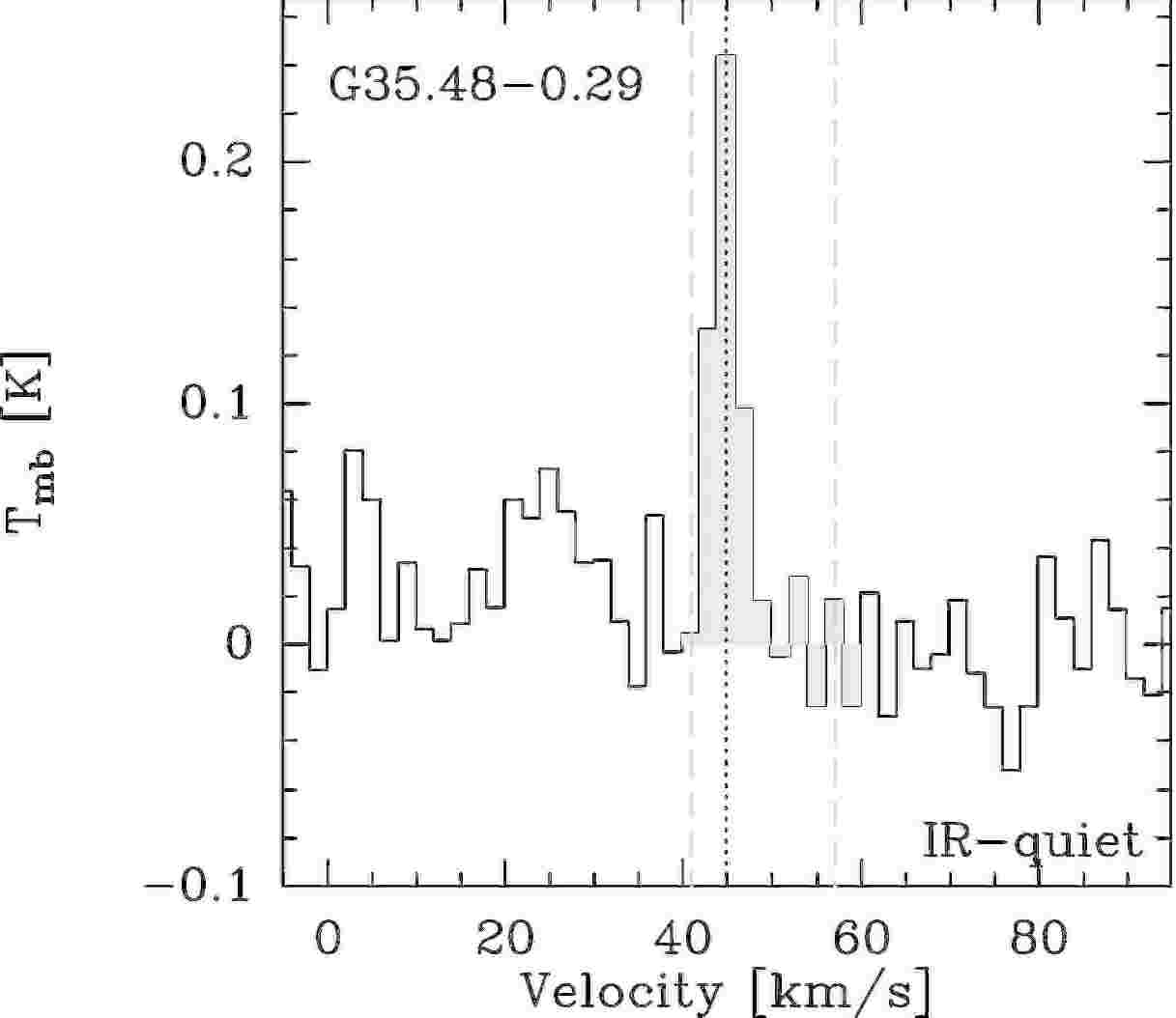} 
  \includegraphics[width=5.6cm,angle=0]{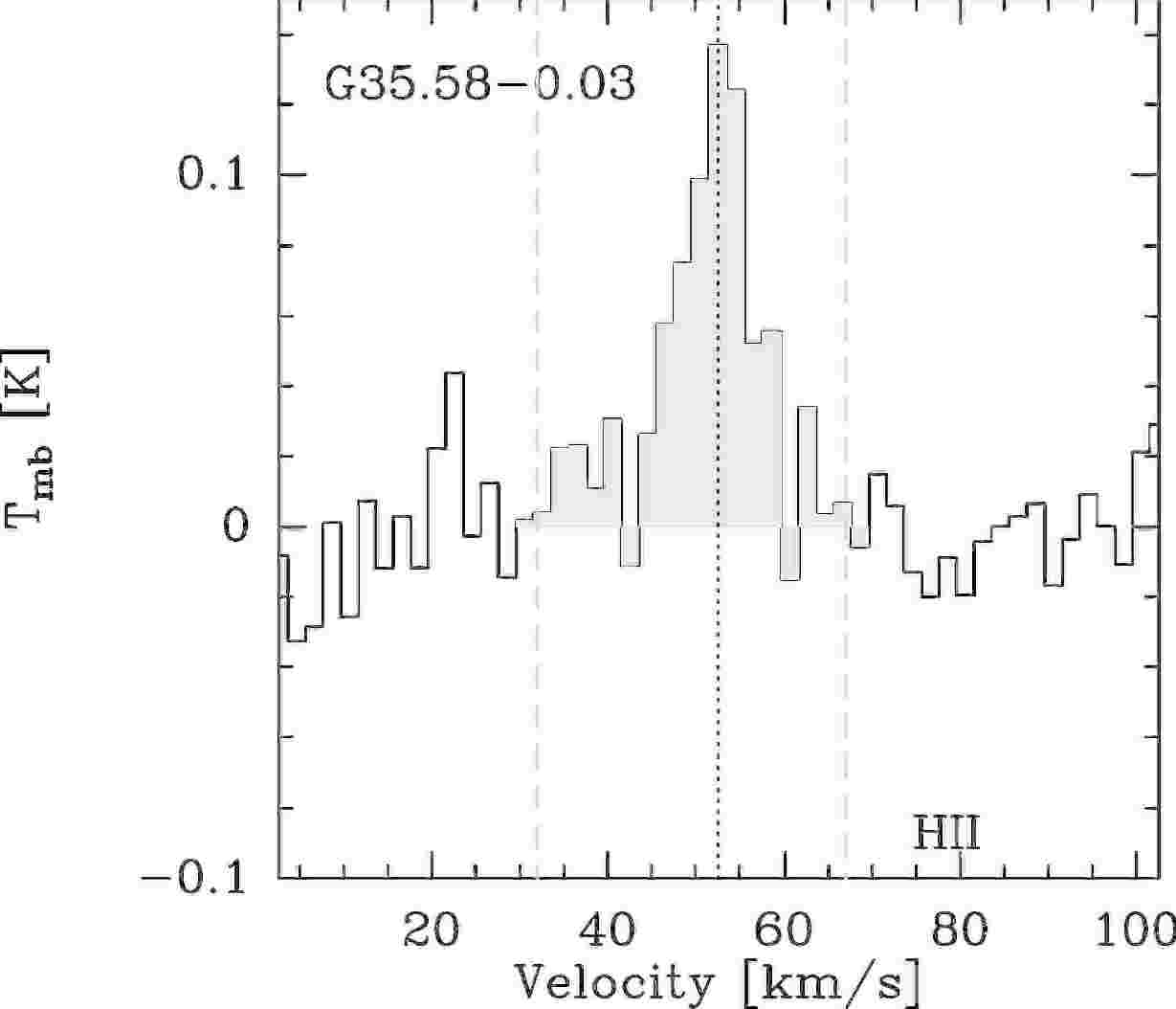} 
  \includegraphics[width=5.6cm,angle=0]{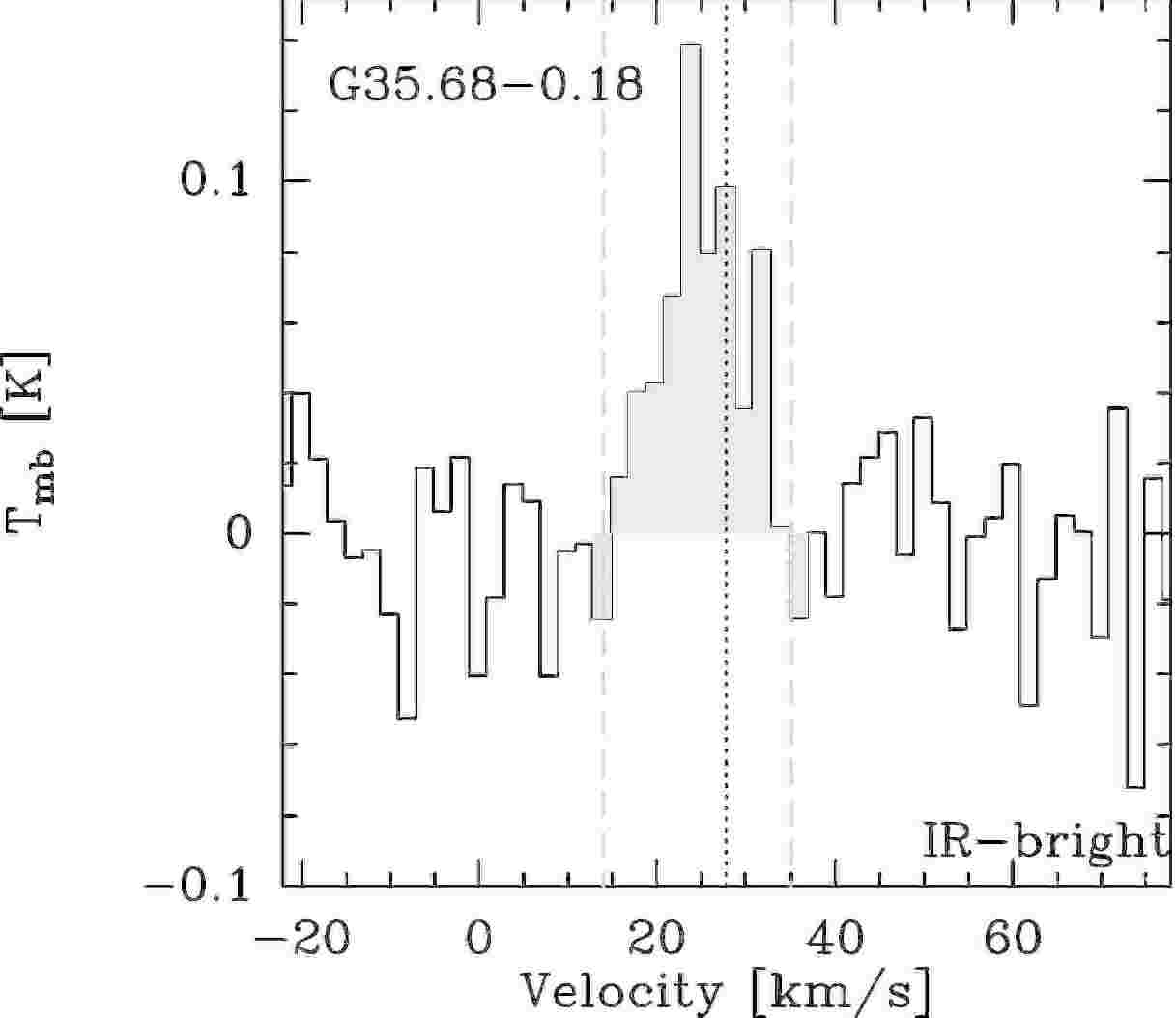} 
  \includegraphics[width=5.6cm,angle=0]{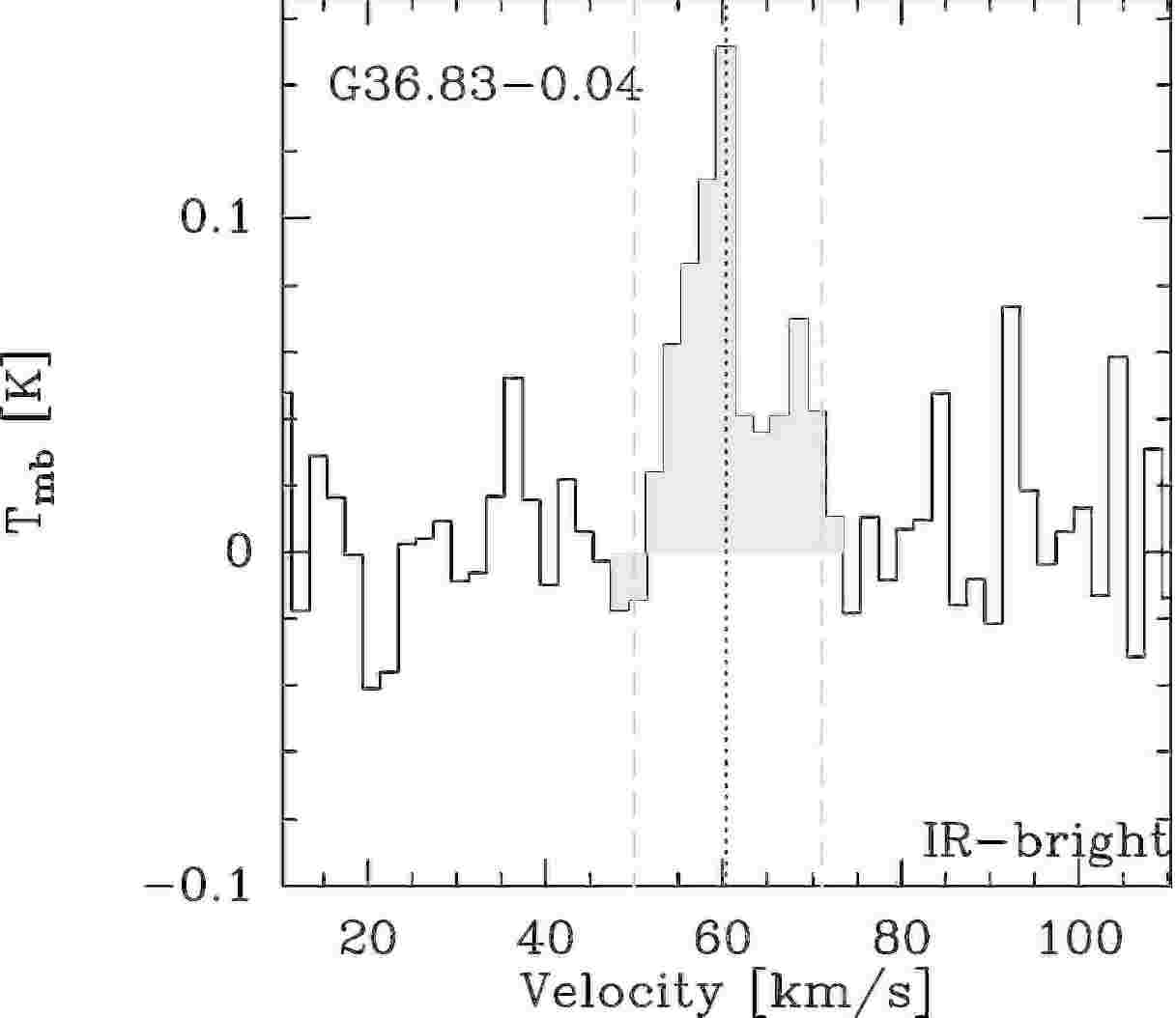} 
  \includegraphics[width=5.6cm,angle=0]{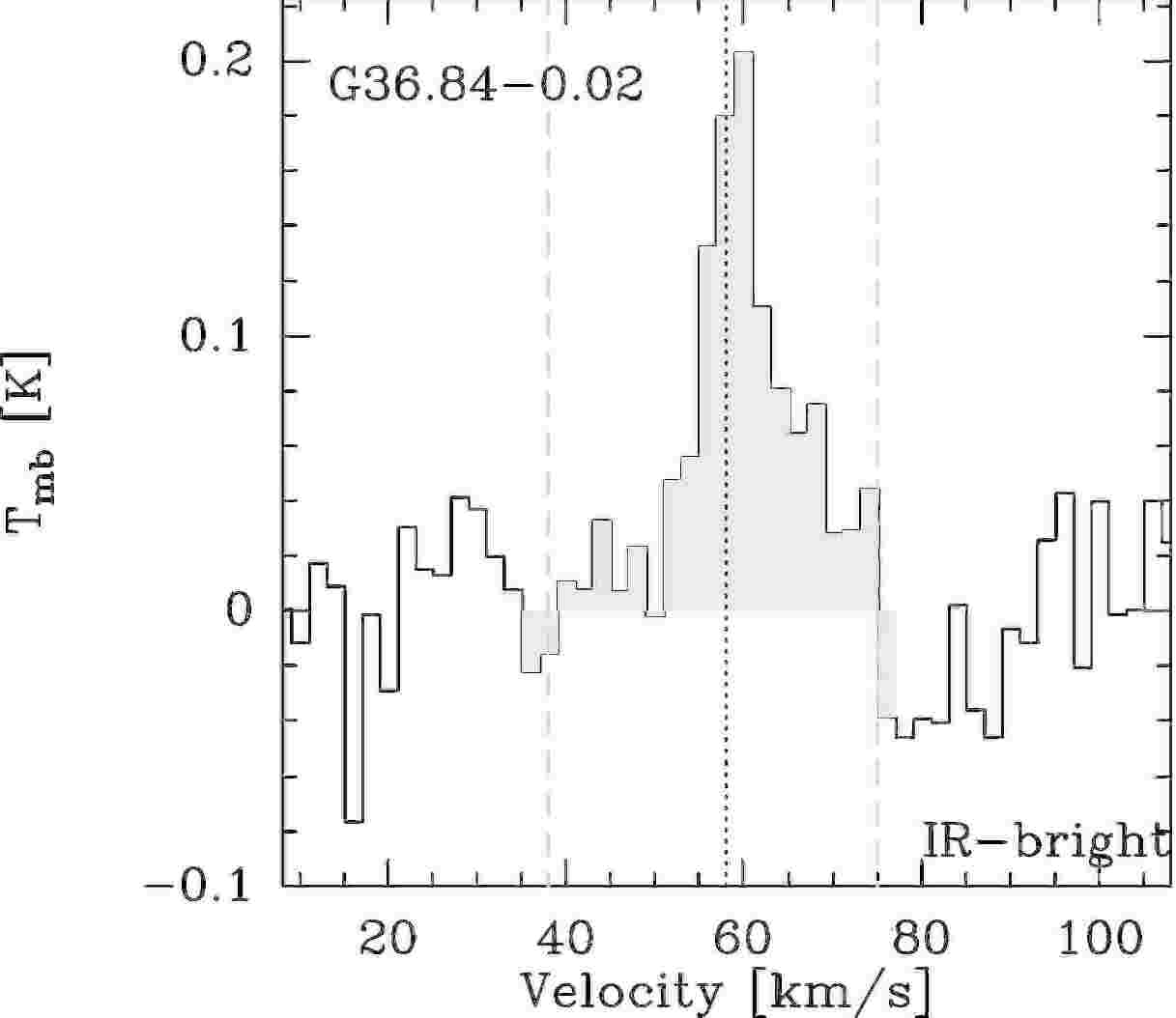} 
  \includegraphics[width=5.6cm,angle=0]{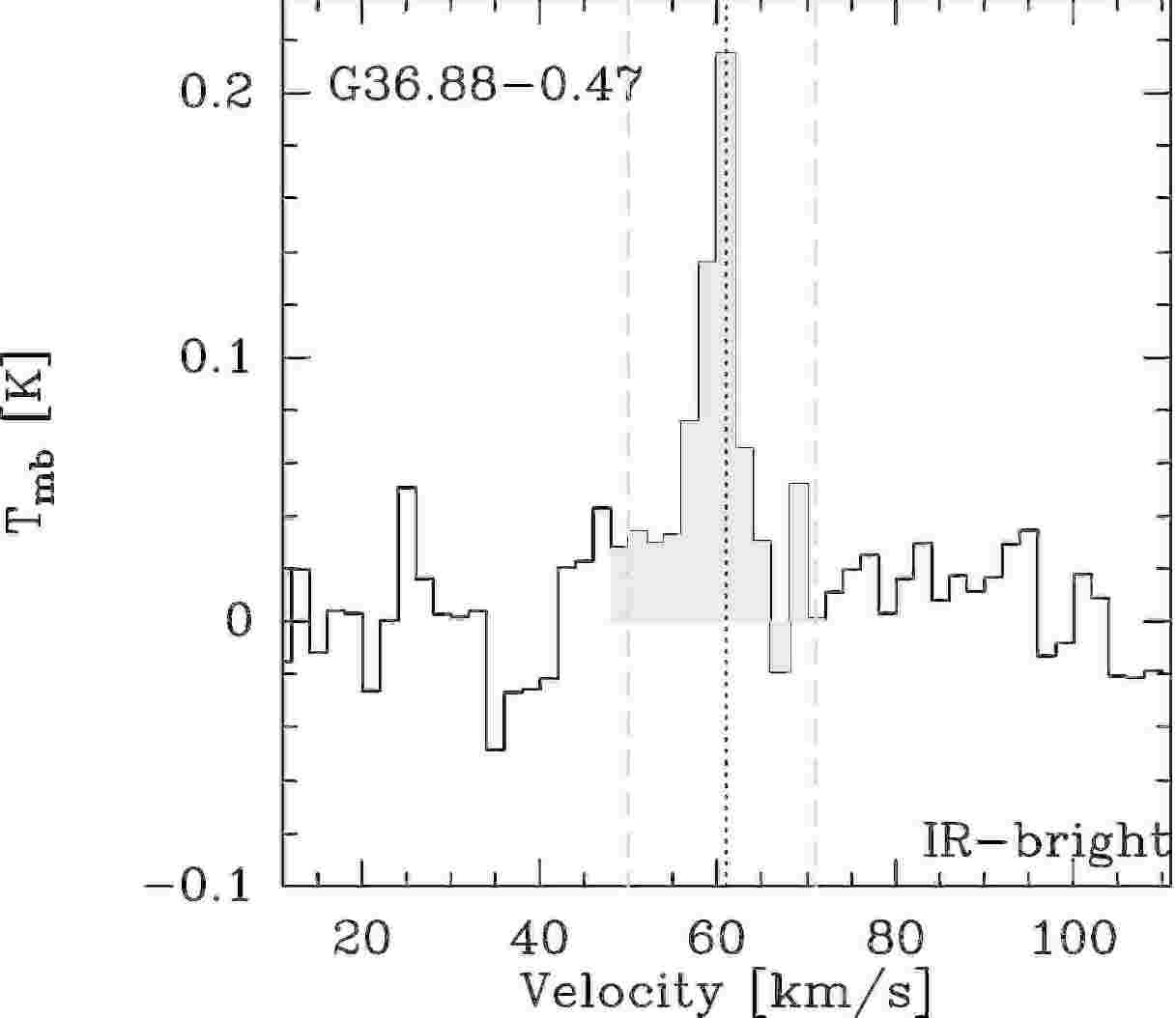} 
  \includegraphics[width=5.6cm,angle=0]{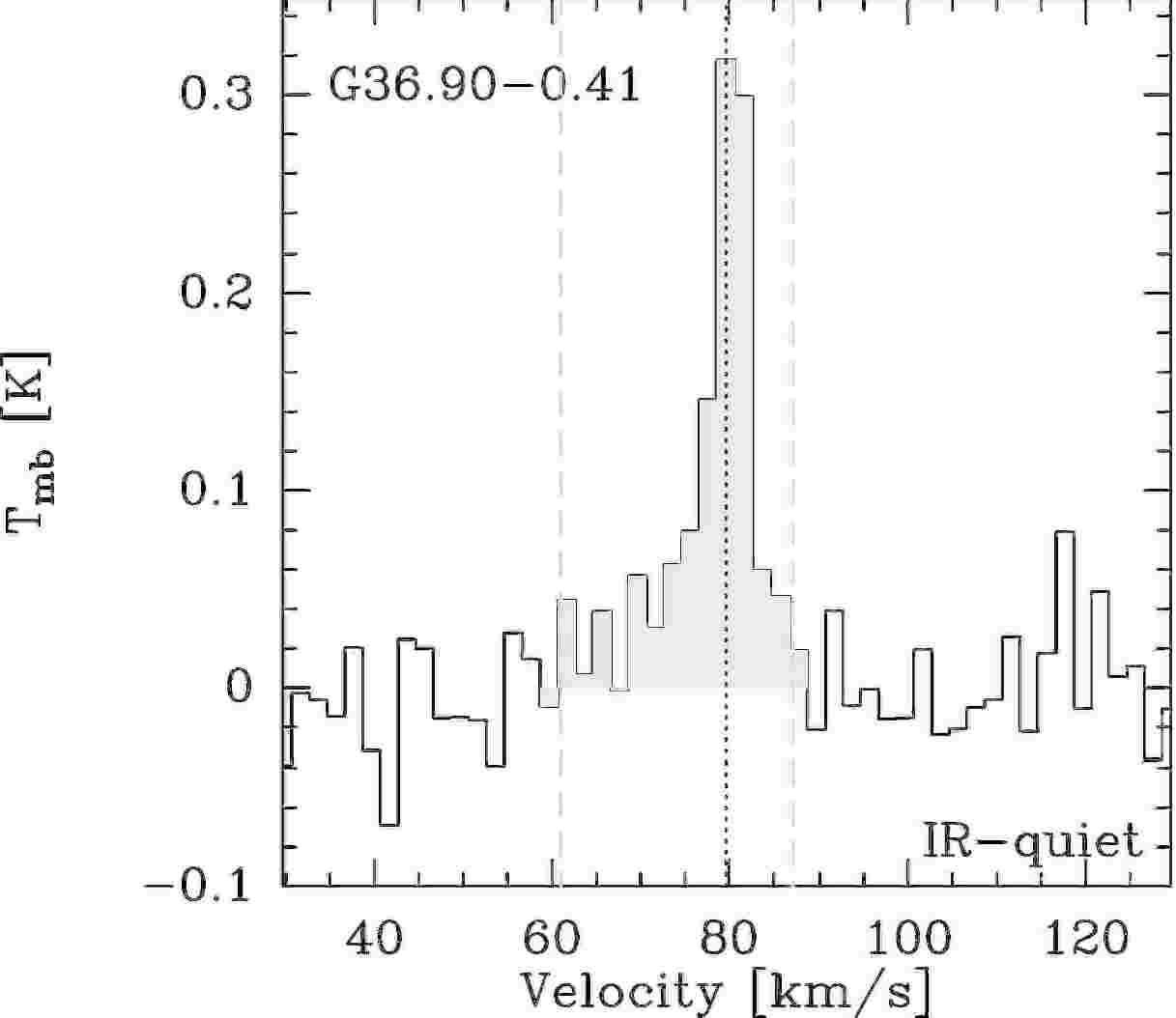} 
  \includegraphics[width=5.6cm,angle=0]{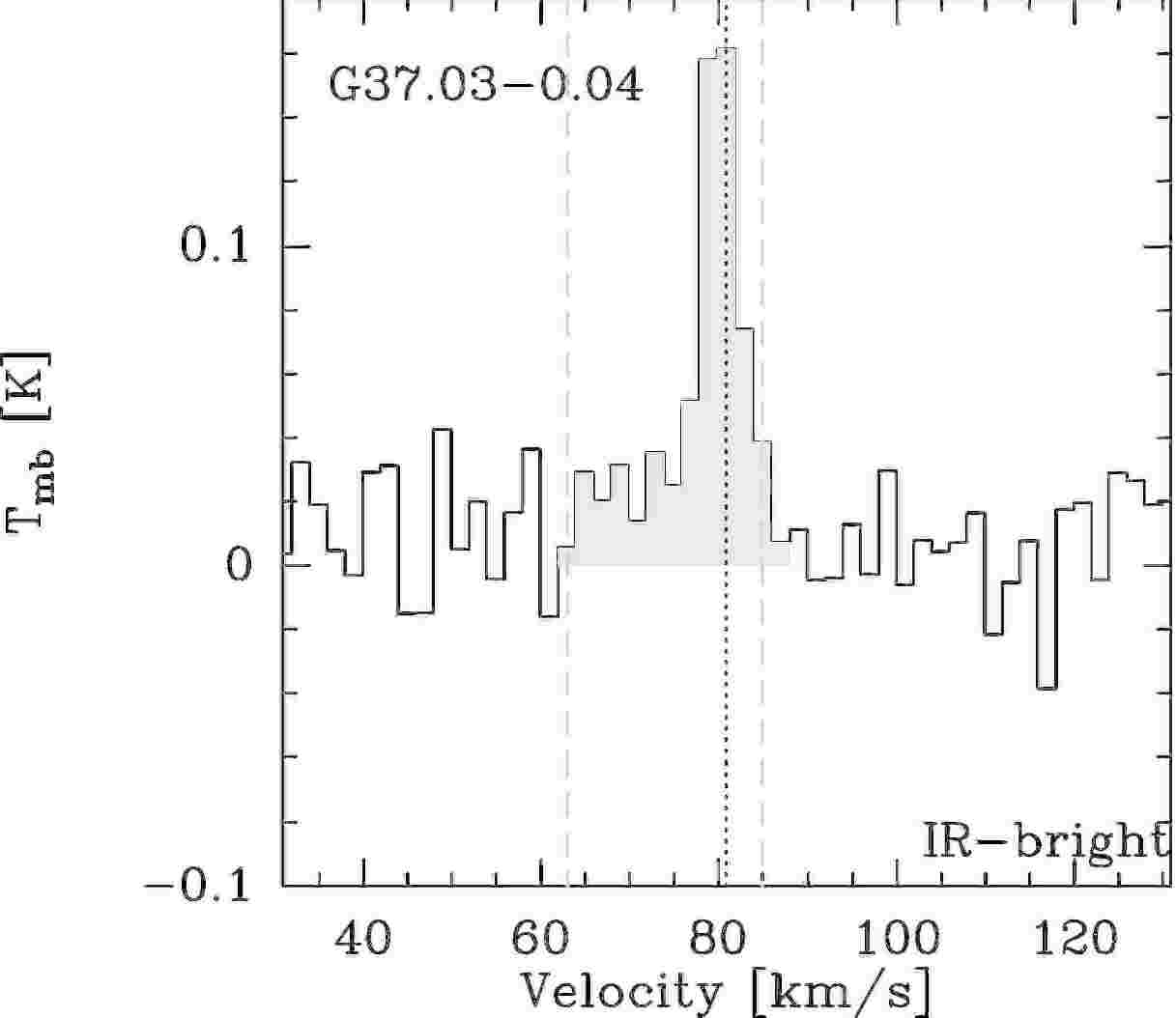} 
  \caption{Continued.}
\end{figure}
\end{landscape}

\begin{landscape}
\begin{figure}
\centering
\ContinuedFloat
  \includegraphics[width=5.6cm,angle=0]{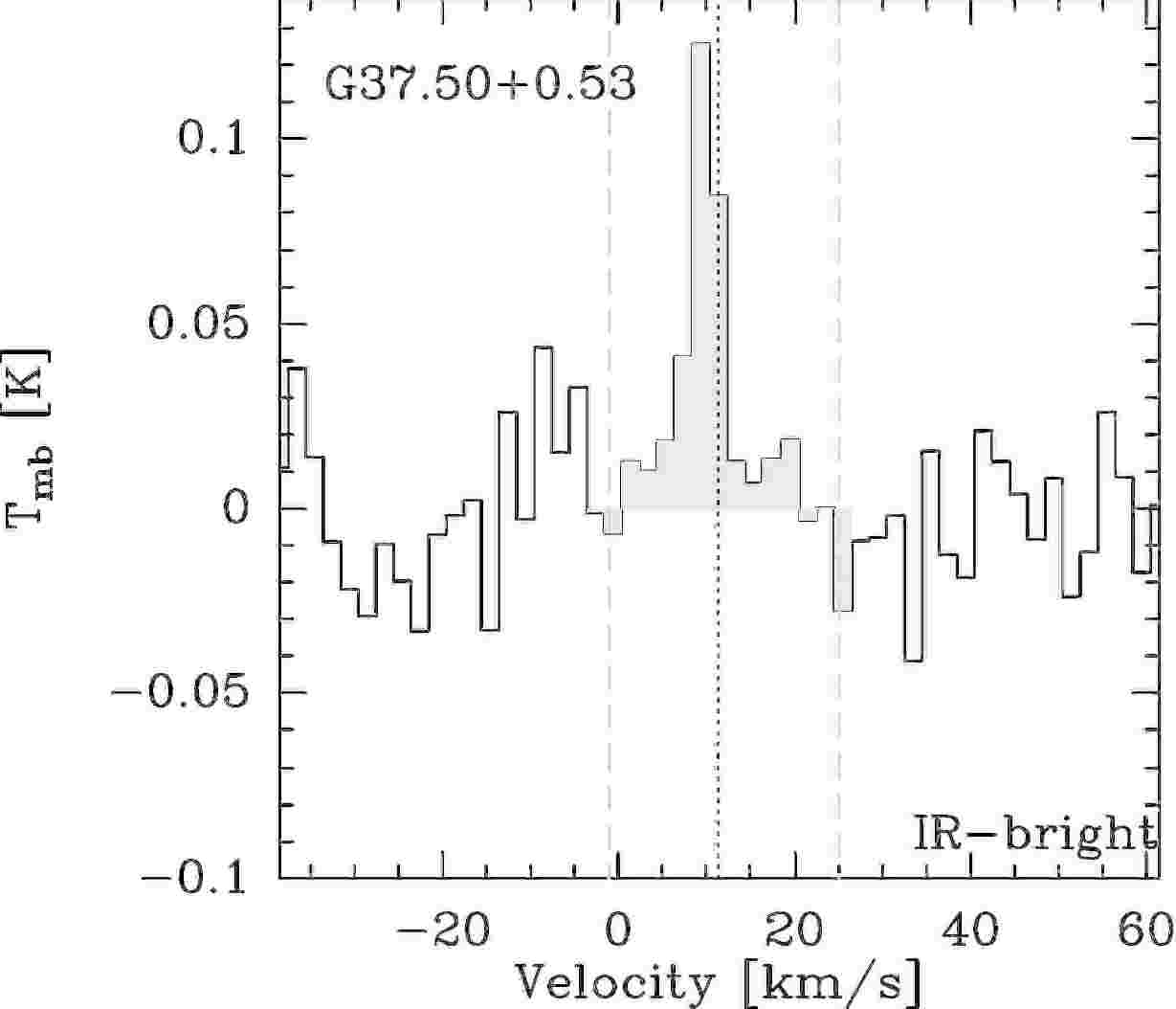} 
  \includegraphics[width=5.6cm,angle=0]{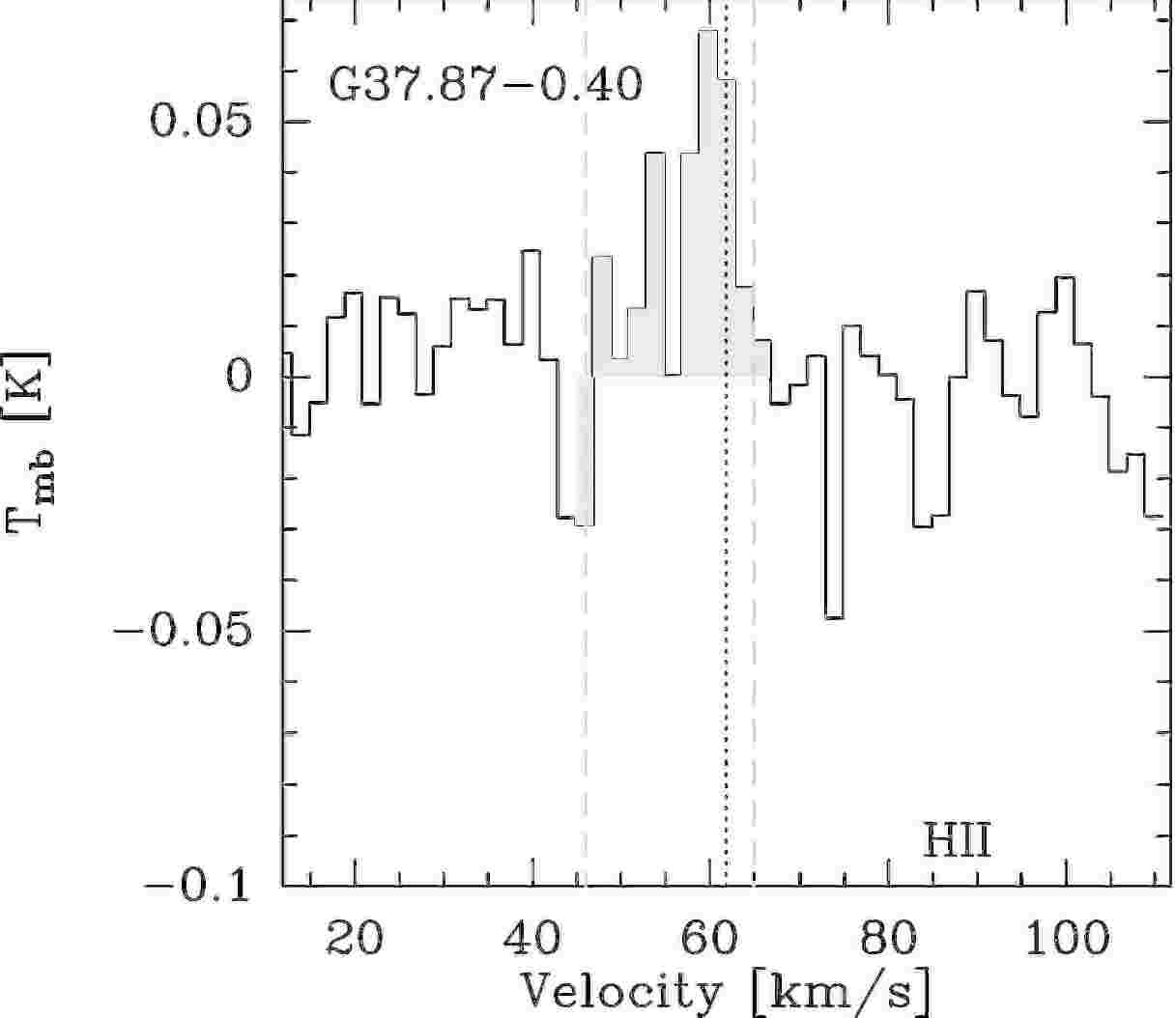} 
 \includegraphics[width=5.6cm,angle=0]{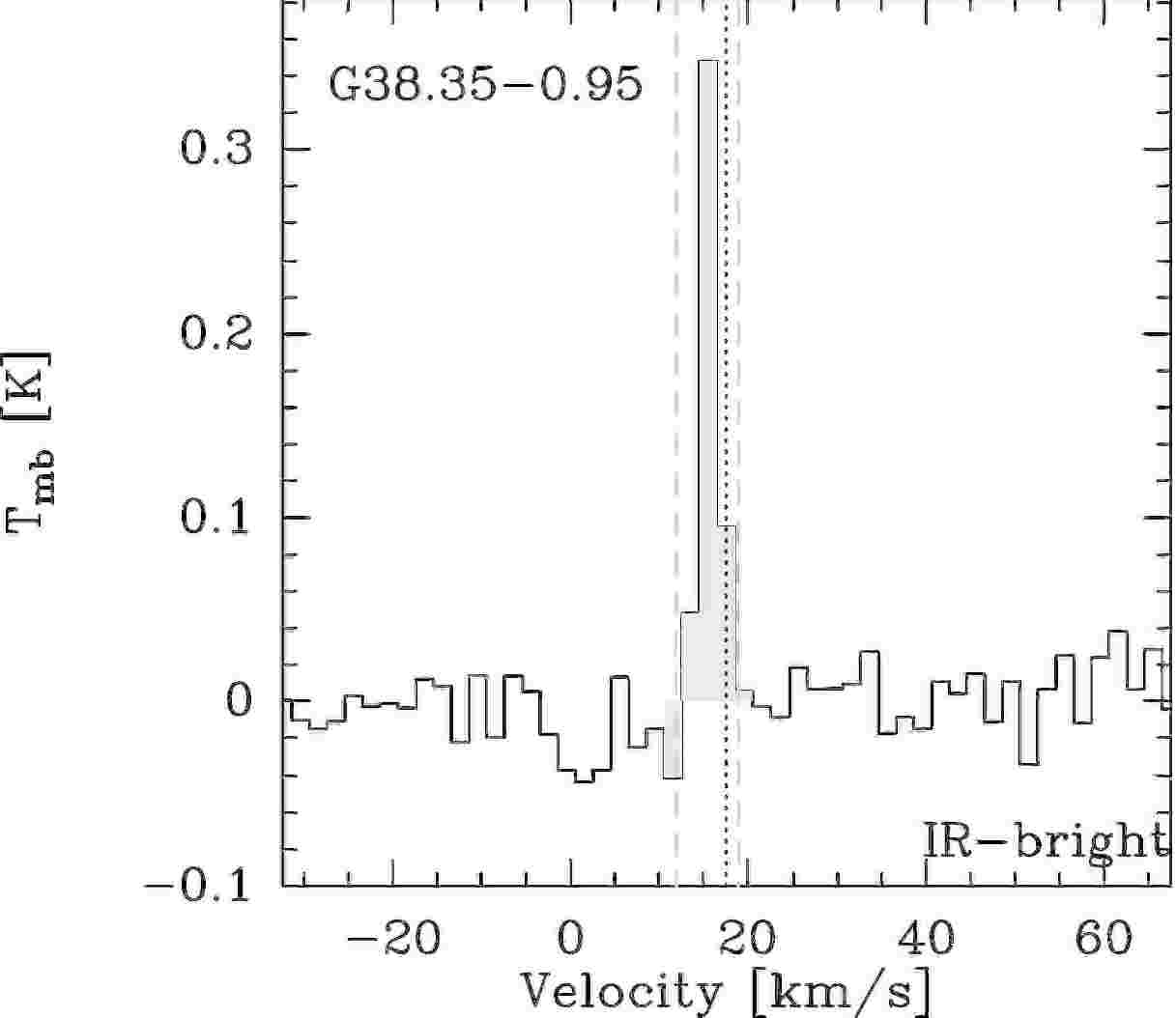} 
  \includegraphics[width=5.6cm,angle=0]{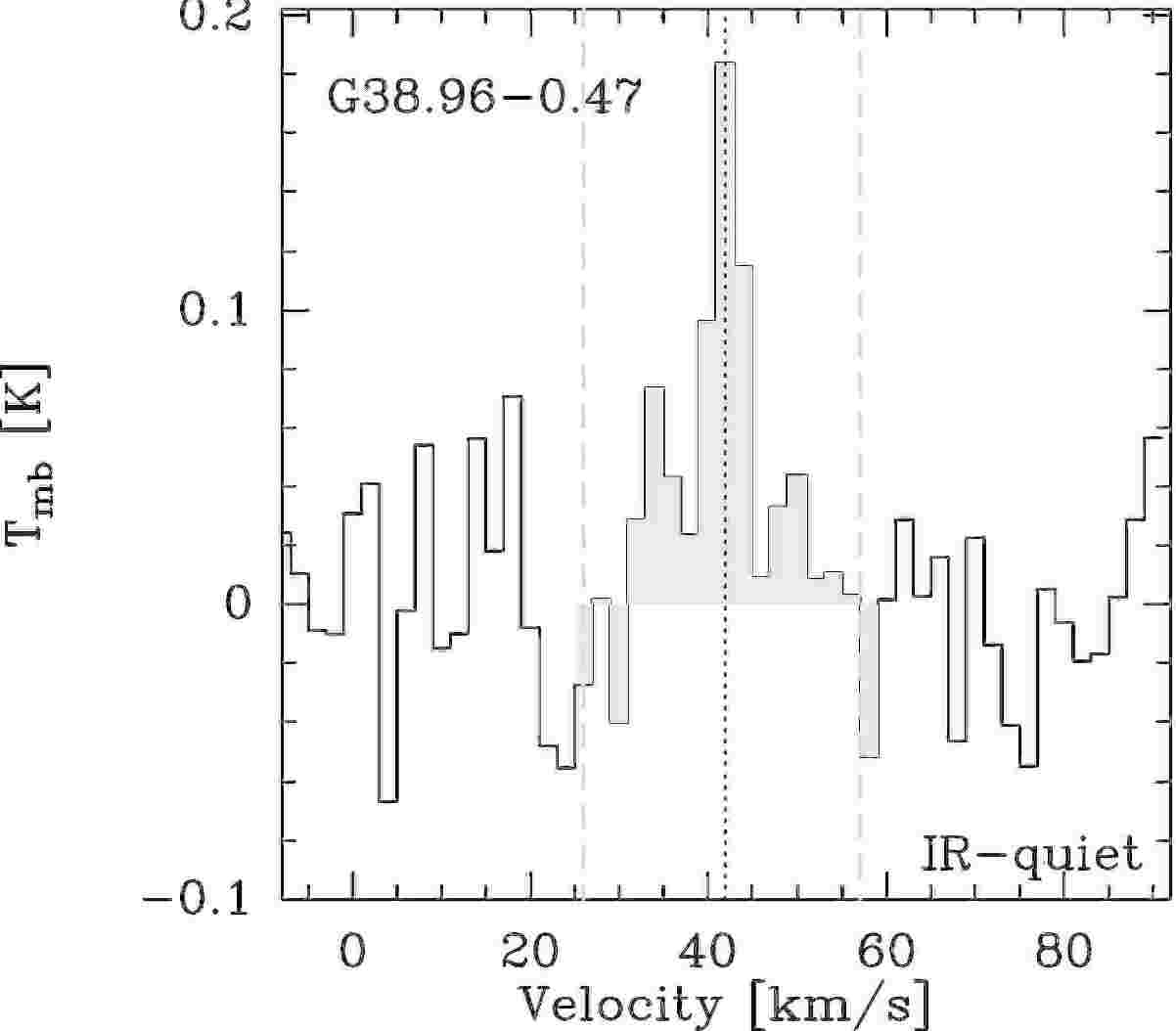} 
  \includegraphics[width=5.6cm,angle=0]{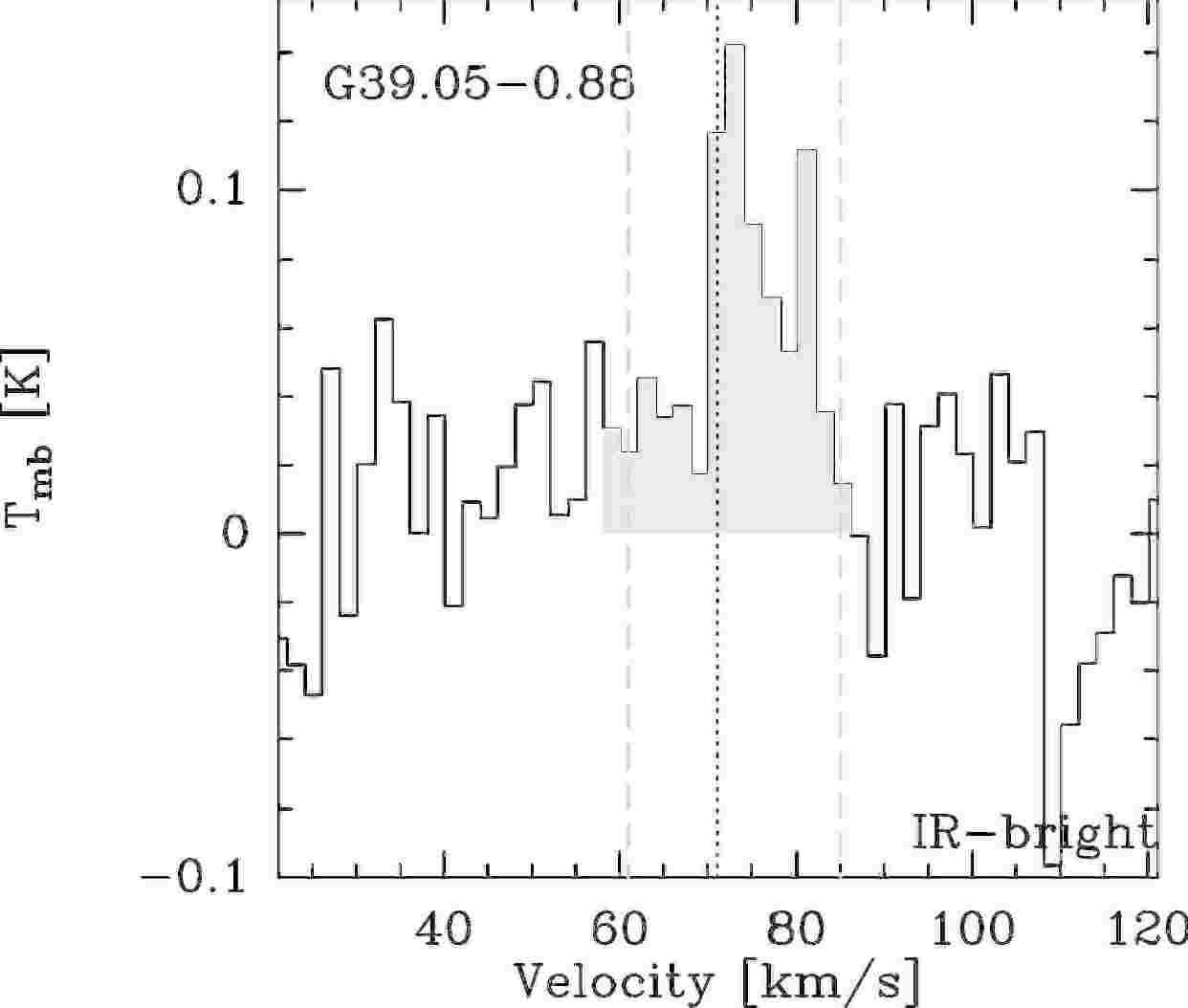} 
  \includegraphics[width=5.6cm,angle=0]{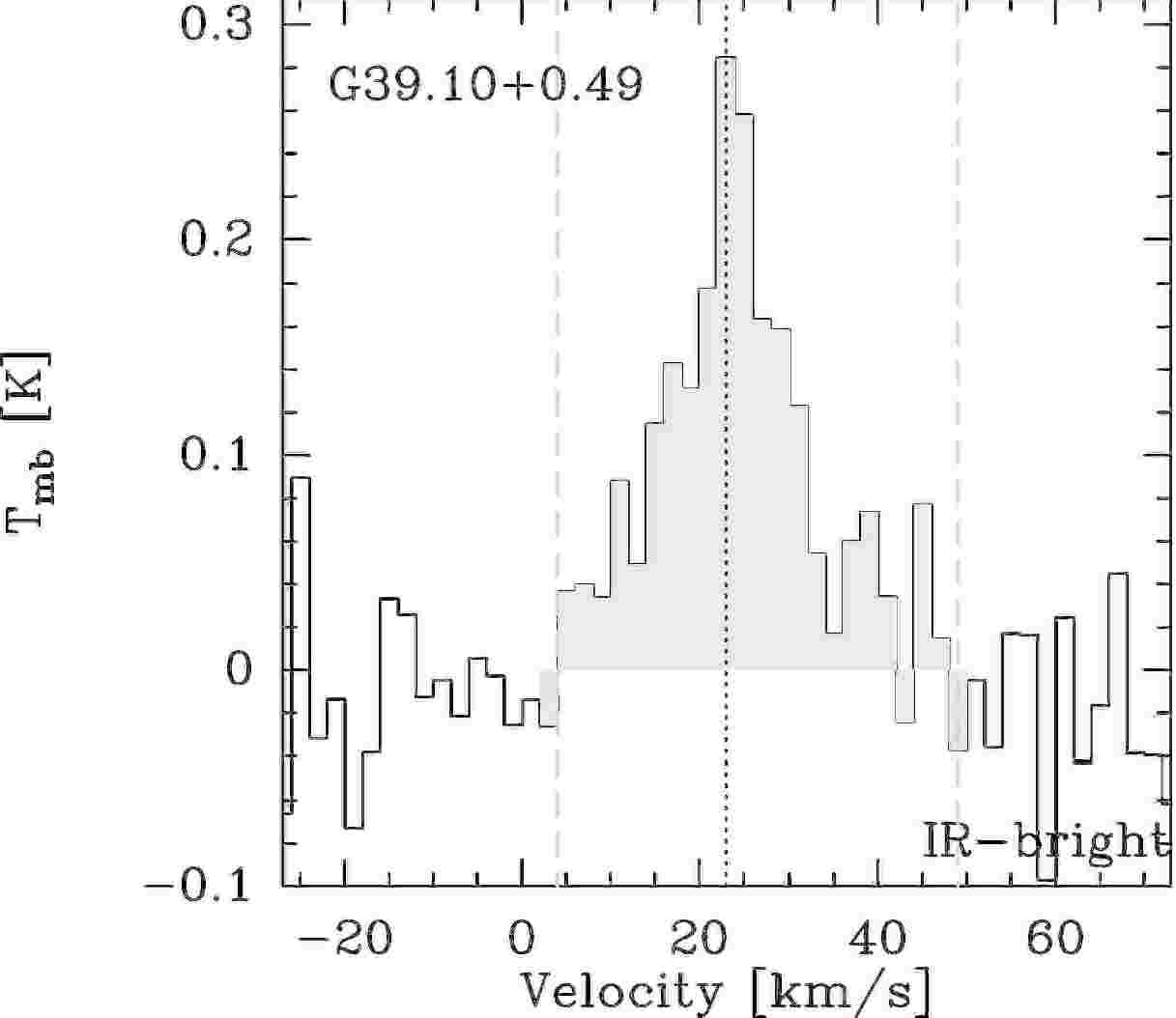} 
  \includegraphics[width=5.6cm,angle=0]{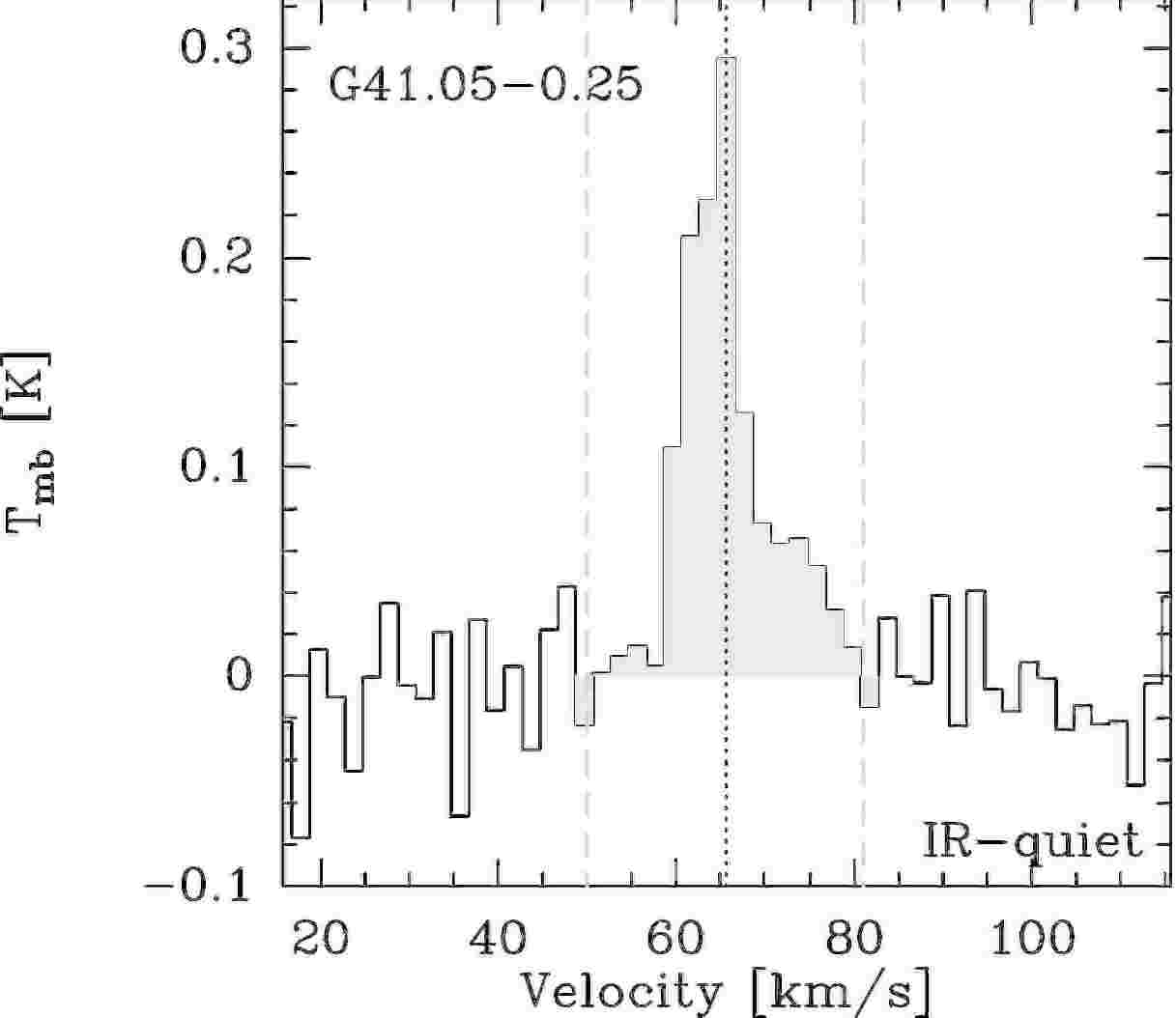} 
  \includegraphics[width=5.6cm,angle=0]{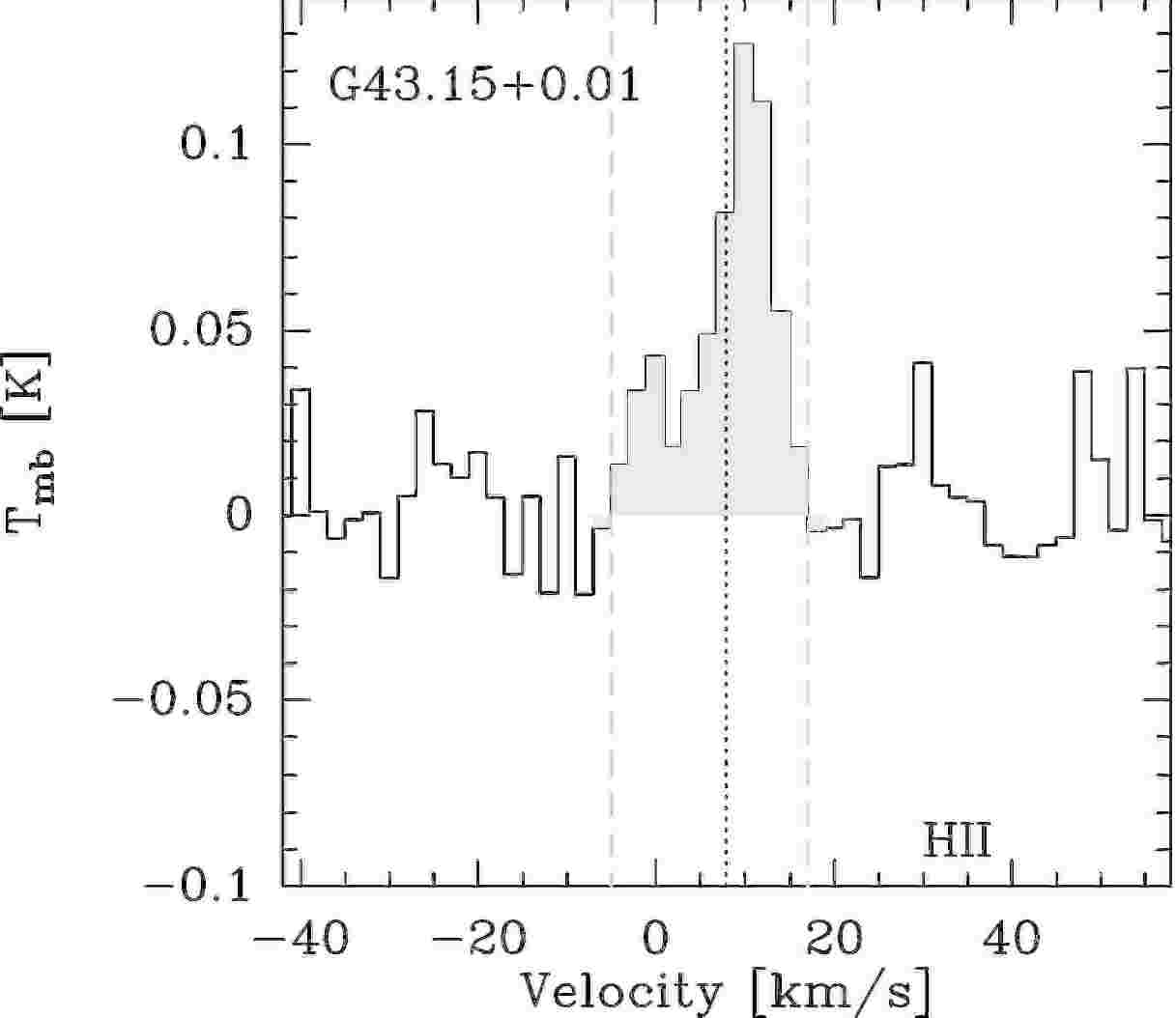} 
  \includegraphics[width=5.6cm,angle=0]{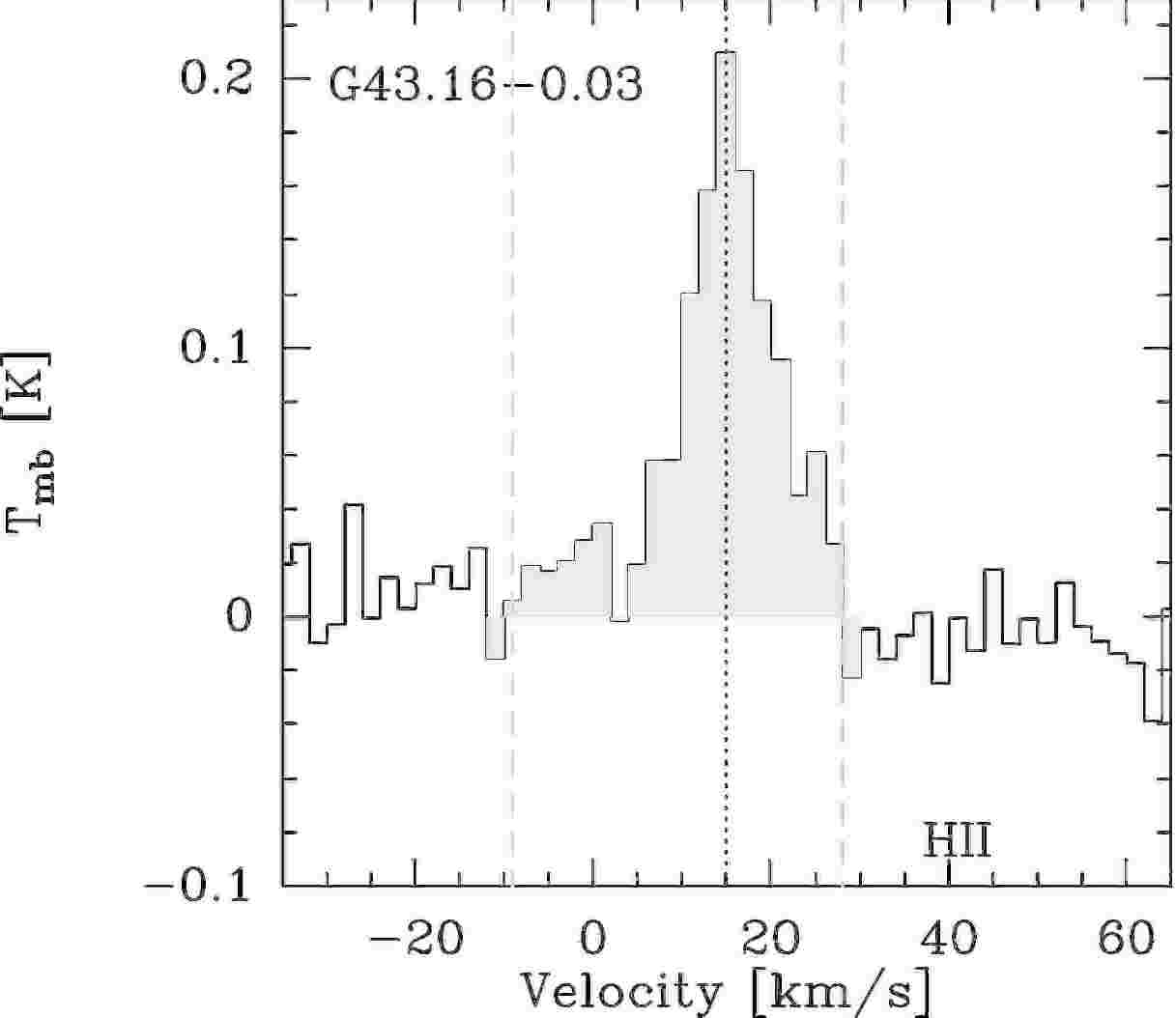} 
   \includegraphics[width=5.6cm,angle=0]{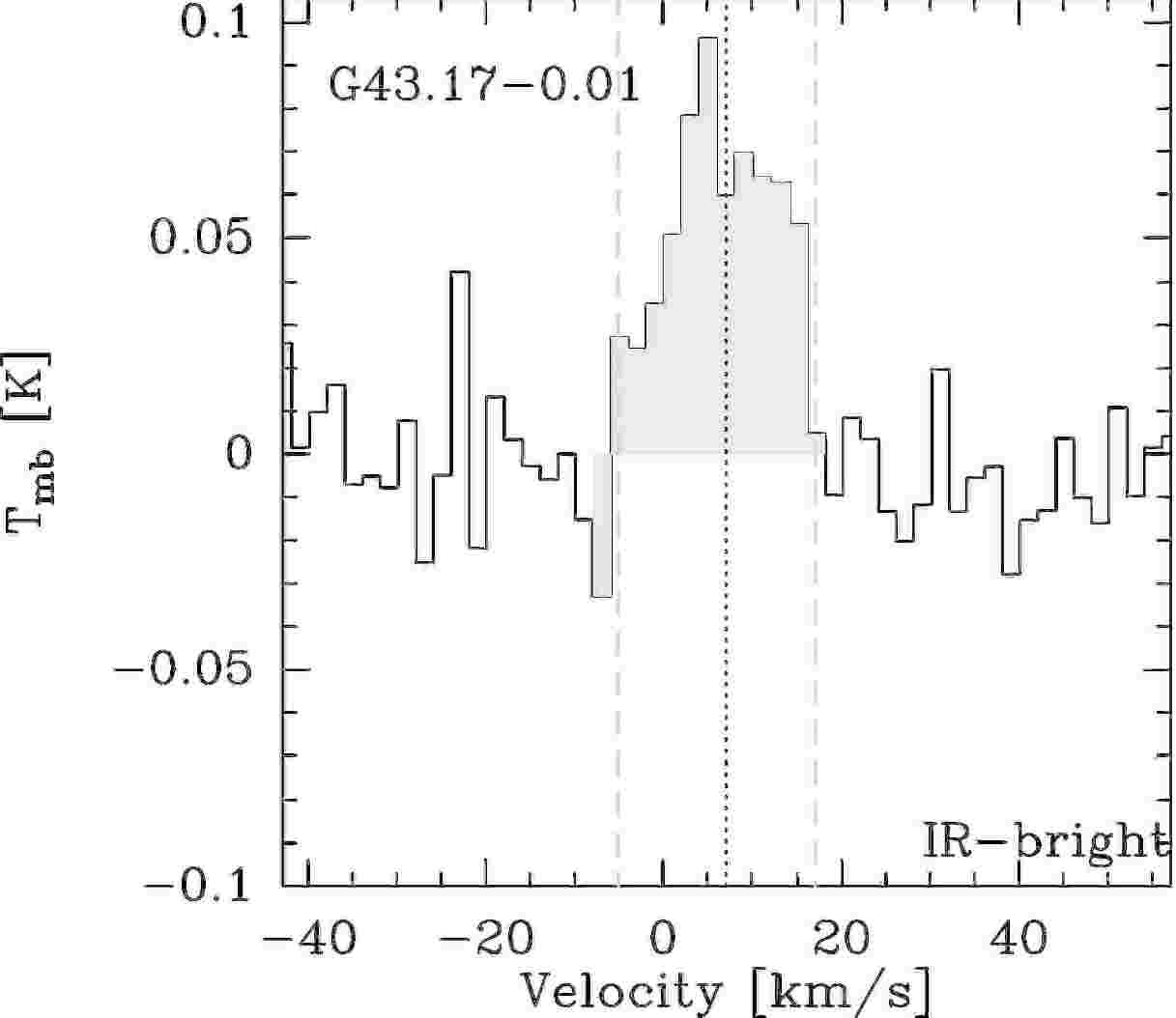} 
  \includegraphics[width=5.6cm,angle=0]{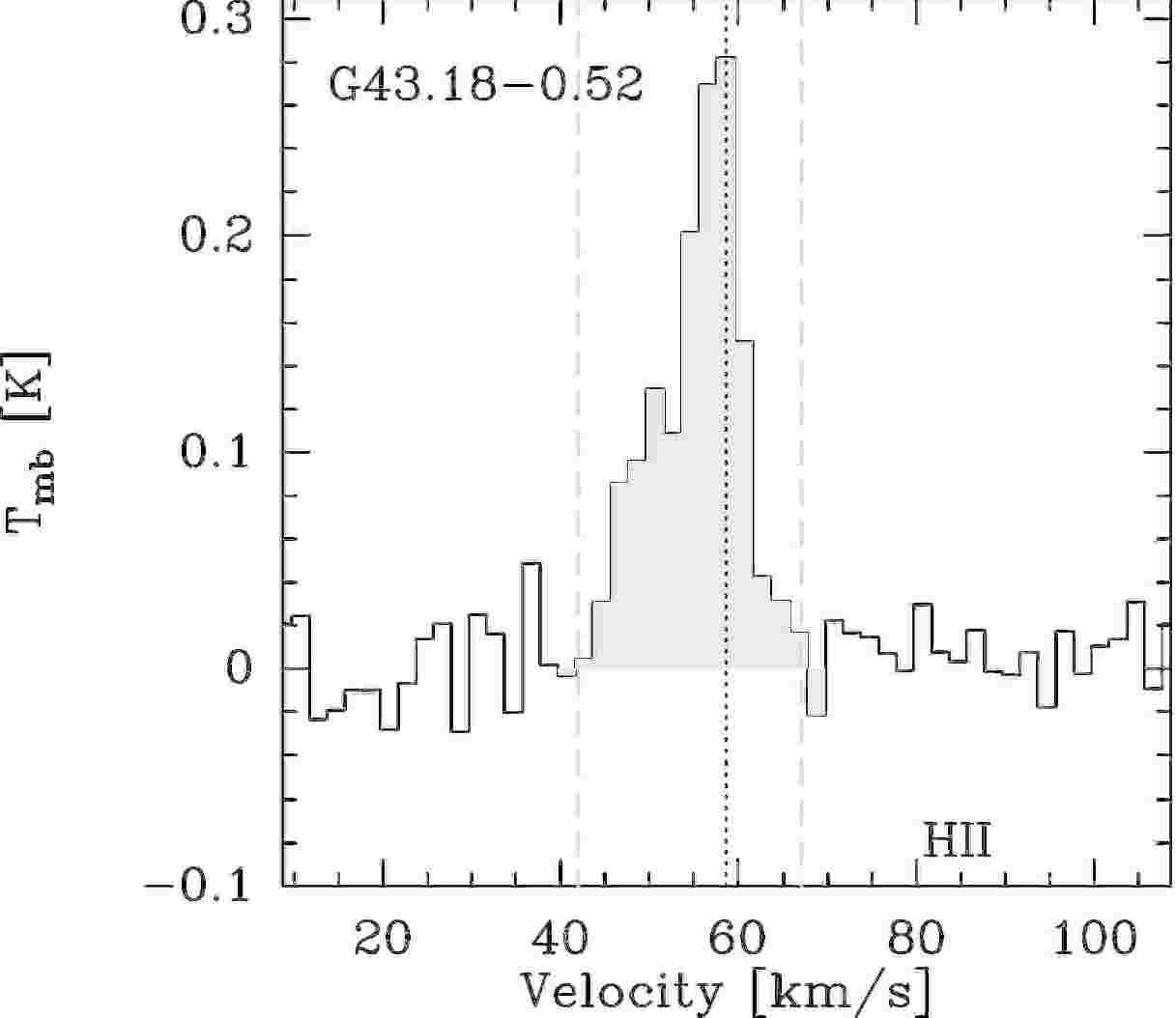} 
  \includegraphics[width=5.6cm,angle=0]{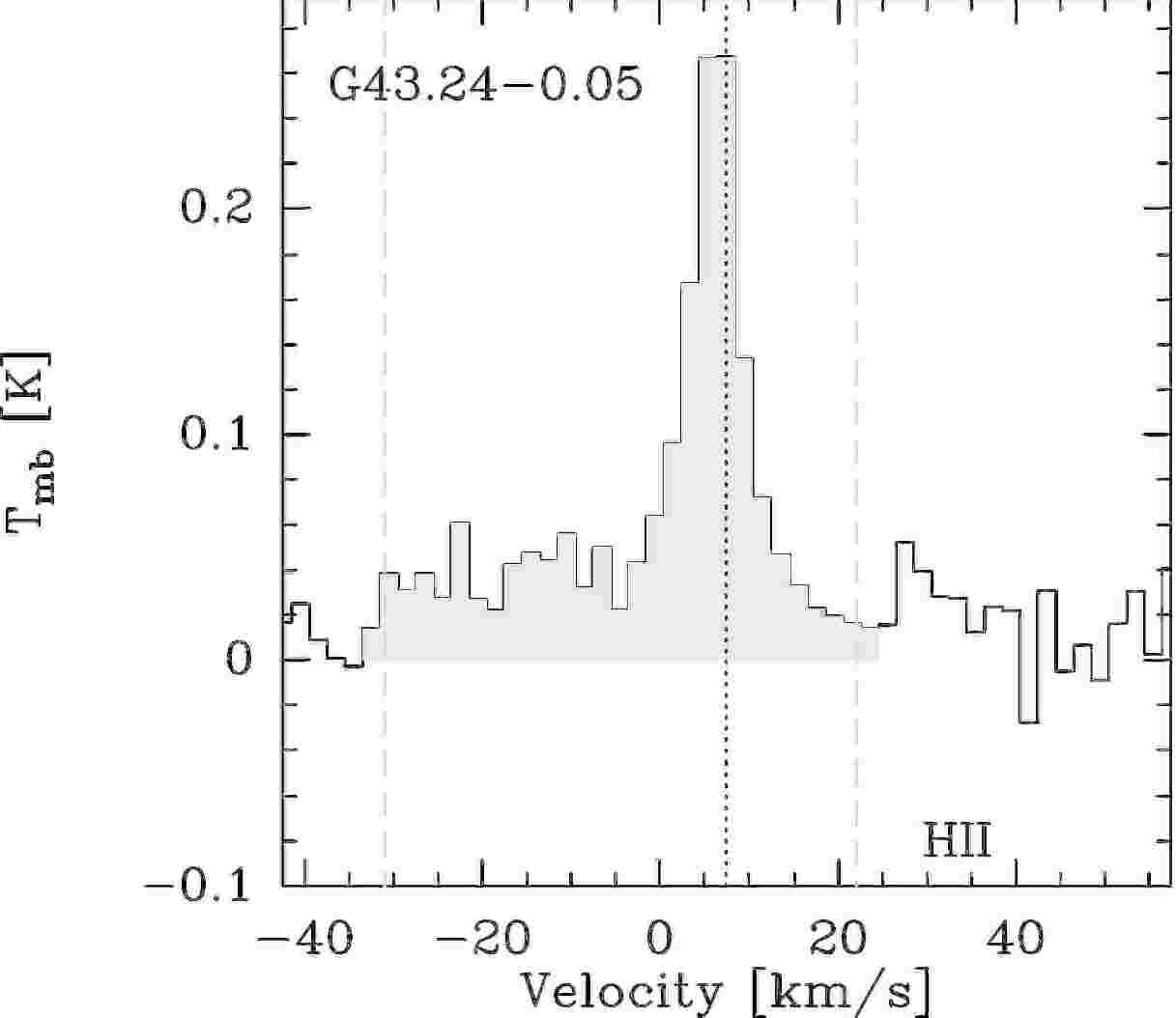} 
\caption{Continued.}
\end{figure}
\end{landscape}

\begin{landscape}
\begin{figure}
\centering
\ContinuedFloat
  \includegraphics[width=5.6cm,angle=0]{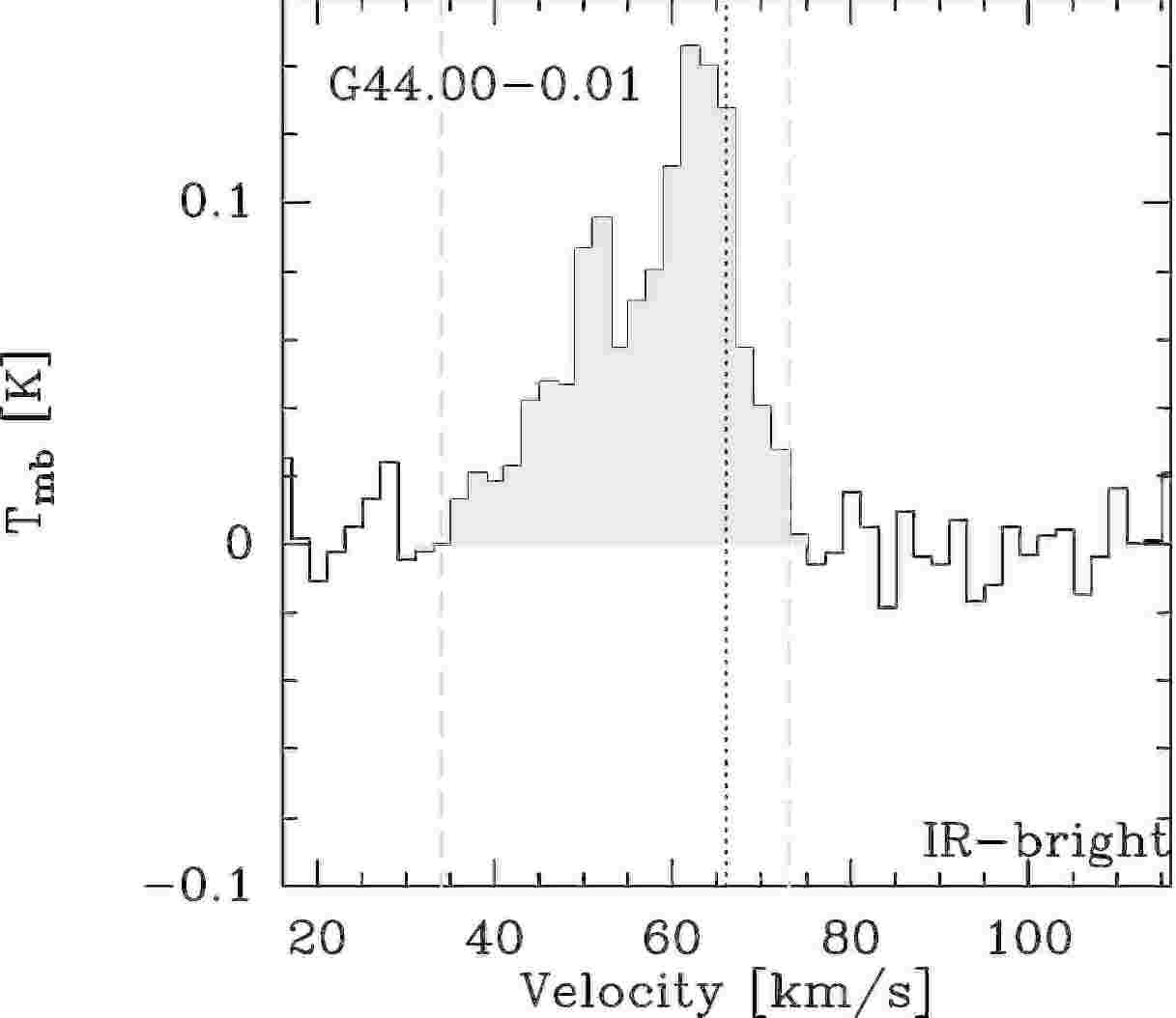} 
  \includegraphics[width=5.6cm,angle=0]{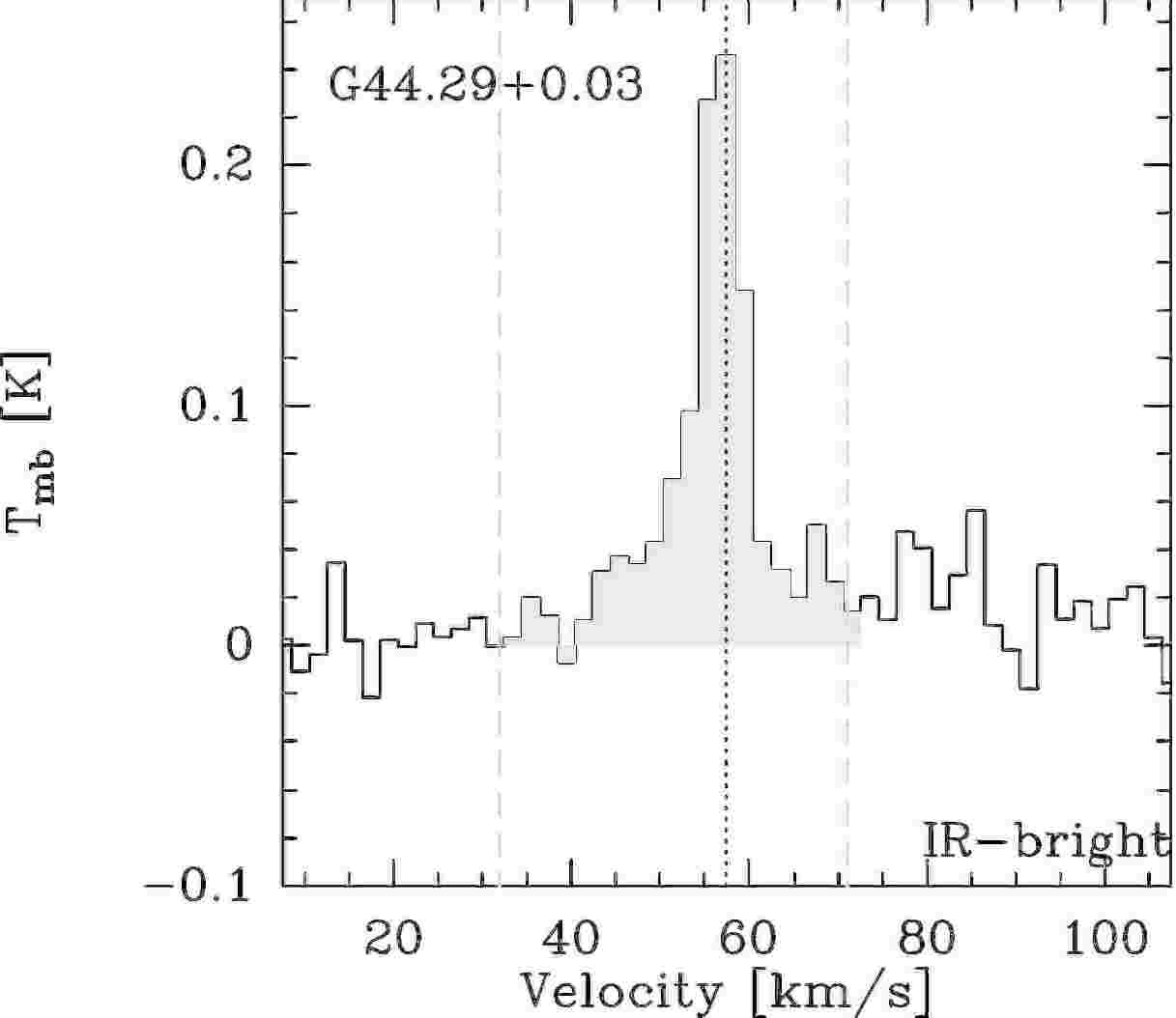} 
  \includegraphics[width=5.6cm,angle=0]{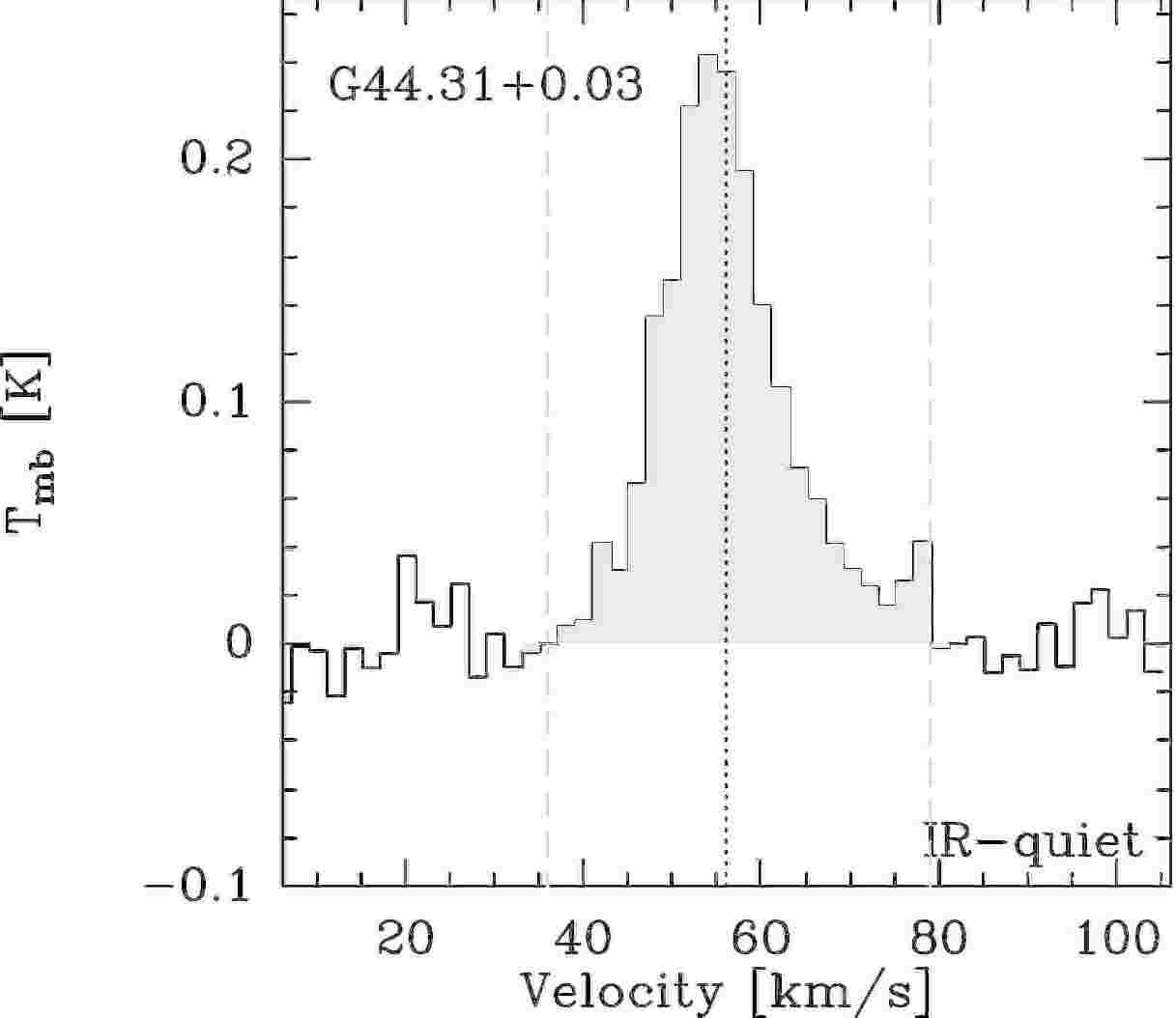} 
  \includegraphics[width=5.6cm,angle=0]{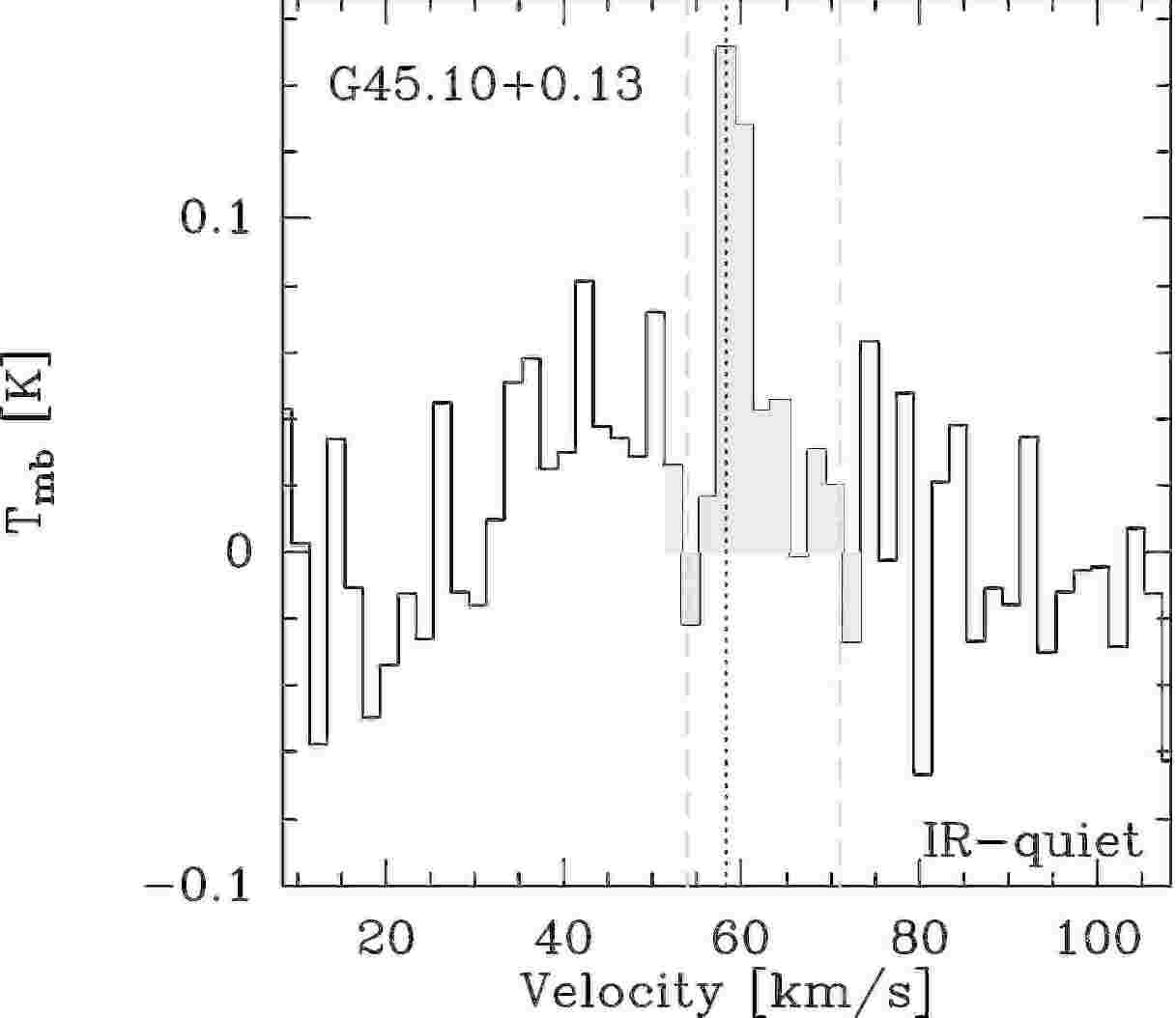} 
  \includegraphics[width=5.6cm,angle=0]{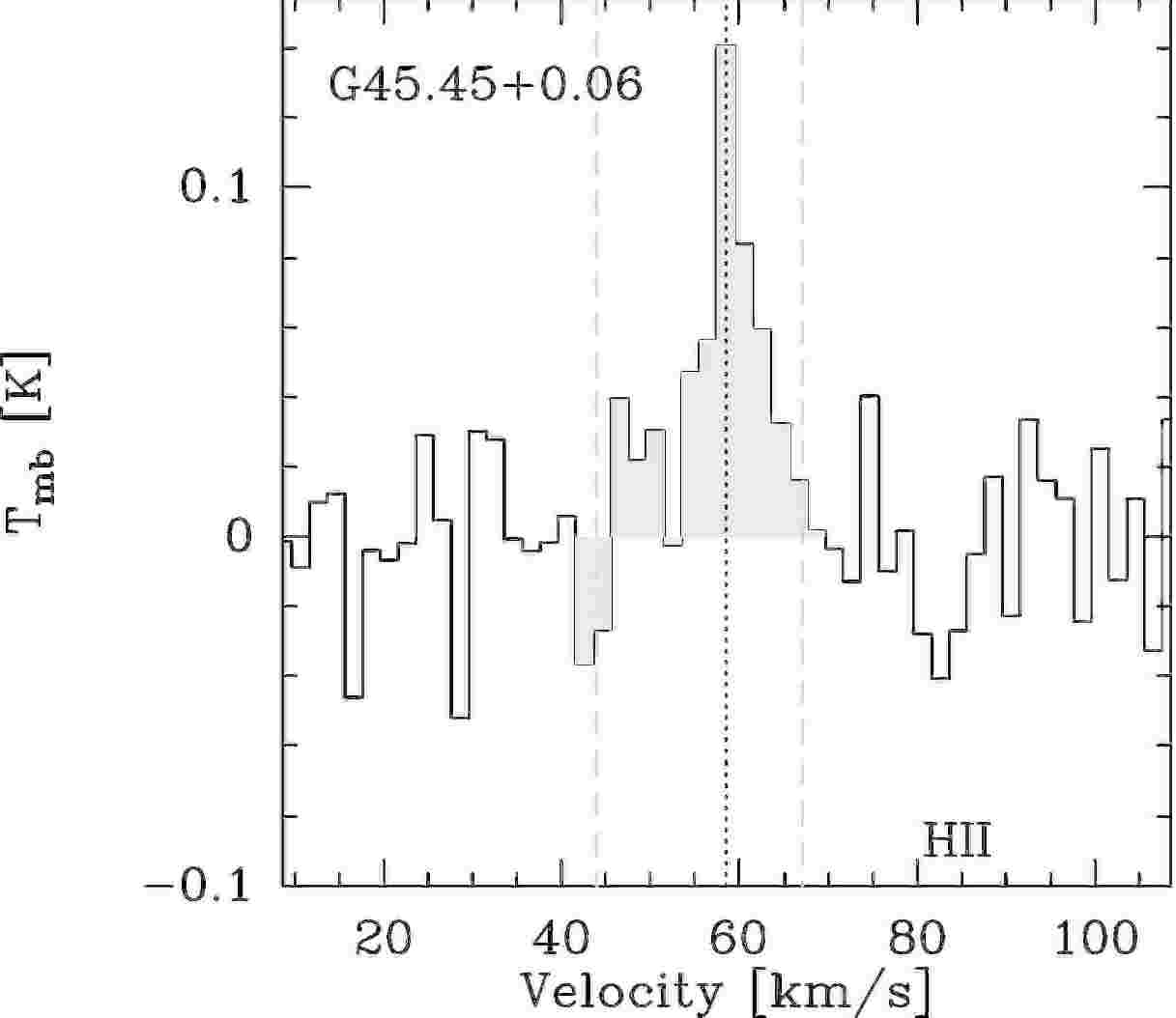} 
  \includegraphics[width=5.6cm,angle=0]{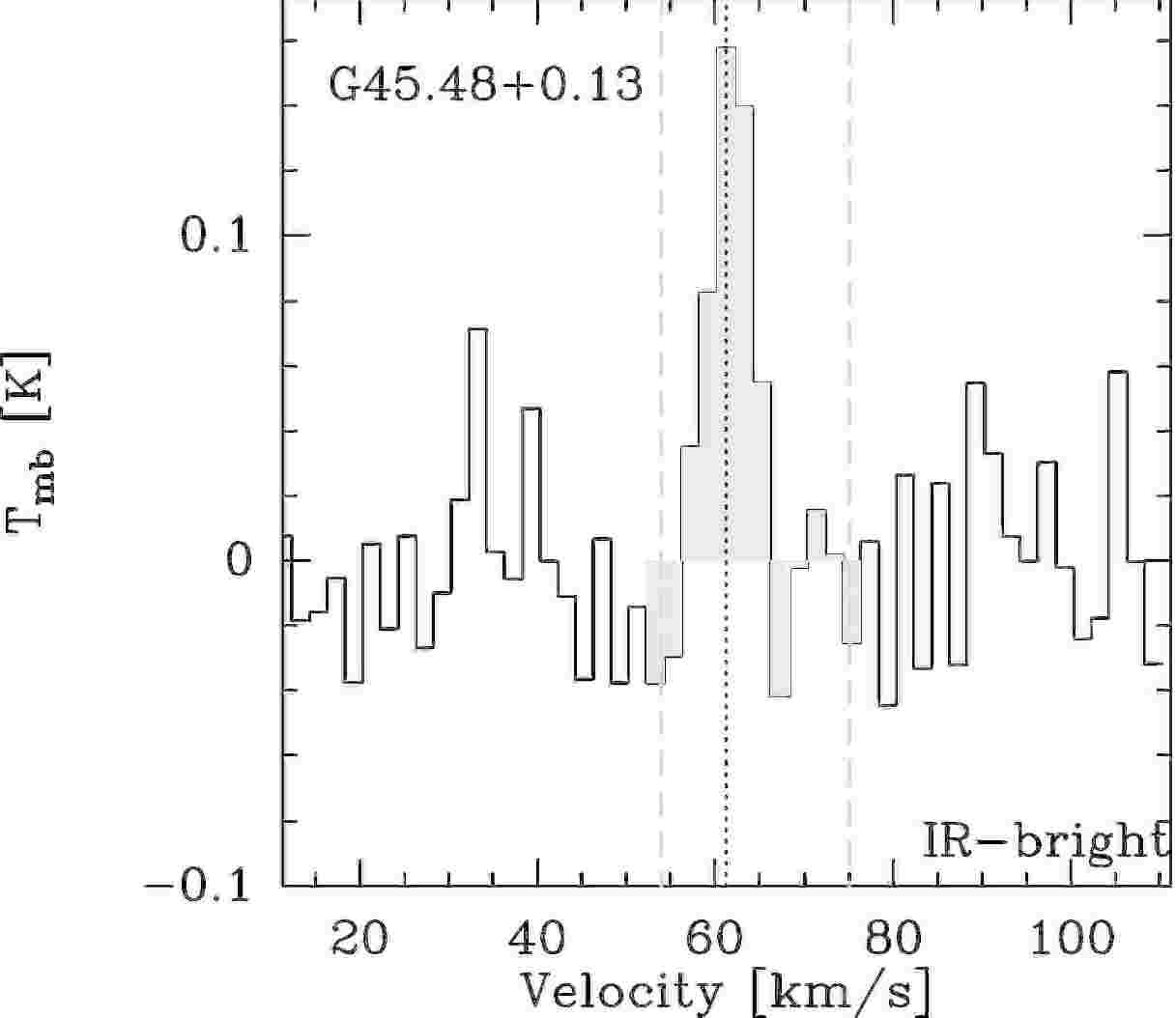} 
  \includegraphics[width=5.6cm,angle=0]{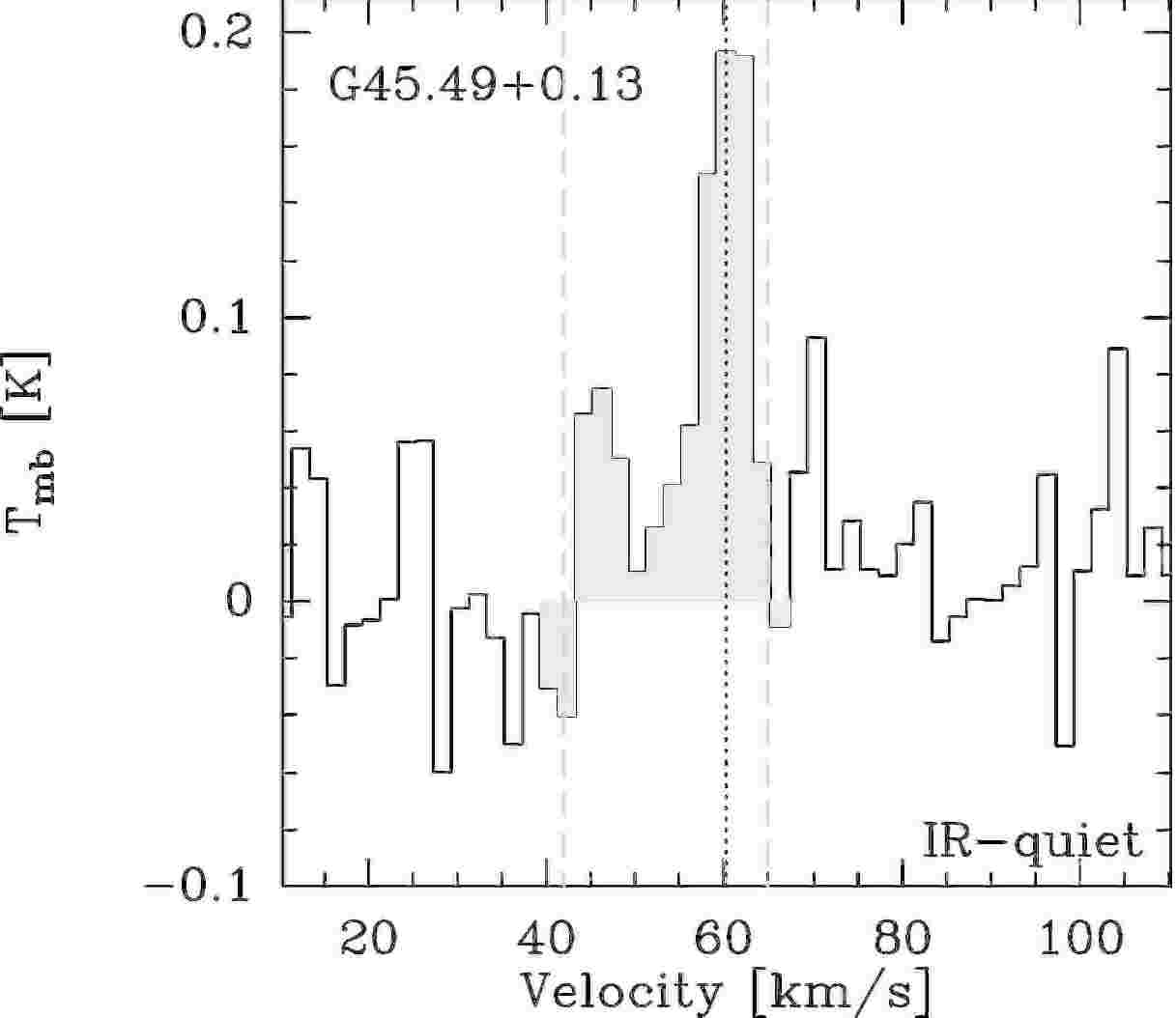} 
  \includegraphics[width=5.6cm,angle=0]{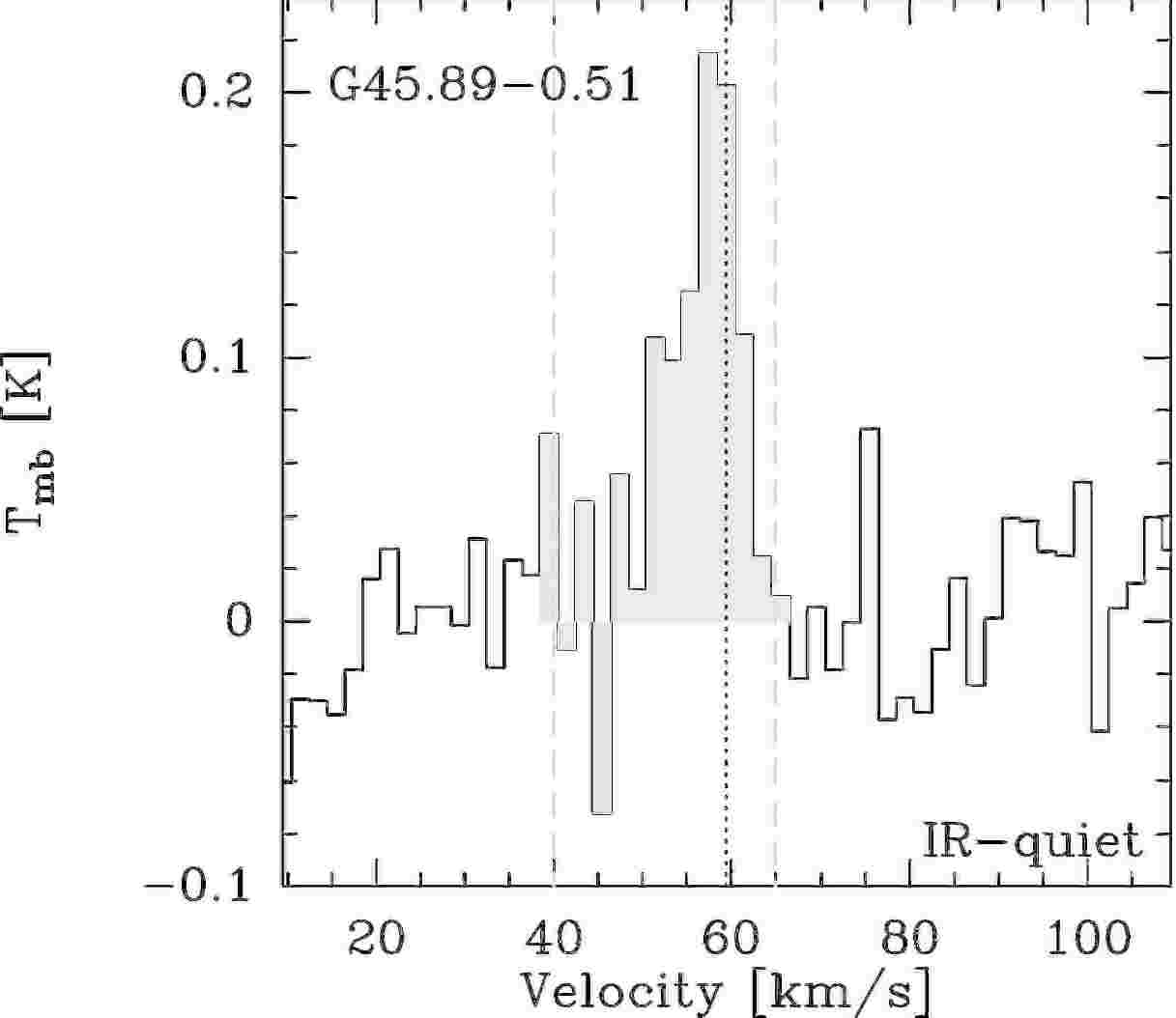} 
  \includegraphics[width=5.6cm,angle=0]{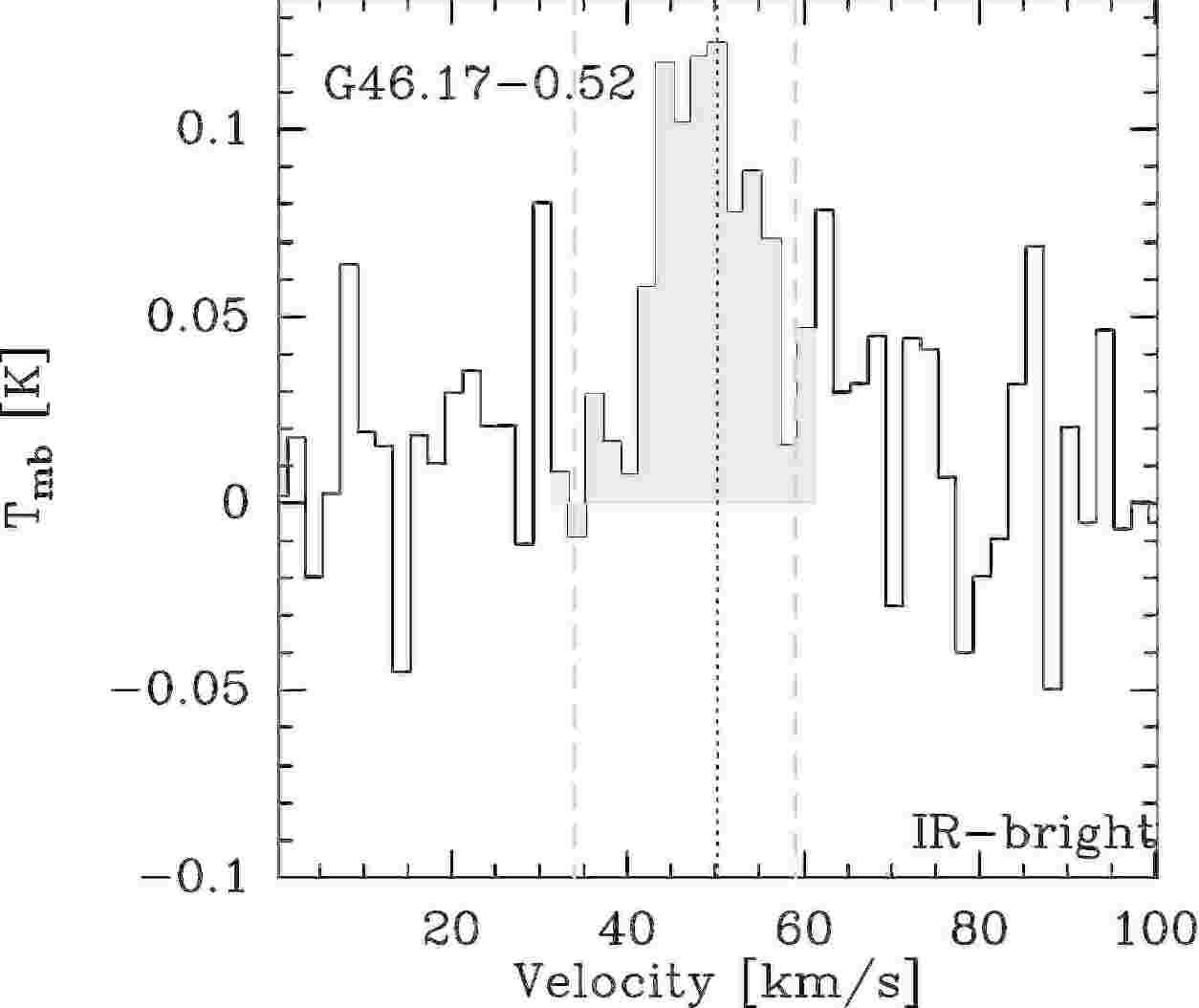} 
  \includegraphics[width=5.6cm,angle=0]{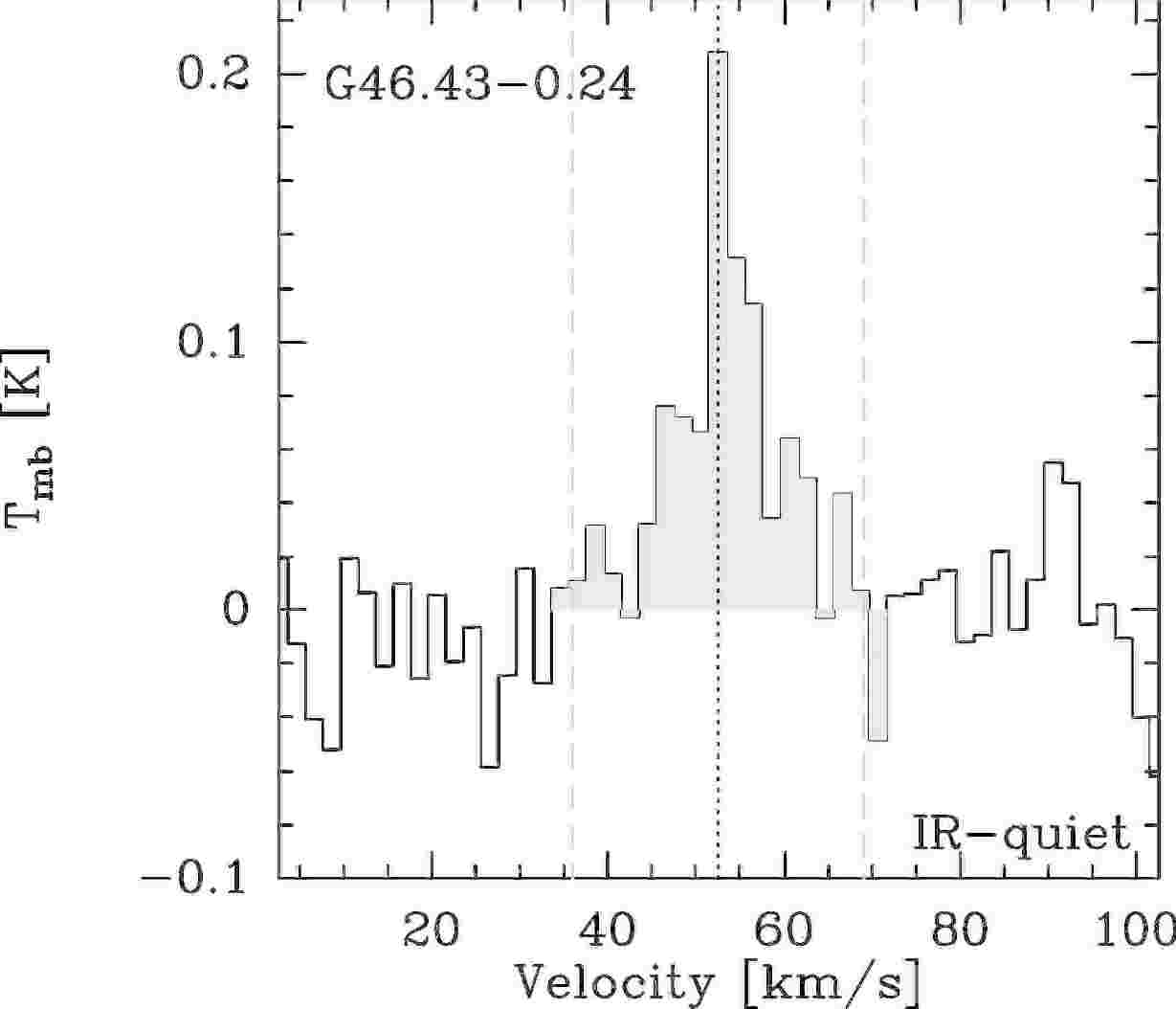} 
  \includegraphics[width=5.6cm,angle=0]{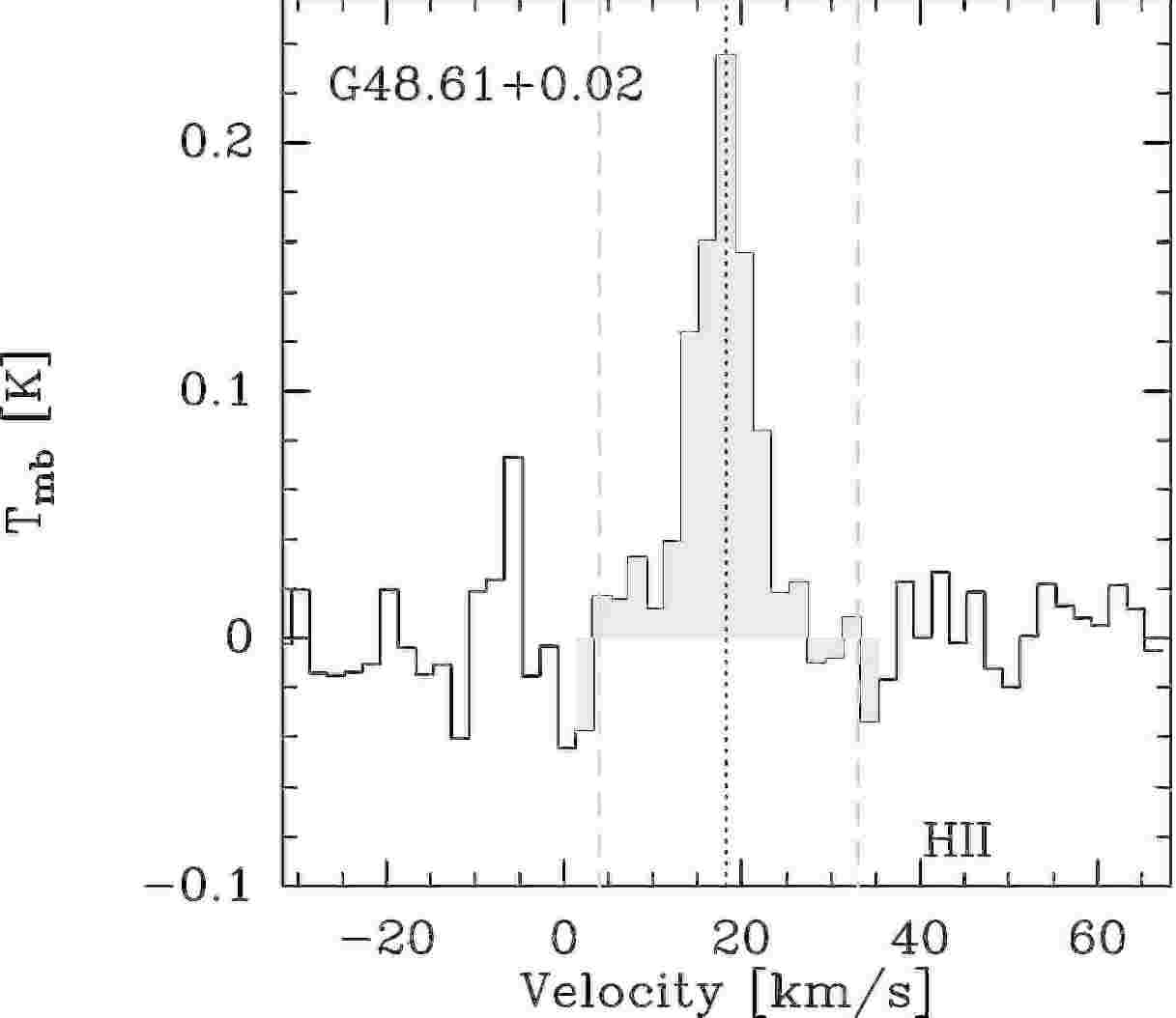} 
  \includegraphics[width=5.6cm,angle=0]{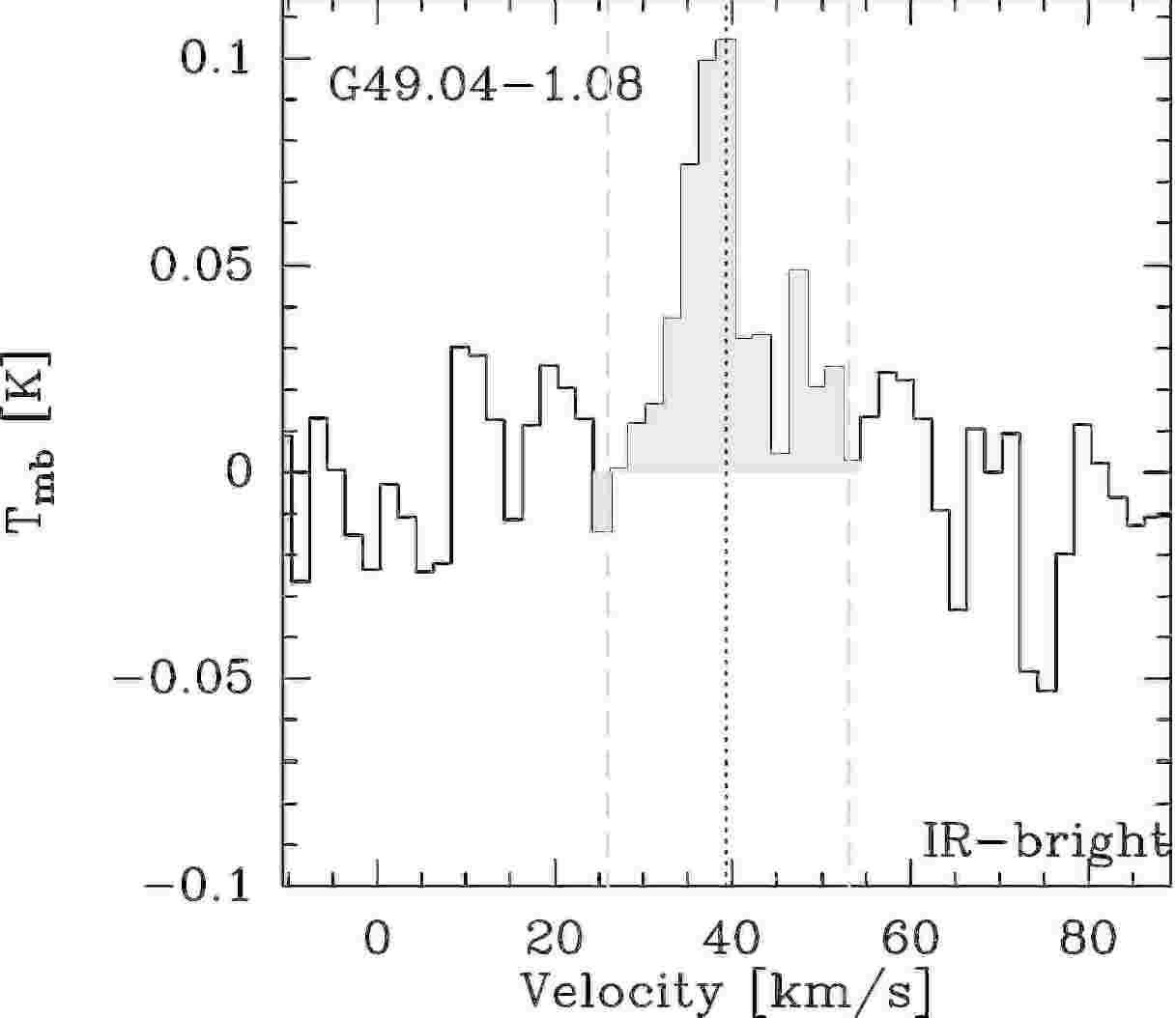} 
  \caption{Continued.}
\end{figure}
\end{landscape}

\begin{landscape}
\begin{figure}
\centering
\ContinuedFloat
 \includegraphics[width=5.6cm,angle=0]{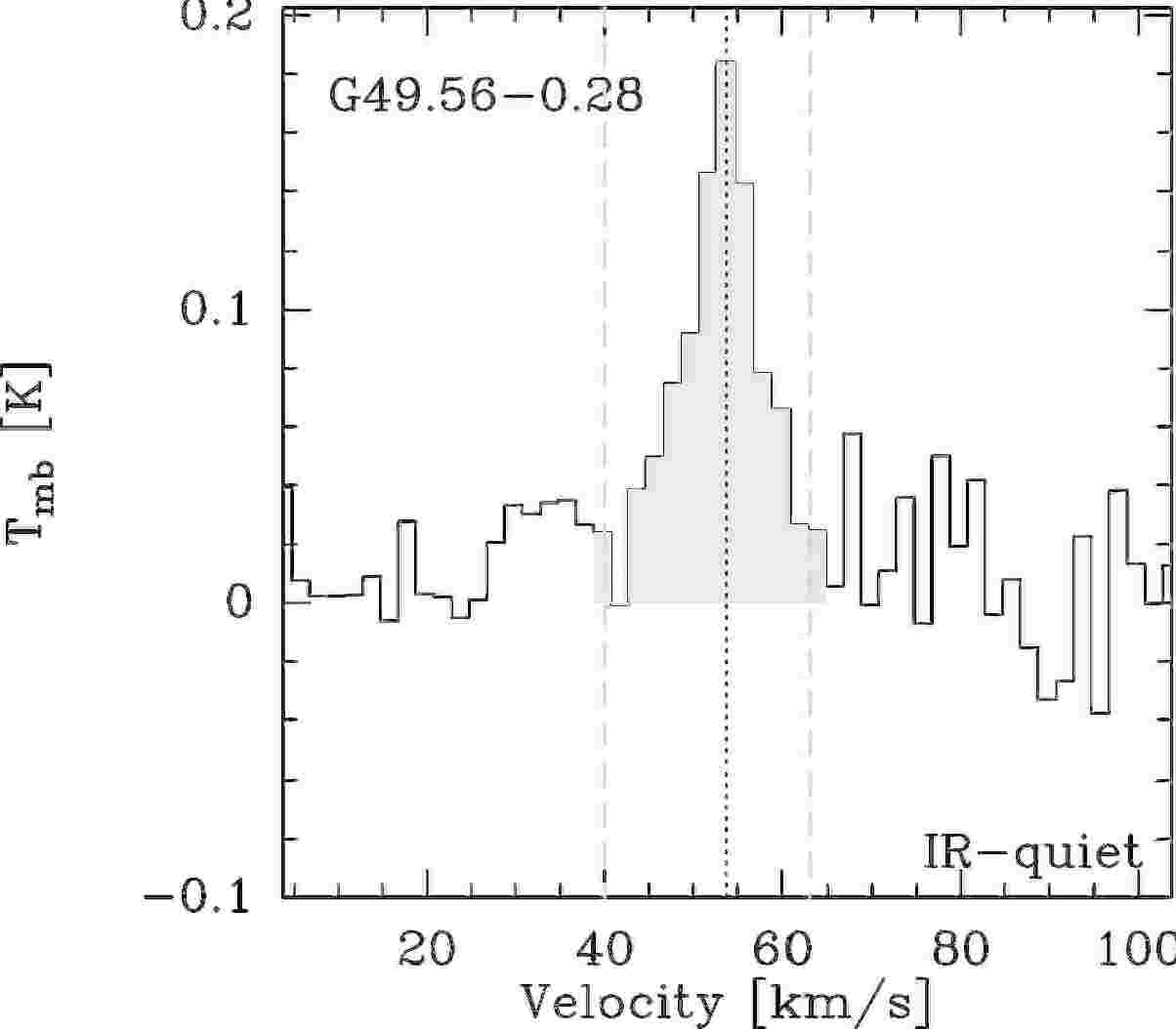} 
  \includegraphics[width=5.6cm,angle=0]{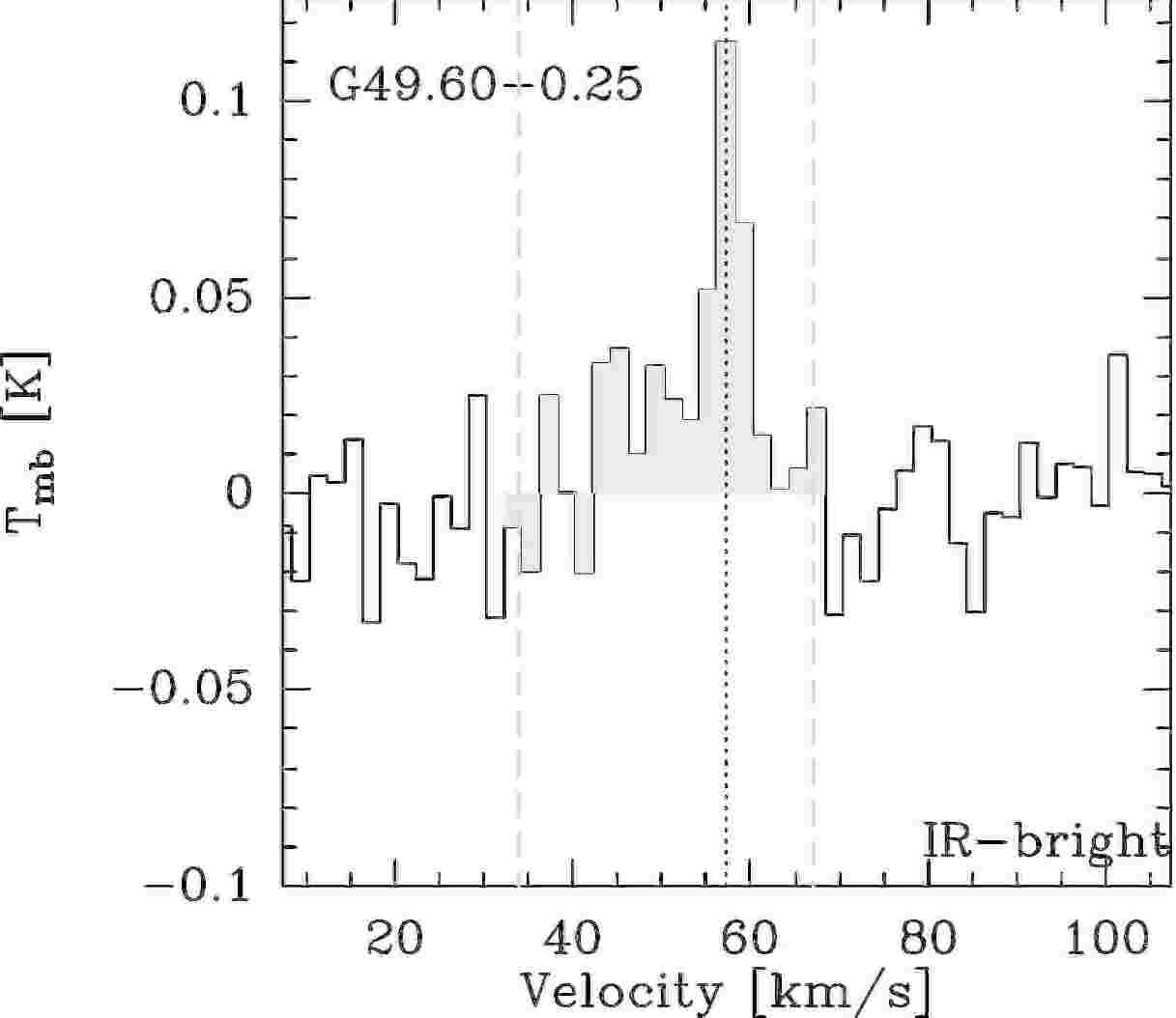} 
  \includegraphics[width=5.6cm,angle=0]{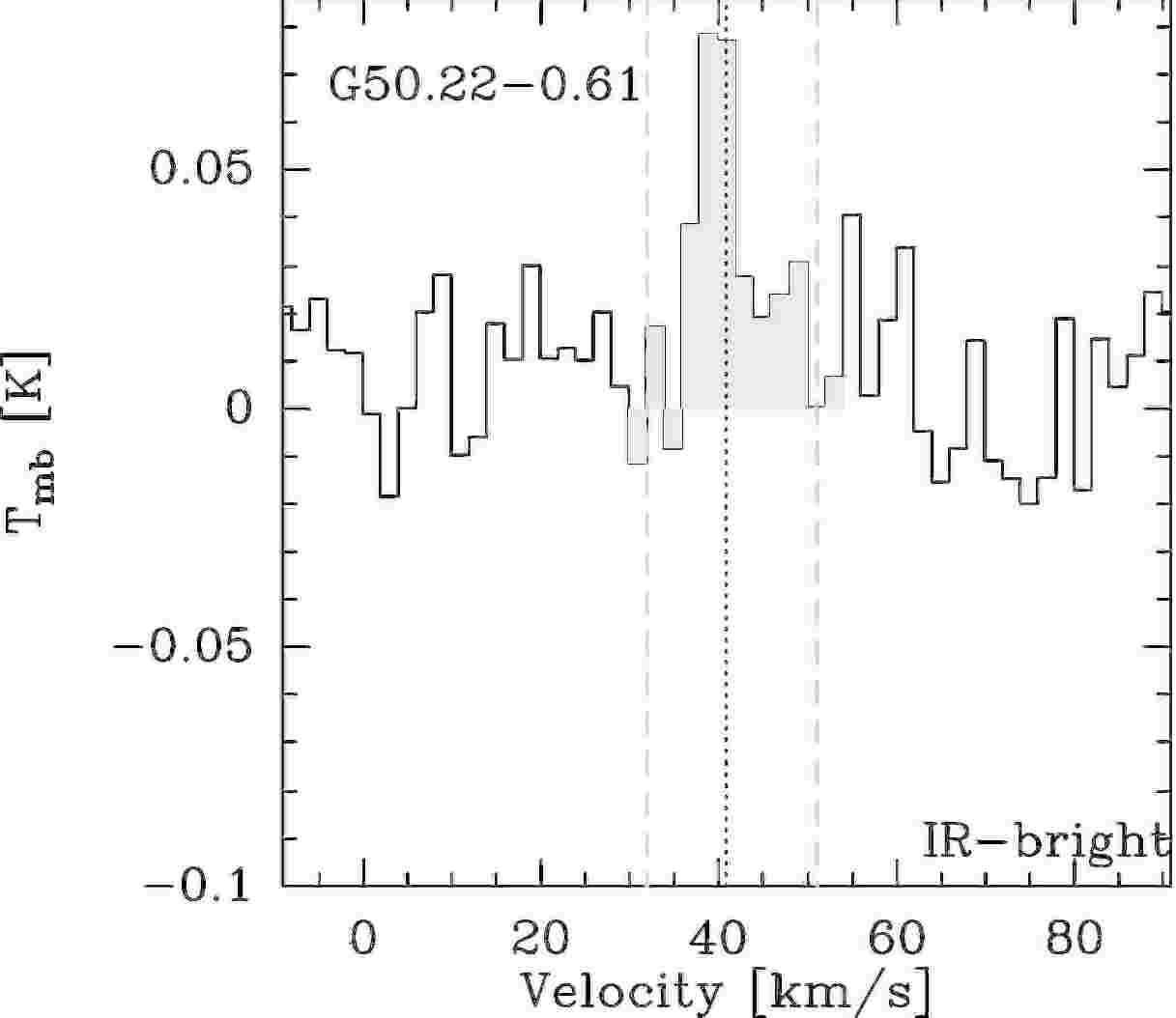} 
  \includegraphics[width=5.6cm,angle=0]{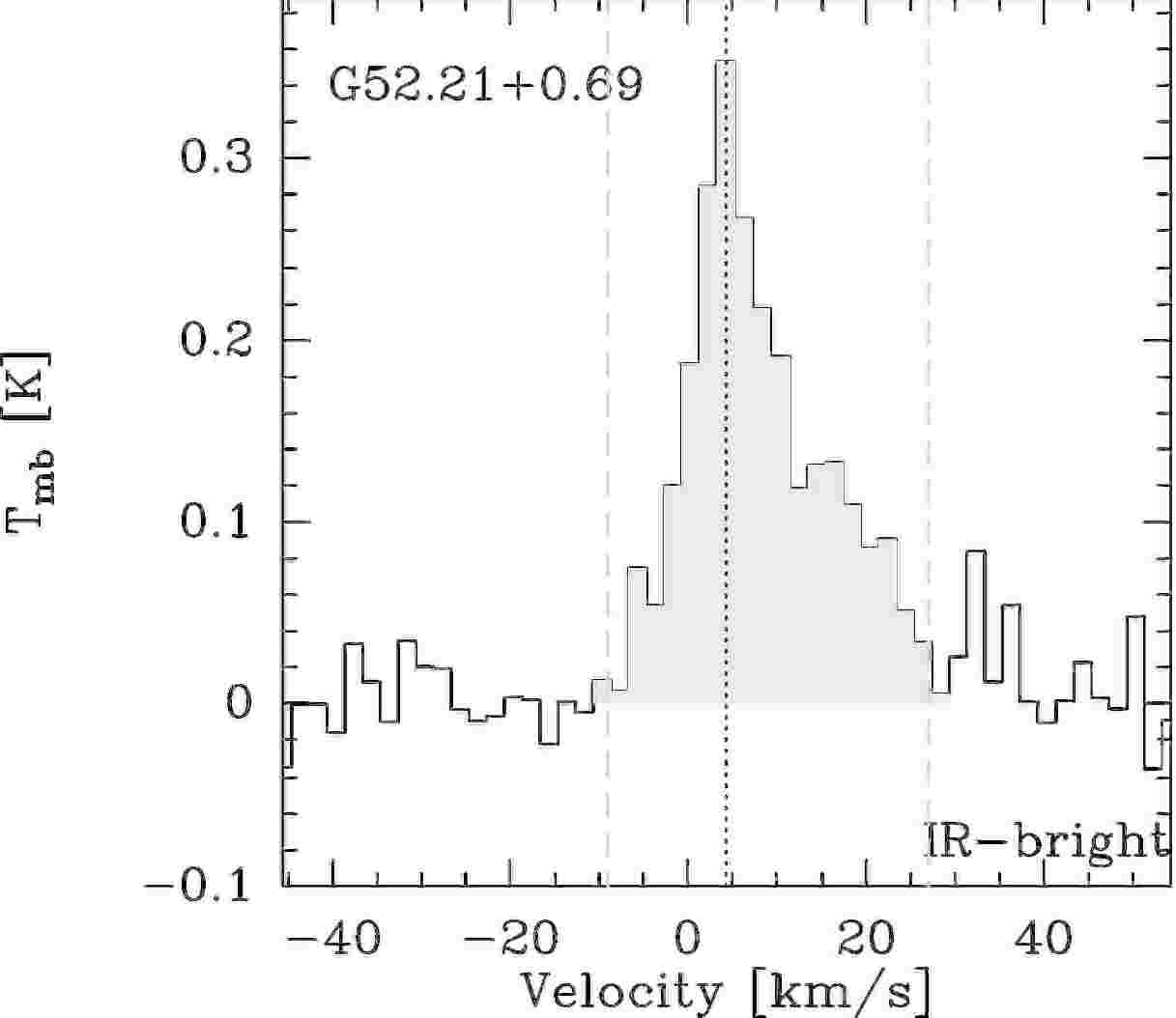} 
  \includegraphics[width=5.6cm,angle=0]{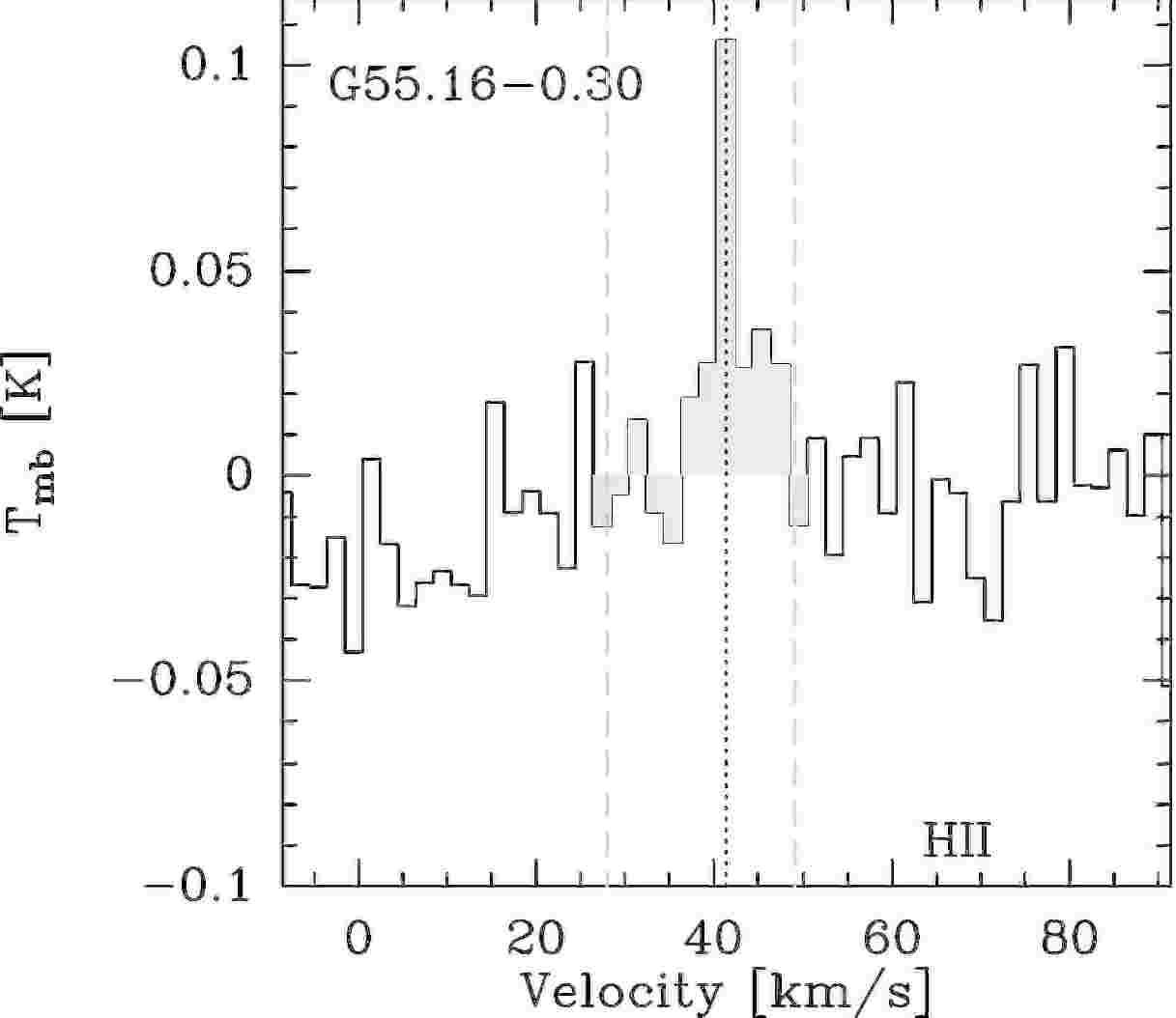} 
  \includegraphics[width=5.6cm,angle=0]{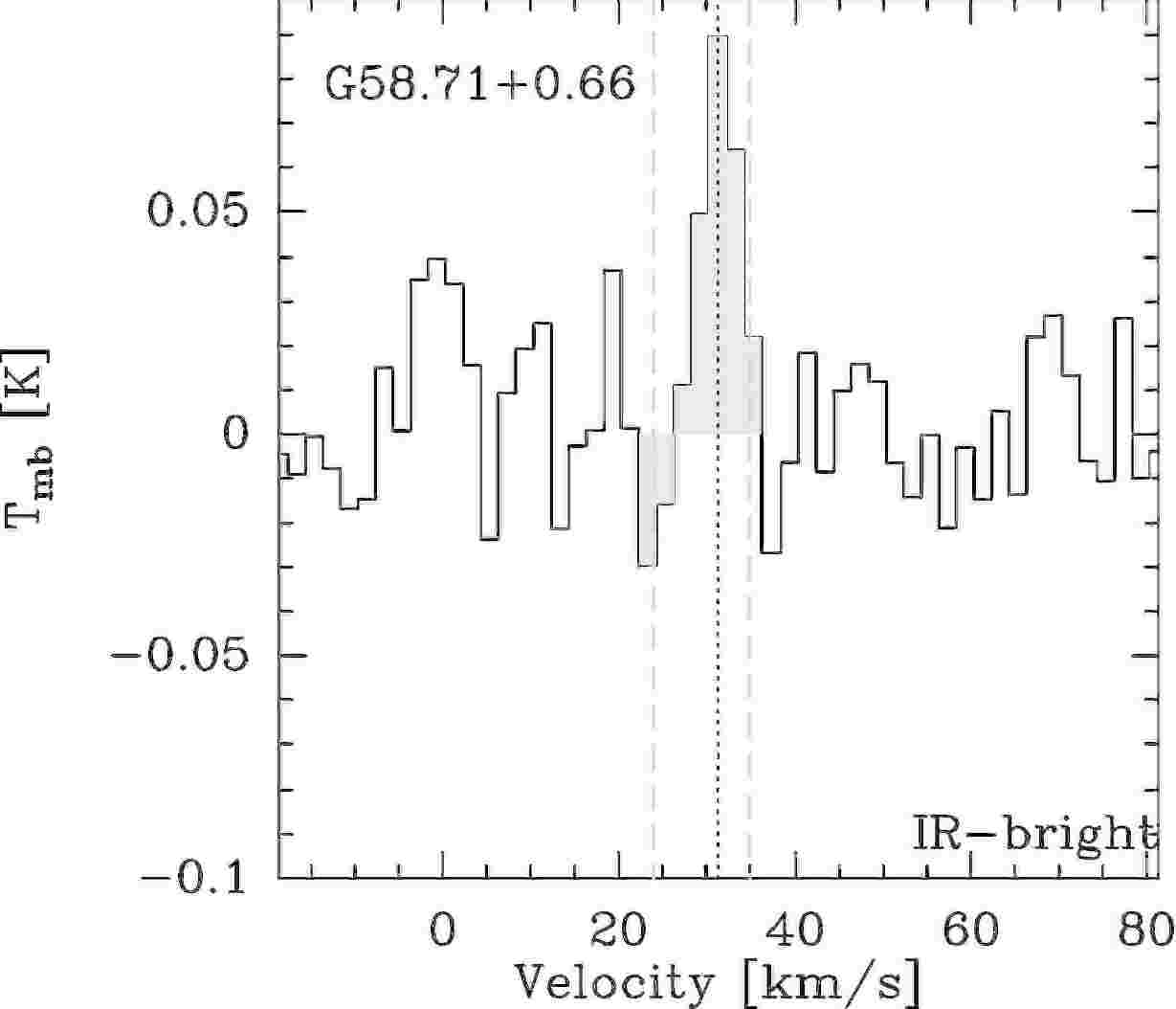} 
  \includegraphics[width=5.6cm,angle=0]{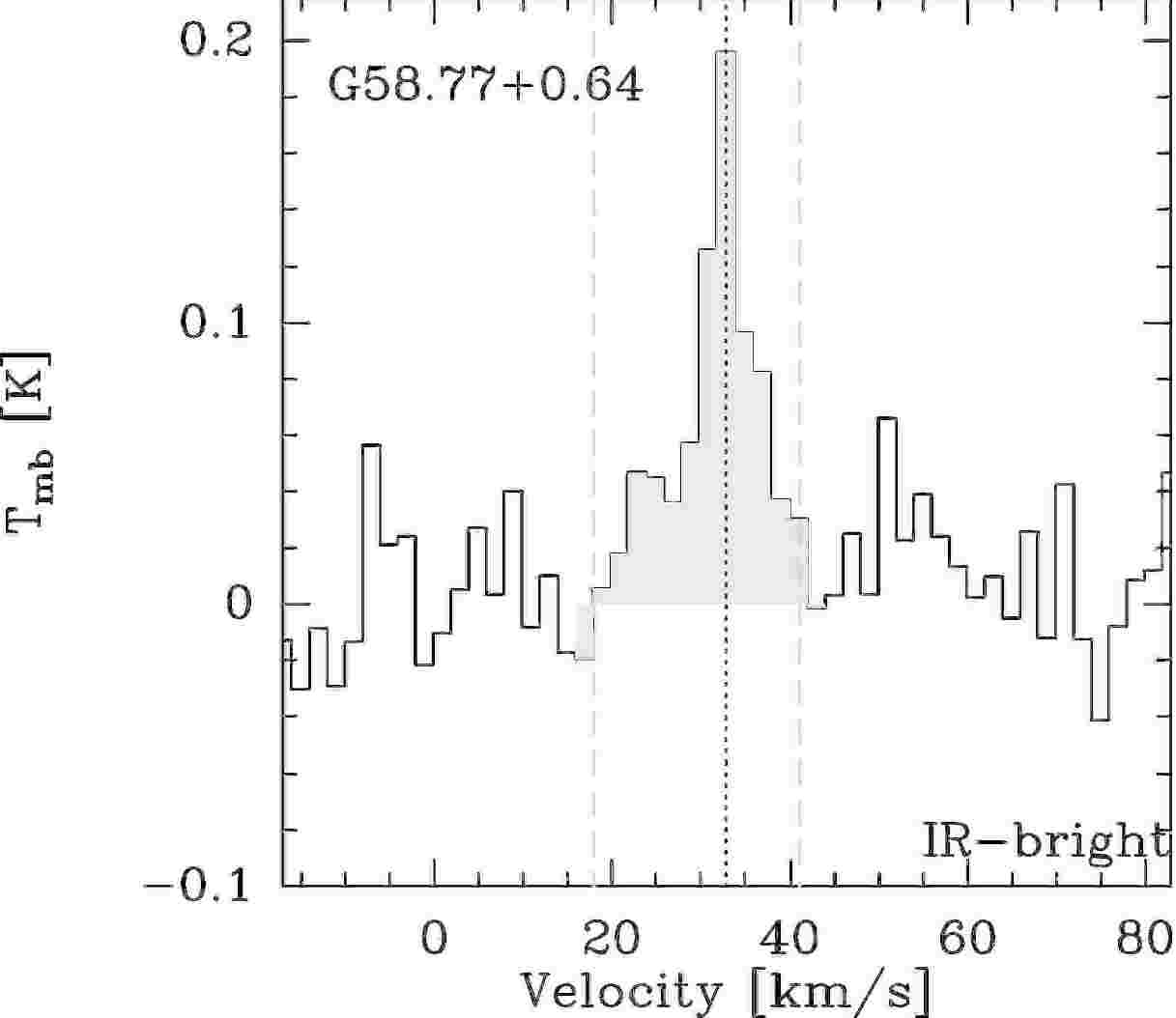} 
  \includegraphics[width=5.6cm,angle=0]{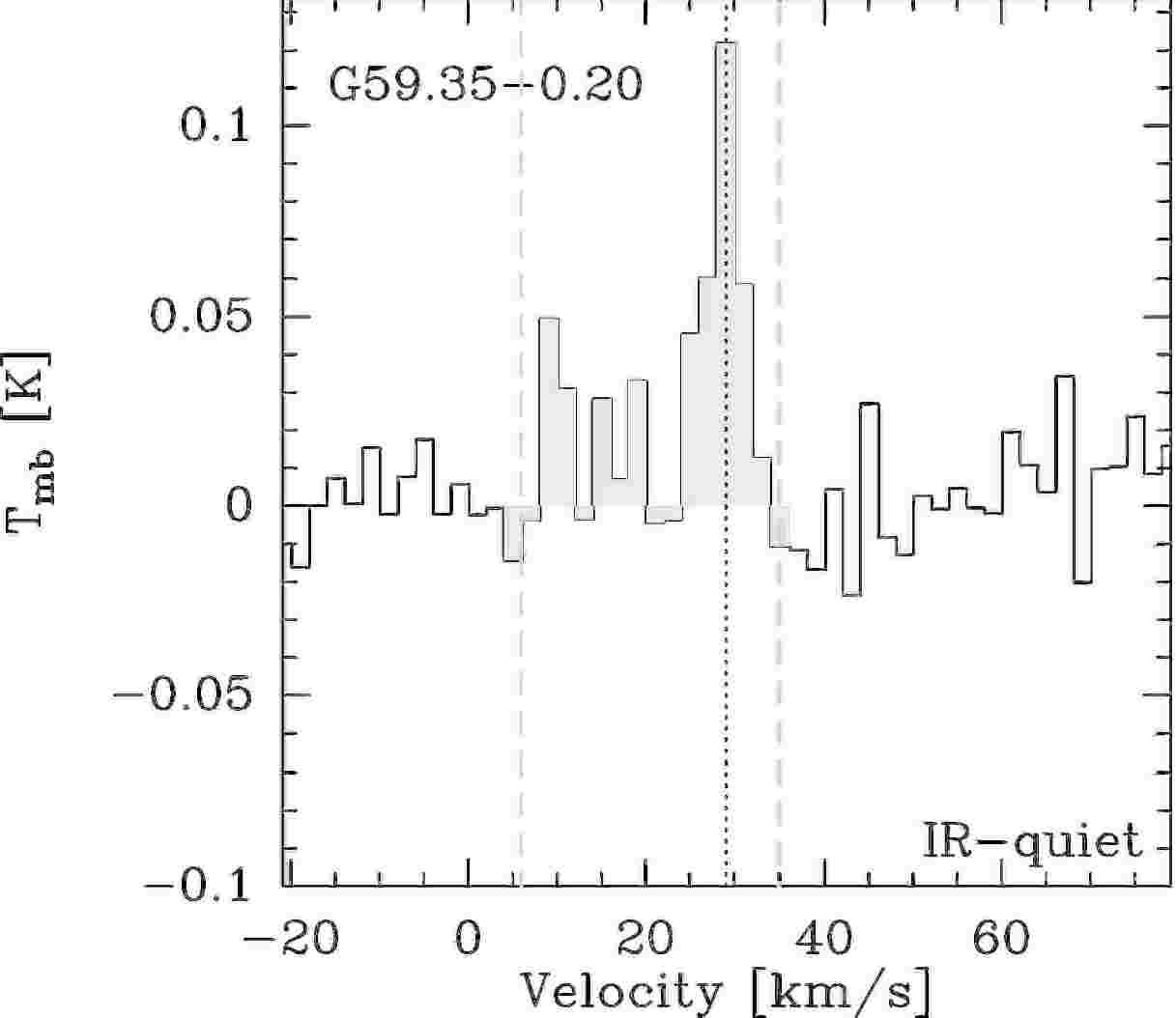} 
  \includegraphics[width=5.6cm,angle=0]{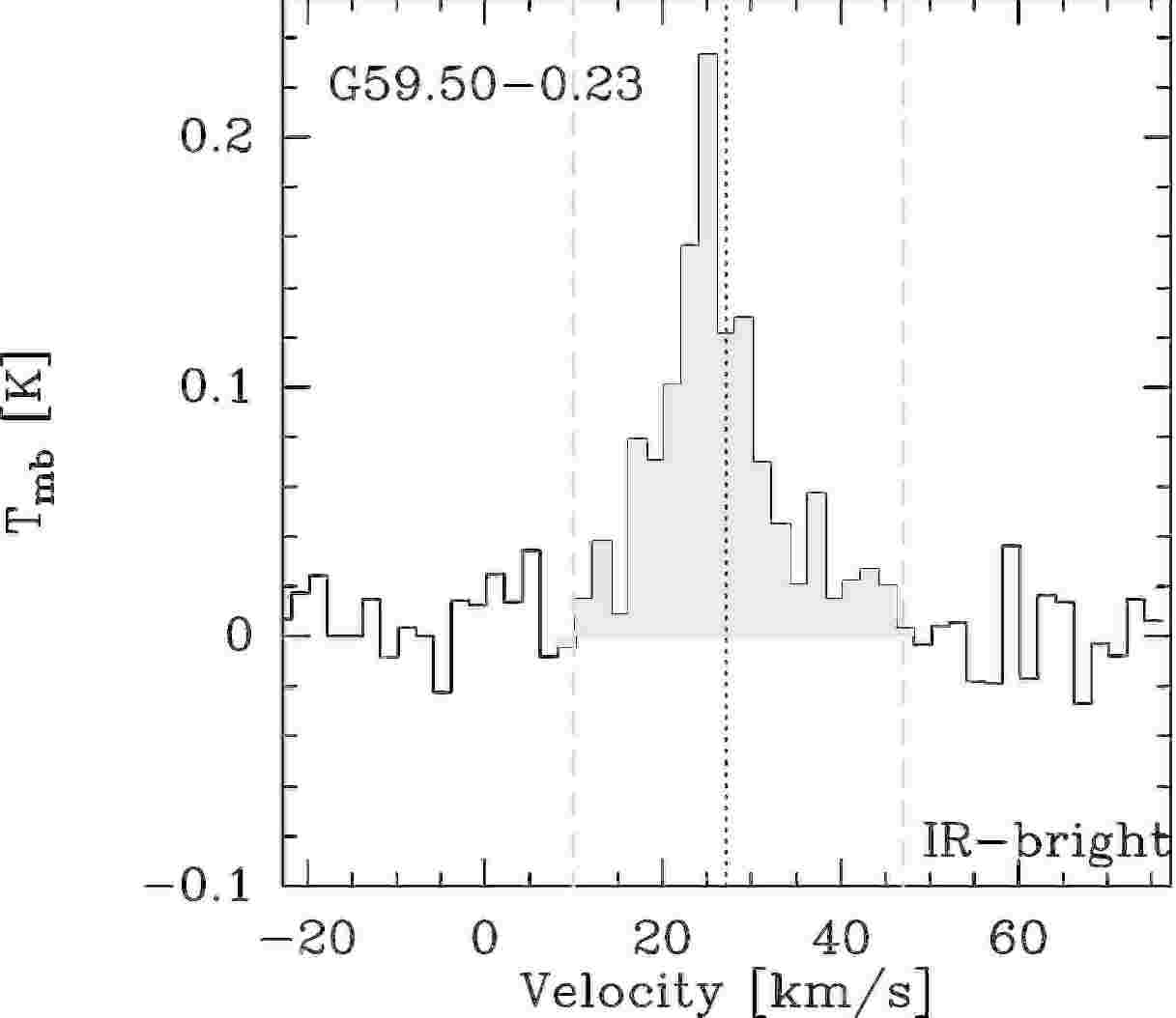} 
  \includegraphics[width=5.6cm,angle=0]{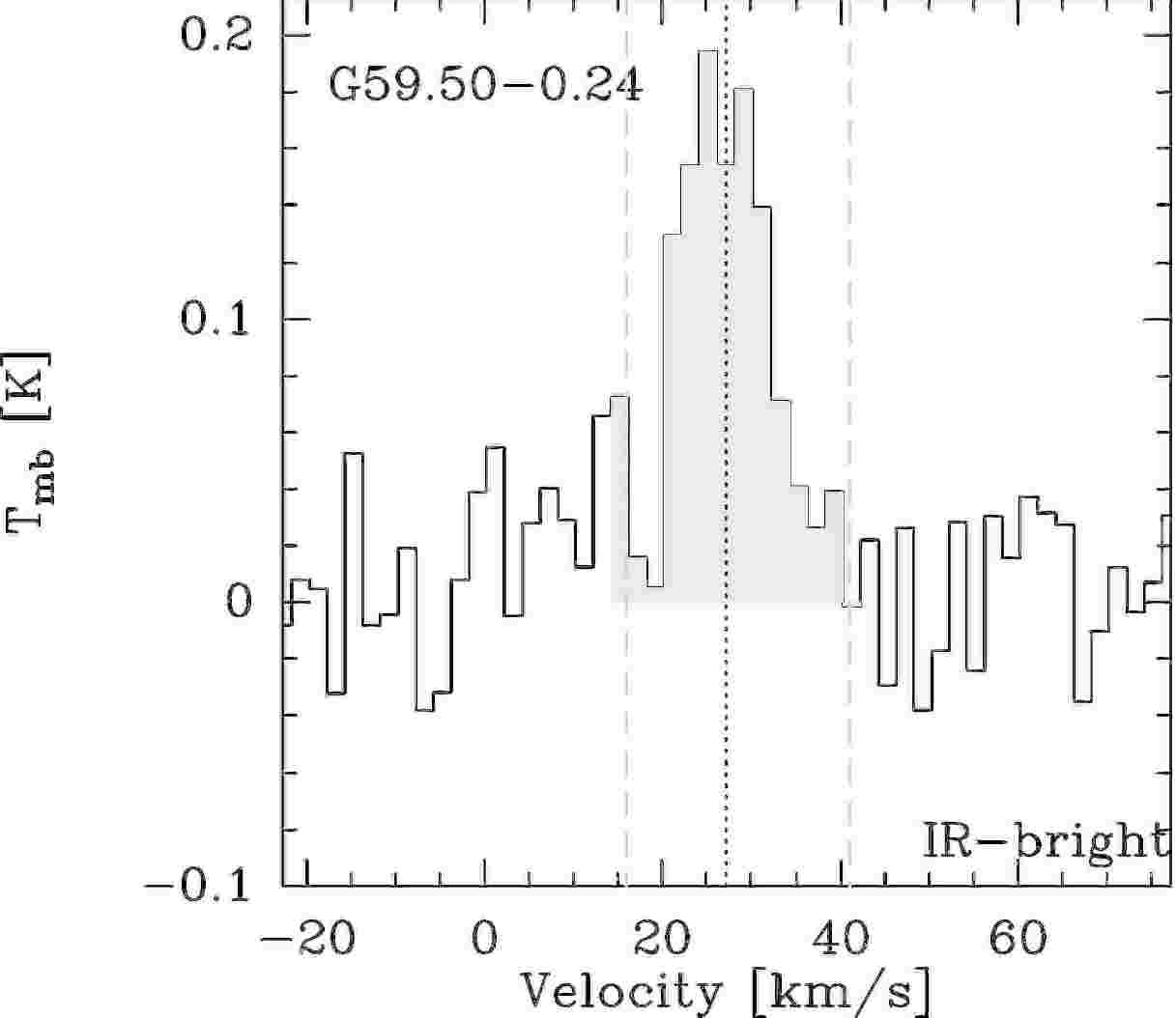} 
  \includegraphics[width=5.6cm,angle=0]{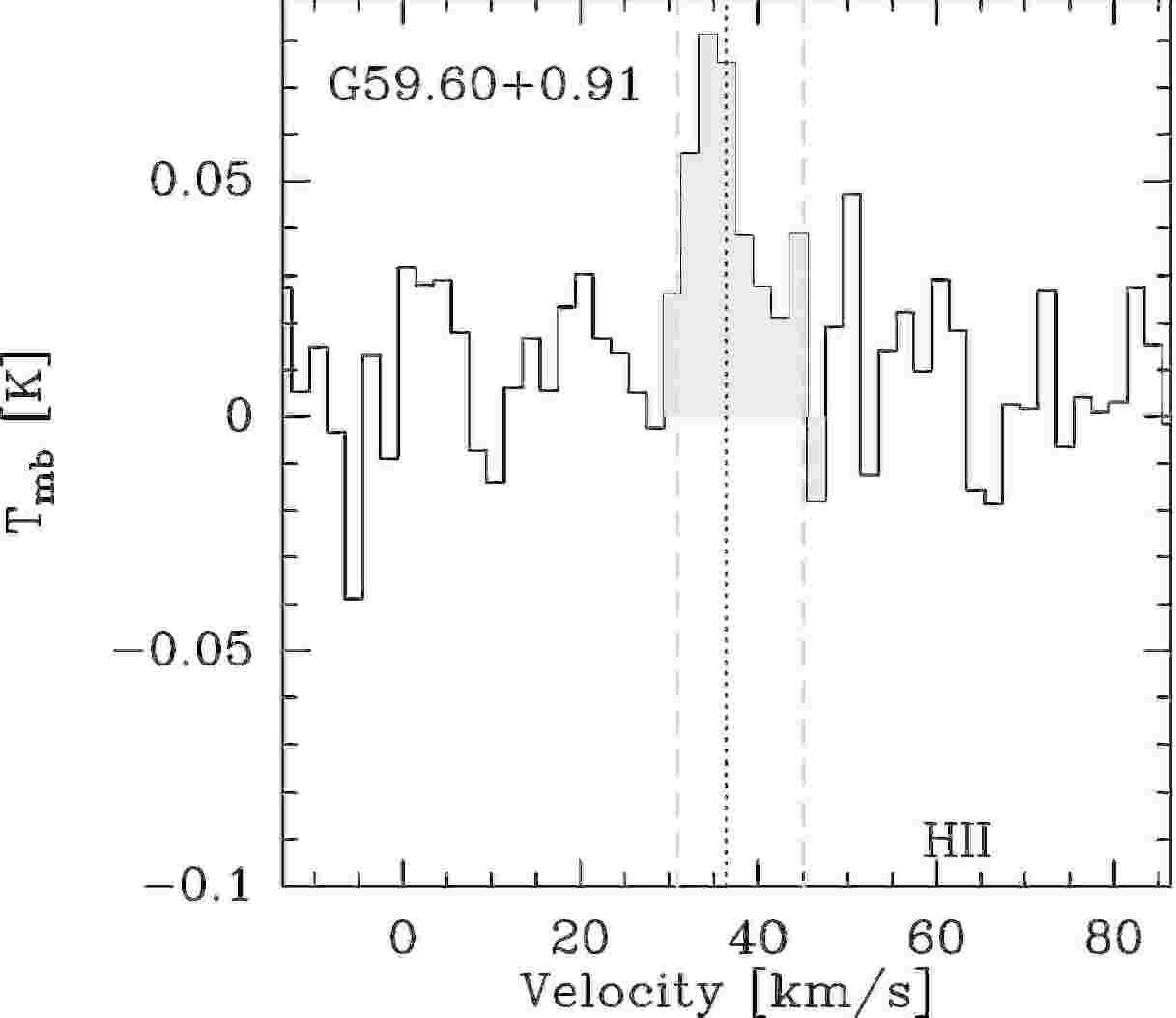} 
  \includegraphics[width=5.6cm,angle=0]{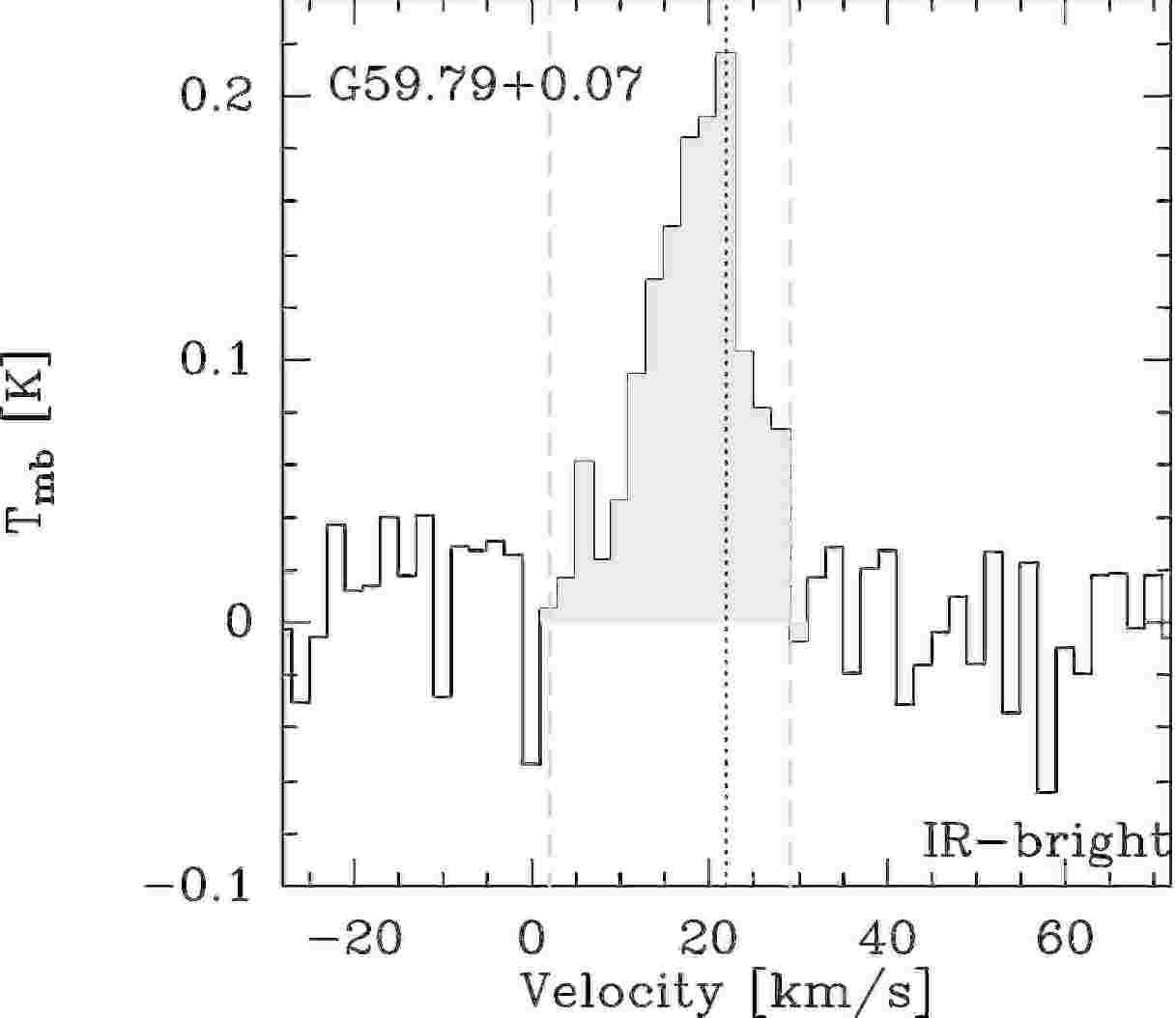} 
 \caption{Continued.}
\end{figure}
\end{landscape}
\begin{landscape}
\begin{figure}
\centering
\ContinuedFloat
  \includegraphics[width=5.6cm,angle=0]{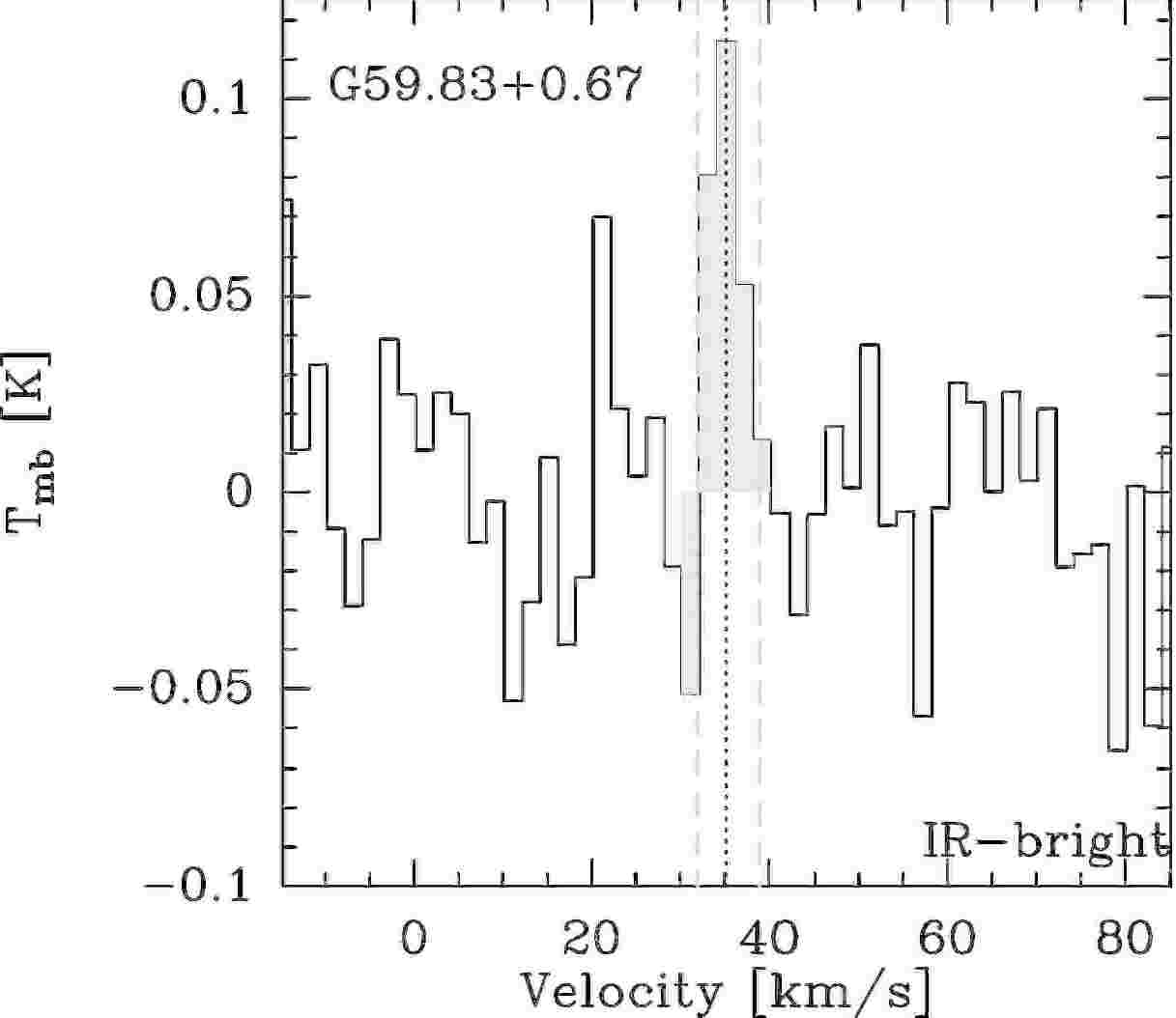} 
 \caption{Continued.}
\end{figure}
\end{landscape}

\clearpage
\subsection{Spectra of sources with only SiO ($2-1$) observations and non-detections}

\begin{landscape}
\begin{figure}
\centering
  \includegraphics[width=5.2cm,angle=90]{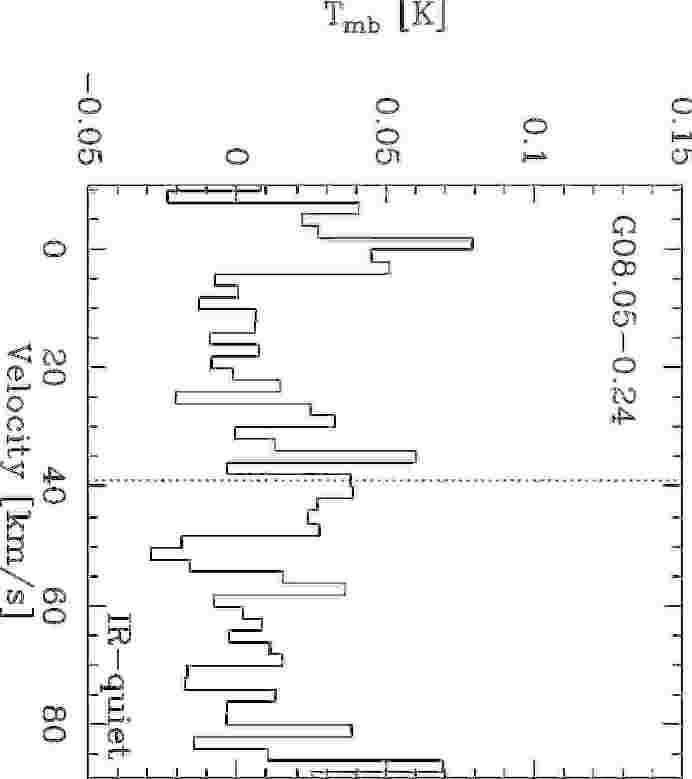} 
  \includegraphics[width=5.2cm,angle=90]{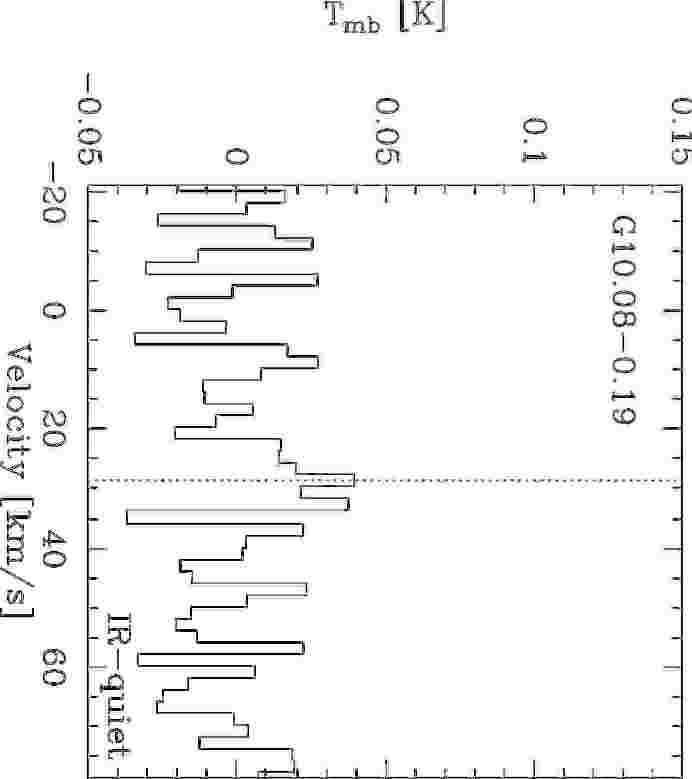} 
  \includegraphics[width=5.2cm,angle=90]{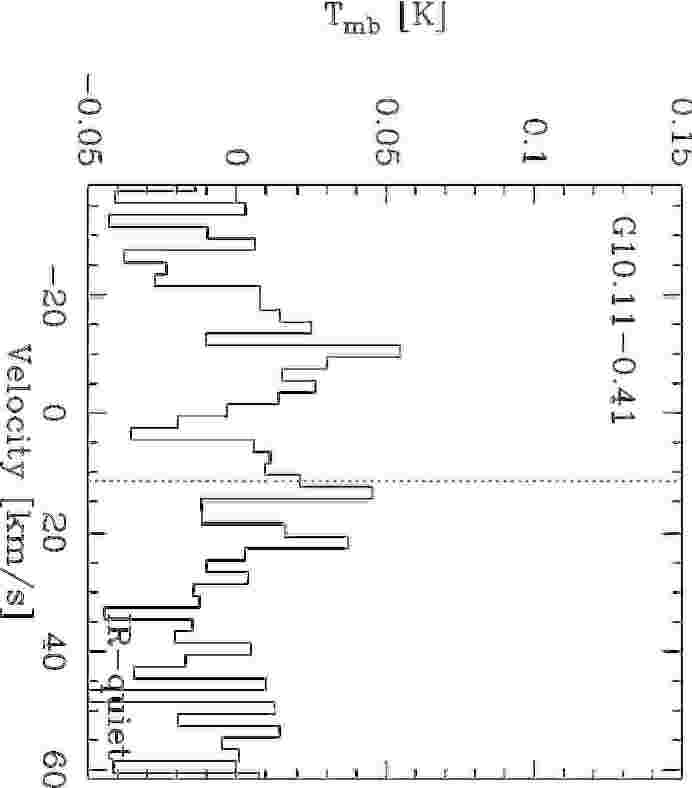} 
  \includegraphics[width=5.2cm,angle=90]{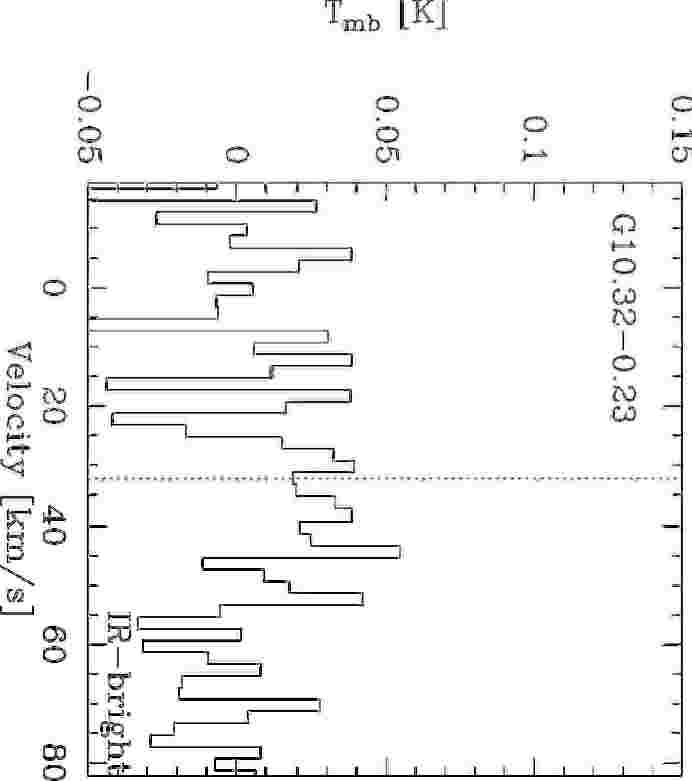} 
  \includegraphics[width=5.2cm,angle=90]{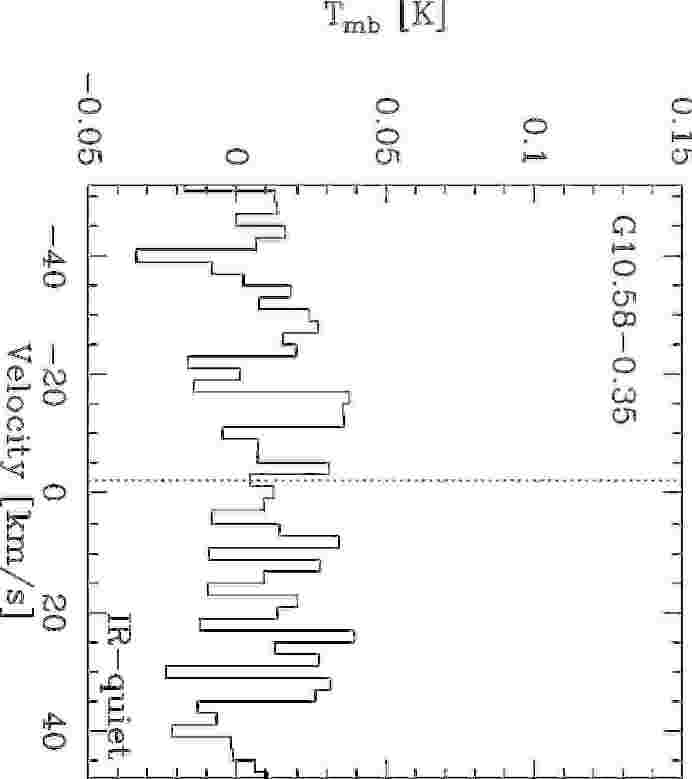} 
  \includegraphics[width=5.2cm,angle=90]{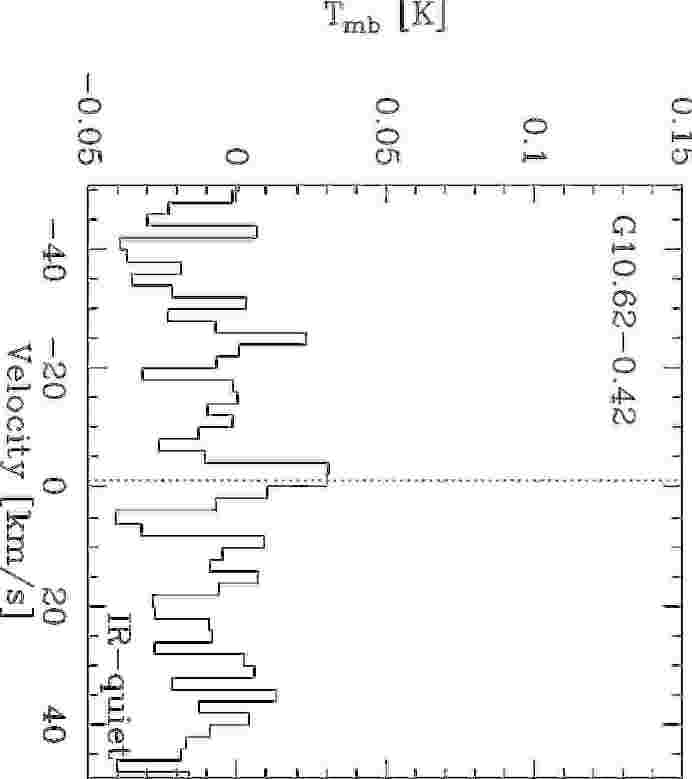} 
  \includegraphics[width=5.2cm,angle=90]{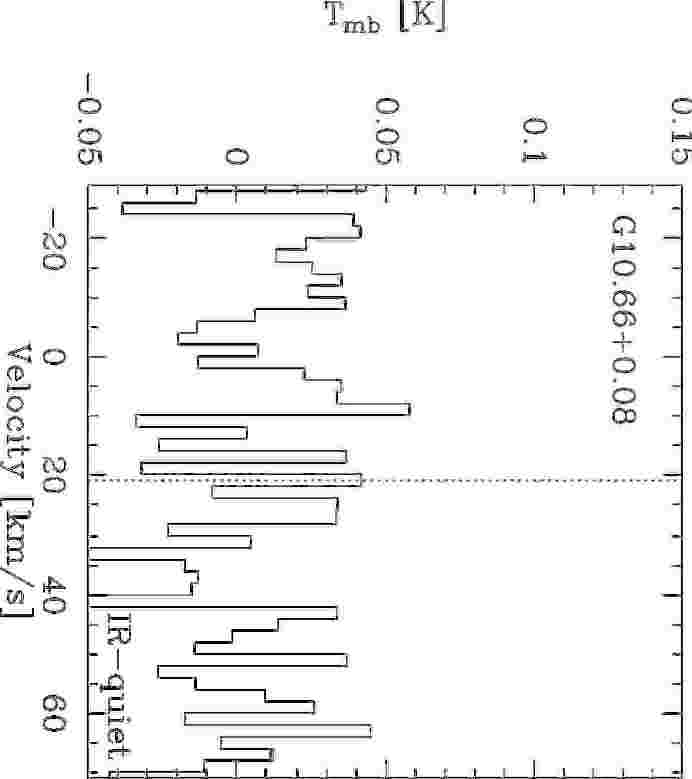} 
  \includegraphics[width=5.2cm,angle=90]{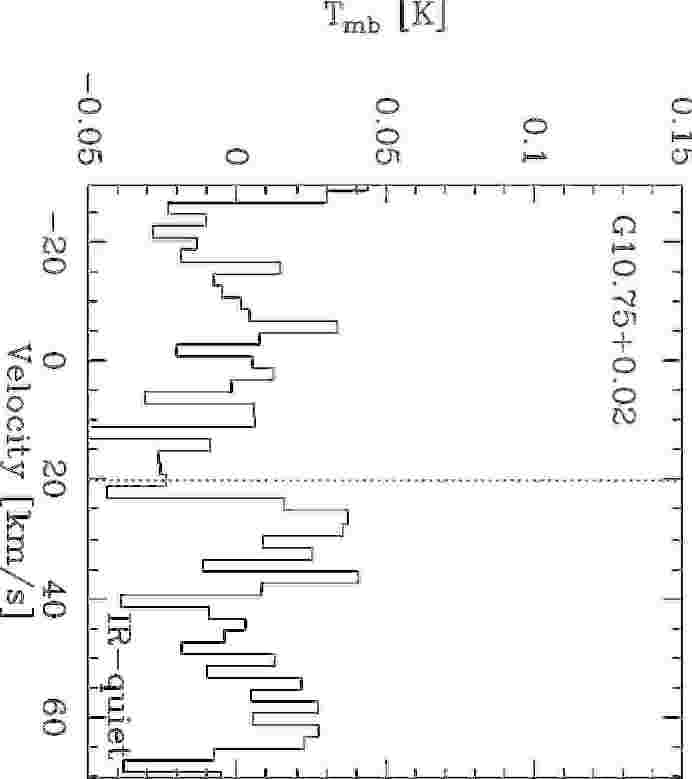} 
  \includegraphics[width=5.2cm,angle=90]{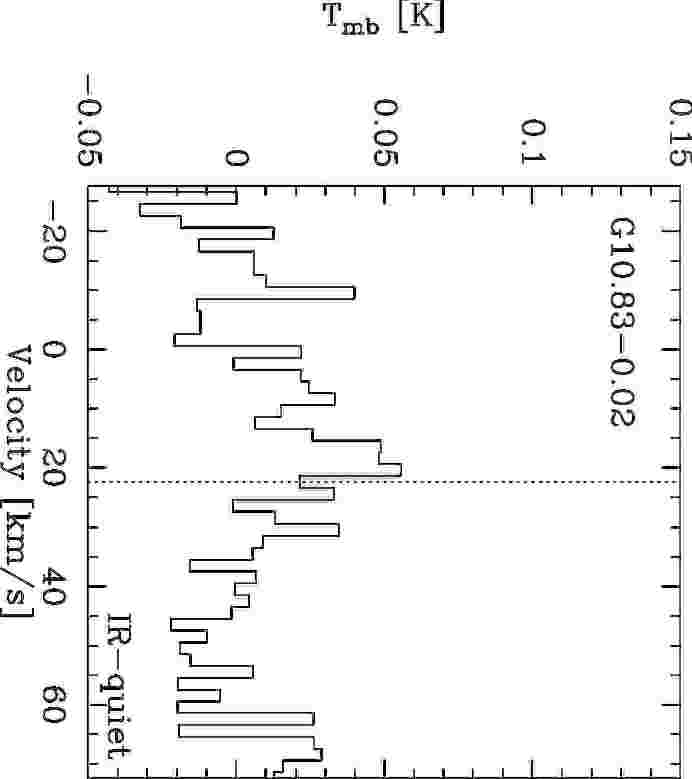} 
  \includegraphics[width=5.2cm,angle=90]{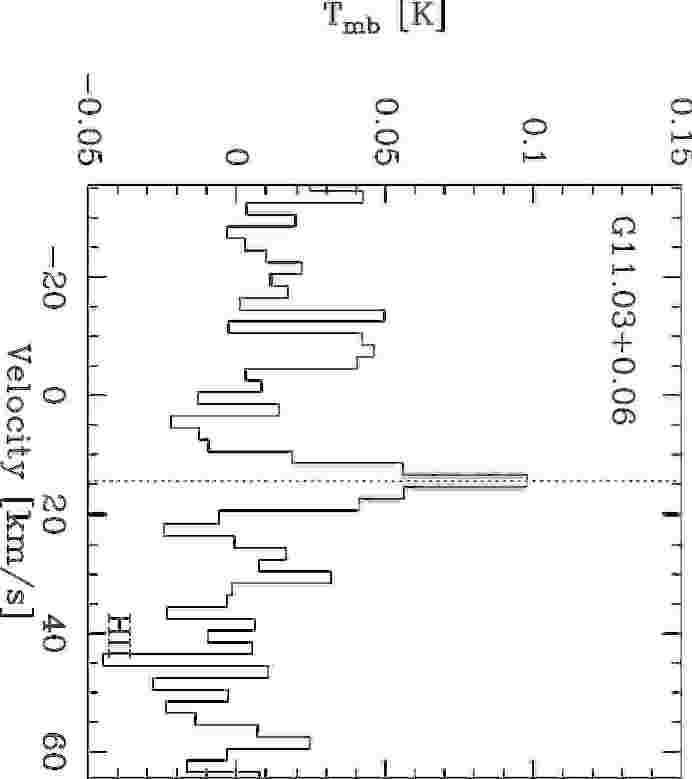} 
  \includegraphics[width=5.2cm,angle=90]{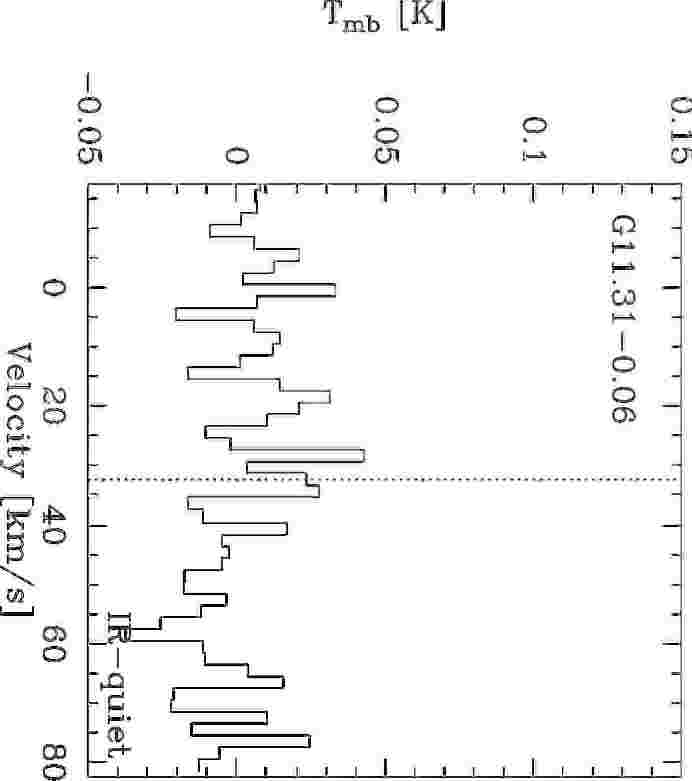} 
  \includegraphics[width=5.2cm,angle=90]{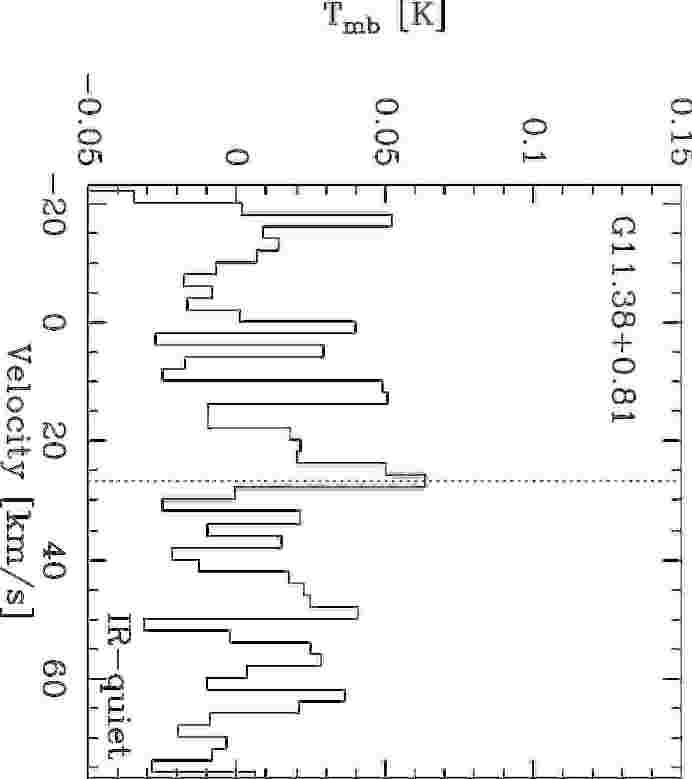} 
 \caption{SiO (2--1) non-detections.}
\end{figure}
\end{landscape}

\begin{landscape}
\begin{figure}
\ContinuedFloat
\centering
  \includegraphics[width=5.2cm,angle=90]{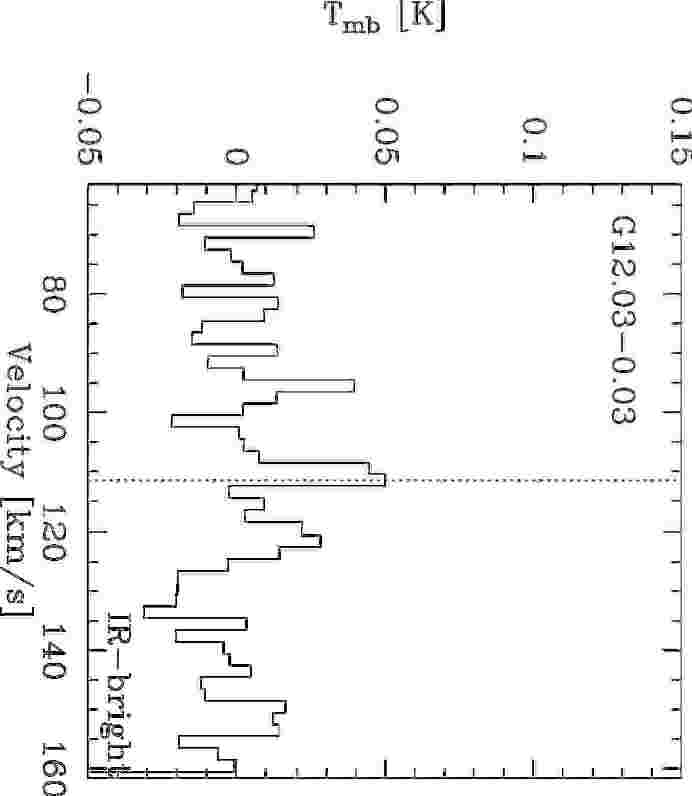} 
  \includegraphics[width=5.2cm,angle=90]{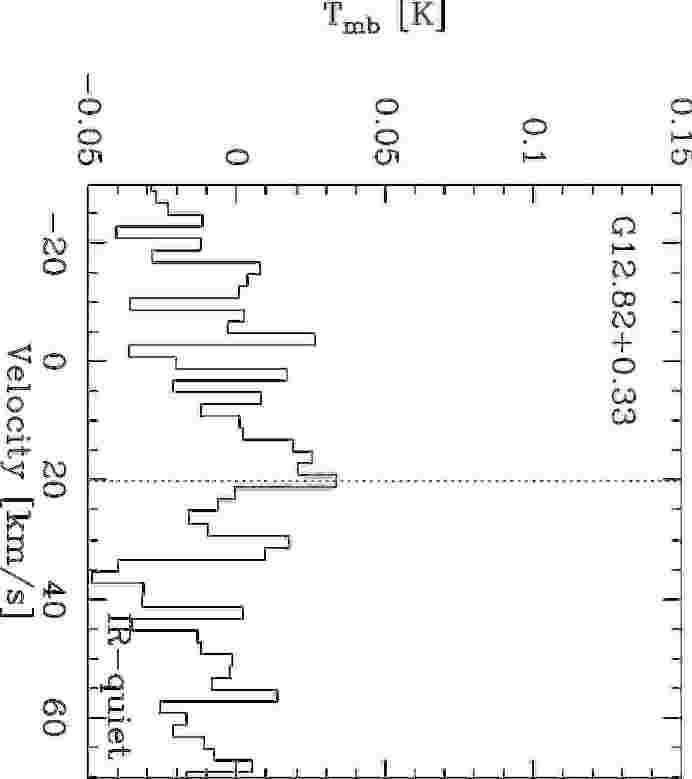} 
  \includegraphics[width=5.2cm,angle=90]{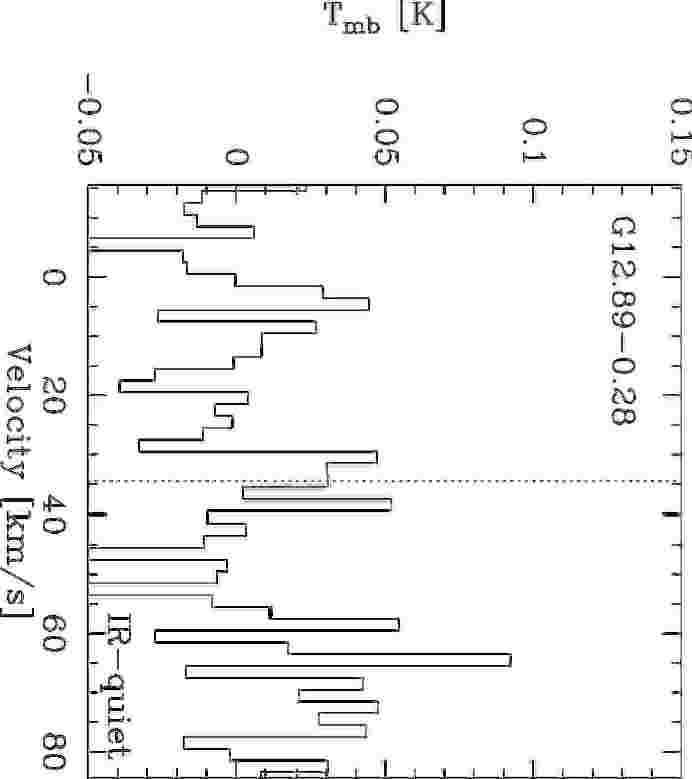} 
  \includegraphics[width=5.2cm,angle=90]{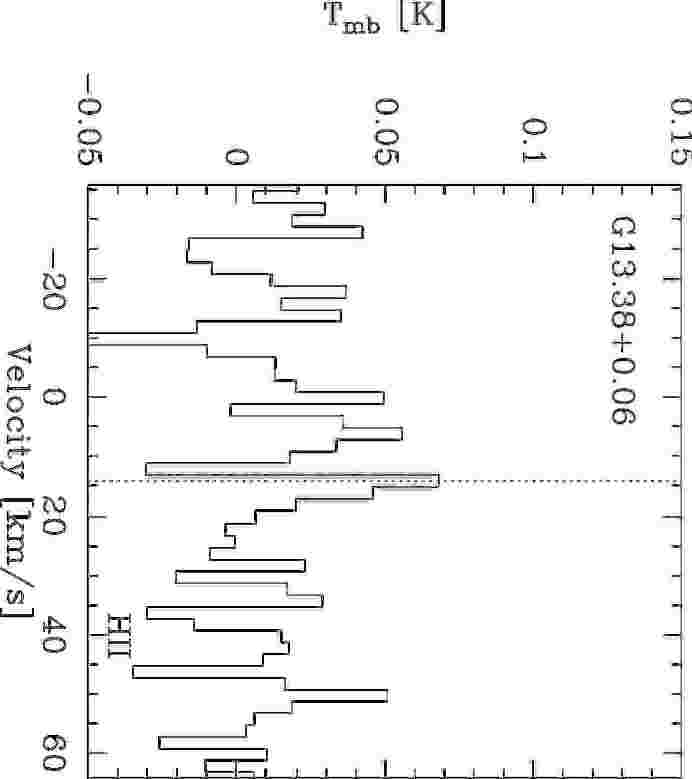} 
  \includegraphics[width=5.2cm,angle=90]{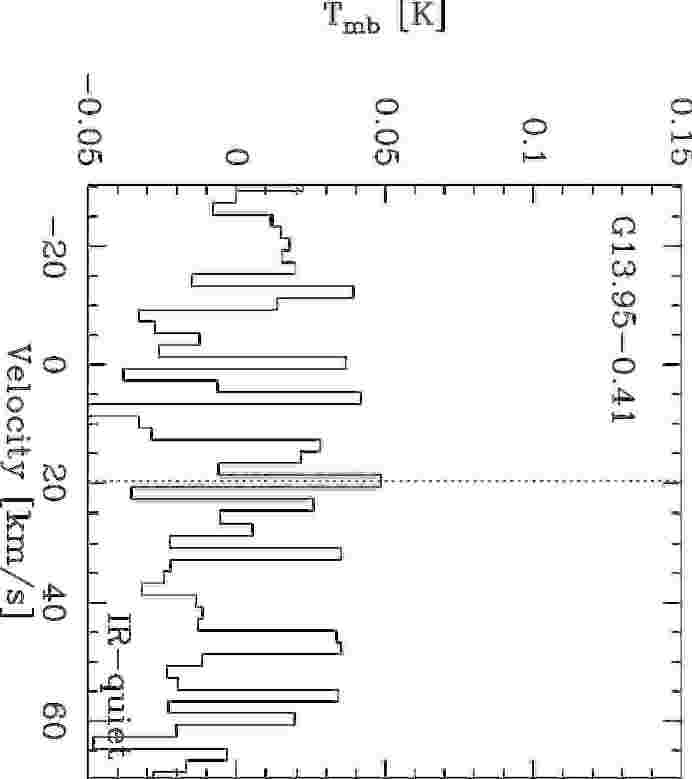} 
  \includegraphics[width=5.2cm,angle=90]{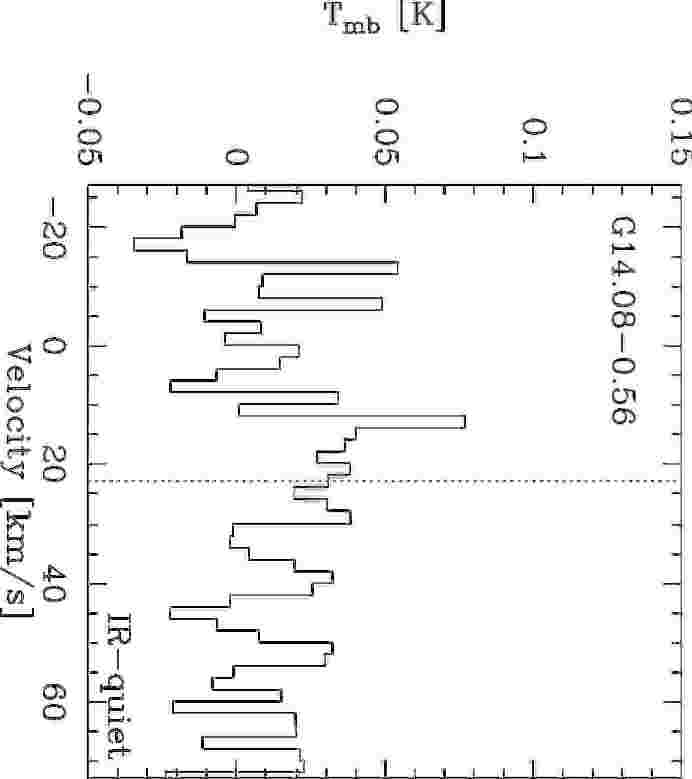} 
  \includegraphics[width=5.2cm,angle=90]{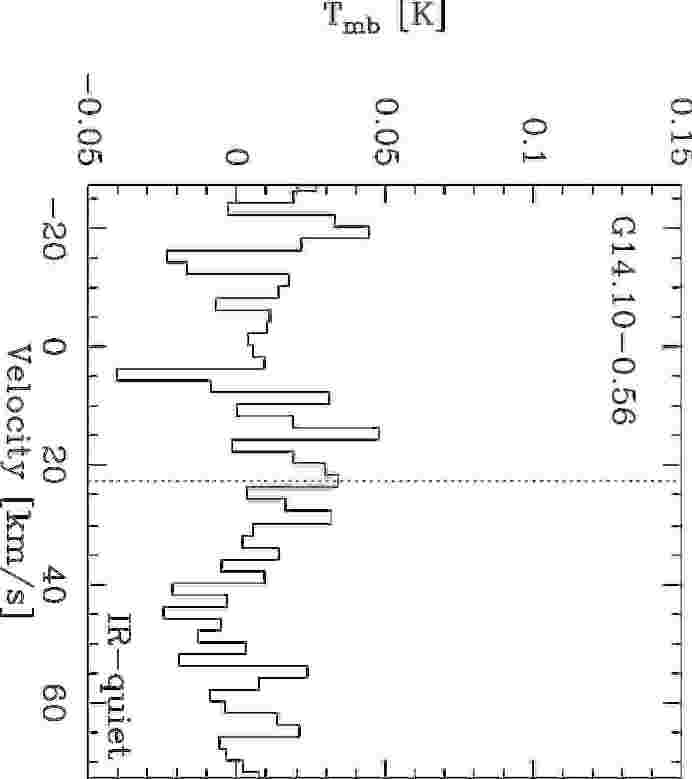} 
  \includegraphics[width=5.2cm,angle=90]{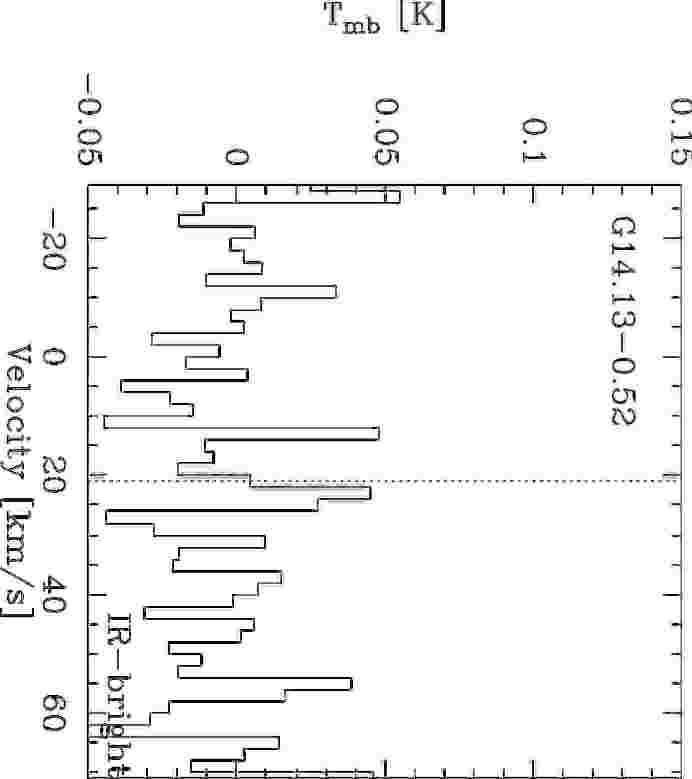} 
  \includegraphics[width=5.2cm,angle=90]{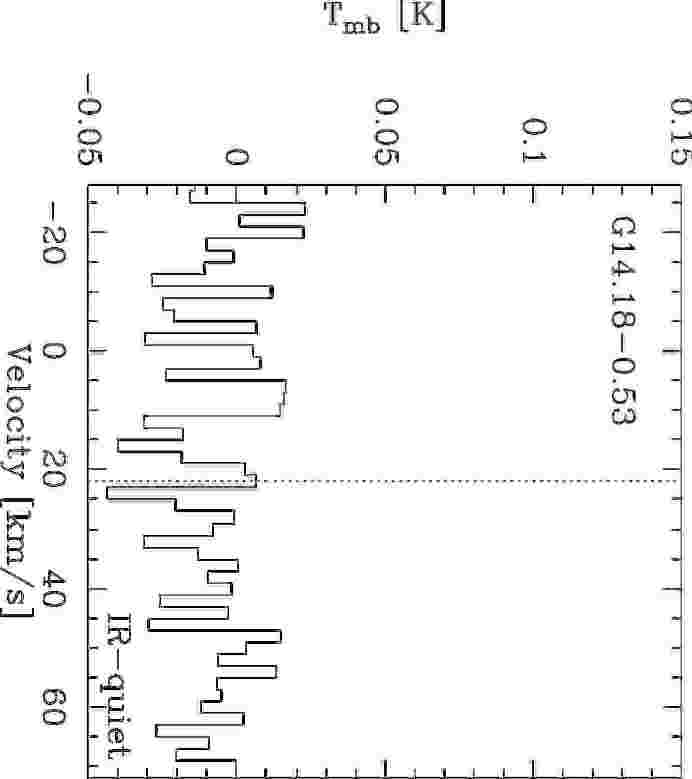} 
  \includegraphics[width=5.2cm,angle=90]{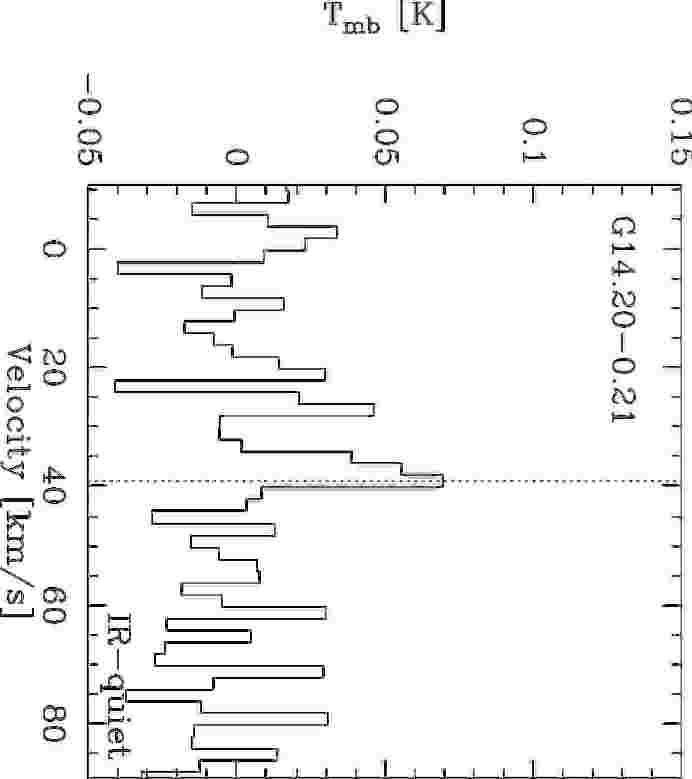} 
  \includegraphics[width=5.2cm,angle=90]{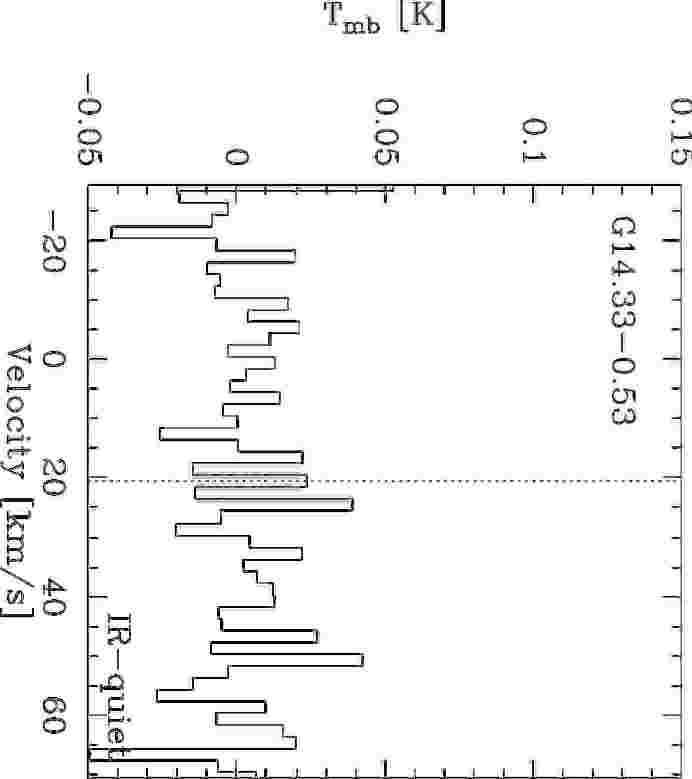} 
  \includegraphics[width=5.2cm,angle=90]{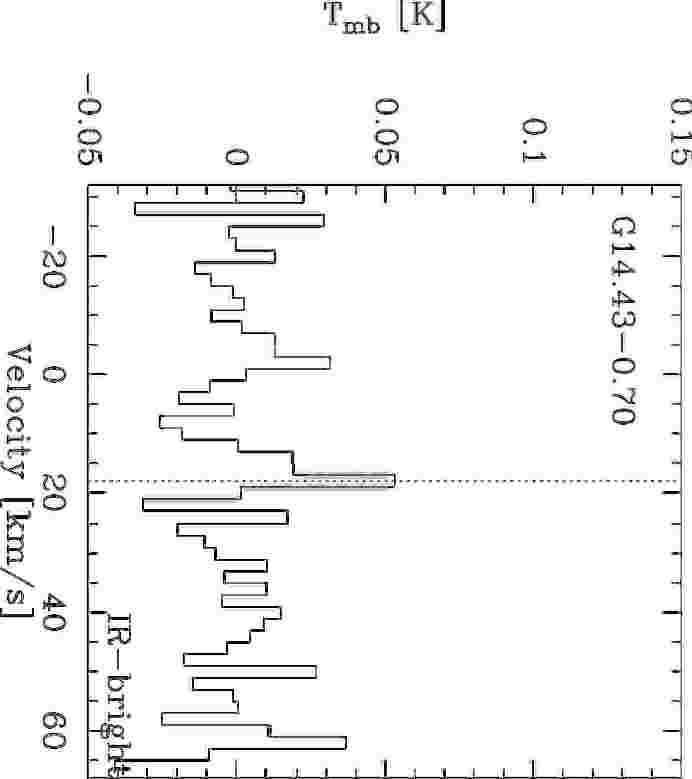} 
 \caption{Non-det.}
\end{figure}
\end{landscape}

\begin{landscape}
\begin{figure}
\ContinuedFloat
\centering
  \includegraphics[width=5.2cm,angle=90]{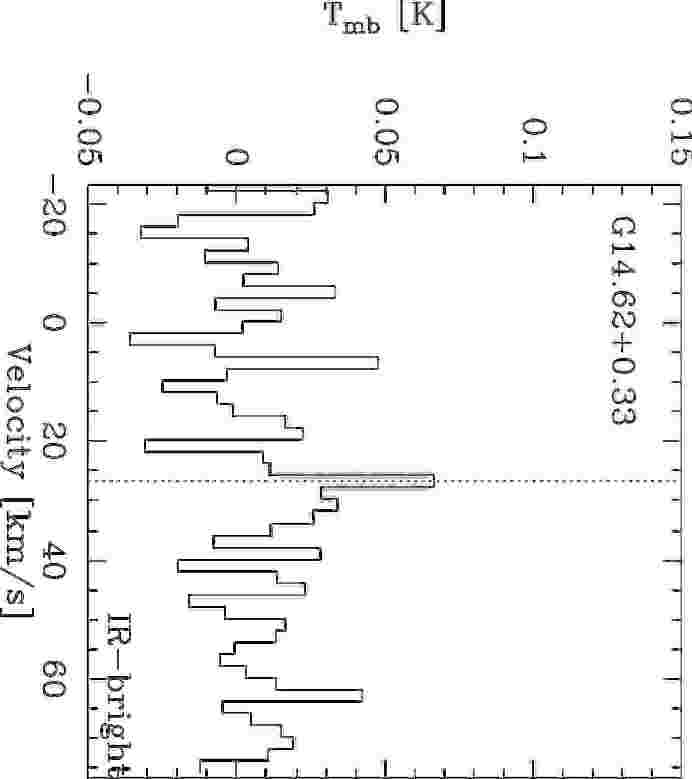} 
  \includegraphics[width=5.2cm,angle=90]{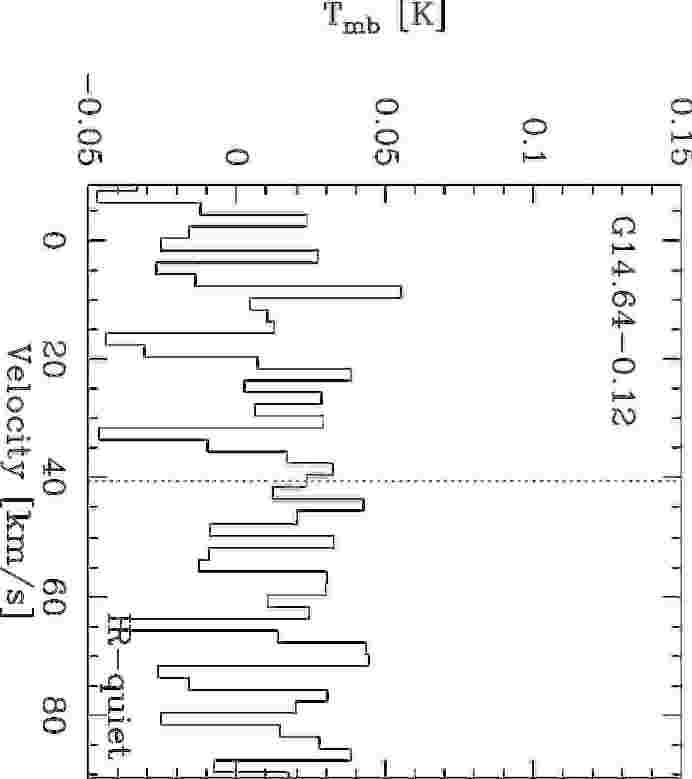} 
  \includegraphics[width=5.2cm,angle=90]{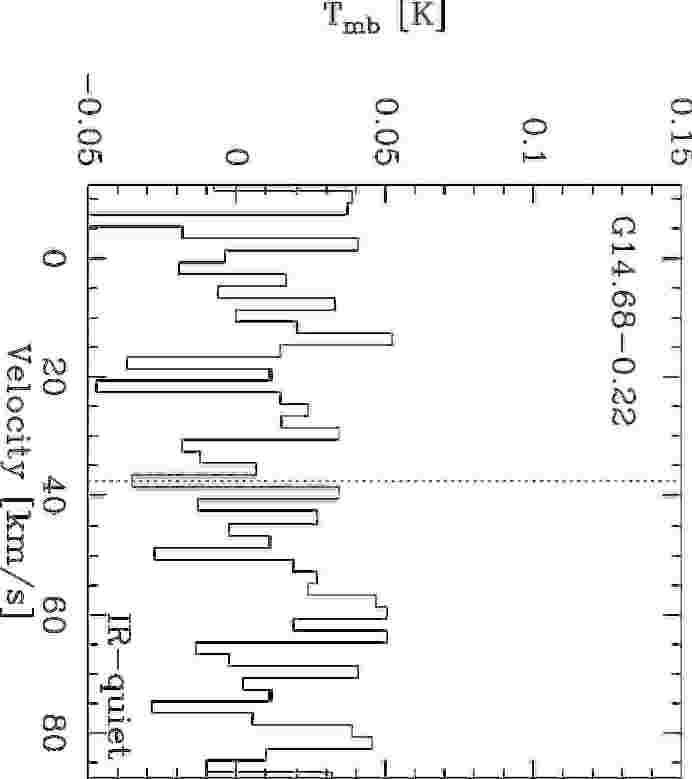} 
  \includegraphics[width=5.2cm,angle=90]{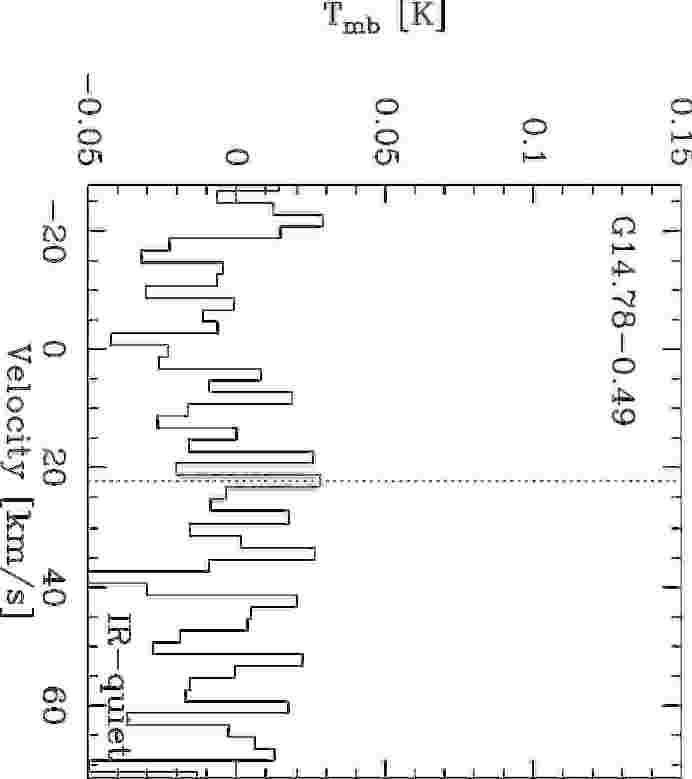} 
  \includegraphics[width=5.2cm,angle=90]{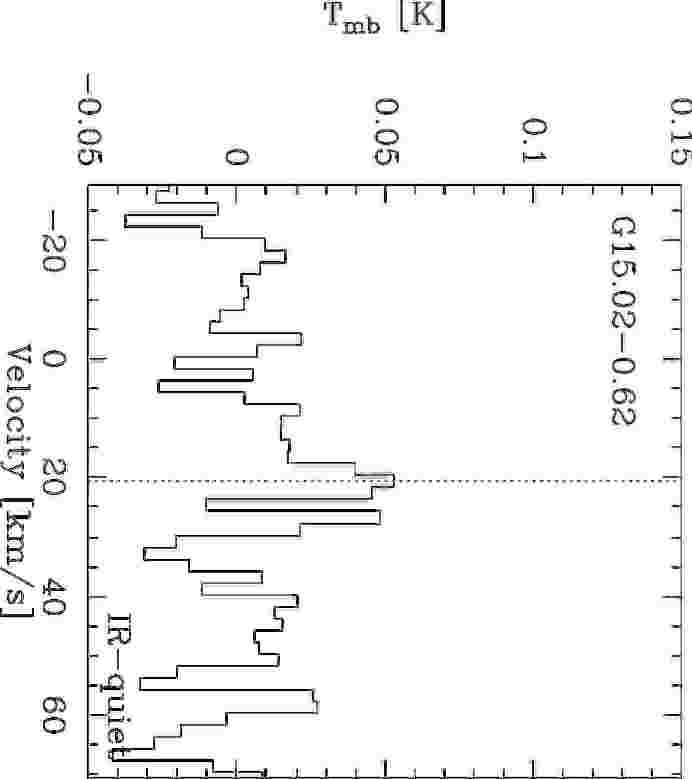} 
  \includegraphics[width=5.2cm,angle=90]{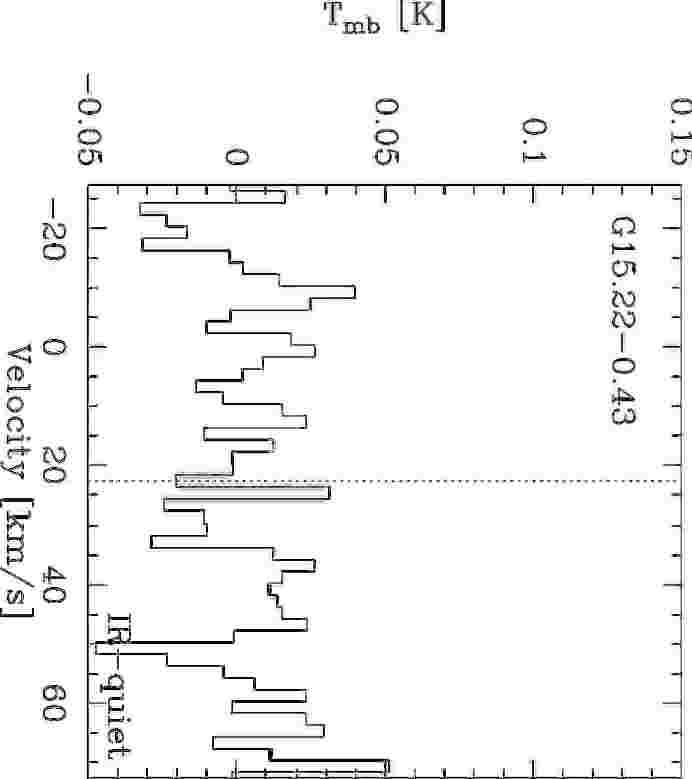} 
  \includegraphics[width=5.2cm,angle=90]{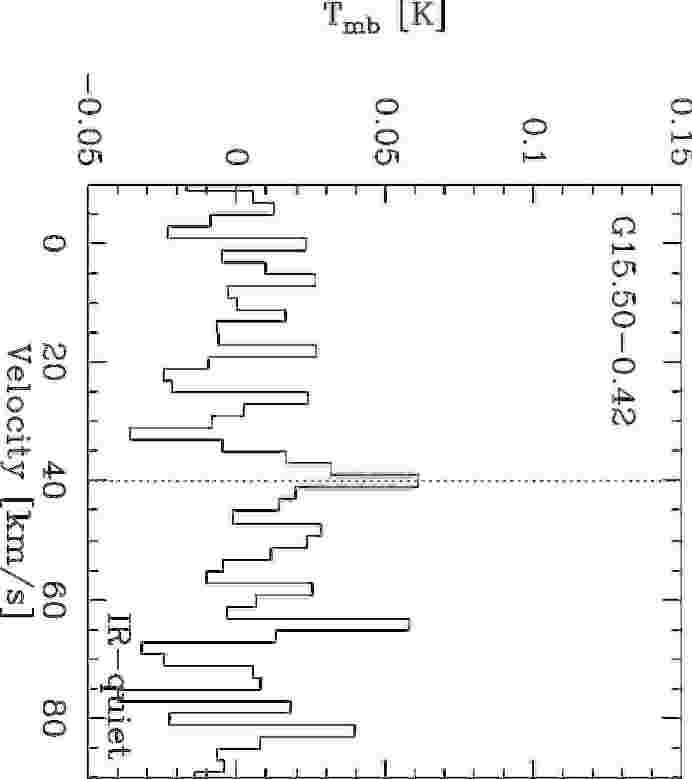} 
  \includegraphics[width=5.2cm,angle=90]{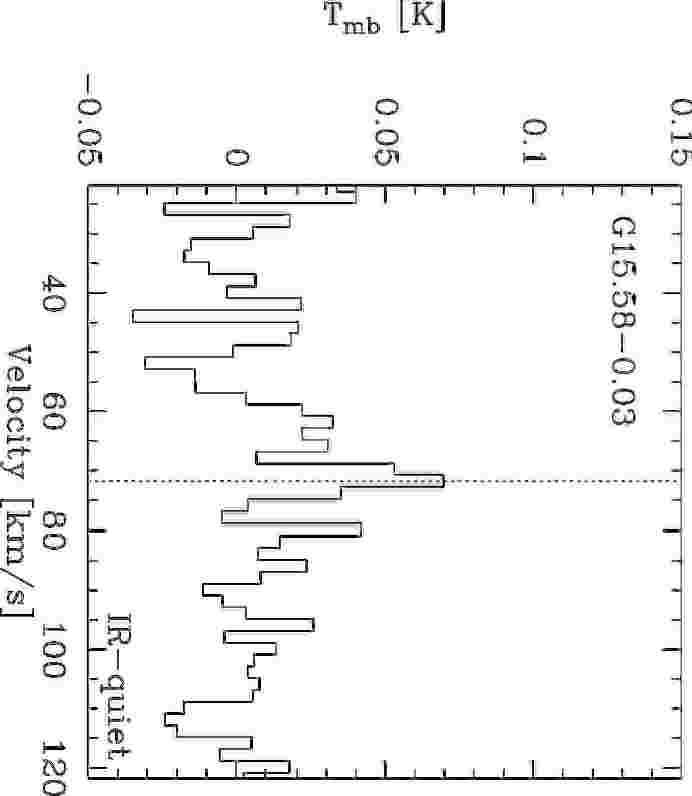} 
  \includegraphics[width=5.2cm,angle=90]{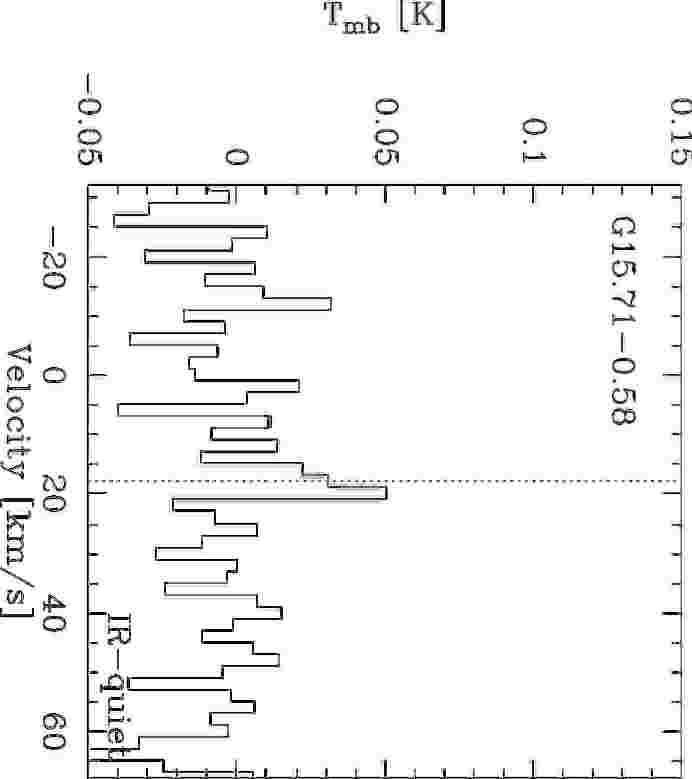} 
  \includegraphics[width=5.2cm,angle=90]{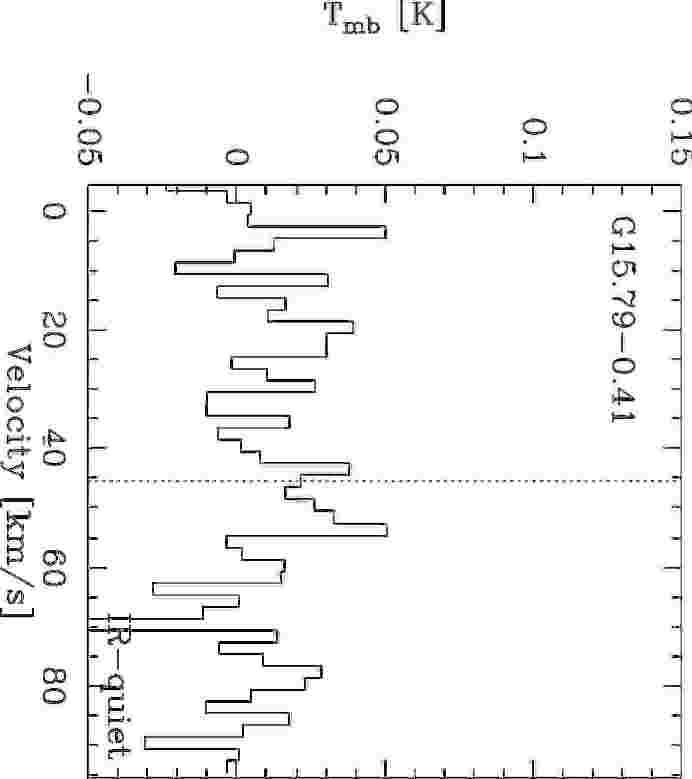} 
  \includegraphics[width=5.2cm,angle=90]{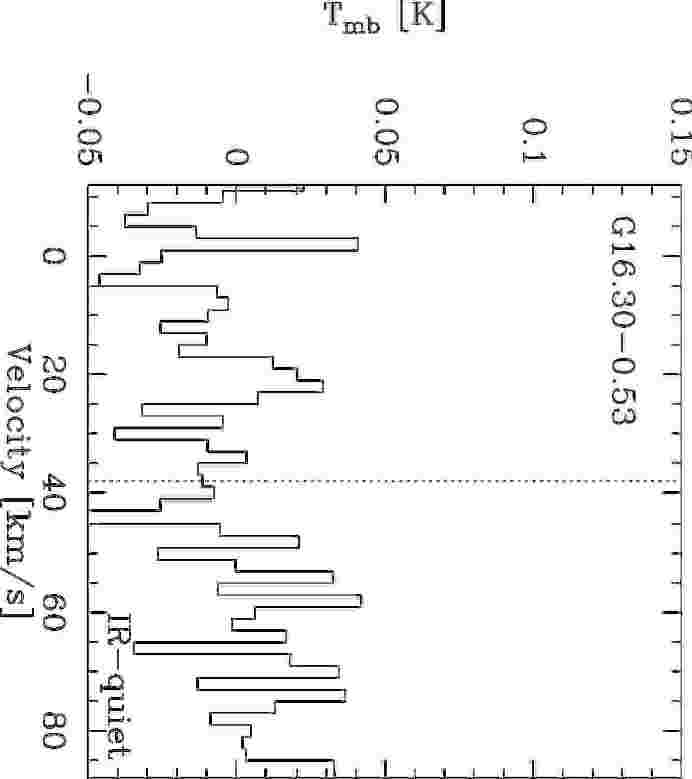} 
  \includegraphics[width=5.2cm,angle=90]{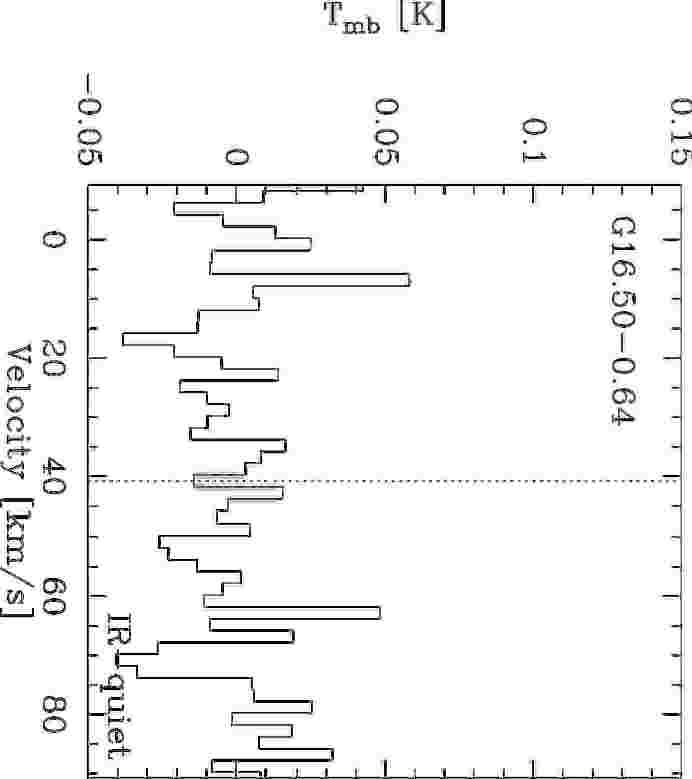} 
 \caption{Non-det.}
\end{figure}
\end{landscape}

\begin{landscape}
\begin{figure}
\ContinuedFloat
\centering
  \includegraphics[width=5.2cm,angle=90]{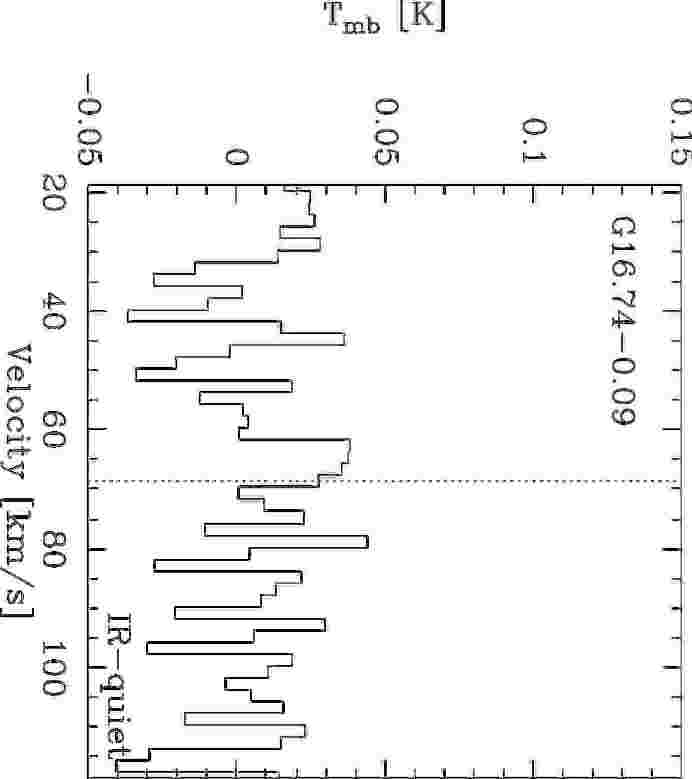} 
  \includegraphics[width=5.2cm,angle=90]{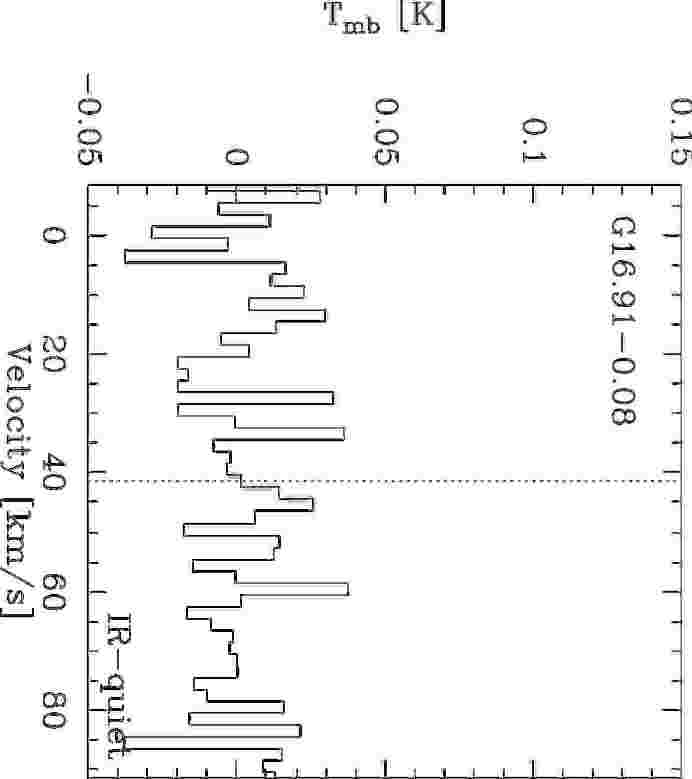} 
  \includegraphics[width=5.2cm,angle=90]{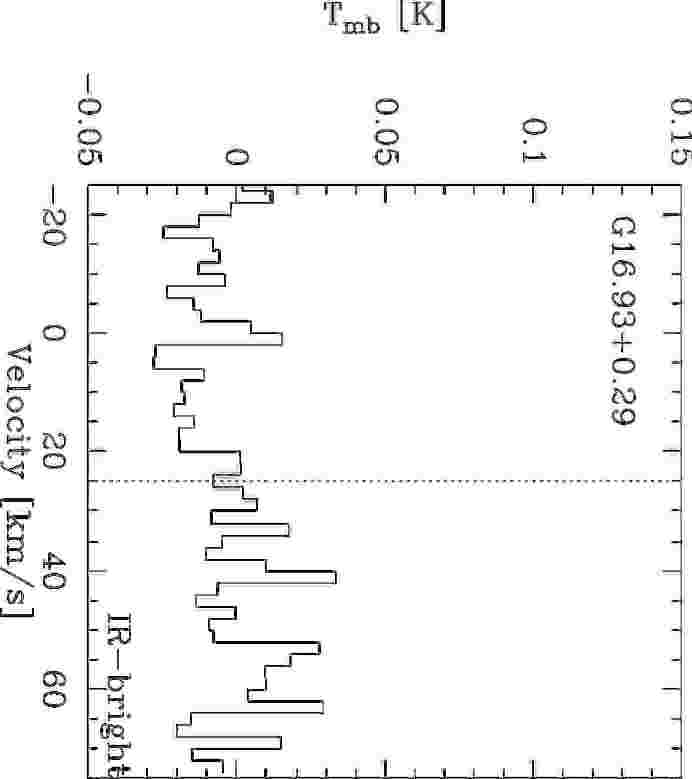} 
  \includegraphics[width=5.2cm,angle=90]{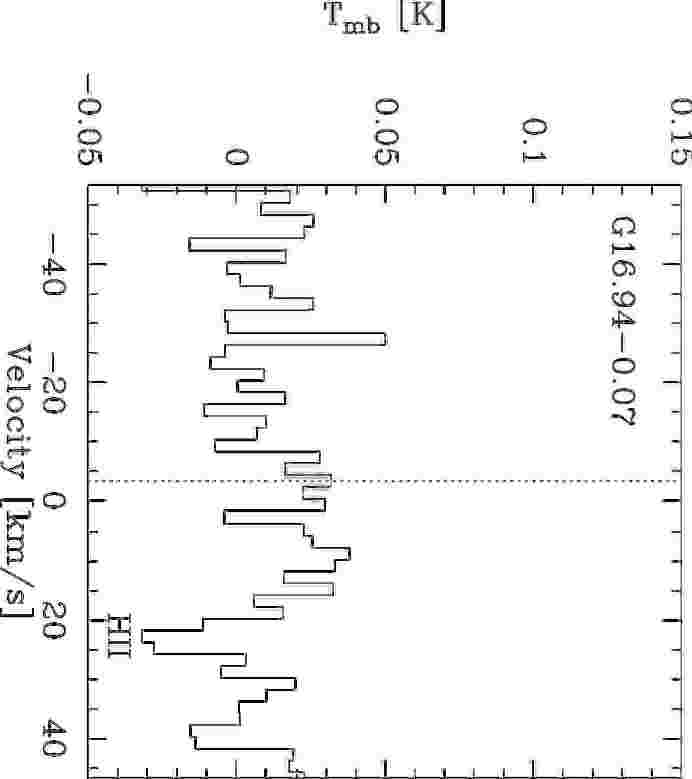} 
  \includegraphics[width=5.2cm,angle=90]{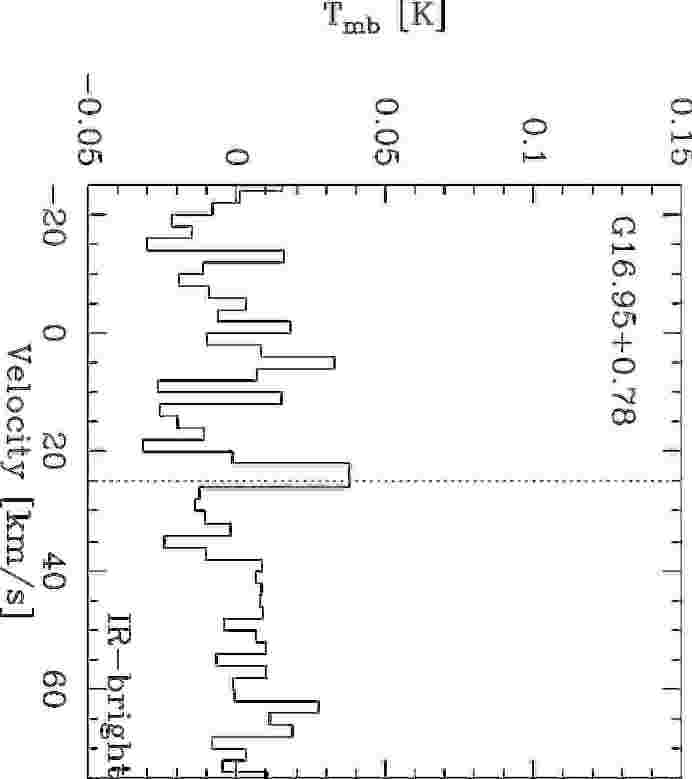} 
  \includegraphics[width=5.2cm,angle=90]{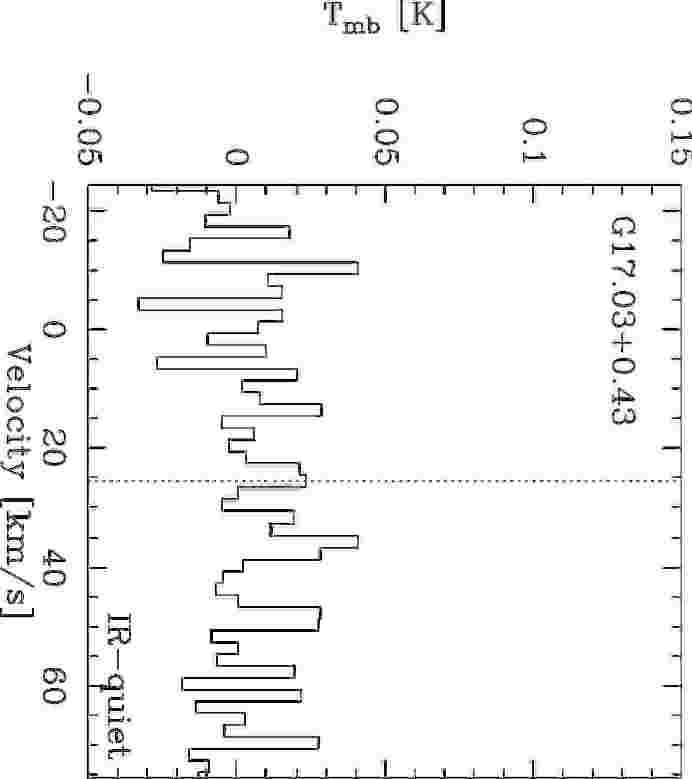} 
  \includegraphics[width=5.2cm,angle=90]{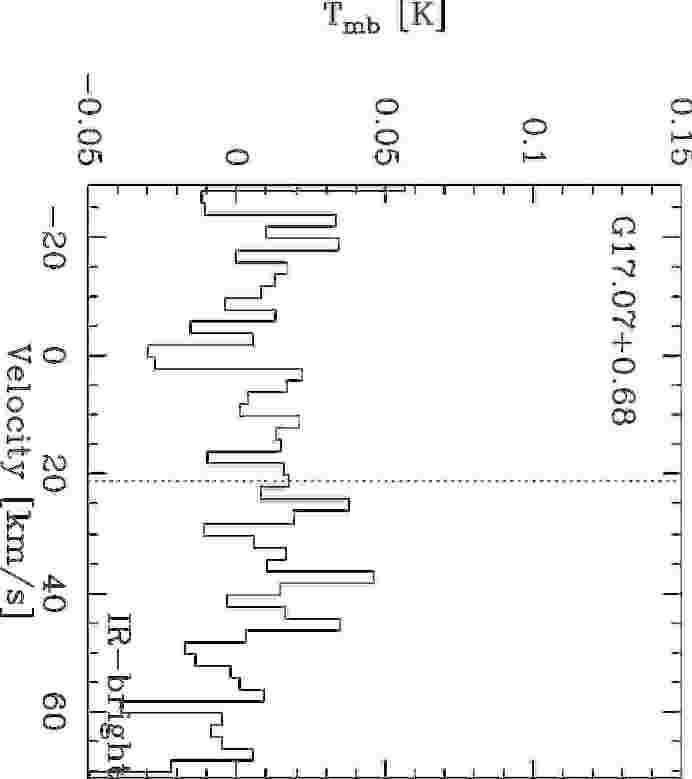} 
  \includegraphics[width=5.2cm,angle=90]{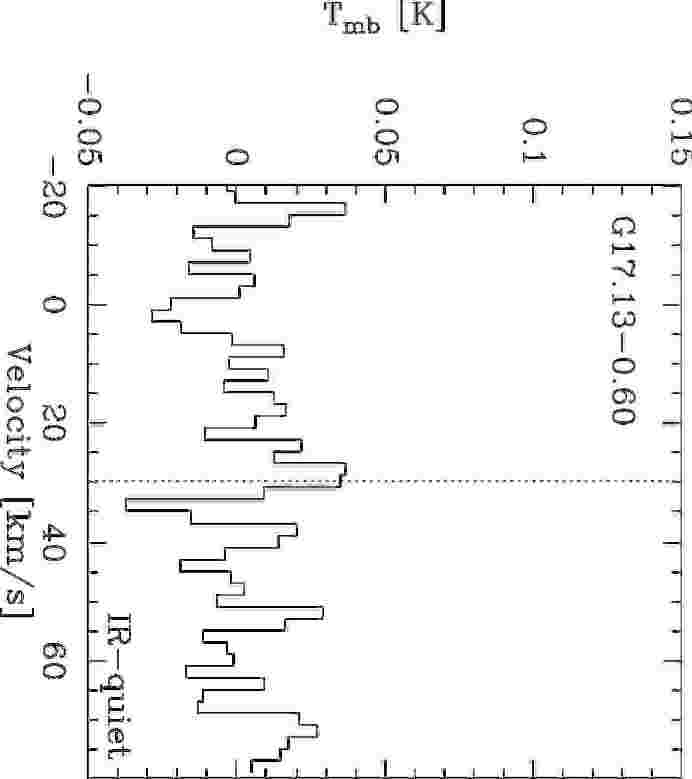} 
  \includegraphics[width=5.2cm,angle=90]{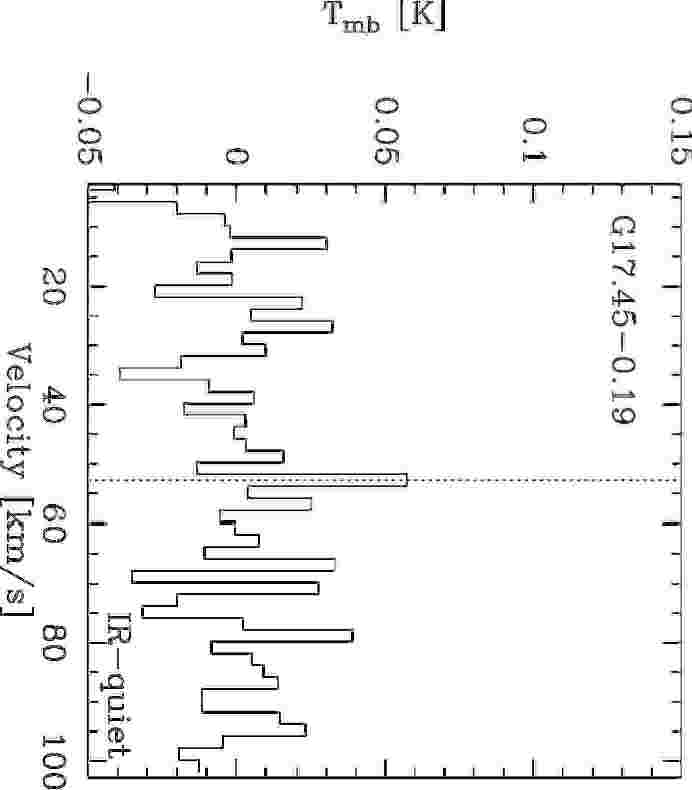} 
  \includegraphics[width=5.2cm,angle=90]{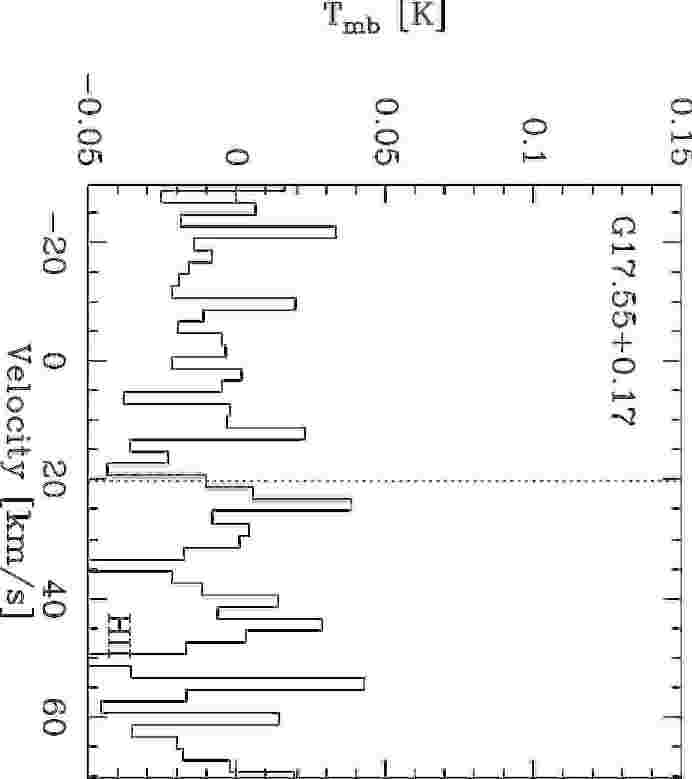} 
  \includegraphics[width=5.2cm,angle=90]{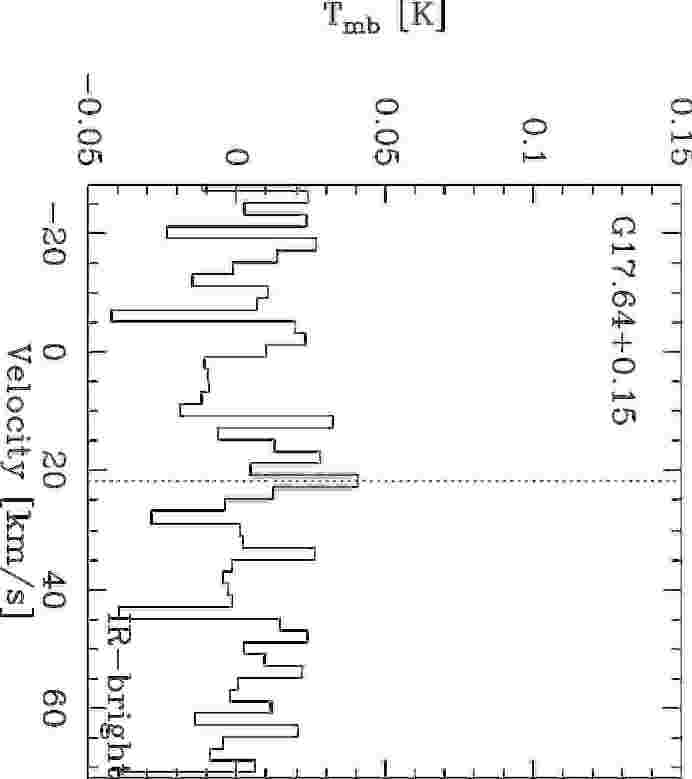} 
  \includegraphics[width=5.2cm,angle=90]{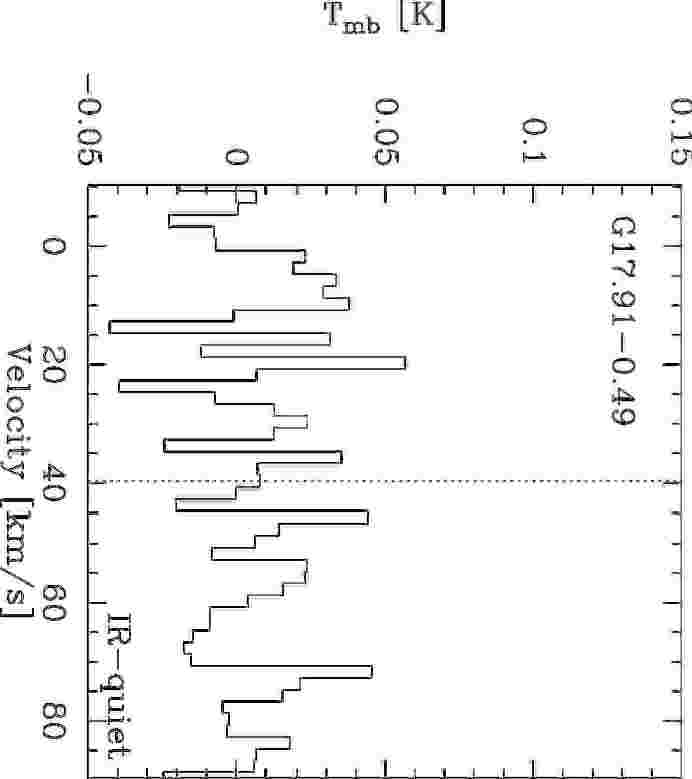} 
 \caption{Non-det.}
\end{figure}
\end{landscape}

\begin{landscape}
\begin{figure}
\ContinuedFloat
\centering
  \includegraphics[width=5.2cm,angle=90]{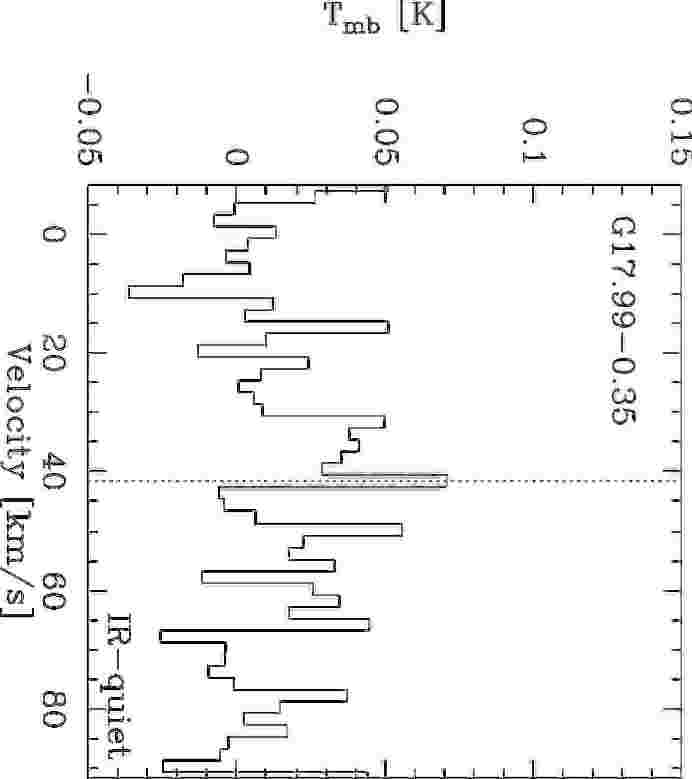} 
  \includegraphics[width=5.2cm,angle=90]{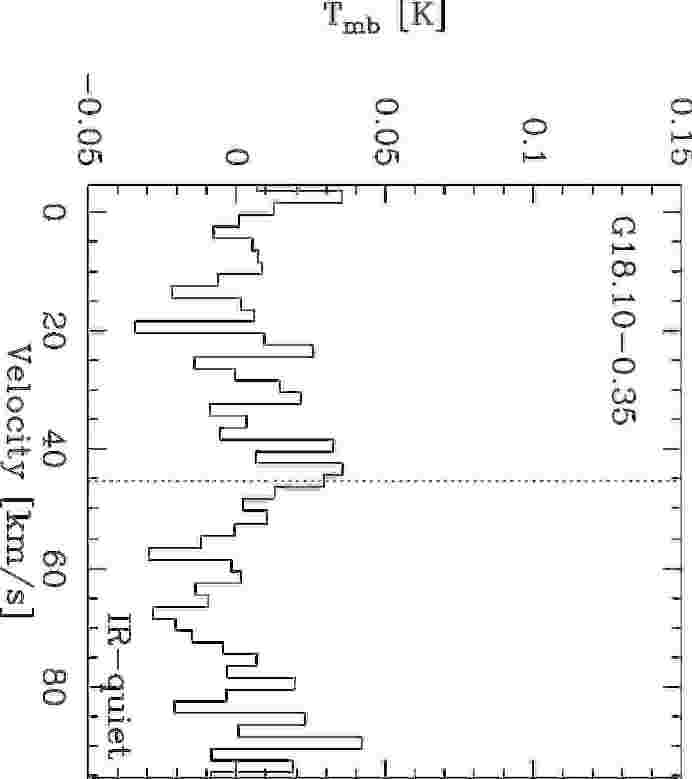} 
  \includegraphics[width=5.2cm,angle=90]{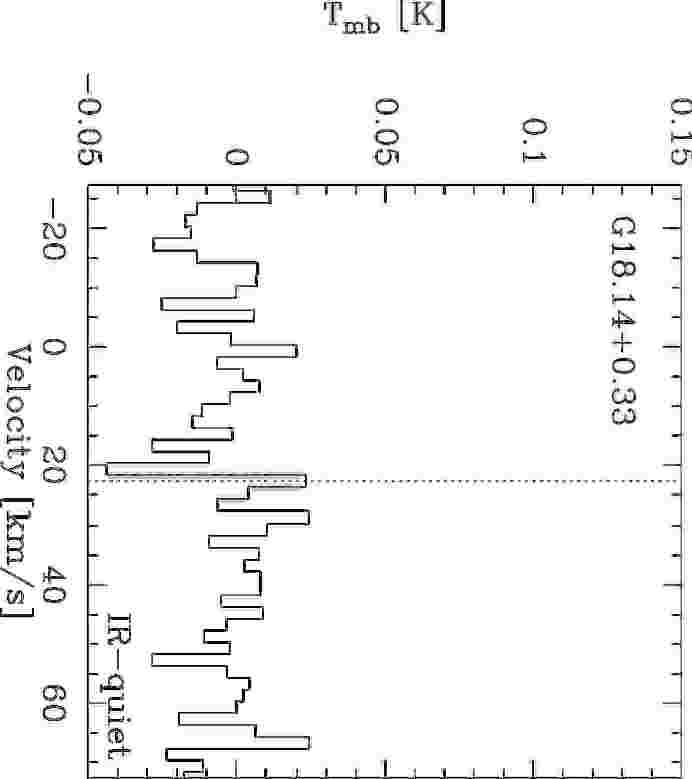} 
  \includegraphics[width=5.2cm,angle=90]{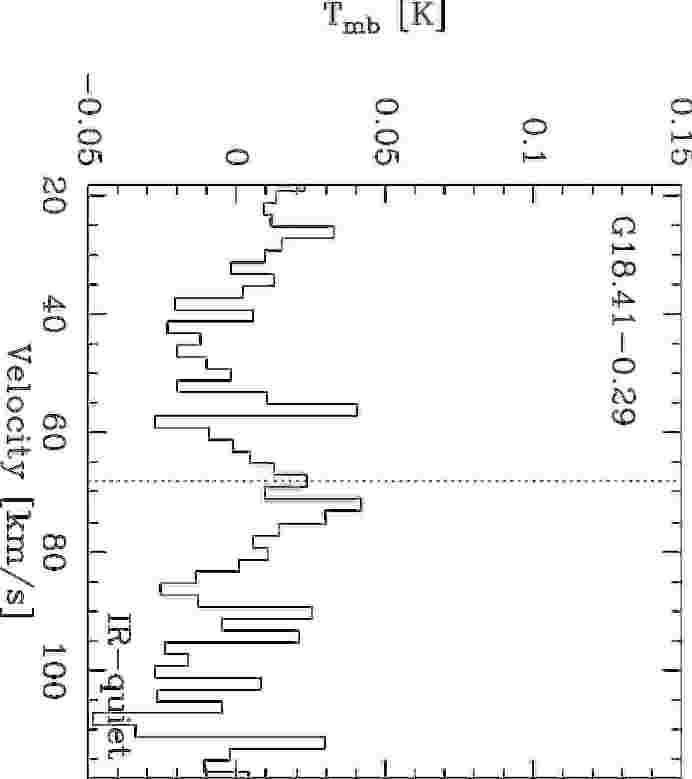} 
  \includegraphics[width=5.2cm,angle=90]{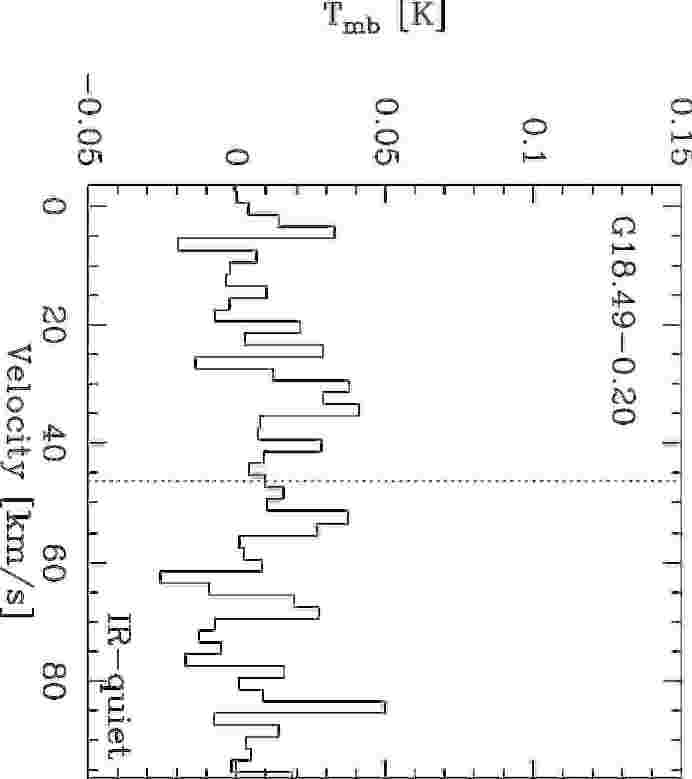} 
  \includegraphics[width=5.2cm,angle=90]{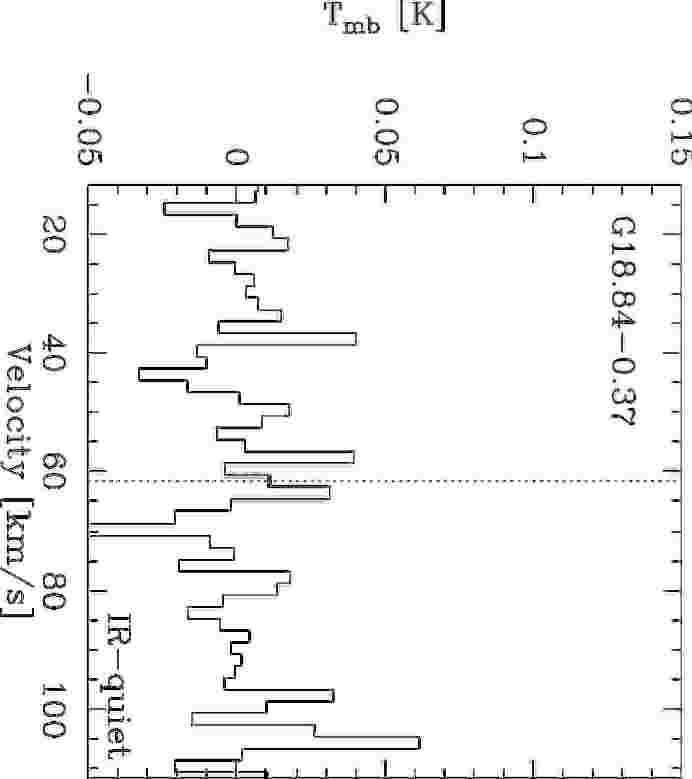} 
  \includegraphics[width=5.2cm,angle=90]{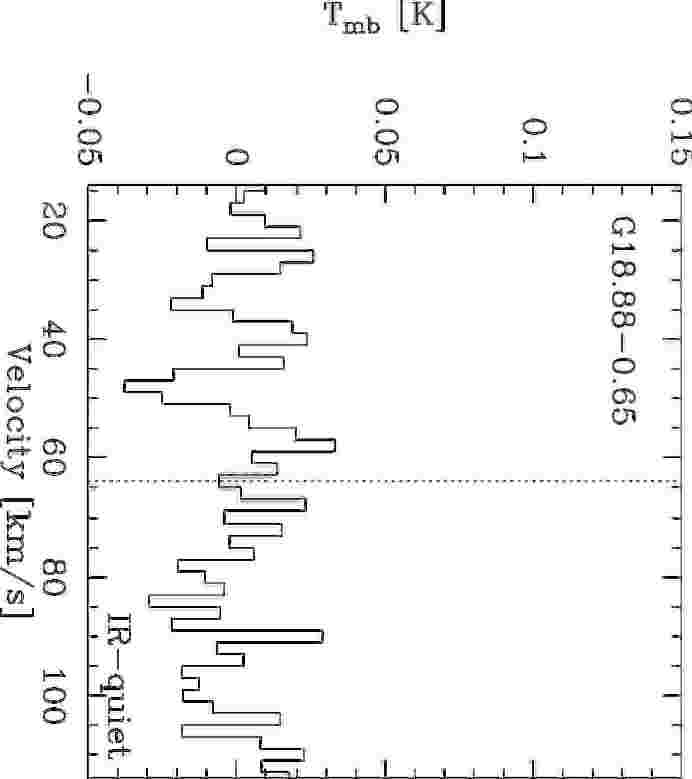} 
  \includegraphics[width=5.2cm,angle=90]{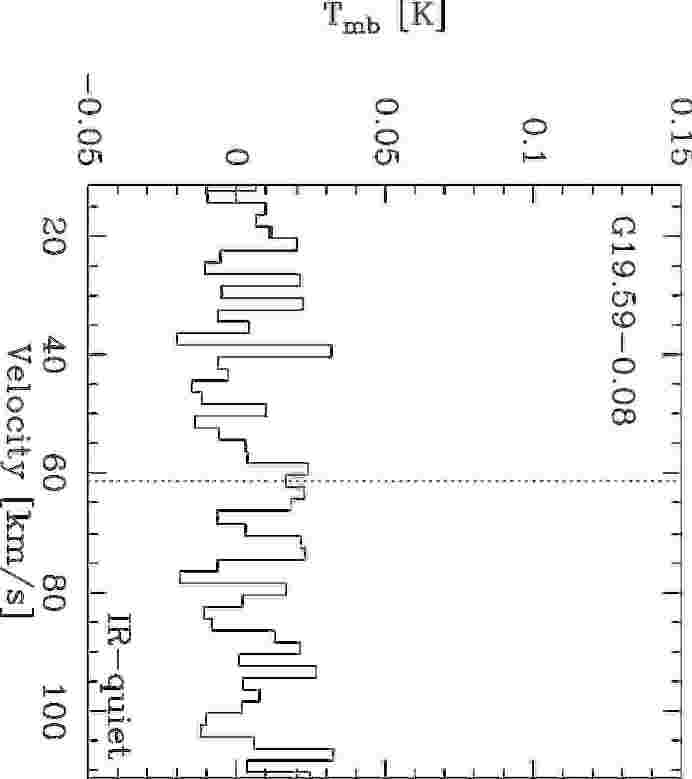} 
  \includegraphics[width=5.2cm,angle=90]{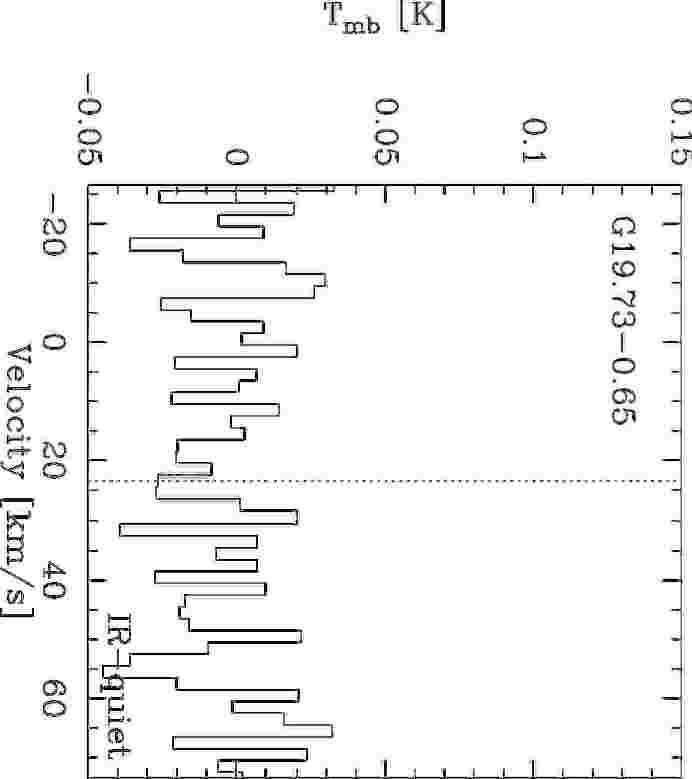} 
    \includegraphics[width=5.2cm,angle=90]{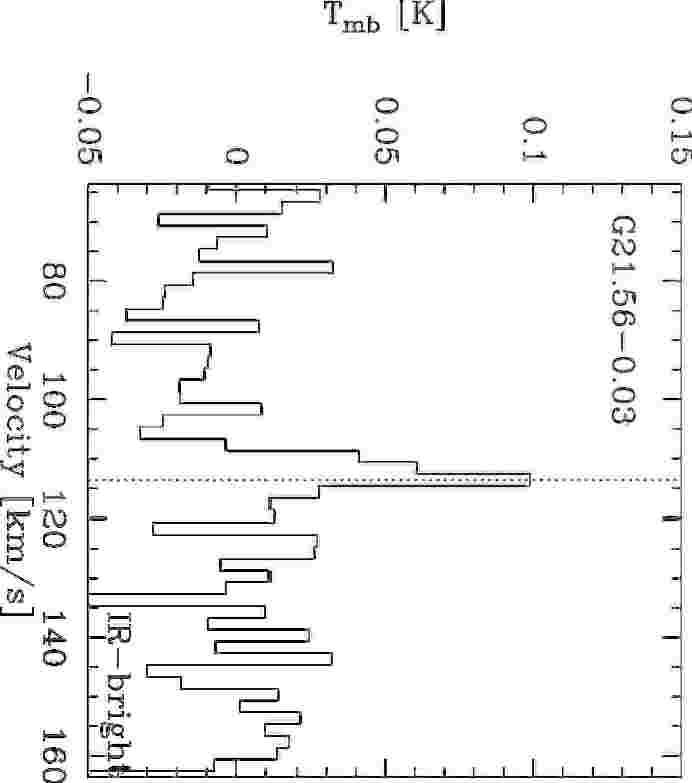} 
  \includegraphics[width=5.2cm,angle=90]{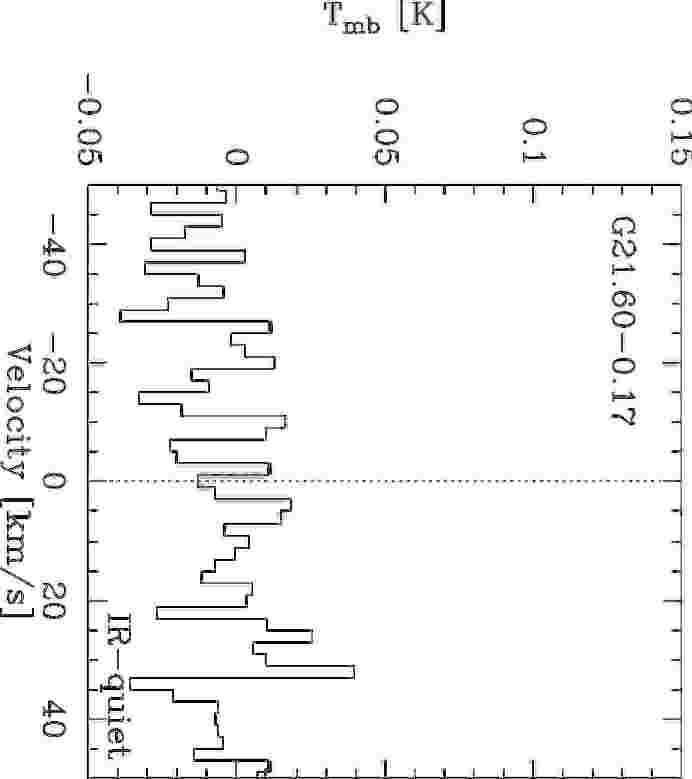} 
  \includegraphics[width=5.2cm,angle=90]{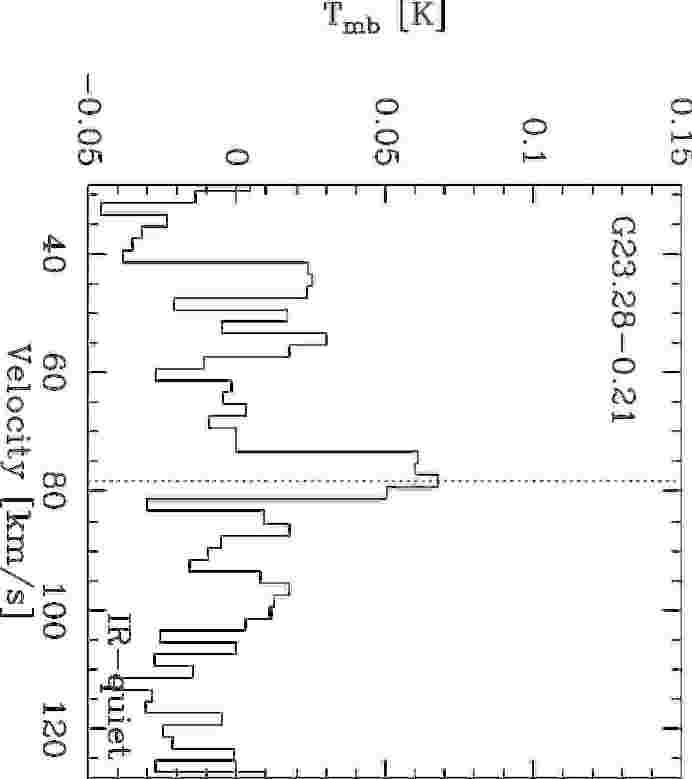} 
 \caption{Non-det.}
\end{figure}
\end{landscape}

\begin{landscape}
\begin{figure}
\ContinuedFloat
\centering
  \includegraphics[width=5.2cm,angle=90]{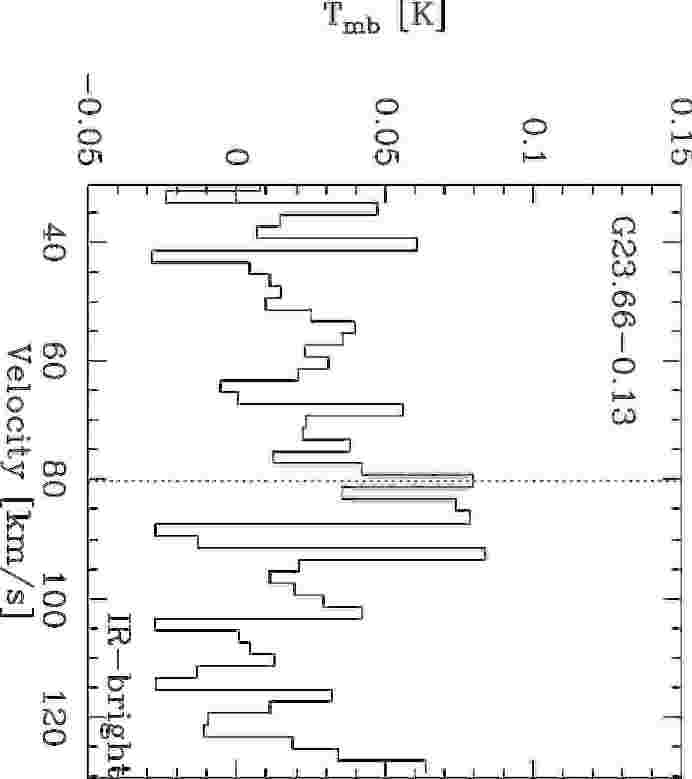} 
  \includegraphics[width=5.2cm,angle=90]{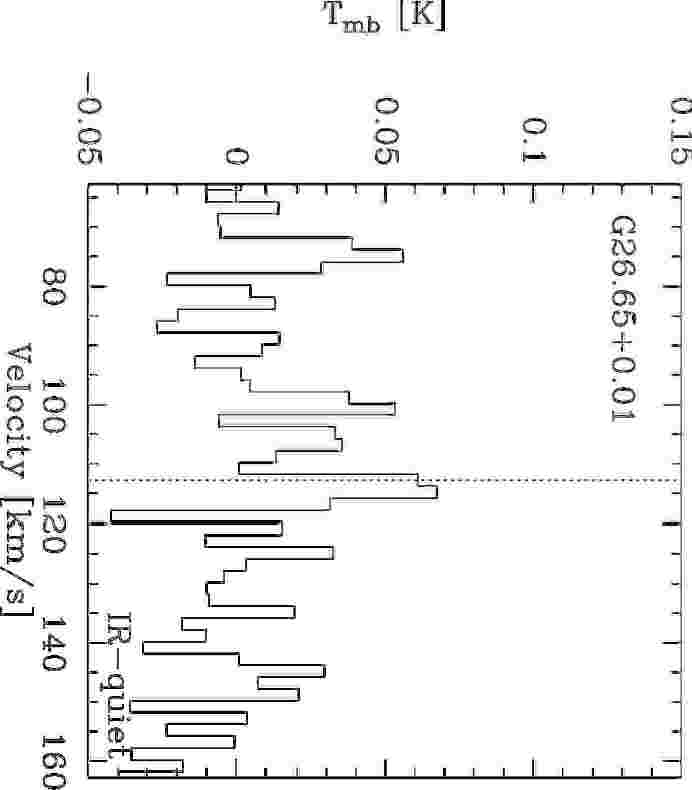} 
  \includegraphics[width=5.2cm,angle=90]{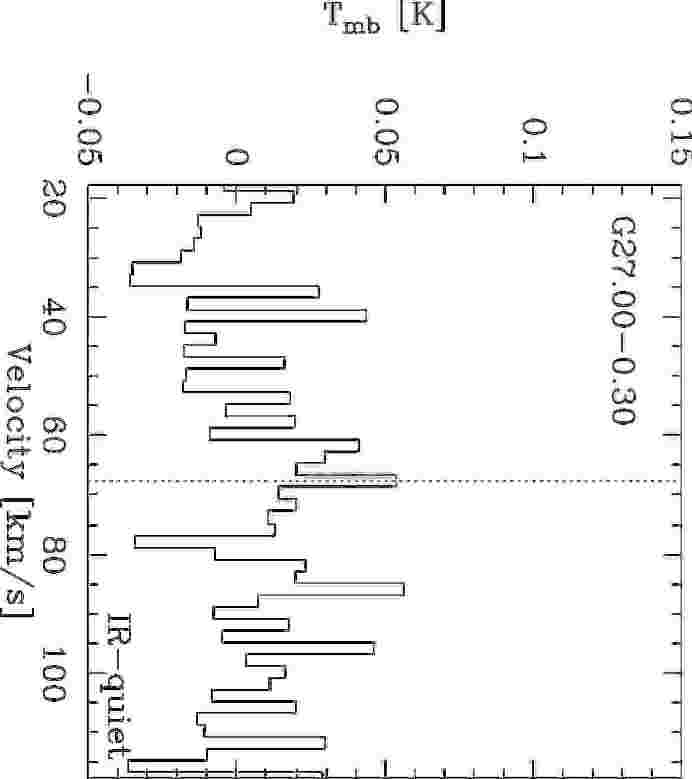} 
  \includegraphics[width=5.2cm,angle=90]{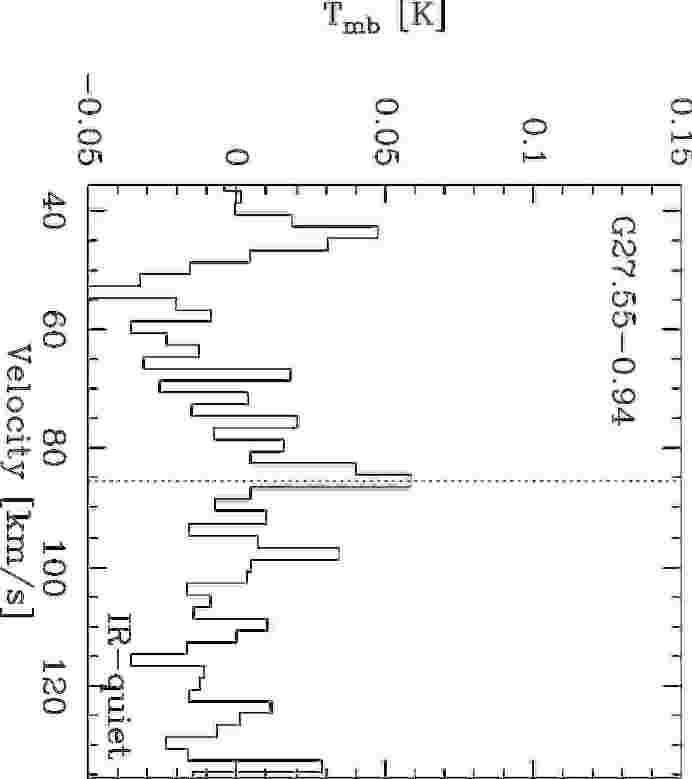} 
  \includegraphics[width=5.2cm,angle=90]{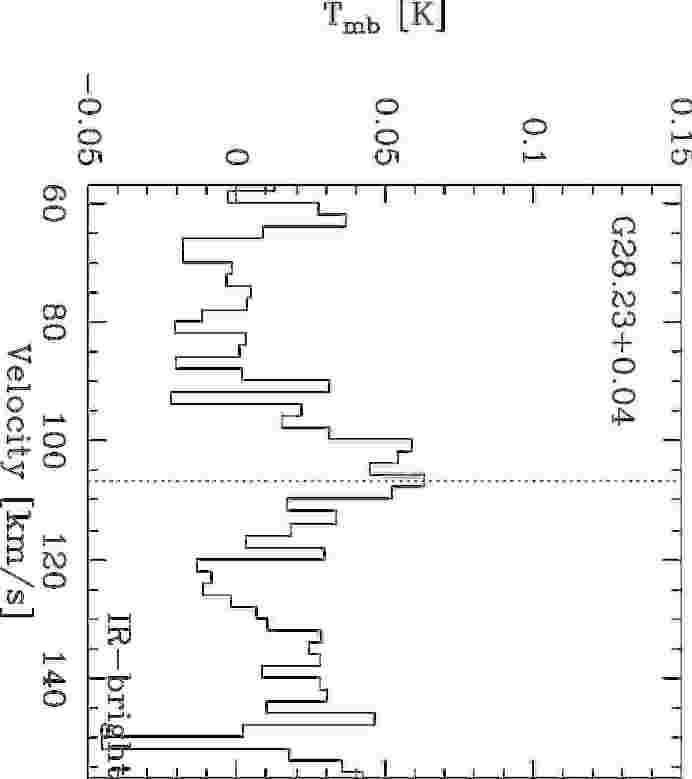} 
  \includegraphics[width=5.2cm,angle=90]{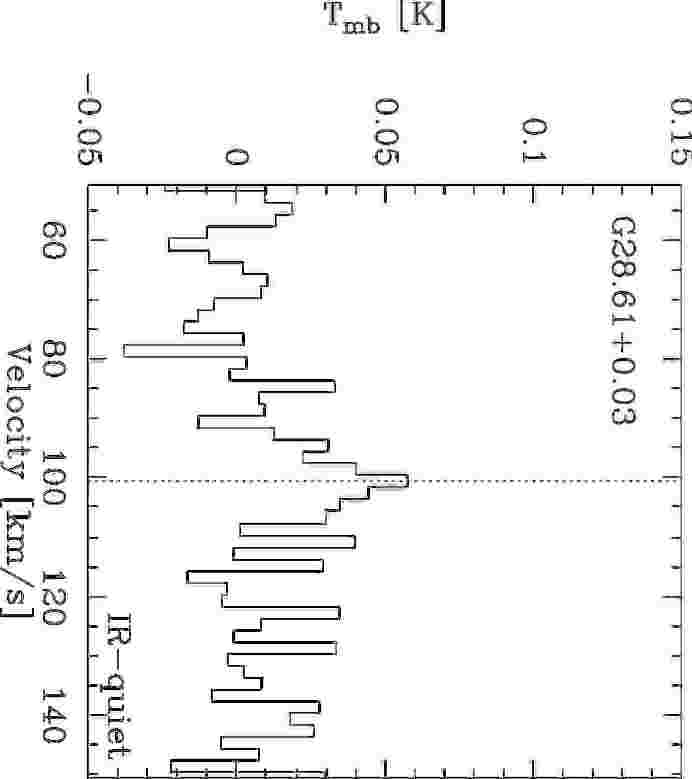} 
  \includegraphics[width=5.2cm,angle=90]{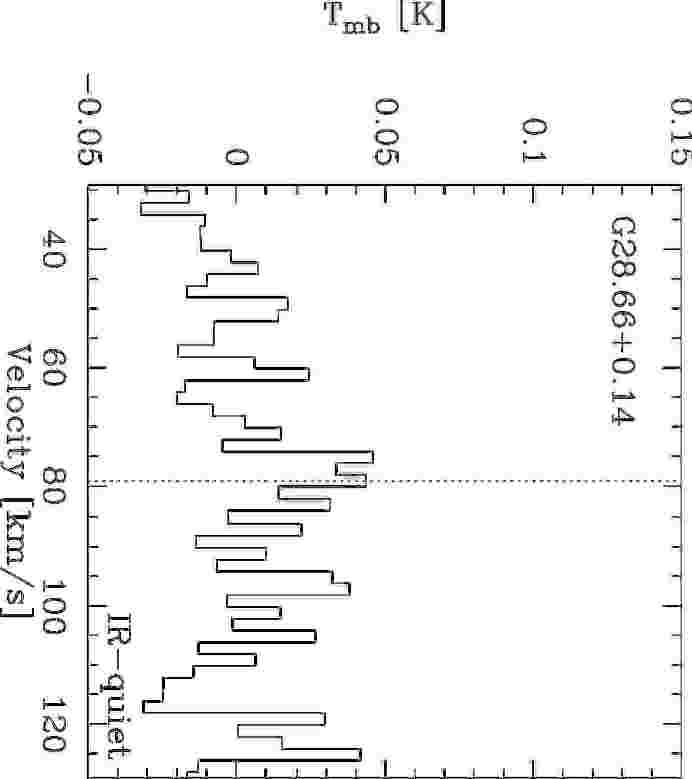} 
  \includegraphics[width=5.2cm,angle=90]{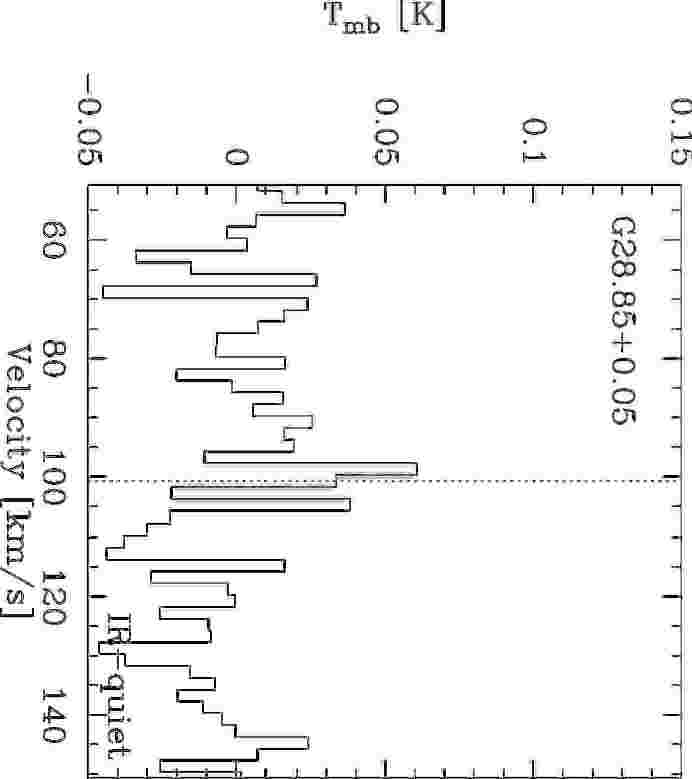} 
  \includegraphics[width=5.2cm,angle=90]{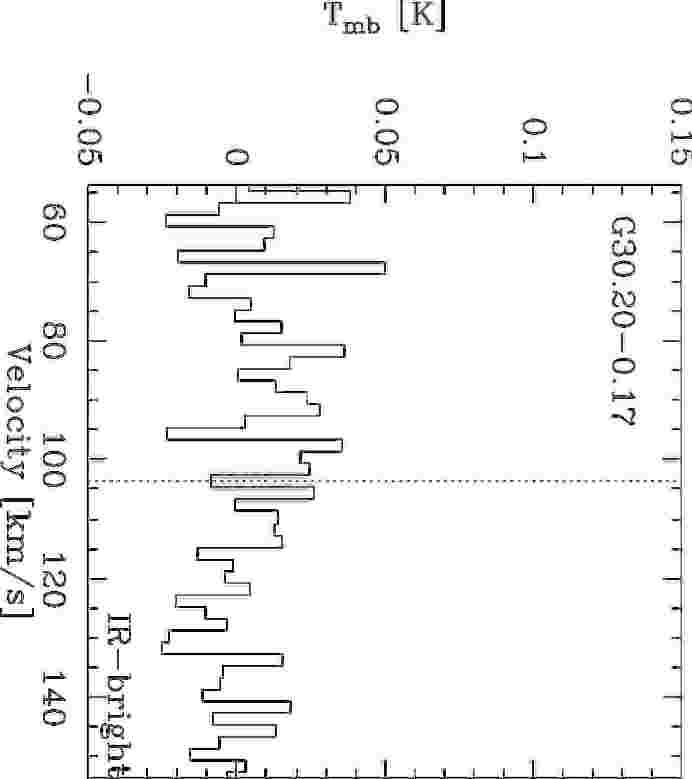} 
  \includegraphics[width=5.2cm,angle=90]{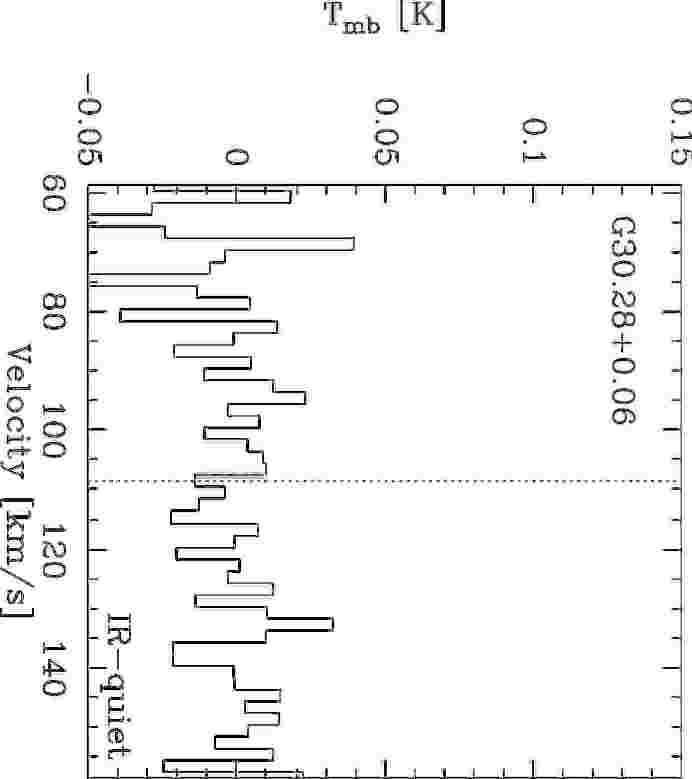} 
  \includegraphics[width=5.2cm,angle=90]{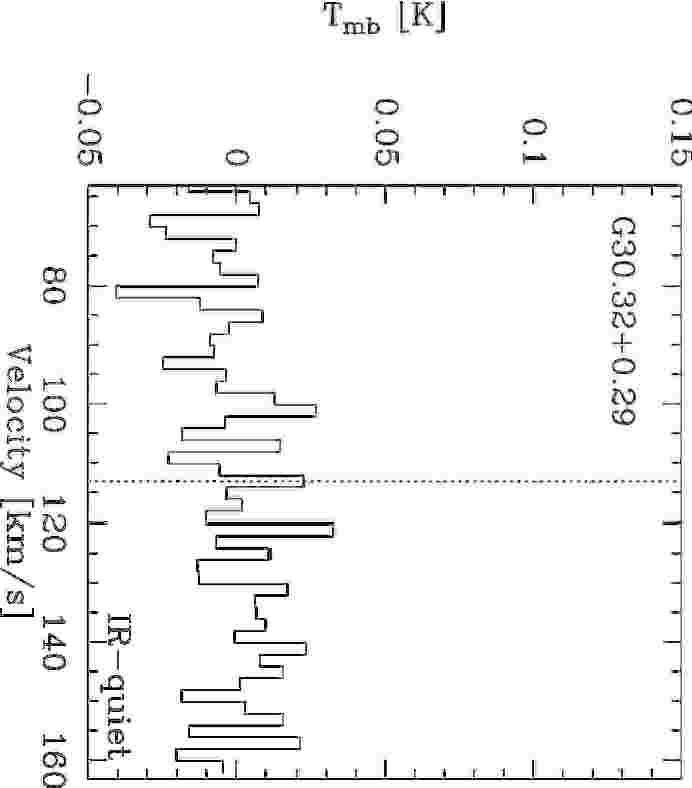} 
  \includegraphics[width=5.2cm,angle=90]{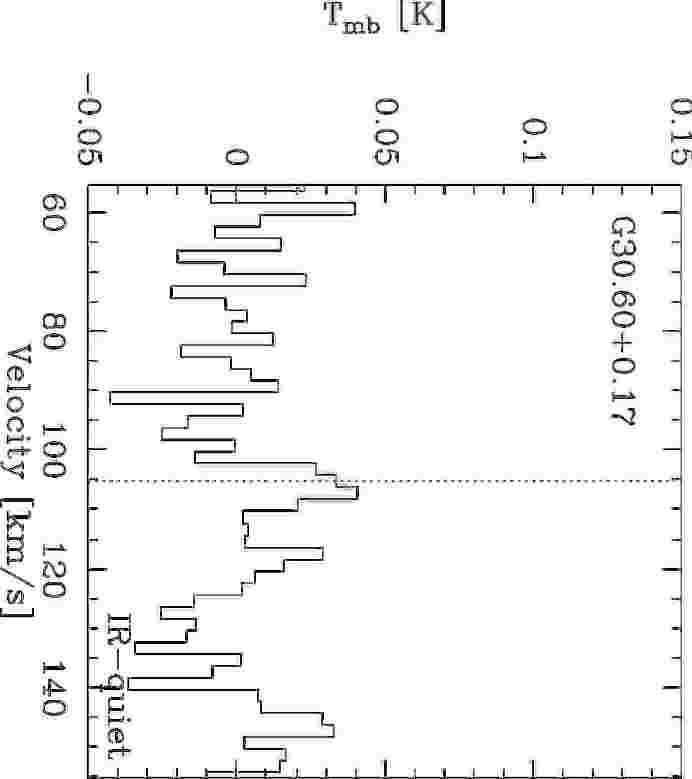} 
 \caption{Non-det.}
\end{figure}
\end{landscape}

\begin{landscape}
\begin{figure}
\ContinuedFloat
\centering
  \includegraphics[width=5.2cm,angle=90]{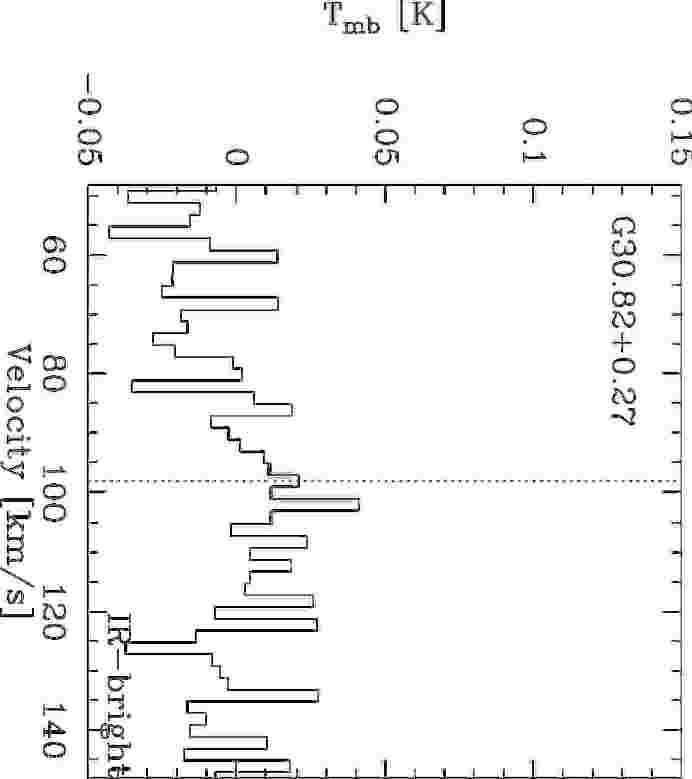} 
  \includegraphics[width=5.2cm,angle=90]{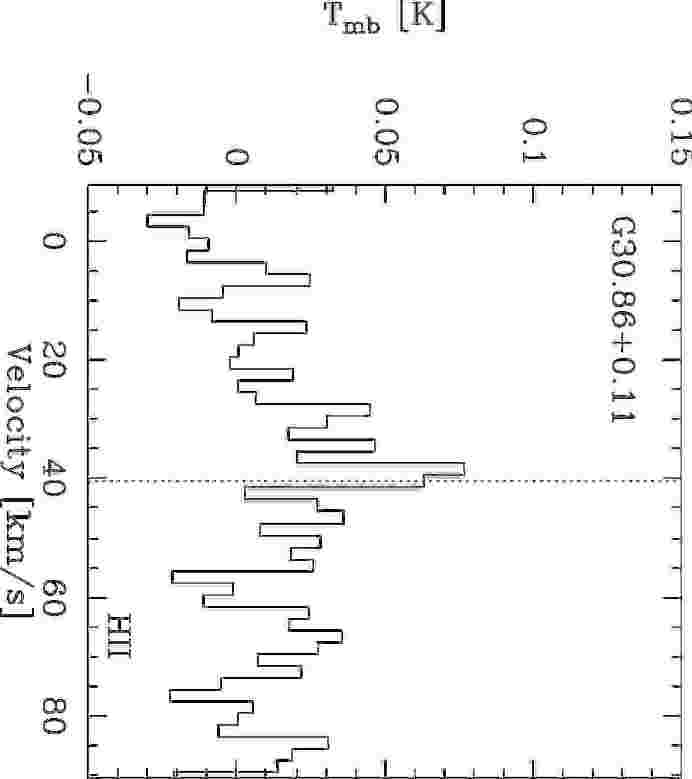} 
  \includegraphics[width=5.2cm,angle=90]{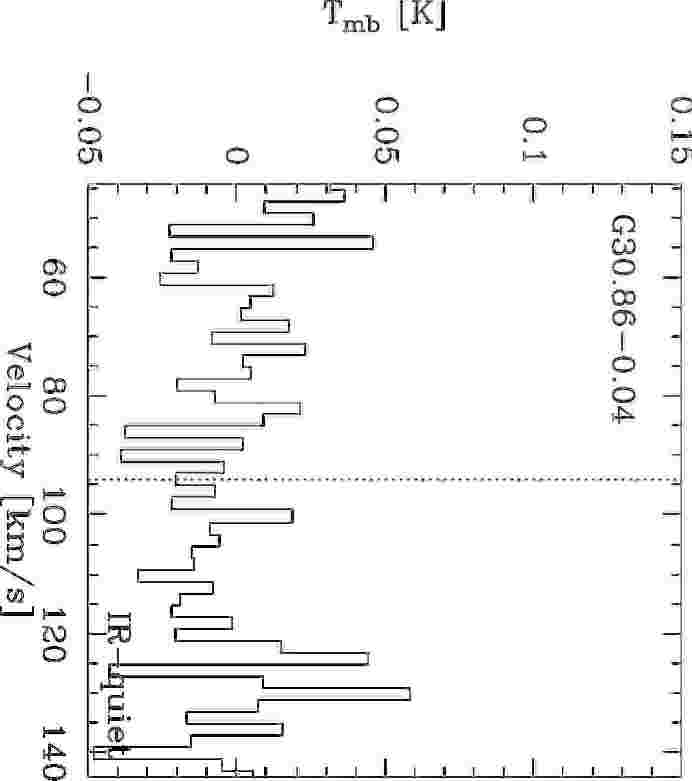} 
  \includegraphics[width=5.2cm,angle=90]{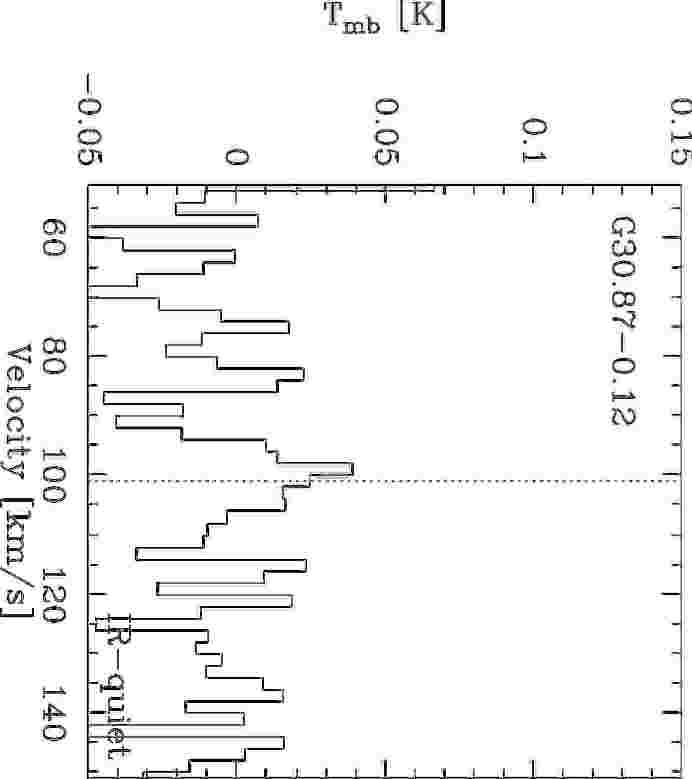} 
   \includegraphics[width=5.2cm,angle=90]{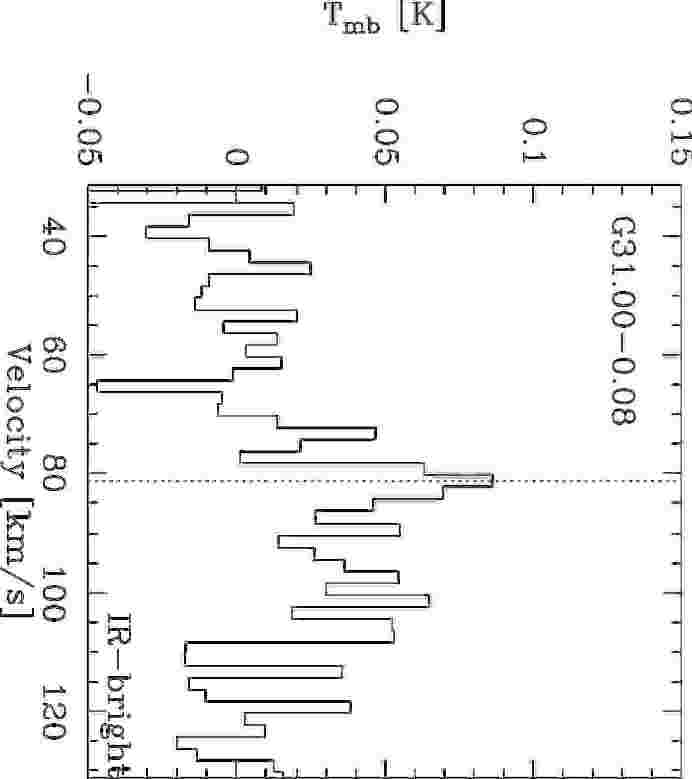} 
  \includegraphics[width=5.2cm,angle=90]{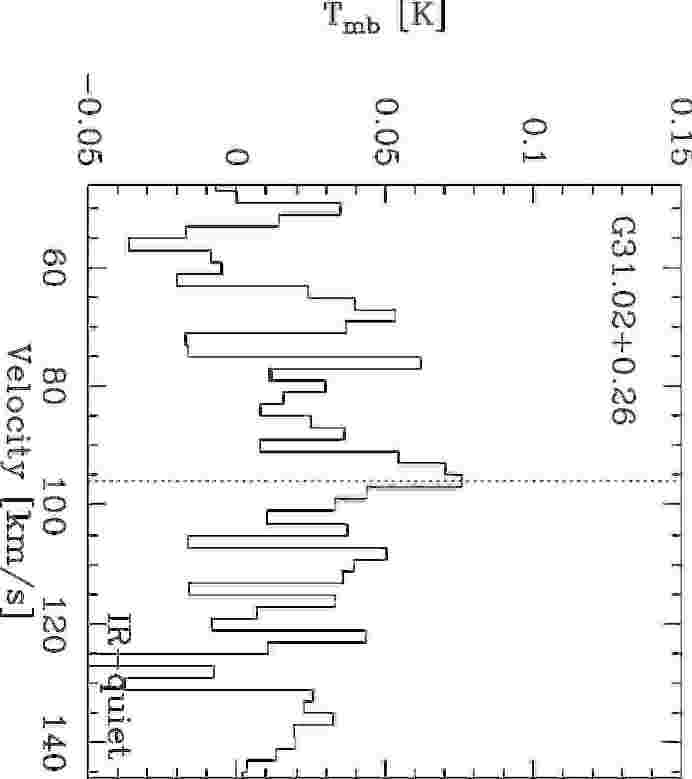} 
  \includegraphics[width=5.2cm,angle=90]{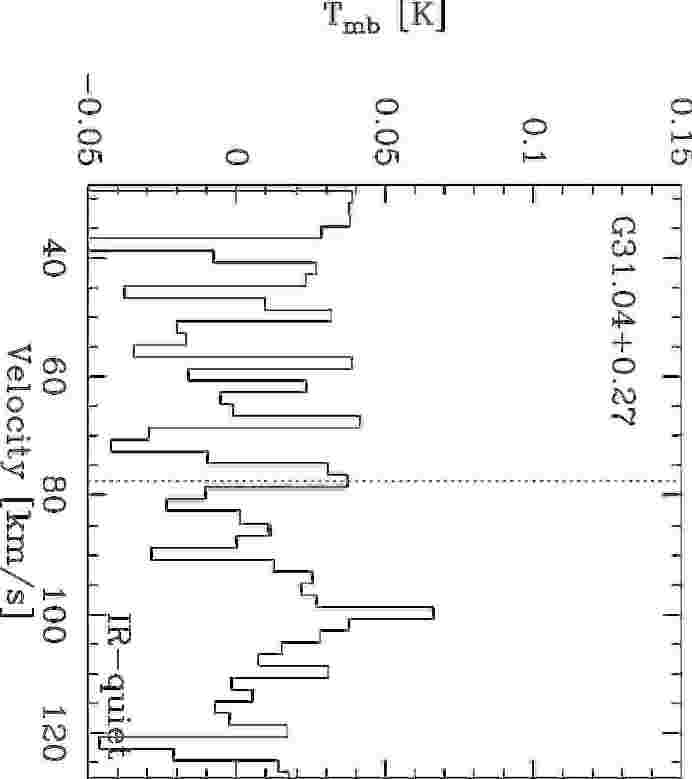} 
  \includegraphics[width=5.2cm,angle=90]{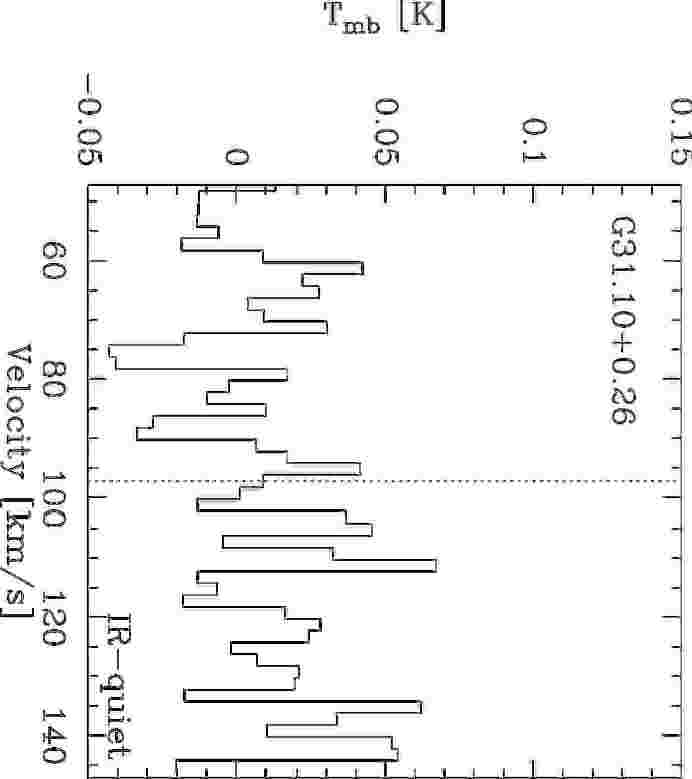} 
  \includegraphics[width=5.2cm,angle=90]{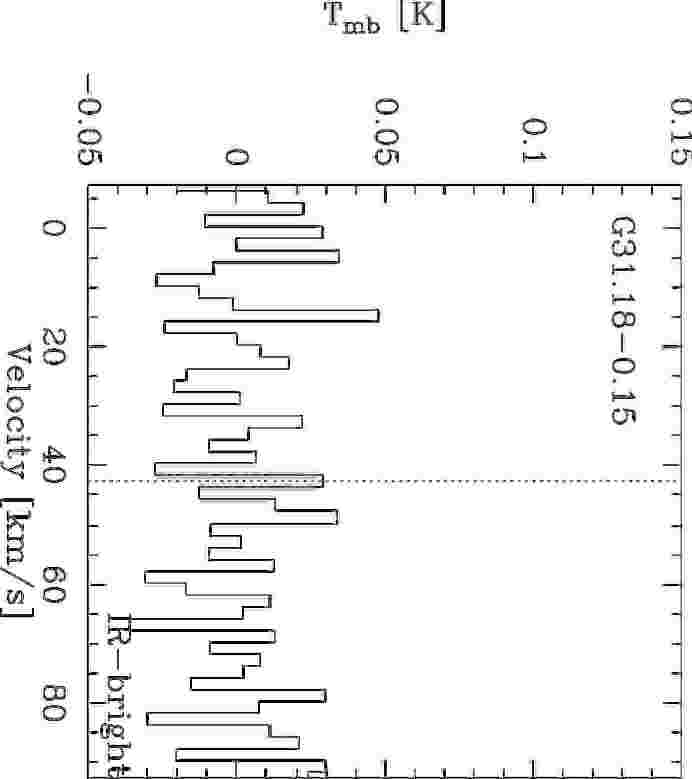} 
  \includegraphics[width=5.2cm,angle=90]{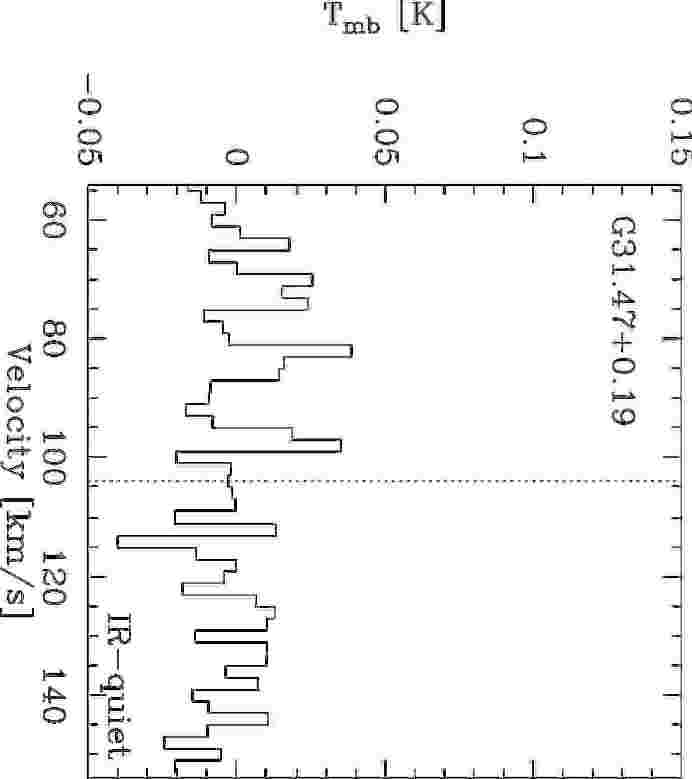} 
  \includegraphics[width=5.2cm,angle=90]{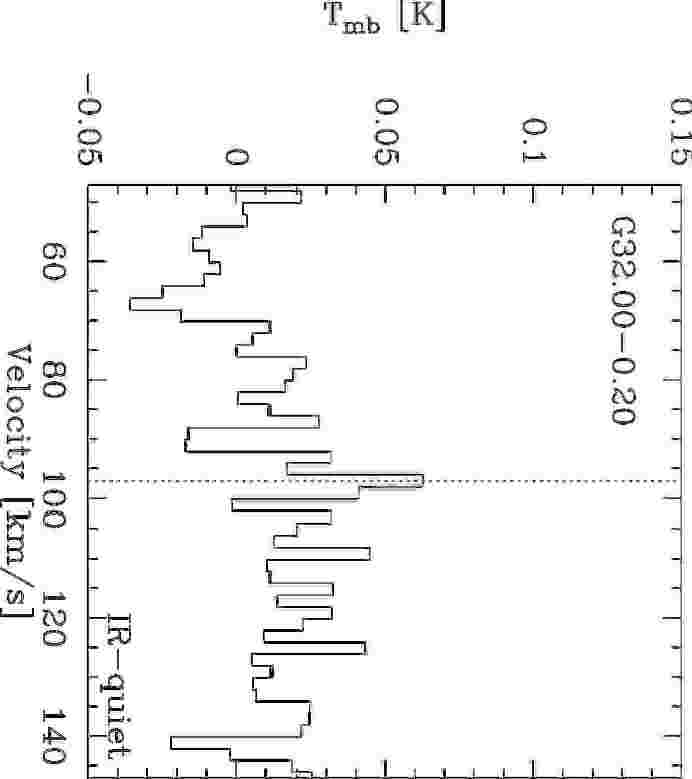} 
  \includegraphics[width=5.2cm,angle=90]{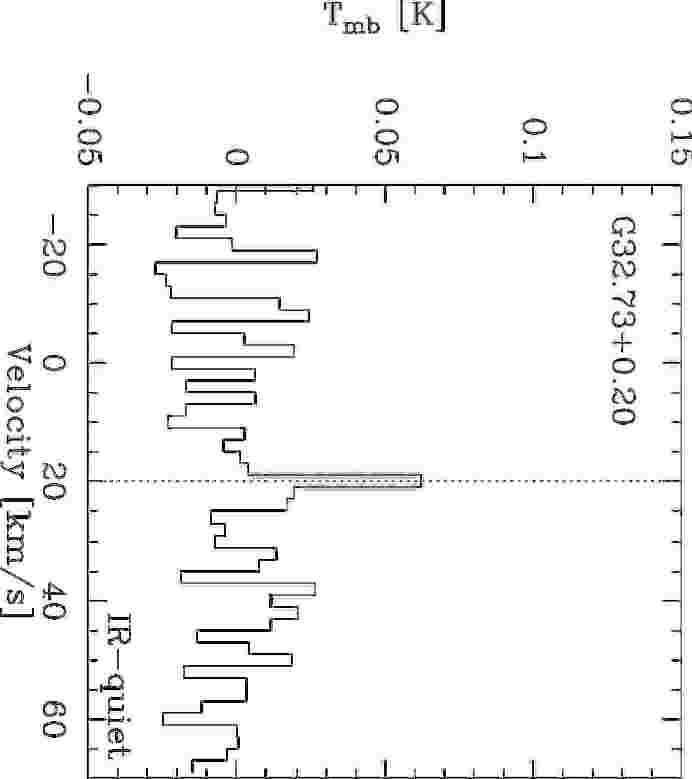} 
 \caption{Non-det.}
\end{figure}
\end{landscape}

\begin{landscape}
\begin{figure}
\ContinuedFloat
\centering
  \includegraphics[width=5.2cm,angle=90]{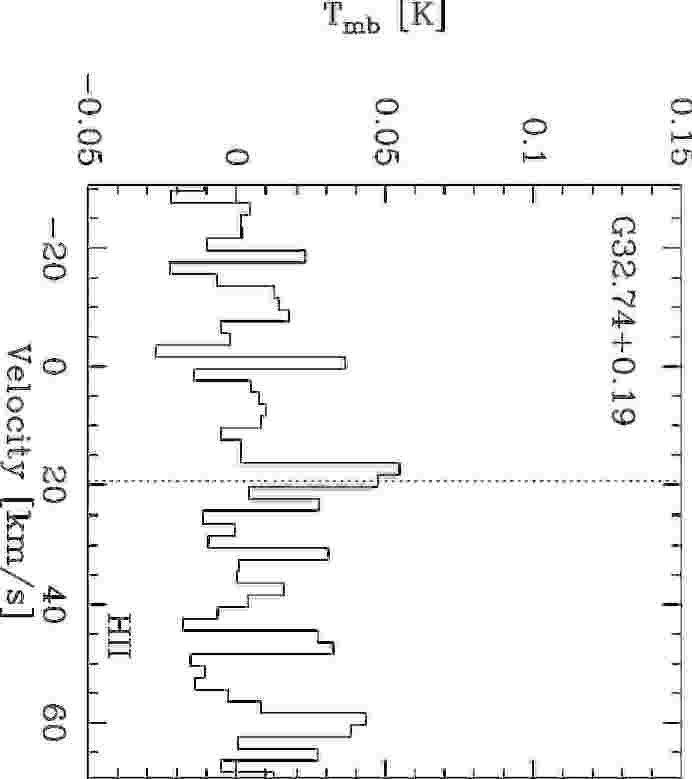} 
  \includegraphics[width=5.2cm,angle=90]{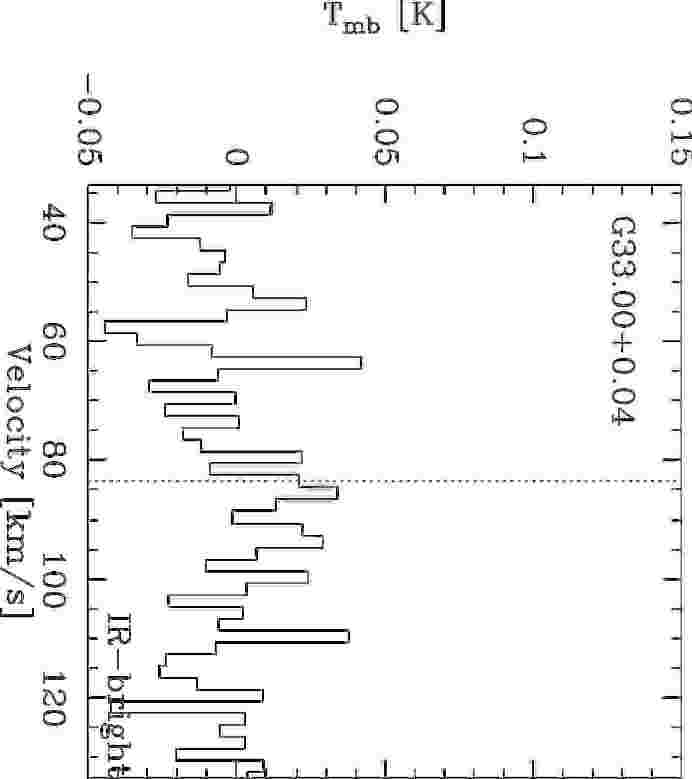} 
  \includegraphics[width=5.2cm,angle=90]{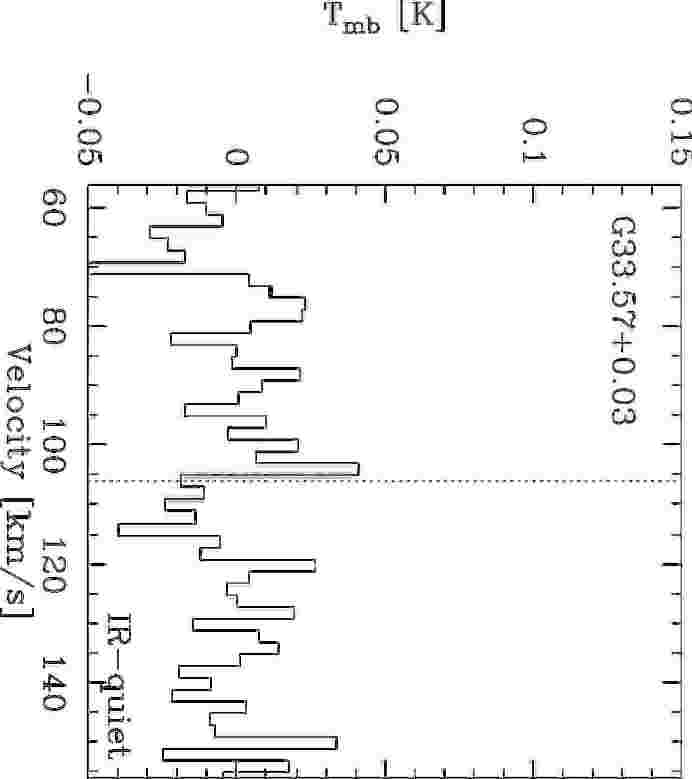} 
  \includegraphics[width=5.2cm,angle=90]{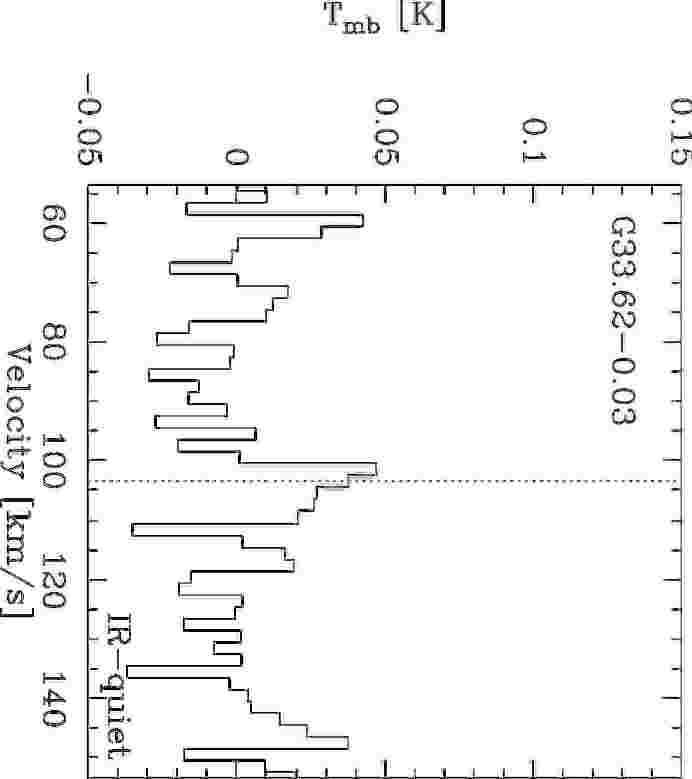} 
  \includegraphics[width=5.2cm,angle=90]{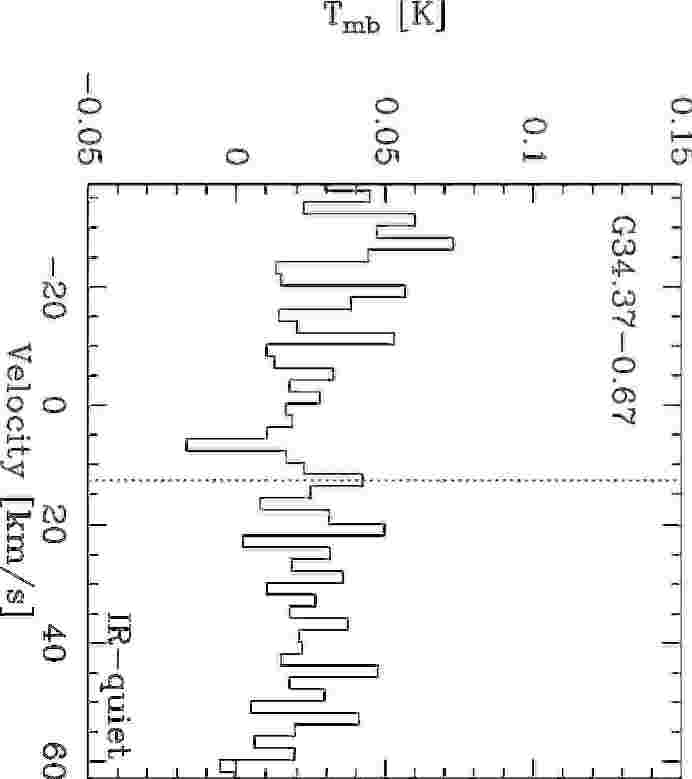} 
  \includegraphics[width=5.2cm,angle=90]{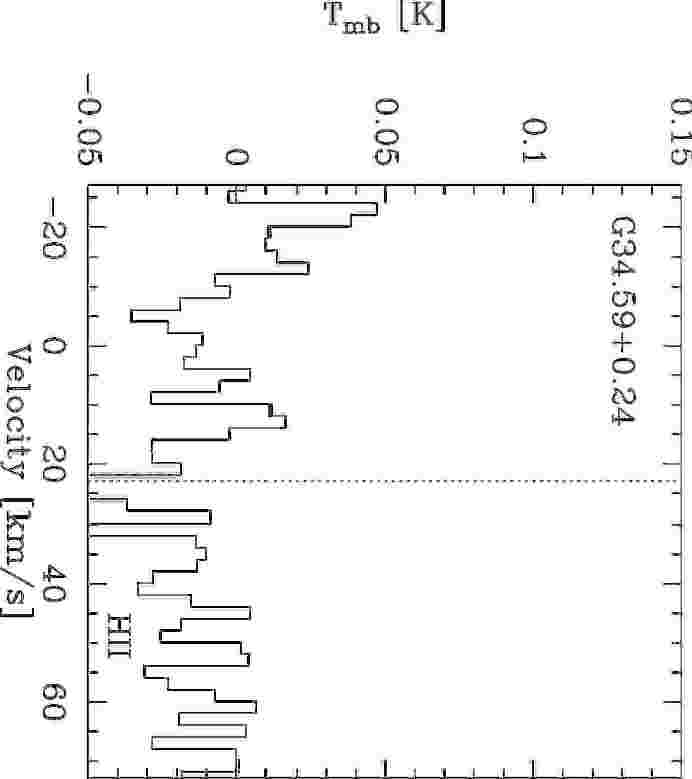} 
  \includegraphics[width=5.2cm,angle=90]{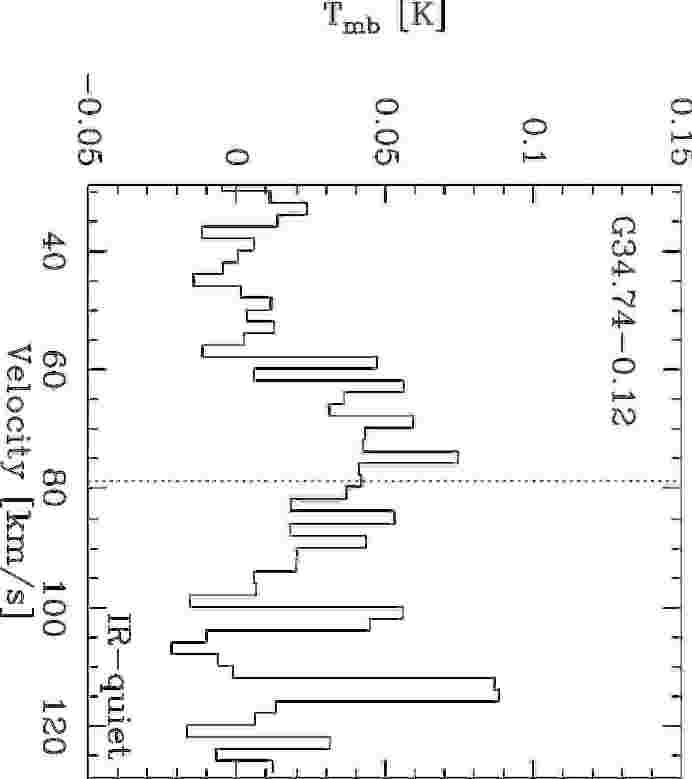} 
  \includegraphics[width=5.2cm,angle=90]{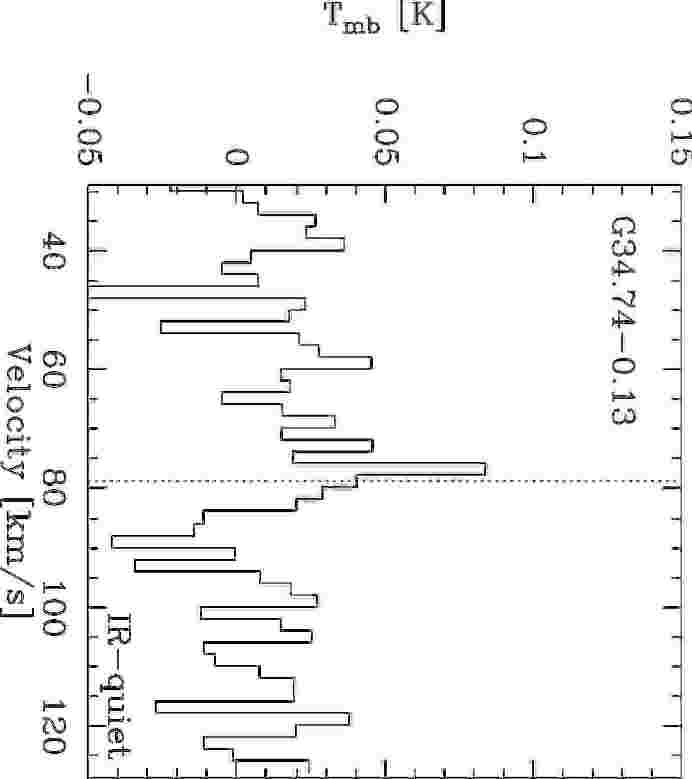} 
  \includegraphics[width=5.2cm,angle=90]{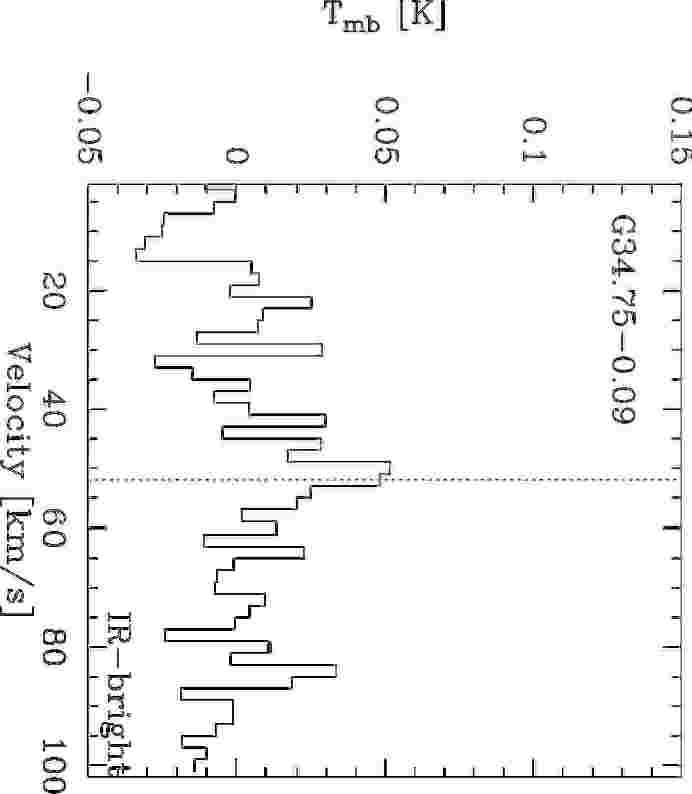} 
  \includegraphics[width=5.2cm,angle=90]{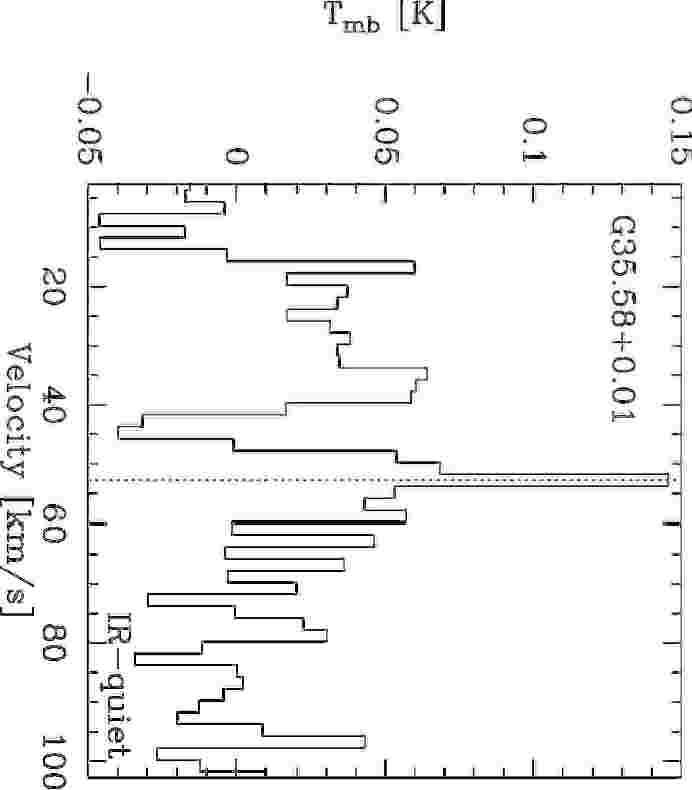} 
  \includegraphics[width=5.2cm,angle=90]{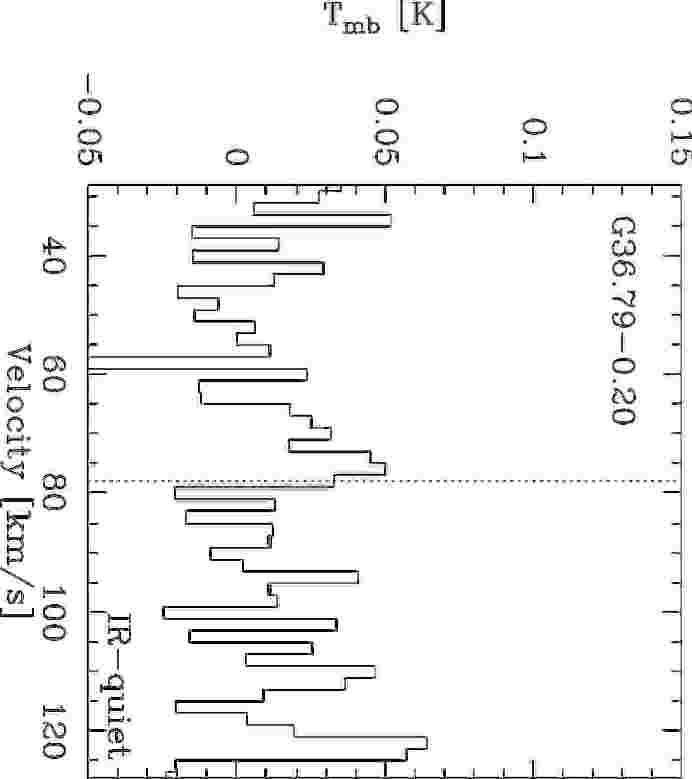} 
  \includegraphics[width=5.2cm,angle=90]{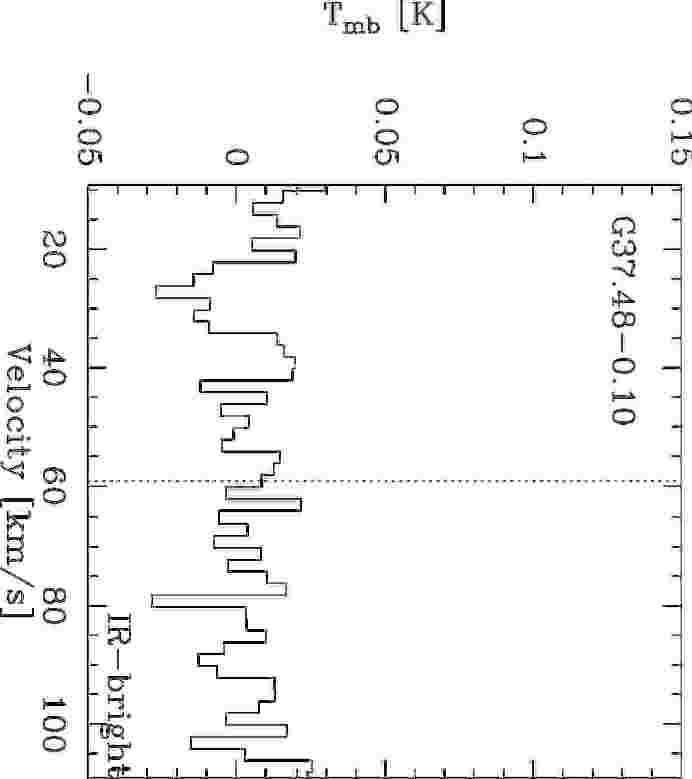} 
 \caption{Non-det.}
\end{figure}
\end{landscape}

\begin{landscape}
\begin{figure}
\ContinuedFloat
\centering
  \includegraphics[width=5.2cm,angle=90]{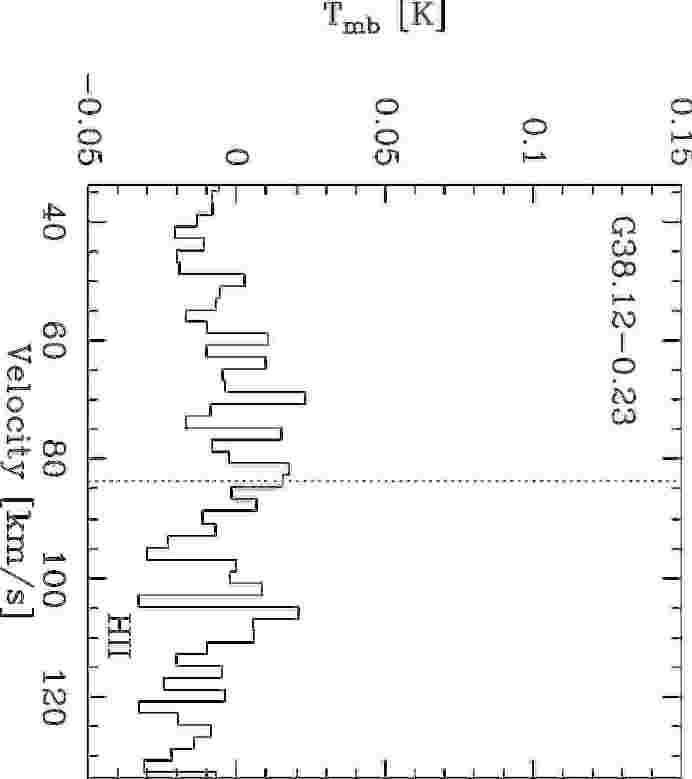} 
  \includegraphics[width=5.2cm,angle=90]{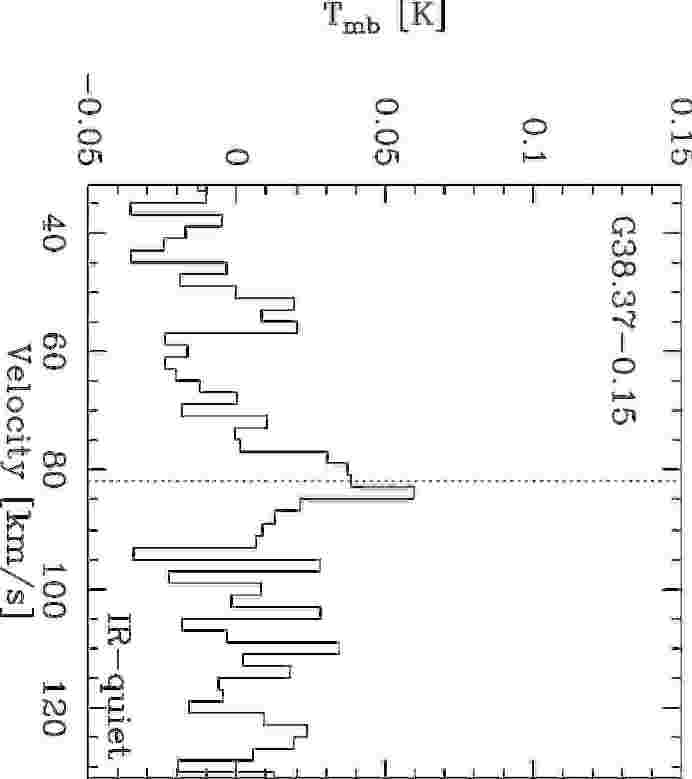} 
  \includegraphics[width=5.2cm,angle=90]{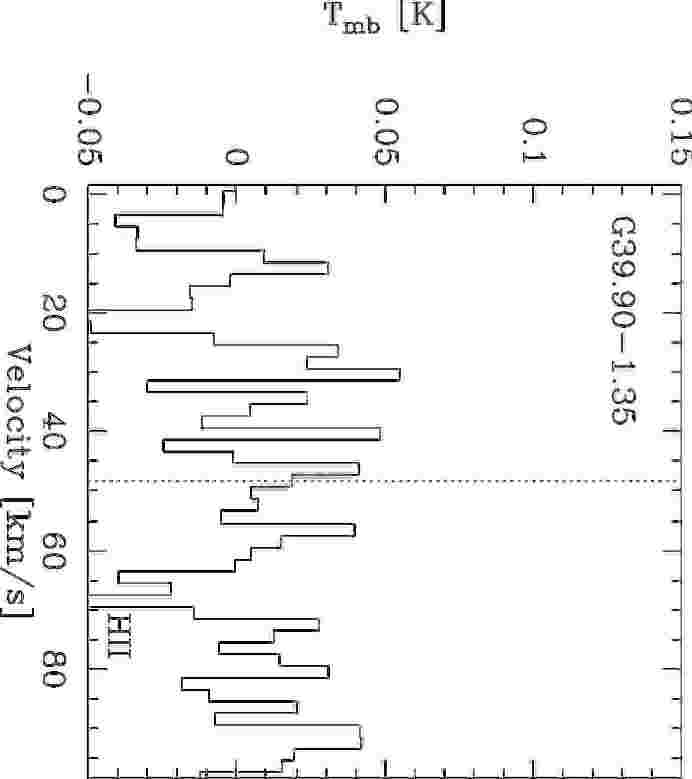} 
  \includegraphics[width=5.2cm,angle=90]{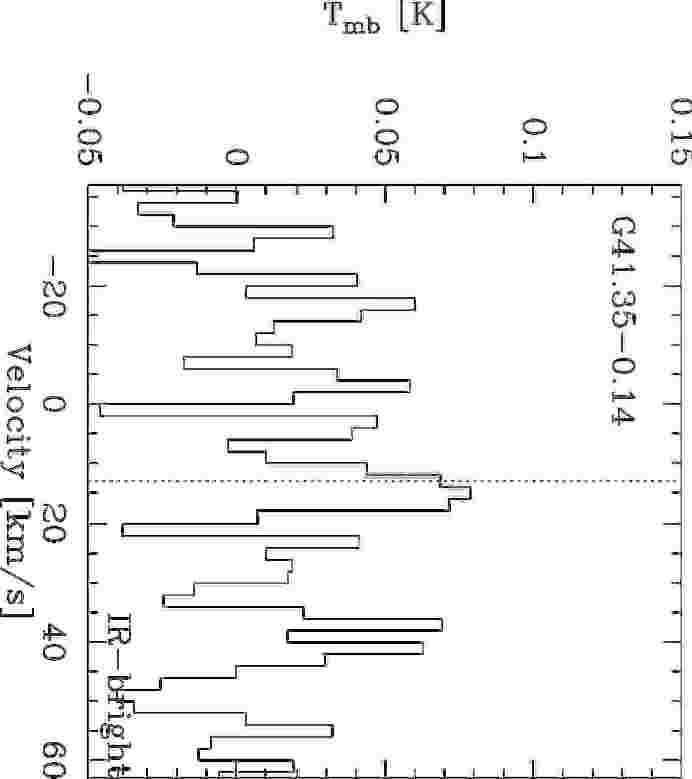} 
  \includegraphics[width=5.2cm,angle=90]{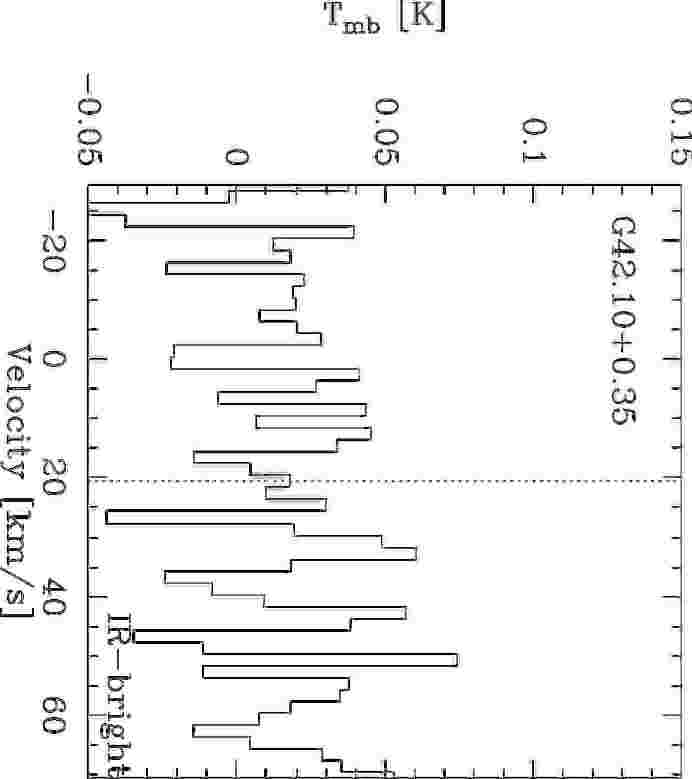} 
  \includegraphics[width=5.2cm,angle=90]{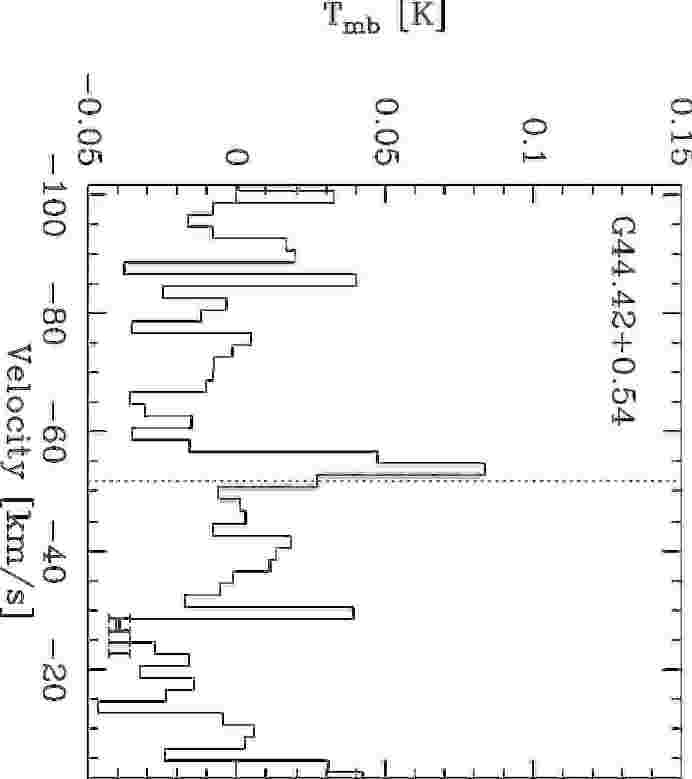} 
  \includegraphics[width=5.2cm,angle=90]{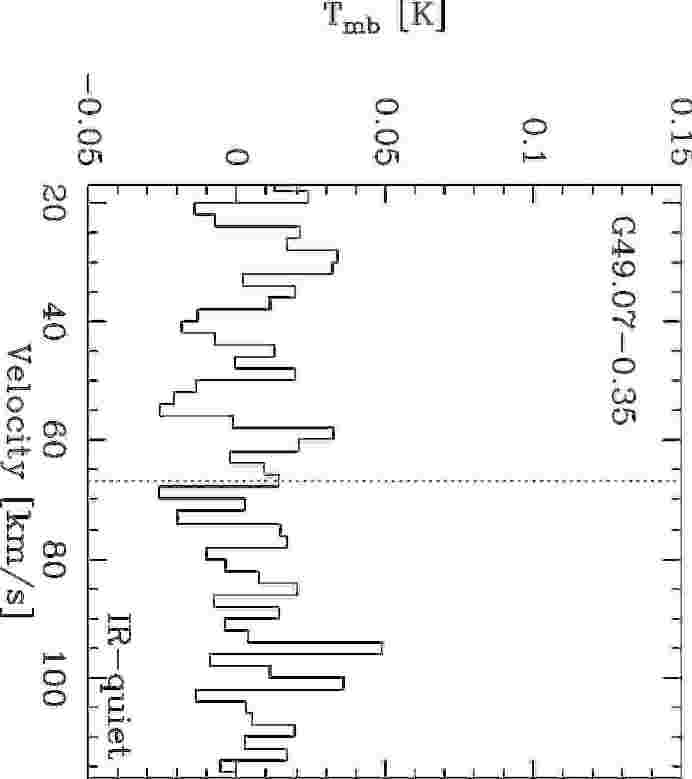} 
  \includegraphics[width=5.2cm,angle=90]{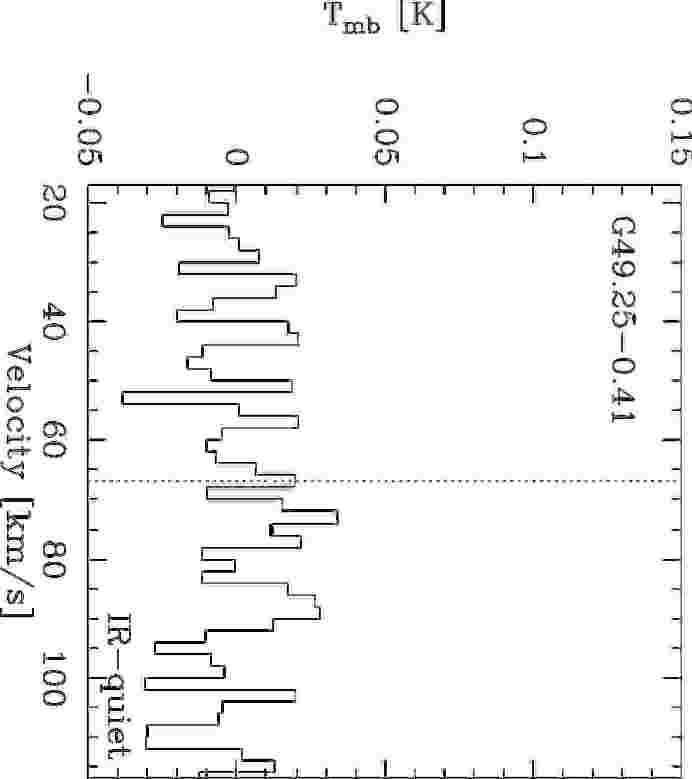} 
  \includegraphics[width=5.2cm,angle=90]{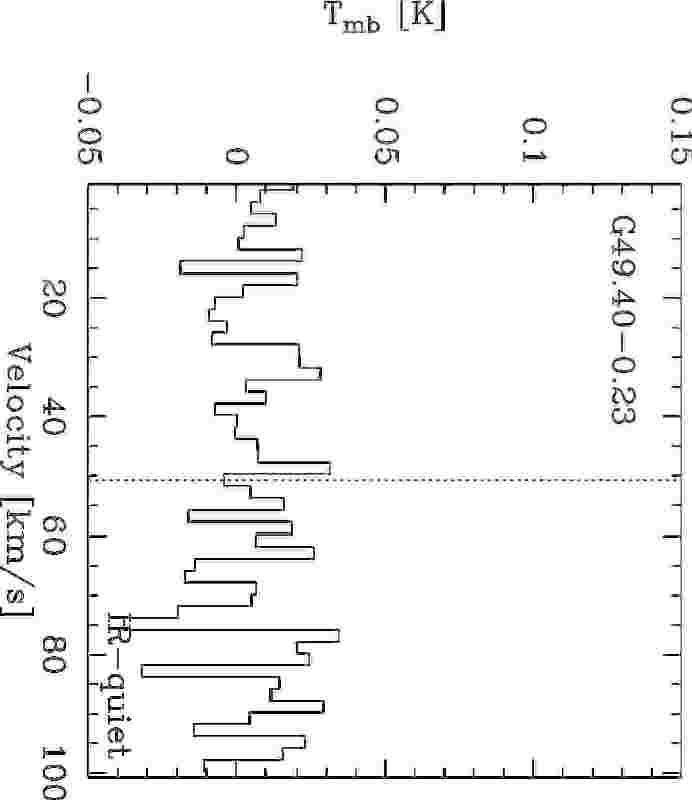} 
  \includegraphics[width=5.2cm,angle=90]{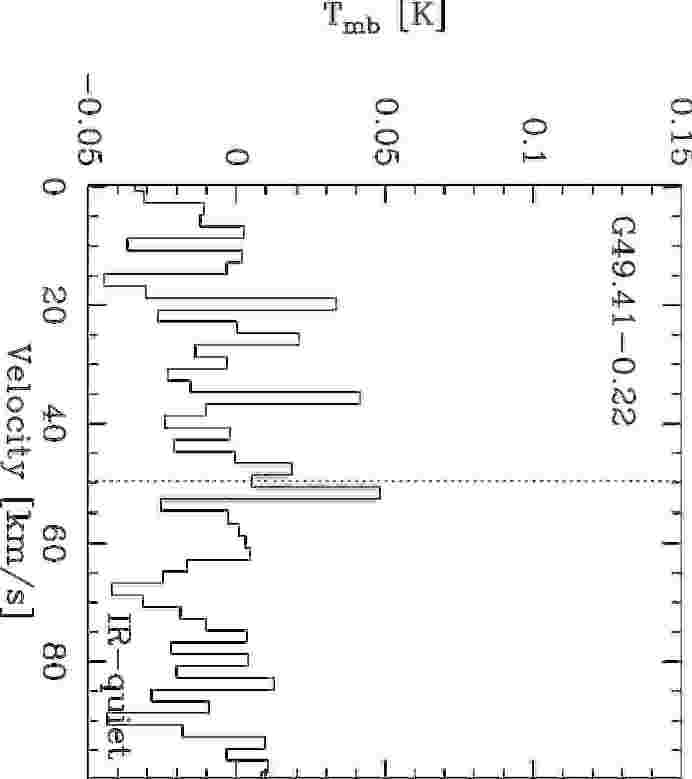} 
  \includegraphics[width=5.2cm,angle=90]{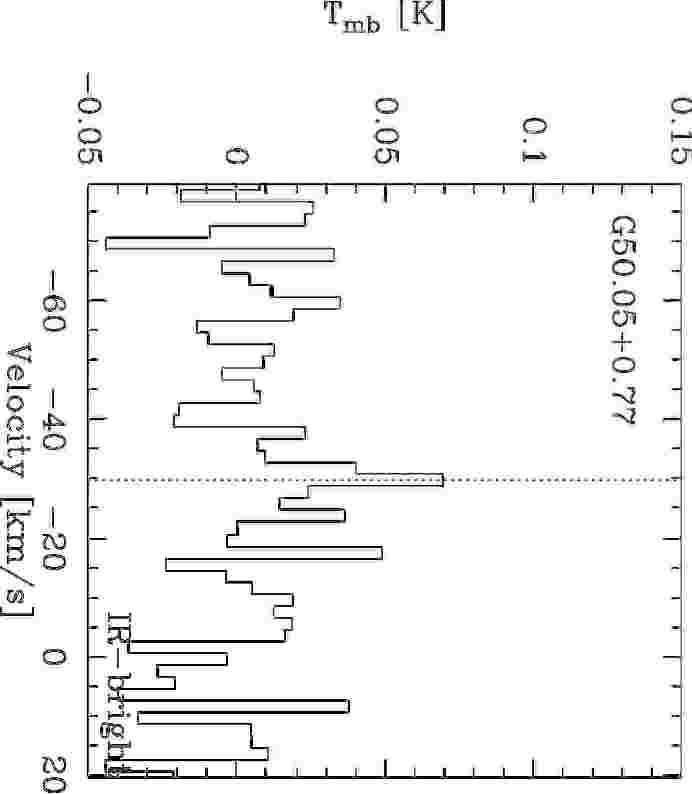} 
  \includegraphics[width=5.2cm,angle=90]{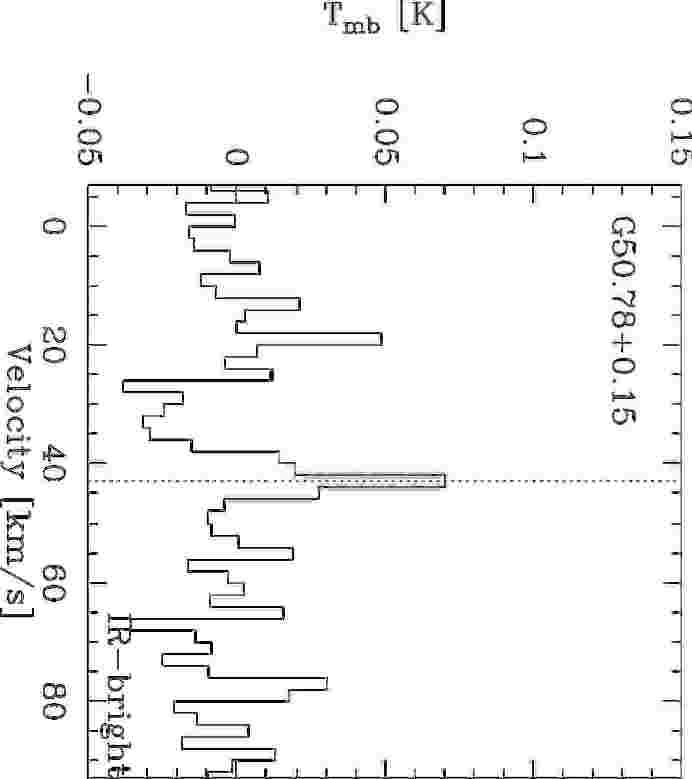} 
 \caption{Non-det.}
\end{figure}
\end{landscape}

\begin{landscape}
\begin{figure}
\ContinuedFloat
\centering
  \includegraphics[width=5.2cm,angle=90]{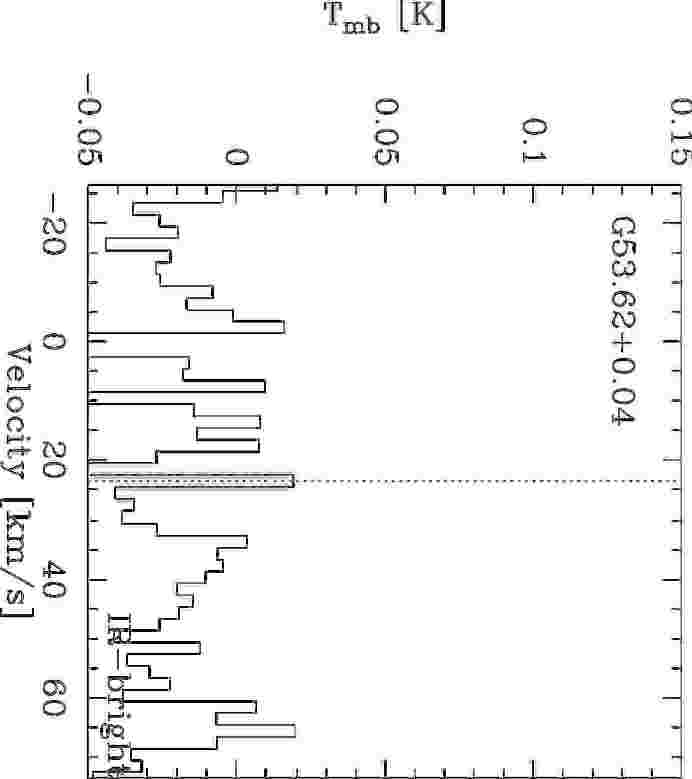} 
  \includegraphics[width=5.2cm,angle=90]{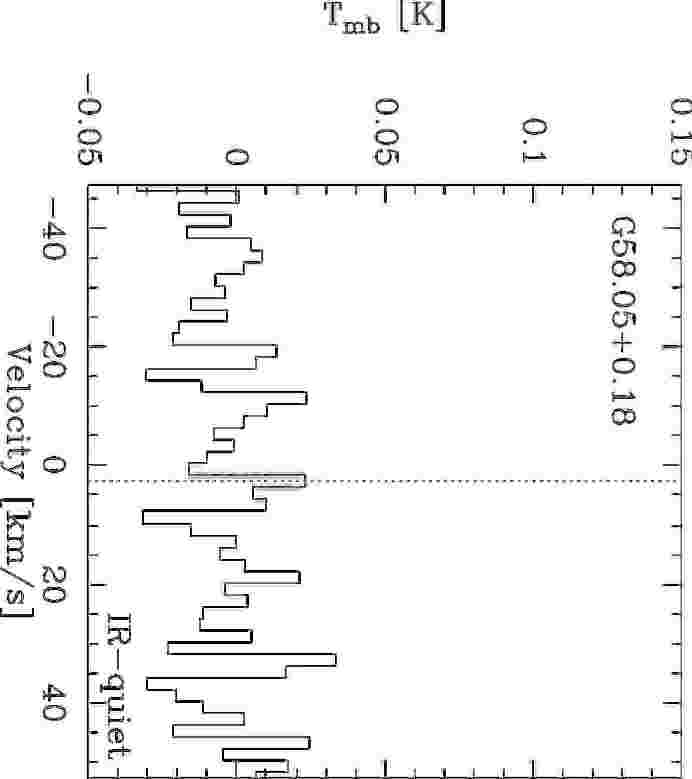} 
  \includegraphics[width=5.2cm,angle=90]{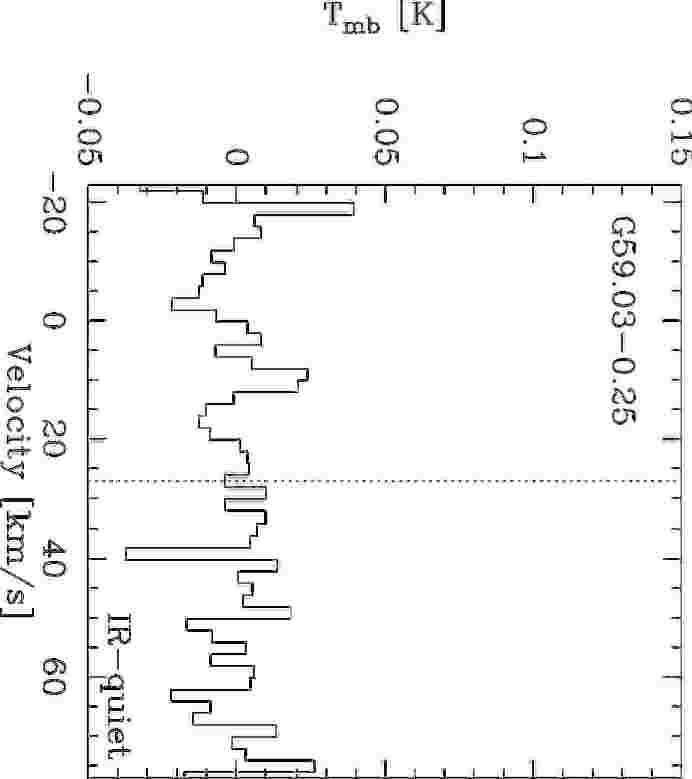} 
  \includegraphics[width=5.2cm,angle=90]{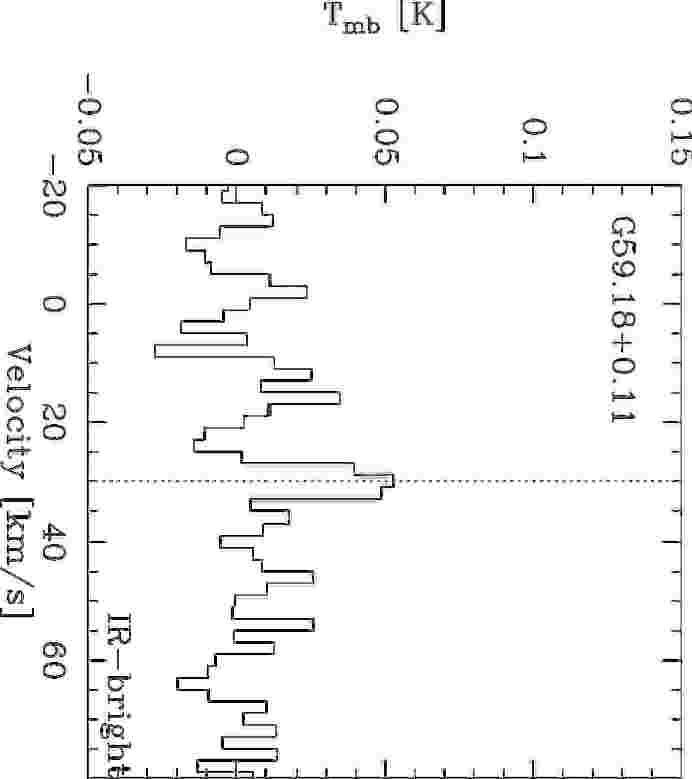} 
  \includegraphics[width=5.2cm,angle=90]{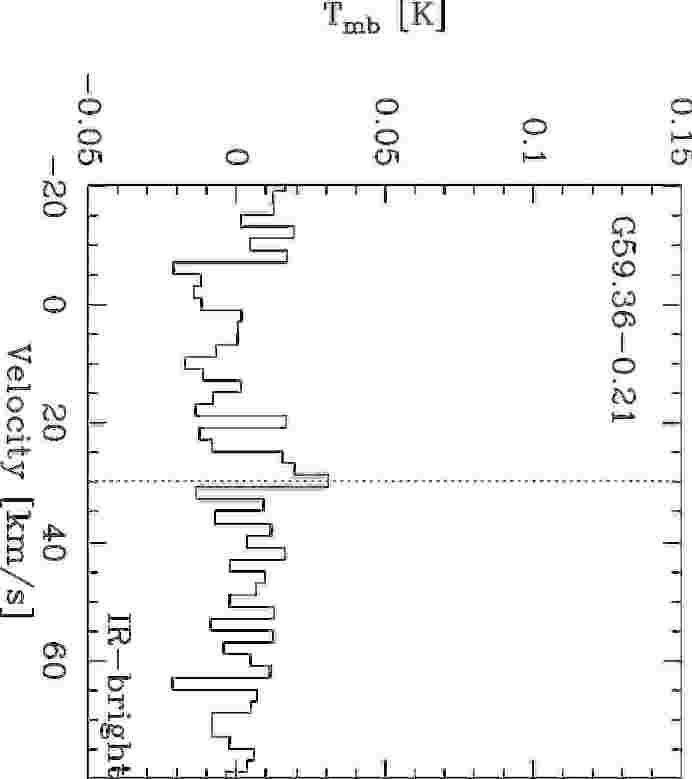} 
  \includegraphics[width=5.2cm,angle=90]{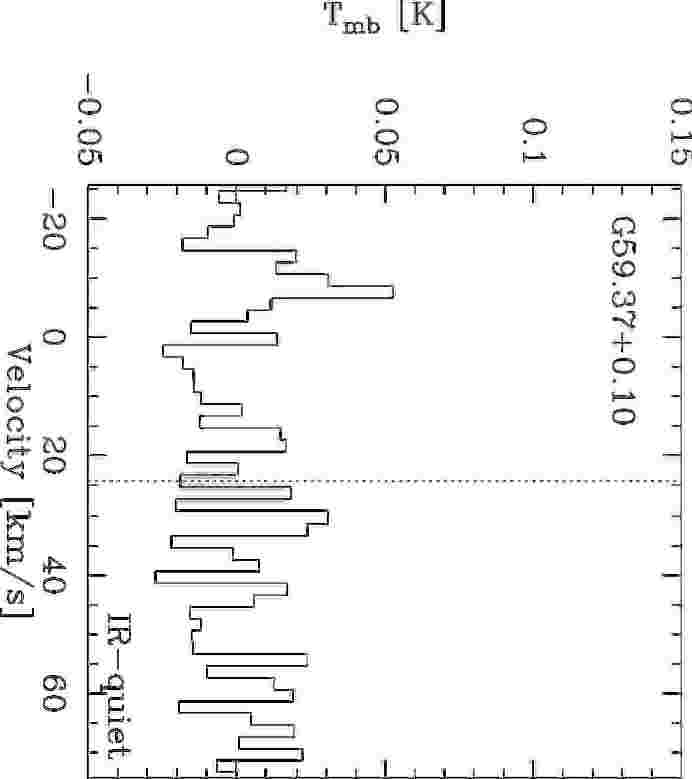} 
  \includegraphics[width=5.2cm,angle=90]{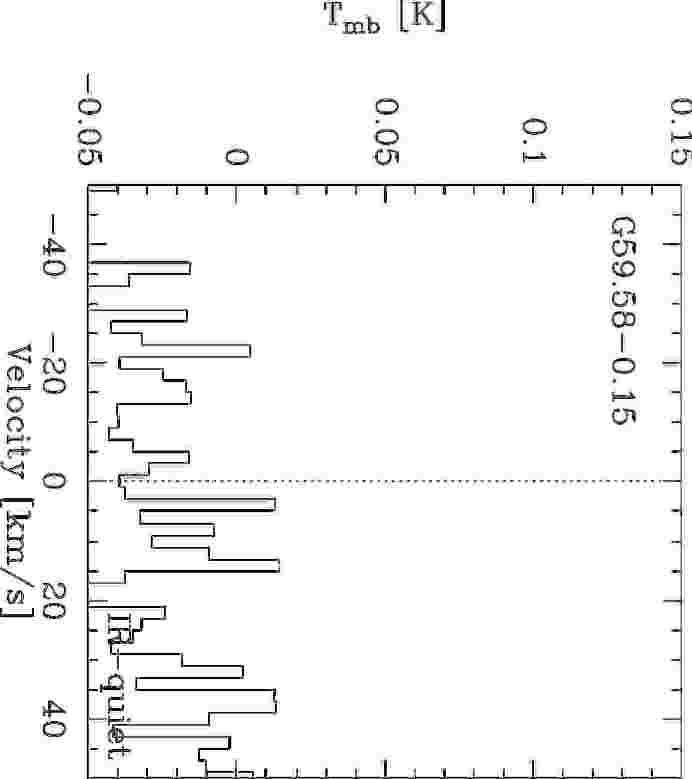} 
  \includegraphics[width=5.2cm,angle=90]{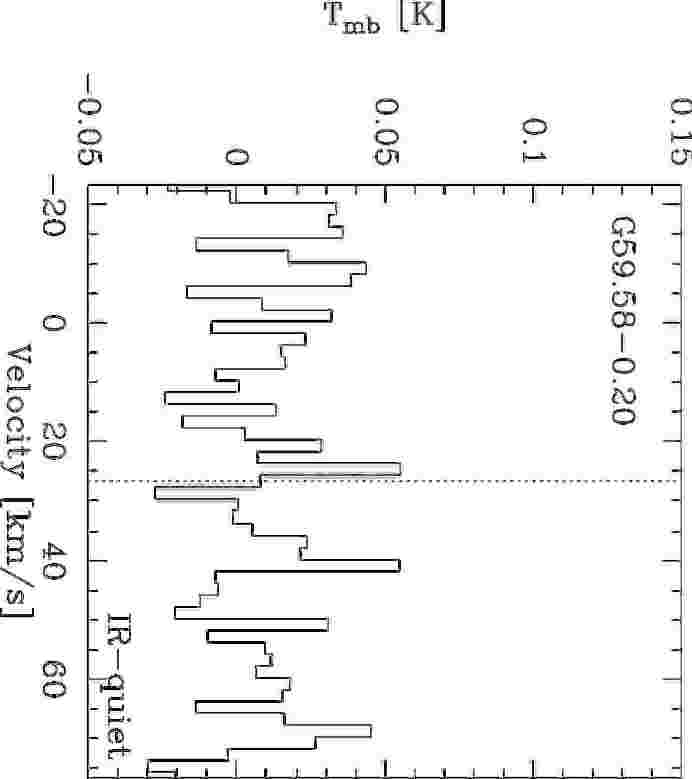} 
  \includegraphics[width=5.2cm,angle=90]{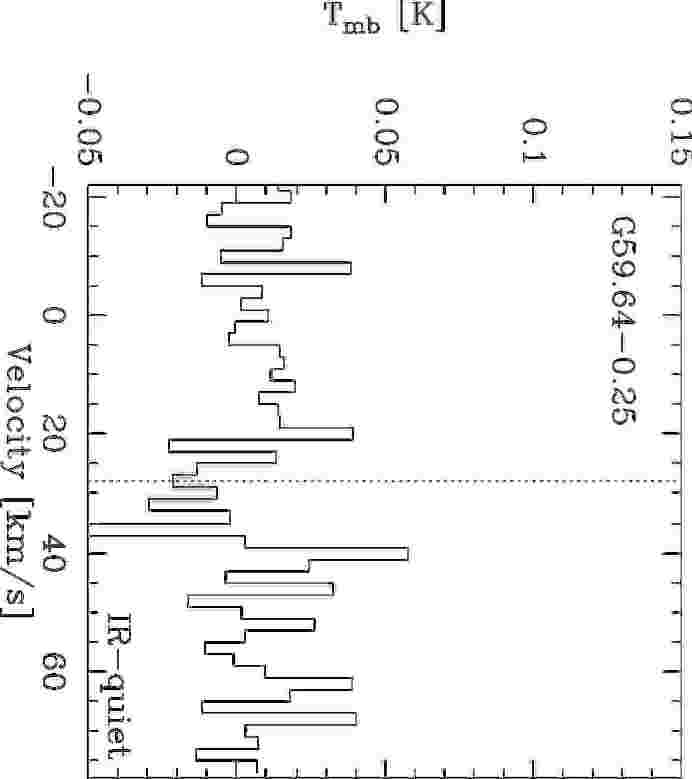} 
  \includegraphics[width=5.2cm,angle=90]{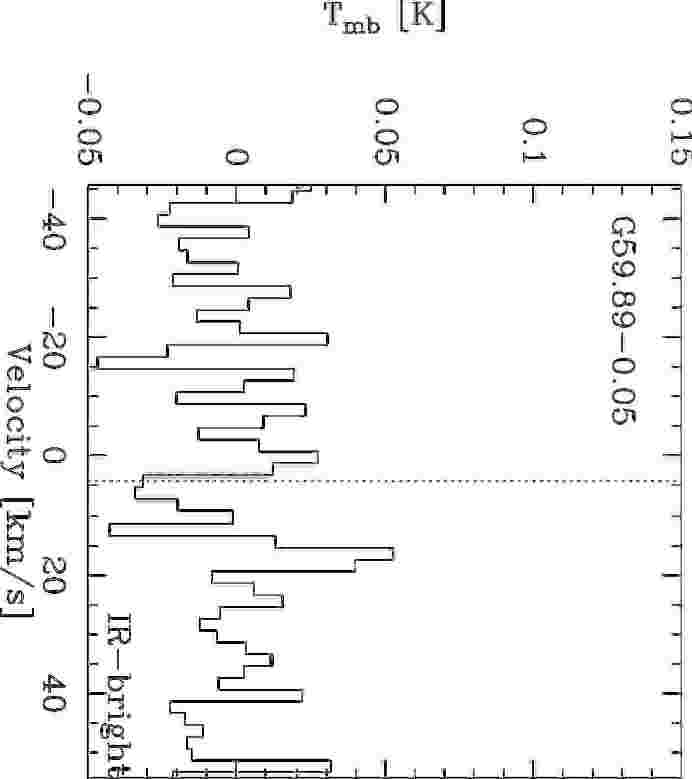} 
 \caption{Non-det.}
\end{figure}
\end{landscape}
}

%
\end{appendix}
\end{document}